\documentclass[10pt]{amsbook}

\usepackage[text={475pt,660pt},centering]{geometry}

\usepackage{color}
\usepackage{esint,amssymb}
\usepackage{graphicx}
\usepackage{MnSymbol}
\usepackage{amsmath, mathtools}
\usepackage[colorlinks=true, pdfstartview=FitV, linkcolor=blue, citecolor=blue, urlcolor=blue,pagebackref=false]{hyperref}
\usepackage{microtype}

\usepackage{bm}
\usepackage{scalerel} 
\usepackage{dsfont}
\usepackage{mathrsfs}
\usepackage[font={footnotesize}]{caption}
\usepackage{stmaryrd}

\definecolor{darkgreen}{rgb}{0,0.5,0}
\definecolor{darkblue}{rgb}{0,0,0.7}
\definecolor{darkred}{rgb}{0.9,0.1,0.1}

\makeatletter
\newtheorem*{rep@theorem}{\rep@title}
\newcommand{\newreptheorem}[2]{%
\newenvironment{rep#1}[1]{%
 \def\rep@title{#2 \ref{##1}}%
 \begin{rep@theorem}}%
 {\end{rep@theorem}}}
\makeatother

\newtheorem{theorem}{Theorem}
\newreptheorem{theorem}{Theorem}
\newreptheorem{proposition}{Proposition}
\newreptheorem{lemma}{Lemma}
\newtheorem{proposition}{Proposition}
\newtheorem{lemma}[proposition]{Lemma}
\newtheorem{corollary}[proposition]{Corollary}

\theoremstyle{remark}

\theoremstyle{definition}
\newtheorem{definition}[proposition]{Definition}
\newtheorem{remark}[proposition]{Remark}

\numberwithin{equation}{section}
\numberwithin{proposition}{section}

\newcommand{\Z}{\mathbb{Z}}
\newcommand{\N}{\mathbb{N}}
\newcommand{\R}{\mathbb{R}}

\newcommand{\E}{\mathbb{E}}
\renewcommand{\P}{\mathbb{P}}
\newcommand{\T}{\mathbb{T}}

\newcommand{\Zd}{\mathbb{Z}^d}
\newcommand{\Rd}{{\mathbb{R}^d}}

\newcommand{\di}{\mathrm{d}}
\newcommand{\inte}[1]{%
  {\kern0pt#1}^{\mathrm{o}}%
}

\newcommand{\ep}{\varepsilon}

\renewcommand{\a}{\mathbf{a}}
\newcommand{\ahom}{{\overbracket[1pt][-1pt]{\a}}}  

\newcommand{\per}{\mathrm{per}}

\renewcommand{\subset}{\subseteq}

 \newcommand{\cu}{\square}

\DeclareMathOperator{\dist}{dist}

\DeclareMathOperator*{\esssup}{ess\,sup}

\DeclareMathOperator*{\osc}{osc}
\DeclareMathOperator{\var}{var}
\DeclareMathOperator{\cov}{cov}
\DeclareMathOperator{\diam}{diam}

\DeclareMathOperator{\supp}{supp}

\renewcommand{\bar}{\overline}
\renewcommand{\tilde}{\widetilde}

\newcommand{\indc}{\mathds{1}}

\newcommand{\C}{C}
\newcommand{\Pa}{\mathcal{P}}

\newcommand{\Qa}{\mathcal{Q}}

\newcommand{\G}{\mathcal{G}}

\newcommand{\X}{\mathcal{X}}

\newcommand{\f}{\mathbf{f}}
\renewcommand{\L}{\mathcal{L}}
\newcommand{\Lspat}{\mathcal{L}_{\mathrm{spat}}}

\begin{document}

\sloppy

\title{Massless Phases for the Villain model in $d\geq 3$}
\author{Paul Dario, Wei Wu}
\address[P. Dario]{School of Mathematical Sciences, Tel Aviv University
Ramat Aviv, Tel Aviv 69978, Israel}
\email{paul.dario@ens.fr}

\address[W. Wu]{Statistics department, University of Warwick, Coventry CV4 7AL, UK}
\email{w.wu.9@warwick.ac.uk}

\keywords{}
\subjclass[2010]{}
\date{\today}

\maketitle
\begin{abstract}
We consider the classical Villain rotator model in $\Zd, d\geq 3$ at sufficiently low temperature, and prove that the truncated two-point function decays asymptotically as $|x|^{2-d}$, with an algebraic rate of convergence. We also obtain the same asymptotic decay separately for the transversal two-point functions. This quantifies the spontaneous magnetization result for the Villain model at low temperature, and rigorously establishes the Gaussian spin-wave conjecture in dimension $d\ge 3$. We believe that our method extends to finite range interactions and to other abelian spin systems and abelian gauge theory in $d\geq 3$.

\end{abstract}

\setcounter{tocdepth}{1}
\tableofcontents

\chapter{Introduction}

\section{General overview and statement of the main result} In this paper, we obtain precise low temperature properties of certain classical rotator models in the Euclidean lattice $\Zd$, $d\geq3$. A canonical model of rotator is the XY model, which assigns to each $x\in\Zd$ a spin taking value in $S^1$ with the corresponding angle $\theta(x) \in (-\pi,\pi]$. The XY model is defined formally as a Gibbs measure with Hamiltonian given by
\begin{equation*}
H_{XY} := -\sum_{x\sim y} \cos(\theta(x) - \theta(y)).
\end{equation*}

The classical Villain model is another canonical two-component spin model which is closely related to the XY model \cite{Vil}. Given a finite cube $\cu \subset \Zd$, we denote by $\cu^\circ$ its interior, $\partial\cu$ its boundary and  $E(\cu)$ its edge set. The Villain model on the cube $\cu$ with zero boundary condition is given by the following Gibbs measure
\begin{equation} \label{e.finitedirichlet}
    d\mu^V_{\beta,\cu,0}(\theta):= Z_{\cu,0}^{-1} \prod_{(x,y) \subset E(\cu)} v_\beta(\theta(x) - \theta(y)) \prod_{x\in \partial \cu}\delta_0(\theta(x)) \prod_{x\in \cu^\circ} \indc_{[-\pi,\pi)} (\theta(x)) \,d\theta
\end{equation}
where 
\begin{equation*}
v_\beta(\theta) = \sum_{m\in \Z} \exp{\left(-\frac{\beta}{2} (\theta+ 2\pi m)^2\right)}
\end{equation*}
is the heat kernel on $S^1$ and $Z_{\cu,0}$ is the normalization constant that makes \eqref{e.finitedirichlet} a probability measure. 
The expectation with respect to the Gibbs measure \eqref{e.finitedirichlet} is denoted by $\left\langle \cdot \right\rangle_{\mu^V_{\beta,\cu,0}}$. We also define the spin variable, which takes value on the unit circle, by $S_x= (\cos{\theta(x)},\sin{\theta(x)})$. By the $\theta \to - \theta$ symmetry, we have
\begin{equation*}
\left\langle S_0 \cdot S_x \right\rangle_{\mu^V_{\beta,\cu,0}} = \left\langle e^{i (\theta(0) - \theta(x))}\right\rangle_{\mu^V_{\beta,\cu,0}}.
\end{equation*}

It is known that, as a consequence of correlation inequalities \cite{Gi, BFL, MMP}, the thermodynamic limit of the measures \eqref{e.finitedirichlet} exists as $\cu \to \Z^d$. We denote by $\mu^V_\beta$ the corresponding infinite volume Gibbs measure. It is clear that the quantity 
\begin{equation*}
    \left\langle S_0 \cdot S_x \right\rangle_{\mu^V_\beta} = \left\langle e^{i (\theta(0) - \theta(x))}\right\rangle_{\mu^V_\beta}
\end{equation*}
is invariant under the rotations $\theta \to \theta + \alpha  \mod{2\pi}$. 

There has been long standing interests in the low temperature behavior of the XY and the Villain model. A simple heuristics suggests that, as the temperature goes to zero, the spins tend to align with each other so as to minimize the Hamiltonian. Since $\exp(\beta \cos(\delta \theta)) \approx \sum_{m\in \Z} \exp{\left(-\frac{\beta}{2} (\delta\theta+ 2\pi m)^2\right)}$ and $\cos 2\pi\left(
\delta \theta \right) \approx 1-\left( \delta \theta \right) ^{2}/2$, it is expected that at low temperature, both the XY and the Villain Gibbs
measures on large scales behave like a massless free field. This idea
originated in \cite{Dy} (see also \cite{MW}) and was referred to as the
Gaussian spin-wave approximation. By further making connections between
the rotational symmetry of these models and the recurrence/transience property
of simple random walks, it was proved in~\cite{MW} that for $d\leq 2$, there
is no spontaneous magnetization at any strictly positive temperature. A
related argument was applied in~\cite{FSS} to show that for $d\geq 3$, with
appropriate boundary conditions, the $SO(2)$ symmetry is broken at low temperature, i.e., there is spontaneous magnetization $\left\langle \cos \theta_0 \right\rangle_{\mu^V_{\beta}} >0$,
but there is no spontaneous magnetization at high temperature (and thus there is
a phase transition). 

Quantitative improvements of the spontaneous magnetization result were established in the 1980s. If we believe that the low temperature behavior of the rotator models is like the one of a Gaussian free field, then a Gaussian computation yields that for $d\geq 3$,
\begin{equation}
\label{e.free}
 \left\langle e^{i (\theta(0) - \theta(x))}\right\rangle_{\mu^V_\beta} 
 \approx \exp \left( \delta_0 - \delta_x, -\frac1{2\beta} \Delta^{-1}  (\delta_0 - \delta_x) \right)
 = C_1 (\beta) + \frac{C_2 (\beta) }{|x|^{d-2}}+ o \left(\frac{ 1}{|x|^{d-2}}\right).
\end{equation} 
For $d=2$, however, a Gaussian computation yields that for large $\beta$,
\begin{equation}
\label{e.free2d}
 \left\langle e^{i (\theta(0) - \theta(x))}\right\rangle_{\mu^V_\beta} 
 \approx 
 C |x|^{-\frac 1 {2\pi\beta}}+ o \left(|x|^{-\frac 1 {2\pi\beta}}\right).
\end{equation} 
The algebraic decay of correlation indicates the  so-called Kosterlitz-Thouless
transition in two dimensions~\cite{KT}. In $d=2$, the algebraic decay of correlations is expected to be valid for all temperatures
below the critical temperature corresponding to the K-T transition, and above which one expects exponential decay of correlations. The Gaussian spin-wave approximation \eqref{e.free2d} for $d=2$ remains largely open. For $\beta$ sufficiently large,  rigorous (but non-optimal) power law upper and lower bounds for the two-point function were established in the celebrated works \cite{MS} and \cite{FrSp}.

For  $d\ge 3$, Fr{\"o}hlich and Spencer \cite{FS4d} observed that the classical Villain model in $\Zd$ can be mapped, via duality, to a statistical mechanical model of lattice Coulomb gas, with local neutrality constraints. They further  employed a one step renormalization argument and gave the following next order description of the two-point function at low temperature. 

\begin{proposition}[\cite{FS4d}]
\label{p.FS}
Let $\mu^V_{\beta}$ be the thermodynamic limit of the Villain model in $\Z^d$, for $d\geq3$.  There exist constants $\beta_0=\beta_0(d)$, $c_0= c_0(\beta,d)$, such that for all $\beta>\beta_0$,
\begin{equation*}
    \left\langle S_0 \cdot S_x \right\rangle_{\mu^V_{\beta}} =
    c_0 + O\left(\frac{1}{|x|^{d-2}}\right) 
\end{equation*} 
Moreover, as $\beta \to \infty$,
\begin{equation*}
   \exp \left( \delta_0 - \delta_x, -\frac1{2\beta} \Delta^{-1}  (\delta_0 - \delta_x) \right) \geq
    \left\langle S_0 \cdot S_x \right\rangle_{\mu^V_{\beta}} 
\geq 
\exp \left( \delta_0 - \delta_x, \left(-\frac1{2\beta}+o\left(\frac1\beta\right)\right) \Delta^{-1}  (\delta_0 - \delta_x) \right).
\end{equation*}
\end{proposition}

This suggests that the truncated two-point function may be related to a massless free field in $\Rd$, which corresponds to the emergence of a (conjectured) Goldstone Boson. Similar results were also obtained for the Abelian gauge theory in $d=4$ (see~\cite{FS4d, Gu}). Kennedy and King in~\cite{KK} obtained a  similar low temperature expansion for the Abelian Higgs model, which couples an XY model with a gauge fixing potential. Their proofs rely on a different approach, via a transformation introduced by \cite{BBIJ} and a polymer expansion.

It is also of much interest to justify the spin-wave conjecture separately for the longitudinal and transversal two-point functions of the rotator models. The best known result is due to Bricmont et. al. \cite{BFLLS},  where they perform a low temperature expansion of the truncated two-point function of the XY model and obtain that there exist $c_1>c_2>0$, such that for sufficiently large $\beta$,

\begin{align*}
\frac{c_2}{\beta |x|^{d-2}} \leq \langle \sin\theta(0)  \sin\theta(x)  \rangle_{\mu^{XY}_\beta}
&\leq \frac{c_1}{\beta |x|^{d-2}}.
\end{align*}
The proof of~\cite{BFLLS} relies on a combination of the infrared bound, a Mermin-Wagner type argument, and correlation inequalities, and is therefore restricted to the cubic lattice with nearest neighbor interactions.

The main result of our paper, stated below, resolves the long standing spin-wave conjecture for the Villain model in $d\ge 3$, by obtaining the precise asymptotics of the two-point functions at low temperature.
\begin{theorem}
\label{t.main}
Let $\mu^V_{\beta}$ be the thermodynamic limit of the Villain model in $\Z^d$, for $d\geq3$. There exist constants $\beta_0=\beta_0(d)$, $c_0= c_0(\beta,d), c_1= c_1(\beta,d), c_2= c_2(\beta,d)$, and $\alpha=\alpha(d)>0$ such that for all $\beta>\beta_0$,
\begin{equation} \label{e.2ptfirstfirst}
\langle \sin\theta(0)  \sin\theta(x)  \rangle_{\mu^{V}_\beta}
= \frac{c_2}{|x|^{d-2}} + O\left(\frac{1}{|x|^{d-2+\alpha}}\right),
\end{equation}
and 
\begin{equation}
\label{e.2ptfirst}
    \left\langle S_0 \cdot S_x \right\rangle_{\mu^{V}_\beta} =
    c_0 + \frac{c_1}{|x|^{d-2}}+ O\left(\frac{1}{|x|^{d-2+\alpha}}\right).
  \end{equation}  
\end{theorem}

 It will be clear from the proof of Theorem \ref{t.main} (see the presentation in Chapters \ref{chap:chap3} and \ref{section3.4}), that the constant $c_1$ is a small correction to the prediction $C_2$ from the free field computation \eqref{e.free} (and satisfies $|c_1 - C_2|\leq e^{-c\beta^{1/2}}$). Indeed, as observed in \cite{FS4d}, the emergent massless free field that leads to the $|x|^{2-d}$ term is contributed from a free field \eqref{e.free}, plus a vortex correction (see Chapter~\ref{chap:chap3}, in particular \eqref{e.dual2pt} there). Our result gives the covariance structure of the emergent massless free field. In fact, we see that $c_0 = \left\langle \cos \theta_0 \right\rangle_{\mu^{V}_\beta}$. We may characterize $c_1$ from the homogenization of a family of elliptic PDEs (see Chapter~\ref{section3.4}), and thus identify the correlation structure of the continuum free field.
 
The proof in this paper is robust and can be adapted to general non-cubic lattices and general finite range interactions. It also applies to boundary conditions other than Dirichlet. For example, we may consider the Gibbs states arising as the infinite volume limit of the Villain models with Neumann or periodic boundary conditions. The same proof applies (except for several changes in the boundary conditions in Chapter~\ref{chap:chap3} when we apply the duality) and we can also prove Theorem~\ref{t.main} for these Gibbs states. With the same approach, but more elaborate estimates, it is possible to further generalize our result and prove that the spin field $\cos \theta_x - \left\langle \cos \theta_x \right\rangle_{\mu^{V}_\beta}$ converges in distribution to a Gaussian free field, with the covariance structure given in~\eqref{e.2ptfirst}. Finally, the method extends to more general models such as the higher dimensional abelian spin models and abelian lattice gauge theorems, and these may lead to several subsequent works. 

\section{Strategy of the proof}

\subsection{Sine-Gordon representation and polymer expansion} \label{section2.11457}

The spin wave computation \eqref{e.free} is only heuristic and does not give the correct constants $C_1, C_2$. The main problem for the spin wave is that it ignores the formation of  vortices, that are defined on the faces of $\Zd$. Kosterlitz and Thouless \cite{KT} gave a heuristic argument, that indicates that the vortices are interacting like a neutral Coulomb gas, taking integer valued charges. 

Our proof of Theorem \ref{t.main} starts from an insight of Fr{\"o}hlich and Spencer \cite{FS4d}, which makes this observation rigorous. In particular, the two-point function of the Villain model in $\Zd$, $d\ge 3$ can be mapped, via duality, to a statistical mechanical model with integer valued and locally neutral charges on 2-forms $\Lambda^2(\Zd)$. By performing a Fourier transform with respect to the charge variable, a classical random field representation of the Coulomb gas, known as the sine-Gordon representation is obtained. When the temperature is low enough, the charges are essentially supported on short range dipoles, therefore a one-step renormalization argument and a cluster expansion can be applied, following the presentation of~\cite{Bau}, in order to reduce the effective activity of the charges. This leads to an effective, vector-valued random interface model with infinite range and uniformly convex potential. The question of the asymptotic behavior of the two-point function is thus reduced to the question of the quantitative understanding of the large-scale properties of the random interface model. 

\subsection{Hellfer-Sj\"{o}strand PDE and regularity}

The study of the large-scale properties of the random interface model starts from the insight of Naddaf and Spencer \cite{NS} that the fluctuations of the field are closely related to an elliptic homogenization
problem for the Helffer-Sj{\"o}strand equation \cite{HS, Sj}. This approach has been used by Giacomin, Olla and Spohn in~\cite{GOS} to prove that the large-scale space-time fluctuations of the field is described by an infinite dimensional Ornstein-Uhlenbeck process and by Deuschel, Giacomin and Ioffe to establish concentration properties and large deviation principles on the random surface (see also Sheffield~\cite{Sh}, Biskup and Spohn~\cite{BS11}, Cotar and Deuschel~\cite{CD12} and Adams, Buchholz, Koteck{\`y} and M{\"u}ller \cite{ABKM} for an extension of these results to some non-convex potentials). An important component of the strategy presented in many of the aforementioned articles relies on a probabilistic approach: one can, through the Helffer-Sj\"{o}strand representation, reduce the problem to a question of random walk in dynamic random environment, and then prove properties on this object, e.g. invariance principles, using the results Kipnis and Varadhan~\cite{KV86}, or annealed upper bounds on the heat kernel (see Delmotte and Deuschel~\cite{DD05}). However, the results obtained so far using this probabilistic approach are not quantitative, and it cannot be easily implemented to study nonlinear functionals of the field. A more analytical approach was developed by Armstrong and Wu in~\cite{AW}, where they extend and quantify the homogenization argument of Naddaf and Spencer~\cite{NS}, resolved an open question posed by Funaki and Spohn~\cite{FS} regarding the $C^2$ regularity of surface tension, and also positively resolve the fluctuation-dissipation conjecture of~\cite{GOS}. 

The approach followed in this article is the analytical one, and fits into the program developed in \cite{NS, AW} on homogenization for the random interface models. Since the sine-Gordon representation and the polymer expansion give a random interface model valued in the vector space $\R^{\binom d2}$ (which corresponds to the dimension of the space of discrete $2$-forms on $\Zd$ used to derive the dual Villain model as explained in Section~\ref{section2.11457} and Chapter~\ref{chap:chap3}) with long range and uniformly convex potential, an application of the strategy of Naddaf and Spencer~\cite{NS} to this model gives a Helffer-Sj\"{o}strand operator of the form
\begin{equation} \label{introHS}
    \Delta_\phi + \mathcal{L},
\end{equation}
which is an infinite dimensional elliptic operator acting on functions defined in the space $\Omega \times \Zd$ (see \eqref{e.HSop} in Chapter~\ref{chap:chap3} for the precise definition of this operator), where $\Omega$ is the space of functions from $\Zd$ to $\R^{\binom d2}$ in which the vector-valued random interface considered in this article is valued. The operator $\Delta_\phi$ is the (infinite dimensional) Laplacian computing derivatives with respect to the height of the random surface and $\mathcal{L}$ is an operator associated to a uniformly elliptic \emph{system of equations} with infinite range (and with exponential decay on the size of the long range coefficients) on the discrete lattice $\Zd$. The analysis of these systems requires to overcome some difficulties; a number of properties which are valid for elliptic equations, and used to study the random interface models, are known to be false for elliptic systems. It is for instance the case for the maximum principle, which implies there is no the random walk representation for this model, the De Giorgi-Nash-Moser regularity theory for uniformly elliptic and parabolic PDE (see~\cite{NA58, DG1},~\cite[Chapter 8]{GT01} and the counterexample of De Giorgi~\cite{DeG}) and the Nash-Aronson estimate on the heat kernel (see~\cite{Ar}).

To resolve this lack of regularity, we rely on a perturbative argument and make use of ideas from \emph{Schauder theory} (see~\cite[Chapter 3]{QL}); we leverage on the fact that the inverse temperature $\beta$ is chosen very large so that the elliptic operator $\mathcal{L}$ can be written
\begin{equation*}
    \mathcal{L} := - \frac{1}{2\beta} \Delta + \mathcal{L}_{\mathrm{pert}},
\end{equation*}
where the operator $\mathcal{L}_{\mathrm{pert}}$ is a perturbative term; its typical size is of order $\beta^{-\frac 32} \ll\beta^{-1}$. One can thus prove that any solution $u$ of the equation~\eqref{introHS} is well-approximated on every scale by a solution $\bar u$ of the equation $\Delta_\phi - \frac{1}{2\beta} \Delta$ for which the regularity can be easily established. It is then possible to borrow the strong regularity properties of the function $\bar u$ and transfer it to the solution of~\eqref{introHS}. This strategy is implemented in Chapter~\ref{section:section4} and allows us to prove the $C^{0 , 1-\ep}$-regularity of the solution of the Helffer-Sj\"{o}strand equation, and to deduce from this regularity property various estimates on other quantities of interest (e.g, decay estimates on the heat kernel in dynamic random environment, decay and regularity for the Green's function associated to the Helffer-Sj\"{o}strand operator). The regularity exponent $\ep$ depends on the dimension $d$ and the inverse temperature $\beta$, and tends to $0$ as $\beta$ tends to infinity; in the perturbative regime, the result turns out to be much stronger than the $C^{0 , \alpha}$-regularity provided by the De Giorgi-Nash-Moser theory (for some tiny exponent $\alpha >0$) in the case of elliptic equations, and implies sufficient mixing of the solution to the Helffer-Sj\"{o}strand equation.

\subsection{Stochastic homogenization} The main difficulty in the establishment of Theorem~\ref{t.main} is that since the Villain model is not exactly solvable, the dependence  of the constants $c_1$ and $c_2$ on the dimension $d$ and the inverse temperature $\beta$ is highly non explicit; one does not expect to have a simple formula for these coefficients but it is necessary to analyze them in order to prove the expansions~\eqref{e.2ptfirstfirst} and~\eqref{e.2ptfirst}. This is achieved by using tools from the theory of quantitative stochastic homogenization.

This theory is typically interested in the understanding of the large-scale behavior of the solutions of the elliptic equation
\begin{equation} \label{eq:TVintro1802}
    - \nabla \cdot \mathbf{a}(x) \nabla u = 0,
\end{equation}
where $\mathbf{a}$ is a random, uniformly elliptic coefficient field that is stationary and ergodic. The general objective is to prove that, on large scales, the solutions of~\eqref{eq:TVintro1802} behave like the solutions of the elliptic equation
\begin{equation} \label{eq:TV11450502}
    - \nabla \cdot \ahom \nabla u = 0,
\end{equation}
where $\ahom$ is a constant uniformly elliptic coefficient called \emph{the homogenized matrix}.
The theory was initially developed in the 80's, in the works of Kozlov~\cite{K1}, Papanicolaou and Varadhan~\cite{PV1} and Yurinski\u\i~\cite{Y1}. Dal Maso and Modica~\cite{DM1, DM2} extended these results a few years later to nonlinear equations using variational arguments inspired by $\Gamma$-convergence. All of these results rely on the use of the ergodic theorem and are therefore purely qualitative.

The main difficulty in the establishment of a quantitative theory is the question of the transfer of the quantitative ergodicity encoded in the coefficient field $\mathbf{a}$ to the solutions of the equation. This problem was addressed in a satisfactory fashion for the first time by Gloria and Otto in~\cite{GO1, GO2}, where, building upon the ideas of~\cite{NS}, they used spectral gap inequalities (or concentration inequalities) to transfer the quantitative ergodicity of the coefficient field to the solutions of~\eqref{eq:TVintro1802}. These results were then further developed in~\cite{GO15, GO115} and also in collaboration with Neukamm in~\cite{GNO} and ~\cite{GNO14}.

Another approach, which is the one pursued in this article, was initiated by Armstrong and Smart in~\cite{AS}, who extended the techniques of Avellaneda and Lin~\cite{AL1, AL2}, the ones of Dal Maso and Modica~\cite{DM1, DM2} and obtained an algebraic, suboptimal rate of convergence for the homogenization error of the Dirichlet problem associated to the nonlinear version of the equation~\eqref{eq:TVintro1802}. These results were then improved in~\cite{AKM1, AKMInvent, AKM} to obtain optimal rates. Their approach relies on mixing conditions on the coefficient fields and on the quantification of the subadditivity defect of dual convex quantities (see Chapter~\ref{section5}). The paper \cite{AS} also addressed for the first time a large-scale regularity theory for the solutions of~\eqref{eq:TVintro1802}, which was later improved in the works of ~\cite{GNO14} and \cite{AKMInvent}. An extension of the techniques of~\cite{AKM} to the setting of differential forms (which also appear in this article in the dual Villain model) can be found in~\cite{Daf18}, and to the uniformly convex gradient field model in~\cite{Dar}.

To prove Theorem~\ref{t.main}, we apply the techniques of~\cite{AKM} to the Helffer-Sj\"{o}strand equation to prove the quantitative homogenization of the mixed derivative of the Green's function associated to this operator. The strategy can be decomposed into two steps.

The first one relies on the variational structure of the Helffer-Sj\"{o}strand operator and is the main subject of Chapter~\ref{section5}: following the arguments of~\cite[Chapter 2]{AKM}, we define two subadditive quantities, denoted by $\nu$ and $\nu^*$. The first one corresponds to the energy of the Dirichlet problem associated to the Hellfer-Sj\"{o}strand operator~\eqref{introHS} in a domain $U \subseteq \Zd$ and subject to affine boundary condition, the second one corresponds to the energy of the Neumann problem of the same operator with an affine flux. Each of these two quantities depends on two parameters: the domain of integration $U$ and the slope of the affine boundary condition, denoted by $p$ (for $\nu$) and $p^*$ (for $\nu^*$). These energies are quadratic, uniformly convex with respect to the variables $p$ and $p^*$, and satisfy a subadditivity property with respect to the domain $U$; in particular, an application of Fekete's Lemma shows that these quantities converge as the size of the domain tends to infinity
\begin{equation*}
    \nu \left(  U , p\right) \underset{|U| \to \infty}{\longrightarrow} \frac 12 p \cdot \ahom p \hspace{5mm} \mbox{and} \hspace{5mm} \nu^* \left(  U , p^*\right) \underset{|U| \to \infty}{\longrightarrow} \frac 12 p^* \cdot \ahom_* p^*.
\end{equation*}
The coefficient $\ahom$ obtained this way plays a similar role as the homogenized matrix in~\eqref{eq:TV11450502}; in the case of the present random interface model, it gives the coefficient obtained in the continuous (homogenized) Gaussian free field which describes the large-scale behavior of the random surface as established by Naddaf and Spencer in~\cite{NS}. The objective of the proofs of Chapter~\ref{section5} is to quantify this convergence and to obtain an algebraic rate: we show that there exists an exponent $\alpha>0$ such that for any cube $\cu \subseteq \Zd$ of size $R> 0$,
\begin{equation} \label{eq:TV11240502}
     \left| \nu \left(  U , p\right) - \frac 12 p \cdot \ahom p \right| + \left| \nu \left(  U , p^* \right) - \frac 12 p^* \cdot \ahom_* p^* \right| \leq C R^{-\alpha}.
\end{equation}
The strategy to prove the quantitative rate~\eqref{eq:TV11240502} is to use that the maps $p \mapsto \nu \left( U , p \right)$ and $p^* \mapsto \nu^* \left( U , p^* \right)$ are approximately convex dual. We use a multiscale argument to prove that, by passing from one scale to another, the convex duality defect must contract (in particular, it is equal to $0$ in the infinite volume limit, i.e., $\ahom_* = \ahom^{-1}$). More precisely we show that the convex duality defect can be controlled by the subadditivity defect, and then iterate the result over all the scales from $1$ to $R$ to obtain~\eqref{eq:TV11240502}. As a byproduct of the proof, we obtain a quantitative control on the sublinearity of the finite-volume corrector defined as the solution of the Dirichlet problem: given an affine function $l_p$ of slope $p$ and a cube $\cu \subseteq \Zd$ of size $R$,
\begin{equation*}
    \left\{ \begin{aligned} 
    \left(\Delta_\phi + \mathcal{L}\right)\left( l_p + \chi_{\cu , p} \right) & = 0 ~\mbox{in}~ \Omega \times \cu, \\
    \chi_{\cu , p} & = 0 ~\mbox{on}~ \Omega \times \partial \cu.
    \end{aligned}
    \right.
\end{equation*}
This estimate takes the following form
\begin{equation} \label{eq:TV14530502}
    \frac1R \left\|  \chi_{\cu , p}  \right\|_{\underline{L}^2 \left( \cu , \mu_\beta\right)} \leq \frac{C}{R^\alpha},
\end{equation}
where the average $L^2$-norm is considered over both the spatial variable and the random field. 

We note that, contrary to the case of the homogenization of the elliptic equation~\eqref{eq:TVintro1802}, the subadditive quantities are deterministic objects and are applied to the operator~\eqref{introHS} which is essentially infinite dimensional. While the proofs of~\cite[Chapter 2]{AKM} rely on a finite range dependence assumption to quantify the ergodicity of the coefficient field, we rely here on the regularity properties of the Helffer-Sj\"{o}strand operator to prove sufficient decorrelation estimates on the field to obtain the algebraic rate of convergence stated in~\eqref{eq:TV11240502}. The same issues were addressed in the work of Armstrong and Wu~\cite{AW}, to study the $\nabla \phi$-model and prove $C^{2}$-regularity of the surface tension conjectured by Funaki and Spohn \cite{FS}; the arguments presented there are somewhat similar to ours but with a distinct difference: they rely on couplings based on the probabilistic interpretation of the equation to obtain sufficient decorrelation of the gradient field. In the present paper, we rely on the observation that the differential with respect to the field $\partial u$, where $u$ is a solution to the Helffer-Sj\"{o}strand equation, solves a differentiated Helffer-Sj\"{o}strand equation introduced in Section~\ref{sec.section4.5} of Chapter~\ref{section:section4}, and the decorrelation follows from the regularity theory for the differentiated Helffer-Sj\"{o}strand operator.

The second step in the argument, which extends the results of~\cite{AW}, is to prove quantitative homogenization of the mixed derivative of the Green's function associated to the Helffer-Sj\"{o}strand operator~\eqref{introHS}. In the setting of the divergence from elliptic operator~\eqref{eq:TVintro1802}, the properties of the Green's function are well-understood; moment bounds on the Green's function, its gradient and mixed derivative are proved in~\cite{DD05},~\cite{BG15} and~\cite{CGO17}. Quantitative homogenization estimates are proved in~\cite[Chapters 8 and 9]{AKM} and in~\cite{BGO17}. The argument relies on a common strategy in stochastic homogenization: the two-scale expansion. It is implemented as follows: given a function $\mathbf{f}: \Omega \to \R$ which depends on the field and satisfies some suitable regularity assumptions (e.g. $\mathbf{f}$ depends on finitely many height variables of the random surface, is smooth and compactly supported), the large-scale behavior of the fundamental solution $\G_\mathbf{f}: \Omega \times \Zd \to \R^{\binom d2 \times \binom d2}$ of the $\binom d2$-dimensional elliptic system
\begin{equation*}
        \left( \Delta_\phi + \mathcal{L} \right) \G_\mathbf{f} = \mathbf{f}  \delta_0,
\end{equation*}
is described by the (deterministic) fundamental solution $\bar G : \Zd \to \R^{\binom d2 \times \binom d2}$ of the homogenized elliptic system
\begin{equation*}
    - \nabla \cdot \ahom \nabla \bar G = \left\langle \mathbf{f}\right\rangle_{\mu_\beta} \delta_0,
\end{equation*}
where the notation $\mu_\beta$ is used to denote the law of the random surface and $\left\langle \mathbf{f}\right\rangle_{\mu_\beta}$ denotes the expectation of the function $\mathbf{f}$ with respect to the probability measure $\mu_\beta$. The proof of this result relies on a two-scale expansion for systems of equations: we select a suitable cube $\cu \subseteq \Rd$ and define the function
\begin{equation*}
\mathcal{H}_{\cdot k} := \bar G_{\cdot k} + \sum_{i = 1}^d \sum_{j = 1}^{\binom d2}  \chi_{\cu , e_{ij}} \nabla_i \bar G_{j k}.
\end{equation*}
We then compute the value of $\left( \Delta_\phi + \mathcal{L} \right) \mathcal{H}$ and prove, by using the quantitative information obtained on the corrector~\eqref{eq:TV14530502}, that this value is small in a suitable functional space. This argument shows that the function $\mathcal{H}$ (resp. its gradient) is quantitatively close to the functions $\mathcal{G}_{\mathbf{f}}$ (resp. its gradient). Once this is achieved, we can iterate the argument to obtain a quantitative homogenization result for the mixed derivative of the Green's function following the description given at the beginning of Chapter~\ref{section5}.
The overall strategy is similar to the one in the case of the divergence form elliptic equations~\eqref{eq:TVintro1802} but a number of technicalities need to be treated along the way:
\begin{itemize}
    \item The infinite dimensional Laplacian $\Delta_\phi$ needs to be taken into account in the analysis;
    \item The elliptic operator $\L$ given in the model has infinite range;
    \item One needs to homogenize an elliptic system instead of an elliptic PDE.
\end{itemize}
While the first point has been successfully addressed by \cite{AW}, the last two points are intrinsic to the Coulomb gas representation of the dual Villain model. Overcoming these difficulties  requires new adaptation of the methods developed in \cite{AKM} and \cite{AW}.

\section{First order expansion of the two point-functions}

The first order expansion of the two-point function is obtained by post-processing all the arguments above. We first use the sine-Gordon representation and the polymer expansion to reduce the problem to the understanding of the large scale behavior of a vector-valued random surface model, whose Hamiltonian is a small perturbation of the one of a Gaussian free field. We then use the ideas of Naddaf and Spencer~\cite{NS} and Schauder regularity theory (through a perturbative argument) to obtain a precise understanding of the correlation structure of the random field. Unfortunately, the regularity obtained this way is not strong enough to establish Theorem~\ref{t.main} and to obtain a sharp control on the large-scale behavior of the solutions of the Helffer-Sj\"{o}strand equation, we adapt the recent techniques in quantitative stochastic homogenization developed in~\cite{AKM}. We can then combine this result with the regularity theory and the rotation and symmetry invariance of the Villain model to prove Theorem~\ref{t.main}. The proof of this result requires to analyze a number of terms to isolate the leading order terms and to estimate quantitatively the lower order ones. It is rather technical and is split into two chapters: in Chapter~\ref{section3.4} we present a detailed sketch of the argument, isolate the leading order from the lower order terms and state the estimates on each of these terms; Chapter~\ref{section7} is devoted to the proof of the technical estimates and the estimates of the various terms obtained in Chapter~\ref{section3.4} using the results proved in Chapters~\ref{section:section4},~\ref{section5} and~\ref{sec:section6}.

\section{Open questions}

Finally, we discuss some open questions regarding the low temperature phase of the classical XY model in $\Zd$, $d\geq 3$. The Gaussian spin-wave approximation predicts that the two-point function of the XY model also admits a low temperature expansion stated in Theorem \ref{t.main}. It is believable that our method can be adapted to resolve this conjecture for the XY model. The main challenge is that, unlike the Villain model, when passing to the dual model the two-point function cannot be factorized as a Gaussian contribution and a vortex contribution (see Chapter~\ref{chap:chap3}, \eqref{e.dual2pt}), thus requires a new idea for renormalization. 

\section{Organization of the paper}

This paper is organized as follows. In the next chapter we introduce some preliminary notations. In Chapter~\ref{chap:chap3}, we recall the dual formulation of the Villain model in terms of a vector-valued random interface model, based on the ideas of Fr{\"o}hlich and Spencer~\cite{FS4d} and following the presentation of Bauerschmidt~\cite{Bau}. We also establish the existence of a thermodynamic limit for the dual Villain model by coupling the Langevin dynamics of the finite volume models following the technique of Funaki and Spohn~\cite{FS}. We then derive the Helffer-Sj{\"o}strand equation for the dual model and state the main regularity estimates on the Green's function proved in Chapter~\ref{section:section4} and the quantitative homogenization of the mixed derivative of the Green's function proved in Chapters~\ref{section5} and~\ref{sec:section6}. In Chapter~\ref{section3.4}, we sketch the proof of the main theorem, assuming the $C^{0,1-\ep}$ regularity for the solutions of the Helffer-Sj{\"o}strand equation (established in Chapter~\ref{section:section4}), and the quantitative homogenization of the mixed derivative of the Green's function (established in Chapter~\ref{sec:section6}). In Chapter~\ref{section5}, we introduce the subadditive energy quantities and show by a multiscale iterative argument that they converge at an algebraic rate. Finally in Chapter~\ref{section7}, we give detailed proofs of the claims in Chapter~\ref{section3.4}. 

\medskip

\textbf{Acknowledgments.} P.D. is supported by the Israel Science Foundation grants 861/15 and 1971/19 and by the European Research Council starting grant 678520 (LocalOrder). W.W. is supported in part by the EPSRC grant EP/T00472X/1.
We thank T. Spencer for many insightful discussions that
inspired the project, R. Bauerschmidt for kindly explaining the arguments in \cite{Bau}, and S. Armstrong for many helpful discussions. We also thank S. Armstrong and J.-C. Mourrat for helpful feedbacks on a previous version of the paper.

\chapter{Preliminaries} \label{Chap:chap2}

\section{Notations and assumptions} \label{SecNotandprelim}

\subsection{General notations and assumptions}
We work on the Euclidean lattice $\Zd$ in dimension $d \geq 3$. We say that two points $x , y \in \Zd$ are neighbors, and denote it by $x \sim y$, if $|x - y|_1 = 1$. We denote by $e_1 , \ldots , e_k$ the canonical basis of $\Rd$. We denote by $\left| \cdot \right|$ the standard Euclidean norm on the lattice $\Zd$. For each integer $k \in \{ 1 , \ldots, d \}$, a $k$-cell of the lattice $\Zd$ is a set of the form, for a subset $\{ i_1 , \ldots, i_{k} \} \subseteq \left\{ 1 , \ldots , d \right\}$ and a point $x \in \Zd$,
\begin{equation*}
    \left\{x + \sum_{l = 1}^k \lambda_l e_{i_l} \in \Rd \, : \, 0 \leq \lambda_1 , \ldots ,  \lambda_k \leq 1 \right\}.
\end{equation*}
We equip the set of $k$-cells with an orientation induced by the canonical orientation of the lattice $\Zd$ and denote by $\Lambda^k(\Zd)$ the set of oriented $k$-cells of the lattice $\Zd$. Given a $k$-cell $c_k$, we denote by $\partial c_k$ the boundary of the cell; it can be decomposed into a disjoint union of $(k-1)$-cells. The values $k = 0 , 1 , 2$ are of specific interest to us; they correspond to the set of vertices, edges and faces of the lattice $\Zd$. We will denote these spaces by $V(\Zd)$, $E(\Zd)$ and $F(\Zd)$ respectively.

Given a subset $U \subseteq \Zd$ we define its interior $U^\circ$ and its boundary $\partial U$ by the formulas
\begin{equation*}
    U^\circ := \left\{ x \in U \, : \, x \sim y \implies y \in U \right\} \hspace{5mm} \mbox{and} \hspace{5mm} \partial U := U \setminus U^\circ.
\end{equation*}
If the subset $U \subseteq \Zd$ is finite, we denote by $|U|$ its cardinality and refer to this quantity as the volume of $U$. We denote by $\diam U$ the diameter of $U$ defined by the formula $\diam U := \sup_{x , y \in U} |x - y|$. Given a point $x \in \Zd$ and a radius $r>0$, we denote by $B(x,r)$ the discrete euclidean ball of center $x$ and radius $r$. We frequently use the notation $B_r$ to mean $B(0 , r)$.

A discrete cube $\cu$ of $\Zd$ is a subset of the form
\begin{equation*}
        \cu := x + \left[ -N , N \right]^d \cap \Zd ~\mbox{with}~ x \in \Zd ~\mbox{and}~ N \in \N.
\end{equation*}
We refer to the point $x$ as the center of the cube $\cu$ and to the integer $2N +1$ as its length. Given a parameter $r > 0$, we use the nonstandard convention of denoting by $r\cu$ the cube 
\begin{equation*}
    r \cu := x + \left[ -r N , rN \right]^d \cap \Zd.
\end{equation*}
We denote by $\Lambda^k(\cu)$ the set of oriented $k$-cells qhich are included in the cube $\cu$. In the specific cases $k = 1 ,2$ and $3$, we denote by $V(\cu)$, $E(\cu)$ and $F(\cu)$ the set of vertices, edges and faces of the cube $\cu$ respectively.

For each integer $i \in \{ 1 , \ldots , d \}$,we denote by $h_i$ the reflection of the lattice $\Zd$ with respect to the hyperplane $\left\{ z \in \Zd \, : \, z_i = 0 \right\}$, i.e.,
\begin{equation*}
    h_i := \left\{ \begin{aligned}
    \Zd &\to  \Zd \\
    (z_1 , \ldots, z_d) & \mapsto  (z_1 , \ldots, - z_i , \ldots,  z_d).
    \end{aligned}
    \right.
\end{equation*}
For each pair of integers $i,j \in \{ 1 , \ldots , d \}$ with $i < j$, we denote by $h_{ij}$ the map
\begin{equation*}
    h_i := \left\{ \begin{aligned}
    \Zd &\to  \Zd \\
    (z_1 , \ldots, z_d) & \mapsto  (z_1 , \ldots, z_j, \ldots, z_i , \ldots,  z_d).
    \end{aligned}
    \right.
\end{equation*}
We define $H$ the group of lattice preserving transformation to be the group of linear maps generated by the collections of functions $\left( h_i \right)_{1 \leq i \leq d}$ and $\left( h_{ij}\right)_{1 \leq i < j \leq d}$ with respect to the composition law.

Given three real numbers $X , Y \in \R$ and $\kappa \in [0 , \infty)$, we write
\begin{equation*}
    X = Y + O (\kappa) \hspace{3mm} \mbox{if and only if} \hspace{3mm} \left| X - Y \right| \leq \kappa.
\end{equation*}

\subsection{Notations for vector-valued functions}
For each integer $k \in \N$, we let $\mathcal{F} \left( \Zd , \R^k\right)$ be the set of functions defined on $\Zd$ and taking values in $\R^k$. Given a function $g \in \mathcal{F} \left( \Zd , \R^k\right)$, we denote by $g_1, \ldots, g_k$ its components on the canonical basis of $\R^k$ and write $g = (g_1 , \ldots, g_k)$. We define the support of the function $g$ to be the set
\begin{equation*}
    \supp g := \left\{ x \in \Zd \, : \, g(x) \neq 0 \right\}.
\end{equation*}
The oscillation of a function $g$ over a set $U \subseteq \Zd$ is defined by the formula
\begin{equation*}
    \osc_U g := \sup_U g - \inf_U g.
\end{equation*}
For each exponent $\alpha > 0$, we define the $C^{0,\alpha}$-H\"{o}lder seminorm of the function $g$ over the set $U$ by
\begin{equation*}
    \left\| g \right\|_{C^{0,\alpha} (U)} := \sup_{x , y \in U, x\neq y} \frac{\left| g(x) - g(y) \right|}{|x - y|^\alpha}.
\end{equation*}
For each integer $i \in \{ 1 , \ldots, d \}$, we define its discrete $i$-th derivative $\nabla_i g : \Zd \to \R^{k}$ by the formula, for each $x \in \Zd$,
\begin{equation*}
    \nabla_i g (x) :=  g (x + e_i) - g(x),
\end{equation*}
and its gradient $\nabla g: \Zd \to \R^{d \times k}$ by the formula
\begin{equation} \label{def.gradg1600}
    \nabla g(x) = \left( \nabla_i g (x) \right)_{1 \leq i \leq d} = \left( \nabla_i g_j (x) \right)_{ 1 \leq i \leq d,  1 \leq j \leq k }.
\end{equation}
We denote by $\nabla_i^*$ the adjoint gradient defined by the formula $\nabla_i^* g(x) = g(x-e_i) - g(x)$.
A property of the discrete setting is that the $L^{\infty}$-norm of the gradient of a function $g$ is bounded from above by the $C^{0,\alpha}$-H\"{o}lder seminorm of this function: we have, for any $U \subseteq \Zd$ and any exponent $\alpha > 0$,
\begin{equation*}
     \sup_{x \in U} \left| \nabla g(x) \right| \leq \left\| g \right\|_{C^{0,\alpha} (U)}.
\end{equation*}
The Laplacian $\Delta g : \Zd \to \R^{k}$ is defined by the formula, for each point $x \in \Zd$,
\begin{equation} \label{eq:TV11322901}
    \Delta g(x) = \sum_{y \sim x} \left( g(y) - g(x) \right).
\end{equation}
For each integer $n \in \N$, one can consider the iteration $\nabla^n$ on the gradient and $\Delta^n$ of the Laplacian. We note that these discrete operators have range $n$ and $2n$ respectively, i.e., given a point $x \in \Zd$ and a function $u: \Zd \to \R^k$ one can compute the value of $\nabla^n u(x)$ (resp. $\Delta^n u(x)$) by knowing only the values of $u$ inside the ball $B(x,n)$ (resp. $B(x , 2n)$). For each function $g : \Zd \times \Zd \to \R^k$, we denote by $\nabla_x$ and $\nabla_y$ the gradients with respect to the first and second variable respectively, i.e., for each point $(x, y ) \in \Zd \times \Zd$, we write
\begin{equation*}
    \nabla_x g(x , y) = \left( g_j (x + e_i,y) - g_j(x,y) \right)_{ 1 \leq i \leq d,  1 \leq j \leq k } ~\mbox{and}~\nabla_y g(x , y) = \left( g_j (x ,y+e_i) - g_j(x,y) \right)_{ 1 \leq i \leq d,  1 \leq j \leq k }.
\end{equation*}
We similarly define the $i$-th derivatives $\nabla_{i,x}$ and $\nabla_{i,y}$ and the Laplacians $\Delta_x$ and $\Delta_y$ with respect to the first and second variables.

Given two functions $f , g : \Zd \to \R^k$ and a point $x \in \Zd$, we define the scalar product $f(x) \cdot g(x) := \sum_{i=1}^d f_k(x) g_k(x)$. To ease the notation, we may write $f(x)g(x)$ to mean $f(x) \cdot g(x)$. We define the $L^2$-scalar product $(\cdot , \cdot)$ according to the formula
\begin{equation} \label{eq:TV07309}
    \left( f , g \right) = \sum_{x \in \Zd} f(x) g(x),
\end{equation}
We restrict this scalar product to a set $U \subseteq \Zd$ and define, for any pair of functions $f , g : U \to \R^k$,
\begin{equation} \label{eq:VEr161902}
    \left( f , g \right)_U := \sum_{x \in U} f(x)  g(x).
\end{equation}

We define the divergence operator $\nabla \cdot$ on vector valued functions: given a function $F : \Zd \to \R^{d \times k}$, we denote by $\nabla \cdot F : \Zd \to \R^{k}$ the unique function which satisfies, for each compactly supported function $g : \Zd \to \R^k$, 
\begin{equation*}
    \left( F , \nabla g \right) = - \left( \nabla \cdot F , g \right).
\end{equation*}
If we use the definition~\eqref{def.gradg1600} and denote by $\left( F_{ij} \right)_{1 \leq i \leq d, 1\leq j \leq k }$ the components of the function $F$, then we have the identity $\nabla \cdot F = \left( \sum_{i=1}^d \nabla_i^* F_{ij} \right)_{1 \leq j \leq k}$.

Given an integer $l \in \N$ and a function $h :  \Zd \to \R^{l}$, we denote by $g \otimes h : \Zd \to \R^{k \times l}$ the tensor product between the two functions $h$ and $g$; it is defined by the formula, for each point $x \in \Zd$,
\begin{equation} \label{eq:TV11011622}
    g \otimes h(x) := \left( g_i(x) h_j (x) \right)_{ 1 \leq i \leq k,  1 \leq j \leq l}.
\end{equation}
This notation allows to expand gradients of products of functions: for each function $u : \Zd \to \R$, one has
\begin{equation} \label{eq:TV16450801}
    \nabla (ug)(x) = \nabla u \otimes g(x) + u(x) \nabla g(x).
\end{equation}
Given a bounded subset $U \subseteq \Zd$, we define the average of $g$ over the set $U$ by the formula
\begin{equation*}
    \left( g \right)_U := \frac{1}{|U|} \sum_{x \in U} g(x) \in \R^k. 
\end{equation*}
For each real number $p \in [1 , \infty)$ and each subset $U \subseteq \Zd$, we define the $L^p \left( U \right)$-norm
\begin{equation*}
    \left\| g \right\|_{L^p \left( U \right)} := \left( \sum_{x \in U} |g(x)|^p\right)^{\frac 1p} \hspace{5mm} \mbox{and} \hspace{5mm} \left\| g \right\|_{L^\infty \left( U \right)} := \sup_{x \in U} \left| g(x) \right|.
\end{equation*}
where the notation $|\cdot|$ denotes the euclidean norm on $\R^k$. Given a bounded subset $U \subseteq \Zd$, we denote by $\underline{L}^p(U)$ the normalized norms
\begin{equation*}
    \left\| g \right\|_{\underline{L}^p \left( U \right)} := \left( \frac{1}{|U|} \sum_{x \in \Zd} |g(x)|^p\right)^{\frac 1p}.
\end{equation*}
We introduce the normalized Sobolev norms $\underline{H}^1(U)$ and $\underline{H}^{-1}(U)$ by the formulas
\begin{equation*}
    \left\| g \right\|_{\underline{H}^1 (U)} := \frac{1}{\diam U} \left\| g \right\|_{\underline{L}^2 \left( U \right)} +  \left\| \nabla g \right\|_{\underline{L}^2 \left( U \right)} \hspace{5mm} \mbox{and} \hspace{5mm} \left\| g \right\|_{\underline{H}^{-1} (U)} := \left\{\left( f , g\right)_U \, : \, f : U \to \R^k, \, \left\| f \right\|_{\underline{H}^1 \left( U \right)} \leq 1 \right\}.
\end{equation*}
We note that for each bounded connected subset $U \subseteq \Zd$, the Poincar\'e inequality implies the estimate
\begin{equation*}
    \left\| g  - \left( g \right)_U \right\|_{\underline{L}^2 (U)} \leq C \left( \diam U \right) \left\| \nabla g \right\|_{\underline{L}^2 \left( U \right)}.
\end{equation*}
We denote by $H^1_0(U)$ the set of functions from $U$ to $\R^k$ which are equal to $0$ outside the set $U$ (by analogy to the Sobolev space). We implicitly extend the functions of $H^1_0(U)$ by the value $0$ to the entire lattice $\Zd$. For $p, q \in [1 , \infty]$, we need to consider linear operators from $L^p(\Zd)$ into $L^q(\Zd)$; we introduce the operator norm on this space according to the formula, for each $A : L^p(\Zd) \to L^q(\Zd)$,
\begin{equation*}
    ||| A |||_{L^p(\Zd) \to L^q(\Zd)} := \sup \left\{ \left\| Au \right\|_{L^q \left( \Zd \right)} \, : \, u \in L^p \left( \Zd \right), \, \left\| u \right\|_{L^p \left( \Zd \right)} \leq 1 \right\}.
\end{equation*}
Note that, for each $p \in [1 , \infty]$, the discrete gradient and Laplacian have a finite operator norm in the space $L^p \left( \Zd \right)$.

We frequently consider functions defined from $\Zd$ an d valued in $\R$ of the form $x \to |x|^{-k}$. We implicitly extend these functions at the point $x = 0$ by the value $1$ so that they are defined on the entire lattice $\Zd$. For instance, we may write, given two integers $k , l \in \N$ such that $k + l > d$ and a point $y \in \Zd$,
\begin{equation*}
    \sum_{x \in \Zd} \frac{1}{|x|^k} \cdot \frac{1}{|x - y|^l} \hspace{5mm} \mbox{to mean} \hspace{5mm} \sum_{x \in \Zd, x \notin \{ 0,y\}} \frac{1}{|x|^k} \cdot \frac{1}{|x - y|^l} + \frac{1}{|y|^k} + \frac{1}{|y|^l}.
\end{equation*}
Finally Section~\ref{sec:chap8.4} of Chapter~\ref{section7} requires to work with Fourier analysis, we thus introduce the Schwartz space of rapidly decreasing functions of $\Rd$ by the formula
\begin{equation*}
    \mathcal{S} \left( \Rd \right) := \left\{ f \in C^\infty \left( \Rd \right) \, : \, \forall \alpha , \beta \in \N, ~ \sup_{x \in \Rd} |x|^\alpha \left| \nabla^\beta f \right| < \infty \right\}
\end{equation*}
as well as the set of tempered distributions $\mathcal{S}' \left( \Rd \right)$ to be its topological dual. Given a function $g \in \mathcal{S} \left( \Rd \right)$, we define its Fourier transform $\hat{g} : \Rd \to \R$ by the formula
\begin{equation*}
    \hat{g}(\xi) := \int_{\Rd} g(x) e^{- i \xi \cdot x} \, \di x,
\end{equation*}
and extend the Fourier transform to the space of tempered distributions following the standard procedure.

\subsection{Notations for matrix-valued functions}
Given an pair of integers $k,l \in \N$, we may identify the vector space $\R^{k \times l}$ with the space of $(k \times l)$-matrices with real coefficients. Given a map $F : \Zd \to \R^{k \times l}$, we denote its components by $\left( F_{ij}\right)_{1 \leq i \leq k, 1 \leq j \leq l}$. For each integer $i \in \left\{ 1 , \ldots, k  \right\}$, we denote by $F_{i\cdot}$ the map
\begin{equation*}
    F_{i \cdot} : \left\{ \begin{aligned}
    \Zd & \to \R^{l}, \\
    x & \mapsto \left( \sum_{j=1}^k F_{ij}(x) \right)_{1 \leq j \leq l}.
    \end{aligned} \right.
\end{equation*}
We similarly define the map $F_{\cdot j}$ for each integer $j \in \{ 1 , \ldots, l \}$. We define the product between two maps $F : \Zd \to \R^{l \times k}$ and $g: \Zd \to \R^{k}$ by the formula
\begin{equation*}
    F g : \left\{ \begin{aligned}
    \Zd & \to \R^{l}, \\
    x & \mapsto \left( \sum_{j=1}^k F_{ij}(x) f_j(x) \right)_{1 \leq i \leq l}.
    \end{aligned} \right.
\end{equation*}
We may abuse notations and write $gF$ instead of $Fg$.

\subsection{Notations for the parabolic problem} \label{Secchap2parapb}
In Chapter~\ref{section:section4}, we need to study solutions of parabolic equations. We introduce in this section a few definitions and notations pertaining to this setting. For $s > 0$ and $t \in \R$, we define the time intervals $I_s := \left( -s , 0 \right]$ and $I_s(t) := \left( -s + t , t \right]$. Given a point $x \in \Zd$ and a radius $r > 0$, we denote the parabolic cylinder by $Q_{r}(t , x) := I_{r^2}(t) \times B(x , r)$ (where $B(x , r)$ is the discrete ball). To simplify the notation, we write $Q_r$ to mean $Q_r(0,0)$. Given a function $u :  Q_{r}(t , x)  \to \R$, we define its average over the parabolic cylinder $\left( u \right)_{Q_r(t,x)}$ by the formula
\begin{equation*}
    \left( u \right)_{Q_r(t,x)}  := \frac{1}{r^2 \left| B_r \right|} \int_{-r^2}^0 \sum_{x \in B_r} u(t,x) \, dt.
\end{equation*}
Given a finite subset $V \subseteq \Zd$ or a bounded open set $V \subset \Rd$, we denote by $\partial_{\sqcup}(I_r \times V)$ the parabolic boundary of the cylinder $I_r \times V$ defined by the formula
\begin{equation*}
    \partial_{\sqcup}(I_r \times V) := \left( I_r \times \partial V \right) \cup \left( \{ -r^2\} \times V \right).
\end{equation*}

\subsection{Notations pertaining to Gibbs measures} \label{sec:notGIbbsmeasure}

Given a cube $\cu \subset\Zd$, we let $\Omega_0(\cu)$ be the set of vector-valued functions $\phi: \cu\to \R^{\binom d2}$ such that $\phi=0$ on $\partial\cu$. We often drop the dependence on the domain when it is clear from the context. We also let $\Omega_r$ be the set of vector-valued functions $\phi: \Zd\to \R^{\binom d2}$ satisfying the growth condition $\sum_{x\in\Zd} |\phi(x)|^2 e^{-|x|} <\infty$.
Given $z\in \Zd$, we define $\tau_z: \Omega \to \Omega$ to be a translation of elements in $\Omega$: $\tau_z \phi(\cdot)= \phi(z+\cdot)$.
\smallskip

Given an inverse temperature $\beta$, a probability measure $\mu_\beta$ on $\Omega$ and measurable function $X : \Omega \to \R$ which is either nonnegative or integrable with respect to the measure $\mu_\beta$, we denote its expectation and variance by
\begin{equation*}
    \left\langle X \right\rangle_{\mu_\beta} := \int_{\Omega} X(\phi) \mu_\beta(\di \phi) \hspace{5mm} \mbox{and}  \hspace{5mm} \var_{\mu_\beta} \left[ X \right] = \int_{\Omega} \left| X(\phi) - \left\langle X \right\rangle_{\mu_\beta} \right|^2 \mu_\beta(\di \phi).
\end{equation*}
For each real number $p \in [1 , \infty)$, we define the $L^p \left( \mu_\beta \right)$-norm of the random variable $X$ according to the formula
\begin{equation*}
    \left\| X \right\|_{L^p \left( \mu_\beta \right)} : = \left( \int_{\Omega} \left| X(\phi) \right|^p \mu_\beta(\di \phi) \right)^{\frac 1p} \hspace{5mm} \mbox{and}  \hspace{5mm} \left\| X \right\|_{L^\infty \left( \mu_\beta \right)} := \esssup_{\phi \in \Omega} |X(\phi)|.
\end{equation*}
For each point $x \in \Zd$ and each integer $i \in \left\{  1 , \ldots , \binom d2 \right\}$, we let $\omega_{x,i}$ be the function
\begin{equation*}
    \omega_{x,i}(y) := \left\{ 
    \begin{aligned}
    e_i & ~\mbox{if}~ x=y \\
    0 & ~\mbox{if}~ x\neq y,
    \end{aligned}
    \right.
\end{equation*}
where $\left( e_1 , \ldots , e_{\binom d2} \right)$ is the canonical basis of $\R^{\binom d2}$. We define the differential operators $\partial_{x,i}$ and $\partial_x$ by the formulas 
\begin{equation*}
    \partial_{x,i} u (\phi) := \lim_{h \to 0} \frac {u(\phi + h\omega_{x,i}) - u(\phi)}h \hspace{5mm} \mbox{and} \hspace{5mm} \partial_{x} u (\phi) = \left(\partial_{x,1} u , \ldots , \partial_{x, \binom d2} u \right). 
\end{equation*}
We let $C_{\mathrm{loc}}^\infty(\Omega)$ be the set of smooth, local and compactly supported functions of the set $\Omega$. We define the space $H^1 \left( \mu_\beta \right)$ to be the closure of the space $C_{\mathrm{loc}}^\infty(\Omega)$ with respect to the norm (rescaled with respect to the inverse temperature $\beta$)
\begin{equation*}
    \left\| u \right\|_{H^1 \left( \mu_\beta \right)} := \left\| u \right\|_{L^2 \left( \mu_\beta \right)} + \left( \beta \sum_{x \in \Zd} \left\| \partial_x u \right\|_{L^2 \left( \mu_\beta \right)}^2\right)^{\frac 12}.
\end{equation*}
For any subset $U \subseteq \Zd$, we let $L^2 \left( U , \mu_\beta \right)$ to be the set of measurable functions $u : \Zd \times \Omega \to \R^k$ which satisfy
\begin{equation*}
    \left\| u \right\|_{L^2(U , \mu_\beta) } := \left( \sum_{x \in U} \left\| u(x , \cdot) \right\|_{L^2 \left( \mu_\beta \right)}^2 \right)^{\frac 12} < \infty.
\end{equation*}
When the set $U$ is finite, we define the normalized $\underline{L}^2 \left( U , \mu_\beta \right)$-norm by the formula
\begin{equation*}
    \left\| u \right\|_{\underline{L}^2(U , \mu_\beta) } := \left( \frac 1{|U|}\sum_{x \in U} \left\| u(x , \cdot) \right\|_{L^2 \left( \mu_\beta \right)}^2 \right)^{\frac 12},
\end{equation*}
as well as the space-field average
\begin{equation*}
    \left( u \right)_{U , \mu_\beta} := \frac1{|U|} \sum_{x \in U} \left\langle u(x , \cdot)\right\rangle_{\mu_\beta}.
\end{equation*}
More generally, for $p,q \in [1 , \infty]$ and a set $U \subseteq \Zd$, we introduce the $L^p \left( U , L^q \left( \mu_\beta\right) \right)$-norm by the formula
\begin{equation*}
    \left\| u \right\|_{L^p \left(U ,L^q \left( \mu_\beta \right)\right) } := \left( \sum_{x \in U} \left\| u(x , \cdot) \right\|_{L^q \left( \mu_\beta \right)}^p \right)^{\frac 1p}.
\end{equation*}
We define the norm $H^1(U , \mu_\beta)$ by the formula
\begin{equation*}
    \left\| u \right\|_{H^1(U , \mu_\beta) } := \left( \sum_{x \in U} \left\| u(x , \cdot) \right\|_{H^1(\mu_\beta)}^2 + \left\| \nabla u \right\|_{L^2 \left( U , \mu_\beta \right)}^2 \right)^{\frac 12},
\end{equation*}
as well as the normalized $\underline{H}^1(U , \mu_\beta)$-norm
\begin{equation*}
     \left\| u \right\|_{\underline{H}^1(U , \mu_\beta) } := \left(   \frac1{\left( \diam U\right)^2|U|}\sum_{x \in U} \left\|  u(x , \cdot) \right\|_{L^2(\mu_\beta)}^2 + \frac{\beta}{|U|} \sum_{x,y \in U} \left\| \partial_y u(x , \cdot) \right\|_{L^2(\mu_\beta)}^2 + \frac1{|U|} \left\| \nabla u \right\|_{L^2 \left( U , \mu_\beta \right)}^2 \right)^{\frac 12}.
\end{equation*}
We define the subset $H^1_0 \left(U , \mu_\beta \right)$ to be the subset of functions of $H^1_0 \left(U , \mu_\beta \right)$ which are equal to $0$ on the boundary $\partial U \times \Omega$. We implicitly extend these functions by the value $0$ to the space $\Zd$. In particular, we always think of elements of $H^1_0\left(U , \mu_\beta \right) $ as functions defined on the the entire space. We introduce the seminorm
\begin{equation*}
    \left\llbracket u \right\rrbracket_{\underline{H}^1(U , \mu_\beta) } := \left( \frac{\beta}{|U|}\sum_{x \in U, y \in \Zd} \left\| \partial_y u(x , \cdot) \right\|_{L^2(\mu_\beta)}^2 + \frac1{|U|} \left\| \nabla u \right\|_{L^2 \left( U , \mu_\beta \right)}^2 \right)^{\frac 12}.
\end{equation*}
We define the $H^{-1}(U , \mu)$-norm by the formula
\begin{equation*}
    \left\| u \right\|_{H^{-1}\left( U , \mu_\beta \right)} := \sup \left\{ \frac{1}{|U|}\sum_{x \in U} \left\langle u(x , \cdot) v(x , \cdot) \right\rangle_{\mu_\beta} \, : \, v \in H^1_0 \left( U , \mu_\beta\right), ~ \left\| v \right\|_{\underline{H}^1 \left(U , \mu_\beta \right)} \leq 1   \right\}.
\end{equation*}

We next state a Poincar\'e inequality for~$H^1(U,\mu_\beta)$. We give two statements, one for functions which vanish on the boundary of~$U$ and another for zero-mean functions in the case~$U$ is a cube.

\begin{lemma}[{Poincar\'e inequality for $H^1(U,\mu_\beta)$}]
\label{l.spectralgap.U}
Let $\cu_L$ be a cube of size $L$. There exists $C(d,\beta)<\infty$ such that:
\begin{enumerate}
\item[(i)] For every subset $U\subseteq \cu_L$ and $w \in H^1_0(U,\mu_\beta)$, 
\begin{equation}
\left\| w \right\|_{L^2(U,\mu_\beta)}
\leq 
CL \left\llbracket w \right\rrbracket_{H^1(U,\mu_\beta)}.
\end{equation}
\item[(ii)] For every $L\in\N$, every cube $\cu'\subseteq \cu_L$ and $w \in H^1(\cu',\mu_\beta)$, 
\begin{equation}
\left\| w - \left( w \right)_{\cu'} \right\|_{L^2(\cu',\mu_\beta)}
\leq 
CL \left\llbracket w \right\rrbracket_{H^1(\cu',\mu_\beta)}.
\end{equation}
\end{enumerate}
\end{lemma}
\begin{proof}
 In the case of~(i), since $w$ vanishes on $\partial U$, the (discrete) Poincar\'e inequality on $U$ yields
\begin{equation*}
\left\langle\sum_{x\in U}
|w(x,\cdot)|^2  \right\rangle_{\mu_\beta}
\leq 
CL^2 \sum_{e\in E(U)} \left\langle \left| \nabla w(e,\cdot) \right|^2 \right\rangle_{\mu_\beta}.
\end{equation*}
In the case of~(ii), we may suppose without loss of generality that~$\left( w \right)_{\cu'}=0$ and then apply the (discrete) Poincar\'e inequality for mean-zero functions on~$\cu'$ to obtain
\begin{equation*}
\left\langle\sum_{x\in \cu'} 
w(x,\cdot)^2  \right\rangle_{\mu_\beta}
\leq 
\left\langle C\diam(\cu')^2 
\sum_{e\in E(\cu')} \left( \nabla w(e,\cdot)\right)^2  \right\rangle_{\mu_\beta}
\leq 
CL^2 \sum_{e\in E(\cu')} \left\langle \left( \nabla w(e,\cdot) \right)^2 \right\rangle_{\mu_\beta     
 }.
\end{equation*}
\end{proof}

\section{Discrete differential forms} \label{secDefbaspropdifffo}

\subsection{Definitions and basic properties} \label{subsecDefbaspropdifffo} Given an integer $k \in \{ 0 , \ldots, d \}$, we denote by $\Lambda^k(\Zd)$ the set of oriented $k$-cells of the hypercubic lattice $\Zd$; 

For each $k$-cell $c_k$, we denote by $c_k^{-1}$ the same $k$-cell as $c_k$ with reverse orientation and by $\partial c_k$ the boundary this cell. A $k$-form $u$ is a mapping from $\Lambda^k(\cu)$ to $\R$ such that $u\left(c_k^{-1}\right) = - u\left(c_k\right).$

Given a $k$-form $u$, we define its exterior derivative $\di u$ according to the formula, for each oriented $(k+1)$-cell~$c_{k+1}$,
\begin{equation} \label{eq:TVd}
\di u \left( c_{k+1} \right) = \sum_{c_k \subset \partial c_{k+1}} u(c_k),
\end{equation}
where the orientation of the face $c_k$ is given by the orientation of the $(k+1)$-cell $c_{k+1}$; we set the convention $\di u = 0$ for any $d$-form $u$. We define the codifferential~$\di^*$ according to the formula, for each $(k-1)$-cell $c_{k-1}$ and each $k$-form $u : \Lambda^k \left( \cu \right) \to \R$,
\begin{equation} \label{eq:TV15557}
\di^* u \left(c_{k-1}\right) := \sum_{\partial c_k \ni c_{k-1}} u(c_k). 
\end{equation}
Clearly, $\di u $ is a $(k+1)$-form and $\di^* u$ is a $(k-1)$-form; we set $\di^* u = 0$ for any $0$-form $u$. One also verifies the properties, for each $k$-form $u : \Lambda^k(\cu) \to \R$, $\di \di u = 0$ and $\di^* \di^* u = 0.$ For arbitrary $k$-forms $u, v : \Lambda^k(\Zd) \to \R $ with finite support, we define the scalar product $(\cdot , \cdot)$ by the formula
\begin{equation} \label{eq:TV07319}
    (u , v) = \sum_{c_k \in \Lambda^k(\Zd)} u(c_k) v(c_k).
\end{equation}
We restrict the scalar product $(\cdot , \cdot)$ to forms which are only defined in a cube $\cu$; we denote the corresponding scalar product by $(\cdot , \cdot )_{\cu}$. It is defined by the formula, for each pair of forms $k$-forms $u, v : \Lambda^k(\cu) \to \R $,
\begin{equation*}
    (u , v) = \sum_{c_k \in \Lambda^k(\cu)} u(c_k) v(c_k).
\end{equation*}

The codiferential $\di^*$ is the formal adjoint of the exterior derivative $\di$ with respect to this scalar product: Given a $k$-form $u: \Lambda^k(\Zd) \to \R$ and a $(k+1)$-form $v: \Lambda^{k+1}(\Zd) \to \R$ with finite supports, one has the identity
\begin{equation} \label{eq:TV10687}
    \left( \di u , v  \right) =  \left( u , \di^* v  \right).
\end{equation}

For an integer $k \in \{ 0 , \ldots , d -1 \}$ and a cube $\cu \subseteq \Zd$, we define the tangential boundary of the cube $\partial_{k, \mathbf{t}} \cu$ to be the set of all the $k$-cells which are included in the boundary of the cube $\cu$. Given a $k$-form $u: \Lambda^k (\cu) \to \R$, we define its tangential trace $\mathbf{t}u$ to be the restriction of the form $u$ to the set $\partial_{k, \mathbf{t}} \cu$. One has the formula, for each $k$-form $u : \Lambda^k(\cu) \to \R$ such that $\mathbf{t} u = 0$ and each $(k+1)$-form $v: \Lambda^k(\cu) \to \R$,
\begin{equation*}
\left( \di u , v \right)_\cu = \left(  u , \di^* v \right)_\cu.
\end{equation*}

We will need the following lemma.

\begin{lemma}[Poincar\'e] \label{Poincforms}
Let $\cu \subseteq \Zd$ be a cube of the lattice $\Zd$ of sidelength $R$ and $k$ be an integer in the set $\{ 1 , \ldots , d-1 \}$. For each $k$-form $f: \Lambda^k (\cu) \to \R$ such that $\di f = 0$ and $\mathbf{t}f = 0$ on the tangential boundary $\partial_{k , \mathbf{t}} \cu$, there exists a $(k-1)$-form $u : \Lambda^{k-1}(\cu) \to \R$ such that $\mathbf{t}u = 0$ on the tangential boundary $\partial_{k , \mathbf{t}} \cu$ and $\di u = f$ in the cube $\cu$. Additionally, one can choose the form $u$ such that
\begin{equation*}
    \frac{1}{R} \left\| u \right\|_{L^2 \left( \cu \right)} + \left\| \nabla u \right\|_{L^2 \left( \cu \right)} \leq C \left\| f \right\|_{L^2 \left( \cu \right)}.
\end{equation*}
\end{lemma}

An important role is played by the set of integer-valued, compactly supported forms $q$ which satisfy $\di q = 0$ and have connected support. We denote by $\mathcal{Q}$ the set of these forms, i.e.,
\begin{equation} \label{def.defQchap2}
    \mathcal{Q} := \left\{ q : \Zd \to \Z \, : \, \left| \supp q \right| < \infty, ~\supp q ~\mbox{is connected}~\mbox{and}~ \di q = 0 \right\}.
\end{equation}
We may restrict our considerations to the charges of $\mathcal{Q}$ whose support is included in a cube $\cu \subseteq \Zd$; to this end, we introduce the notation
\begin{equation*}
    \mathcal{Q}_\cu := \left\{ q : \Zd \to \Z \, : \, \supp q \subseteq \cu, ~\supp q ~\mbox{is connected}~\mbox{and}~ \di q = 0  \right\}.
\end{equation*}
We will need to use the following version of Lemma~\ref{Poincforms} for the forms of the set $\mathcal{Q}$.

\begin{lemma}[Poincar\'e for integer valued forms] \label{eq:TVPoincaforint}
Let $k$ be an integer of the set $\{ 1 , \ldots , d-1 \}$ and $q$ be a $k$-form with values in $\Z$ such that $\di q = 0$, then there exists a $(k-1)$-form $n_q$ with values in $\Z$ such that $q = \di n_q$. Moreover, $n_q$ can be chosen such that $\supp n_q$ is contained in the smallest hypercube containing the support of $q$ and such that
\begin{equation*}
\left\| n_q \right\|_{L^\infty} \leq C \left\| q \right\|_1. 
\end{equation*}
\end{lemma}
As it is useful in the article, we record a series of inequalities satisfied by the charges $q \in \mathcal{Q}$,
\begin{equation} \label{ineqcharge}
    \left\{ \begin{aligned}
    \left\| q \right\|_{L^\infty} \leq \left\| q \right\|_1,\\
    \diam q \leq \left| \supp q \right| \leq  \left\| q \right\|_1, \\
    \left\| n_q \right\|_{L^\infty} \leq C \left\| q \right\|_1, \\
    \diam n_q \leq C \left\| q \right\|_1, \\
    \left| \supp n_q \right| \leq C \left\| q \right\|_1^d, \\
    \left\| n_q \right\|_{L^1} \leq \left| \supp n_q \right| \left\| n_q \right\|_{L^\infty} \leq  C \left\| q \right\|_1^{d+1}, \\
    \left\| n_q \right\|_{L^2} \leq \left\| n_q \right\|_{L^1}^{\frac12} \left\| n_q \right\|_{L^\infty}^{\frac12} \leq  C \left\| q \right\|_1^{\frac d2 +1}.
    \end{aligned}
    \right.
\end{equation}
The proofs of these results use~\eqref{eq:TVPoincaforint} and the fact that the charges are valued in the set $\Z$; they are left to the reader. Given a point $(x , y)\in \Zd \times \Zd$, we denote by $\mathcal{Q}_x$ and $\mathcal{Q}_{x,y}$ the set of charges $q \in \mathcal{Q}$ such that the point $x$ and the points $x , y$ belong to the support of $n_q$ respectively, i.e.,
\begin{equation*}
    \mathcal{Q}_x := \left\{ q \in \mathcal{Q} \,:\, x \in \supp n_q \right\} \hspace{5mm} \mbox{and} \hspace{5mm} \mathcal{Q}_{x,y} := \left\{ q \in \mathcal{Q} \,:\, x \in \supp n_q ~ \mbox{and}~ y \in \supp n_q \right\}.
\end{equation*}
Similarly we also define
\begin{equation*}
    \mathcal{Q}_{\cu,x} := \left\{ q \in \mathcal{Q_\cu} \,:\, x \in \supp n_q \right\} \hspace{5mm} \mbox{and} \hspace{5mm} \mathcal{Q}_{\cu,x,y} := \left\{ q \in \mathcal{Q_\cu} \,:\, x \in \supp n_q ~ \mbox{and}~ y \in \supp n_q \right\}.
\end{equation*}
We also record two inequalities involving the sum of charges: for each pair of points $(x , y) \in \Zd$, each integer $k \in \N$ and each constants $c > 0$ and $\beta \geq 1$
\begin{equation} \label{eq:sumcharges}
    \left\{ \begin{aligned}
    \sum_{q \in \mathcal{Q}_x} \left\| q \right\|_1^k e^{-c \sqrt{\beta} \left\| q \right\|_1} \leq C e^{-c_0 \sqrt{\beta}}  \\
    \sum_{q \in \mathcal{Q}_{x , y}} \left\| q \right\|_1^k e^{-c \sqrt{\beta} \left\| q \right\|_1} \leq C e^{-c_0 \sqrt{\beta} |x - y|} 
    \end{aligned}
    \right.
\end{equation}
where the constants $C$, $c_0$ depend on $k$, $c$ and the dimension $d$. The proofs of these inequalities rely on the assumption that the charges are integer-valued; they are fairly elementary and left to the reader.

\subsection{Differential forms as vector-valued functions} \label{sec:diffform=vectfct}

Given a subset $I = (i_1 , \ldots , i_k) \subseteq \{ 1 , \ldots , d \}$ of cardinality $k$. We denote by $\Lambda^k_I(\Zd)$ the set of oriented $k$-cells of the hypercubic lattice $\Zd$ which are parallel to the vectors $(e_{i_1} , \ldots , e_{i_k})$. This set can be characterized as follows: if we let $c_{I}$ be the $k$-cell defined by the formula
\begin{equation*}
    c_{I} := \left\{\sum_{l = 1}^k \lambda_l e_{i_l} \in \Rd \, : \, 0 \leq \lambda_1 , \ldots ,  \lambda_k \leq 1 \right\},
\end{equation*}
then we have
\begin{equation} \label{eq:TV14077}
    \Lambda^k_I(\Zd) =  \left\{  x + c_{I} \, : \, x \in \Zd \right\}.
\end{equation}
The identity~\eqref{eq:TV14077} allows to identify the vector space of $k$-forms to the vector space of functions defined on $\Zd$ and valued in $\R^{\binom dk}$ according the procedure described below. Note that there are $\binom dk$ subsets of $\{ 1 , \ldots , d \}$ of cardinality $k$ and consider an arbitrary enumeration $I_1 , \ldots, I_{\binom dk}$ of these sets. To each $k$-form $\hat u : \Lambda^k (\Zd) \to \R$, we can associate a vector-valued function $ u : \Zd \to \R^{\binom dk}$ defined by the formula, for each point $x \in \Zd$,
\begin{equation} \label{eq:TV15027}
     u (x) = \left( \hat u \left(x + c_{I_1} \right) , \ldots, \hat u \left(x + c_{I_{\binom dk}} \right) \right).
\end{equation}
This identification is enforced in most of the article; in fact, except in Section~\ref{Chap3sec1} of Chapter~\ref{chap:chap3}, we always work with vector-valued functions instead of differential forms. We use the identification~\eqref{eq:TV15027} to extend the formalism described in Section~\ref{SecNotandprelim} to differential forms; we may for instance refer to the gradient of a form, or the Laplacian of a form etc. Reciprocally, we extend the formalism described in Section~\ref{subsecDefbaspropdifffo} to vector-valued functions; given a function $u : \Zd \to \R^{\binom dk}$, we may refer to the exterior derivative, the codifferential and the tangential trace of the function $u$, which we still denote by $\di u$, $\di^* u$ and $\mathbf{t}u$ respectively. We note that the two definitions of the scalar products~\eqref{eq:TV07309} for vector valued functions and~\eqref{eq:TV07319} for differential forms coincide through the identification~\eqref{eq:TV15027}.

\smallskip

From the definition of the exterior derivative $\di$ and the codifferential $\di^*$ given in~\eqref{eq:TVd} and~\eqref{eq:TV15557} and the identification~\eqref{eq:TV15027}, one sees that the differential operators $\di$ and $\di^*$ are linear functionals of the gradient $\nabla$: for each integer $k \in \{ 1 , \ldots, d\}$, there exist linear maps $L_{k , \di} : \R^{d \times \binom dk} \to \R^{ \binom d{k+1}}$ and $L_{k , \di^*} : \R^{d \times \binom dk} \to \R^{\binom d{k-1}}$ such that, for each function $u :  \Zd \to \R^{\binom dk}$ and each point $x \in \Zd$,
\begin{equation} \label{eq:formLkdde}
    \di u(x) = L_{k , \di} \left( \nabla u(x) \right) \hspace{5mm} \mbox{and} \hspace{5mm} \di^* u(x) = L_{k , \di^*} \left( \nabla u(x) \right).
\end{equation}
Using that linear maps on finite dimensional vector spaces are continuous, we obtain the estimates, for each point $x \in \Zd$,
\begin{equation*}
     \left| \di u(x) \right| \leq C \left| \nabla u(x) \right| \hspace{5mm} \mbox{and} \hspace{5mm} \left| \di^* u(x) \right| \leq C \left| \nabla u(x) \right|,
\end{equation*}
for some constant $C$ depending only on the dimension $d$.

\smallskip

This article frequently deals with functions defined on the space $\Zd \times \Omega \times \Zd$ (resp. $\Z^{2d} \times \Omega \times \Z^{2d}$) and valued in $\R^{\binom d2 \times \binom d2}$ (resp. $\R^{\binom d2^4}$) since these maps correspond to the fundamental solutions of the Hellfer-Sj\"{o}strand operator (resp. differentiated Hellfer-Sj\"{o}strand operator) associated to the dual Villain model introduced in Section \ref{s.HSrep} of Chapter~\ref{chap:chap3}. Given a map $F : \Zd \times \Zd \times \Omega \to \R^{\binom d2 \times \binom d2}$, we denote by $\di_x F : \Zd \times \Omega \to \R^{\binom d2 \times  \binom d{3}} $, $\di_y F : \Zd \times \Omega \to \R^{\binom d{3} \times \binom d2} $, $\di_x^* F : \Zd \times \Omega \to \R^{d \times \binom d2} $ and $\di_y^* F : \Zd \times \Omega \to \R^{\binom d2 \times d} $ the exterior derivative with respect to the first, second variable and the codifferential with respect to the first and second variable respectively. They are defined by the formulas, for each triplet $(x , y , \phi) \in \Zd \times \Zd \times \Omega$ and each integer $k \in \left\{ 1 , \ldots , \binom d2 \right\}$,
\begin{equation*}
    \left( \di_x F(x , y , \phi) \right)_{\cdot k} = L_{2,\di} \left( \nabla_x F_{\cdot k}(x , y , \phi) \right), \hspace{5mm} \left( \di_y F(x , y , \phi) \right)_{ k \cdot} = L_{2,\di} \left( \nabla_y F_{ k \cdot}(x , y , \phi) \right)
\end{equation*}
and
\begin{equation*}
    \left( \di_x^* F(x , y , \phi) \right)_{\cdot k} = L_{2,\di^*} \left( \nabla_x F_{\cdot k}(x , y , \phi) \right), \hspace{5mm} \left( \di_y^* F(x , y , \phi) \right)_{ k \cdot} = L_{2,\di^*} \left( \nabla_y F_{ k \cdot}(x , y , \phi) \right).
\end{equation*}
Similarly, given a function $F : \Z^{2d} \times \Omega \times \Z^{2d} \to \R^{\binom d2^4}$, we define, for each $(x , y , \phi, x_1 , y_1) \in \Zd \times \Zd \times \Omega \times \Zd \times \Zd$, each field $\phi \in \Omega$ and each triplet of integers $i,j,k \in \left\{ 1 , \ldots , \binom d2 \right\}$
\begin{equation*}
    \left\{ \begin{aligned}
        \left( \di_x F(x , y , \phi , x_1 , y_1) \right)_{\cdot ijk} &= L_{2,\di} \left( \nabla_x F_{ \cdot i jk}(x , y , \phi , x_1 , y_1) \right), \\
        \left( \di_y F(x , y , \phi , x_1 , y_1) \right)_{ i \cdot jk} &= L_{2,\di} \left( \nabla_x F_{i \cdot j k}(x , y , \phi , x_1 , y_1) \right), \\
        \left( \di_{x_1} F(x , y , \phi , x_1 , y_1) \right)_{ ij \cdot k} &= L_{2,\di} \left( \nabla_x F_{ij \cdot k}(x , y , \phi , x_1 , y_1) \right), \\
        \left( \di_{y_1} F(x , y , \phi , x_1 , y_1) \right)_{ ijk \cdot} &= L_{2,\di} \left( \nabla_x F_{ijk \cdot}(x , y , \phi , x_1 , y_1) \right),
    \end{aligned} \right.
\end{equation*}
and similarly
\begin{equation*}
    \left\{ \begin{aligned}
        \left( \di_x^* F(x , y , \phi , x_1 , y_1) \right)_{\cdot ijk} &= L_{2,\di^*} \left( \nabla_x F_{ \cdot i jk}(x , y , \phi , x_1 , y_1) \right), \\
        \left( \di_y^* F(x , y , \phi , x_1 , y_1) \right)_{ i \cdot jk} &= L_{2,\di^*} \left( \nabla_x F_{i \cdot j k}(x , y , \phi , x_1 , y_1) \right), \\
        \left( \di_{x_1}^* F(x , y , \phi , x_1 , y_1) \right)_{ ij \cdot k} &= L_{2,\di^*} \left( \nabla_x F_{ij \cdot k}(x , y , \phi , x_1 , y_1) \right), \\
        \left( \di_{y_1}^* F(x , y , \phi , x_1 , y_1) \right)_{ ijk \cdot} &= L_{2,\di^*} \left( \nabla_x F_{ijk \cdot}(x , y , \phi , x_1 , y_1) \right).
    \end{aligned} \right.
\end{equation*}
We extend these definitions so that we can consider \emph{mixed derivatives}; for instance, we may use the notation $\di_y^* \di_x^* F$ (or any other combination of exterior derivatives and codifferentials). It is clear that as long as the derivatives involve different variables, they commute: we have for instance $\di_y^* \di_x^* F = \di_x^* \di_y^*  F$.
\smallskip

We record the following identity which relates the Laplacian $\Delta$ to the exterior derivative $\di$ and the codifferential $\di^*$,
\begin{equation} \label{eq:TV11272901}
-\Delta = \di \di^* + \di^* \di.
\end{equation}
Using the identity $\di \circ \di = 0$, one obtains that the Laplacian commutes with the exterior derivative: we have
\begin{equation*}
    \di \Delta = -\di \left( \di \di^* + \di^* \di \right) = -\di \di^* \di = -\left( \di \di^* \di + \di^* \di \right) \di = \Delta \di.
\end{equation*}
Similarly, using this time the identity $\di^* \circ \di^* = 0$, we obtain that the Laplacian commutes with the codifferential: we have
\begin{equation*}
    \di^* \Delta = -\di^* \left( \di \di^* + \di^* \di \right) = -\di^* \di \di^* = -\left( \di \di^* \di + \di^* \di \right) \di^* = \Delta \di^*.
\end{equation*}

We complete this section by recording the Gaffney-Friedrichs inequality which provides an upper bound on the $L^2$-norm of the gradient of a form in terms of the $L^2$-norm of its exterior derivative and the codifferential assuming that the tangential trace of the form vanishes. 

\begin{proposition}[Gaffney-Friedrichs inequality for cubes] \label{GFCD} 
Let $\cu$ be a cube of $\Zd$. Then there exists a constant $C := C(d) < \infty$ such that for each $k$-form $u: \Lambda^k(\cu) \to \R$ with vanishing tangential trace, we have
\begin{equation*}
\| \nabla u \|_{\underline{L}^2(\cu)} \leq C \left( \| \di u \|_{\underline{L}^2 (\cu)} + \| \di^* u \|_{L^2 (\cu)} \right).
\end{equation*}
\end{proposition}
The proof of the continuous version of this inequality can be found in~\cite{G51, F55} or in the monograph~\cite[Proposition 2.2.3]{S06}. We complete this section by proving the solvability of a boundary value problem involving discrete differential forms used in Section~\ref{Chap3sec1} of Chapter~\ref{chap:chap3}.

\begin{proposition} \label{prop2.4chap2}
For any integer $k \in \{ 1 , \ldots , d-1\}$ and any cube $\cu \in \Zd$ and any $k$-form $q := (q_1, \ldots , q_{\binom dk}): \cu \to \R^{\binom dk}$ such that 
$\di q = 0$ in the cube $\cu$ and $\mathbf{t}q = 0$ on the boundary $\partial \cu$, there exists a unique solution to the boundary value problem
\begin{equation} \label{eq:VER192002}
\left\{ \begin{aligned}
\di \di^* w &= q ~\mbox{in}~ \cu, \\
\di w& = 0 ~\mbox{in}~ \cu, \\
\mathbf{t} w &= 0 ~\mbox{on}~ \partial \cu, \\
\mathbf{t} \di^* w &= 0 ~\mbox{on}~ \partial \cu.
\end{aligned} \right.
\end{equation}
If we denote by $w_1 , \ldots , w_{\binom dk}$ the coordinates of the map $w$, then they solve the following boundary value problem: for each $i \in \left\{ 1 , \ldots, \binom dk  \right\}$, if we denote by $\partial_{I_i} \cu$ the subset of faces of the boundary $\partial \cu$ which are parallel to the cell $c_{I_i}$, then we have 
\begin{equation} \label{eq:TV10042901}
    \left\{ \begin{aligned}
-\Delta w_i &= q_i ~\mbox{in}~ \cu, \\
w_i & = 0 ~\mbox{in}~ \partial_{I_i} \cu, \\
\nabla w_i \cdot \mathbf{n} &= 0 ~\mbox{on}~\partial \cu \setminus \partial_{I_i} \cu.
\end{aligned} \right.
\end{equation}
\end{proposition}

\begin{remark}
The boundary condition~\eqref{eq:TV10042901} is a combination of the Dirichlet and Neumann boundary conditions: given an integer $i \in \{ 1 , \ldots, \binom dk \}$, we assign Dirichlet boundary condition on the faces which are parallel to the cell $c_{I_{i}}$ and Neumann boundary condition on the faces which are orthogonal to the cell $c_{I_{i}}$.
\end{remark}

\begin{proof}
 The boundary value problem~\eqref{eq:VER192002} admits a variational formulation which we can use to prove existence and uniqueness of solutions. We first define the set of $k$-forms
 \begin{equation*}
     C^{k}_0 (\cu) := \left\{ u : \cu \to \R^{\binom dk} \, : \, \di u = 0 ~\mbox{in} \, \cu \, \mbox{and}\,  \mathbf{t} u = 0 \,\mbox{on}\, \partial \cu \right\}.
 \end{equation*}
 We then define the energy functional $J_q : C^{k}_0 (\cu) \to \R $ according to the formula
 \begin{equation*}
     J_q (u) := \frac 12\left\| \di^*u \right\|_{L^2 \left( \cu \right)} - \left(q , u \right)_\cu.
 \end{equation*}
 To prove the solvability of the problem~\eqref{eq:VER192002}, we prove that there exists unique minimizer to the variational problem
 \begin{equation*}
     \inf_{u \in C^{k}_0 (\cu)} J(u).
 \end{equation*}
We first use that, by Lemma~\ref{Poincforms}, there exists a $(k-1)$-form $n_q : \cu \to \R^{\binom{d}{k-1}}$ such that $\mathbf{t} n_q = 0$ on $\partial \cu$ and $\di n_q = q$ in the cube $\cu$. We then perform an integration by parts to write
 \begin{equation*}
     J_q (u) = \frac 12\left\| \di^* u \right\|_{L^2 \left( \cu \right)} - \left(n_q , \di^* u \right)_\cu.
 \end{equation*}
 The technique then follows the standard strategy of the calculus of variations. The energy functional $J_q$ is bounded from below and we consider a minimizing sequence $\left( w_n\right)_{n \in \N}$. It is clear that the norms $\left\| \di^* w_n \right\|_{L^2 \left( \cu \right)}$ are uniformly bounded in $n \in \N$. Using that $\di w_n = 0$ and the Gaffney-Friedrich inequality stated in Proposition~\ref{GFCD}, we obtain that the norms $\left\| \nabla w_n \right\|_{L^2 \left( \cu \right)}$ and $\left\| w_n \right\|_{L^2 \left( \cu \right)}$ are uniformly bounded in $n$. We can thus extract a subsequence which converges in the discrete space $L^2 \left( \cu \right)$ and verify that the limit is solution to the problem~\eqref{eq:VER192002}. The uniqueness is a consequence of the uniform convexity of the functional $J_q$.
 
 To prove~\eqref{eq:TV10042901}, note that the condition $\di w=0$ and the identity~\eqref{eq:TV11272901} imply that $-\Delta w = q$ in the cube $\cu$. Using the definition of the Laplacian for vector-valued function (stated in~\eqref{eq:TV11322901}), we have that for each integer $i \in \{ 1 , \ldots, \binom dk \}$, $ -\Delta w_i = q_i$ in the cube $\cu$. The boundary condition $\mathbf{t}w = 0$ implies that $w_i$ is equal to $0$ on each face which is parallel to the cell $c_{I_i}$; the condition $\mathbf{t} \di^* w = 0$ implies that the function $w_i$ satisfies a Neumann boundary condition on the faces of the boundary $\partial \cu$ which are orthogonal to the cell $c_{I_i}$.
\end{proof}

\section{Convention for constants and exponents} Throughout this article, the symbols $c$ and $C$ denote positive constants which may vary from line to line. These constants may depend only on the dimension $d$ and the inverse temperature $\beta$. We use the symbols $\alpha, \, \beta, \, \gamma, \, \delta$ to denote positive exponents which depend only on the dimension $d$. Usually, we use the letter $C$ for large constants (whose value is expected to belong to $[1, \infty)$) and $c$ for small constants (whose value is expected to be in $(0,1]$). The values of the exponents $\alpha, \, \beta, \, \gamma, \, \delta$ are always expected to be small. When the constants and exponents depend on other parameters, we write it explicitly and use the notation $C := C(d , \beta , t)$ to mean that the constant $C$ depends on the parameters $d , \beta$ and $t$.

When the constants depend on the charges $q \in \mathcal{Q}$ (see~\eqref{def.defQchap2}), we frequently need to keep track of their dependence in this parameter; more specifically we need that the growth of the constant $C$ is at most algebraic in the parameter $\left\| q \right\|_1$. We usually denote by $C_q$ a constant which depends on the parameter $d , \beta$ and $q$ and which satisfies the growth condition $C_q \leq C \left\| q \right\|_1^k$, for some $C := C(d , \beta) < \infty$ and $k : = k(d) <\infty$. We allow the values of $C$ and $k$ to vary from line to line and we may write
\begin{equation*}
    C_q + C_q \leq C_q \hspace{5mm} \mbox{or} \hspace{5mm} C_q C_q \leq C_q.
\end{equation*}

We usually do not keep track of the dependence of the constants in the inverse temperature $\beta$ (even though we believe it should be possible with our techniques) except in Chapters~\ref{section:section4} and~\ref{section5}. In these two chapters, we assume that the constants depend only on the dimension $d$ and make it explicit if they depend on the inverse temperature $\beta$.

\chapter{Duality and Helffer-Sj{\"o}strand representation } \label{chap:chap3}

\section{From Villain model to solid on solid model} \label{Chap3sec1}
In this section we recall the duality relation between the Villain model in $\Zd$ and  a statistical mechanical model of lattice Coulomb gas, with integer valued and locally neutral charges (which can also be viewed as a solid-on-solid model) defined on $\Lambda^2(\Zd)$, as observed in \cite{FS4d}. One may then perform a Fourier transform with respect to the charge variable, and obtain a classical random field representation of the Coulomb gas, known as the sine-Gordon representation. When the temperature is low enough, we may apply a one-step renormalization argument, following the presentation of Bauerschmidt \cite{Bau} (see also \cite{FS4d}), to reduce the effective activity of the charges, thus obtain an effective, real valued random interface model on $2$-forms with a convex action.

Recall that the partition function for the Villain model in a cube $\cu\subset \Zd$ with zero boundary condition is given by
\begin{equation*}
Z_{\cu,0} := \int \prod_{e \subseteq E(\cu)} \sum_{m \in \Z} \exp \left( -\frac{\beta}{2} \left( \nabla \theta(e) - 2 \pi m \right)^2 \right) \prod_{x \in \partial \cu }\delta_{0} \left( \theta (x) \right) \prod_{x \in \cu^\circ} \indc_{[-\pi,\pi)} (\theta(x)) \, \di \theta(x).
\end{equation*}
Since we need to use the formalism of discrete differential forms later in this chapter, we note that the function $\theta : \cu \mapsto \R$ can be seen as a $0$-form, in that case the discrete gradient $\nabla \theta$ can be seen as a $1$-form and is equal to the exterior derivative $\di \theta$. We may thus rewrite
\begin{equation*}
Z_{\cu,0} := \int \prod_{e \subseteq E(\cu)} \sum_{m \in \Z} \exp \left( -\frac{\beta}{2} \left( \di \theta(e) - 2 \pi m \right)^2 \right) \prod_{x \in \partial \cu }\delta_{0} \left( \theta (x) \right) \prod_{x \in \cu^\circ} \indc_{[-\pi,\pi)} (\theta(x)) \, \di \theta(x).
\end{equation*}
Permuting the sum with the product and the integral, we obtain
\begin{equation} \label{eq:TV17030701}
Z_{\cu,0}  = \sum_{\mathbf{m} \in \Z^{E(\cu)}_{\mathbf{t}  = 0}} \int  \prod_{e \subseteq E(\cu)} \exp \left(  -\frac{\beta}{2} \left( \di \theta(e) - 2 \pi \mathbf{m}(e) \right)^2 \right)  \prod_{x \in \partial \cu }\delta_{0} \left( \theta (x) \right)  \prod_{x \in \cu^\circ} \indc_{[-\pi,\pi)} (\theta(x)) \, \di \theta(x),
\end{equation}
where we have used the notation
\begin{equation*}
    \Z^{E(\cu)}_{\mathbf{t}  = 0} := \left\{ \mathbf{m} : E(\cu) \mapsto \Z \, : \, \mathbf{t}\mathbf{m} = 0 ~\mbox{on}~ \partial \cu \right\}.
\end{equation*}
Observe that we may split the sum according to
\begin{equation} \label{eq:TV17040701}
\sum_{\mathbf{m} \in \Z^{E(\cu)}_{\mathbf{t}  = 0}} = \sum_{q \in \Z^{F(\cu)}_{\mathbf{t}  = 0}, \di q = 0} \sum_{\mathbf{m} \in  \Z^{E(\cu)}_{\mathbf{t}  = 0},  \di \mathbf{m} = q},
\end{equation}
where we have set
\begin{equation*}
    \Z^{F(\cu)}_{\mathbf{t}  = 0} := \left\{ q : F(\cu) \mapsto \Z \, : \, \mathbf{t}q = 0 ~\mbox{on}~ \partial \cu \right\}.
\end{equation*}
A combination of~\eqref{eq:TV17030701} and~\eqref{eq:TV17040701} yields
\begin{equation*}
Z_{\cu,0}  = \sum_{q \in \Z^{F(\cu)}_{\mathbf{t}  = 0}, \di q = 0} \sum_{\mathbf{m} \in  \Z^{E(\cu)}_{\mathbf{t}  = 0},  \di \mathbf{m} = q}  \int  \prod_{e \subseteq E(\cu)} \exp \left( -\frac{\beta}{2} \left( \di \theta(e) - 2 \pi \mathbf{m}(e) \right)^2 \right)  \prod_{x \in \partial \cu }\delta_{0} \left( \theta (x) \right)  \prod_{x \in  \cu^\circ} \indc_{[-\pi,\pi)} (\theta(x)) \, \di \theta(x).
\end{equation*}
Here $q:F(\cu) \to \Z$ is the ``vortex charge" on each plaquette of $\cu$, which arises, informally, from
\begin{equation*}
    \oint_F \, \di \theta(e) = 2\pi q(F).
\end{equation*}
For each $q \in \Z^{F(\cu)}_{\mathbf{t}  = 0}$, satisfying $\di q = 0$, we denote by $\mathbf{n}_q$ an element of $\Z^{E(\cu)}_{\mathbf{t}  = 0}$ such that $\di \mathbf{n}_q = q$, chosen arbitrarily among all the possible candidates, note that the set of candidates is not empty by Proposition~\ref{eq:TVPoincaforint} of Chapter~\ref{Chap:chap2}. Using that each $1$-form $\mathbf{m} \in  \Z^{E(\cu)}_{\mathbf{t}  = 0}$ satisfying $\di \mathbf{m} = 0$ can be uniquely written $\di w$, for some $w: \cu \mapsto \Z $ satisfying $w = 0$ on the boundary $\partial \cu$, one can rewrite the previous display according to 
\begin{equation*}
Z_{\cu,0}  = \sum_{q \in \Z^{F(\cu)}_{\mathbf{t}  = 0}, \di q = 0} \sum_{w \in  \Z^{\cu}_0}  \int  \prod_{e \subseteq E(\cu)} \exp \left(  -\frac{\beta}{2} \left( \di \theta(e) - 2 \pi \left( \mathbf{n}_q + \di w \right)(e) \right)^2 \right)  \prod_{x \in \partial \cu }\delta_{0} \left( \theta (x) \right)  \prod_{x \in  \cu^\circ} \indc_{[-\pi,\pi)} (\theta(x)) d \theta(x),
\end{equation*}
where we have set
\begin{equation*}
     \Z^{\cu}_0 := \left\{ w: \cu \mapsto \Z  \, : \, w=0 ~\mbox{on}~ \partial \cu \right\}.
\end{equation*}
Using the change of variable $\phi := \theta + 2\pi w$, and summing over all the maps $w \in  \Z^{\cu}_0$, one obtains
\begin{equation*}
Z_{\cu,0}  = \sum_{q \in \Z^{F(\cu)}_{\mathbf{t}  = 0}, \di q = 0}  \int_{\R^\cu} \prod_{e \subseteq E(\cu)} \exp \left(  -\frac{\beta}{2} \left( \di \phi(e) - 2 \pi \mathbf{n}_q(e) \right)^2 \right)  \prod_{x \in \partial \cu }\delta_{0} \left( \phi (x) \right)  \prod_{x \in  \cu^\circ} d \phi(x).
\end{equation*}
Using Proposition~\ref{prop2.4chap2} of Chapter~\ref{Chap:chap2}, we denote by $\left( \di \di^* \right)^{-1} q$ the (unique) solution of the problem
\begin{equation*}
\left\{ \begin{aligned}
\di \di^* w &= q ~\mbox{in}~ \cu, \\
\di w& = 0 ~\mbox{in}~ \cu, \\
\mathbf{t} w &= 0 ~\mbox{on}~ \partial \cu, \\
\mathbf{t} \di^* w &= 0 ~\mbox{on}~ \partial \cu.
\end{aligned} \right.
\end{equation*}
It is then clear that the charge $\mathbf{n}_q - \di^* \left( \di \di^* \right)^{-1} q$ satisfies 
$\di \left( \mathbf{n}_q - \di^* \left( \di \di^* \right)^{-1} q \right) = q - q = 0$ in $\cu$ and $\mathbf{t} \left( \mathbf{n}_q - \di^* \left( \di \di^* \right)^{-1} q \right) = 0$ on $\partial \cu$.
Applying Proposition~\ref{Poincforms} of Chapter~\ref{Chap:chap2}, one can write 
\begin{equation} \label{eq:TV23200701}
\mathbf{n}_q = \di \phi_{\mathbf{n}_q} + \di^* \left( \di \di^* \right)^{-1} q,
\end{equation}
for some $\phi_{\mathbf{n}_q} : \cu \mapsto \R$ with $\phi_{\mathbf{n}_q} = 0$ on $\partial \cu$. As a remark, note that the operator $\left( \di \di^* \right)^{-1}$ depends on the cube $\cu$, since this cube is fixed through the proof, we omit the dependence in the cube $\cu$ in the notation. Then using the translation invariance of the Lebesgue measure, we obtain
\begin{equation*}
Z_{\cu,0}  = \sum_{q \in \Z^{F(\cu)}_{\mathbf{t}  = 0}, \di q = 0}  \int_{\R^\cu} \prod_{e \subseteq E(\cu)} \exp \left(  -\frac{\beta}{2} \left( \di \phi(e) - 2 \pi \di^* \left( \di \di^* \right)^{-1} q(e) \right)^2 \right)  \prod_{x \in \partial \cu }\delta_{0} \left( \phi (x) \right)  \prod_{x \in  \cu} \, \di \phi(x).
\end{equation*}
The previous identity can be simplified
\begin{align} \label{eq:TV15302801}
    Z_{\cu,0} & = Z_{GFF} \times Z(0) \\
    & :=  \int_{\R^\cu} \exp \left( -\frac{\beta}{2}  \left( \di \phi , \di \phi  \right) \right)  \prod_{x \in \partial \cu }\delta_{0} \left( \phi (x) \right)  \prod_{x \in  \cu} \di \phi(x) \times \sum_{q \in \Z^{F(\cu)}_{\mathbf{t}  = 0}, \di q = 0} \exp \left( - 2 \pi^2 \beta \left( q,  \left( \di \di^* \right)^{-1} q \right) \right). \notag
\end{align}
Using the identity $\di \phi = \nabla \phi$ (valid for $0$-forms), we see that the first term in the left hand side of~\eqref{eq:TV15302801} is the partition function of the discrete Gaussian free field in the cube $\cu$ with Dirichlet boundary condition. In other words, the Villain partition function factorizes into the partition function of a Gaussian free field in $F(\cu)$, and the vortex charges that form a (neutral) Coulomb gas.

One can use the same argument to study the two-point functions
\begin{equation*}
\left\langle e^{i \left( \theta(x) - \theta(0) \right)} \right\rangle_{\mu^V_{\beta,\cu,0}},
\end{equation*}
 For any point $x \in \cu$, consider the string observable $h_{0,x} : E(\cu) \mapsto \Z$ be such that $\di^* h_{0,x} = \indc_x - \indc_0$ and $h_{x} : E(\cu) \mapsto \Z$ such that $\di^* h_{x} = \indc_x$, with the same computation, we obtain 
\begin{equation}
\label{e.dual2pt}
\left\langle e^{i \left( \theta(x) - \theta(0) \right)} \right\rangle_{\mu^V_{\beta,\cu,0}} = \left\langle e^{i \left( \phi(x) - \phi(0) \right)} \right\rangle_{GFF} \left\langle e^{-2i\pi  \left( q , ( \di \di^*)^{-1} \di h_{0,x} \right)  }\right\rangle_{\mu_C(\beta)}
\end{equation}
and
\begin{equation*}
\left\langle e^{i \theta(x) } \right\rangle_{\mu^V_{\beta,\cu,0}} = \left\langle e^{i \phi(x)} \right\rangle_{GFF} \left\langle e^{-2i\pi  \left( q , ( \di \di^*)^{-1} \di h_{x} \right)  }\right\rangle_{\mu_C(\beta)}.
\end{equation*}
Here 
\begin{equation*}
    \left\langle e^{i \left( \phi(x) - \phi(0) \right)} \right\rangle_{GFF}
    := Z_{GFF}^{-1} \times \int_{\R^{\cu}} e^{i \left( \phi(x) - \phi(0) \right)} \exp \left( -\frac{\beta}{2}  \left( \nabla \phi , \nabla \phi  \right) \right)  \prod_{x \in \partial \cu }\delta_{0} \left( \phi (x) \right)  \prod_{x \in  \cu^\circ} d \phi(x) 
\end{equation*}
and 
\begin{equation*}
    \left\langle e^{-2i\pi  \left( q , ( \di \di^*)^{-1} \di h_{0,x} \right)  }\right\rangle_{\mu_C(\beta)}
    = Z(0)^{-1} \times \sum_{q \in \Z^{F(\cu)}_{\mathbf{t}  = 0}, \di q = 0} e^{- 2 \pi^2 \beta \left( q,  \left( \di \di^* \right)^{-1} q \right) } e^{-2i\pi  \left( q , ( \di \di^*)^{-1} \di h_{0,x} \right)} .
\end{equation*}
Following \cite{Bau}, we define the notations $\sigma_x := ( \di \di^*)^{-1} \di h_{x}$ and $\sigma_{0x} := ( \di \di^*)^{-1} \di h_{0,x}$. Note that for each $q \in \Z^{E(\cu)}_{\mathbf{t}=0}$ satisfying $\di q = 0$, one has
\begin{equation} \label{eq:TV23100701}
\left( q , \sigma_x - \sigma_0 \right)  = \left( q , \sigma_{0x} \right) ~ \mbox{mod}~\Z.
\end{equation}
To justify the identity~\eqref{eq:TV23100701}, we note that, with the same argument as in~\eqref{eq:TV23200701}, we may write
\begin{equation*}
    h_x - h_0 - h_{0x} = \di \phi + \di^* \sigma_x - \di^* \sigma_0 - \di^* \sigma_{0x},
\end{equation*}
for some field $\phi : \cu \mapsto \R$ satisfying $\phi = 0$ on the boundary $\partial \cu$. Taking the scalar product with the $1$-form $\di \phi$ and performing integrations by parts, we obtain that
\begin{equation*}
    \left( \di \phi , h_x - h_0 - h_{0x}\right) =  \left( \phi , \di^* h_x - \di^* h_0 - \di^* h_{0x}\right) = \left( \phi , \indc_x  - \indc_0 - \left( \indc_x - \indc_0 \right) \right) = 0
\end{equation*}
and
\begin{equation*}
    \left( \di \phi + \di^* \sigma_x - \di^* \sigma_0 - \di^* \sigma_{0x}, \di \phi \right) = \left( \di \phi, \di \phi \right) + \left(\sigma_x -  \sigma_0 - \sigma_{0x}, \di \di \phi  \right) = \left( \di \phi, \di \phi \right).
\end{equation*}
A combination of the two previous displays implies $\di \phi = 0$ and thus  $h_x - h_0 - h_{0x} =  \di^* \sigma_x - \di^* \sigma_0 - \di^* \sigma_{0x}$. We then use that $q = \di n_q$ for some $n_q \in \Z^{E(\cu)}_{\mathbf{t} = 0}$ to write
\begin{equation*}
    \left( q,  \sigma_x - \sigma_0 -  \sigma_{0x} \right) = \left( n_q,  \di^* \sigma_x - \di^* \sigma_0 - \di^* \sigma_{0x} \right) = \left( n_q,  h_x - h_0 - h_{0x} \right) \in \Z.
\end{equation*}
This is~\eqref{eq:TV23100701}. A consequence of~\eqref{eq:TV23100701} is that for each $q \in \Z^{E(\cu)}_{\mathbf{t}=0}$ satisfying $\di q = 0$, 
\begin{equation*}
e^{-2i\pi \left( q , \sigma_x - \sigma_0 \right) } = e^{-2i\pi  \left( q , \sigma_{0x} \right)}.
\end{equation*}
 For later use, we note that, by Proposition~\ref{prop2.4chap2} of Chapter~\ref{Chap:chap2}, the maps $\sigma_0, \, \sigma_x$ and $\sigma_{0x}$ can be equivalently defined by the formulas
\begin{equation*}
    \sigma_x = \left( - \Delta \right)^{-1} \di h_x,~ \sigma_0 = \left( - \Delta \right)^{-1} \di h_x ~\mbox{and}~ \sigma_0 = \left( - \Delta \right)^{-1} \di h_x ~\mbox{in}~\cu
\end{equation*}
where the Laplacian is subject to the boundary condition stated in~\eqref{eq:TV10042901}. In particular, using that the Laplacian commutes with the operators $\di$ and $\di^*$, we formally obtain
\begin{equation} \label{eq:TV10330102}
    \di^* \sigma_0 = \left( - \Delta \right)^{-1} \di^* \di h_x = -h_x - \left( - \Delta \right)^{-1} \di \di^* h_0 = -h_0 - \left( - \Delta \right)^{-1} \di \indc_0 = -h_0 - \nabla G, 
\end{equation}
where $G$ is the standard random walk Green's function on the lattice $\Zd$ and where we have used the identity $\di = \nabla$ valid for any function defined on $\Zd$. A consequence of the identity~\eqref{eq:TV10330102} is the equality
\begin{equation} \label{eq:TV10490102}
    e^{-2i\pi \left( q , \sigma_0 \right) } = e^{-2i\pi \left( n_q , \nabla G \right) }.
\end{equation}
While the identity~\eqref{eq:TV10330102} is not exactly true in finite volume (since we cannot a priori commute the operators $\di$, $\di^*$ and $\left( - \Delta \right)^{-1}$), it becomes true by taking the infinite volume limit (i.e., sending the volume of the cube $\cu$ to $\infty$). To avoid further technicalities, we will assume that the identity~\eqref{eq:TV10490102} holds in the rest of this chapter.
Similar statements hold for the maps $\sigma_x$ and $\sigma_{0x}$ and we may write
\begin{equation*}
    e^{-2i\pi \left( q , \sigma_x \right) } = e^{-2i\pi \left( n_q , \nabla G_x \right) } ~\mbox{and}~ e^{-2i\pi \left( q , \sigma_{0x} \right) } = e^{-2i\pi \left( n_q , \nabla G_x  - \nabla G \right) } .
\end{equation*}
where we have used the notation $G_x := G(\cdot - x)$. We set the notation, for each $\sigma : F(\cu) \mapsto \R$,
\begin{equation}
\label{e.Zsigma}
Z \left( \sigma \right) := \sum_{q \in \Z^{F(\cu)}_{\mathbf{t}  = 0}, \di q = 0} e^{- 2 \pi^2 \beta \left( q,  \left( \di \di^* \right)^{-1} q \right) } e^{-2i\pi  \left( q , \sigma \right)}.
\end{equation}
So that
\begin{equation*}
\frac{Z \left( \sigma_x \right)}{Z \left( 0 \right)} = \left\langle e^{-2i\pi  \left( q , \sigma_x \right)  }\right\rangle_{\mu_C(\beta)} ~\mbox{and}~\frac{Z \left( \sigma_{0x} \right)}{Z \left( 0 \right)}= \left\langle e^{-2i\pi  \left( q , \sigma_{0x}\right)  }\right\rangle_{\mu_C(\beta)}.
\end{equation*}
Let $\phi_{1} , \ldots, \phi_{\binom d2}$ be independent, real-valued, Gaussian free fields in the cube $\cu$ with boundary conditions given by~\eqref{eq:TV10042901} of Chapter~\ref{Chap:chap2}. We denote by $\phi := \left( \phi_1 , \ldots, \phi_{\binom dk} \right)$ the corresponding vector-valued Gaussian field, it is valued in the space
\begin{equation*}
    C (\cu) := \left\{ w := \left( w_1 , \ldots, w_{\binom d2} \right) : \cu \to \R^{\binom d2} \, : \, \forall i \in \left\{ 1 , \ldots , \binom d2 \right\},~ w_i = 0 ~\mbox{on}~ \partial \cu \setminus \partial_{I_i} \cu \right\}.
\end{equation*}
By Proposition~\ref{prop2.4chap2} of Chapter~\ref{Chap:chap2}, if we denote by $(-\Delta)^{-1} q$ the solution of the boundary value problem~\eqref{eq:TV10042901}, then one has the identity $(-\Delta)^{-1}q = \left( \di \di^* \right)^{-1} q$. This observation implies that for each $q \in \Z^{E(\cu)}$ satisfying $\di q = 0$ and $\mathbf{t} q = 0$,
\begin{equation*}
\E \left[ e^{2i\pi \left( q, \phi \right)} \right] =   e^{- 2 \pi^2 \beta \left( q,  \left(-\Delta \right)^{-1} q \right) } =  e^{- 2 \pi^2 \beta \left( q,  \left( \di \di^* \right)^{-1} q \right) }.
\end{equation*}
Consequently,
\begin{equation*}
Z \left( \sigma_{0x} \right) =  \sum_{q \in \Z^{F(\cu)}_{\mathbf{t}  = 0}, \di q = 0} \E \left[ e^{ - 2i\pi \left( q, \phi + \sigma_{0x} \right)} \right].
\end{equation*}
Thus the partition function of this lattice Coulomb gas can be represented in terms of a characteristic function with respect to a Gaussian measure. We then claim that for $\beta$ sufficiently large, a one-step renormalization maps the Coulomb gas model to an effective one with very small effective activity. 
Using that the discrete Laplacian is bounded from above, one has that $\left( - \Delta \right)^{-1} \geq c$, for some $c := c(d) > 0$. We then choose the inverse temperature $\beta$ larger than then value $c^{2}$ and decompose the Gaussian field $\phi$ as the sum of two independent Gaussian fields $\phi_1 + \phi_2$, such that $\phi_1$ and $\phi_2$ have covariance matrices $\beta \left( (- \Delta)^{-1} - \beta^{-\frac12}Id\right)$ and $\beta^{\frac12}Id$. We can thus write
\begin{equation*}
Z \left( \sigma_{0x} \right) = 
\sum_{q \in \Z^{F(\cu)}_{\mathbf{t}  = 0}, \di q = 0} \E \left[ e^{ - 2i\pi \left( q, \phi_1+\phi_2 + \sigma_{0x} \right)} \right]
=\sum_{q \in \Z^{F(\cu)}_{\mathbf{t}  = 0}, \di q = 0} e^{-  \pi^2 \beta^{1/2} \left( q,   q \right) }  \E_{\mu_1} \left[ e^{ - 2i\pi \left( q, \phi_1 + \sigma_{0x} \right)} \right],
\end{equation*}
where $\mu_1$ is a Gaussian measure on $C(\cu)$, given by
\begin{equation*}
    d\mu_1(\phi_1)= \text{Const} \times \exp{\left(-\frac12 \left( \phi_1, \frac{1}{\beta} \left( (- \Delta)^{-1} - \beta^{-\frac12}Id\right)^{-1} \phi_1\right)\right) } \indc_{\phi_1 \in C(\cu)} \, d\phi_1.
\end{equation*}
For $\beta$ sufficiently large, we may expand $\left( (- \Delta)^{-1} - \beta^{-\frac12}Id\right)^{-1}$ into a convergent sum
\begin{equation*}
    \left( (- \Delta)^{-1} - \beta^{-\frac12}Id\right)^{-1} = -\Delta+ \sum_{n\geq 1}  \frac{1}{\beta^{n/2}} (-\Delta)^{n+1}.
\end{equation*}
Thus
\begin{equation*}
    d\mu_1(\phi_1)= Z_1^{-1} \times \exp{\left( \frac{1}{2\beta} (\phi_1, \Delta \phi_1) - \sum_{n\geq 1} \frac{1}{2\beta} \frac{1}{\beta^{n/2}}(\phi_1, (-\Delta)^{n+1} \phi_1)  \right)}\indc_{\phi_1\in C(\cu)} \, d\phi_1.
\end{equation*}

Following \cite{Bau}, (especially see Lemmas 5.14 and 5.15 there), since $e^{-  \pi^2 \beta^{1/2} \left( q,   q \right) }$ decays to zero rapidly in $\|q\|_1:= \sum_{x\in F(\cu)} |q(x)|$, we may apply a standard cluster expansion to conclude that for $\beta$ large enough, one can re-sum $Z \left( \sigma_{0x} \right)$ as
\begin{equation*}
Z (\sigma_{0x}) =  \E_{\mu_1} \left[ \exp{\left( \sum_{q \in \mathcal{Q}_\cu} z(\beta, q)e^{- 2i\pi \left( q, \phi_1 + \sigma_{0x} \right)}\right)} \right]
= \E_{\mu_1} \left[ \exp{\left( \sum_{q \in \mathcal{Q}_\cu} z(\beta, q)\cos{ 2\pi \left( q, \phi_1 + \sigma_{0x} \right)}\right)} \right],
\end{equation*}
where the sum is over all lattice animals $q \in \mathcal{Q}_\cu$ with connected support satisfying $\di q = 0$ and $\mathbf{t} q = 0$ (see~\eqref{def.defQchap2} of Chapter~\ref{Chap:chap2}), and $z (\beta,q)$ is a real number satisfying the estimate
\begin{equation}
    \label{e.zest}
    |z(\beta , q)| \leq e^{-c \beta^{1/2} \|q\|_1}, \quad \mbox{for some } c :=c(d) > 0.
\end{equation}
Similarly, 
\begin{equation*}
Z (0) =  \E_{\mu_1} \left[ \exp{\left( \sum_{q \in \mathcal{Q}_\cu} z(\beta, q)e^{- 2i\pi \left( q, \phi_1  \right)}\right)} \right]= \E_{\mu_1} \left[ \exp{\left( \sum_{q \in \mathcal{Q}_\cu} z(\beta, q)\cos{ 2\pi \left( q, \phi_1  \right)}\right)} \right].
\end{equation*}
Using the trigonometric identity $$\cos{ 2\pi \left( q, \phi_1 + \sigma_{0x} \right)} = \cos{ 2\pi \left( q, \phi_1  \right)} \cos{ 2\pi \left( q, \sigma_{0x}  \right)} - \sin{ 2\pi \left( q, \phi_1  \right)}\sin{ 2\pi \left( q, \sigma_{0x}  \right)},$$ we may write 
\begin{equation}
\label{e.ratio}
    \frac{Z (\sigma_{0x}) }{Z (0) }=  \left\langle \exp \left( \sum_{q \in \mathcal{Q}_\cu} z(\beta , q) \sin 2\pi (\phi , q) \sin 2\pi(\sigma_{0x} , q) + \sum_{q \in \mathcal{Q}_\cu} z(\beta , q) \cos 2\pi(\phi , q) \left( \cos 2\pi(\sigma_{0x} , q) - 1  \right) \right) \right\rangle_{\mu_{\beta,\cu}}.
\end{equation}
Here $\mu_{\beta,\cu}$ is defined as a measure on the space $C(\cu)$ by
\begin{equation*}
    d\mu_{\beta,\cu}(\phi) := \text{Const} \times \exp{\left( \frac{1}{2\beta} (\phi, \Delta \phi) -\sum_{n\geq 1} \frac{1}{2\beta} \frac{1}{\beta^{n/2}}(\phi, (-\Delta)^{n+1} \phi)  + \sum_{q \in \mathcal{Q}_\cu} z(\beta, q)\cos{ 2\pi \left( q, \phi  \right)} \right)} \indc_{\phi \in C(\cu)} \, \di \phi.
\end{equation*}

Combining \eqref{e.dual2pt} and \eqref{e.ratio}, we have the following dual representation for the two-point function of the Villain model. Define $G_{\cu}$ be the solution of the problem 
\begin{equation}
\label{e.Gcu}
\left\{ \begin{aligned}
- \Delta G_\cu(x,\cdot) &= \delta_x ~\mbox{in}~ \cu, \\
G_\cu(x,\cdot) &= 0 ~\mbox{on}~ \partial \cu.
\end{aligned} \right.
\end{equation}

\begin{proposition}
\label{p.2ptfinite}
Let $G_\cu$ be defined as above. For $\beta$ sufficiently large, we have 
\begin{multline}
\label{e.2ptfinite}
      \left\langle e^{i \left( \theta(x) - \theta(0) \right)} \right\rangle_{\mu^{V}_{\beta,\cu,0}}
    \exp \left(\frac{1}{2\beta}G_\cu(0,x)\right) \\
    = \left\langle \exp \left( \sum_{q \in \mathcal{Q}_\cu} z(\beta , q) \sin 2\pi (\phi , q) \sin 2\pi(\sigma_{0x} , q) + \sum_{q \in \mathcal{Q}_\cu} z(\beta , q) \cos 2\pi(\phi , q) \left( \cos 2\pi(\sigma_{0x} , q) - 1  \right) \right) \right\rangle_{\mu_{\beta,\cu}}.
\end{multline}

\end{proposition}

Following the same argument we also obtain the dual representation for $\left\langle e^{i \left( \theta(x) + \theta(0) \right)} \right\rangle_{\mu^{V}_{\beta,\cu,0}}$. Define $\bar\sigma_{0x} := \sigma_0 + \sigma_x$. We then have

\begin{multline}
      \left\langle e^{i \left( \theta(x) + \theta(0) \right)} \right\rangle_{\mu^{V}_{\beta,\cu,0}}
    \exp \left(\frac{1}{2\beta}G_\cu(0,x)\right) \\
    = \left\langle \exp \left( \sum_{q \in \mathcal{Q}_\cu} z(\beta , q) \sin 2\pi (\phi , q) \sin 2\pi(\bar\sigma_{0x} , q) + \sum_{q \in \mathcal{Q}_\cu} z(\beta , q) \cos 2\pi(\phi , q) \left( \cos 2\pi(\bar\sigma_{0x} , q) - 1  \right) \right) \right\rangle_{\mu_{\beta,\cu}}.
\end{multline}
In view of~\eqref{e.dual2pt}, to study the two-point function of the (finite volume) Villain model, it suffices to compute the expectation of a nonlinear functional~\eqref{e.2ptfinite} with respect to the Gibbs measure $\mu_{\beta,\cu}$. Notice that for $\beta$ large, the exponential smallness of $z(\beta,q)$ implies that $\mu_{\beta,\cu}$ is a perturbation of a Gaussian measure. The neutrality condition $\di q =0$ indicates $\mu_{\beta,\cu}$ is a measure of \emph{gradient-type}, i.e., the Hamiltonian only depends on $\di \phi$. In later sections, we combine the Helffer-Sj{\"o}strand representation with quantitative homogenization to show that, for sufficiently large $\beta$, on large scales (i.e., $|x|\to \infty$ and $|\cu| \to \infty$) the measure $\mu_{\beta,\cu}$ behaves like an {\it effective} Gaussian free field, with the covariance matrix depending on $\beta$. This shows, along the line of the Gaussian heuristics, that the subleading order of~\eqref{e.ratio} (and therefore, the truncated two-point function) decays asymptotically as $C|x|^{2-d}$ where the constant depends on $\beta$.

\remark To see that the Gaussian heuristics implies that for $\beta $ sufficiently large, \eqref{e.2ptfinite} has an asymptotic power law $|x|^{2-d}$, we begin by noting that the assumption \eqref{e.zest} implies that for $\beta \gg 1$, charges in $\mathcal{Q}_\cu$ are essentially supported on dipoles, i.e., of the form $q = z(\beta)(\delta_x - \delta_{x+e_i})$, for $i=1,\ldots,d$. Thus the right side of \eqref{e.2ptfinite} is approximately 
\begin{multline*}
    \left\langle \exp \left( \sum_{e \in E(\cu)} z(\beta ) \sin 2\pi (\nabla \phi(e) ) \sin 2\pi(\nabla G(e) - \nabla G_x(e)) \right) \right. \\ \left. \times \exp \left( \sum_{e \in E(\cu)} z(\beta ) \cos 2\pi(\nabla \phi(e)) \left( \cos 2\pi(\nabla G(e) - \nabla G_x(e)) - 1  \right) \right) \right\rangle_{\mu_{\beta,\cu}}.
\end{multline*}
Since $|\cos 2\pi(\nabla G(e) - \nabla G_x(e)) - 1 |\leq C(\nabla G(e) - \nabla G_x(e))^2$ decays fast away from $0$ and $x$, let us assume for now that the term $\sum_{e \in E(\cu)} z(\beta ) \cos 2\pi(\nabla \phi(e)) \left( \cos 2\pi(\nabla G(e) - \nabla G_x(e)) - 1  \right)$ only contributes to the lower order. By further making the approximation $\sin a \approx a $ for small $a$,  we may further approximate the expression above by 
\begin{equation*}
     \left\langle \exp \left( \sum_{e \in E(\cu)} z(\beta )  2\pi (\nabla \phi(e) )  2\pi(\nabla G(e) - \nabla G_x(e)) \right) \right\rangle_{\mu_{\beta,\cu}}.
\end{equation*}
Using an integration by parts, this equals to $\left\langle \exp( 4\pi^2 (\phi(0) - \phi(x)))\right\rangle_{\mu_{\beta,\cu}}.$ Note that for $\beta$ sufficiently large, $\mu_{\beta,\cu}$ is a small perturbation of a Gaussian free field, we may conclude 
\begin{equation*}
   \left\langle \exp( 4\pi^2 (\phi(0) - \phi(x)))\right\rangle_{\mu_{\beta,\cu} }
   \approx \exp\left( \frac12 \var_{\mu_{\beta,\cu}}(4\pi^2 (\phi(0) - \phi(x))) \right)
   \approx
   C_0(d,\beta) + C_1(d,\beta)|x|^{2-d}.
\end{equation*}

We  remark that the computation above is only heuristical and the constants $C_0, C_1$ obtained are not the right constants. Indeed, the nonlocal charges in $\mathcal{Q}_\cu$, the nonlinear function $\sin x$, and the non-Gaussian field $\mu_{\beta,\cu}$ contribute to a nontrivial correction of these constants. Such corrections can be obtained rigorously through the homogenization of the Helffer-Sj\"ostrand PDE.

\remark The preceding derivation is for the measure with Dirichlet boundary condition. For the Villain model with Neumann and periodic boundary conditions, similar dual representation holds. To state the result, we fix a base point $x^*\in \partial \cu$ or $x^*\in  \T$. Define
\begin{equation*}
    d\mu_{\beta,\cu, x^*}(\phi) := \text{Const} \times \exp{\left( \frac{1}{2\beta} (\phi, \Delta \phi) -\sum_{n\geq 1} \frac{1}{2\beta} \frac{1}{\beta^{n/2}}(\phi, (-\Delta)^{n+1} \phi)  + \sum_{q \in \mathcal{Q}_\cu} z(\beta, q)\cos{ 2\pi \left( q, \phi  \right)} \right)} \indc_{\phi(x^*)=0} \, d\phi
\end{equation*}
and 
\begin{equation*}
    d\mu_{\beta,\T, x^*}(\phi) := \text{Const} \times \exp{\left( \frac{1}{2\beta} (\phi, \Delta \phi)_\T -\sum_{n\geq 1} \frac{1}{2\beta} \frac{1}{\beta^{n/2}}(\phi, (-\Delta)^{n+1} \phi)_\T  + \sum_{q \in \mathcal{Q}_\cu} z(\beta, q)\cos{ 2\pi \left( q, \phi  \right)} \right)} \indc_{\phi(x^*)=0} \, d\phi
\end{equation*}
We then have
\begin{multline*}
      \left\langle e^{i \left( \theta(x) - \theta(0) \right)} \right\rangle_{\mu^{V}_{\beta,\cu,\f}}
    \exp \left(\frac{1}{2\beta}G_{\cu,x^*}(0,x)\right) \\
    = \left\langle \exp \left( \sum_{q \in \mathcal{Q}_\cu} z(\beta , q) \sin 2\pi (\phi , q) \sin 2\pi(\sigma_{0x} , q) + \sum_{q \in \mathcal{Q}_\cu} z(\beta , q) \cos 2\pi(\phi , q) \left( \cos 2\pi(\sigma_{0x} , q) - 1  \right) \right) \right\rangle_{\mu_{\beta,\cu,x^*}}
\end{multline*}
and
\begin{multline*}
      \left\langle e^{i \left( \theta(x) - \theta(0) \right)} \right\rangle_{\mu^{V}_{\beta,\T}}
    \exp\left(\frac{1}{2\beta}G_{\T,x^*}(0,x)\right) \\
    = \left\langle \exp \left( \sum_{q \in \mathcal{Q}_\cu} z(\beta , q) \sin 2\pi (\phi , q) \sin 2\pi(\sigma_{0x} , q) + \sum_{q \in \mathcal{Q}_\cu} z(\beta , q) \cos 2\pi(\phi , q) \left( \cos 2\pi(\sigma_{0x} , q) - 1  \right) \right) \right\rangle_{\mu_{\beta,\T,x^*}},
\end{multline*}
where $G_{\cu,x^*}$ (and $G_{\T,x^*}$) are  Green's function in $\cu$ (and $\T$), with zero boundary condition at $x^*$, defined by 
\begin{equation*}
\left\{ \begin{aligned}
- \Delta G_{\cu,x^*}(x,\cdot) &= \delta_x - \delta_{x^*} ~\mbox{in}~ \cu, \\ G_{\cu,x^*}(x,x^*) &= 0, \\
\nabla G_{\cu,x^*}(x,\cdot) \cdot \mathbf{n} &= 0 ~\mbox{on}~ \partial \cu
\end{aligned} \right.
\end{equation*}
and 
\begin{equation*}
\left\{ \begin{aligned}
-\Delta G_{\T,x^*}(x,\cdot) &= \delta_x - \delta_{x^*} ~\mbox{in}~ \T,\\
G_{\T,x^*}(x,x^*) &= 0 .
\end{aligned} \right.
\end{equation*}

\section{Brascamp-Lieb inequality}

For $L \geq 1$ we denote by $\cu_L := [-L,L]^d \cap \Zd$ and write $\mu_{\beta, \cu_L}$ as $\mu_{\beta,L}$. As we discussed, when $\beta$ is sufficiently large, the measure $\mu_{\beta,L}$ is a small perturbation of the Gaussian measure, and is therefore log-concave. We next present the Brascamp-Lieb inequality \cite{BL, BobLed}, which states that the variance of observables  with respect to a log-concave measure is dominated by that of a Gaussian measure.  
Denote, for any $x\in\Zd$, $\partial_x := \left(\partial_{x,1},\cdots, \partial_{x,\binom d2}\right)$ and $\partial: = (\partial_x)_{x\in\Zd}$. We let $G_{C(\cu)}$ be the Green's function defined by the formula
\begin{equation*}
    \left\{\begin{aligned}
    - \Delta G_{C(\cu)}(x , \cdot) = \delta_x ~\mbox{in}~ \cu\\
    G_{C(\cu)}(x , \cdot) \in C(\cu),
    \end{aligned} \right.
\end{equation*}
we denote its components by $ G_{C(\cu)}$, $1 \leq i \leq \binom d2$.

\begin{proposition}[Brascamp-Lieb inequality for $\mu_{\beta,\cu}$]
\label{p.BL} 
Let $\beta$ be sufficiently large. For every $F\in H^1(\mu_{\beta,\cu})$, there exists $C = C(d, \beta) < \infty$ such that 
\begin{equation}
\label{e.BL.var}
\var_{\mu_{\beta,\cu}} \left[ F \right] 
\leq 
C
\sum_{x,y \in \cu^\circ} \sum_{i=1}^{\binom d2}  G_{C(\cu),i}(x,y) \left\langle \left( \partial
_{x,i}F \right) \left( \partial _{y,i}F\right) \right\rangle_{\mu_{\beta,\cu}}.
\end{equation}
\end{proposition}

Given $f: \cu \to \R^{\binom d2}$, recall that we denote by $(f,\phi) := \sum_{x\in\cu}  f(x) \phi(x)$ the linear functional of $\phi$. Here we follow the notation in Chapter~\ref{Chap:chap2} and omit the $\cdot$ when taking the scalar product for two vector valued functions. As a direct consequence of Proposition \ref{p.BL}, we obtain the following variance bound for linear functionals. 

\begin{corollary} 
\label{c.BL}
Let $\beta$ be sufficiently large. For every $f: \cu \to \R^{\binom d2}$, there exists $C = C(d, \beta) < \infty$ such that 
\begin{equation}
\label{e.BL.var2}
\var_{\mu_{\beta,\cu}} \left[ (f,\phi) \right] \leq 
C
\sum_{x,y \in \cu^\circ} |f(x)| \left| G_{C(\cu)}(x,y)\right| |f(y)|.
\end{equation}
Moreover, for any $t\in \R$, 
\begin{equation*}
\left\langle \exp \left[t (f,\phi) \right] \right\rangle_{\mu_{\beta,\cu}}\leq 
\exp\left(Ct^2
\sum_{x,y \in \cu^\circ} |f(x)| \left| G_{C(\cu)}(x,y)\right| |f(y)| \right).
\end{equation*}
\end{corollary}
\begin{proof}
The variance bound is a direct consequence of \eqref{e.BL.var}. To prove the bound for exponential moments, we differentiate the quantity 
\begin{equation}
\label{e.difflog}
\frac{\partial^2}{\partial t^2} \log \left\langle \exp \left( t (f,\phi) \right)\right\rangle_{\mu_{\beta,\cu} }
= 
\var_{\mu_t}\left[ (f,\phi) \right],
\end{equation}
where $\mu_t$ satisfies
\begin{equation*}
\frac{d\mu_t}{d\mu_{\beta,\cu}} = Const \times \exp \left( t (f,\phi) \right).
\end{equation*}
In other words, $\mu_t$ is obtained from $\mu_{\beta,\cu}$ by adding a linear tilt, and is therefore log concave. The variance estimate \eqref{e.BL.var2} for measure $\mu_t$ in place of $\mu$ can be obtained without any changes to the arguments.
The claim thus follows from integrating \eqref{e.BL.var2} for $\mu_t$. 
\end{proof}

\begin{corollary}
\label{c.BLexp}
Let $\beta$ be sufficiently large. For every $F: \R \to \R$, $F'' \in L^\infty(\R)$ and $|g(\beta, q)| \leq \exp(-\beta^{1/2}\|q\|_1)$, we have 
\begin{equation*}
 \left\langle \exp \left( \sum_{q \in \mathcal{Q}_\cu} g(\beta , q) F(\phi , q) \right)\right\rangle_{\mu_{\beta,\cu} } \leq
  \exp \left(2 \var_{\mu_{GFF,\cu}}\left( \sum_{q \in \mathcal{Q}_\cu} g(\beta , q) F(\phi , q) \right) \right),
 \end{equation*}
 where 
 \begin{equation*}
    d\mu_{GFF,\cu}= Z_{GFF,\cu} \times \exp{\left( -\frac{1}{2\beta} (\phi, \Delta \phi)  \right)}\indc_{\phi \in C \left( \cu \right)} \, d\phi.
\end{equation*}
\end{corollary}
\begin{proof}
Similar to \eqref{e.difflog}, we have 
\begin{equation}
\frac{\partial^2}{\partial t^2} \log \left\langle \exp \left( \sum_{q \in \mathcal{Q}_\cu} g(\beta , q) F(\phi , q) \right)\right\rangle_{\mu_{\beta,\cu} } 
= 
\var_{\mu_t} \left( \sum_{q \in \mathcal{Q}_\cu} g(\beta , q) F(\phi , q) \right),
\end{equation}
where $\mu_t$ is defined via density
\begin{equation*}
\frac{d\mu_t}{d\mu_{\beta,\cu}} = Const \times \exp \left( t \sum_{q \in \mathcal{Q}_\cu} g(\beta , q) F(\phi , q)\right).
\end{equation*}
Denote by $H_{\beta, \cu} $ and $H_t$ the Hamiltonian associated with the Gibbs measures $\mu_{\beta,\cu}$ and $\mu_t$. We then have for all $x,y\in \cu$,
\begin{equation*}
\partial_x \otimes \partial_y H_t = \partial_x \otimes \partial_y H_{\beta, \cu} + \sum_{q \in \mathcal{Q}_{\cu,x,y} } z(\beta, q) F''(\phi , q) q(x) \otimes q(y),
\end{equation*}
where $\otimes$ denotes the tensor product of two vectors (see~\eqref{eq:TV11011622} of Chapter~\ref{Chap:chap2}). 
Using the estimate $|z(\beta, q)| \leq \exp(-\beta^{1/2}\|q\|_1)$, we see that for $\beta$ sufficiently large, 
\begin{equation*}
\left|  \sum_{q \in \mathcal{Q}_{\cu,x,y}} z(\beta, q) F''(\phi , q) q(x) \otimes q(y) \right| \leq
C_1 e^{-c_2 \beta^{1/2}}.
\end{equation*}
Therefore for $t\in [0,1]$, $\partial_x \otimes \partial_y H_t  \geq \beta^{-1} \Delta$. The claim thus follows from integrating $t$. 
\end{proof}

\section{Coupling the finite volume Gibbs measures}

 The finite-volume Gibbs measures $\mu_{\beta,L}$ can be realized as the invariant measure of a Markov process, known as the Langevin dynamics. Consider the diffusion process $\{ \phi_t\}: \cu \times \R \to \R^{\binom d2}$ evolving according to
\begin{equation}
\label{e.dynamics.muL}
\left\{
\begin{aligned}
& d\phi_t
= \left[\frac{1}{2\beta} \Delta\phi_t - \sum_{n\geq 1} \frac{1}{2\beta} \frac{1}{\beta^{n/2}}(-\Delta)^{n+1} \phi_t -\sum_{q \in \mathcal{Q}} 2\pi z(\beta, q) q \sin{ 2\pi \left( q, \phi_t  \right)}   \right] \,dt  + \sqrt{2} \,dB_t,
\\ & 
\phi_t \in C \left( \cu \right),
\end{aligned} 
\right.  
\end{equation}
where $\left( B_t \right)_{t \geq 0}$ is a brownian motion on the space $C \left( \cu \right)$ (equipped with the standard $L^2$ scalar product). 
The infinitesimal generator of this process 
is the operator~$-\Delta_\phi$ defined by
\begin{equation} \label{eq:TV13000102}
-\Delta_\phi F (\phi) 
: = 
\sum_{x\in \cu^\circ} \partial_x^2 F(\phi) 
- 
\sum_{x\in \cu^\circ} 
 \left[\frac{1}{2\beta} \Delta\phi(x) - \sum_{n\geq 1} \frac{1}{2\beta} \frac{1}{\beta^{n/2}}(-\Delta)^{n+1} \phi(x)-\sum_{q \in \mathcal{Q}} 2\pi z(\beta, q) q(x) \sin{ 2\pi \left( q, \phi  \right)}   \right] \partial_xF(\phi),
\end{equation}
where the notation $\partial_x^2$ means $\sum_{i=1}^{\binom d2} \partial_{x,i}^2$ and we implicitly take the scalar product between the two terms in the right side of~\eqref{eq:TV13000102}.
The domain of the operator~$\Delta_\phi$ includes the space of twice differentiable compactly supported functions on the set $C \left( \cu \right)$ denoted by $C^2_c(C \left( \cu \right))$. Notice that we can write $\Delta_\phi$ as
\begin{equation*}
\Delta_\phi F = \sum_{x\in \cu^\circ} \partial_x^* \cdot \partial_x F, 
\end{equation*}
where~$\partial_x^*$ denotes the formal adjoint of~$\partial_x$ with respect to $\mu_{\beta,L}$, given by 
\begin{equation*}
\partial_x^* w := -\partial_x w 
+
\left[\frac{1}{2\beta} \Delta\phi(x) - \sum_{n\geq 1} \frac{1}{2\beta} \frac{1}{\beta^{n/2}}(-\Delta)^{n+1} \phi(x)-\sum_{q \in \mathcal{Q}} 2\pi z(\beta, q) q(x) \sin{ 2\pi \left( q, \phi  \right)}   \right]  w(\phi).
\end{equation*}
The operator $\Delta_\phi$ is thus symmetric with respect to the measure $\mu_{\beta,L}$, and we define the Dirichlet form
\begin{equation*}
\mathcal{E}\left( F , G \right) := \left\langle F \Delta_\phi  G \right \rangle_{\mu_{\beta,L}} 
=
\left\langle G \Delta_\phi  F \right \rangle_{\mu_{\beta,L}} 
=
\sum_{x\in\Zd} \langle \partial_x F, \partial_x G \rangle_{\mu_{\beta,L}},
\quad \forall F,G \in C^2_c(C \left( \cu \right)).
\end{equation*}
In particular, 
\begin{equation}
\label{e.GLmuLF.bnd}
\left| \left\langle G \Delta_\phi  F \right \rangle_{\mu_{\beta,L}} \right| 
\leq 
\left\| F \right\|_{H^1(\mu_{\beta,L})}\left\| G \right\|_{H^1(\mu_{\beta,L})}, 
\quad \forall F,G \in C^2_c(C \left( \cu \right)),
\end{equation}
where we define the norm~$\| \cdot\|_{H^1(\mu_{\beta,L})}$ by 
\begin{equation*}
\| F \|_{H^1(\mu_{\beta,L})} :=
\left\langle |F|^2 \right\rangle_{\mu_{\beta,L}}^{\frac12}
+
\left( \sum_{x\in \cu^\circ} \left\langle |\partial_x F|^2 \right\rangle_{\mu_{\beta,L}} \right)^{\frac12}.
\end{equation*}

Let $H^1(\mu_{\beta,L})$ be the completion of $C^2_c(C(\cu))$ with respect to the norm~$\| \cdot\|_{H^1(\mu_{\beta,L})}$. By~\eqref{e.GLmuLF.bnd} and a density argument, the domain of the Dirichlet form $\mathcal{E}$ can be extended so that it includes the space $H^1(\mu_{\beta,L})$, and we have
\begin{equation}
\label{e.weakform.L.QL}
\left\langle G \Delta_\phi F \right \rangle_{\mu_{\beta,L}} 
=
\sum_{x\in \cu} \langle \partial_x F, \partial_x G \rangle_{\mu_{\beta,L}},
\quad \forall F,G \in H^1({\mu_{\beta,L}}).
\end{equation}

\subsection{Dynamical coupling}
The Langevin dynamics provides a convenient way to construct couplings between different finite volume Gibbs measures. In the context of the gradient Gibbs measures with uniformly convex potential, this coupling technique was first used by Funaki and Spohn to prove the uniqueness of the infinite volume Gibbs state \cite{FS}, and later used by \cite{Mil} to prove the CLT in finite domains and by \cite{AW} to obtain quantitative rate of convergence for the Hessian of finite volume surface tensions.  We will use this technique to obtain estimates on the difference of the~$\nabla\phi$ fields corresponding to different underlying Gibbs measures $\mu_{\beta,L}$ and $\mu_{\beta,M}$ for different $M,L\in\N$.

The basic idea is that we can couple the measures by driving the dynamics in~\eqref{e.dynamics.muL} with the same family~$\{  B_t(x) \}$ of Brownian motions and estimating the difference of the solutions of the system of SDEs with the aid of parabolic estimates. Specifically, we will apply the $C^{0,1-\ep}$-regularity estimate for solutions of parabolic equations with small ellipticity contrast, proved in  \eqref{e.paraboliccouple} of Chapter~\ref{section:section4}.

We denote by $\P'_{L,\phi}$ and $\P'_{M,\tilde{\phi}}$ the laws of the solution to~\eqref{e.dynamics.muL} in the cubes $\cu_L$ and $\cu_M$, starting from the initial data $\phi$ and $\tilde{\phi}$ respectively. We use the symbol $\otimes$ to denote the product of measures.

\begin{proposition}[{Dynamic coupling of $\mu_{\beta, L}$ and $\mu_{\beta, M}$}]
\label{p.coupling}
There exist constants $\beta_0 := \beta_0 (d)$ such that the following statement holds for all $\beta>\beta_0$. Let $L,M \in\N$ with  $4 \leq L\leq M \leq e^L$ and let the finite volume measures $\mu_{\beta, L}$ and $\mu_{\beta, M}$ be defined as above.  There exists a random element $(\nabla \phi,\nabla \tilde{\phi})$ of $C(\R^+;\Omega_0(\cu_L))\times C(\R^+;\Omega_0(\cu_{M}))$ with law $\Theta$ such that:
\begin{equation}
\label{e.lawyes}
\mbox{the law of~$\nabla \phi$ is $\mu_{\beta, L} \otimes \P'_{L,\phi} $,}
\end{equation}
\begin{equation}
\label{e.lawyestilde}
\mbox{the law of~$\nabla \tilde{\phi}$ is $\mu_{\beta, M}\otimes \P'_{M,\tilde{\phi}}$,}
\end{equation}
and a constant ~$\varepsilon \in \left(0,\tfrac12 \right]$, such that for all $L>4$, there exists $C= C(d)<\infty$ such that 
\begin{equation}
\label{e.couplingbound}
\E_\Theta\left[ \sup_{x\in \cu_{L/2}} \left| \nabla \phi(x) - \nabla \tilde{\phi}(x) \right| \right]
\leq
CL^{-1+\ep}.
\end{equation}
\end{proposition}

\begin{proof}
Let  $\P'_{L,\phi_0}$ and $\P'_{M,\tilde{\phi}_0}$ be law of the Langevin dynamics \eqref{e.dynamics.muL} in the cubes $\cu_L$ and $\cu_M$, starting from the initial data $\phi_0$ and $\tilde{\phi_0}$ respectively.
We may couple these measures by requiring that the family $\{  B_t(x) \,:\, x\in \cu_L^\circ \}$ of Brownian motions driving the dynamics are the same. 

\smallskip

We let $\P^*_{(\phi_0,\tilde\phi_0)}$ be the resulting coupled measure of the joint process $(\phi,\tilde\phi)$. In other words, $\P^*_{(\phi_0,\tilde\phi_0)}$ is the law on the set of trajectories $(\phi_t,\tilde{\phi}_t)$ satisfying the coupled set of equations
\begin{equation}
\label{e.dynamics.coupling}
\left\{
\begin{aligned}
& d\phi_t 
= \left[\frac{1}{2\beta} \Delta\phi_t - \sum_{n\geq 1} \frac{1}{2\beta} \frac{1}{\beta^{n/2}}(-\Delta)^{n+1} \phi_t -\sum_{q \in \mathcal{Q}_{\cu_L}} 2\pi z(\beta, q) q \sin{ 2\pi \left( q, \phi_t  \right)}  \right] \,dt  + \sqrt{2} \,dB_t, 
\\  & 
d\tilde{\phi}_t 
= \left[\frac{1}{2\beta} \Delta\tilde\phi_t - \sum_{n\geq 1} \frac{1}{2\beta} \frac{1}{\beta^{n/2}}(-\Delta)^{n+1} \tilde\phi_t-\sum_{q \in \mathcal{Q}_{\cu_M}} 2\pi z(\beta, q) q \sin{ 2\pi \left( q, \tilde \phi_t  \right)}   \right] \,dt + \sqrt{2} \,dB_t,
\\ &
\phi_t \in C \left( \cu_L \right),
\\ & 
\tilde{\phi}_t \in C \left( \cu_M \right),
\end{aligned} 
\right.  
\end{equation}
with initial data $(\phi_0,\tilde{\phi}_0)$. Let us sample the initial data with $\mu_{\beta,L} \times \mu_{\beta,M}$ itself by setting 
\begin{equation}
\Theta':= \left( \mu_{\beta,L} \times \mu_{\beta,M} \right) \otimes \P^*_{(\phi,\tilde\phi)}. 
\end{equation}
In other words, $\Theta'$ is the law of the pair $(\phi_t,\tilde{\phi}_t)$ of trajectories obtained by first sampling $\phi_0$ and $\tilde{\phi}_0$ according to the measures~$\mu_{\beta,L}$ and~$\mu_{\beta,M}$, respectively, and then running the dynamics~\eqref{e.dynamics.coupling}. 

\smallskip

It is clear, by the invariance of the Gibbs measures with respect to the dynamics, that at any time $t$, the law of $\phi_t$ is $\mu_{\beta,L}$ and the law of $ \tilde{\phi}_t$ is $\mu_{\beta,M}$. We will eventually take the measure~$\Theta$ as in the statement of the proposition to be the law of~$( \phi_t ,\tilde{\phi}_t)$ at a given time~$t_*$ which will be selected below. This ensures that~\eqref{e.lawyes} and~\eqref{e.lawyestilde} are satisfied. It remains therefore to show that we can select $t_*$ in such a way that the bound~\eqref{e.couplingbound} is satisfied. 

\smallskip

Consider the difference 
\begin{equation}
u(t,x) := \phi_t(x) - \tilde{\phi}_t(x), \quad (t,x) \in (0,\infty) \times \cu_L.
\end{equation}
Observe that $u$ satisfies the parabolic equation, for any $(t, x) \in (0,\infty) \times \cu_{\frac 34 L}$,
\begin{multline}
\label{e.paraboliccouple}
\partial_t u(t,x) - \frac{1}{2\beta} \Delta u(t,x) - \sum_{n\geq 1} \frac{1}{2\beta} \frac{1}{\beta^{n/2}}(-\Delta)^{n+1} u(t,x) \\ -\sum_{q \in \mathcal{Q}_{\cu_L}}\nabla_q^* \cdot \hat{\a}_q\nabla_q u(t , \cdot)
- \sum_{q \in \mathcal{Q}_{\cu_M,x}\setminus\mathcal{Q}_{\cu_L} } 2\pi z(\beta, q) q(x) \sin{ 2\pi \left( q, \tilde \phi_t  \right)}
=  0 ,
\end{multline}
where $\nabla_q^* \cdot \hat{\a}_q\nabla_q$  is defined by
\begin{equation}
\nabla_q^* \cdot \hat{\a}_q\nabla_q u
:=
 (2\pi)^2 z(\beta, q) q(x)(u,q) \int_0^1 \cos{ \left(2\pi s\left( q, \phi_t  \right)+ 2\pi(1-s)(q, \tilde \phi_t)\right)}\,ds  ,
\end{equation} 
Denote, for $r>0$ and $(t,x)\in (0,\infty) \times \Zd$,  the parabolic cylinder 
\begin{equation*}
Q_r (t,x) 
:= 
(t,x) + (-r^2,0] \times \cu_r.
\end{equation*}
We first notice that the term involving $q\in \mathcal{Q}_{\cu_M,x}\setminus\mathcal{Q}_{\cu_L}$ is exponentially small in $L$. Indeed, since the number of lattice animal with diameter $r$ grows exponentially in $r$, using the estimate $|z(\beta,q)|\leq e^{-\beta^{1/2} \|q\|_1}$, we may take $\beta>\beta_0(d)$ such that 
\begin{equation*}
    \left|\sum_{q \in \mathcal{Q}_{\cu_M,x}\setminus\mathcal{Q}_{\cu_L} } 2\pi z(\beta, q) q(x) \sin{ 2\pi \left( q, \tilde \phi_t  \right)} \right|
    \leq C \sum_{r=\frac L2}^{2M} e^{-r}
    \leq e^{-c L}.
\end{equation*}
Using the estimate $|z(\beta,q)|\leq e^{-c \beta^{1/2} \|q\|_1}$, we see that for large $\beta$, \eqref{e.paraboliccouple} is a small perturbation of the heat equation $\partial_t u + \frac{1}{2\beta}\Delta u =0$, and therefore the solution possess very strong regularity. 
By the $C^{0,1-\ep}$ regularity estimate for the parabolic equation \eqref{e.paraboliccouple}, which can deduced from the arguments presented Proposition~\ref{prop:prop4.6} and~\eqref{eq:C^1-epregcal} of Chapter~\ref{section:section4} with a minor adaptation to include the case of nonzero but small right-hand side, we obtain that if $\beta$ is chosen large enough, then there exist $\ep\in \left(0,\tfrac12 \right]$ and $C<\infty$ such that for every $t\in [\frac{L^2}{4},\infty)$,
\begin{equation} 
    L^{1-\ep}\left[ u \right]_{C^{0, 1-\ep}\left(Q_{\frac  L2}(t,0)\right)} \leq C \left\| u - \left(u \right)_{Q_{\frac34 L}(t,0)} \right\|_{\underline{L}^2 \left( Q_{\frac34 L}(t,0)\right)} + \int_{-\frac{L^2}{4}+t}^t \sum_{y \in \Zd} e^{-c \left(\ln \beta\right)  \left( \frac{L}{2} \vee |y| \right)} \left| u(s , y) \right|^2 \, ds 
    + e^{-\frac{c L}{2}}
\end{equation}
We claim that the second term on the right side acts as a small perturbation when $\beta$ is large. To see this, observe that for $\beta$ large enough, 
\begin{equation*}
    \int_{-\frac{L^2}{4}+t}^t \sum_{y \in \cu_L} e^{-c \left(\ln \beta\right)  \left( \frac{L}{2} \vee |y| \right)} \left| u(s , y) \right|^2 \, ds
    \leq 
    e^{-c L} \left\| u  \right\|_{\underline{L}^2 \left( Q_L(t,0)\right)}
\end{equation*}
In particular, since we are on discrete lattice, one may divide by $L^{1-\ep}$ and apply the triangular inequality to obtain
\begin{align} \label{e.nash}
&
\frac12 \sup_{x\in \cu_{\lceil L/2\rceil}} 
\left| \nabla \phi(x) - \nabla\tilde{\phi}(x) \right| 
\leq
\frac12 \left[ u \right]_{C^{0,1-\ep}\left(Q_{\lceil L/2\rceil} (t,0) \right)} 
\\ & 
\leq 
CL^{-1+\ep} \left( \left\| \phi \right\|_{\underline{L}^2( Q_{L}(t,0) )} + \left\| \tilde{\phi} \right\|_{\underline{L}^2( Q_{L}(t,0) )} +  \right)  +CL^{-1+\ep} \int_{-\frac{L^2}{4}+t}^t \sum_{y \in \cu_M\setminus\cu_L} e^{-c \left(\ln \beta\right)  |y| } \left| \tilde \phi(s , y) \right|^2 \, ds + e^{-c L} \notag
\end{align}
Applying Lemma~\ref{l.oscillation.dyn} below, we obtain for some $C<\infty$, all $L>4$, all $t > 1$ and all $s > 1$,
\begin{equation*}
\P_\Theta \left[ \left\| \phi \right\|_{\underline{L}^2( Q_L(t,0) )} + \left\| \tilde{\phi}  \right\|_{\underline{L}^2( Q_{L}(t,0) )} > Cs\sqrt{\log Lt} \right]\leq \exp(-cs^2 \log(Lt)).
\end{equation*}
We then use the exponential smallness of $e^{-c \left(\ln \beta\right) L}$ to absorb any polynomial in $L$, and obtain
\begin{multline*}
\P_\Theta\left[ \left| \int_{-\frac{L^2}{4}+t}^t \sum_{y \in \cu_M\setminus\cu_L} e^{-c \left(\ln \beta\right)  |y| } \left| \tilde \phi(s , y) \right|^2 \, ds \right|> Cs\frac{\sqrt{\log M}}{\sqrt{L}} 
\right] \\
\leq \P_\Theta\left[ e^{-c \left(\ln \beta\right)L} |\cu_M| \max_{(t,x)\in [0,\frac{L^2}{4}]\times \cu_M} |\tilde\phi(t,x)| > Cs\frac{\sqrt{\log M}}{\sqrt{L}}.
\right]
\end{multline*}
Since $M<e^L$, we may choose $\beta(d)$ large enough so that $e^{-c \left(\ln \beta\right)L} |\cu_M|<\frac{1}{\sqrt{L}}$, apply Lemma~\ref{l.oscillation.dyn} again to $|\cu_M|$ we see that the above probability is bounded by $\exp(-cs^2\log(Mt))$. Since $\frac{\sqrt{\log M}}{\sqrt{L}} <1$, 
taking $t= L^2$ and take expectation with respect to $\Theta$ in \eqref{e.nash}, we conclude the proposition. 
\end{proof}

The following lemma is a direct consequence of the Brascamp-Lieb inequality. 

\begin{lemma}
\label{l.oscillation}
Let $\beta$ be sufficiently large. There exists~$C(d,\beta)<\infty$, such that, for every $s \geq  C$ and $L\in\N$,
\begin{equation}
\label{e.oscillation}
\mu_{\beta,L} \left( 
\max_{x\in \cu_L} 
| \phi(x)| 
> C s \sqrt{\log L}
 \right) 
\leq
\exp\left( -s^2 \log L \right).
\end{equation}
\end{lemma}

\begin{proof}
We will prove~\eqref{e.oscillation} by estimating the exponential moments of $|\phi(x)|$ for each $x\in \cu_L$, , which we do by an application of the Brascamp-Lieb inequality (Corollary~\ref{c.BL}), and then take a union bound over~$x$. 
We obtain, for a constant~$C(d,\beta)<\infty$, and all $s \in \R$,
\begin{align*}
\max_{x\in Q_L} \left\langle \exp (s |\phi(x)|) \right \rangle_{\mu_{\beta,L}} 
\leq
\exp\left( cs^2\max_{x\in Q_L} \left ( G_{\cu_L} (x,x) \right) \right) \leq \exp(Cs^2).
\end{align*}
Applying the Chebyshev inequality and optimize over $s$, we obtain, for a constant~$C_1(d,\beta)< \infty$ and every $s >0$,
\begin{equation*}
\max_{x\in Q_L} \mu_{\beta,L} \left\{ |\phi (x)| > C_1 s \sqrt{\log L} \right\}
\leq \exp \left(- s^2 \log L \right).
\end{equation*}
The claim follows by taking a union bound over all $x$. 
\end{proof}

We next give an estimate on the oscillations of the dynamical field $|\phi_{t}|$. 

\begin{lemma}
\label{l.oscillation.dyn}
 Let $\beta$ be sufficiently large. There exists $C(d,\beta)<\infty$  such that, for every $T,s\in (1,\infty)$ and $L >1$,
\begin{equation}
\left( \mu_{\beta,L} \otimes \P'_{L,\phi}  \right) 
\left[
\max_{(t,x) \in (0,T] \times \cu_L} 
| \phi_{t}(x)| 
> C s \sqrt{ \log (LT)}
\right]
\leq
\exp\left( -s^2 \left(  \log (LT) \right)  
\right).
\end{equation}
\end{lemma}
\begin{proof}
 Since the time parameter is continuous, we prove the claim in two steps. First we discretize the time into intervals of length $(\log L)^{-1}$, and define the corresponding comb set by $\mathcal{C} := \{(t,x)\in  (0,T] \times \cu_L, t\log L \in \Z \}$. A union bound over the tail estimate proved in Lemma \ref{l.oscillation} controls the maximum of $|\phi_{t}|$ over $(t,x)\in \mathcal{C}$. Then we use continuity of the Brownian motion to bound $|\phi_{t}(x) - \phi_{t_0}(x)|$, whenever $|t-t_0| <(\log L)^{-1}$.

\smallskip

We first discuss the continuity estimates in $t$. The dynamics \eqref{e.dynamics.muL} imply, for every~$e = (x,y) \in E(\cu_L)$, 
\begin{align*}
d \nabla \phi_t(e) = \\
& \left[\frac{1}{2\beta} \Delta\nabla \phi_t(e) - \sum_{n\geq 1} \frac{1}{2\beta} \frac{1}{\beta^{n/2}}(-\Delta)^{n+1} \nabla \phi_t(e) - \left( \sum_{q \in \mathcal{Q}_{\cu,x}} - \sum_{q \in \mathcal{Q}_{\cu,y}}  \right) 2\pi z(\beta, q) q(x) \sin{ 2\pi \left( q, \phi_t  \right)} \right] \,dt \\ & 
+ 2 \,dB_t(e),
\end{align*}
where $B_t(e): = \frac{1}{\sqrt{2}} (B_t(y) - B_t(x))$ is a standard $\R^{\binom d2}$-valued Brownian motion. Let $G_t := \max_{e\in E(\cu_L)}| \nabla \phi_{t}(e)| $ and $M :=\max_{e\in E(\cu_L)} \max_{t\in (0, (\log L)^{-1}]} |B_t(e)| $, we use the fact that $\Delta $ is a bounded operator on $\Zd$, $\sum q(x) =0$ (so that $ \left( q, \phi_t  \right)$ is a linear combination of $\nabla \phi_t$) and the estimate $z(\beta, q) \leq \exp(-\beta^{1/2} \|q\|_1 )$ to conclude that for all large $\beta$, there exists $C_1=C_1(d,\beta)$ so that 
\begin{equation*}
G_t \leq C_1 \int_0^t  G_s\,ds + 2 M.
\end{equation*}
We apply Gronwall inequality to obtain for $t\in (0, (\log L)^{-1}]$ 
\begin{equation*}
G_t \leq 2(M+1) + 2C_1 \int_0^t (M+1) \exp\left( C_1(t-s) \right) \,ds.
\end{equation*}
That is, 
\begin{equation*}
G_t \leq C (M+1)\exp\left( C_1 t\right).
\end{equation*}
We now bound $|\phi_{t}|$ by a comparison with independent Brownian motions. Denote by $\Psi_{t} := \phi_{t} - (\sqrt{2} B_t +\phi_{0} )$. We then have for all $x \in \cu_L$,  there exists $C_2 = C_2(d,\beta)$ such that
\begin{align*}
\left|\frac{d\Psi_{t} (x)}{d t}\right| &\leq 
C_2(M+1) \exp\left( C_1 t\right).
\end{align*}
Integrating over $ t\in (0, (\log L)^{-1}]$, we have the following inequality in law:
\begin{equation*}
\max_{t\in (0, (\log L)^{-1}]} |\Psi_{t} (x)| \leq C_2( M +1).
\end{equation*}

We are now ready to finish the proof of the lemma. Given $t\in (0,T]$, take $t^* \in \frac{1}{\log L}\Z$ such that $t-t^* \in (0,(\log L)^{-1}] $. Using the stationarity of $\phi_{t}$  in time, we have the following inequalities in law:
\begin{align}
\label{e.cont}
\lefteqn{
\max_{(t,x) \in (0,T] \times \cu_L} 
| \phi_{t}(x)| 
} \quad & 
\\ & \notag
\leq \max_{(t,x) \in \mathcal{C}} | \phi_{t}(x)|  + \max_{(t,x) \in (0,T] \times \cu_L} |\phi_{t}(x)- \phi_{t^*}(x)| 
\\ & \notag
\leq  \max_{(t,x) \in \mathcal{C}} | \phi_{t}(x)| + \max_{(t^*,x) \in \mathcal{C}} \max_{t\in (0, (\log L)^{-1}]}|\phi_{t+t^*}(x)- \phi_{t^*}(x)| 
\\ & \notag
\leq  \max_{(t,x) \in \mathcal{C}} | \phi_{t}(x)| + \max_{(t^*,x) \in \mathcal{C}} \max_{t\in (0, (\log L)^{-1}]} |\Psi_{t} (x)|  
+ 2\max_{(t^*,x) \in \mathcal{C}} \max_{t\in (0, (\log L)^{-1}]} |B_t(x)| \notag
\\ & \notag 
\leq  \max_{(t,x) \in \mathcal{C}} | \phi_{t}(x)| + C_2M+ 2\max_{(t^*,x) \in \mathcal{C}} \max_{t\in (0, (\log L)^{-1}]} |B_t(x)|.
\end{align}
Applying Lemma~\ref{l.oscillation} and taking a union bound over~$t\in (\log L)^{-1}\Z$, we find, for~$L > 1$,
\begin{equation*}
\left( \mu_{\beta,L} \otimes \P'_{L,\phi}  \right) 
\left[
\max_{(t,x) \in \mathcal{C}} 
| \phi_{t}(x)| 
> C s \sqrt{ \log (LT)}
\right]
\leq T\log L \exp\left(-s^2 \log(LT)\right) 
\leq
\exp\left(-\frac{s^2}{2} \log(LT)\right).
\end{equation*}
Applying a union bound and then Doob's inequality, we obtain 
\begin{align*}
\lefteqn{ 
\left( \mu_{\beta,L} \otimes \P'_{L,\phi}  \right)  \left[ M > s \log(LT) \right]
} \qquad & 
\\ & 
\leq |Q_L| \left( \mu_{L,\xi,\per} \otimes \P'_{L,\xi,\per,\phi}  \right) \left[ \max_{t\in (0, (\log L)^{-1}]}|B_t(0)| \geq \log L \right]
\\ & 
\leq 
|Q_L| \exp \left(-\frac12{(\log L)^3}\right) \leq \exp \left(-\frac13{(\log L)^3}\right). 
\end{align*}
Taking a union bound over $(t^*,x)\in \mathcal{C}$ then yields
\begin{equation*}
\left( \mu_{\beta , L} \otimes \P'_{L,\phi}  \right) \left[ \max_{(t^*,x) \in \mathcal{C}} \max_{t\in (0, (\log L)^{-1}]} |B_t(x)| > Cs \log(LT) \right] \leq\exp \left(-\frac14 (\log L)^3\right). 
\end{equation*}
Combining \eqref{e.cont} with the last three inequalities we conclude the lemma. 
\end{proof}

\subsection{Thermodynamic limit}
The coupling result from the previous subsection allows us to prove the thermodynamic limit of the measure $\mu_{\beta,L}$ as $L\to \infty$. For $\beta<\infty$, define the infinite volume Gibbs measure formally by

\begin{equation}
     d\mu_{\beta}(\phi)= \text{Const} \times \exp{\left( \frac{1}{2\beta} (\phi, \Delta \phi) -\sum_{n\geq 1} \frac{1}{2\beta} \frac{1}{\beta^{n/2}}(\phi, (-\Delta)^{n+1} \phi)  + \sum_{q \in \mathcal{Q}} z(\beta, q)\cos{ 2\pi \left( q, \phi  \right)}\right) }.
\end{equation}
Since $\mu_\beta$ is a translation-invariant, ergodic Gibbs measure that only depends on $\nabla \phi$, we uniquely determine the Gibbs state by requiring that $\left\langle \phi(0) \right\rangle_{\mu_\beta}=0$. 

Apply Proposition~\ref{p.coupling}, we see that for any $F \in C_c^\infty(\Omega)$,  the sequence of random variable $\left\langle F(\nabla \phi)\right\rangle_{\mu_{\beta, L}}$, $L\in\N$, forms a Cauchy sequence, thus converges as $L\to \infty$. Since $\left\langle \phi(0) \right\rangle_{\mu_{\beta,L}}=0$ for all $\beta$, it follows that the law of $\nabla \phi$ under $\mu_{\beta, L}$, viewed as an element of $\Omega$, converges weakly as $L\to \infty$ to $\mu_\beta$.

Sending $|\cu| \to \infty $ in Corollary \ref{c.BL}, and notice that all constants in the statement do not depend on the volume, we obtain the Brascamp-Lieb inequality for the infinite volume measure $\mu_\beta$. Let $G$ be the simple random walk Green's function in $\Zd$.

\begin{corollary} 
\label{c.BLZd}
Let $\beta$ be sufficiently large. For every $f: \Zd \to \R^{\binom d2}$, there exists $C = C(d, \beta) < \infty$ such that 
\begin{equation}
\var_{\mu_{\beta}} \left[ (f,\phi) \right] \leq 
C
\sum_{x,y \in \Zd} |f(x)| G(x,y) |f(y)|.
\end{equation}
Moreover, for any $t\in \R$, 
\begin{equation*}
\left\langle \exp \left[t (f,\phi) \right] \right\rangle_{\mu_{\beta}}\leq 
\exp\left(Ct^2
\sum_{x,y \in \Zd} |f(x)| G(x,y) |f(y)| \right).
\end{equation*}
\end{corollary}

An application of the Brascamp-Lieb inequality implies that for $\beta$ sufficiently large, $\var_{\mu_\beta}[|\phi(0)|] \leq c_\beta \Delta^{-1}(0,0) < \infty$, thus $\mu_\beta$ is a well-defined $\phi$-Gibbs measure.  
Combining with the thermodynamic limit results for the Villain model \cite{BFL,Gi}, we are now ready to state the following dual representation in infinite volume.
\begin{proposition}
\label{p.2pt}
Let $G$ be the simple random walk Green's function in $\Zd$. For $\beta$ sufficiently large, we have 
\begin{multline}
\label{e.2pt}
      \left\langle e^{i \left( \theta(x) - \theta(0) \right)} \right\rangle_{\mu^V_{\beta}}
    \exp \left(\frac{1}{2\beta}G(0,x) \right) \\
    = \left\langle \exp \left( \sum_{q \in \mathcal{Q}} z(\beta , q) \sin 2\pi (\phi , q) \sin 2\pi(\sigma_{0x} , q) + \sum_{q\in \mathcal{Q}} z(\beta , q) \cos 2\pi(\phi , q) \left( \cos 2\pi(\sigma_{0x} , q) - 1  \right) \right) \right\rangle_{\mu_{\beta}}
\end{multline}
    and 
\begin{multline}
\label{e.2ptplus}
      \left\langle e^{i \left( \theta(x) + \theta(0) \right)} \right\rangle_{\mu^V_{\beta}}
    \exp \left(\frac{1}{2\beta}G(0,x) \right) \\
    = \left\langle \exp \left( \sum_{q \in \mathcal{Q}} z(\beta , q) \sin 2\pi (\phi , q) \sin 2\pi(\bar\sigma_{0x} , q) + \sum_{q\in \mathcal{Q}} z(\beta , q) \cos 2\pi(\phi , q) \left( \cos 2\pi(\bar\sigma_{0x} , q) - 1  \right) \right) \right\rangle_{\mu_{\beta}}.
\end{multline}
\end{proposition}
\begin{proof}
We give the proof of \eqref{e.2pt} below, \eqref{e.2ptplus} follows from the same argument. 
The thermodynamic limit of the Villain model implies $\left\langle e^{i \left( \theta(x) - \theta(0) \right)} \right\rangle_{\mu^V_{\beta,L} }\to \left\langle e^{i \left( \theta(x) - \theta(0) \right)} \right\rangle_{\mu^V_{\beta}}$ as $L\to \infty$  \cite{BFL, Gi}, and we also have $G_{\cu_L}(0,x) \to G(0,x)$. Apply Proposition \ref{p.2ptfinite}, it suffices to show the weak convergence of $\mu_{\beta,L}$ to $\mu_{\beta}$ implies the right side of \eqref{e.2ptfinite} converges to that of \eqref{e.2pt}.

To simplify the notation, denote by 
\begin{equation*}
    X:=  \sum_{q \in \mathcal{Q}} z(\beta , q) \sin 2\pi (\phi , q) \sin 2\pi(\sigma_{0x} , q) + \sum_{q\in \mathcal{Q}} z(\beta , q) \cos 2\pi(\phi , q) \left( \cos 2\pi(\sigma_{0x} , q) - 1  \right).
\end{equation*}
and for any $R<\infty$, define the truncation of $X$ in $\cu_R$ by
\begin{equation*}
    X_R:=  \sum_{q \in \mathcal{Q}_{\cu_R}} z(\beta , q) \sin 2\pi (\phi , q) \sin 2\pi(\sigma_{0x} , q) + \sum_{q\in \mathcal{Q}_{\cu_R}} z(\beta , q) \cos 2\pi(\phi , q) \left( \cos 2\pi(\sigma_{0x} , q) - 1  \right).
\end{equation*}
We then have for all $|x|\ll R\ll L$,
\begin{equation*}
\left\langle \exp(X_L) \right\rangle_{\mu_{\beta,L} } 
= \left\langle \exp(X_{R}) \right\rangle_{\mu_{\beta,L} } 
+\left\langle \exp(X_{R}) \left(\exp(X_L-X_{R}) -1 \right) \right\rangle_{\mu_{\beta,L} }
\end{equation*}
and 
\begin{equation*}
\left\langle \exp(X) \right\rangle_{\mu_{\beta} } 
= \left\langle \exp(X_{R}) \right\rangle_{\mu_{\beta} } 
+\left\langle \exp(X_{R}) \left(\exp(X_L-X_{R}) -1 \right) \right\rangle_{\mu_{\beta}}. 
\end{equation*}
In particular, by applying Proposition \ref{p.coupling} to the geometric scales $2^k L, k\in\N$, using the fact that $(\phi,q)$ is a linear functional of $\nabla \phi$ and the estimate $|z(\beta,q)|\leq \exp(-\beta^{1/2}\|q\|_1)$, we see that as $L\to \infty$,
\begin{align*}
    \left|  \left\langle \exp(X_{R}) \right\rangle_{\mu_{\beta,L} } 
    - \left\langle \exp(X_{R}) \right\rangle_{\mu_{\beta} }  \right| &
    \leq 
    \sum_{k\geq 0}  \left|  \left\langle \exp(X_{R}) \right\rangle_{\mu_{\beta,2^kL} } 
    - \left\langle \exp(X_{R}) \right\rangle_{\mu_{\beta,2^{k+1}L} }  \right| \\
    & \leq
    \sum_{k\ge 0} \frac{e^{CR^d}}{2^{(1 - \ep) k}L^{1 - \ep}} \\ & \leq   \frac{C e^{CR^d}}{L^{(1 - \ep)/2}},
\end{align*}
which tends to $0$ as $L$ tends to infinity. We apply the H\"older and Brascamp-Lieb inequalities (Corollary \ref{c.BL}, \ref{c.BLexp} and \ref{c.BLZd}) to obtain 
\begin{align*}
    \left\langle \exp(X_{R}) \left(\exp(X_L-X_{R})-1\right)\right\rangle_{\mu_{\beta,L} } 
    &\leq 
    \left\langle \exp(2X_{R}) \right\rangle_{\mu_{\beta,L} }^{1/2}
    \left\langle \left(\exp(X_L-X_{R}) -1 \right)^2 \right\rangle_{\mu_{\beta,L} }^{1/2} \\
    & \leq
    \exp(C\var_{\mu^G_{\beta,L} } X_{R}) 
     \left\langle \left(\exp(X_L-X_{R}) -1 \right)^2 \right\rangle_{\mu_{\beta,L} } .
\end{align*}
We claim that the Brascamp-Lieb inequality implies 
\begin{equation*}
\left\langle \left(\exp(X_L-X_{R}) -1 \right)^2 \right\rangle_{\mu_{\beta,L} } \leq
C\var_{\mu^G_{\beta,L} } (X_L- X_{R}).
\end{equation*}
Indeed, we can Taylor expand the left side and obtain
 \begin{equation*}
LHS = \left\langle \left( \sum_{k\ge 1} \frac{1}{k!} (X_L- X_R)^k \right)^2 \right\rangle_{\mu_{\beta,L} }
=
\sum_{k\ge 1} \left\langle \sum_{j=1}^{2k} \frac{2}{j! (2k-j)!} (X_L -X_R)^{2k}\right\rangle_{\mu_{\beta,L} }.
\end{equation*}
We may then apply the exponential Brascamp-Lieb inequality to even moments of $X_L -X_R$ and the Wick Theorem for the Gaussian measures to conclude 
\begin{equation*}
    \left\langle (X_L -X_R)^{2k}\right\rangle_{\mu_{\beta,L} }
    \leq C^k  \left\langle (X_L -X_R)^{2k}\right\rangle_{\mu_{\beta,L}^G }
    \leq  C^k \frac{(2k)!}{k!2^k} \left( \var_{\mu^G_{\beta,L} } (X_L- X_{R})\right)^k.
\end{equation*}
Therefore 
\begin{align*}
    \left\langle \left(\exp(X_L-X_{R}) -1 \right)^2 \right\rangle_{\mu_{\beta,L} } \leq
    \sum_{k\ge 1} \frac{2^{k+1}}{k!} C^k \left( \var_{\mu^G_{\beta,L} } (X_L- X_{R})\right)^k
    & \leq \exp(C\var_{\mu^G_{\beta,L} } (X_L- X_{R}) ) -1 \\ & \leq  C\var_{\mu^G_{\beta,L} } (X_L- X_{R}) .
\end{align*}
The same computation yields
\begin{equation*}
     \left\langle \exp(X_{R}) \left(\exp(X-X_{R})-1\right)\right\rangle_{\mu_{\beta} } 
       \leq
    C \exp(C\var_{\mu^G_{\beta} } X_{R}) 
     \var_{\mu^G_{\beta} } (X- X_{R}).
\end{equation*}
We claim that for $R\gg |x|$ (e.g., $R=e^{|x|}$), both $\left\langle \exp(X_{R}) \left(\exp(X-X_{R})-1\right)\right\rangle_{\mu_{\beta} }  $ and $ \left\langle \exp(X_{R}) \left(\exp(X_L-X_{R})-1\right)\right\rangle_{\mu_{\beta,L} } $ are bounded by $R^{2 + \ep - d} $ for some $\ep \ll 1$.

By a Gaussian computation we see that $\var_{\mu^G_{\beta,L} } X_{R} < \infty$ and $\var_{\mu^G_{\beta} } X_{R} < \infty$. Also, for $R\gg |x|$ and $r>R$  we have for every charge $q \in \mathcal{Q}$ such that $ \supp q \cap \cu_{2r}\setminus \cu_r \neq \emptyset$,
\begin{equation}
\label{e.estsin}
     |z(\beta, q)\sin 2\pi(\sigma_{0x} , q) | \leq  2\pi\exp(-\beta \|q\|_1)|(\sigma_{0x} , q) |
     \leq C \left(\frac{1}{r^{d-2}}- \frac{1}{(r-|x|)^{d-2}} \right) 
     \leq \frac{C}{r^{d-1-\ep}}
\end{equation}
and similarly.
\begin{equation}
\label{e.estcos}
     |z(\beta, q)\left(\cos 2\pi(\sigma_{0x} , q) - 1 \right) | \leq  C\exp(-\beta \|q\|_1)(\sigma_{0x} , q)^2 
     \leq \frac{C}{r^{2d-2-\ep}}.
\end{equation}
To estimate $\var_{\mu^G_{\beta} } (X- X_{R})$, we decompose  
\begin{equation*}
    X- X_R = \sum_{k\geq 0} \Delta X_{2^k R},
\end{equation*}
where
\begin{align*}
      \Delta X_{2^k R} & := \sum_{q\in \mathcal{Q}_{\cu_{2^{k+1}R}} \setminus \mathcal{Q}_{\cu_{2^{k}R}}} z(\beta , q) \sin 2\pi (\phi , q) \sin 2\pi(\sigma_{0x} , q)  \\
    & \qquad + \sum_{q\in \mathcal{Q}_{\cu_{2^{k+1}R}} \setminus \mathcal{Q}_{\cu_{2^{k}R}}} z(\beta , q) \cos 2\pi(\phi , q) \left( \cos 2\pi(\sigma_{0x} , q) - 1  \right).
\end{align*}
 Using the estimates \eqref{e.estsin} and \eqref{e.estcos}, we conclude that 
\begin{equation*}
    \var_{\mu^G_{\beta} } ( \Delta X_{2^k R}) \leq
    \frac{1}{(2^k R)^{d-2-2\ep}}.
\end{equation*}
Therefore 
\begin{equation*}
    \var_{\mu^G_{\beta} }^\frac 12 (X- X_{R})
    \leq \sum_{k\in \mathbb{N}} \var_{\mu^G_{\beta} }^\frac 12 \Delta X_{2^k R}
    \leq \frac{C}{R^{d-2-2\ep}}.
\end{equation*}
Combining the estimates above we obtain 
\begin{equation*}
     \left\langle \exp(X_{R}) \exp(X-X_{R})\right\rangle_{\mu_{\beta} } 
       \leq
  \frac{1}{R^{d-2-2\ep}}
\end{equation*}
and similarly,
\begin{equation*}
     \left\langle \exp(X_{R}) \exp(X_L-X_{R})\right\rangle_{\mu_{\beta,L} } 
       \leq
  \frac{1}{R^{d-2-2\ep}}.
\end{equation*}
So that we conclude the proof.
\end{proof}

\remark By the same proof, one may obtain analogues of Proposition \ref{p.coupling} for the Gibbs measures $\mu_{\beta,\cu, x^*}$ and $\mu_{\beta,\T, x^*}$. One also observes that they converge to the same thermodynamic limit $\mu_\beta$. Therefore \eqref{e.2pt} also holds for the infinite volume measure $\mu^V_{\beta,\f}$, and the proofs in the rest of the paper applies to the Villain model with Neumann or periodic boundary conditions. 

\section{The Helffer-Sj{\"o}strand PDE} \label{secchap3HSPDE}
Following the idea of \cite{NS} (which was in turn inspired by the works \cite{HS} and \cite{Sj}) and \cite{GOS},
we will show in this section that the elliptic operator $\L$, defined in \eqref{e.HSop} below, where
\begin{multline*}
\Delta_\phi F (\phi)  := \\ 
\sum_{x\in F(\Zd)} \partial_x^2 F(\phi) 
- 
\sum_{x\in \Zd} 
 \left[\frac{1}{2\beta} \Delta\phi(x) - \sum_{n\geq 1} \frac{1}{2\beta} \frac{1}{\beta^{n/2}}(-\Delta)^{n+1} \phi(x)-\sum_{q \in \mathcal{Q}} 2\pi z(\beta, q) q(x) \sin{ 2\pi \left( q, \phi  \right)}   \right] \partial_xF(\phi),
\end{multline*}
arises naturally when one considers the variance of certain
observables with respect to the Gibbs measure~$\mu_\beta$. 
In Section \ref{s.HSrep} we derive the Helffer-Sj{\"o}strand representation, which identifies the variance of certain
observables under the measure $\mu_\beta$ with the energy density of the Helffer-Sj{\"o}strand PDE. In Section \ref{s.HSwellpose} we give variational characterizations for the Helffer-Sj{\"o}strand PDEs in the finite and infinite volume. The variational characterization gives the solvability of the equation, and will be crucial to prove the main homogenization result for the Green's function, which we state in Section \ref{s.homog}. The proof here combines the ideas in \cite{NS}, \cite{GOS} and \cite{AW}, and extend them to the setting of differential forms and long range operators.

\subsection{Helffer-Sj{\"o}strand representation} \label{s.HSrep}

The main result of this section is the Helffer-Sj{\"o}strand representation for the Gibb measure $\mu_\beta$, stated as Proposition \ref{p.HSrep}. To state the result, we introduce the Helffer-Sj{\"o}strand operator

\begin{equation}
\label{e.HSop}
\L : = \Delta_\phi - \frac{1}{2\beta} \Delta + \frac{1}{2\beta}\sum_{n \geq 1} \frac{1}{\beta^{ \frac n2}} \left(-\Delta\right)^{n+1} + \sum_{q \in \mathcal{Q}} \nabla_q^* \cdot \a_q \nabla_q
\end{equation}
where we introduce the notation
\begin{equation*}
 \nabla_q^* \cdot \a_q \nabla_q u = 4\pi^2 z\left( \beta , q \right) \cos 2\pi\left( \phi , q \right) \left( u , q \right) q.
\end{equation*}

\begin{proposition} \label{p.HSrep}
Fix $F,G \in H^1( \mu_\beta)$ and assume that there exist $f , g : \Zd \times \Omega \to \R^{d \times d}$ which belong to the space $L^2 \left( \Zd , \mu_\beta \right)$ such that $\partial_x F = \nabla \cdot f (x)$ and $\partial_x G = \nabla \cdot g (x)$. Then we have
\begin{equation} \label{eq:TV13060102}
\cov_{\mu_\beta} (F; G)
= \left\langle (\partial F, \L^{-1} \partial G )\right\rangle_{\mu_\beta} = \left\langle (f, \nabla \L^{-1} \nabla \cdot g )\right\rangle_{\mu_\beta}.
\end{equation}
Equivalently, we may write
\begin{equation*}
    \cov_{\mu_\beta} (F; G)
= \sum_{x\in\Zd}  \left\langle \partial_x F, \nabla u(x, \cdot) \right\rangle_{\mu_\beta},
\end{equation*}
where $u$ is the solution to the PDE
\begin{equation*}
    \L u  = \nabla \cdot g ~\mbox{in}~\Zd \times \Omega.
\end{equation*}
\end{proposition}

\begin{remark}
    The assumption that $\partial F$ and $\partial G$ are divergence of $L^2(\Zd , \mu_\beta)$-vector fields is not essential to prove the Helffer-Sj\"{o}strand representation formula. Nevertheless, it simplifies the proof of this formula and is sufficient for the purposes of this article.
\end{remark}
By polarization, it suffices to prove the identity for variances in the above proposition.
We would like to apply the integration by parts and obtain
\begin{equation*}
\left\langle \left( F - \langle F \rangle_{\mu_\beta} \right)^2 \right\rangle_{\mu_\beta}
=
\left\langle \left( F - \langle F \rangle_{\mu_\beta} \right) \Delta_\phi (\Delta_\phi)^{-1}  \left( F - \langle F \rangle_{\mu_\beta} \right)\right\rangle_{\mu_\beta}
=
\sum_{x\in \Zd} \left\langle ( \partial_x F ) \left( \partial_x \left(  \Delta_\phi^{-1} \left( F - \langle F \rangle_{\mu_\beta}  \right)\right) \right) \right\rangle_{\mu_\beta}.
\end{equation*}
However, it is not clear that $ \Delta_\phi^{-1} \left( F - \langle F \rangle_{\mu_\beta}  \right)$ is well-defined in the infinite volume. Therefore we will prove Proposition \ref{p.HSrep} by first solving a PDE with a mass term $\lambda$ and then send $\lambda\to 0$.
  
We are motivated to study solutions of the equation
\begin{equation}
\label{e.Lmu}
\Delta_\phi F + \lambda F = H,
\end{equation}
where $\langle G \rangle_{\mu_\beta}=0$ and $\lambda >0$. We have the 
unique variational solvability of~\eqref{e.Lmu} for any right-hand side~$G\in L^2(\mu_\beta)$. This is stated in the next lemma.

\begin{lemma}
\label{l.solvability}
Let  $G\in L^2(\mu_\beta)$. Then there exists a solution $F \in H^1(\mu_\beta)$ of the equation 
\begin{equation}
\label{e.HSmass}
\Delta_\phi F + \lambda F = H.
\end{equation}
Moreover the solution~$F$ of~\eqref{e.HSmass} is unique, and there exists a constant $C(\lambda,\beta, d)<\infty$ such that 
\begin{equation*}
\left\| F - \left\langle F \right\rangle_{\mu_\beta} \right\|_{H^1(\mu_\beta)} 
\leq C \left\| H \right\|_{L^2(\mu_\beta)}.
\end{equation*}
\end{lemma}
\begin{proof}
This result can be obtained by an application of the Lax-Milgram lemma, or, alternatively, by considering the variational problem 
\begin{equation*}
\inf
_{w\in H^1({\mu_\beta})} 
\left( \frac 12 \sum_{x\in \Zd} \left\langle ( \partial_x w)^2 \right\rangle_{\mu_\beta} + \frac\lambda 2 \left\langle w^2 \right\rangle_{\mu_\beta} - \left\langle Hw \right\rangle_{\mu_\beta} \right). 
\end{equation*}
In either case, we just use uniform coercivity with respect to the~$H^1(\mu_\beta)$ norm (and the coercivity depends on~$\lambda$). This problem can be solved equivalently by using tools of spectral theory as follows. The operator $\Delta_\phi$ is well-defined on the space of smooth compactly supported functions defined on the space $\Omega$ and depending on finitely many variables. Following the arguments of~\cite{NS, GOS}, we may extend this operator into a closed, self adjoint operator of $L^2 \left( \mu_\beta\right)$ which we still denote by $\Delta_\phi$. We denote by $\left( e^{-t \Delta_\phi}\right)_{t \geq 0}$ the $L^2(\mu_\beta)$-semigroup generated by this operator. Additionally, it follows from standard arguments (see~\cite{FS}), in view of the growth condition imposed on the elements of $\Omega$ in Chapter~\ref{Chap:chap2}, that for any initial condition $\phi \in \Omega$, the infinite volume Langevin dynamics
 \begin{equation}
 \label{e.LangZd}
 \left\{ \begin{aligned}
  d\phi_t(x) 
& = \left[\frac{1}{2\beta} \Delta\phi_t(x) - \sum_{n\geq 1} \frac{1}{2\beta} \frac{1}{\beta^{n/2}}(-\Delta)^{n+1} \phi_t(x) -\sum_{q \in \mathcal{Q}} 2\pi z(\beta, q) q(x) \sin{ 2\pi \left( q, \phi_t  \right)}  \right] \,dt  + \sqrt{2} \,dB_t(x), \quad x\in \Zd, \\
\phi_0 & = \phi
\end{aligned} \right.
\end{equation}
is well-posed. We denote by $\left( P_t\right)_{t \geq 0}$ the probability transition semigroup associated to this dynamics. By the arguments of~\cite[Section 3 and Theorem 4.2]{fritz1982infinite} (as mentioned in~\cite{GOS}), the two semigroup $\left( e^{-t \Delta_\phi}\right)_{t \geq 0}$ $\left
(P_t \right)_{t \geq 0}$ coincide. In particular, we have the identities
\begin{equation} \label{eq:TV13480102}
    w = \int_{0}^\infty e^{-\lambda t} e^{- t \Delta_\phi} H \di t = \int_{0}^\infty e^{-\lambda t} P_t H \di t.
\end{equation}
\end{proof}
Applying the previous lemma and the integration by parts, we obtain for all $\lambda>0$,
\begin{multline}
\label{e.HSlambda}
\left\langle F^2 \right\rangle_{\mu_\beta}
=
\left\langle F (\Delta_\phi + \lambda)(\Delta_\phi+ \lambda)^{-1}  F \right\rangle_{\mu_\beta} 
= 
 \sum_{x\in \Zd} \left\langle ( \partial_x F ) \left( \partial_x   (\Delta_\phi+ \lambda)^{-1}  F \right)\right\rangle_{\mu_\beta}
+ \lambda \left\langle F (\Delta_\phi+ \lambda)^{-1} F \right\rangle_{\mu_\beta}.
\end{multline}
Using the fact that 
\begin{equation*}
[\partial_x, \Delta_\phi] = \sum_{y \in \Zd} \partial_x \partial_y H
= \frac{1}{2\beta} \Delta - \frac{1}{2\beta}\sum_{n \geq 1} \frac{1}{\beta^{ \frac n2}} (-\Delta)^{n+1} - \sum_{q\in \mathcal{Q}_x} \nabla_q^* \cdot \a_q \nabla_q,
\end{equation*}
we may rewrite the first term on the right side of \eqref{e.HSlambda} formally as 
\begin{equation*}
 \sum_{x\in \Zd} \left\langle ( \partial_x F ) \left( \L+ \lambda)^{-1} \left(\partial F \right)\left( x , \cdot \right) \right)\right\rangle_{\mu_\beta},
\end{equation*}
where $u =  \left( \L+ \lambda\right)^{-1} \left(\partial F\right)$ is the solution to the problem
\begin{equation} \label{eq:TV13240102}
 \Delta_\phi u(x , \cdot) + \frac{1}{2\beta} \Delta u(x , \cdot) - \frac{1}{2\beta}\sum_{n \geq 1} \frac{1}{\beta^{ \frac n2}} (-\Delta)^{n+1} u(x , \cdot) -\sum_{q\in \mathcal{Q}_x} \nabla_q^* \cdot \a_q \nabla_q u+ \lambda u(x , \cdot)
 = \partial_x F(\cdot)
\quad \text{in } \Zd \times \Omega.
\end{equation}
We refer to~\cite{GOS} for the rigorous justification of this step. In Lemma~\ref{l.wellposeHS.Zd} below, we use a variational argument to prove the solvability of the equation~\eqref{eq:TV13240102}.

By monotone convergence theorem, we see that the $\lambda \to 0$ limit of $ \sum_{x\in \Zd} \left\langle ( \partial_x F ) \left( \L+ \lambda)^{-1} \partial_x F \right)\right\rangle_{\mu_\beta}$ exists. We denote the limit by $\sum_{x\in \Zd} \left\langle ( \partial_x F ) \L^{-1} \left(\partial F\right)(x , \cdot)\right\rangle_{\mu_\beta}$, and posit the existence of this object to Lemma~\ref{l.wellposeHS.Zd}.

We next claim the second term in the right side of \eqref{e.HSlambda} converges to $\langle F\rangle^2_{\mu_\beta}$ as $\lambda \to 0$. Combining with the previous argument implies 
\begin{equation}
\label{e.HSvar}
\var_{\mu_\beta} F 
= - \sum_{x\in \Zd} \left\langle ( \partial_x F ) \L^{-1} \left(\partial F\right)(x , \cdot) \right\rangle_{\mu_\beta},
\end{equation}
and the proof of Proposition \ref{p.HSrep} is complete. 

By the identity~\eqref{eq:TV13480102} and the change of variable $t \to \lambda t$, we obtain
\begin{align*}
\lambda (\Delta_\phi+ \lambda)^{-1} F
&= \int_0^\infty \lambda e^{-(\Delta_\phi + \lambda)t} F \,dt \\
&= \int_0^\infty \lambda e^{-\lambda t} P_t F \,dt
= \int_0^\infty e^{-t} P_{t/\lambda} F \,dt.
\end{align*}
It suffices to prove the right side converges in $L^2(\mu_\beta) $ to $\langle F\rangle_{\mu_\beta}$. Indeed, 
\begin{align*}
\left\langle  \left( \int_0^\infty e^{-t} P_{t/\lambda} F \,dt -\langle F\rangle_{\mu_\beta}  \right)^2  \right\rangle_{\mu_\beta}
&= \left\langle  \left( \int_0^\infty e^{-t} \left(P_{t/\lambda} F  -\langle F\rangle_{\mu_\beta} \right)\,dt  \right)^2  \right\rangle_{\mu_\beta} \\
&\leq C \int_0^\infty e^{-t} \left\langle  \left(P_{t/\lambda} F  -\langle F\rangle_{\mu_\beta} \right)^2 \right\rangle_{\mu_\beta} \,dt.
\end{align*}
Notice that for all $s>0$, $\left\langle  \left(P_s F  -\langle F\rangle_{\mu_\beta} \right)^2 \right\rangle_{\mu_\beta}  = \left\langle (P_s F)^2 \right\rangle_{\mu_\beta}  - \langle F\rangle_{\mu_\beta}^2 \leq \var_{\mu_\beta} F$, which follows from the fact that 
\begin{equation*}
\frac{d}{ds} \left\langle (P_s F)^2 \right\rangle_{\mu_\beta}  
= 2\left\langle (P_s F) \frac{d}{ds} (P_s F)\right\rangle_{\mu_\beta}  = -2 \left\langle (P_s F) \Delta_\phi (P_s F)\right\rangle_{\mu_\beta}  \leq 0.
\end{equation*}
Using the ergodicity of the Langevin dynamics, we obtain the almost sure convergence $P_t F \to  \langle F\rangle_{\mu_\beta}$ as $t \to \infty$. We can thus apply the dominated convergence theorem and conclude that $\left\langle  \left( \int_0^\infty e^{-t} P_{t/\lambda} F \,dt -\langle F\rangle_{\mu_\beta}  \right)^2  \right\rangle_{\mu_\beta} \to 0$. This concludes \eqref{e.HSvar}, and therefore Proposition \ref{p.HSrep}.

\subsection{Well-posedness for  the Helffer-Sj{\"o}strand equation}
\label{s.HSwellpose}
In this section, we study the solvability of the Helffer-Sj\"{o}strand equation in finite and infinite volume. The equations are solved variationally; this approach is one of the main techniques used in this article as it allows to prove quantitative homogenization estimates on the solutions of the Helffer-Sj\"{o}strand PDE (see Theorem~\ref{thm:homogmixedderchap4} below).

\subsection{Solvability in infinite volume}
The goal of this section is to prove that the equation 
\begin{equation*}
  -\Delta_\phi u + \frac{1}{2\beta} \Delta u- \frac{1}{2\beta}\sum_{n \geq 1} \frac{1}{\beta^{ \frac n2}} (-\Delta)^{n+1} u - \sum_{q\in \mathcal{Q}_x} \nabla_q^* \cdot \a_q \nabla_q u + \lambda u
 = \partial_x F
\quad \text{in } \Zd \times \Omega,
\end{equation*}
for all $\lambda \geq 0$ admits a variational characterization and a unique solution. We recall that we assumed $ \partial_x F = \nabla \cdot f$ for some $f \in L^2(\Zd, \mu_\beta)$. 

We now state the well-posedness for the Helffer-Sj\"ostrand equation in the infinite volume.

\begin{lemma}
\label{l.wellposeHS.Zd}
Assume that $\beta$ is sufficiently large and select $\lambda > 0$. There exists a unique solution $u_\lambda \in H^1(\Zd,\mu_\beta)$ of the equation 
\begin{equation}
\label{e.HS.eqn.Zd}
 \Delta_\phi u_\lambda - \frac{1}{2\beta} \Delta u_\lambda +  \frac{1}{2\beta}\sum_{n \geq 1} \frac{1}{\beta^{ \frac n2}} (-\Delta)^{n+1} u_\lambda + \sum_{q\in \mathcal{Q}} \nabla_q^* \cdot \a_q \nabla_q u_\lambda + \lambda u_\lambda
 = \nabla \cdot f
\quad \text{in } \Zd \times \Omega,
\end{equation}
which satisfies, for a constant $C(d,\beta)<\infty$, the estimate
\begin{equation}
\label{e.HSsolest}
\lambda \left\| u_\lambda \right\|_{L^2(\Zd,\mu_\beta)}^2 + \sum_{x \in \Zd} \left\| \partial_x u_{\lambda}  \right\|_{L^2(\Zd,\mu_\beta)}^2 + \left\| \nabla u_{\lambda} \right\|_{L^2 \left( \Zd , \mu_\beta\right)}^2
\leq 
C \left\| f \right\|_{ L^2(\Zd,\mu_\beta)}^2. 
\end{equation}
For each point $x \in \Zd$, the map $u_\lambda$ converges weakly in the spaces $L^2 \left( \mu_\beta \right)$ as $\lambda \to 0$. The weak limit $u := \mathcal{L} \nabla \cdot g$ is the unique solution (up to a constant) of the equation
\begin{equation*}
    \Delta_\phi u - \frac{1}{2\beta} \Delta u +  \frac{1}{2\beta}\sum_{n \geq 1} \frac{1}{\beta^{ \frac n2}} (-\Delta)^{n+1} u + \sum_{q\in \mathcal{Q}} \nabla_q^* \cdot \a_q \nabla_q u
    = -\nabla \cdot f
\quad \text{in } \Zd \times \Omega,
\end{equation*}
\end{lemma}
\begin{proof}
The proof is variational. A function $u_{\lambda}\in H^1(\Zd, \mu_\beta)$ is a solution of~\eqref{e.HS.eqn.Zd} if and only if  
\begin{multline}
\label{e.HSsolbvchar}
\sum_{y\in \Zd} \sum_{x\in \Zd} \left\langle (\partial_y u_{\lambda}(x,\cdot) )(\partial_y w (x,\cdot) ) \right\rangle_{\mu_\beta}
+
\frac{1}{2\beta}\sum_{x\in \Zd} 
\left\langle \nabla u_{\lambda}(x,\cdot) \nabla w(x,\cdot)  \right\rangle_{\mu_\beta} 
\\ + \frac{1}{2\beta}\sum_{n \geq 1} \frac{1}{\beta^{ \frac n2}}\sum_{x\in \Zd} 
\left\langle  \nabla^{n+1}  u_{\lambda}(x,\cdot), \nabla^{n+1} w(x,\cdot)  \right\rangle_{\mu_\beta}  
+ \sum_{x\in \Zd} \lambda \left\langle  u_{\lambda}(x,\cdot),w (x,\cdot)  \right\rangle_{\mu_\beta}
+
\sum_{q\in \mathcal{Q}}\left\langle \nabla_q u_{\lambda} \cdot \a_q \nabla_q w \right\rangle_{\mu_\beta}  
\\ =
-\sum_{x\in \Zd}  \left\langle f(x,\cdot) \nabla w(x,\cdot) \right\rangle_{\mu_\beta}, \quad \forall w\in H^1(\Zd,\mu_\beta). 
\end{multline}
Using the estimate $|\a_q| \leq C e^{-\beta \|q\|_1}$, if $\beta$ is sufficiently large, the symmetric bilinear form on the left side of the previous display is coercive with respect to the $H^1(\Zd,\mu_\beta)$ norm. The Lax-Milgram lemma therefore yields the existence of a unique solution~$u\in H^1(\Zd,\mu_\beta)$. We see that this function satisfies~\eqref{e.HSsolest} by taking $w=u_{\lambda}$ in~\eqref{e.HSsolbvchar}.

By the Gagliardo-Nirenberg-Sobolev inequality, we see that for any $\lambda > 0$ and any field $\phi \in \Omega$,
\begin{equation*}
    \left\| u_\lambda(x , \cdot) \right\|_{L^2 \left( \mu_\beta\right)}^2 \leq \left\langle \left( \sum_{x \in \Zd }\left\| u_\lambda(x, \phi) \right\|^{\frac{2d}{d-2}} \right)^{\frac{d-2}{d}} \right\rangle_{\mu_\beta} \leq \sum_{x \in \Zd }\left\| \nabla u_\lambda \right\|^2_{L^2 \left( \mu_\beta\right)} \leq C.
\end{equation*}
Combining the previous estimate with~\eqref{e.HSsolest}, we obtain that (up to extraction) for each point $x \in \Zd$, the map $ u_\lambda(x , \cdot)$ converges weakly to a function $u(x , \cdot) \in L^2 \left( \mu_\beta \right)$ and that (still up to extraction) the functions $\nabla u_\lambda$ and $\partial u_\lambda$ converge weakly to $\nabla u$ and $\partial u$ in the spaces $L^2 \left( \Zd , \mu_\beta\right)$ and $L^2 \left( \Zd \times \Zd, \mu_\beta\right)$ respectively. From~\eqref{e.HSsolbvchar}, we deduce that $u$ satisfies the identity: for each map $w \in H^1 \left( \Zd , \mu_\beta \right)$
\begin{multline*}
\sum_{y\in \Zd} \sum_{x\in \Zd} \left\langle (\partial_y u(x,\cdot) )(\partial_y w (x,\cdot) ) \right\rangle_{\mu_\beta}
+
\frac{1}{2\beta}\sum_{x\in \Zd} 
\left\langle \nabla u(x,\cdot) \nabla w(x,\cdot)  \right\rangle_{\mu_\beta} 
\\ + \frac{1}{2\beta}\sum_{n \geq 1} \frac{1}{\beta^{ \frac n2}}\sum_{x\in \Zd} 
\left\langle  \nabla^{n+1}  u(x,\cdot), \nabla^{n+1} w(x,\cdot)  \right\rangle_{\mu_\beta}  
+
\sum_{q\in \mathcal{Q}}\left\langle \nabla_q u \cdot \a_q \nabla_q w \right\rangle_{\mu_\beta}  
 =
-\sum_{x\in \Zd}  \left\langle f(x,\cdot) \nabla w(x,\cdot) \right\rangle_{\mu_\beta}. 
\end{multline*}
\end{proof}

\subsection{Solvability of the Dirichlet problem}

We next present the well-posedness of the Dirichlet boundary value problem for the Helffer-Sj\"ostrand equation in a cube $\cu \subset \Zd$. This will be used to prove the regularity of the solutions in Chapter~\ref{section:section4} and to establish the convergence of subadditive quantities associated with the equations in Chapter~\ref{section5}.

\begin{lemma}
\label{l.wellposeHS.dir}
Assume that $\beta$ is sufficiently large. Let~$\cu \subseteq \Zd$ be a cube of size $R$, $h \in L^2(\cu, \mu_\beta)$ and $\bar u \in H^1 \left( \cu, \mu_\beta \right)$. There exists a unique solution $u \in H^1(\cu,\mu_\beta)$ of the boundary-value problem 
\begin{equation}
\label{e.HS.eqn.dir}
\left\{ 
\begin{aligned}
& -\Delta_\phi u + \frac{1}{2\beta} \Delta u- \frac{1}{2\beta}\sum_{n \geq 1} \frac{1}{\beta^{ \frac n2}} (-\Delta)^{n+1} u - \sum_{q \in \mathcal{Q} ,\, \supp q \cap \cu \neq \emptyset } \nabla_q^* \cdot \a_q \nabla_q u
 = h
& \mbox{in} & \ \cu^\circ \times \Omega, 
\\ & 
u = \bar u & \mbox{on} & \ \partial \cu \times\Omega,
\end{aligned}
\right.
\end{equation}
which satisfies, for a constant $C(d,\beta)<\infty$, the estimate
\begin{equation}
\left\| u \right\|_{\underline{H}^1(\cu,\mu_\beta)}
\leq 
C \left( R \left\| h \right\|_{ \underline{L}^2(\cu,\mu_\beta)} + \left\| \bar u \right\|_{\underline{H}^1(\cu,\mu_\beta)} \right). 
\end{equation}
\end{lemma}
\begin{proof}
 A function $u\in \bar u + H^1_0(\cu, \mu_\beta)$ is a solution of~\eqref{e.HS.eqn.dir} if and only if  
\begin{multline}
\sum_{y\in \cu} \sum_{x\in \cu} \left\langle (\partial_y u(x,\cdot) )(\partial_yw (x,\cdot) ) \right\rangle_{\mu_\beta}
+
\frac{1}{2\beta}\sum_{x\in \cu} 
\left\langle \nabla u(x,\cdot) \nabla w(x,\cdot)  \right\rangle_{\mu_\beta} 
+ \frac{1}{2\beta}\sum_{n \geq 1} \frac{1}{\beta^{ \frac n2}}\sum_{x\in \cu} 
\left\langle   \nabla^{n+1} u(x,\cdot), \nabla^{n+1} w(x,\cdot)  \right\rangle_{\mu_\beta}  
\\
+
\sum_{q \in \mathcal{Q} ,\, \supp q \cap \cu \neq \emptyset }\left\langle \nabla_q u \cdot \a_q \nabla_q w \right\rangle_{\mu_\beta}  
=
\sum_{x\in \cu}  \left\langle h(x,\cdot) w(x,\cdot) \right\rangle_{\mu_\beta}, \quad \forall w\in H^1_0(\cu,\mu_\beta). 
\end{multline}
For large $\beta$,  the symmetric bilinear form on the left side of the previous display is is a small perturbation of the Laplacian term, and therefore $\left\llbracket u \right\rrbracket_{\underline{H}^1(\cu,\mu_\beta)}
\leq 
C \left( R \left\| h \right\|_{ \underline{L}^2(\cu,\mu_\beta)} + \left\| \bar u \right\|_{ \underline{H}^1(\cu,\mu_\beta)} \right)$. Apply Lemma \ref{l.spectralgap.U} (i), we see that the bilinear form is coercive with respect to the $H^1(\cu,\mu_\beta)$ norm. The
Lax-Milgram lemma therefore yields the existence of a unique solution $u\in H^1_0(\cu,\mu_\beta)$. It also follows that the solution to \eqref{e.HS.eqn.dir} admits a variational characterization: $u$ minimizes the energy 
\begin{multline*}
  \sum_{y \in \Zd} \left\| \partial_y u \right\|_{L^2 \left( \cu , \mu_\beta \right)}^2 +  \frac1{2\beta}  \left\| \nabla  u\right\|_{L^2 \left( \cu , \mu_\beta \right)}^2 + \sum_{n \geq 1} \frac1{\beta^{\frac n2}}\left\| \nabla^{n+1}  u \right\|_{L^2 \left( \Zd, \mu_\beta \right)}^2 \\  +  \sum_{q\in \mathcal{Q}, \supp q \cap \cu \neq \emptyset}  \left\langle \nabla_q u \cdot \a_q  \nabla_q u   \right\rangle_{\mu_{\beta}}
  - \sum_{x\in \cu}  \left\langle h(x,\cdot) \nabla u(x,\cdot) \right\rangle_{\mu_\beta},
\end{multline*}
among all the functions in the space $\bar u + H^1_0 \left( \cu , \mu_\beta \right).$
\end{proof}

\subsection{Green's matrix: existence, decay and homogenization}
\label{s.homog}
In this section, we present record some properties of the Green's matrix associated to the elliptic Helffer-Sj\"{o}strand operator $\L$. We highlight that since the operator $\L$ is an elliptic system, the fundamental solution is a matrix; this object is used repeatedly in the following chapters as it allows to decompose the solution of the Helffer-Sj\"{o}strand equation stated in~\eqref{e.HS.eqn.Zd}. 
We fix an exponent $p \in [1, \infty ]$, and a function $\f \in L^p \left( \mu_\beta \right)$ and define the elliptic Green's matrix $\G_\f:  \Zd \times \Omega \times \Zd \mapsto \R^{\binom d2 \times \binom d2}$ by the formula 
\begin{equation}
\label{e.Green}
\mathcal{L} \G_\f(x  , \phi ; y) = \f( \phi) \delta_y(x) ~\mbox{in}~ \Zd \times \Omega,
\end{equation}
such that $\left\| \G_\f(x ,\cdot ; y) \right\|_{L^p \left( \mu_\beta \right)}$ tends to $0$ as $x$ tends to infinity. To be slightly more precise in the definition, we see the Dirac $\delta_y$ to be the diagonal matrix $\delta_y(x) := \left( \indc_{x = y} \cdot \indc_{i=j} \right)_{1 \leq i, j \leq \binom d2}$. To solve the equation~\eqref{e.Green}, we fix a column in the matrix $\mathbf{f} \delta_y$, solve the system~\eqref{e.Green} (with this specific column) and obtain a function valued in the space $\R^{\binom d2}$. We then perform the same operation on the $\binom d2 -1$ columns and use the $\binom d2$ solutions obtained this way to define the matrix $\G_\f(\cdot ,\cdot ; y)$.

In the case $p=2$, we can solve \eqref{e.Green} variationally, by applying the Gagliardo-Nirenberg-Sobolev inequality. The solvability in the general case relies on the Feynman-Kac formula and tools from spectral theory as presented in the introduction of Chapter~\ref{section:section4} (following the ideas of~\cite[Section 2.2.2]{NS}) 

\begin{lemma}
\label{l.Greenvariation}
There exists $\beta_0 = \beta_0(d)$, such that for all $\beta>\beta_0$, there exists $C=C(d,\beta)$, such that the Green's function $\G_\f$ defined from \eqref{e.Green} satisfies $\left\| \G_\f \right\|_{L^{2^*}(\Zd,L^2 \left(\mu_\beta)\right)} \leq C \|\f\|_{L^2(\mu_\beta)}$. 
\end{lemma}

\begin{proof}
The proof follows from the same argument as Lemma \ref{l.wellposeHS.Zd}. We test \eqref{e.Green} using $\G_\f$, and observing that for large $\beta$, $\L$ is a small perturbation of $\frac{1}{2\beta}\Delta$, to obtain 
 \begin{equation*}
 \left\| \nabla \G_\f \right\|_{L^{2}(\Zd,\mu_\beta)}^2 \leq C \langle \G_\f(y) \f(\cdot) \rangle_{\mu_\beta}
 \leq C \left\| \G_\f \right\|_{L^{2^*}\left(\Zd,L^2 \left(\mu_\beta\right) \right)} \left\| \f \right\|_{L^{2}(\mu_\beta)}.
 \end{equation*}
 Applying the Gagliardo-Nirenberg-Sobolev inequality, we have $\left\| \G_{\mathbf{f}} \right\|_{L^{2^*}\left(\Zd,L^2 \left( \mu_\beta\right) \right)} \leq C(d)\left\| \nabla \G_{\mathbf{f}} \right\|_{L^{2}(\Zd,\mu_\beta)} $.
\end{proof}
 
The following proposition quantifies the solvability lemma and prove asymptotic decays on the $L^p\left( \mu_\beta \right)$-norm of the Green's matrix, its gradient and its mixed derivative. The proof is based on regularity theory for the Helffer-Sj\"{o}strand operator and is one of the main subject of Chapter~\ref{section:section4}.

\begin{proposition} \label{prop.prop4.11chap4}
For any regularity exponent $\ep > 0$, there exists an inverse temperature $\beta_0 (d , \ep) < \infty$ such that the following result holds. For any $\beta > \beta_0$, there exists a constant $C(d ) < \infty$ such that for any $x \in \Zd$, one has the estimates
\begin{equation*}
\left\| \G_\f (x , \cdot;y ) \right\|_{L^p \left( \mu_\beta \right)} \leq \frac{C \beta \left\| \f \right\|_{L^p \left( \mu_\beta \right)}}{|x-y|^{d-2}},
\end{equation*}
and the regularity estimates on the gradient and the mixed derivative
\begin{equation*}
\left\| \nabla_x \G_\f (x, \cdot ; y ) \right\|_{L^p\left( \mu_\beta \right)}  \leq \frac{C \beta \left\| \f \right\|_{L^p \left( \mu_\beta \right)} }{|x-y|^{d - 1 - \ep}} \hspace{10mm} \mbox{and} \hspace{10mm} \left\| \nabla_x \nabla_y \G_\f (x ,\cdot ; y) \right\|_{L^p\left( \mu_\beta \right)}  \leq \frac{C \beta \left\| \f \right\|_{L^p \left( \mu_\beta \right)}}{|x-y|^{d - \ep}}.
\end{equation*}
\end{proposition}

\subsection{Homogenization of the mixed derivative of the Green's matrix}
In this section, we state a quantitative result which establishes homogenization for the mixed gradient of the Green's matrix associated to the Helffer-Sj\"ostrand operator. The proof of the theorem below will be presented in Chapters~\ref{section5} and~\ref{sec:section6}. Naturally, one expects that the Green's matrix associated to the Helffer-Sj\"ostrand operator \eqref{e.HSop}, defined by 
\begin{equation*}
   \L G = \delta_0 ~\mbox{in}~\Zd \times \Omega
\end{equation*}
 homogenizes to the Green's matrix $\bar G$ associated to the Laplacian operator $\nabla \cdot \ahom_\beta \nabla $: 
\begin{equation}
\label{e.Greenhom}
    -\nabla \cdot \ahom_\beta \nabla \bar G = \delta_0 ~\mbox{in}~\Zd,
\end{equation}
where $\ahom_\beta$ is a positive definite matrix which is a small perturbation of the matrix $\frac{1}{2\beta} I_d$. It is defined in Chapters~\ref{section5} and~\ref{sec:section6} as the limit of the energy associated to the Dirichlet problem with affine boundary condition (see Definition~\ref{def.defnunustar} and Corollary~\ref{cor:coro5.13} in Chapter~\ref{section5} and the introduction of Chapter~\ref{sec:section6}); it is deterministic and depends only on the dimension $d$ and the inverse temperature $\beta$. The solvability of the equation~\eqref{e.Greenhom} is ensured by the fact that $\ahom_\beta$ is a perturbation of a diagonal matrix and the arguments of Chapter~\ref{section:section4}.

When applying to the Villain model (see computations in Chapter~\ref{section3.4}) we need a largely nonlinear generalization of the homogenization result \eqref{e.Greenhom} and also the convergence of the mixed gradient of the Green's matrix. We present as Theorem \ref{thm:homogmixedderchap4} below and the proof is deferred to Chapters~\ref{section5} and~\ref{sec:section6}.

 For each pair of integers $(i,j) \in \{ 1 , \ldots, d \} \times \left\{1 , \ldots, \binom d2 \right\}$, we let $l_{e_{ij}}$ be the linear function defined by the formula
    \begin{equation*}
    l_{e_{ij}} := \left\{\begin{aligned}
    \R^{d} &\to \R^{\binom d2}, \\
    x &\to \left( 0, \ldots, x \cdot e_i, \ldots , 0 \right),
    \end{aligned} \right.
\end{equation*}
where the term $ x \cdot e_i$ appears in the $j$-th position.
We denote by $\nabla \chi_{ij}$ the gradient of the infinite-volume corrector, which is the unique stationary solution of the Helffer-Sj\"{o}strand equation
\begin{equation*}
    \mathcal{L} \left( l_{e_{ij}} + \chi_{ij} \right) = 0 ~\mbox{in}~ \Zd \times \Omega.
\end{equation*}
It is constructed as the infinite volume limit of finite volume correctors, the later measures the homogenization error for the Helffer-Sj\"ostrand equation with affine boundary condition. For a precise definition, see Proposition~\ref{prop5.26} of Chapter~\ref{section5}. Once equipped with the gradient of the corrector, we can define the exterior derivative $\di^* \chi_{ij}$ by using that the codifferential $\di^*$ is a linear functional of the gradient (see \eqref{eq:formLkdde}). The following theorem proves a quantitative homogenization result for a version of the mixed derivative of the Green's function~\eqref{e.Green}, the specific form of the function~\eqref{eq:defmathcalUchap4} is justified by the fact that it is the correct object to consider in order to prove Theorem~\ref{t.main} in Chapter~\ref{section3.4}. We mention that the techniques developed in Chapters~\ref{section5} and~\ref{sec:section6} can be adapted to prove quantitative homogenization of more general solutions of the Helffer-Sj\"{o}strand equation.

\begin{theorem}[Homogenization of the mixed derivative of the Green's matrix] \label{thm:homogmixedderchap4}
We fix a charge $q_1 \in \mathcal{Q}$ such that $0$ belongs to the support of $n_{q_1}$, let $\mathcal{U}_{q_1}$ be the solution of the Helffer-Sj{\"o}strand equation
\begin{equation} \label{eq:defmathcalUchap4}
\L \mathcal{U}_{q_1} = \cos 2\pi\left( \phi , q_1\right) q_1 ~\mbox{in}~\Zd \times \Omega,
\end{equation}
and let $\bar G_{q_1} := \left( \bar G_{q_1, 1} , \ldots , \bar G_{q_1, \binom d2} \right)$ be the map defined by the formula, for each integer $k \in \left\{ 1 , \ldots , \binom d2 \right\}$,
\begin{equation}
\label{e.Gq}
    \bar G_{q_1, k} = \sum_{1 \leq i \leq d} \sum_{1 \leq j\leq \binom d2} \left\langle \cos 2\pi\left( \phi , q_1\right) \left( n_{q_1} , \di^* l_{e_{ij}} + \di^* \chi_{ij} \right)   \right\rangle_{\mu_\beta} \nabla_i \bar G_{jk}.
\end{equation}
There exist an inverse temperature $\beta_0 := \beta_0(d) <\infty$, an exponent $\gamma := \gamma(d) >0$ and a constant $C_{q_1}$ which satisfies the estimate
$\left| C_{q_1} \right| \leq  C \left\| q_1 \right\|_1^{k}$ for some $C := C(d,\beta) <\infty$ and $k := k(d) < \infty$, such that for each $\beta \geq \beta_0$ and each radius $R \geq 1$, one has the inequality
\begin{equation} \label{eq:TV19253.main}
    \left\| \nabla \mathcal{U}_{q_1} - \sum_{1 \leq i \leq d} \sum_{1 \leq j\leq \binom d2} \left( e_{ij} + \nabla \chi_{ij}\right) \nabla_i \bar G_{q_1, j}\right\|_{\underline{L}^2 \left( 
    B_{2R}  \setminus B_R, \mu_\beta \right)} \leq \frac{C_{q_1}}{R^{d + \gamma}}.
\end{equation}
\end{theorem}

\begin{remark}
    The functions $\nabla \mathcal{U}_{q_1}$ and $\nabla_i \bar G_{q_1}$ behaves like mixed derivative of Green's matrices, in particular, they should decay like the map $x \to |x|^{-d}$. Theorem~\ref{thm:homogmixedderchap4} states that their difference is quantitatively closer than the typical size of the two functions: we obtain an algebraic rate of convergence with additional exponent $\gamma > 0$ in the right side of~\eqref{eq:TV19253.main}.
\end{remark}

\begin{remark} \label{remark4.914470302}
For the purposes of Chapter~\ref{section3.4}, we record here that the statement of Theorem~\ref{thm:homogmixedderchap4} can be simplified by using the formalism of discrete differential forms. To this end, we recall the definition of the operator $L_{2 , \di^*}$ introduced in Section~\ref{sec:diffform=vectfct} of Chapter~\ref{Chap:chap2} record the following properties:
\begin{itemize}
    \item The operator $- \nabla \cdot \ahom_\beta \nabla$ can be written  
    \begin{equation} \label{eq:TV11490302}
        - \nabla \cdot \ahom_\beta \nabla = \frac{1}{2\beta} \left( \di^* \di + \left(1 + \bar \lambda_\beta \right) \di \di^* \right),
    \end{equation}
    where $\bar \lambda_\beta$ is a real coefficient which is small tends to $0$ as $\beta$ tends to infinity. This property is stated in Remark~\ref{remark1.1110320302} of Chapter~\ref{section5};
    \item The gradient of the infinite volume corrector only depends on the value of the codifferential $\di^* l_{e_{ij}}$ (in particular, it is equal to $0$ if $\di^* l_{e_{ij}} = 0$) as mentioned in Remarks~\ref{remark4.20302} and~\ref{remark09110302} of Chapter~\ref{section5}. We use the notation of  Remark~\ref{remark09110302}: given an integer $k \in \{ 1 , \ldots , d\}$, we let select a vector $p:= \sum_{1 \leq i \leq d} \sum_{1 \leq j\leq \binom d2} p_{ij} e_{ij}$ such that $\di l_{p} = e_k$ and denote by $\nabla \chi_k := \sum_{1 \leq i \leq d} \sum_{1 \leq j\leq \binom d2} p_{ij} \nabla \chi_{ij}$.
\end{itemize}
Using these ingredients, we can rewrite the definition of the map $\bar G_{q_1 , k}$ stated in~\eqref{e.Gq}: we have
\begin{equation*}
    \bar G_{q_1, k} = \sum_{1 \leq i \leq d} \left\langle \cos 2\pi\left( \phi , q_1\right) \left( n_{q_1} , e_i + \di^* \chi_{i} \right)   \right\rangle_{\mu_\beta} \left( \di^* \bar G_{\cdot k} \right) \cdot e_i.
\end{equation*}
We then use that, by definition, the map $\bar G_{\cdot , k}$ solves the equation $- \nabla \cdot \ahom_\beta \nabla \bar G = \delta_0$ and the identities $ - \Delta = \di \di^* + \di^* \di$, $\di \circ \di = 0$ and $\di^* \circ \di^* = 0$ to write
\begin{align*}
    - \left(1 + \bar \lambda_\beta\right) \Delta \di^* \bar G =   \left(1 + \bar \lambda_\beta\right) \left( \di \di^* + \di^* \di \right) \di^* \bar G_{\cdot k} & =  \left(1 + \bar \lambda_\beta\right)  \di^* \di  \di^* \bar G_{\cdot k} \\
    & = \di^* \left( \di^*  \di \bar G_{\cdot k} +  \left(1 + \bar \lambda_\beta\right) \di  \di^* \right) \bar G_{\cdot k} \\
    & = \di^* \left( - \nabla \cdot \ahom_\beta \nabla \bar G_{\cdot , k} \right) \\
    & = \di^* \delta_0.
\end{align*}
The exterior derivative $\di^* \bar G$ can thus be explicitly computed in terms of the gradient of the Green's function associated to the operator $- \left(1 + \bar \lambda_\beta\right) \Delta$ which is equal to the the standard random walk Green's function on the lattice $\Zd$ multiplied by the value $\left( 1 + \bar \lambda_\beta\right)^{-1}$.
\end{remark}

\chapter{First-order expansion of the two-point function} \label{section3.4}

In this chapter we show that by combining Theorem~\ref{thm:homogmixedderchap4}, which gives a quantitative rate of convergence of the mixed gradients of the Helffer-Sj{\"o}strand Green's matrix, with a regularity theory for the Helffer-Sj{\"o}strand operator, implies the convergence of the two-point function stated in Theorem~\ref{t.main}.  The proof relies on the regularity theory that is developed in Chapter \ref{section:section4}.

The objective of this chapter is to prove Theorem~\ref{t.main}. To this end, by Proposition~\ref{p.2pt} of Chapter~\ref{chap:chap3}, it is enough to prove the expansion stated in the following theorem.
\begin{theorem} \label{thm.expdualVil}
There exist constants $\beta_0 := \beta_0 (d), \, c_0 := c_0 \left( \beta , d \right),  c_1 \left( \beta , d \right)$ and an exponent $\gamma' := \gamma'(d) >0$ such that for every $\beta > \beta_0$, and every $x \in \Zd$,
\begin{equation*}
    \frac{Z \left( \sigma_{0x} \right)}{Z(0)} = c_0 + \frac{c_1}{|x|^{d-2}} + O \left( \frac{C}{|x|^{d-2+\gamma'}} \right),
\end{equation*}
and 
\begin{equation*}
    \frac{Z \left(\bar \sigma_{0x} \right)}{Z(0)} = c_0 + \frac{\bar c_1}{|x|^{d-2}} + O \left( \frac{C}{|x|^{d-2+\gamma'}} \right),
\end{equation*}
\end{theorem}
The proof of Theorem~\ref{thm.expdualVil} requires to use the following statements established in Chapter~\ref{section:section4}, Chapter~\ref{section5} and Chapter~\ref{sec:section6}:
\begin{itemize}
    \item We need to use the quantitative homogenization of the mixed derivative of the Green's function associated to the Helffer-Sj{\"o}strand equation $\L$. The precise statement we need to use is given in Theorem~\ref{thm:homogmixedderchap4}. The proof of this theorem is the subject of Chapters~\ref{section5} and~\ref{sec:section6};
    \item We need to use the $C^{0,1-\ep}$ regularity theory established in Chapter~\ref{section:section4}; more specifically, we need to use the regularity estimates for the Green's function $\mathcal{G}$ associated to the operator $\L$ stated in Proposition~\ref{prop.prop4.11chap4} of Chapter \ref{chap:chap3} and on the Green's function $\G_{\mathrm{der}}$ stated in Proposition~\ref{cor:corollary4.14} of Chapter~\ref{section:section4}. We additionally make the assumption that the regularity exponent $\ep$ is very small compared to the exponent $\gamma$ which appears in the statement of Theorem~\ref{thm:homogmixedderchap4} of Chapter~\ref{chap:chap3} (for instance, we assume that the ratio $\frac{\gamma}{\ep}$ is larger than $100d$). This condition can always be ensured by increasing the inverse temperature $\beta$.
\end{itemize}
Apart from these three results, the proof of Theorem~\ref{thm.expdualVil}, which is contained in this chapter (and Chapter~\ref{section7} for the technical estimates) is largely independent from Chapters~\ref{section:section4},~\ref{section5} and~\ref{sec:section6}. 

Finally, some computations presented in this chapter requires to prove estimates on terms of the form
\begin{equation*}
    \sum_{x \in \Zd} \frac{1}{|x|^{\alpha}} \frac{1}{|x - y|^{\beta}},
\end{equation*}
for some exponents $\alpha, \beta > 0$ satisfying $\alpha + \beta > d$. We refer to Appendix~\ref{app.appC} for the proof of the upper bounds and directly write the results in the sections below.

This chapter is organized as follows. We first set up the argument and introduce some preliminary notations in Section~\ref{subsec3.4.1}. We then simplify the expression~\eqref{eq:TV11102} below in a series of technical lemmas stated in Sections~\ref{sec:chap4.1},~\ref{chap4sec2Rcc} and~\ref{chap4sec3De}. In particular, in Sections ~\ref{chap4sec2Rcc} and~\ref{chap4sec3De}, we sketch the argument that one can decouple the Helffer-Sj{\"o}strand Green's matrix from the exponential terms arising from the dual model in Chapter~\ref{chap:chap3}. The proofs of these lemmas relies on the $C^{0,1-\ep}$-regularity theory established in Chapter~\ref{section:section4}, we give an outline of the arguments and postpone the proof to Chapter~\ref{section7}. The core of the proof of Theorem \ref{thm.expdualVil} (thus Theorem~\ref{t.main}) is given in Section~\ref{Chap4sec4Roetpf}. This section is decomposed into two subsections. We first write an outline of the argument in Section~\ref{subsec3.4.1} and then present the details of the proof in Section~\ref{sectionsection4.2789}. 

\section{Preliminary notations} \label{subsec3.4.1}

We first recall that we have the identity
    \begin{equation} \label{eq:TV11102}
    \frac{Z_\beta (\sigma_{0x})}{Z_\beta(0)} =  \left\langle \exp \left( \sum_{q \in \mathcal{Q}} z(\beta , q) \sin 2\pi (\phi , q) \sin 2\pi(\sigma_{0x} , q) + \sum_{q\in \mathcal{Q}} z(\beta , q) \cos 2\pi(\phi , q) \left( \cos 2\pi(\sigma_{0x} , q) - 1  \right) \right) \right\rangle_{\mu_{\beta}}.
\end{equation}
We also recall that, by the definition of the function $\sigma_{0x}$ given in Section~\ref{Chap3sec1} of Chapter~\ref{chap:chap3}, we have the equality
\begin{align*}
    \di^* \sigma_{0x}  = \di^* \di \left( - \Delta\right)^{-1} h_{0x}
                        =  h_{0,x} - \di \di^* \left( - \Delta\right)^{-1} h_{0x} 
                         & = h_{0,x} - \di  \left( - \Delta\right)^{-1} \di^* h_{0x} \\
                         & = h_{0,x} - \di  \left( - \Delta\right)^{-1} \left(\indc_x - 
                        \indc_0 \right) \\
                        & = h_{0,x} + \nabla G - \nabla G_x.
\end{align*}
We then use the identity $q = \di n_q$, that the maps $q$, $n_q$ and $h_{0,x}$ are valued in $\Z$, and the periodicity of the sine and the cosine to deduce that
\begin{equation*}
    \sin 2\pi(\sigma_{0x} , q) =  \sin 2\pi(\nabla G - \nabla G_x , n_q) \hspace{5mm} \mbox{and} \hspace{5mm}  \cos 2\pi(\sigma_{0x} , q)  =  \cos 2\pi(\nabla G - \nabla G_x , n_q).
\end{equation*}
One can then expand the sine and the cosine by using the trigonometric formulas. We obtain the identities
\begin{multline} \label{eq:TV11112}
    \sin 2\pi(\nabla G - \nabla G_x , n_q) =  \sin 2\pi(\nabla G , n_q) - \sin 2\pi(\nabla G_x , n_q) \\ + \left( \cos 2\pi(\nabla G_x , n_q) -1 \right) \sin 2\pi(\nabla G , n_q) - \left( \cos 2\pi(\nabla G , n_q) -1 \right) \sin 2\pi(\nabla G_x , n_q),
\end{multline}
and
\begin{multline}  \label{eq:TV11122}
    \cos 2\pi(\nabla G - \nabla G_x , n_q) - 1 = \left( \cos 2\pi(\nabla G , n_q)  - 1  \right) \left( \cos 2\pi(\nabla G_x , n_q)  - 1  \right) \\ + \left( \cos 2\pi(\nabla G , n_q) - 1 \right) + \left(\cos 2\pi(\nabla G_x , n_q) - 1 \right)+ \sin 2\pi(\nabla G , n_q) \sin 2\pi(\nabla G_x , n_q). 
\end{multline}
We then combine the identities~\eqref{eq:TV11112} and~\eqref{eq:TV11122} with the right side of~\eqref{eq:TV11102}. To ease the notation, we introduce the following random variables
\begin{equation} \label{eq:moneq}
    \left\{ \begin{aligned}
    X_{x} &:= \exp \left( -\sum_{q \in \mathcal{Q}} z(\beta , q) \left( \sin 2\pi(\phi , q) \sin 2\pi(\nabla G_x  , n_q)  - \frac12\cos 2\pi(\phi , q) \left( \cos 2\pi(\nabla G_x , n_q) - 1 \right)  \right)\right), \\  
    Y_{0} &:= \exp \left(  \sum_{q \in \mathcal{Q}} z(\beta , q) \left( \sin 2\pi(\phi , q)\sin 2\pi(\nabla G  , n_q)  + \frac12 \cos 2\pi(\phi, q) \left( \cos 2\pi(\nabla G , n_q) - 1 \right)\right) \right), \\
            Y_{x} &:= \exp \left( \sum_{q \in \mathcal{Q}} z(\beta , q) \left( \sin 2\pi(\phi , q)\sin 2\pi(\nabla G_x  , n_q)  + \frac12 \cos 2\pi(\phi, q) \left( \cos 2\pi(\nabla G_x , n_q) - 1 \right)\right) \right), \\
    X_{\sin \cos} & := \exp \left( - \sum_{q \in \mathcal{Q}} z(\beta , q) \sin 2\pi(\phi , q) \sin 2\pi( \nabla G_x  , n_q) \left( \cos 2\pi(\nabla G  , n_q) - 1 \right) \right) \\ & \quad \times \exp \left( \sum_{q \in \mathcal{Q}} z(\beta , q) \sin 2\pi(\phi , q)  \sin 2\pi(\nabla G  , n_q) \left( \cos 2\pi(\nabla G_x  , n_q) - 1 \right)  \right),\\
    X_{\cos \cos} & := \exp \left(\sum_{q \in \mathcal{Q}} z(\beta , q) \cos 2\pi(\phi , q)   \left( \cos 2\pi (\nabla G , n_q) - 1 \right) \left(\cos 2\pi(\nabla G_x , n_q) -1 \right) \right),\\
    X_{\sin \sin} & := \exp \left( \sum_{q \in \mathcal{Q}} z(\beta , q) \cos 2\pi(\phi , q) \sin 2\pi(\nabla G , n_q) \sin 2\pi(\nabla G_x , n_q) \right).
    \end{aligned} \right.
\end{equation}
In this notation we have
\begin{equation} \label{eq:TV11282}
    \frac{Z_\beta (\sigma_{0x})}{Z_\beta(0)} = \left\langle Y_0X_x X_{\sin \cos}  X_{\cos \cos}X_{\sin \sin}  \right\rangle_{\mu_\beta}.
\end{equation}
Our aim is then to simplify the identity~\eqref{eq:TV11282} and then to apply Theorem~\ref{thm:homogmixedder}.

\section{Removing the terms \texorpdfstring{$X_{\sin \cos},$ $X_{\cos \cos}$ and $X_{\sin \sin}$}{1}} \label{sec:chap4.1}
We first show that the terms $X_{\sin \cos},$ $X_{\cos \cos}$ and $X_{\sin \sin}$ are lower order terms which can be removed from the analysis. We prove the following lemma.

\begin{replemma}{lem.lemma7.1}
There exist constants $\beta_0 := \beta_0 (d) < \infty$, $c := c(d , \beta)$ and $C := C(d, \beta)$ such that for each $\beta > \beta_0$,
\begin{equation} \label{eq.TV12012}
     \frac{Z_\beta (\sigma_{0x})}{Z_\beta(0)} = \left\langle Y_0 X_x \right\rangle_{\mu_\beta} + \frac{c \left\langle Y_0 X_x \right\rangle_{\mu_\beta}}{|x|^{d-2}} + O \left( \frac{C}{|x|^{d-1}} \right).
\end{equation}
A consequence of the identity~\eqref{eq.TV12012} is the equivalence
\begin{equation*}
    \exists c_1 , c_2 \in \R, \frac{Z_\beta (\sigma_{0x})}{Z_\beta(0)} = c_1 + \frac{c_2}{|x|^{d-2}} + O \left( \frac{C}{|x|^{d-2 + \gamma'}} \right) \iff  \exists c_1, c_2 \in \R, \left\langle Y_0 X_x \right\rangle_{\mu_\beta} = c_1+ \frac{c_2}{|x|^{d-2}} + O \left( \frac{C}{|x|^{d-2 + \gamma'}} \right).
\end{equation*}
\end{replemma}

This lemma is technical and its proof is not the core of the argument; the proof is thus deferred to Chapter~\ref{section7}. We provide here a sketch of the argument.

\begin{proof}[Sketch of the proof of Lemma~\ref{lem.lemma7.1}]
To prove the identity~\eqref{eq.TV12012}, we first record four standard inequalities, for each $y \in \Zd$, and each $a \in \R$,
\begin{equation} \label{eq:TV12472}
    \left| \nabla G(y) \right| \leq \frac{C}{|y|^{d-1}}, \hspace{3mm}  \left| \nabla G_x(y) \right| \leq \frac{C}{|y - x|^{d-1}}, \hspace{3mm} \left| \sin a \right| \leq |a| \hspace{3mm} \mbox{and} \hspace{3mm} \left| \cos a -1 \right| \leq \frac12 |a|^2.
\end{equation}
Using the estimates~\eqref{eq:TV12472} and the exponential decay of the coefficient $z(\beta, q)$, we prove the following estimates:
\begin{itemize}
    \item[(i)] The random variables $X_{\sin \cos}$ and $X_{\cos \cos}$ belongs to the space $L^\infty \left( \mu_\beta \right)$ and satisfy the estimates
    \begin{equation} \label{eq:TV18392}
    \left\{ \begin{aligned}
    \left\| X_{\sin \cos} - 1 \right\|_{L^\infty} \leq \frac{C}{|x|^{d-1}}, \\
    \left\| X_{\cos \cos} - 1 \right\|_{L^\infty} \leq \frac{C}{|x|^{d-1}}.
    \end{aligned} \right.
    \end{equation}
    \item[(ii)] We prove that the random variable $X_{\sin \sin}$ also belongs to the space $L^\infty \left( \mu_\beta \right)$ and that its fluctuations around the value $1$ are of order $|x|^{2-d}$. This is larger than the fluctuations of the random variables $X_{\sin \cos}$ and $X_{\cos \cos}$ and one needs to be more precise in the analysis: we prove the following estimates on the expectation and the variance $X_{\sin \sin}$
        \begin{equation} \label{eq:TV19202}
    \left\{ \begin{aligned}
    \var_{\mu_\beta} X_{\sin \sin} & \leq \frac{C}{|x|^{2d-2}}, \\
    \left\langle X_{\sin \sin} \right\rangle_{\mu_\beta} & = 1 + \frac{c}{|x|^{d-2}} + O \left( \frac{C}{|x|^{d-1}} \right).
    \end{aligned} \right.
    \end{equation}
    The variance is estimated thanks to the Brascamp-Lieb inequality and the expectation is estimated thanks to the estimates~\eqref{eq:TV12472} and a Taylor expansion of the exponential.
\end{itemize}
A combination of the estimates~\eqref{eq:TV18392} and~\eqref{eq:TV19202} is then sufficient to prove Lemma~\ref{lem.lemma7.1}.
\end{proof}

\remark The same proof also yields 
\begin{equation} \label{eq.TV12012plus}
     \frac{Z_\beta (\bar \sigma_{0x})}{Z_\beta(0)} = \left\langle Y_0 Y_x \right\rangle_{\mu_\beta} + \frac{\bar c \left\langle Y_0 Y_x \right\rangle_{\mu_\beta}}{|x|^{d-2}} + O \left( \frac{C}{|x|^{d-1}} \right).
\end{equation}
In general, $\bar c \neq c $ since the $O(|x|^{d-2})$ term above is contributed by $\langle X_{\sin \sin}^{-1}\rangle_{\mu_\beta}$ instead of $\langle X_{\sin \sin}\rangle_{\mu_\beta}$. 

\section{Removing the contributions of the cosines} \label{chap4sec2Rcc}
From Lemma~\ref{lem.lemma7.1}, we see that to prove Theorem~\ref{t.main}, it is sufficient to obtain the following expansion
\begin{equation} \label{eq:TV10583}
    \exists c_1, c_2 \in \R, \hspace{3mm} \left\langle Y_0 X_x \right\rangle_{\mu_\beta} = c_1+ \frac{c_2}{|x|^{d-2}} + O \left( \frac{C}{|x|^{d-2 + \gamma'}} \right).
\end{equation}
We then note that, by the translation invariance of the measure $\mu_\beta$, the expectation of the random variable $X_x$ does not depend on the point $x$: we have, for each $x \in \Zd$, $\left\langle X_x \right\rangle_{\mu_\beta} = \left\langle X_0 \right\rangle_{\mu_\beta}$.
A consequence of this observation is that to prove~\eqref{eq:TV10583}, it is sufficient to show
\begin{equation} \label{eq:TV10593}
    \cov \left[ X_x , Y_0 \right] = \frac{c_2}{|x|^{d-2}} + O \left( \frac{C}{|x|^{d-2 + \gamma'}} \right).
\end{equation}
Indeed, the expansion~\eqref{eq:TV10593} implies~\eqref{eq:TV10583} with the value $c_1 = \left\langle Y_0 \right\rangle_{\mu_\beta} \left\langle X_0 \right\rangle_{\mu_\beta}$.
To prove the identity~\eqref{eq:TV10583}, we use the Helffer-Sj{\"o}strand representation formula and write the covariance in the following form
\begin{equation} \label{eq:V07043}
     \cov \left[ X_x , Y_0 \right] = \sum_{y \in \Zd} \left\langle \left(\partial_y  X_x\right) \mathcal{Y}(y, \cdot ) \right\rangle_{\mu_\beta},
\end{equation}
where $\mathcal{Y}$ is the solution of the Helffer-Sj{\"o}strand equation, for each $(y , \phi) \in \Zd \times \Omega$,
\begin{equation} \label{def.wVill}
    \mathcal{L}\mathcal{Y}(y , \phi) = \partial_y Y_0(\phi).
\end{equation}
For each point $x \in \Zd$, we introduce the notation $Q_x$ to denote the (random) charge: for pair each $(y , \phi) \in \Zd \times \Omega$,
\begin{equation} \label{eq:TVdefcapitQ}
     Q_x(y , \phi) :=  2\pi \sum_{q \in \mathcal{Q}} z(\beta , q) \cos 2\pi(\phi , q) \sin 2\pi(\nabla G_x  , n_q)  q(y).
\end{equation}
These charges are defined so as to have the identities, for each $y \in \Zd$,
\begin{equation}\label{e.partialY}
    \partial_y Y_0(\phi) =   \left(Q_0(y, \phi)  - \frac12 2\pi \sum_{q\in \mathcal{Q}} z \left( \beta , q \right) \sin2\pi(\phi, q) \left( \cos 2\pi(\nabla G , n_q) - 1 \right) q(y) \right) Y_0(\phi)
\end{equation}
and
\begin{equation} \label{eq:TV14213}
    \partial_y X_x(\phi) = -\left(Q_x(y, \phi)  + \frac12 2\pi \sum_{q\in \mathcal{Q}}  z \left( \beta , q \right) \sin2\pi(\phi, q) \left( \cos 2\pi(\nabla G_x , n_q) - 1 \right) q(y) \right) X_x(\phi).
\end{equation}
We also define the random charges $n_{Q_x}$ according to the formula
\begin{equation} \label{eq:TV14223}
     n_{Q_x} :=  \sum_{q \in \mathcal{Q}} 2\pi z(\beta , q) \left( \cos2\pi(\phi , q) \sin2\pi(\nabla G_x  , n_q)  \right) n_q \hspace{5mm} \mbox{so that} \hspace{5mm} \di  n_{Q_x} = Q_x.
\end{equation}
We note that by the exponential decay $\left| z\left( \beta , q\right) \right| \leq C e^{-c \sqrt{\beta} \left\| q\right\|_1}$, the decay of the gradient of the Green's function stated in~\eqref{eq:TV12472} and the inequality $|\sin a| \leq |a|$, the random charges $Q_x$ and $n_{Q_x}$ satisfy the $L^\infty \left( \mu_\beta\right)$-estimate: for each $y \in \Zd$, 
\begin{equation} \label{eq:TV15583}
        \left\| Q_x(y, \cdot) \right\|_{L^\infty \left( \mu_\beta \right)} \leq \frac{C}{|y-x|^{d-1}} ~\mbox{and}~
        \left\| n_{Q_x}(y, \cdot) \right\|_{L^\infty \left( \mu_\beta \right)} \leq \frac{C}{|y- x|^{d-1}}.
\end{equation}
By a similar argument, but this time using the inequality $|\cos a - 1| \leq \frac 12 |a|^2$, one obtains the inequality, for each $y \in \Zd$,
\begin{equation} \label{eq:TV14183}
    \left| \sum_{q\in \mathcal{Q}}  z \left( \beta , q \right) \sin2\pi(\phi, q) \left( \cos 2\pi(\nabla G_x , n_q) - 1 \right) q(y) \right| \leq \frac{C}{|y - x|^{2d-2}}
\end{equation}
and
\begin{equation} \label{eq:TV14193}
    \left| \sum_{q\in \mathcal{Q}}  z \left( \beta , q \right) \sin2\pi(\phi, q) \left( \cos 2\pi(\nabla G_x , n_q) - 1 \right) n_q(y) \right| \leq \frac{C}{|y - x|^{2d-2}}.
\end{equation}
The reason we record the inequalities~\eqref{eq:TV14183} and~\eqref{eq:TV14193} is that, since $2d - 2 > d-1$, the function $x \mapsto |x|^{2d-2}$ decays faster than $x \mapsto |x|^{d-1}$. From this observation, we expect that the terms $Q_0(y)Y_0$ and $Q_x(y)X_x$ are the leading order terms in the identities \eqref{e.partialY} and~\eqref{eq:TV14213}  and that the terms involving the cosine of the gradient of the Green's functions (left sides of~\eqref{eq:TV14183} and~\eqref{eq:TV14193}) are lower order terms which can be removed from the analysis. We prove this result in the following lemma.

\begin{replemma}{p.removecos}[Removing the contributions of the cosines]
One has the identity
\begin{equation} \label{eq:TV14563}
    \cov \left[ X_x , Y_0 \right] = \sum_{y \in \Zd} \left\langle X_x Q_x(y) \mathcal{V}(y, \cdot ) \right\rangle_{\mu_\beta} + O \left( \frac{C}{|x|^{d - 1 - \ep}} \right),
\end{equation}
where $\mathcal{V}$ is the solution of the Helffer-Sj{\"o}strand equation, for each pair $(y, \phi ) \in \Zd \times \Omega$,
\begin{equation*}
    \mathcal{L} \mathcal{V}(y , \phi) = Q_0(y, \phi) Y_0(\phi).
\end{equation*}
A consequence of the identity~\eqref{eq:TV14563} is the equivalence
\begin{equation*}
\exists c_2 \in \R, ~\cov \left[ X_x , Y_0 \right] = \frac{c_2}{|x|^{d-2}} + O \left( \frac{C}{|x|^{d-2 + \gamma'}} \right) \hspace{2mm} \iff \hspace{2mm} \exists c_2 \in \R, ~ \sum_{y \in \Zd} \left\langle X_x Q_x(y) \mathcal{V}(y, \cdot ) \right\rangle_{\mu_\beta} = \frac{c_2}{|x|^{d-2}} + O \left( \frac{C}{|x|^{d-2 + \gamma'}} \right).
\end{equation*}
\end{replemma}

\begin{remark}
We recall that $\ep$ is the regularity exponents for the gradient and the mixed derivatives of the Green's function $\G$ and $\G_{\mathrm{der}}$ stated in Proposition~\ref{prop.prop4.11chap4} of Chapter \ref{chap:chap3} and Proposition~\ref{cor:corollary4.14} of Chapter~\ref{section:section4} respectively.
\end{remark}

The proof of this result is again technical and does not represent the core of the argument; it is thus deferred to Chapter~\ref{section7}. The argument relies on two ingredients:
\begin{itemize}
    \item[(i)] We use the decay estimates for the Green's function associated to the Helffer-Sj{\"o}strand operator $\mathcal{L}$ and its mixed derivative stated in Proposition~\ref{prop.prop4.11chap4} of Chapter~\ref{chap:chap3};
    \item[(ii)] We use the decay estimates~\eqref{eq:TV15583} and~\eqref{eq:TV14193} and take advantage of the fact that the function $x \mapsto |x|^{2d-2}$ decays faster than the map $x \mapsto |x|^{d-1}$.
\end{itemize}

We complete this section by recording that we may also prove 
\begin{equation}
    \exists c_1, c_2 \in \R, \hspace{3mm} \left\langle Y_0 Y_x \right\rangle_{\mu_\beta} = c_1+ \frac{c_2}{|x|^{d-2}} + O \left( \frac{C}{|x|^{d-2 + \gamma'}} \right)
\end{equation}
by showing
\begin{equation}
    \cov \left[ Y_x , Y_0 \right] = \frac{c_2}{|x|^{d-2}} + O \left( \frac{C}{|x|^{d-2 + \gamma'}} \right).
\end{equation}
Indeed, we have the following analogue of \eqref{eq:TV14563}
\begin{equation*} 
    \cov \left[ Y_x , Y_0 \right] = \sum_{y \in \Zd} \left\langle Y_x Q_x(y) \mathcal{V}(y, \cdot ) \right\rangle_{\mu_\beta} + O \left( \frac{C}{|x|^{d - 1 - \ep}} \right).
\end{equation*}
The proof of this identity is almost the same as  \eqref{eq:TV14563} with only notational changes, and is therefore omitted.

\section{Decoupling the exponentials} \label{chap4sec3De}
The next (and final) technical step consists in removing the exponential terms $X_x$ and $Y_0$ from the computation. To this end, we prove the decorrelation estimate stated in the following lemma.

\begin{replemma}{p.decoupleexp}[Decoupling the exponential terms]
One has the following estimate
\begin{equation} \label{eq:TV17423}
    \cov \left[ X_x , Y_0 \right]  = \left\langle Y_0 \right\rangle_{\mu_\beta} \left\langle X_0 \right\rangle_{\mu_\beta}  \sum_{y \in \Zd} \left\langle Q_x(y, \cdot) \mathcal{U}(y, \cdot ) \right\rangle_{\mu_\beta} + O \left( \frac{C}{|x|^{d-1+\ep}} \right),
\end{equation}
and 
\begin{equation} \label{eq:TV17423plus}
    \cov \left[ Y_x , Y_0 \right]  = \left\langle Y_0 \right\rangle_{\mu_\beta}^2 \sum_{y \in \Zd} \left\langle Q_x(y,\cdot) \mathcal{U}(y, \cdot ) \right\rangle_{\mu_\beta} + O \left( \frac{C}{|x|^{d-1+\ep}} \right).
\end{equation}
where the function $\mathcal{U}$ is the solution of the Helffer-Sj{\"o}strand equation $\mathcal{L} \mathcal{U} = Q_0$ in $\Zd \times \Omega$. The identity~\eqref{eq:TV17423} implies the equivalence
\begin{multline*}
    \exists c_2 \in \R, ~ \sum_{y \in \Zd} \left\langle X_x Q_x(y,\cdot) \mathcal{V}(y, \cdot ) \right\rangle_{\mu_\beta} = \frac{c_2}{|x|^{d-2}} + O \left( \frac{C}{|x|^{d-2 + \gamma'}} \right)  \\ \iff \hspace{3mm}  \exists c_2 \in \R, ~ \sum_{y \in \Zd} \left\langle Q_x(y) \mathcal{U}(y, \cdot ) \right\rangle_{\mu_\beta} = \frac{c_2}{|x|^{d-2}} + O \left( \frac{C}{|x|^{d-2 + \gamma'}} \right).
\end{multline*}
\end{replemma}

\begin{remark} \label{rem:rem3.13}
The function $\mathcal{U}$ can be decomposed according to the following procedure: if for each charge $q_1 \in \mathcal{Q}$, we denote by $\mathcal{U}_{q_1}$ the solution of the Helffer-Sj{\"o}strand equation
\begin{equation} \label{eq:TV09004}
\L \mathcal{U}_{q_1} = \cos 2\pi\left( \phi , q_1\right) q_1 ~\mbox{in}~\Zd \times \Omega,
\end{equation}
then we have the identity
\begin{equation} \label{eq:TV09555}
\mathcal{U} =  2\pi \sum_{q_1 \in \mathcal{Q}} z(\beta , q_1) \sin 2\pi(\nabla G  , n_{q_1})  \mathcal{U}_{q_1}.
\end{equation}
\end{remark}

\begin{remark} \label{rem:rem3.3}
By writing $q_1 = \di n_{q_1}$, we can rewrite the equation~\eqref{eq:TV09004} in the following form
\begin{equation*}
    \L \mathcal{U}_{q_1} = \di \left(\cos 2\pi\left( \cdot , q_1\right) n_{q_1} \right) ~\mbox{in}~\Zd \times \Omega.
\end{equation*}
As a consequence the function $\mathcal{U}_{q_1}$ can be expressed in terms of the Green's function $\mathcal{G}_{\cos 2\pi\left( \cdot , q_1\right) }$ according to the formula, for each pair $(y , \phi) \in \Zd \times \Omega$,
\begin{equation} \label{eq:VER192130}
    \mathcal{U}_{q_1}(y , \phi) = \sum_{z \in \supp n_{q_1}} \di^*_z \mathcal{G}_{\cos 2\pi\left( \cdot , q_1\right)} (y , \phi ; z) n_{q_1}(z).
\end{equation}
Using the decay estimate on the gradient and mixed derivative of the Green's function given in Proposition~\ref{prop.prop4.11chap4} of Chapter \ref{chap:chap3}, we obtain that the map $\mathcal{U}_{q_1}$ satisfies the upper bounds, for each $y \in \Zd$,
\begin{equation} \label{eq:TV10154}
    \left\| \mathcal{U}_{q_1}(y , \cdot)  \right\|_{L^\infty \left( \mu_\beta \right)} \leq \frac{C_{q_1}}{|y - z|^{d-1-\ep}} \hspace{3mm} \mbox{and} \hspace{3mm}  \left\| \nabla \mathcal{U}_{q_1}(y , \cdot)  \right\|_{L^\infty \left( \mu_\beta \right)} \leq \frac{C_{q_1}}{|y - z|^{d - \ep}},
\end{equation}
where $z$ is a point which belongs to the support of the charge $n_{q_1}$ (chosen arbitrarily).
\end{remark}

\begin{remark} \label{rem:rem3.4}
A consequence of the estimate~\eqref{eq:TV10154} is that by using the exponential decay of the coefficient $z \left( \beta, q \right)$ (see \eqref{e.zest} of Chapter~\ref{chap:chap3}) and the inequality, for each charge $q_1 \in \mathcal{Q}$,
\begin{equation*}
     \left| \sin 2\pi(\nabla G  , n_{q_1}) \right| \leq 2 \pi \left|(\nabla G  , n_{q_1}) \right| \leq 2 \pi \left\| \nabla G \right\|_{L^2 \left( \supp n_{q_1} \right)} \left\| n_{q_1} \right\|_{2} \leq \frac{C_q}{| z|^{d-1}},
\end{equation*}
where $z$ is a point in the support of $n_q$ (chosen arbitrarily), we deduce the inequality, for each point $y \in \Zd$,
\begin{align*}
    \left\| \mathcal{U}(y,\cdot)  \right\|_{L^\infty \left( \mu_\beta \right)} & \leq  2\pi \sum_{z \in \Zd} \sum_{q_1 \in \mathcal{Q}_z} \left| z(\beta , q_1) \sin 2\pi(\nabla G  , n_{q_1}) \right| \left\| \mathcal{U}_{q_1}(y,\cdot) \right\|_{L^\infty(\mu_\beta)} \\
    & \leq \sum_{z \in \Zd}  \sum_{q_1 \in \mathcal{Q}_z} e^{-c \sqrt{\beta} \left\| q \right\|_1} \frac{C_q}{|z|^{d-1} \times |y - z|^{d-1-\ep}}
    \\ & \leq C \sum_{z \in \Zd}  \frac{1}{|z|^{d-1} \times |y - z|^{d-1-\ep}} \\
    & \leq   \frac{C}{|y|^{d-2-\ep}}.
\end{align*}
where we used the exponential decay of the term $e^{-c \sqrt{\beta} \left\| q \right\|_1}$ to absorb the algebraic growth of the term $C_q \leq C \left\| q \right\|_1^k$ in the third inequality.
The same argument also yields the estimate
\begin{equation*}
    \left\| \nabla \mathcal{U}(y, \cdot)  \right\|_{L^\infty \left( \mu_\beta \right)} \leq \frac{C}{|y|^{d-1-\ep}}.
\end{equation*}
\end{remark}

We now give an heuristic argument explaining why we expect the decoupling estimate~\eqref{eq:TV17423} to hold.
\begin{proof}[Heuristic of the proof of Lemma~\ref{p.decoupleexp}]
The strategy of the proof is to first decouple the exponential term $X_x$ and then decouple the exponential term $Y_0$; to decouple the term $X_x$, we prove the expansion
\begin{equation} \label{eq:TV14164}
     \sum_{y \in \Zd} \left\langle X_x Q_x(y,\cdot) \mathcal{V}(y, \cdot ) \right\rangle_{\mu_\beta} = \left\langle X_0 \right\rangle_{\mu_\beta}\sum_{y \in \Zd} \left\langle  Q_x(y,\cdot) \mathcal{V}(y, \cdot ) \right\rangle_{\mu_\beta} + O \left( \frac{C}{|x|^{d - 1 - \ep}} \right).
\end{equation}
A heuristic reason justifying why one can expect the expansion~\eqref{eq:TV14164} to hold is the following. By the definition of the random variable $X_x$ given in~\eqref{eq:moneq} and the decay of the gradient of the Green's function $\nabla G_x$ stated in~\eqref{eq:TV12472}, we expect the random variable $X_x$ to essentially depend on the value of the gradient of the field around the point $x$. The statement is voluntarily vague; one could give a mathematical meaning to it arguing that if one considers a large constant $C$ depending only on the dimension $d$, then the conditional expectation of the random variable $X_x$ with respect to the sigma-algebra generated by the fields $\left( \nabla \phi(y) \right)_{y \in B(x , C)}$ is a good approximation of the random variable $X_x$ in the space $L^2 \left( \mu_\beta \right)$.  \smallskip

Additionally, using similar arguments to the one presented in Remarks~\ref{rem:rem3.3} and~\ref{rem:rem3.4}, but using the $L^2 \left( \mu_\beta\right)$-estimate $\left\| Y_0 \right\|_{L^2 \left( \mu_\beta \right)} \leq C$ instead of the (stronger) pointwise upper bound $\left| \cos 2\pi\left(\phi , q \right) \right| \leq 1$, one obtains the $L^2(\mu_\beta)$-estimate, for each $y \in \Zd$,
\begin{equation} \label{eq:TV16284}
    \left\| \nabla \mathcal{V}(y , \cdot) \right\|_{L^2 \left( \mu_\beta \right)} \leq \frac{C}{|y|^{d- 1 - \ep}}.
\end{equation}
While we can prove the estimate~\eqref{eq:TV16284} using Proposition~\ref{prop.prop4.11chap4} of Chapter~\ref{chap:chap3}, we expect that its real decay is of order $|y|^{1-d}$, and make this assumption for the rest of the argument. We use an integration by parts to write, for each field $\phi \in \Omega$,
\begin{equation*}
    \sum_{y \in \Zd} Q_x(y , \phi) \mathcal{V}(y, \phi ) = \sum_{y \in \Zd} n_{Q_x}(y , \phi) \di^* \mathcal{V}(y, \phi ).
\end{equation*}
Since we expect the random charge $n_{Q_x}(y)$ to decay like $|y - x|^{1-d}$ (see the estimate~\eqref{eq:TV15583}) and the random variable $ \di^* \mathcal{V}(y , \cdot)$ to decay $|y|^{1-d}$ (since the codifferential $\di^*$ is a linear functional of the gradient $\nabla$), we have
\begin{equation} \label{eq:TV17554}
    \sum_{y \in \Zd} n_{Q_x}(y , \phi) \di^* \mathcal{V}(y, \phi )  \simeq  \sum_{y \in \Zd} \frac{1}{|y - x|^{d-1}} \times \frac{1}{|y |^{d-1}} \simeq \frac{1}{|x|^{d-2}}.
\end{equation}
The point of the identity~\eqref{eq:TV17554} is that while we expect the sum $\sum_{y \in \Zd} n_{Q_x}(y , \phi) \di^* \mathcal{V}(y, \phi )$ to be of order $|x|^{2-d}$, its restriction to the ball $B(x , C)$ is of lower-order since we have
\begin{equation*}
    \sum_{y \in B(x , C)} n_{Q_x}(y , \phi) \di^* \mathcal{V}(y, \phi )  \simeq  \sum_{y \in B(x , C)} \frac{1}{|y - x|^{d-1}} \times \frac{1}{|y |^{d-1}} \simeq \frac{1}{|x|^{d-1}}.
\end{equation*}
A consequence of this result is that we expect the main contribution of the sum $\sum_{y \in \Zd} n_{Q_x}(y , \phi) \di^* \mathcal{V}(y, \phi )$ to come mostly from the points $y$ outside the ball $B(x , C)$. 

\smallskip

To summarize the heuristic explanation, one should expect that:
\begin{itemize}
    \item The random variable $X_x$ depends mostly on the gradient of the field inside a ball $B(x , C)$ for some large but fixed constant $C$ depending only on the dimension;
    \item The random variable $\sum_{y \in \Zd} n_{Q_x}(y , \phi) \di^* \mathcal{V}(y, \phi )$ depends mostly on the value of the gradient of the field outside the ball $B(x , C)$.
\end{itemize}
Since the gradient of the field decorrelates, we expect the random variable $\sum_{y \in \Zd} n_{Q_x}(y , \phi) \di^* \mathcal{V}(y, \phi )$ and $X_x$ to decorrelate; this is what is proved by~\eqref{eq:TV14164}.

 Once we have proved that the identity~\eqref{eq:TV14164}, we can prove the expansion~\eqref{eq:TV14563} by applying the same argument and the symmetry of the Helffer-Sj{\"o}strand operator $\L$ to decorrelate the random variable $Y_0$.
 
 The previous paragraph describes a heuristic argument explaining why the expansion~\eqref{eq:TV17423} is plausible; the rigorous proof of the result is technical and deferred to Chapter~\ref{section7}. It relies on the Helffer-Sj{\"o}strand formula: if we fix a point $y \in \Zd$, then we can write
 \begin{equation*}
     \left\langle X_x Q_x(y) \mathcal{V}(y, \cdot ) \right\rangle_{\mu_\beta} - \left\langle X_0 \right\rangle_{\mu_\beta}\left\langle  Q_x(y) \mathcal{V}(y, \cdot ) \right\rangle_{\mu_\beta} = \cov \left[  X_x , \mathcal{V}(y, \cdot )  \right] = \sum_{z \in \Zd} \left\langle \mathcal{X}_x(z , \cdot) \partial_z \mathcal{V}(y, \cdot )  \right\rangle_{\mu_\beta},
 \end{equation*}
 where the function $\mathcal{X}_x$ is the solution of the Helffer-Sj{\"o}strand equation, for each pair $(z , \phi) \in \Zd \times \Omega$,
 \begin{equation*}
     \L \mathcal{X}_x(z , \phi) = \partial_z X_x (\phi). 
 \end{equation*}
 The map $\mathcal{X}_x(z , \phi)$ can then be explicitly written in terms of the Green's function associated to the Helffer-Sj{\"o}strand operator $\L$ described in Proposition~\ref{prop.prop4.11chap4} of Chapter~\ref{chap:chap3}. The derivative $\partial_z \mathcal{V}(y, \cdot ) $ is studied using properties of the Green's function associated to the differentiated Helffer-Sj{\"o}strand operator $\L_{\mathrm{der}}$ stated in Proposition~\ref{cor:corollary4.14} of Chapter~\ref{section:section4}. The strategy relies on the fact that, since we have assumed the inverse temperature $\beta$ to be large, a $C^{0,1-\ep}$-regularity theory holds for the operators $\L$ and $\L_{\mathrm{der}}$, for some small regularity exponent $\ep : = \ep (d , \beta) > 0$. If the exponent $\ep$ is small enough (or equivalently, if $\beta$ is chosen large enough), this regularity theory is precise enough to describe the behaviour of the Green's functions $\mathcal{G}$, $\mathcal{G}_{\mathrm{der}}$ accurately and we are able to prove that the absolute value of the covariance $\cov \left[  X_x , \mathcal{V}(y, \cdot )  \right]$ is bounded from above by the value $C |x|^{1 - d + \ep}$, which gives the expansion~\eqref{eq:TV14164}.
 
 The expansion~\eqref{eq:TV17423} can then be deduced from~\eqref{eq:TV14164} by applying the same argument and using the symmetry of the Helffer-Sj{\"o}strand operator $\mathcal{L}$. The expansion \eqref{eq:TV17423plus} can be proved by the same argument.
\end{proof}

\section{First order expansion of the two-point function} \label{Chap4sec4Roetpf}
Once the Lemmas~\ref{lem.lemma7.1},~\ref{p.removecos} and~\ref{p.decoupleexp} are established, we have showed that, to prove Theorem~\ref{thm.expdualVil}, it is enough to obtain the expansion
\begin{equation} \label{eq:TV09355}
\exists c \in \R, ~ \sum_{y \in \Zd} \left\langle Q_x(y) \mathcal{U}(y, \cdot ) \right\rangle_{\mu_\beta} = \frac{c}{|x|^{d-2}} + O \left( \frac{C}{|x|^{d-2 + \gamma'}} \right).
\end{equation}

This section is devoted to the proof of~\eqref{eq:TV09355}. We first give a sketch of the proof in Section~\ref{chap4:sect4.1heuristic} and provide the details of the argument in Section~\ref{sectionsection4.2789}.

\subsection{Heuristic argument} \label{chap4:sect4.1heuristic}
In this section, we present a heuristic argument for the proof of the expansion~\eqref{eq:TV09355}. A large part of of the proof is concerned with the treatment of the technicalities inherent to the dual Villain model (sum over all the charges $q \in \mathcal{Q}$, presence of a sine etc.). In order to highlight the main ideas of the argument, we make the following simplifications:
\begin{itemize}
    \item We assume that for $\beta$ large enough, one may essentially reduce the charges to the collection of dipoles $\left( \di \indc_{\{y , y + e_i\}} \right)_{y \in \Zd, 1 \leq i \leq d}$. The exponential decay on the coefficient $z \left( \beta , q \right)$ constraints the $L^1$-norm of the charge $q$ to be small. One can thus assume that only the charges $q \in \mathcal{Q}$ which minimize the value $\left\| q \right\|_1$ are involved in the sum; this leads us to considering the dipoles $\left( \di \indc_{\{x , x + e_i\}} \right)_{x \in \Zd, 1 \leq i \leq d}$. An important, but mostly technical, part of the argument presented in Section~\ref{sectionsection4.2789} is devoted to proving that this dipole approximation yields the correct picture. Under this assumption, one has the simplifications
    \begin{equation*}
        Q_x  = \sum_{i=1}^d \sum_{y \in \Zd} 2\pi \sin \left(2\pi \nabla_i G(y) \right) \di \indc_{\{ y , y+ e_i\}}
        \hspace{3mm} \mbox{in} \hspace{3mm}
        \mathcal{U} = \sum_{i=1}^d \sum_{y \in \Zd} 2\pi \sin \left(2\pi \nabla_i G_x(y) \right) \mathcal{U}_{y,i},
    \end{equation*}
    where the function $\mathcal{U}_{y,i}$ is the solution of the Helffer-Sj{\"o}strand equation $$\L \mathcal{U}_{y,i} =  \di \left(\cos 2\pi\left( \di^*\phi(y)\cdot e_i\right) \indc_{\{y , y + e_i\}} \right) \hspace{3mm} \mbox{in} \hspace{3mm} \Zd \times \Omega. $$
    \item Since the gradients of the Green's functions $\nabla_i G(y)$ are usually small, we consider the first-order expansion of the sine and replace the value $\sin \left(2\pi \nabla_i G_x(y) \right)$ by $ 2\pi \nabla_i G_x(y)$. With this assumption, we have
    \begin{equation*}
        Q_x  = (2\pi)^2 \sum_{i=1}^d \sum_{y \in \Zd}  \nabla_i G(y) \di \indc_{\{ y , y+ e_i\}}
        \hspace{3mm} \mbox{and} \hspace{3mm}
        \mathcal{U} =  (2\pi)^2 \sum_{i=1}^d \sum_{y \in \Zd} \nabla_i G_x(y) \mathcal{U}_{y,i}.
    \end{equation*}
\end{itemize}
Using these simplifications, we compute
\begin{equation} \label{eq:TV09299}
    \sum_{y \in \Zd} \left\langle Q_x(y) \mathcal{U}(y, \cdot ) \right\rangle_{\mu_\beta}  = (4 \pi^2)^2 \sum_{i,j=1}^d \sum_{y,y_1 \in \Zd} \nabla_i G(y) \nabla_j G_x(y_1) \left\langle \cos(\di^* \phi(y_1) \cdot e_i)  \di^* \mathcal{U}_{y , j}(y_1 , \phi) \cdot e_i \right\rangle_{\mu_\beta}.
\end{equation}
Using the translation invariance of the measure $\mu_\beta$, one has the the identity, for each pair of points $y , y_1 \in \Zd$,
\begin{equation} \label{eq:TV09309}
    \left\langle \cos(\di^* \phi(y_1) \cdot e_i)  \di^* \mathcal{U}_{y , j}(y_1 , \phi) \cdot e_i \right\rangle_{\mu_\beta} = \left\langle \cos(\di^* \phi(y_1-y) \cdot e_i)  \di^* \mathcal{U}_{0 , j}( y_1-y , \phi) \cdot e_i \right\rangle_{\mu_\beta}.
\end{equation}
Putting the identity~\eqref{eq:TV09309} into the equality~\eqref{eq:TV09299}  and performing the change of variable $\kappa := y_1 - y$, we obtain
\begin{equation} \label{eq:TV08389}
    \sum_{y \in \Zd} \left\langle Q_x(y) \mathcal{U}(y, \cdot ) \right\rangle_{\mu_\beta}  = (4 \pi^2)^2 \sum_{i,j=1}^d \sum_{y,\kappa \in \Zd} \nabla_i G(y) \nabla_j G_x(\kappa-y) \left\langle \cos(\di^* \phi(\kappa) \cdot e_i)  \di^* \mathcal{U}_{0 , j}(\kappa , \phi) \cdot e_i \right\rangle_{\mu_\beta}.
\end{equation}
The strategy is then to simplify the right side of~\eqref{eq:TV08389} by arguing that the term $\di^* \mathcal{U}_{0 , j}$ behaves like the mixed derivative of a deterministic Green's function. Proving a quantitative version of this result is the subject of Theorem~\ref{thm:homogmixedderchap4} which is proved in Chapters~\ref{sec:section6} and~\ref{section7}; in this setting, it can be stated as follows: there exists an exponent $\gamma := \gamma (d) > 0$ and, for each pair of integers $i,j \in \{ 1 , \ldots , d \}$, there exists a deterministic constants $c_{i,j} := c_{i,j}(d , \beta)$ such that for each radius $R \geq 1$,
\begin{equation} \label{eq:TV09439}
    \sum_{\kappa \in B_{2R} \setminus B_R} \left|  \left\langle \cos(\di^* \phi(\kappa) \cdot e_i)  \di^* \mathcal{U}_{0 , j}(\kappa , \phi) \cdot e_i  \right\rangle_{\mu_\beta} - \sum_{i_1,j_1=1}^d c_{i,i_1} c_{j , j_1} \nabla_{i_1} \nabla_{j_1} G(\kappa) \right| \leq \frac{C}{R^{\gamma}}.
\end{equation}
Once equipped with this estimate, we let $\mathcal{E}_{i,j} : \Zd \mapsto \R$ be the error term defined according to the formula, for each $\kappa \in \Zd$,
\begin{equation*}
    \mathcal{E}_{i,j} (\kappa) := \left\langle \cos(\di^* \phi(\kappa) \cdot e_i)  \di^* \mathcal{U}_{0 , j}(\kappa , \phi) \cdot e_i  \right\rangle_{\mu_\beta} - \sum_{i_1,j_1=1}^d c_{i,i_1} c_{j , j_1} \nabla_{i_1} \nabla_{j_1} G(\kappa).
\end{equation*}
According to the regularity estimate on the gradient of the Green's matrix associated to the Helffer-Sj\"{o}strand operator $\L$ stated in Proposition~\ref{prop.prop4.11chap4} of Chapter~\ref{chap:chap3} (via the formula~\eqref{eq:VER192130}) and the homogenization estimate~\eqref{eq:TV09439}, this term satisfies the $L^1$ and pointwise estimates
\begin{equation} \label{eq:TV1934717}
        \forall R \geq 1, \hspace{2mm} R^{-d} \sum_{\kappa \in B_{2R} \setminus B_R} \left| \mathcal{E}_{i , j}(\kappa) \right| \leq \frac{C}{R^{d+\gamma}} \hspace{3mm} \mbox{and} \hspace{3mm} \forall \kappa \in \Zd, \hspace{2mm} \left| \mathcal{E}_{i , j}(\kappa) \right| \leq  \frac{C}{\left| \kappa \right|^{d - \ep}}.
    \end{equation}
We can use the definition of the term $\mathcal{E}_{i,j}$ to rewrite the identity~\eqref{eq:TV08389}. We obtain
\begin{align} \label{eq:TV10199}
    \sum_{y \in \Zd} \left\langle Q_x(y) \mathcal{U}(y, \cdot ) \right\rangle_{\mu_\beta} & = (4 \pi^2)^2 \sum_{i, i_1,j, j_1=1}^d c_{i,i_1} c_{j , j_1} \sum_{y,\kappa \in \Zd} \nabla_i G(y) \nabla_j G_x( \kappa -y)  \nabla_{i_1} \nabla_{j_1} G(\kappa) \\ 
    & \qquad + \sum_{i,j = 1}^d \nabla_i G(y) \nabla_j G_x(\kappa-y) \mathcal{E}_{i,j}(\kappa). \notag
\end{align}
The right side of the identity~\eqref{eq:TV10199} can then be refined. First using the estimates~\eqref{eq:TV1934717} on the error term $\mathcal{E}_{i,j}$ and Proposition~\ref{prop:propB1} proved in Section~\ref{sec:chap8.5} of Chapter~\ref{section7}, we can show the following expansion: there exists an exponent $\gamma' := \gamma' (d) > 0$ such that
\begin{equation} \label{eq:TV10399}
    \sum_{i,j = 1}^d \sum_{y,\kappa \in \Zd} \nabla_i G(y) \nabla_j G_x(\kappa-y) \mathcal{E}_{i,j}(\kappa) = \sum_{i,j = 1}^d K_{i,j} \sum_{y,\kappa \in \Zd} \nabla_i G(y) \nabla_j G_x(\kappa-y) + O \left( \frac{C}{|x|^{d-2+\gamma'}}\right),
\end{equation}
where the constant $K_{i,j}$ are obtained from the error term $\mathcal{E}_{i,j}$ according to the formula
\begin{equation*}
     K_{i,j} := \left( 4\pi^2 \right)^2 \sum_{\kappa \in \Zd} \mathcal{E}_{i,j}(\kappa),
\end{equation*}
which, by the estimate~\eqref{eq:TV1934717}, is well-defined. A combination of the identity~\eqref{eq:TV10199} with the expansion~\eqref{eq:TV10399} then shows
\begin{align} \label{eq:TV07030}
    \sum_{y \in \Zd} \left\langle Q_x(y) \mathcal{U}(y, \cdot ) \right\rangle_{\mu_\beta} & = 4 \pi^2 \sum_{i, i_1,j, j_1=1}^d c_{i,i_1} c_{j , j_1} \sum_{y,\kappa \in \Zd} \nabla_i G(y) \nabla_j G_x(\kappa - y)  \nabla_{i_1} \nabla_{j_1} G(\kappa) \\ 
    & \qquad + \sum_{i,j = 1}^d K_{i,j} \sum_{y,\kappa \in \Zd} \nabla_i G(y) \nabla_j G_x(\kappa - y) + O \left( \frac{C}{|x|^{d-2+\gamma'}}\right). \notag
\end{align}
This expansion does not give the result~\eqref{eq:TV09355} directly and we need to exploit the symmetries of the dual Villain model to conclude. The argument relies on the following observation: since the Villain and dual Villain model are invariant under the action of the group $H$ of the lattice preserving transformations introduced in Chapter~\ref{Chap:chap2}, the same property holds for the two-point function and thus for the map $x \mapsto \sum_{y \in \Zd} \left\langle Q_x(y) \mathcal{U}(y, \cdot ) \right\rangle_{\mu_\beta}$.

One can then use this invariance property together with the expansion~\eqref{eq:TV07030} to prove that this expansion must take the simpler form
\begin{equation} \label{eq:TV18282701}
    \sum_{y \in \Zd} \left\langle Q_x(y) \mathcal{U}(y, \cdot ) \right\rangle_{\mu_\beta} = \frac{c}{|x|^{d-2}} +  O \left( \frac{C}{|x|^{d-2+\gamma'}}\right).
\end{equation}
This is achieved by using the property of the discrete Green's function and relies on tools from Fourier analysis. The proof can be found in Section~\ref{sec:chap8.4} of Chapter~\ref{section7}. The expansion~\eqref{eq:TV18282701} is exactly~\eqref{eq:TV09355}; the proof is thus complete.

\subsection{Proof} \label{sectionsection4.2789}

We first write $Q_x = \di n_{Q_x}$, perform an integration by parts and use the identities~\eqref{eq:TV14223} and~\eqref{eq:TV09555} to expand the sum $\sum_{y \in \Zd} \left\langle Q_x(y) \mathcal{U}(y, \cdot ) \right\rangle_{\mu_\beta}$. We obtain
\begin{align} \label{eq:TV13505}
    \lefteqn{\sum_{y \in \Zd} \left\langle Q_x(y) \mathcal{U}(y, \cdot ) \right\rangle_{\mu_\beta} } \qquad & \\ & = \sum_{y \in \Zd} \left\langle n_{Q_x}(y) \di^* \mathcal{U}(y, \cdot ) \right\rangle_{\mu_\beta} \notag \\ & = 4\pi^2 \sum_{y \in \Zd} \sum_{q_1 , q_2 \in \mathcal{Q}} z(\beta , q_1) z(\beta , q_2) \sin 2\pi(\nabla G  , n_{q_2}) \sin 2\pi(\nabla G_x  , n_{q_1})  \left\langle \cos2\pi(\phi , q_1)  \di^* \mathcal{U}_{q_2}(y , \phi) \right\rangle_{\mu_\beta} n_{q_1}(y). \notag
\end{align}
To simplify the sum over all the charges $q_1 , q_2$, we introduce an equivalence class on the set of charges $\mathcal{Q}$: we say that two charges $q$ and $q'$ are equivalent, and denote it by $q \sim q'$, if and only if one is the translation of the other, i.e.,
\begin{equation*}
    q \sim q' \iff \exists z \in \Zd, \hspace{2mm} q(z + \cdot) = q'.
\end{equation*}
We denote this quotient space by $\mathcal{Q}/\Zd$ and for each charge $q \in \mathcal{Q}$, we denote by $[q]$ its equivalence class. For each equivalence class $[q] \in \mathcal{Q}/\Zd$, we select a charge $q \in \mathcal{Q}$ such that $0$ belongs to the support of $n_q$ (if there is more than one candidate, we break ties by using an arbitrary criterion). We note that, for each charge $q \in \mathcal{Q}$, by the definition of the charge $n_q$ and the coefficient $z \left( \beta , q \right)$, we have the identities, for each point $z \in \Zd$,
\begin{equation} \label{eq:TV13485chap4}
    z \left( \beta , q \right) = z \left( \beta , q(\cdot - z ) \right), \hspace{5mm} n_{q(\cdot - z)} = n_q (\cdot - z) \hspace{5mm} \mbox{and} \hspace{5mm} \left( n_{q(\cdot - z)} \right) = \left( n_q \right).
\end{equation}
We also note that, by using the translation invariance of the measure $\mu_\beta$ and the definition of the function $\mathcal{U}_{q_2}$ given in~\eqref{eq:TV09004}, we have the equality, for each pair of points $(y , z) \in \Zd$,
\begin{equation*}
    \left\langle \cos2\pi(\phi , q_1)  \di^* \mathcal{U}_{q_2(\cdot - z)}(y , \phi) \right\rangle_{\mu_\beta} =  \left\langle \cos2\pi(\phi , q_1(\cdot+ 
    z))  \di^* \mathcal{U}_{q_2}(y - z , \phi) \right\rangle_{\mu_\beta}.
\end{equation*}
Additionally, we can decompose the sum over the charges $q \in \mathcal{Q}$ along the equivalence classes, i.e., we can write, for any summable function $F : \mathcal{Q} \to \R$,
\begin{equation} \label{eq:TV13495chap4}
    \sum_{q \in \mathcal{Q}} F(q) = \sum_{[q] \in \mathcal{Q}/\Zd} \sum_{z \in \Zd} F(q \left( \cdot - z\right)),
\end{equation}
where the charge $q$ in the right side is the element of the equivalence class $[q] \in \mathcal{Q}/\Zd$ chosen such that $0$ belongs to the support of the charge $n_q$.

Combining the identities~\eqref{eq:TV13485chap4} and~\eqref{eq:TV13495chap4}, we can rewrite the equality~\eqref{eq:TV13505},
\begin{multline} \label{eq:TV15345}
    \sum_{y \in \Zd} \left\langle Q_x(y) \mathcal{U}(y, \cdot ) \right\rangle_{\mu_\beta} = 4\pi^2 \sum_{[q_1], [q_2] \in \mathcal{Q}/\Zd} z \left( \beta , q_1 \right) z \left( \beta , q_2 \right) \\ \times \left[ \sum_{z_1 , z_2 , y \in \Zd}  \sin 2\pi(\nabla G  , n_{q_2}( \cdot  - z_2)) \sin 2\pi(\nabla G_x  , n_{q_1}(\cdot - z_1))  \left\langle \cos2\pi(\phi , q_1(\cdot - z_1 + z_2))  \di^* \mathcal{U}_{q_2}(y - z_2 , \phi) \right\rangle_{\mu_\beta} n_{q_1}( y - z_1) \right].
\end{multline}
We first rearrange the identity~\eqref{eq:TV15345}. We use the identities $(\nabla G_x  , n_{q_1}(\cdot - z_1)) = (\nabla G_x(\cdot + z_1)  , n_{q_1}) $, $(\nabla G  , n_{q_2}(\cdot - z_2)) = (\nabla G(\cdot + z_2)  , n_{q_2}) $ and perform the change of variable $y := y - z_1$. We obtain
\begin{multline} \label{eq:TV15335}
    \sum_{y \in \Zd} \left\langle Q_x(y) \mathcal{U}(y, \cdot ) \right\rangle_{\mu_\beta} = 4\pi^2 \sum_{[q_1], [q_2] \in \mathcal{Q}/\Zd} z \left( \beta , q_1 \right) z \left( \beta , q_2 \right) \\ \times \underbrace{\left[ \sum_{z_1 , z_2 , y \in \Zd}  \sin 2\pi(\nabla G(\cdot + z_2)  , n_{q_2}) \sin 2\pi(\nabla G_x(\cdot + z_1)  , n_{q_1})  \left\langle \cos2\pi(\phi , q_1(\cdot - z_1 + z_2))  \di^* \mathcal{U}_{q_2}(y + z_ 1 - z_2 , \phi) \right\rangle_{\mu_\beta} n_{q_1}( y ) \right].}_{\eqref{eq:TV15335}-(q_1, q_2)}
\end{multline}
The rest of the proof is decomposed into two steps:
\begin{enumerate}
    \item In the first step, we use Theorem~\ref{thm:homogmixedderchap4} and the regularity estimates established in Proposition~\ref{prop.prop4.11chap4} of Chapter~\ref{chap:chap3} to prove the following result: there exists an exponent $\gamma' := \gamma' (d)>0$ such that for each pair of charges $q_1 , q_2 \in \mathcal{Q}$ and each pair of integers $(i,j) \in \left\{ 1 , \ldots, d \right\}^2$, there exist constants $K_{q_1, q_2} := K_{q_1, q_2}(q_1 , q_2 , d , \beta)$, $C_{q_1, q_2} := C_{q_1, q_2}(q_1 , q_2 , d , \beta)$, $c_{ij}^{q_1}:= c_{ij}^{q_1}(i , j , q_1 , d , \beta) $ such that the term $\eqref{eq:TV15335}-(q_1, q_2)$ satisfies the expansion
\begin{multline} \label{eq:TV16245}
    \eqref{eq:TV15335}-(q_1, q_2) = \sum_{i,j,k,l = 1}^d c_{ij}^{q_1} c_{kl}^{q_2} \sum_{z_1,z_2 \in \Zd} \nabla_i G (z_1) \nabla_j \nabla_k G (z_1 - z_2) \nabla_l G (x - z_2) \\  + K_{q_1 , q_2} \sum_{z_1  \in \Zd}  \nabla G(z_1) \cdot (n_{q_2}) \times \nabla G_x(x - z_1) \cdot (n_{q_1}) + O \left( \frac{C_{q_1, q_2}}{|x|^{d-2+\gamma'}} \right).
\end{multline}
We recall that the vectors $(n_{q_1})$ and $(n_{q_2})$ belongs to $\Rd$ and are defined by the formulas
\begin{equation*}
    (n_{q_1}) := \sum_{y \in \Zd} n_{q_1}(y) \hspace{5mm}\mbox{and}  \hspace{5mm} (n_{q_2}) := \sum_{y \in \Zd} n_{q_2}(y).
\end{equation*}
These two quantities belong to the space $\Rd$ (or more precisely $\Zd$). We also record that all the constants $K_{q_1,q_2}$, $c_{ij}^{q_1,q_2}$ and $C_{q_1, q_2}$ grow at most algebraically fast in the values $\left\| q_1 \right\|_1$ and $\left\| q_2 \right\|_1$, i.e., there exist an exponent $k := k(d) < \infty$ and a constant $C := C(d , \beta) < \infty$ such that one has the estimates
\begin{equation} \label{eq:TV16475}
    \left| c_{ij}^{q_1} \right| \leq C  \left\| q_1 \right\|_1^k, \hspace{5mm} \left| K_{q_1,q_2} \right| \leq C  \left\| q_1 \right\|_1^k \left\| q_2 \right\|_1^k \hspace{5mm}\mbox{and} \hspace{5mm} \left| C_{q_1,q_2} \right| \leq C  \left\| q_1 \right\|_1^k \left\| q_2 \right\|_1^k .
\end{equation}
    \item In the second step, we use the symmetry and rotation invariance of the dual Villain model to prove that the expansion~\eqref{eq:TV16245} implies the expansion~\eqref{eq:TV09355}.
\end{enumerate}
We first give the proof of the second item of \eqref{eq:TV16245}, as the argument is simpler and less technical. We first sum the expansion~\eqref{eq:TV16245} over all the equivalence classes $[q_1], [q_2] \in \mathcal{Q}/\Zd$ and use the exponential decay of the coefficients $z \left( \beta , q_1 \right)$ and $z \left( \beta , q_2 \right)$ to absorb the algebraic growth of the constants $c_{ij}^{q_1}$, $c_{ij}^{q_2}$ and $C_{q_1,q_2}$. We obtain
\begin{align} \label{eq:TV17335}
    \sum_{y \in \Zd} \left\langle Q_x(y) \mathcal{U}(y, \cdot ) \right\rangle_{\mu_\beta} & = 4\pi^2 \sum_{[q_1], [q_2] \in \mathcal{Q}/\Zd} z \left( \beta , q_1 \right) z \left( \beta , q_2 \right) \times  \eqref{eq:TV15335}-(q_1, q_2) \\
    & = 4\pi^2 \sum_{[q_1], [q_2] \in \mathcal{Q}/\Zd} \sum_{i,j,k,l = 1}^d z \left( \beta , q_1 \right) z \left( \beta , q_2 \right)   c_{ij}^{q_1} c_{kl}^{q_1} \sum_{z_1,z_2 \in \Zd} \nabla_i G (z_1) \nabla_j \nabla_k G (z_1 - z_2) \nabla_l G (x - z_2) \notag \\
    & \qquad +  4\pi^2 \sum_{[q_1], [q_2] \in \mathcal{Q}/\Zd} z \left( \beta , q_1 \right) z \left( \beta , q_2 \right) K_{ q_1 , q_2}  \sum_{z_1  \in \Zd}  \nabla G(z_1) \cdot (n_{q_2}) \times \nabla G(x - z_1) \cdot (n_{q_1}) \notag
    \\ & \qquad + 4\pi^2 \sum_{[q_1], [q_2] \in \mathcal{Q}/\Zd}  z \left( \beta , q_1 \right) z \left( \beta , q_2 \right) O \left( \frac{C_{q_1, q_2}}{|x|^{d-2+\gamma'}} \right) \notag \\
    & =  \sum_{i,j,k,l = 1}^d c_{ij} c_{kl} \sum_{z_1,z_2 \in \Zd} \nabla_i G (z_1) \nabla_j \nabla_k G (z_1 - z_2) \nabla_l G (x - z_2) \notag \\ & \qquad + \sum_{i,j = 1}^d K_{i,j}  \sum_{z_1  \in \Zd}  \nabla_i G(z_1) \nabla_j G(x - z_1) + O \left( \frac{C}{|x|^{d-2+\gamma'}} \right), \notag
\end{align}
where we have set
\begin{equation*}
     c_{ij} := 4\pi^2 \sum_{[q] \in \mathcal{Q}/\Zd}  z \left( \beta , q \right) c_{ij}^{q} \hspace{5mm} \mbox{and} \hspace{5mm}  K_{i,j} := 4\pi^2 \sum_{[q_1] , [q_2] \in \mathcal{Q}/\Zd} z \left( \beta , q_1 \right) z \left( \beta , q_2 \right) K_{q_1,q_2} \left[ \left(n_{q_1} \right) \cdot e_i \right] \times \left[\left(n_{q_1} \right) \cdot e_j \right],
\end{equation*}
which are well-defined by the estimate $\left| z(\beta , q)\right| \leq e^{-c \sqrt{\beta} \left\| q\right\|_1}$ and~\eqref{eq:TV16475}.

We then simplify the expansion~\eqref{eq:TV17335} by noting that, since the measure $\mu_\beta$ is invariant under the symmetries and rotations of the lattice $\Zd$, the function $x \mapsto \sum_{y \in \Zd} \left\langle Q_x(y) \mathcal{U}(y, \cdot ) \right\rangle_{\mu_\beta}$ is also invariant over the symmetries and rotations of the lattice. It is proved in Proposition~\ref{prop:propB1} in Chapter~\ref{section7} that this invariance property combined with the expansion~\eqref{eq:TV17335} implies that there exists a constant $c := c(d , \beta)$ such that
\begin{equation*}
    \sum_{y \in \Zd} \left\langle Q_x(y) \mathcal{U}(y, \cdot ) \right\rangle_{\mu_\beta} = \frac{c}{|x|^{d-2}} + O \left( \frac{C}{|x|^{d-2+\gamma'}} \right).
\end{equation*}
This is precisely the expansion~\eqref{eq:TV09355}. We have thus proved that the expansion~\eqref{eq:TV16245} implies the expansion~\eqref{eq:TV09355}.

The rest of the demonstration is devoted to the proof of~\eqref{eq:TV16245}. We first simplify the term~$\eqref{eq:TV15335}-(q_1, q_2)$ by removing the sine. To this end, we use the following ingredients:
 \begin{itemize}
     \item We use the inequality, $\left| \sin a - a \right| \leq \frac{1}{6} a^3$, valid for any real number $a \in \R$ and the inequality, for each charge $q \in \mathcal{Q}_0$ and each point $z \in \Zd$,
     \begin{equation*}
         \left| \left( \nabla G , n_q(\cdot - z) \right) \right| \leq \left\| \nabla G(z + \cdot) \right\|_{L^2 \left( \supp n_q \right)} \left\| n_q \right\|_{2} \leq \frac{C_q}{|z|^{d-1}}.
     \end{equation*}
     We deduce that, for each pair of charges $q_1 , q_2 \in \mathcal{Q}$ and each pair of points $z_1, z_2 \in \Zd$,
     \begin{equation} \label{eq:TV07196}
         \left| \sin 2\pi(\nabla G_x(\cdot + z_2)  , n_{q_2}) - 2\pi(\nabla G(\cdot + z_2)  , n_{q_2}) \right| \leq \frac{C_{q_2}}{|z_2|^{3d-3}}
     \end{equation}
     and
     \begin{equation} \label{eq:TV07206}
     \left| \sin 2\pi(\nabla G_x(\cdot + z_1)  , n_{q_1}) - 2\pi(\nabla G_x(\cdot + z_1)  , n_{q_1}) \right| \leq \frac{C_{q_1}}{|z_1 - x|^{3d-3}};
     \end{equation}
     \item We further simplify the terms $ 2\pi(\nabla G  , n_{q_1}(\cdot - z_2))$ and $ 2\pi(\nabla G_x  , n_{q_1}(\cdot - z_1))$. We use that the double gradient of the Green's function decays like $|z|^{-d}$ and the assumption that $0$ belongs to the support of the charges $n_{q_1}$ and $n_{q_2}$ to write
     \begin{align} \label{eq:TV09466}
         \left| 2\pi(\nabla G_x(\cdot  + z_1)  , n_{q_1}) - 2 \pi \left( n_{q_1} \right) \cdot \nabla G_x(z_1) \right| & = \left| 2\pi(\nabla G_x(z_1 + \cdot) -  \nabla G_x(z_1) , n_{q_1})  \right| \\
         & \leq C \left\| n_q \right\|_2 \left\| \nabla G_x(z_1 + \cdot) -  \nabla G_x(z_1) \right\|_{L^2 \left( \supp n_{q_2}\right)} \notag \\
         & \leq C \left\| n_{q_2} \right\|_2 \left| \supp n_{q_2} \right|^\frac 12 \sup_{z \in \supp n_q} \left| \nabla \nabla G (z + z_2 - x) \right| \notag \\
         & \leq \frac{C_{q_2}}{|z_2 - x|^d}. \notag
     \end{align}
     A similar argument shows the estimate
     \begin{equation} \label{eq:TV09476}
          \left| 2\pi(\nabla G  , n_{q_2}( \cdot  - z_2)) - 2 \pi \left( n_{q_2} \right) \cdot \nabla G(z_2) \right| \leq \frac{C_{q_2}}{|z_2|^d}.
     \end{equation}
     We then combine the inequalities~\eqref{eq:TV07196} and~\eqref{eq:TV09466} on the one hand and~\eqref{eq:TV07206} and
     ~\eqref{eq:TV09476} on the other hand and use the inequality $3d - 3 > d$. We obtain the two estimates
         \begin{equation} \label{eq:TV09486}
         \left| \sin 2\pi(\nabla G_x  , n_{q_2}( \cdot  - z_2)) - 2 \pi \left( n_{q_2} \right) \cdot \nabla G_x(z_2) \right| \leq \frac{C_{q_2}}{|x - z_1|^{d}}
     \end{equation}
     and
     \begin{equation} \label{eq:TV09496}
     \left| \sin 2\pi(\nabla G_x  , n_{q_1}(\cdot - z_1)) -2 \pi \left( n_{q_1} \right) \cdot \nabla G(z_1) \right| \leq \frac{C_{q_1}}{|z_1|^{d}};
     \end{equation}
     \item We recall the estimate~\eqref{eq:TV10154} which reads for each $y \in \Zd$, $ \left\| \nabla \mathcal{U}_{q_2}(y , \cdot)  \right\|_{L^\infty \left( \mu_\beta \right)} \leq \frac{C_{q_2}}{|x - y|^{d - \ep}}$. From this inequality, we deduce that for each pair of charges $q_1, q_2 \in \mathcal{Q}$ such that $0$ belongs to the supports of $n_{q_1}$ and $n_{q_2}$ and for each point $y \in \supp n_{q_1},$ 
     \begin{equation*}
      \left| \left\langle \cos2\pi(\phi , q_1(\cdot - z_1 + z_2))  \di^* \mathcal{U}_{q_2}(y + z_1 - z_2 , \phi) \right\rangle_{\mu_\beta} \right| \leq \frac{C_{q_1, q_2}}{|z_1 - z_2|^{d - \ep}}.
      \end{equation*}
     \item We have the inequalities, for each point $x \in \Zd$,
     \begin{equation} \label{eq:TV14460901}
         \sum_{z_1 , z_2 \in \Zd} \frac{1}{|x - z_1|^d} \times \frac{1}{|z_1 - z_2|^{d - \ep}} \times  \frac{1}{|z_2|^{d-1}}  \leq \frac{C\ln \left| x \right|}{|x|^{d - 1 - \ep}} \hspace{2mm} \mbox{and}  \sum_{z_1 , z_2 \in \Zd} \frac{1}{|x - z_1|^{d-1}} \times \frac{1}{|z_1 - z_2|^{d - \ep}} \hspace{2mm} \times  \frac{1}{|z_2|^{d}}  \leq \frac{C}{|x|^{d - 1 - \ep}}.
     \end{equation}
     The proof of these estimates are postponed to Proposition~\ref{coro.appCcoro} of Appendix~\ref{app.appC}.
 \end{itemize}
 A combination of the the four items described above implies the following expansion
 \begin{multline} \label{eq:TV12276}
     \eqref{eq:TV15335}-(q_1, q_2) \\ = \underbrace{4 \pi^2 \sum_{z_1 , z_2 , y \in \Zd}  \nabla G(z_2) \cdot (n_{q_2}) \nabla G_x(z_1) \cdot (n_{q_1})  \left\langle \cos2\pi(\phi , q_1(\cdot - z_1 + z_2))  \di^* \mathcal{U}_{q_2}(y + z_ 1 - z_2 , \phi) \right\rangle_{\mu_\beta} n_{q_1}( y )}_{\eqref{eq:TV12276}-(q_1,q_2)} \\ + O \left(\frac{C_{q_1 , q_2}}{|x|^{d - 1 -\ep}} \right).
 \end{multline}
 A consequence of the identity~\eqref{eq:TV12276} is that to prove the expansion~\eqref{eq:TV16245}, it is enough to prove the following result
 \begin{multline} \label{eq:TV12506}
     \eqref{eq:TV12276}-(q_1,q_2) = \sum_{i,j,k,l = 1}^d c_{ij}^{q_1} c_{kl}^{q_2} \sum_{z_1,z_2 \in \Zd} \nabla_i G (z_1) \nabla_j \nabla_k G (z_1 - z_2) \nabla_l G (x - z_2) \\ + 4\pi^2 K_{q_1 , q_2} \sum_{z_1 , y \in \Zd}  \nabla G(z_2) \cdot (n_{q_2}) \nabla G_x(z_2 + \kappa) \cdot (n_{q_1}) + O \left( \frac{C_{q_1, q_2}}{|x|^{d-2+\gamma}} \right).
 \end{multline}
The rest of the argument is devoted to the proof of~\eqref{eq:TV12506} and relies on the homogenization of the mixed derivative associated to the Helffer-Sj{\"o}strand equation (Theorem~\ref{thm:homogmixedderchap4}).

We first consider the term $\eqref{eq:TV12276}-(q_1,q_2)$ and perform the change of variable $\kappa := z_1 - z_2$. We obtain
 \begin{multline} \label{eq:TV17267}
     \eqref{eq:TV12276}-(q_1,q_2) \\ = 4 \pi^2 \sum_{z_1 , \kappa , y \in \Zd}  \nabla G(z_2) \cdot (n_{q_2}) \nabla G_x(z_2 + \kappa) \cdot (n_{q_1})  \left\langle \cos2\pi(\phi , q_1(\cdot - \kappa))  \di^* \mathcal{U}_{q_2}(y + \kappa , \phi) \right\rangle_{\mu_\beta} n_{q_1}( y ).
 \end{multline}
We then apply the homogenization theorem for the solution of the Helffer-Sj{\"o}strand equation stated in Theorem~\ref{thm:homogmixedderchap4}.

We post-process the result of Theorem~\ref{thm:homogmixedderchap4} so that it can be used to estimate the term~$\eqref{eq:TV12276}-(q_1,q_2)$; the objective is to prove the estimate~\eqref{eq:TV17347} below. We first use that the codifferential $\di^*$ is a linear functional of the gradient to deduce from Theorem~\ref{thm:homogmixedderchap4} that, for each radius $R \geq 1$,
\begin{equation} \label{eq:TV22306}
    \left\| \di^* \mathcal{U}_{q_2} - \sum_{1 \leq i \leq d} \sum_{1 \leq j\leq \binom d2} \left( \di^* l_{e_{ij}} + \di^* \chi_{ij}\right) \nabla_i  \bar G_{q_2,j}\right\|_{\underline{L}^2 \left( B_{2R} \setminus B_R , \mu_\beta \right)} \leq \frac{C_{q_2}}{R^{d + \gamma}}.
\end{equation}
To ease the notations, we let $A_R:= B_{2R} \setminus B_R$. Using the arguments and notations introduced in Remark~\eqref{remark4.914470302} of Chapter~\ref{chap:chap3}, we obtain the identity
\begin{equation*}
   \sum_{1 \leq i \leq d} \sum_{1 \leq j\leq \binom d2} \left( \di^* l_{e_{ij}} + \di^* \chi_{ij}\right) \nabla_i  \bar G_{q_2,j} = \sum_{1 \leq i \leq d} \left( e_i + \di^* \chi_{i} \right) \left( \di^* \bar G_{q_2} \cdot e_i \right).
\end{equation*}
The estimate~\eqref{eq:TV22306} then implies, by using the stationarity of the gradient of the infinite-volume corrector and the Cauchy-Schwarz inequality
\begin{multline} \label{eq:TV20126}
    \frac 1{R^d}\sum_{\kappa \in A_R}\left| \left\langle \cos2\pi(\phi , q_1(\cdot - \kappa))  \di^* \mathcal{U}_{q_2}( \kappa , \phi) \right\rangle_{\mu_\beta} - \sum_{1 \leq i \leq d}  \left\langle \cos2\pi(\phi , q_1) \left( e_i + \di^* \chi_{i}(0 , \phi)\right) \right\rangle_{\mu_\beta} \left( \di^* \bar G_{q_2}(\kappa) \cdot e_i \right) \right| \\ \leq \frac{C_{q_2}}{R^{d + \gamma} }.
\end{multline}
We then generalize the inequality~\eqref{eq:TV20126} and prove the following result: for each point $y \in \Zd$, one has the estimate
\begin{multline} \label{eq:TV08157}
\frac{1}{R^d}\sum_{\kappa \in A_R}\left| \left\langle \cos2\pi(\phi , q_1(\cdot - \kappa))  \di^* \mathcal{U}_{q_2}( y + \kappa , \phi) \right\rangle_{\mu_\beta} - \sum_{1 \leq i \leq d} \left\langle \cos2\pi(\phi , q_1) \left( e_i + \di^* \chi_{i}(y , \phi)\right) \right\rangle_{\mu_\beta}   \left( \di^* \bar G_{q_2}(\kappa) \cdot e_i \right)  \right| \\ \leq \frac{C_{q_2} ( 1 +\left| y \right|^{2d+\gamma})}{R^{d + \gamma}}.
\end{multline}
To prove this result, we distinguish two cases, whether the norm of $y$ is larger or smaller than $\frac{R}{2}$.

\medskip

\textit{Case 1. The norm of $y$ is smaller than $\frac{R}{2}$.} In that case, we note that the set $y + A_R$ is included in the annuli $B_{2R} \setminus B_{\frac{R}{2}}$. We can use the identity~\eqref{eq:TV22306} with the two annulus $A_R$ and $A_{\frac R2}$ to deduce that
\begin{multline} \label{eq:TV07167}
\frac{1}{R^d}\sum_{\kappa \in A_R}\left| \left\langle \cos2\pi(\phi , q_1(\cdot - \kappa))  \di^* \mathcal{U}_{q_2}( y + \kappa , \phi) \right\rangle_{\mu_\beta} - \sum_{1 \leq i \leq d}  \left\langle \cos2\pi(\phi , q_1) \left( e_i + \di^* \chi_{i}(y , \phi)\right) \right\rangle_{\mu_\beta}  \left( \di^* \bar G_{q_2}(y +\kappa) \cdot e_i \right)  \right| \\ \leq \frac{C_{q_2}}{R^{d + \gamma}}.
\end{multline} 
We then use the definition of the function $G_{q_2}$ stated in~\eqref{e.Gq} combined with the fact that the triple gradient of the Green's function $G$ decays like $z \mapsto |z|^{d+1}$. We deduce that, for each point $\kappa \in A_R$,
\begin{equation} \label{eq:TV07217}
    \left|   \left( \di^* \bar G_{q_2}(y + \kappa) \cdot e_i \right) -  \nabla_i   \left( \di^* \bar G_{q_2}(\kappa) \cdot e_i \right)  \right| \leq |y| \times \supp_{z \in B(\kappa , y)} \left| \nabla^2 \bar G_{q_2}(\kappa) \right| \leq \frac{C_{q_2} |y|}{R^{d+1}}.
\end{equation}
A combination of the inequalities~\eqref{eq:TV07167} and~\eqref{eq:TV07217} implies the estimate~\eqref{eq:TV08157} in Case 1. 

\medskip

\textit{Case 2. The norm of $y$ is larger than $\frac{R}{2}$.} In that case, we use the following (rough) estimates: for each pair of points~$y , \kappa \in \Zd$,
\begin{equation} \label{eq:TV08227}
    \left|\left\langle \cos2\pi(\phi , q_1(\cdot - \kappa))  \di^* \mathcal{U}_{q_2}( y + \kappa , \phi) \right\rangle_{\mu_\beta} \right| \leq C_{q_2}
\end{equation}
and 
\begin{equation} \label{eq:TV08227bis}
    \sum_{1 \leq i \leq d} \left| \left\langle \cos2\pi(\phi , q_1) \left( e_i + \di^* \chi_{i}(y , \phi)\right) \right\rangle_{\mu_\beta} \left( \di^* \bar G_{q_2}(y +\kappa) \cdot e_i \right) \right| \leq C_{q_2}.
\end{equation}
Using the estimates~\eqref{eq:TV08227} and~\eqref{eq:TV08227bis}, the fact that the volume of the annulus $A_R$ is of order $R^d$ and the assumption $|y| \geq \frac R2$, we deduce that
\begin{align*}
    \lefteqn{ \sum_{\kappa \in A_R}\left| \left\langle \cos2\pi(\phi , q_1(\cdot - \kappa))  \di^* \mathcal{U}_{q_2}( y + \kappa , \phi) \right\rangle_{\mu_\beta} - \sum_{1 \leq i \leq d} \left\langle \cos2\pi(\phi , q_1) \left( e_i + \di^* \chi_{i}(y , \phi)\right) \right\rangle_{\mu_\beta}  \left( \di^* \bar G_{q_2}(y +\kappa) \cdot e_i \right) \right| } \qquad \hspace{130mm} & \\ & \leq C_{q_2} R^d \\ & \leq  \frac{2^{2d + \gamma} C_{q_2}|y|^{2d + \gamma}}{R^{d + \gamma}}.
\end{align*}
The proof of the estimate~\eqref{eq:TV08157} is complete.

We then consider the estimate~\eqref{eq:TV08157} for a point $y$ in the support of the charge $n_{q_1}$, consider the scalar product with the vector $n_{q_1}(y)$ and the sum over all the points $y$ in the support of $n_{q_1}$. We additionally use the inequalities $\left| n_{q_2} \right| \leq C_{q_2}$, $\left| \supp n_{q_2} \right| \leq C_{q_2}$ and the fact that, since $0$ belongs to the support of $n_{q_2}$, one has the estimate, for each point $y$ in the support of $n_{q_2}$, $|y| \leq \diam n_{q_2} \leq C_{q_2}$. We obtain
\begin{multline*}
\sum_{\kappa \in A_R}\left| \left\langle \cos2\pi(\phi , q_1(\cdot - \kappa))  n_{q_1}(y)\di^* \mathcal{U}_{q_2}( y + \kappa , \phi) \right\rangle_{\mu_\beta} - \sum_{1 \leq i \leq d} \left\langle \cos2\pi(\phi , q_1) n_{q_1}(y) \left( e_i + \di^* \chi_{i}(y , \phi)\right) \right\rangle_{\mu_\beta}  \left( \di^* \bar G_{q_2}(\kappa) \cdot e_i \right)  \right|\\ \leq \frac{C_{q_1, q_2}}{R^{\gamma}}.
\end{multline*}
We sum over all the points $y$ in the support of the charge $n_{q_1}$ and obtain
\begin{multline} \label{eq:TV17347}
    \sum_{\kappa \in A_R}\left|  \left\langle \cos2\pi(\phi , q_1(\cdot - \kappa))  \left( n_{q_1}, \di^* \mathcal{U}_{q_2}( \cdot + \kappa , \phi) \right) \right\rangle_{\mu_\beta} - \sum_{1 \leq i \leq d} \left\langle \cos2\pi(\phi , q_1) \left( n_{q_1},  e_i + \di^* \chi_{i}(y , \phi)\right) \right\rangle_{\mu_\beta}  \left( \di^* \bar G_{q_2}(\kappa) \cdot e_i \right) \right| \\ \leq \frac{C_{q_1,q_2}}{R^{ \gamma}}.
\end{multline}
We then focus on the term~$\eqref{eq:TV12276}-(q_1,q_2)$ (and more specifically on the right side of~\eqref{eq:TV17267}) and use the inequality~\eqref{eq:TV17347} to simplify it.
To ease the notation, we introduce the following definitions:
\begin{itemize}
    \item We let $\mathcal{E}_{q_1 , q_2}$ be the map from $\Zd$ to $\R$ defined according to the formula, for each point $\kappa \in \Zd$,
    \begin{equation*}
        \mathcal{E}_{q_1 , q_2}(\kappa) :=  \left\langle \cos2\pi(\phi , q_1(\cdot - \kappa))  \left(n_{q_1}, \di^* \mathcal{U}_{q_2}( \cdot + \kappa , \phi)\right) \right\rangle_{\mu_\beta} - \sum_{1 \leq i \leq d} \left\langle \cos2\pi(\phi , q_1) \left( n_{q_1}, e_i + \di^* \chi_{i}\right) \right\rangle_{\mu_\beta} \left( \di^* \bar G_{q_2}(\kappa) \cdot e_i \right),
    \end{equation*}
    It is an error term which is small; in view of the estimate~\eqref{eq:TV17347}, Remark~\ref{rem:rem3.3} and the definition of the map $G_{q_2 , j}$ stated in~\eqref{e.Gq} of Theorem~\ref{thm:homogmixedderchap4}, it satisfies the inequalities
    \begin{equation} \label{eq:TV19347}
        \forall R \geq 1, \hspace{2mm} \sum_{\kappa \in A_R} \left| \mathcal{E}_{q_1 , q_2}(\kappa) \right| \leq \frac{C_{q_1 , q_2}}{R^{\gamma}} \hspace{3mm} \mbox{and} \hspace{3mm} \forall \kappa \in \Zd, \hspace{2mm} \left| \mathcal{E}_{q_1 , q_2}(\kappa) \right| \leq  \frac{C_{q_1,q_2}}{\left| \kappa \right|^{d - \ep}};
    \end{equation}
    \item We recall the definition of the coefficient $\bar \lambda_\beta$ stated in Remark~\ref{rem:rem3.3} of Chapter~\ref{chap:chap3}. For each pair of integers $i,j \in \left\{ 1 , \ldots , d \right\}^2$, we define the coefficient $c_{ij}^{q}$ according to the formula
    \begin{equation*}
        c_{ij}^{q} := 2\pi \left(1 + \bar \lambda_\beta \right)^{-\frac 12} \left[ \left( n_{q} \right) \cdot e_i \right] \times \left\langle \cos2\pi(\phi , q_1) \left( n_{q_1}, \di^* l_{e_{ij}} + \di^* \chi_{ij}\right) \right\rangle_{\mu_\beta}.
    \end{equation*}
\end{itemize}
Using these notations together with Remark~\ref{remark4.914470302} of Chapter~\ref{chap:chap3} and an explicit computation (which we omit here), we obtain the formula
\begin{multline*}
    \sum_{1 \leq i \leq d} \left\langle \cos2\pi(\phi , q_1) \left( n_{q_1},  e_i + \di^* \chi_{i}(y , \phi)\right) \right\rangle_{\mu_\beta}  \left( \di^* \bar G_{q_2}(\kappa) \cdot e_i \right) \\ = \left( 1 + \bar \lambda_\beta\right)^{-1} \sum_{1 \leq i, j  \leq d} \left\langle \cos 2\pi\left( \phi , q_1\right) \left( n_{q_1} , e_i + \di^* \chi_{i} \right)   \right\rangle_{\mu_\beta} \left\langle \cos 2\pi\left( \phi , q_1\right) \left( n_{q_1} , e_i + \di^* \chi_{i} \right)   \right\rangle_{\mu_\beta}  \nabla_i \nabla_j G.
\end{multline*}
The term $\eqref{eq:TV12276}-(q_1,q_2)$ then becomes
\begin{multline*}
    \eqref{eq:TV12276}-(q_1,q_2) = \sum_{i,j,k,l = 1}^d c_{ij}^{q_1} c_{kl}^{q_2} \sum_{z_1,z_2 \in \Zd} \nabla_i G (z_1) \nabla_j \nabla_k G (z_1 - z_2) \nabla_l G (x - z_2) \\ + 4 \pi^2 \sum_{z_2, \kappa \in \Zd}  \nabla G(z_2) \cdot (n_{q_2}) \nabla G(z_2 + \kappa - x) \cdot (n_{q_1})  \mathcal{E}_{q_1,q_2}(\kappa).
\end{multline*}
To prove the estimate~\eqref{eq:TV12506}, it is thus sufficient to prove that there exists a constant $K_{q_1 , q_2}$ such that one has the expansion
\begin{multline*}
    4\pi^2\sum_{z_2 , \kappa  \in \Zd}  \nabla G(z_2) \cdot (n_{q_2}) \nabla G_x(z_2 + \kappa) \cdot (n_{q_1})  \mathcal{E}_{q_1,q_2}(\kappa) \\= K_{q_1 , q_2} \sum_{z_1  \in \Zd}  \nabla G(z_2) \cdot (n_{q_2}) \nabla G_x(z_2) \cdot (n_{q_1}) + O \left(\frac{C_{q_1 , q_2}}{R^{d + \gamma'}} \right).
\end{multline*}
The proof of this result relies on the estimates~\eqref{eq:TV19347}; it is the subject of Proposition~\ref{prop:propB1} and is deferred to Chapter~\ref{section7}. We note that the argument gives an explicit value for the constant $K_{q_1 , q_2}$ which is given by
\begin{equation*}
    K_{q_1 , q_2} =  4\pi^2 \sum_{\kappa \in \Zd} \mathcal{E}_{q_1,q_2}(\kappa).
\end{equation*}
By the estimates~\eqref{eq:TV19347}, the constant $K_{q_1 , q_2}$ is well-defined and grows at most algebraically fast in the parameters $\left\| q_1 \right\|_1$ and $\left\| q_2 \right\|_1$ as required.

\chapter{Regularity theory for low temperature dual Villain model} \label{section:section4}

This chapter is devoted to the study of the solutions of the Helffer-Sj{\"o}strand operator
\begin{equation} \label{eq:theHSoperator}
    \mathcal{L} := \Delta_\phi - \frac{1}{2\beta} \Delta + \frac{1}{2\beta}\sum_{n \geq 1} \frac{1}{\beta^{ \frac n2}} (-\Delta)^{n+1} + \sum_{q \in \mathcal{Q}} \nabla_q^* \cdot \a_q \nabla_q,
\end{equation}
where we recall the notation, for each integer-valued closed and compactly supported charge $q \in \mathcal{Q}$,
\begin{equation} \label{eq:TV18290702}
 \nabla_q^* \cdot \a_q \nabla_q u = z\left( \beta , q \right) \cos 2\pi\left( \phi , q \right) \left( u , q \right) q.
\end{equation}
This operator acts on the space $\Zd \times \Omega$. We decompose it into two terms: the operator $\Delta_\phi$ which acts on the field $\phi$ and the spatial term $\mathcal{L}_{\mathrm{spat}}$ defined by the formula
\begin{equation*}
    \mathcal{L}_{\mathrm{spat}} := - \frac{1}{2\beta} \Delta + \frac{1}{2\beta}\sum_{n \geq 1} \frac{1}{\beta^{ \frac n2}} (-\Delta)^{n+1} + \sum_{q \in \mathcal{Q}} \nabla_q^* \cdot \a_q \nabla_q.
\end{equation*}
The operator $\mathcal{L}_{\mathrm{spat}}$ is uniformly elliptic. The purpose of this chapter is to apply the techniques from the theory of elliptic regularity to understand the behavior of the solutions of the equations $\Lspat u =0$ and $\L u =0$. We study three types of objects:
\begin{itemize}
    \item In Sections~\ref{sec:Caccineq} and~\ref{sec:section4.2}, we study the solutions of the equation $\L u =0$. We establish a Caccioppoli inequality (Proposition~\ref{Caccio.ineq}) and $C^{0, 1- \ep}$-regularity estimates (Proposition~\ref{prop:prop4.6});
        \item In Section~\ref{sec:section4.3}, we study the Green's matrix $\mathcal{G}_{\mathbf{f}}$ and the heat kernel $\mathcal{P}_{\mathbf{f}}$. We prove Gaussian bounds on the heat kernel, decay estimates on the Green's matrix and $C^{0,1-\ep}$-regularity estimates;
    \item Section~\ref{sec.section4.5} is devoted to the study of the Green's matrix $\mathcal{G}_{\mathrm{der},\mathbf{f}}$ and the heat kernel $\mathcal{P}_{\mathrm{der},\mathbf{f}}$ associated to the differentiated Helffer-Sj{\"o}strand $\mathcal{L}_{\mathrm{der}}$ (see \eqref{e.Lder}); we prove Gaussian bounds on the heat kernel and decay estimates on the Green's matrix and $C^{0,1-\ep}$-regularity estimates.
\end{itemize}

Let us give a few comments and heuristic of the proofs presented in this chapter. The demonstrations rely on two main ingredients:
\begin{itemize}
    \item If we write
    \begin{equation} \label{eq:defLpert}
        \mathcal{L} = \underbrace{\Delta_\phi  - \frac{1}{2\beta} \Delta }_{\mathcal{L}_0} + \underbrace{\frac{1}{2\beta}\sum_{n \geq 1} \frac{1}{\beta^{ \frac n2}} (-\Delta)^{n+1}  + \sum_{q \in \mathcal{Q}} \nabla_q^* \cdot \a_q \nabla_q }_{\mathcal{L}_{\mathrm{pert}}},
    \end{equation}
    then the operator $\mathcal{L}_{\mathrm{pert}}$ is a perturbation of $\mathcal{L}_0$ if the inverse temperature $\beta$ is large enough. The operator $\mathcal{L}_0$ has properties similar to the Laplacian and a complete regularity theory is available. The strategy to obtain regularity estimates relies on \emph{Schauder theory} (see~\cite[Chapter 3]{QL}): since the operator $\mathcal{L}_{\mathrm{pert}}$ is a perturbation of the operator $\mathcal{L}_0$, one can prove that each solution of the equation $\mathcal{L} u = 0$ is well-approximated on every scale by a solution $\bar u$ of the equation $\mathcal{L}_{0} \bar u = 0$. One can then borrow the regularity of the function $\bar u$ and transfer it to the function $u$. This process causes a deterioration of the regularity for the function $u$: one obtains a $C^{0, 1-\ep}$-regularity theory for the solutions of the system $\mathcal{L} u = 0$, for some strictly positive exponent $\ep$. The size of the exponent $\ep$ depends on the inverse temperature $\beta$ and tends to $0$ as $\beta$ tends to infinity.
    \item The second ingredient is the Feynman-Kac formula which is used to study heat kernel and Green's matrix and is described in the following paragraph. Given a real number $p \in [1 , \infty)$ and a  function $\mathbf{f} \in L^p \left( \mu_\beta\right)$, we let $\mathcal{P}_{\mathbf{f}}$ be the solution of the parabolic equation
    \begin{equation} \label{eq:TVdefmatcapPf}
        \left\{ \begin{aligned} 
        \partial_t \mathcal{P}_{\mathbf{f}} + \L \mathcal{P}_{\mathbf{f}} = 0 ~\mbox{in}~ (0, \infty) \times \Omega \times \Zd, \\
        \mathcal{P}_{\mathbf{f}} (0 , \cdot , \cdot) = \mathbf{f}(\cdot) \delta_0 ~\mbox{in}~ \Omega \times \Zd.
        \end{aligned}
        \right.
    \end{equation}
    Given a field $\phi \in \Omega$ which satisfies the growth condition $\sum_{x \in \Zd} \left| \phi(x)\right| e^{-r|x|} < \infty$ for some $r > 0$, we let $(\phi_t)_{t \geq 0}$ be the diffusion process evolving according to the Langevin dynamics
    \begin{equation} \label{def.phi_t}
        \left\{ \begin{aligned}
        d \phi_t(x) & =  \frac{1}{2\beta} \Delta \phi_t (x) dt - \frac{1}{2\beta}\sum_{n \geq 1} \frac{1}{\beta^{ \frac n2}} (-\Delta)^{n+1} \phi_t (x) dt + \sum_{q\in \mathcal{Q}} \left(\nabla_q^* \cdot \a_q(\phi_t) \nabla_q \phi_t\right) (x) dt + d B_t(x), \\
        \phi_0 (x) & = \phi(x),
        \end{aligned} \right.
    \end{equation}
    where $\left( B_t(x) \right)_{x \in \Rd}$ is a collection of normalized $\R^{\binom d2}$-valued independent Brownian motions. We denote by $\P_\phi$ the law of the dynamics $(\phi_t)_{t \geq 0}$ starting from $\phi$ and by $\E_\phi$ the expectation with respect to the measure $\P_\phi$. The solvability of the SDE~\eqref{def.phi_t} is guaranteed by standard arguments (see~\cite[Section 2.1.3]{GOS} or~\cite[Section 2.2]{FS}). The solution $\mathcal{P}_{\mathbf{f}}$ of the parabolic system~\eqref{eq:TVdefmatcapPf} is related to the diffusion process $(\phi_t)_{t \geq 0}$ through the Feynman-Kac representation formula which reads as follows, for each $(x , \phi) \in \Zd \times \Omega$, one has the identity
    \begin{equation} \label{eq:FeynKac}
         \mathcal{P}_{\mathbf{f}} \left( t , x , \phi ; y\right) = \E_{\phi} \left[ \mathbf{f}(\phi_t) P^{\phi_\cdot}(t , x , y) \right],
    \end{equation}
    where $P^{\phi_\cdot}$ is the solution of the parabolic system
    \begin{equation*}
        \left\{ \begin{aligned}
        \partial_t P^{\phi_\cdot}\left(t,\cdot , y\right) - \frac{1}{2\beta} \Delta P^{\phi_\cdot}\left(t,\cdot\right) + \frac{1}{2\beta}\sum_{n \geq 1} \frac{1}{\beta^{ \frac n2}} (-\Delta)^{n+1} P^{\phi_\cdot}\left(t,\cdot\right) + \sum_{q\in \mathcal{Q}} \left(\nabla_q^* \cdot \a_q(\phi_t) \nabla_q P^{\phi_\cdot}\left(t,\cdot \right)\right) =0 ~\mbox{in}~ \Zd, \\
        P^{\phi_\cdot}\left(0,\cdot ,y \right) = \delta_y ~\mbox{in}~\Zd.
        \end{aligned} \right.
    \end{equation*}
    The rigorous justification of the Feynman-Kac formula requires to use tools from spectral theory. The argument in the case of the dual Villain model is identical to the one presented for the uniformly elliptic $\nabla \phi$-model which can be found in the articles of Naddaf and Spencer~\cite{NS} and Giacomin, Olla and Spohn~\cite{GOS} and not presented here.
\end{itemize}
While most of the ideas and techniques come from the theory of elliptic and parabolic regularity, one needs to treat the infinite range of the elliptic operator $\mathcal{L}_{\mathrm{spat}}$; for each integer $n \in \N$, the iteration of the Laplacian $\Delta^n$ has range $2n$ and the sum over all the charges $q \in \mathcal{Q}$ involves charges with arbitrarily large support. Nevertheless one has exponential decay for the long range terms, due to the exponent $\beta^{\frac n2}$ and to the estimate \eqref{e.zest} of Chapter~\ref{chap:chap3} on the coefficient $z \left( \beta , q \right)$; this allows to prove that the contribution of these terms is negligible and to obtain the desired results.

We complete this section by mentioning that we need to keep track of the dependence of the constants in the inverse temperature $\beta$; our objective is to prove that the regularity exponent $\ep$ tends to $0$ as the inverse temperature $\beta$ tends to infinity. We thus only allow the constants to depend on the dimension.
    
\section{Caccioppoli inequality for the solutions of the Helffer-Sj{\"o}strand equation} \label{sec:Caccineq}

In this section, we prove a Caccioppoli inequality for the operator $\mathcal{L}$, the proof follows the standard technique but some technical difficulties have to be taken into account due the infinite range of the operator~$\mathcal{L}$. In particular, the result obtained is slightly different from the one of the standard Caccioppoli inequality: there is a long range term in the right sides of~\eqref{est:Caccio.HS} and~\eqref{est:Caccio.HSbis} which takes into consideration the infinite range of the Helffer-Sj{\"o}strand operator. Since the coefficients of the operator~$\mathcal{L}$ decay exponentially fast, the long range term in the right sides of~\eqref{est:Caccio.HS} and~\eqref{est:Caccio.HSbis} exhibits the same decay.

\begin{proposition}[Caccioppoli inequality] \label{Caccio.ineq}
Fix a radius $R\geq 1$ and let $u : \Zd \rightarrow \R^{\binom d2}$ be a solution of the Helffer-Sj{\"o}strand equation
\begin{equation} \label{HSforreg}
 \mathcal{L} u = 0  ~\mbox{in}~ B_{2R} \times \Omega.
\end{equation}
Then there exist a constant $C:= C(d) < \infty$ and an exponent $c:=c(d) > 0$ such that the following estimates hold
\begin{equation} \label{est:Caccio.HS}
\beta \sum_{y \in \Zd} \left\| \partial_y u \right\|_{\underline{L}^2 \left(B_R , \mu_\beta \right)} + \left\|\nabla u \right\|_{\underline{L}^2\left( B_R , \mu_\beta \right) } \leq \frac CR \left\| u \right\|_{\underline{L}^2\left( B_{2R} , \mu_\beta \right) }  + \sum_{x \in \Zd \setminus B_{2R}} e^{-c \left( \ln \beta \right) |x|}   \left\| u(x , \cdot) \right\|_{L^2\left( \mu_\beta \right) },
\end{equation}
and
\begin{equation} \label{est:Caccio.HSbis}
\left\|\nabla u \right\|_{\underline{L}^2\left( B_R , \mu_\beta \right) } \leq \frac CR \left\| u - \left( u \right)_{B_{2R}} \right\|_{\underline{L}^2\left( B_{2R} , \mu_\beta \right) }  + \sum_{x \in \Zd} e^{-c \left( \ln \beta \right) \left(R \vee |x|\right)}   \left\| u(x , \cdot) \right\|_{L^2\left( \mu_\beta \right) }.
\end{equation}
\end{proposition}

\begin{remark}
The two long range terms in the right sides of~\eqref{est:Caccio.HS} and~\eqref{est:Caccio.HSbis} are error terms which are small and are caused by the infinite range of the operator $\L_{\mathrm{spat}}$. They decay exponentially fast and are typically of order $e^{-R}$.
\end{remark}

\begin{proof}
The strategy of the proof follows the standard outline of the proof of the Caccioppoli inequality. We let $\eta: \Zd \rightarrow \R$ be a cutoff function satisfying the following properties
\begin{equation} \label{eq:propetaCaccio}
0 \leq \eta \leq 1, \hspace{3mm} \eta = 1 ~\mbox{in}~ B_R, \hspace{3mm} \eta = 0 ~\mbox{in}~ \Zd \setminus  B_{2R} ~\mbox{and}~ \left| \nabla \eta \right| \leq \frac CR.
\end{equation}
We note that the properties on the function $\eta$ imply, for each charge $q \in \mathcal{Q}$,
\begin{equation} \label{eq:supinfetaCacc}
    \sup_{\supp n_q} \eta \leq \inf_{\supp n_q} \eta + \frac{C \diam n_q}{R} \leq \inf_{\supp n_q} \eta + \frac{C_q}{R}.
\end{equation}
By testing the function $\eta u$ in the equation~\eqref{HSforreg}, we obtain
\begin{multline} \label{eq:Caccioexpanded}
\beta \sum_{x , y \in \Zd} \eta(x)^2 \left\langle \left( \partial_y u(x , \phi) \right)^2 \right\rangle_{\mu_\beta} + \frac 1{2}\sum_{x \in \Zd}  \left\langle \nabla u(x, \phi) \cdot \nabla \left( \eta^2 u \right)(x, \phi) \right\rangle_{\mu_\beta} \\ +  \underbrace{ \beta \sum_{q \in \mathcal{Q}}  \left\langle \nabla_q u \cdot \a_q \nabla_q \left( \eta^2 u \right) \right\rangle_{\mu_\beta}}_{\eqref{eq:Caccioexpanded}-(i)} +  \frac 1{2} \underbrace{\sum_{n \geq 1} \sum_{x \in \Zd}  \frac{1}{\beta^{ \frac n2}} \left\langle  \nabla^{n+1} u(x, \phi) \cdot \nabla^{n+1} \left( \eta^2 u \right)(x,\phi) \right\rangle_{\mu_\beta}}_{\eqref{eq:Caccioexpanded}-(ii)} = 0.
\end{multline}
The terms in the first line can be estimated with the standard arguments of the Caccioppoli inequality. The terms in the second line require specific considerations.

To estimate the term~\eqref{eq:Caccioexpanded}-(i), we first fix a charge $q \in \mathcal{Q}$ and write $$\nabla_q \left( \eta^2  u \right) = \left( \eta^2 u , q \right) = \left( \di^* \left( \eta^2 u\right) , n_q \right).$$ We then expand the codifferential of the product $\di^* \left( \eta^2 u\right)$ thanks to the formula~\eqref{eq:formLkdde} of Chapter~\ref{Chap:chap2}. We obtain
\begin{align*}
    \di^* \left( \eta^2 u\right)(x) =  L_{k , \di^*} \left( \nabla \left( \eta^2 u \right)(x, \phi) \right) & = L_{k , \di^*} \left(   \eta^2(x) \nabla u(x, \phi) \right) 
    +  L_{k , \di^*} \left(   \nabla \eta^2(x) \otimes  u(x, \phi) \right) \\ & =  \eta^2(x) L_{k , \di^*} \left(   \nabla u(x, \phi) \right) 
    +  L_{k , \di^*} \left( 2 \eta(x)  \nabla \eta(x) \otimes u(x, \phi) \right) \\
    & =  \eta^2(x) \di^* u(x, \phi) 
    +  2 \eta(x) L_{k , \di^*} \left(   \nabla \eta(x) \otimes u(x, \phi) \right).
\end{align*}
We use the estimate~\eqref{eq:propetaCaccio} on the gradient of $\eta$ and use that the linear map $L_{k , \di^*}$ is bounded to deduce the following inequality
\begin{equation*}
    \left| \nabla_q \left( \eta^2  u \right) - \left( \eta^2 \di^* u , n_q \right) \right| \leq \frac{C  \left\| n_q\right\|_{L^\infty}}{R} \sum_{z \in \supp q} |\eta(z) u(z,\phi)| \leq \frac{C_q}{R}  \sum_{z \in \supp q} |\eta(z) u(z,\phi)|.
\end{equation*}
We use the estimate~\eqref{eq:propetaCaccio} on the gradient of $\eta$ a second time, the fact that the codifferential is a bounded operator and the estimate~\eqref{eq:supinfetaCacc}, to obtain, for any point $x$ in the support of the charge $n_q$,
\begin{align} \label{eq:TV20360401}
    \left| \left( \eta^2 \di^* u , n_q \right) - \eta^2(x) \left(  \di^* u , n_q \right) \right| & = \left| \left( (\eta^2 - \eta^2(x)) \di^* u , n_q \right)  \right| \\
    & \leq \left| \left( (\eta - \eta(x)) (\eta + \eta(x)) \di^* u , n_q \right)  \right| \notag \\
    & \leq \frac{C_q}{R} \sum_{z \in \supp n_q} |(\eta(z) + \eta(x)) u(z,\phi)| \notag\\
    & \leq \frac{C_q}{R} \sum_{z \in \supp n_q} |\eta(z) u(z,\phi)| + \frac{C_q}{R^2} \sum_{z \in \supp n_q} |u(z,\phi)|. \notag
\end{align}
A combination of the two previous displays yields the inequality, for any point $x \in \supp n_q$,
\begin{equation} \label{eq:Caccio1925}
\left| \nabla_q \left( \eta^2  u \right) - \eta(x)^2 \nabla_q   u \right| \leq \frac{C_q}{R}\sum_{z \in \supp n_q} |\eta(z) u(z,\phi)| + \frac{C_q}{R^2} \sum_{z \in \supp n_q} |u(z,\phi)|.
\end{equation}
For each charge $q \in \mathcal{Q}$, we select a point $x_q$ which belongs to its support arbitrarily. Applying the estimate~\eqref{eq:Caccio1925} with the point $x = x_q$, the Cauchy-Schwarz inequality and the estimate
\begin{equation}
\label{e.aest}
|\a_q| \leq |z(\beta,q)|\|q\|_1
\leq C_q e^{ - c \sqrt{\beta} \left\| q\right\|_1},
\end{equation}
 we obtain
\begin{align*}
\left\langle \nabla_q u \cdot \a_q \nabla_q \left( \eta^2 u \right) \right\rangle_{\mu_\beta} &\geq \eta(x_q)^2 \left\langle \nabla_q u \cdot \a_q \nabla_q u \right\rangle_{\mu_\beta} -  \frac{C_q e^{ - c \sqrt{\beta} \left\| q\right\|_1}}{R} \left\| \nabla_q u\right\|_{L^2\left(\mu_\beta\right)} \sum_{x \in \supp n_q} \eta(x) \left\|  u(x , \cdot) \right\|_{L^2 \left( \mu_\beta \right)} \\
&  \qquad -  \frac{C_q e^{ - c \sqrt{\beta} \left\| q\right\|_1}}{R^2} \left\| \nabla_q u\right\|_{L^2\left(\mu_\beta\right)} \sum_{x \in \supp n_q} \left\|  u(x , \cdot) \right\|_{L^2 \left( \mu_\beta \right)}.
\end{align*}
We sum over all the charges $q \in \mathcal{Q}$ such that the support of $\eta$ intersects the support of $n_q$. We obtain the estimate
\begin{align} \label{eq:Cqccio1932}
 \lefteqn{\sum_{q \in \mathcal{Q}}  \left\langle \nabla_q u \cdot \a_q \nabla_q \left( \eta^2 u \right) \right\rangle_{\mu_\beta}} \qquad & \\ & \geq \underbrace{\sum_{q \in \mathcal{Q}} \eta(x_q)^2  \left\langle \nabla_q u \cdot \a_q \nabla_q u \right\rangle_{\mu_\beta} }_{\eqref{eq:Cqccio1932}-(i)} -   \underbrace{\sum_{q \in \mathcal{Q}} \frac{C_qe^{ - c \sqrt{\beta} \left\| q \right\|_1}}{R}  \left\| \nabla_q u\right\|_{L^2\left(\mu_\beta\right)} \sum_{z \in \supp n_q } \eta(z) \left\| u(z, \cdot) \right\|_{L^2 \left( \mu_\beta \right)}}_{\eqref{eq:Cqccio1932}-(ii)} \notag \\
 & \underbrace{- \sum_{q \in \mathcal{Q}} \frac{C_q e^{ - c \sqrt{\beta} \left\| q\right\|_1} \indc_{\{\supp n_q \cap \supp \eta \neq \emptyset\}}}{R^2} \left\| \nabla_q u\right\|_{L^2\left(\mu_\beta\right)} \sum_{x \in \supp n_q} \left\|  u(x , \cdot) \right\|_{L^2 \left( \mu_\beta \right)}}_{\eqref{eq:Cqccio1932}-(iii)}. \notag
\end{align}
We then estimate the three terms in the right side separately. The term~\eqref{eq:Cqccio1932}-(i) can be estimated thanks to the estimate \eqref{e.aest} on the value of the coefficient $\a_q$. We obtain
\begin{align*}
 \sum_{q \in \mathcal{Q}} \eta(x_q)^2 \left| \left\langle \nabla_q u \cdot \a_q \nabla_q u \right\rangle_{\mu_\beta} \right| & \leq \sum_{q \in \mathcal{Q}}C_q e^{-c \sqrt{\beta} \left\| q \right\|_1} \eta(x_q)^2 \left| \left\langle (\di^* u , n_q)^2 \right\rangle_{\mu_\beta}  \right| \\
 & \leq \sum_{q \in \mathcal{Q}} C_q e^{-c \sqrt{\beta} \left\| q \right\|_1} \eta(x_q)^2  \sum_{x \in \supp n_q} \left\| \nabla u(x , \cdot) \right\|_{L^2 \left( \mu_\beta \right)}^2 \\
 & \leq \sum_{q \in \mathcal{Q}} C_q e^{-c \sqrt{\beta} \left\| q \right\|_1} \\ & \quad \times \left( \sum_{x \in \supp n_q} \eta(x)^2 \left\| \nabla u(x , \cdot) \right\|_{L^2 \left( \mu_\beta \right)}^2 + \sum_{x \in \supp n_q} \left( \eta(x_q)^2 - \eta(x)^2 \right) \left\| \nabla u(x , \cdot) \right\|_{L^2 \left( \mu_\beta \right)}^2 \right).
\end{align*}
We use the estimates~\eqref{eq:propetaCaccio} and~\eqref{eq:supinfetaCacc} on the function $\eta$ and a computation similar to the one performed in~\eqref{eq:TV20360401} and the fact that $R$ is always larger than $1$. We obtain
\begin{multline} \label{eq:ineqTV1838}
 \sum_{q \in \mathcal{Q}} \eta(x_q)^2 \left| \left\langle \nabla_q u \cdot \a_q \nabla_q u \right\rangle_{\mu_\beta} \right| \\ \leq \sum_{q \in \mathcal{Q}} C_q e^{-c \sqrt{\beta} \left\| q \right\|_1} \left( \sum_{x \in \supp n_q} \eta(x)^2 \left\| \nabla u(x , \cdot) \right\|_{L^2 \left( \mu_\beta \right)}^2 + \frac{1}{R^2}\sum_{x \in \supp n_q} \indc_{\{\supp n_q \cap \supp \eta \neq \emptyset\}} \left\| \nabla u(x , \cdot) \right\|_{L^2 \left( \mu_\beta \right)}^2 \right).
\end{multline}
We then use that, since the support of the function $\eta$ is included in the ball $B_{2R}$, one has the inequalities, for each point $x \in \Zd$, \begin{equation} \label{eq:Bauest}
    \left\{ \begin{aligned}
    \sum_{q\in \mathcal{Q}_x} C_q \indc_{\{\supp n_q \cap \supp \eta \neq \emptyset\}} e^{-c \sqrt{\beta} \left\| q \right\|_1} &\leq C e^{-c \sqrt{\beta} \dist\left(x , B_{2R} \right)}, \\
    \sum_{q\in \mathcal{Q}_x} C_q e^{-c \sqrt{\beta} \left\| q \right\|_1} &\leq C e^{-c \sqrt{\beta}}.
    \end{aligned} \right.
\end{equation}
These estimates are a consequence of the inequalities~\eqref{eq:sumcharges} in Chapter~\ref{Chap:chap2}. Using~\eqref{eq:Bauest} and the fact that the discrete gradient is a bounded operator, we can simplify the estimate~\eqref{eq:ineqTV1838} and obtain
\begin{equation} \label{eq:TVfinCacc2}
    \sum_{q \in \mathcal{Q}} \eta(x_q)^2 \left\langle \nabla_q u \cdot \a_q \nabla_q u \right\rangle_{\mu_\beta}  \geq - C e^{-c \sqrt{\beta} } \sum_{x \in \Zd} \eta(x)^2 \left\| \nabla u(x , \cdot ) \right\|_{L^2 \left( \mu_\beta\right)}^2 - \frac{C}{R^2}\sum_{x \in \Zd} e^{-c \sqrt{\beta} \dist(x , B_{2R})} \left\| u(x , \cdot) \right\|_{L^2 \left( \mu_\beta \right)}^2.
\end{equation}
This completes the estimate of the term~\eqref{eq:Cqccio1932}-(i).

To estimate the term~\eqref{eq:Cqccio1932}-(ii), we fix a charge $q \in \mathcal{Q}$. We note that only the charges $q'$ such that the support of $n_{q'}$ intersects the support of $\eta$ (or equivalently the ball $B_{2R}$) are counted in the sum; we can thus assume without loss of generality that the support of $n_q$ intersects the support of $\eta$. We have the inequality, for each field $ \phi \in \Omega$, 
$$ \left| \nabla_q u(\cdot , \phi)\right| = (\di^* u (\cdot , \phi), n_q) \leq \left\|  n_q \right\|_{\infty} \sum_{z \in \supp q} |\nabla u(z, \phi) | \leq C_q \sum_{z \in \supp q} |\nabla u(z, \phi)|.$$
Using this estimate, we obtain
\begin{align*}
    \left\| \nabla_q u\right\|_{L^2\left(\mu_\beta\right)} \sum_{z \in \supp n_q } \eta(z) \left\| u(z, \cdot) \right\|_{L^2 \left( \mu_\beta \right)} & \leq C_q \left( \sum_{z \in \supp n_q }  \left\| \nabla u(z, \cdot) \right\|_{L^2 \left( \mu_\beta \right)} \right) \left( \sum_{z \in \supp n_q } \eta(z) \left\| u(z, \cdot) \right\|_{L^2 \left( \mu_\beta \right)} \right)  \notag \\
    & \leq C_q \left(\sup_{\supp n_q} \eta\right) \left( \sum_{z \in \supp n_q } \left\| \nabla u(z, \cdot) \right\|_{L^2 \left( \mu_\beta \right)} \right)\left( \sum_{z \in \supp n_q } \left\| u(z, \cdot) \right\|_{L^2 \left( \mu_\beta \right)} \right).
\end{align*}
We then use the property~\eqref{eq:supinfetaCacc} to deduce that
\begin{align} \label{eq:TV30905}
     \lefteqn{\left\| \nabla_q u\right\|_{L^2\left(\mu_\beta\right)} \sum_{z \in \supp n_q } \eta(z) \left\| u(z, \cdot) \right\|_{L^2 \left( \mu_\beta \right)}} \qquad & \\ & \leq C_q \left(\inf_{\supp n_q} \eta + \frac{C_q}{R}\right) \left( \sum_{z \in \supp n_q } \left\| \nabla u(z, \cdot) \right\|_{L^2 \left( \mu_\beta \right)} \right)\left( \sum_{z \in \supp n_q } \left\| u(z, \cdot) \right\|_{L^2 \left( \mu_\beta \right)} \right) \notag \\
     & \leq C_q \left( \sum_{z \in \supp n_q } \eta(z) \left\| \nabla u(z, \cdot) \right\|_{L^2 \left( \mu_\beta \right)} \right)\left( \sum_{z \in \supp n_q } \left\| u(z, \cdot) \right\|_{L^2 \left( \mu_\beta \right)} \right) \notag \\
     & \quad +  \frac{C_q}{R} \left( \sum_{z \in \supp n_q } \left\| \nabla u(z, \cdot) \right\|_{L^2 \left( \mu_\beta \right)} \right)\left( \sum_{z \in \supp n_q } \left\| u(z, \cdot) \right\|_{L^2 \left( \mu_\beta \right)} \right). \notag
\end{align}
We estimate the first term of the right side of the inequality~\eqref{eq:TV30905} by the Cauchy-Schwarz and Young's inequalities
\begin{multline} \label{eq:TV9322}
    \left( \sum_{z \in \supp n_q } \eta(z) \left\| \nabla u(z, \cdot) \right\|_{L^2 \left( \mu_\beta \right)} \right)\left( \sum_{z \in \supp n_q } \left\| u(z, \cdot) \right\|_{L^2 \left( \mu_\beta \right)} \right) \\ \leq C_q R \sum_{z \in \supp n_q } \eta(z)^2 \left\| \nabla u(z, \cdot) \right\|_{L^2 \left( \mu_\beta \right)}^2 +\frac{C_q}{R} \sum_{z \in \supp n_q } \left\| u(z, \cdot) \right\|_{L^2 \left( \mu_\beta \right)}^2 .
\end{multline}
To estimate the second term in the right side of~\eqref{eq:TV30905}, we use that the discrete gradient is a bounded operator and the Cauchy-Schwarz inequality. We obtain
\begin{equation} \label{eq:TV9321}
    \left( \sum_{z \in \supp n_q } \left\| \nabla u(z, \cdot) \right\|_{L^2 \left( \mu_\beta \right)} \right)\left( \sum_{z \in \supp n_q } \left\| u(z, \cdot) \right\|_{L^2 \left( \mu_\beta \right)} \right) \leq 
    C_q \sum_{z \in \supp n_q } \left\| u(z, \cdot) \right\|_{L^2 \left( \mu_\beta \right)}^2.
\end{equation}
Collecting the estimates~\eqref{eq:TV30905},~\eqref{eq:TV9322},~\eqref{eq:TV9321}, we obtain the upper bound for the term~\eqref{eq:Cqccio1932}-(ii)
\begin{multline} \label{eq:TV95030}
    \left\| \nabla_q u\right\|_{L^2\left(\mu_\beta\right)} \sum_{z \in \supp n_q } \eta(z) \left\| u(z, \cdot) \right\|_{L^2 \left( \mu_\beta \right)} \leq C_q R \sum_{z \in \supp n_q } \eta(z)^2 \left\| \nabla u(z, \cdot) \right\|_{L^2 \left( \mu_\beta \right)}^2 \\
    + \frac{C_q}{R} \sum_{z \in \supp n_q } \left\| u(z, \cdot) \right\|_{L^2 \left( \mu_\beta \right)}^2.
\end{multline}
Multiplying the inequality~\eqref{eq:TV95030} by $e^{-c \sqrt{\beta} \left\| q\right\|_1}$, summing over all the charges $q \in \mathcal{Q}$ and using the estimates~\eqref{eq:Bauest}, we obtain
\begin{multline} \label{eq:TVfinCacc1}
    \frac{1}{R}\sum_{q \in \mathcal{Q}} C_q e^{ - c \sqrt{\beta} \left\| q \right\|_1} \left\| \nabla_q u\right\|_{L^2\left(\mu_\beta\right)} \sum_{z \in \supp n_q } \eta(z) \left\| u(z, \cdot) \right\|_{L^2 \left( \mu_\beta \right)} \\ \leq C e^{-c \sqrt{\beta}} \sum_{x \in \Zd} \eta(x)^2 \left\| \nabla u(x , \cdot ) \right\|_{L^2 \left( \mu_\beta\right)}^2 + \frac{C}{R^2}\sum_{x \in \Zd} e^{-c \sqrt{\beta} \dist(x , B_{2R})} \left\|  u(x , \cdot) \right\|_{L^2 \left( \mu_\beta \right)}^2.
\end{multline}

The term~\eqref{eq:Cqccio1932}-(iii) is estimated following a similar strategy and we omit the details. We obtain
\begin{multline} \label{eq:TVfinCacc1RGHT}
     \sum_{q \in \mathcal{Q}} \frac{C_q e^{ - c \sqrt{\beta} \left\| q\right\|_1} \indc_{\{\supp n_q \cap \supp \eta \neq \emptyset\}}}{R^2} \left\| \nabla_q u\right\|_{L^2\left(\mu_\beta\right)} \sum_{x \in \supp n_q} \left\|  u(x , \cdot) \right\|_{L^2 \left( \mu_\beta \right)} \\ \leq C e^{-c \sqrt{\beta}} \sum_{x \in \Zd} \eta(x)^2 \left\| \nabla u(x , \cdot ) \right\|_{L^2 \left( \mu_\beta\right)}^2 + \frac{C}{R^2}\sum_{x \in \Zd} e^{-c \sqrt{\beta} \dist(x , B_{2R})} \left\|  u(x , \cdot) \right\|_{L^2 \left( \mu_\beta \right)}^2.
\end{multline}

Combining the estimates~\eqref{eq:Cqccio1932},~\eqref{eq:TVfinCacc2},~\eqref{eq:TVfinCacc1},~\eqref{eq:TVfinCacc1RGHT} and choosing the inverse temperature $\beta$ large enough so that the exponential terms $Ce^{ - c \sqrt{\beta}}$ are smaller than $\frac 1{4\beta}$, we obtain the inequality 
\begin{equation} \label{eq:TV1040C}
    \beta \sum_{q \in \mathcal{Q}}  \left\langle \nabla_q u \cdot \a_q \nabla_q \left( \eta^2 u \right) \right\rangle_{\mu_\beta} \geq - \frac 12 \sum_{x \in \Zd} \eta(x)^2 \left\| \nabla u(x , \cdot ) \right\|_{L^2 \left( \mu_\beta\right)}^2 - \frac{C}{R^2}\sum_{x \in \Zd} e^{-c \sqrt{\beta} \dist(x , B_{2R})} \left\|  u(x , \cdot) \right\|_{L^2 \left( \mu_\beta \right)}^2.
\end{equation}
This completes the estimate of the term~\eqref{eq:Caccioexpanded}-(i).

There only remains to estimate the term~\eqref{eq:Caccioexpanded}-(ii) pertaining to the iterations of the Laplacian. The proof follows similar ideas so we do not write down every details. We use the two following ingredients: the discrete gradient is an operator which has a finite operator norm in $L^1(\Zd)$ and the discrete operator $\nabla^n$ has range $n$. By expanding the term $\nabla^n \left( \eta u\right)$, and using the estimate on the gradient of $\eta$ stated in~\eqref{eq:propetaCaccio}, we obtain the following inequality, for each pair $(x,\phi) \in \Zd \times \Omega$,
\begin{equation} \label{eq:TV1549C}
\left| \nabla^n \left( \eta^2 u \right) ( x, \phi) -  \eta^2(x) \nabla^n u  ( x, \phi) \right| \leq \frac{C^n}{R} \sum_{z \in B(x,n)} \eta(z) |u( z, \phi)|.
\end{equation}
Using this inequality, we obtain
\begin{multline} \label{eq:TV29091519}
  \sum_{x \in B_{2R}}   \left\langle  \nabla^n u(x,\cdot ) \cdot \nabla^n \left( \eta^2 u \right)(x , \cdot) \right\rangle_{\mu_\beta} \\ \geq   \sum_{x \in B_{2R}}  \eta(x)^2 \left\|  \nabla^n u(x, \cdot) \right\|^2_{L^2 \left( \mu_\beta \right)} - \frac{C^n}{R}\sum_{x \in B_{2R}} \left\| \nabla^n u(x , \cdot) \right\|_{L^2 \left( \mu_\beta\right)}  \sum_{z \in B(x , n)} \eta(z) \left\|u(z , \cdot) \right\|_{L^2 \left( \mu_\beta \right)}.
\end{multline}
We then use that the discrete gradient is a bounded operator, the properties on the function $\eta$ stated in~\eqref{eq:propetaCaccio} and the fact that the volume of the ball $B_n$ is of order $n^d$ (and has smaller growth than the exponential term $C^n$). We obtain the estimate
\begin{multline} \label{eq:TV290915191}
    \sum_{x \in B_{2R}} \left\| \nabla^n u(x , \cdot) \right\|_{L^2 \left( \mu_\beta\right)}  \sum_{z \in B(x , n)} \eta(z) \left\|u(z , \cdot) \right\|_{L^2 \left( \mu_\beta \right)} \\ \leq \frac{R}{2 C^n} \left(\sum_{x \in B_{2R}} \eta(x)^2 \left\| \nabla^n u(x, \cdot) \right\|_{L^2 \left( \mu_\beta \right)}^2 \right) + \frac{\tilde C^n}{R} \sum_{x \in B_{2R + n}} \left\| u(x, \cdot) \right\|_{L^2 \left( \mu_\beta \right)}^2,
\end{multline}
where $C$ is the constant in~\eqref{eq:TV29091519} and $\tilde C > C$.
Combining the estimates~\eqref{eq:TV29091519} and~\eqref{eq:TV290915191}, we obtain
\begin{equation} \label{eq:TV1216111}
    \sum_{x \in B_{2R}}   \left\langle  \nabla^n u(x,\cdot ) \cdot \nabla^n \left( \eta^2 u \right)(x , \cdot) \right\rangle_{\mu_\beta} \\ \geq  \frac12 \sum_{x \in B_{2R}}  \eta(x)^2 \left\|  \nabla^n u(x, \cdot) \right\|^2_{L^2 \left( \mu_\beta \right)} - \frac{C^n}{R^2}\sum_{x \in B_{2R + n}} \left\|u(x , \cdot) \right\|_{L^2 \left( \mu_\beta \right)}^2.
\end{equation}
Multiplying both sides of the inequality~\eqref{eq:TV1216111} by $\beta^{\frac{n+1}{2}}$ and summing over the integers $ n \in \N$, we obtain the inequality
\begin{multline*} 
    \frac{1}{2\beta}\sum_{n \geq 1} \frac{1}{\beta^{\frac n2}}\sum_{x \in B_{2R}}   \left\langle  \nabla^{n+1} u(x,\cdot ) \cdot \nabla^{n+1} \left( \eta^2 u \right)(x , \cdot) \right\rangle_{\mu_\beta} \\ \geq  \frac1{4\beta} \left( \sum_{n \geq 1} \frac{1}{\beta^{\frac n2}} \right) \sum_{x \in B_{2R}}  \eta(x)^2 \left\|  \nabla^{n+1} u(x, \cdot) \right\|^2_{L^2 \left( \mu_\beta \right)} - \frac{1}{R^2}\sum_{n \geq 1} \left(\frac{C}{\sqrt{\beta}}\right)^n \sum_{x \in B_{2R + n}} \left\|u(x , \cdot) \right\|_{L^2 \left( \mu_\beta \right)}^2.
\end{multline*}
We use that the first term in the right side of~\eqref{eq:TV1216111} is non-negative and assume that the inverse temperature $\beta$ is chosen large enough to rewrite
\begin{align*}
    \frac{1}{2\beta}\sum_{n \geq 1} \frac{1}{\beta^{\frac n2}}\sum_{x \in B_{2R}}   \left\langle  \nabla^{n+1} u(x,\cdot ) \cdot \nabla^{n+1} \left( \eta^2 u \right)(x , \cdot) \right\rangle_{\mu_\beta} & \geq - \frac{1}{R^2}\sum_{n \in \N} \left(\frac{C}{\sqrt{\beta}}\right)^n \sum_{x \in B_{2R + n}} \left\|u(x , \cdot) \right\|_{L^2 \left( \mu_\beta \right)}^2 \\
    & \geq - \frac{C}{R^2\sqrt{\beta}} \sum_{x \in \Zd} \left( \frac{C}{\sqrt{\beta}}\right)^{ \dist\left( x , B_{2R} \right)} \left\|u(x , \cdot) \right\|_{L^2 \left( \mu_\beta \right)}^2.
\end{align*}
If $\beta$ is chosen large enough (for instance larger than $C^4$) then there exists a constant $c > 0$ such that $\frac{C}{\sqrt{\beta}} \leq e^{-c \left( \ln \beta \right)}$. We have obtained
\begin{equation} \label{eq:TV1216}
    \sum_{n \geq 1} \frac{1}{\beta^{\frac n2}}\sum_{x \in B_{2R}}   \left\langle  \nabla^{n+1} u(x,\cdot ) \cdot \nabla^{n+1} \left( \eta^2 u \right)(x , \cdot) \right\rangle_{\mu_\beta} \geq - \frac{C}{R^2\sqrt{\beta}} \sum_{x \in \Zd} e^{-c \left( \ln \beta \right) \dist\left( x , B_{2R} \right)} \left\|u(x , \cdot) \right\|_{L^2 \left( \mu_\beta \right)}^2.
\end{equation}
This completes the estimate of the term~\eqref{eq:Caccioexpanded}-(ii).

\smallskip

We then choose the inverse temperature $\beta$ large enough, combine the identity~\eqref{eq:Caccioexpanded}, the estimates~\eqref{eq:TV1040C},~\eqref{eq:TV1216} and the standard arguments of the proof of the Caccioppoli inequality. We obtain the inequality 
\begin{multline} \label{est:Caccio.HSbisbis}
\beta \sum_{y \in \Zd} \left\| \partial_y u \right\|_{\underline{L}^2 \left(B_R , \mu_\beta \right)} + \left\|\nabla u \right\|_{\underline{L}^2\left( B_R , \mu_\beta \right) } \\ \leq \frac CR \left\| u \right\|_{\underline{L}^2\left( B_{2R} , \mu_\beta \right) }  + \frac{C}{R^{1 + \frac d2}} \sum_{x \in \Zd \setminus B_{2R}} e^{-c \left( \ln \beta \right) \dist(x , B_{2R})}   \left\| u(x , \cdot) \right\|_{L^2\left( \mu_\beta \right) }.
\end{multline}
We finally post-process the estimate~\eqref{est:Caccio.HSbisbis} and use the inequality
\begin{align} \label{eq:TV20103}
    \lefteqn{\frac{C}{R^{1 + \frac d2}} \sum_{x \in \Zd \setminus B_{2R}} e^{-c \left( \ln \beta \right) \dist(x , B_{2R})}   \left\| u(x , \cdot) \right\|_{L^2\left( \mu_\beta \right)}} \qquad & \\ & \leq \frac{C}{R} \left\| u \right\|_{\underline{L}^2\left(B_{4R} \right)} + C\sum_{x \in \Zd \setminus B_{4R}} e^{-c \left( \ln \beta \right) \dist(x , B_{2R})} \left\| u(x , \cdot) \right\|_{L^2\left( \mu_\beta\right)} \notag \\ \notag
    & \leq \frac{C}{R} \left\| u \right\|_{\underline{L}^2\left(B_{4R} \right)} + C\sum_{x \in \Zd \setminus B_{4R}} e^{-c \left( \ln \beta \right) |x|} \left\| u(x , \cdot) \right\|_{L^2\left( \mu_\beta\right)},
\end{align}
where in the second line we used the inequality $|x| \leq C \dist (x , B_{2R})$ valid for any point $x \in \Zd \setminus B_{4R}$. A combination of the estimates~\eqref{est:Caccio.HSbisbis} and~\eqref{eq:TV20103} shows the inequality
\begin{equation} \label{est:Caccio.HSbister}
\beta \sum_{y \in \Zd} \left\| \partial_y u \right\|_{\underline{L}^2 \left(B_R , \mu_\beta \right)} + \left\|\nabla u \right\|_{\underline{L}^2\left( B_R , \mu_\beta \right) } \leq \frac CR \left\| u \right\|_{\underline{L}^2\left( B_{4R} , \mu_\beta \right) }  + C \sum_{x \in \Zd \setminus B_{4R}} e^{-c \left( \ln \beta \right) |x|}   \left\| u(x , \cdot) \right\|_{L^2\left( \mu_\beta \right) }.
\end{equation}
This is the estimate~\eqref{est:Caccio.HS} up to one difference: the right side of~\eqref{est:Caccio.HSbister} involves the ball $B_{4R}$ while the right side of~\eqref{est:Caccio.HS} involves the ball $B_{2R}$. This is a minor issue which can be fixed easily and we skip the details.

\smallskip

The proof of the estimate~\eqref{est:Caccio.HSbis} follows similar lines, except that we need to use the function $ \eta^2 \left( u - \frac{1}{\sum_{x \in \Zd} \eta^2(x)} \sum_{x \in \Zd} \eta^2(x) u(x , \cdot)  \right)$ as a test function in the equation~\eqref{HSforreg}. The proof is similar and the same technicalities (pertaining to the iterations of the Laplacian and to the sum over the charges $q \in \mathcal{Q}$) need to be treated; since the proof does not contain any new element, we omit the details.

\end{proof}

\section{Regularity theory for the Helffer-Sj{\"o}strand operator} \label{sec:section4.2}
The purpose of this section is to prove the $C^{0,1-\ep}$-regularity of the solutions of the Helffer-Sj{\"o}strand equation~\eqref{eq:theHSoperator}. The result is stated in Proposition~\ref{prop:prop4.6}.

The proof relies on Schauder theory; as is explained in~\eqref{eq:defLpert}, the strategy is to decompose the Helffer-Sj{\"o}strand operator $\mathcal{L}$ into two terms, denoted $\mathcal{L}_0$ and $\mathcal{L}_{\mathrm{pert}}$ as recalled below
\begin{equation*}
	\mathcal{L} = \underbrace{\Delta_\phi  - \frac{1}{2\beta} \Delta }_{\mathcal{L}_{0}} + \underbrace{\frac{1}{2\beta}\sum_{n \geq 1} \frac{1}{\beta^{ \frac n2}} (-\Delta)^{n+1}  + \sum_{q\in \mathcal{Q}} \nabla_q^* \cdot \a_q \nabla_q }_{\mathcal{L}_{\mathrm{pert}}}.
\end{equation*}
The operator $\mathcal{L}_{0}$ is the leading order term. For this operator a $C^{0,1}$-regularity theory is available, similar to the one of the Laplacian. This result is stated in Proposition~\ref{prop:prop4.4}; the proof is essentially equivalent to the standard proof of the regularity of the Laplacian.

The second operator $\mathcal{L}_{\mathrm{pert}}$ is a perturbation term; it is small when the inverse temperature $\beta$ is large. The strategy is to argue that any solution $u$ of the Helffer-Sj{\"o}strand equation is well-approximated on every scale by a function $\bar u$ of the equation $\Delta_\phi \bar u  + \frac{1}{2\beta} \Delta \bar u = 0$ and to borrow the regularity of the function $\bar u$ to obtain a $C^{0, 1 - \ep}$-regularity estimate on the solution $u$. This section can be decomposed into three propositions:
\begin{itemize}
	\item Proposition~\ref{prop:prop4.4} establishes a regularity theory for the solutions $\bar u$ of the equation $\mathcal{L}_{0} \bar u = 0$;
	\item Proposition~\ref{prop:prop4.5} states that if a function $u$ is well-approximated, in the sense of the estimate~\eqref{eq:well-approx} below, by a solution of the equation $\Delta_\phi \bar u - \frac 1{2\beta}\Delta \bar u = 0$, then a $C^{0, 1-\ep}$-regularity estimate holds for the function $u$;
	\item Proposition~\ref{prop:prop4.6} establishes the regularity for the solutions of the Helffer-Sj{\"o}strand equation. We prove that any solution $u$ of the equation $\mathcal{L} u = 0$ is well-approximated by a solution $\bar u$ of the equation $\mathcal{L}_{0} \bar u = 0$ and apply Proposition~\ref{prop:prop4.6} to conclude.
\end{itemize}

\subsection{Regularity theory for the operator \texorpdfstring{$\Delta_\phi - \frac1{2\beta}\Delta$}{2}}

In this section, we establish a regularity theory for the operator $\Delta_\phi - \frac1{2\beta}\Delta$.

\begin{proposition}[Regularity theory for the operator $\Delta_\phi  - \frac1{2\beta} \Delta$] \label{prop:prop4.4}
Fix a radius $R > 0$ and let $\bar u : B_{2R} \times \Omega$ be a solution of the equation
\begin{equation*}
\Delta_\phi  \bar u - \frac{1}{2\beta} \Delta \bar u = 0 ~\mbox{in}~ B_{2R} \times \Omega.
\end{equation*}
Then, for any integer $k \in \N$, there exists a constant $C_k < \infty$ depending on the dimension $d$ and the integer $k$ such that the following estimate holds
\begin{equation} \label{eq:regularityL1op}
\sup_{x \in B_R} \left\| \nabla^k \bar u(x , \cdot) \right\|_{L^2 \left(\mu_\beta \right)} \leq \frac{C_k}{R^{k+ \frac d2}} \left\| \bar u - \left( \bar u \right)_{B_{2R}} \right\|_{L^2\left( B_{2R}, \mu_\beta\right)}.
\end{equation}
\end{proposition}

\begin{proof}
The proof is standard and relies on two ingredients: the Caccioppoli inequality and the observation that the spatial gradient $\nabla$ commutes with the two Laplacians $\Delta_\phi$ and $\Delta$. First by the Caccioppoli inequality, which can be deduced from the standard proof as explained in Section~\ref{sec:Caccineq}, one has
\begin{equation*}
\left\|\nabla \bar u \right\|_{\underline{L}^2\left( B_R , \mu_\beta \right) } \leq \frac CR \left\| \bar u - \left( \bar u \right)_{B_{2R}}  \right\|_{\underline{L}^2\left( B_{2R} , \mu_\beta \right) } .
\end{equation*} 
We then note that, since $u$ is a solution of the equation $\mathcal{L}_{0} u = 0$, the gradient of $u$ is also a solution of the equation $\mathcal{L}_{0} \nabla u = 0$. Once can thus apply the Caccioppoli inequality to the gradient of $u$ and deduce 
\begin{equation*}
\left\|\nabla^2 \bar u \right\|_{\underline{L}^2\left( B_R , \mu_\beta \right) } \leq \frac CR \left\| \nabla \bar u \right\|_{\underline{L}^2\left( B_{2R} , \mu_\beta \right) }.
\end{equation*}
An iteration of this argument shows that, for any integer $k \geq 1$, the $ \underline{L}^2\left( B_R , \mu_\beta \right) $-norm of the iterated gradient $\nabla^k u$ is controlled by the $\underline{L}^2\left( B_{2R}, \mu_\beta \right)$-norm of the function $u$ with the appropriate scaling. By an application of the Sobolev embedding theorem (see~\cite[Chapter 4]{AF}), we obtain the regularity estimate~\eqref{eq:regularityL1op}.
\end{proof}

\subsection{Regularity theory for the Helffer-Sj{\"o}strand operator}
The next proposition states that if a map $u$ is well-approximated on every scale by a solution $\bar u$ of the equation $\mathcal{L}_{0} \bar u = 0$, then the function $u$ satisfies a $C^{0,1 - \ep}$ regularity estimate for some exponent $\ep$ depending only on the dimension $d$ and the precision of the approximation. The proof follows a well-known strategy of Campanato (see e.g. \cite{Gia}). The proof written below is an adaptation of the one of Hofmann and Kim~\cite{HK}.

\begin{proposition} \label{prop:prop4.5}
Fix a radius $X \geq 1$, a regularity exponent $\ep > 0$ and a constant $K > 0$. There exists a constant $\delta_\ep> 0$, depending on the parameters $d$ and $\ep$ such that for any radius $R \geq 2X$, and any function $u \in L^{2} \left( B_R , \mu_\beta \right)$ satisfying the property that, for any radius $r \in [X , \frac 12 R]$, there exists a function $\bar u \in  L^{2} \left( B_{2r} , \mu_\beta \right)$, solution of the equation 
\begin{equation} \label{eq:well-approxhomog}
\Delta_\phi \bar u - \frac 1{2\beta} \Delta \bar u = 0 ~\mbox{in}~ B_{2r} \times \Omega,
\end{equation}
satisfying
\begin{equation} \label{eq:well-approx}
\left\| \nabla (u - \bar u)  \right\|_{L^2 \left( B_{r} , \mu_\beta \right)} \leq \delta_\ep \left\| \nabla u \right\|_{L^2 \left( B_{2r} , \mu_\beta \right)} + C K,
\end{equation}
then there exists a constant $C := C(\ep, d) < \infty$ such that for every $r \in [X , R]$,
\begin{equation*}
\left\| \nabla u \right\|_{\underline{L}^2 \left( B_r, \mu_\beta \right)} \leq C \left( \frac{R}{r} \right)^{\ep} \left\| \nabla u \right\|_{\underline{L}^2 \left( B_R , \mu_\beta \right)} + K.
\end{equation*} 
\end{proposition}

Before starting the proof, we record the following lemma, which is a consequence of Giaquinta~\cite[Lemma 2.1]{Gia}.

\begin{lemma} \label{lem:lemma4.6}
Fix two non-negative real numbers $X , R$ such that $R \geq 2 X \geq 2$ and two non-negative constants $C_0, K$. For any regularity exponent $\ep > 0$, there exist two constants $\delta_\ep := \delta_\ep \left( C , \ep ,d \right)$ and $C_1 := C_1 \left(  C , \ep ,d \right)$ such that the following statement holds. If $\phi : \R^+ \to \R$ is a non-negative and non-decreasing function which satisfies the estimate, for each pair of real numbers $\rho , r \in [X , R]$ satisfying $\rho \leq r$,
\begin{equation} \label{eq:ineqlem4.6}
    \phi \left( \rho \right) \leq C_0 \left( \left( \frac{\rho}{r}\right)^{\frac d2} + \delta_\ep \right) \phi(r) + K,
\end{equation}
then one has the estimate, for any $\rho , r \in [X , R]$ satisfying $\rho \leq r$,
\begin{equation} \label{eq:fingiaq}
    \phi \left( \rho \right) \leq C_1 \left( \left( \frac{\rho}{r}  \right)^{\frac d2- \ep} \phi(r) + K \rho^{\frac d2} \right).
\end{equation}
\end{lemma}

\begin{proof}
This lemma can be extracted from \cite[Lemma 2.1 p86]{Gia} by setting $\alpha = \frac{d}2$, $\beta = \frac d2 - \ep$ and by using that the radii $R , r$ are larger than $1.$
\end{proof}

\begin{proof}[Proof of Proposition~\ref{prop:prop4.5}]
We fix a regularity exponent $\ep > 0$, let $\delta_\ep > 0$ be the constant provided by Lemma~\ref{lem:lemma4.6} and fix two radii $\rho , r \in \left[ X , \frac 12 R\right]$. We let $\bar u$ be the solution of the equation~\eqref{eq:well-approxhomog} in the set $B_r \times \Omega$ such that the estimate~\eqref{eq:well-approx} holds. We note that the estimate~\eqref{eq:well-approx} implies the inequality $\left\| \nabla \bar u\right\|_{L^2\left( B_{r} , \mu_\beta \right)} \leq C \left\| \nabla  u\right\|_{L^2 \left( B_{2r} \right)} + K$.  By the regularity theory for the map $\bar u$ established in Proposition~\ref{prop:prop4.4}, we have
\begin{equation} \label{eq:regw}
    \left\| \nabla \bar u \right\|_{L^2 \left( B_\rho, \mu_\beta \right) } \leq C \left( \frac{\rho}{r}\right)^{\frac d2} \left\| \nabla \bar u \right\|_{L^2 \left( B_r, \mu_\beta \right)}.
\end{equation}
By combining the estimates~\eqref{eq:well-approx} and~\eqref{eq:regw} and the estimate on the $L^2$-norm of the gradient of $\bar u$ mentioned above, we compute
\begin{align*}
    \left\| \nabla u \right\|_{L^2 \left( B_\rho , \mu_\beta \right)} & \leq \left\| \nabla \left( u - \bar u \right) \right\|_{L^2 \left( B_\rho , \mu_\beta \right)} + \left\| \nabla \bar u \right\|_{L^2 \left( B_\rho , \mu_\beta \right)} \\
    & \leq \left\| \nabla \left( u - \bar u \right) \right\|_{L^2 \left( B_r , \mu_\beta \right)} + \left( \frac{\rho}{r} \right)^{\frac d2} \left\| \nabla \bar u \right\|_{L^2 \left( B_r , \mu_\beta \right)} \\
    & \leq \delta_\ep \left\| \nabla u \right\|_{L^2 \left( B_{2r} , \mu_\beta \right)} +  K + \left( \frac{\rho}{r}\right)^{\frac d2} \left( C \left\| \nabla u\right\|_{L^2 \left( B_{2r} \right)} + K\right) \\
    & \leq C \left(  \left(\frac{\rho}{r}\right)^{\frac d2} + \delta_\ep \right) \left\| \nabla u \right\|_{L^2 \left( B_{2r} , \mu_\beta \right)} + 2 K.
\end{align*}
We apply Lemma~\ref{lem:lemma4.6} with the function $\phi(\rho) = \left\| \nabla u \right\|_{L^2 \left( B_\rho \right)}$. The inequality~\eqref{eq:fingiaq} with the choice $r = R$ gives, for any radius $\rho \in [X , R]$,
\begin{equation*}
    \left\| \nabla u \right\|_{L^2 \left( B_\rho , \mu_\beta \right)} \leq C_1 \left( \left( \frac{\rho}{R} \right)^{\frac d2- \ep}  \left\| \nabla u \right\|_{L^2 \left( B_R , \mu_\beta \right)}  + 2 K \rho^{\frac d2} \right).
\end{equation*}
Dividing both side of the estimate by $\rho^{\frac d2}$ completes the proof.
\end{proof}

We now use Propositions~\ref{prop:prop4.4} and~\ref{prop:prop4.5} to obtain $C^{0 , 1-\ep}$-regularity for the solutions of the Helffer-Sj{\"o}strand equation.

\begin{proposition}[$C^{0 , 1-\ep}$-regularity theory] \label{prop:prop4.6}
For any regularity exponent $\ep > 0$, there exists an inverse temperature $\beta_0 := \beta_0 \left( d , \ep \right) < \infty$ such that the following statement holds. There exist two constants $C := C(d  , \ep) < \infty$ and $c := c(d)>0$ such that for any radius $R \geq 1$, any inverse temperature $\beta \geq \beta_0$ and any function $u : \Zd \times \Omega \rightarrow \R$ solution of the equation
\begin{equation*}
 \mathcal{L} u = 0  ~\mbox{in}~ B_{R} \times \Omega,
\end{equation*}
one has the estimate
\begin{equation} \label{eq:pointwiseC1-epregest}
\left\| \nabla u(0 , \cdot) \right\|_{L^2 \left( \mu_\beta \right)} \leq \frac{C}{R^{1-\ep}} \left\| u - \left( u \right)_{B_{R}} \right\|_{\underline{L}^2 \left( B_{R}, \mu_\beta \right)} +  \sum_{x \in \Zd} e^{-c (\ln \beta)\left( R \vee |x| \right)}   \left\| u(x , \cdot) \right\|_{L^2\left(\mu_\beta \right) }.
\end{equation}
\end{proposition}

\begin{proof}
The strategy of the proof is to apply Proposition~\ref{prop:prop4.5} to the function $u$ and then to apply the Caccioppoli inequality. We fix a regularity exponent $\ep>0$, a radius $R \geq 1$ and split the argument into two steps:
\begin{itemize}
\item In Step 1, we prove that the map $u$ satisfies the following property: there exist an inverse temperature $\beta_0 \left( \ep , d \right) < \infty$ and a constant $C:= C(d) < \infty$ such that for every $\beta>\beta_0$ and every radius $r \geq \left(\ln R\right)^2$, the following estimate holds
\begin{equation} \label{est:step1regularity}
\left\| \nabla u \right\|_{\underline{L}^2 \left( B_r, \mu_\beta \right)} \leq C \left( \frac{R}{r} \right)^{ \frac{\ep}{2}} \left\| \nabla u \right\|_{\underline{L}^2 \left( B_R , \mu_\beta \right)} +  \sum_{x \in \Zd \setminus B_R} e^{-c  \left( \ln \beta \right) |x|}   \left\| \nabla u(x , \cdot) \right\|_{L^2\left(\mu_\beta \right) }.
\end{equation}
\item In Step 2, we deduce from~\eqref{est:step1regularity} and the Caccioppoli inequality stated in Proposition~\ref{Caccio.ineq}, the pointwise estimate~\eqref{eq:pointwiseC1-epregest}.
\end{itemize} 

\textit{Step 1.} To prove the estimate~\eqref{est:step1regularity}, the strategy is to apply Proposition~\ref{prop:prop4.5}. To this end, we set $X := \left(\ln R\right)^2$, and fix a radius $r \in \left[X, \frac 12 R\right]$. We then define the function $\bar u$ to be the solution of the boundary value problem
\begin{equation} \label{def.wprop4.2}
\left\{ \begin{aligned}
\Delta_\phi \bar u - \frac{1}{2\beta} \Delta \bar u = 0 ~\mbox{in}~ B_{r} \times \Omega, \\
\bar u  = u ~\mbox{on}~ \partial B_{r} \times \Omega.
\end{aligned} \right.
\end{equation}  
We first prove that the map $\bar u$ is a good approximation of the map $u$. Specifically, we prove that there exist two constants $C := C(d) < \infty$ and $c := c(d)>0$ such that
\begin{multline} \label{eq:uwellapprox}
\left\| \nabla (u - \bar u) \right\|_{L^2 \left( B_r , \mu_\beta \right)} \\ \leq \frac{C}{\beta^\frac 12} \left\| \nabla u \right\|_{L^2 \left( B_{2r} , \mu_\beta \right)} + C e^{- c \ln \beta\left( \ln R \right)^2} \left\| \nabla u \right\|_{L^2 \left( B_R , \mu_\beta \right)}  + C  \sum_{x \in \Zd \setminus B(0,R)} e^{-c  \ln \beta |x|}   \left\| \nabla u(x , \cdot) \right\|_{L^2\left(\mu_\beta \right) }.
\end{multline}
To prove the estimate~\eqref{eq:uwellapprox}, we note that the map $\bar u -u$ is a solution of the following system of equations
\begin{equation} \label{eq:w-u}
\left\{ \begin{aligned}
\Delta_\phi (\bar u- u) -  \frac{1}{2\beta}  \Delta (\bar u-u) & =  - \frac{1}{2\beta}\sum_{n \geq 1} \frac{1}{\beta^{ \frac {n}2}} (-\Delta)^{n+1} u - \sum_{q\in \mathcal{Q}} \nabla_q^* \cdot \a_q \nabla_q u ~& \mbox{in}~B_{r} \times \Omega, \\
\bar u  - u &= 0 ~& \mbox{on}~ \partial B_{r} \times \Omega.
\end{aligned} \right.
\end{equation}
We extend the function $(\bar u-u)$ by $0$ outside the ball $B_R$ so that it is defined on the entire space $\Zd$ and use it as a test function in the system~\eqref{eq:w-u}. We obtain
\begin{multline} \label{eq:4.13}
\sum_{y \in \Zd} \left\| \partial_y \left( \bar u - u \right) \right\|_{L^2 \left( B_{r} , \mu_\beta \right)}^2 +  \frac{1}{2\beta} \left\| \nabla \left( \bar u- u \right) \right\|_{L^2 \left( B_{r} , \mu_\beta \right)}^2 \\ =  -\frac{1}{2\beta}\underbrace{\sum_{n \geq 1} \frac{1}{\beta^{ \frac n2}} \sum_{x \in \Zd} \left\langle \nabla^{n+1} u(x, \cdot)\cdot \nabla^{n+1} (\bar u - u)(x , \cdot)  \right\rangle_{\mu_\beta}}_{\eqref{eq:4.13}-(i)} - \underbrace{\sum_{q\in \mathcal{Q}} \left\langle  \nabla_q u \cdot \a_q  \nabla_q (\bar u - u ) \right\rangle_{\mu_\beta}}_{\eqref{eq:4.13}-(ii)}.
\end{multline}
We first focus on the term~\eqref{eq:4.13}-(i) and note that, for each integer $n \in \N$, since the iterated gradient $\nabla^n$ has range $n$ and since the map $\bar u-u$ is equal to $0$ outside the ball $B_r$, the function $\nabla^{n} (\bar u - u) $ is supported in the ball $B_{r  + n}$. By the Cauchy-Schwarz inequality, we obtain
\begin{equation*}
\sum_{x \in \Zd} \left\langle \nabla^{n+1} u(x , \cdot), \nabla^{n+1} (\bar u - u)(x , \cdot)  \right\rangle_{\mu_\beta} \leq  \left\| \nabla^{n+1} u \right\|_{L^2 \left(B_{r  + n} , \mu_\beta \right) }   \left\| \nabla^{n+1} (\bar u-u)\right\|_{L^2 \left(B_{r} , \mu_\beta \right)}. 
\end{equation*} 
Since the discrete gradient $\nabla$ has a finite operator norm on $L^2(\Zd)$, one has the estimate
\begin{align} \label{eq:TV182311}
\sum_{x \in \Zd} \left\langle \nabla^{n+1} u(x , \cdot), \nabla^{n+1} (\bar u - u)(x , \cdot) \right\rangle_{\mu_\beta} & \leq ||| \nabla^{n} |||_{L^2 \to L^2}^2 \left\| \nabla u \right\|_{L^2 \left(B_{r  + n+1} , \mu_\beta \right)}  \left\| \nabla (\bar u -u)\right\|_{L^2 \left(B_{r} , \mu_\beta \right)} \\
& \leq \frac{C^n}{2} \left( \left\| \nabla u \right\|_{L^2 \left(B_{r  + n+1} , \mu_\beta \right)}^2 +  \left\| \nabla (\bar u-u)\right\|_{L^2 \left(B_{r} , \mu_\beta \right)}^2 \right), \notag
\end{align}
where we used the inequality of operator norms $||| \nabla^{n} |||_{L^2 \to L^2}^2 \leq ||| \nabla |||_{L^2 \to L^2}^{2n } \leq C^n$ and Young's inequality in the second line.

\smallskip

Multiplying the inequality~\eqref{eq:TV182311} by $\beta^{-\frac n2}$ and summing over all the integers $n \in \N$, we obtain
\begin{equation*}
    \sum_{n \geq 1} \frac{1}{\beta^{ \frac n2}} \sum_{x \in \Zd} \left\langle \nabla^{n+1} u(x, \cdot), \nabla^{n+1} (\bar u - u)(x , \cdot)  \right\rangle_{\mu_\beta} \leq \sum_{n \geq 1} \frac{C^n}{2 \beta^{ \frac n2}} \left\| \nabla u \right\|_{L^2 \left(B_{r  + n} , \mu_\beta \right)}^2  + \sum_{n \geq 1} \frac{C^n}{2 \beta^{ \frac n2}}  \left\| \nabla (\bar u-u)\right\|_{L^2 \left(B_{r} , \mu_\beta \right)}^2.
\end{equation*}
by choosing the inverse temperature $\beta$ large enough (at larger than the square of the constant $C$), we obtain
\begin{multline} \label{eq:TV23142}
     \sum_{n \geq 1} \frac{1}{\beta^{ \frac n2}} \sum_{x \in \Zd} \left\langle \nabla^{n+1} u(x, \cdot), \nabla^{n+1} (\bar u - u)(x , \cdot)  \right\rangle_{\mu_\beta} \\ \leq \frac{C}{\sqrt{\beta}} \sum_{x \in \Zd} \left( \frac{C}{\sqrt{\beta}} \right)^{\dist(x , B_r)} \left\| \nabla u(x , \cdot) \right\|_{L^2 \left( \mu_\beta \right)}^2 + \frac{C}{\sqrt{\beta}} \left\| \nabla (\bar u-u)\right\|_{L^2 \left(B_{r} , \mu_\beta \right)}^2.
\end{multline}
A similar computation works for the term~\eqref{eq:4.13}-(ii). We decompose the sum over the diameter of the charges
\begin{equation} \label{eq:TV4.30}
  \sum_{q\in \mathcal{Q}} \left\langle  \nabla_q u \cdot \a_q  \nabla_q (\bar u - u ) \right\rangle_{\mu_\beta} = \underbrace{\sum_{\diam q \leq r} \left\langle  \nabla_q u \cdot \a_q  \nabla_q (\bar u - u ) \right\rangle_{\mu_\beta} }_{\eqref{eq:TV4.30}-(i)}+  \underbrace{\sum_{n = r+1}^\infty \sum_{\diam q = n}  \left\langle  \nabla_q u \cdot \a_q  \nabla_q (\bar u - u ) \right\rangle_{\mu_\beta}}_{\eqref{eq:TV4.30}-(ii)}.
\end{equation}
Note that since the function $\bar u-u$ is supported in the ball $B_r$, we can restrict the sums to the charges whose support intersects the ball $B_r$. The term~\eqref{eq:TV4.30}-(i) involving the charges of diameter smaller than $r$ can be estimated by the Cauchy-Schwarz and Young's inequalities. We have
\begin{align*}
\left| \sum_{\diam q \leq r} \left\langle  \nabla_q u \cdot \a_q  \nabla_q (\bar u - u ) \right\rangle_{\mu_\beta} \right| & \leq \sum_{\diam q \leq r} e^{-c \sqrt{\beta} \left\| q \right\|_1} \left\| n_q \right\|_{L^\infty}^2 \left( \sum_{z \in \supp n_q}\left\| \nabla u(z , \cdot) \right\|_{L^2 \left( \mu_\beta \right)} \right) \left(  \sum_{z \in \supp n_q}\left\| \nabla (\bar u - u)(z , \cdot) \right\|_{L^2 \left( \mu_\beta \right)} \right)
\\ & \leq  \sum_{\diam q \leq r} e^{-c \sqrt{\beta} C_q \left\| q \right\|_1}  \left( \sum_{z \in \supp n_q}\left\| \nabla u(z , \cdot) \right\|_{L^2 \left( \mu_\beta \right)} \right)^2 \\ & \quad + \sum_{\diam q \leq r} C_q e^{-c \sqrt{\beta} \left\| q \right\|_1} \left(  \sum_{z \in \supp n_q}\left\| \nabla (\bar u - u)(z , \cdot) \right\|_{L^2 \left( \mu_\beta \right)} \right)^2 \\
& \leq \sum_{\diam q \leq r} e^{-c \sqrt{\beta} C_q \left\| q \right\|_1}  \sum_{z \in \supp n_q}\left\| \nabla u(z , \cdot) \right\|_{L^2 \left( \mu_\beta \right)}^2 \\
& \quad + \sum_{\diam q \leq r} C_q e^{-c \sqrt{\beta} \left\| q \right\|_1}  \sum_{z \in \supp n_q}\left\| \nabla (\bar u - u)(z , \cdot) \right\|_{L^2 \left( \mu_\beta \right)}^2.
\end{align*}
We then absorb the terms $\left\| n_q \right\|_{L^\infty}$ and $\left| \supp n_q \right|$ into the exponential term $e^{-c \sqrt{\beta} \left\| q\right\|_1}$ (by reducing the value of the constant $c$) and use the estimate, for each $z \in \Zd$,
\begin{equation*}
    \sum_{\diam q \leq r} e^{-c \sqrt{\beta} \left\| q \right\|_1} \indc_{\{ z \in \supp n_q \}} \leq \left\{\begin{aligned}
    0 &~\mbox{if}~ z \notin B_{2r},\\
    C e^{-c \sqrt{\beta}} &~\mbox{if}~ z \in B_{2r},
    \end{aligned} \right.
\end{equation*}
where we recall that the sum is restricted to the charges $q \in \mathcal{Q}$ such that the support of $q$ intersects the ball~$B_r$. We obtain the inequality
\begin{equation} \label{eq:TV2240}
    \left| \sum_{\diam q \leq r} \left\langle  \nabla_q u \cdot \a_q  \nabla_q (\bar u - u ) \right\rangle_{\mu_\beta} \right| \leq C e^{-c \sqrt{\beta}} \left\| \nabla (\bar u - u)(z , \cdot) \right\|_{L^2 \left(B_r ,  \mu_\beta \right)}^2 + C e^{-c \sqrt{\beta}} \left\| \nabla u (z , \cdot) \right\|_{L^2 \left(B_{2r} ,  \mu_\beta \right)}^2,
\end{equation}
where we used that the function $\bar u -u$ is supported in the ball $B_r$.

The same computation can be used to estimate the term~\eqref{eq:TV4.30}-(ii). We obtain, for each integer $n \in \N$,
\begin{equation} \label{eq:TV2239}
\left| \sum_{\diam q = n}  \left\langle  \nabla_q u \cdot \a_q  \nabla_q (\bar u - u ) \right\rangle_{\mu_\beta} \right| \leq C e^{-c \sqrt{\beta} n} \left\| \nabla  u \right\|_{L^2 \left( B_{r + n}, \mu_\beta \right)}^2 + C e^{-c \sqrt{\beta} n}   \left\| \nabla \left(\bar u - u \right) \right\|_{L^2 \left( B_{r}, \mu_\beta \right)}^2.
\end{equation}
Combining the identity~\eqref{eq:TV4.30} with the estimates~\eqref{eq:TV2240} and~\eqref{eq:TV2239}, we deduce that
\begin{multline} \label{eq:TV23141}
    \sum_{q\in \mathcal{Q}} \left\langle  \nabla_q u \cdot \a_q  \nabla_q (\bar u - u ) \right\rangle_{\mu_\beta} \\ \leq C e^{-c \sqrt{\beta}} \left\| \nabla (u-\bar u) \right\|_{L^2 \left(B_{r},\mu_\beta\right)}^2  + C e^{-c \sqrt{\beta}} \sum_{x \in \Zd} e^{-c \sqrt{\beta} \dist \left(x , B_{r} \right)} \left\| \nabla u(x , \cdot) \right\|_{L^2 \left(\mu_\beta\right)}^2.
\end{multline}
We now combine the identity~\eqref{eq:4.13} with the estimates~\eqref{eq:TV23142},~\eqref{eq:TV23141} to obtain the inequality
\begin{multline} \label{eq:TV234011}
\sum_{y \in \Zd} \left\| \partial_y \left(\bar u - u \right) \right\|_{L^2 \left( B_{r} , \mu_\beta \right)}^2 +  \frac{1}{2\beta} \left\| \nabla \left(\bar u - u \right) \right\|_{L^2 \left( B_{r} , \mu_\beta \right)}^2 \leq C \left( e^{-c \sqrt{\beta}} + \frac{1}{\beta^{\frac 32}} \right) \left\| \nabla \left( \bar u - u \right) \right\|_{L^2 \left( B_{r}, \mu_\beta \right)}^2 \\ + C\beta^{-\frac 32} \sum_{x \in \Zd} \left( e^{-c \sqrt{\beta} \dist \left(x , B_r \right)} + \left( C\beta^{-\frac12}\right)^{\dist \left(x , B_r \right)}\right)  \left\| \nabla u(x , \cdot) \right\|_{L^2 \left(\mu_\beta \right)}^2.
\end{multline}
We choose the inverse temperature $\beta$ large enough so that the coefficient $ C \left( e^{-c \sqrt{\beta}} + \beta^{-\frac 32} \right)$ is smaller that $\frac{1}{4\beta}$. With this choice, the first term in the right side of the inequality~\eqref{eq:TV234011} can be absorbed in the left side of the same inequality. We obtain the estimate
\begin{equation} \label{eq:right1750}
     \left\| \nabla \left(\bar u - u \right) \right\|_{L^2 \left( B_{r} , \mu_\beta \right)}^2 \leq \frac{C}{\beta^\frac 12} \sum_{x \in \Zd} \left( e^{-c \sqrt{\beta} \dist \left(x , B_r \right)} + \left(C\beta^{-\frac12}\right)^{\dist \left(x , B_r \right)}\right)  \left\| \nabla u(x , \cdot) \right\|_{L^2 \left(\mu_\beta \right)}^2.
\end{equation}
The inequality~\eqref{eq:uwellapprox} can then be deduced from the estimate~\eqref{eq:right1750} thanks to the three ingredients listed below:
\begin{itemize}
    \item By choosing $\beta$ large enough, the exponential term $\left( C\beta^{-\frac12}\right)^{\dist \left(x , B_r \right)}$ is smaller than $1$ for any point $x \in \Zd$. This leads to the estimate in the ball $B_{2r}$
    \begin{equation*}
        \sum_{x \in B_{2r}} \left( e^{-c \sqrt{\beta} \dist \left(x , B_r \right)} + \left( C\beta^{-\frac12}\right)^{\dist \left(x , B_r \right)}\right)  \left\| \nabla u(x , \cdot) \right\|_{L^2 \left(\mu_\beta \right)}^2  \leq 2 \sum_{x \in B_{2r}} \left\| \nabla u(x , \cdot) \right\|_{L^2 \left(\mu_\beta \right)}^2 
         = 2 \left\| \nabla u \right\|_{L^2 \left(B_{2r}, \mu_\beta \right)}^2.
    \end{equation*}
    \item We have the estimate in the annulus $B_R \setminus B_{2r}$,
    \begin{equation} \label{eq:TV083411}
        \sum_{x \in B_R \setminus B_{2r}} \left( e^{-c \sqrt{\beta} \dist \left(x , B_r \right)} + \left(C\beta^{-\frac12}\right)^{\dist \left(x , B_r \right)}\right)  \left\| \nabla u(x , \cdot) \right\|_{L^2 \left(\mu_\beta \right)}^2 \leq \left( e^{-c \sqrt{\beta} r} + \left( C\beta^{-\frac12}\right)^{r}\right) \left\| \nabla u(x , \cdot) \right\|_{L^2 \left(B_{R} \setminus B_{2r}, \mu_\beta \right)}^2.
    \end{equation}
Using the assumption $r \geq \left(\ln R\right)^2$ and choosing $\beta$ large enough, one obtains the estimate
    \begin{align*}
        \sum_{x \in B_R \setminus B_{2r}} \left( e^{-c \sqrt{\beta} \dist \left(x , B_r \right)} + \left( C\beta^{-\frac12}\right)^{\dist \left(x , B_r \right)}\right)  \left\| \nabla u(x , \cdot) \right\|_{L^2 \left(\mu_\beta \right)}^2 & \leq e^{-c r \ln \beta } \left\| \nabla u \right\|_{L^2 \left( B_R \setminus B_{2r}, \mu_\beta \right)}^2 \\
        & \leq e^{-c (\ln R)^2 \ln \beta } \left\| \nabla u \right\|_{L^2 \left( B_R, \mu_\beta \right)}^2.
    \end{align*}
    \item For each point $x \in \Zd \setminus B_R$, we have the estimate $c|x| \leq \dist\left( x , B_{r}\right) \leq C |x|$. This implies
    \begin{equation*}
        \sum_{x \in B_R \setminus B_{2r}} \left( e^{-c \sqrt{\beta} \dist \left(x , B_r \right)} + \left( C\beta^{-\frac12}\right)^{\dist \left(x , B_r \right)}\right)  \left\| \nabla u(x , \cdot) \right\|_{L^2 \left(\mu_\beta \right)}^2 \leq \sum_{x \in  B_R \setminus B_{2r}}e^{-c \left(\ln \beta\right) |x|} \left\| \nabla u(x , \cdot) \right\|_{L^2\left( \mu_\beta\right)}.
    \end{equation*}
\end{itemize}
A combination of the inequality~\eqref{eq:right1750} with the three previous items completes the proof of the estimate~\eqref{eq:uwellapprox}.

We complete Step 1 by proving that the estimate~\eqref{eq:uwellapprox} implies the estimate~\eqref{est:step1regularity}. We consider the regularity exponent $\ep$ fixed at the beginning of the proof and the parameter $\delta_{\frac{\ep}{2}}$ provided by Proposition~\ref{prop:prop4.5} (associated to the exponent $\frac{\ep}{2}$). We let $C := C(d) < \infty$ and $c := c(d)>0$ be the constants which appear in the inequality~\eqref{eq:uwellapprox} and set \begin{equation*}
X := \left( \ln R \right)^2 \mbox{ and } K := C  e^{-c \ln \beta \left( \ln R \right)^2} \left\| \nabla u  \right\|_{L^2 \left(B_R , \mu_\beta \right)} +  C  \sum_{x \in \Zd} e^{-c \left(\ln \beta\right) \left( R \vee |x|\right)} \left\| u(x ,\cdot) \right\|_{L^2\left(\mu_\beta \right)}.
\end{equation*}
An application of Proposition~\ref{prop:prop4.5} shows the inequality: for any radius $r \in [X , R]$,
\begin{equation} \label{eq:TV10430}
\left\| \nabla u \right\|_{\underline{L}^2 \left( B_r, \mu_\beta \right)} \leq C \left( \frac{R}{r} \right)^{\frac{\ep}{2}} \left\| \nabla u \right\|_{\underline{L}^2 \left( B_R , \mu_\beta \right)} +  e^{-c \ln \beta \left( \ln R \right)^2} \left\| \nabla u \right\|_{L^2 \left(B_R , \mu_\beta \right)}  +  C  \sum_{x \in \Zd } e^{-c \left(\ln \beta\right) \left( R \vee |x|\right)} \left\| u(x ,\cdot) \right\|_{L^2\left(\mu_\beta \right)}.
\end{equation}
We then note that the exponential term $ e^{-c \left( \ln \beta \right) \left( \ln R \right)^2} $ decays faster than any power of $R$, so the second term on the right side of~\eqref{eq:TV10430} can be bounded from above by the first term on the right side. This completes the proof of the inequality~\eqref{est:step1regularity}.

\medskip

\textit{Step 2.} We select $r = \left( \ln R \right)^2$, apply the Caccioppoli inequality to estimate the right side of the inequality~\eqref{est:step1regularity} and use that the discrete gradient is a bounded operator to replace the term $\left\| \nabla u (x , \cdot) \right\|_{L^2 \left( \mu_\beta \right)}$ by $\left\| u (x , \cdot) \right\|_{L^2 \left( \mu_\beta \right)}$. We obtain
\begin{equation*}
\left\| \nabla u \right\|_{\underline{L}^2 \left( B_{ \left( \ln R \right)^2}, \mu_\beta \right)} \leq C \left( \frac{R}{ \left( \ln R \right)^2} \right)^{\frac{\ep}{2}} \frac{1}{R} \left\| u - \left( u \right)_{B_R} \right\|_{\underline{L}^2 \left( B_R , \mu_\beta \right)} +  C  \sum_{x \in \Zd } e^{-c \left(\ln \beta\right) \left( R \vee |x| \right)} \left\| u(x ,\cdot) \right\|_{L^2\left(\mu_\beta \right)}.
\end{equation*}
We apply the discrete $L^\infty-L^2$ -estimate
\begin{equation*}
\left\| \nabla u(0) \right\|_{L^2 \left( \mu_\beta \right)} \leq  \left\| \nabla u \right\|_{L^2 \left( B_{ \left( \ln R \right)^2}, \mu_\beta \right)} \leq  \left( \ln R \right)^d \left\| \nabla u \right\|_{\underline{L}^2 \left( B_{ \left( \ln R \right)^2}, \mu_\beta \right)}.
\end{equation*}
We then combine the two previous displays and the estimate $\left( \ln R \right)^{d} \leq C R^{\frac \ep 2}$ to obtain the inequality~\eqref{eq:pointwiseC1-epregest}. The proof of Proposition~\ref{prop:prop4.6} is complete.
\end{proof}

\section{Nash-Aronson estimate and regularity theory for the heat kernel \texorpdfstring{$\Pa_{\f}$}{3}} \label{sec:section4.3}

The main purpose of this section is to prove upper bounds on the heat kernel $\Pa_\f$ and on its spatial derivatives. We introduce the following definition. For each constant $C > 0$, we let $\Phi_C$ be the function defined from $(0, \infty) \times \Zd$ to $\R$ by the formula, for each pair $(t,x) \in (0 , \infty) \times \Zd$,
\begin{equation} \label{def:formPhiC}
\Phi_C(t , x) = \left\{ \begin{aligned}
t^{-\frac d2} \exp \left( - \frac{|x|^2}{Ct} \right) &~\mbox{if}~ |x| \leq t, \\
 \exp \left( - \frac{|x|}{C}\right)  &~\mbox{if} ~ |x| \geq t.
\end{aligned} \right.
\end{equation}

The next proposition is the main result of this section.

\begin{proposition}[Gaussian bounds and  $C^{0,1-\ep}$-regularity for the heat kernel] \label{prop:prop4.7}
For any regularity exponent $\ep > 0$, there exists an inverse temperature $\beta_0(d , \ep) < \infty$ and a constant $C := C(d , \ep) < \infty$ such that for every $\beta > \beta_0$, every exponent $p \in [1 , \infty]$ and every random variable $\f \in L^p \left( \mu_\beta \right)$, the heat kernel $\Pa_\mathbf{f}$ satisfies the following estimate, for each $(t, x, y) \in (0 , \infty) \times \Zd \times \Zd$,
\begin{equation} \label{eq:GaussianP}
\left\| \Pa_\f(t , x , \cdot ; y) \right\|_{L^p \left( \mu_\beta \right)} \leq C \left\| \f \right\|_{L^p(\mu_\beta)} \Phi_C \left( \frac{t}{\beta},x-y \right).
\end{equation}
Moreover, one has the $C^{0 , 1- \ep}$-estimates on the gradient of the heat kernel
\begin{equation} \label{eq:gradGaussianP}
\left\| \nabla_x \Pa_\f(t , x , \cdot ; y) \right\|_{L^p \left( \mu_\beta \right)} \leq C\left\| \f \right\|_{L^p(\mu_\beta)} \left(\frac{\beta}{t}\right)^{\frac 12 - \ep} \Phi_C \left( \frac{t}{\beta},x-y \right),
\end{equation}
and on the mixed derivative of the heat kernel
\begin{equation} \label{eq:gradgradGaussianP}
\left\| \nabla_x \nabla_y \Pa_\f(t , x , \cdot ; y) \right\|_{L^p \left( \mu_\beta \right)} \leq C  \left\| \f \right\|_{L^p(\mu_\beta)} \left(\frac{ \beta}{t}\right)^{1 - \ep} \Phi_C \left( \frac{t}{\beta},x-y \right).
\end{equation}
\end{proposition}

\begin{remark}
The Nash-Aronson type estimate~\eqref{eq:gradGaussianP} is proved by Naddaf and Spencer in~\cite[Section 2.2.2]{NS} in the case of the discrete Ginzburg-Landau interface model.
\end{remark}

\begin{remark}
Due to the discrete setting of the problem and the infinite range of the operator $\mathcal{L}$, the heat kernel does not have Gaussian decay when the value $|x|$ tends to infinity. Instead it decays exponentially fast; this justifies the introduction of the function $\Phi_C$.
\end{remark}

\begin{remark}
For later use, we need to keep track of the dependence of the constants in the inverse temperature $\beta$.
\end{remark}

\begin{proof}
The first ingredient in the proof of Proposition~\ref{prop:prop4.7} is the Feymann-Kac representation formula which is described at the beginning of Chapter~\ref{section:section4} and recalled below. If we let $\left(\phi_t\right)_{t \geq 0}$ be the diffusion process associated to the Langevin dynamics~\eqref{def.phi_t}, then one has the identity
    \begin{equation*}
        \mathcal{P}_{\mathbf{f}} \left( t , x , \phi ; y\right) = \E_{\phi} \left[ \mathbf{f}(\phi_t) P^{\phi_\cdot}(t , x ; y) \right],
    \end{equation*}
where $P^{\phi_\cdot}(\cdot , \cdot \, ; y)$ is the solution of the parabolic system
    \begin{equation} \label{eq:defPhatphi}
        \left\{ \begin{aligned}
        \partial_t P^{\phi_\cdot}\left(\cdot , \cdot \, ; y\right) + \mathcal{L}_{\mathrm{spat}}^{\phi_t}  P^{\phi_\cdot}\left(\cdot , \cdot \, ; y\right) & =0 ~\mbox{in}~ (0 , \infty) \times \Zd, \\
        P^{\phi_\cdot}\left(0,\cdot \, ;y \right) & = \delta_y ~\mbox{in}~\Zd,
        \end{aligned} \right.
    \end{equation}
where $\mathcal{L}_{\mathrm{spat}}^{\phi_t}$ denotes the time-dependent elliptic operator 
\begin{equation*}
\mathcal{L}_{\mathrm{spat}}^{\phi_t} :=  - \frac{1}{2\beta} \Delta +  \frac{1}{2\beta}\sum_{n \geq 1} \frac{1}{\beta^{ \frac n2}} (-\Delta)^{n+1} + \sum_{q \in \mathcal{Q}} \nabla_q^* \cdot \a_q(\phi_t) \nabla_q,
\end{equation*}
The core of the argument is to prove the three following estimates on the heat kernel $P^{\phi_\cdot}$: there exists a constant $C := C(d, \ep) < \infty$ such that for each realization of the diffusion process $\left( \phi_t\right)_{t \geq 0}$ and each triplet $(t,x,y) \in (0, \infty) \times \Zd \times \Zd$,
\begin{equation} \label{eq:3pointwiseest}
    \left\{ \begin{aligned}
    \left| P^{\phi_\cdot}(t , x ; y)\right| \leq C  \Phi_{C} \left( \frac{t}{\beta},x-y \right),  \\
    \left| \nabla_x P^{\phi_\cdot}(t , x ; y)\right| \leq C\left(\frac{\beta}{t}\right)^{\frac 12 - \ep} \Phi_C \left( \frac{t}{\beta},x-y \right), \\
    \left| \nabla_x \nabla_y P^{\phi_\cdot}(t , x ; y)\right| \leq C  \left(\frac{\beta}{t}\right)^{1 - \ep} \Phi_C \left( \frac{t}{\beta},x-y \right).
    \end{aligned} \right.
\end{equation}
The proof of these results is postponed to Propositions~\ref{prop:prop4.8} and~\ref{prop:prop4.9}; we now show how to complete the proof of Proposition~\ref{prop:prop4.7} assuming that the estimates~\eqref{eq:3pointwiseest} hold.

Using that the Gibbs measure $\mu_\beta$ is invariant under the Langevin dynamics~\eqref{def.phi_t}, the inequality~\eqref{eq:GaussianP} is a consequence of the estimates~\eqref{eq:3pointwiseest} and the following computation, for each triplet $(t , x , y) \in (0 , \infty) \times \Zd \times \Zd$,
\begin{align*}
    \left\| \Pa_\f \left( t , x , \cdot ; y \right)\right\|_{L^p \left( \mu_\beta \right)}^p & = \E \left[ \left|\Pa_\f \left( t , x , \phi ; y \right) \right|^p \right] \\
    & = \E \left[  \left| \E_{\phi} \left[ \mathbf{f}(\phi_t) P^{\phi_\cdot}(t , x , y) \right] \right|^p \right] \\
    & \leq \E \left[   \E_{\phi} \left[ \left| \mathbf{f}(\phi_t) P^{\phi_\cdot}(t , x , y) \right|^p \right]  \right] \\
    & \leq \left( C \Phi_C\left( \frac{t}{\beta},x-y \right) \right)^p \E \left[   \E_{\phi} \left[ \left| \mathbf{f}(\phi_t) \right|^p \right]  \right] \\
    & \leq \left( C \Phi_C\left( \frac{t}{\beta},x-y \right) \right)^p \left\| \f \right\|_{L^p \left( \mu_\beta \right)}^p.
\end{align*}
The proofs of the estimates~\eqref{eq:gradGaussianP} and~\eqref{eq:gradgradGaussianP} is similar and we skip the details.
\end{proof}

The rest of Section~\ref{sec:section4.3} is devoted to the statements and proofs of Propositions~\ref{prop:prop4.8} and~\ref{prop:prop4.9}. Gaussian bounds on the heat kernel are usually a consequence of the Nash-Aronson estimate (see~\cite{Ar,FS86}) for uniformly elliptic operators. This result cannot be applied here since the operator $\partial_t + \mathcal{L}_{\mathrm{spat}}^{\phi_t}$ is a parabolic system (see the counter-example of De Giorgi~\cite{DeG} disproving the Liouville property and the $C^{0,\alpha}$-regularity theory for systems of elliptic equations).

To prove Gaussian bounds and regularity on the heat kernel, we proceed differently and organize the proof as follows:
\begin{enumerate}
    \item We use that the elliptic operator $\mathcal{L}_{\mathrm{spat}}^{\phi_t}$ is a perturbation of the Laplacian to establish $C^{0, 1-\ep}$-regularity for the solutions of the system  
    \begin{equation} \label{eq:TV16470}
        \partial_t u + \mathcal{L}_{\mathrm{spat}}^{\phi_t} u = 0;
    \end{equation}
    \item We use the $C^{0 , 1-\ep}$-regularity and an interpolation argument to obtain $L^\infty$-bounds on the solutions of the equation~\eqref{eq:TV16470}. More precisely, we prove that every solution of the system~\eqref{eq:TV16470} in the parabolic cylinder $Q_{2r}$ satisfies the pointwise estimate
    \begin{equation*}
        \left\|  u \right\|_{L^\infty \left( Q_r \right)} \leq C \left\| u \right\|_{\underline{L}^2 \left( Q_{2r}\right)} + \int_{-r^2}^0 \sum_{x \in \Zd \setminus B_r} e^{-c  \left(\ln \beta\right) \left( r \vee |x| \right)} \left| u(t , x) \right|^2 \, dt;
    \end{equation*}
    \item We prove that the solutions of the adjoint of the parabolic operator $\partial_t + \mathcal{L}_{\mathrm{spat}}^{\phi_t}$ satisfies the same pointwise estimate;
    \item We use the pointwise regularity estimates and the technique Fabes and Stroock~\cite{FS86}, which is based on the technique of Davies~\cite{Da1, Da2} (see also the article of Hofmann and Kim~\cite{HK} on which the argument is based) to establish the Gaussian bounds on the heat kernel stated in Proposition~\ref{prop:prop4.8};
    \item We combine the Gaussian bounds on the heat kernel with the $C^{1-\ep}$-regularity theory for the solutions of~\eqref{eq:TV16470} to obtain the upper bounds on the gradient and mixed derivative of the heat kernel stated in Proposition~\ref{prop:prop4.9}.
\end{enumerate}

\subsection{Nash-Aronson estimate for the heat kernel in dynamic environment}
This section is devoted to the statement and proof of Proposition~\ref{prop:prop4.8}; as in Sections~\ref{sec:Caccineq} and~\ref{sec:section4.2}, the infinite range of the operator $\mathcal{L}_{\mathrm{spat}}^{\phi_t}$ has to be taken into consideration in the analysis.

\begin{proposition}[Nash-Aronson type estimate] \label{prop:prop4.8}
There exists an inverse temperature $\beta_0 (d) < \infty$ such that for any point $y \in \Zd$ and any time-dependent continuous field $\phi : \R \times \Zd \mapsto \R$, if we denote by $P^{\phi_\cdot}(\cdot ,\cdot ; y)$ the solution of the parabolic system
\begin{equation*}
\left\{ \begin{aligned}
\partial_t P^{\phi_\cdot}(\cdot , \cdot ; y) + \mathcal{L}_{\mathrm{spat}}^{\phi_t} P^{\phi_\cdot}(\cdot , \cdot ; y) & = 0 ~\mbox{in}~ (0 , \infty) \times \Zd, \\
P^{\phi_\cdot}(0, \cdot ; y)  & = \delta_y  ~\mbox{in}~ \Zd,
\end{aligned} \right.
\end{equation*}  
then there exists a constant $C := C(d) < \infty$ such that one has the estimate, for each pair $(t , x) \in (0, \infty) \times \Zd,$
\begin{equation} \label{eq:NAsystpert}
\left| P^{\phi_\cdot} (t , x ; y) \right| \leq C \Phi_{C} \left( \frac{t}{\beta},x-y \right).
\end{equation}
\end{proposition}

\begin{remark}
Assuming that the field $\phi$ is defined on the entire time line $\R$ is unnecessary; one could assume that it is only define on the interval of positive times $[0 , \infty)$. We make this assumption because it is convenient in the argument and does not cause any loss of generality.
\end{remark}

\begin{proof}
We first simplify the problem by removing some dependence of the parameters in the inverse temperature $\beta$. By the change of variable $t \to \frac{t}{\beta}$, to prove the estimate~\eqref{eq:NAsystpert}, it is sufficient to prove that for every continuous field $\phi : \R \times \Zd \to \R$, the solution of the parabolic system
\begin{equation} \label{eq:TV13592}
\left\{ \begin{aligned}
\partial_t \tilde P^{\phi_\cdot}(\cdot , \cdot ; y) + \beta \mathcal{L}_{\mathrm{spat}}^{\phi_t} \tilde P^{\phi_\cdot}(\cdot , \cdot ; y) & = 0 ~\mbox{in}~ (0 , \infty) \times \Zd, \\
\tilde P^{\phi_\cdot}(0, \cdot ; y)  & = \delta_y  ~\mbox{in}~ \Zd,
\end{aligned} \right.
\end{equation}  
satisfies the estimate
\begin{equation} \label{eq:TV1482}
\left| \tilde P^{\phi_\cdot} (t , x ; y) \right| \leq C \Phi_{C} \left( t,x-y \right).
\end{equation}

We now prove the estimate~\eqref{eq:TV1482} following the sketch of the argument described in Section~\ref{sec:section4.3}. We fix a time-continuous field $\phi : \R \times \Zd \to \R$. 
\medskip

\textbf{Step 1. }We first treat the point (1) and establish the $C^{0,1-\ep}$-regularity of the solutions of the equation~\eqref{eq:TV16470}. More precisely, we prove the following result: for each regularity exponent $\ep > 0$, there exists an inverse temperature $\beta_0 := \beta_0(d, \ep) < \infty$ and constants $C := C(d,\ep) < \infty$ and $c:= c(d ) > 0$  such that for each inverse temperature $\beta \geq \beta_0$, each radius $r >1$, each pair $(t , x) \in \R \times \Zd$ and each function $u : [-r^2 + t ,t ) \times \Zd \to \R^{\binom d2}$ solution of the parabolic system
\begin{equation} \label{eq:TV11512}
    \partial_t u + \beta\mathcal{L}_{\mathrm{spat}}^{\phi_t} u =0 ~\mbox{in}~ Q_r(t , x),
\end{equation}
one has the estimate
\begin{equation} \label{eq:C^1-epregcal}
    \left[ u \right]_{C^{0, 1-\ep}\left(Q_{\frac r2}(t , x)\right)} \leq \frac{C}{r^{1-\ep}} \left\| u - \left(u \right)_{Q_r} \right\|_{\underline{L}^2 \left( Q_r(t , x) \right)} + \int_{-r^2+t}^t \sum_{y \in \Zd} e^{-c \left(\ln \beta\right)  \left( r \vee |y-x| \right)} \left| u(s , y) \right|^2 \, ds.
\end{equation}
We follow the arguments of~\cite{HK} and assume without loss of generality that $t = 0$ and $x = 0$. We decompose the proof of~\eqref{eq:C^1-epregcal} into three substeps:
\begin{itemize}
    \item In Substep 1, we use that the operator $\mathcal{L}_{\mathrm{spat}}^{\phi_t}$ is a perturbation of the Laplacian to prove that the function $u$ is well-approximated by caloric functions. More specifically, we prove the following result: for each parameter $\delta >0$, there exists and inverse temperature $\beta_1 (d , \delta) < \infty$ such that for each $\beta \geq \beta_1$ each radius $r > 1$, there exists a function $\bar u$ caloric on the cylinder $Q_{r}$ such that one has the estimate
    \begin{equation} \label{eq:TV11101}
        \left\| \nabla ( u - \bar u) \right\|_{\underline{L}^2 \left( Q_{\frac{r}{2}}\right)} \leq \delta \left\| \nabla u \right\|_{\underline{L}^2(Q_r)} + \int_{-r^2}^0 \sum_{x \in \Zd } e^{-c  \left(\ln \beta\right)  \left( r \vee |x| \right)} \left| \nabla u(t , x) \right|^2 \, dt.
    \end{equation}
    \item In Substep 2, we use the regularity known on the caloric functions to deduce from Step 1 that for each regularity exponent $\ep >0$, there exists an inverse temperature $\beta_0 := \beta_0(d , \ep) < \infty$ such that for each $\beta \geq \beta_0$ and each pair of radii $r, R \in (1 ,\infty)$ with $r \leq R$,
    \begin{equation} \label{eq:TV11461}
        \left\| u - \left( u \right)_{Q_r} \right\|_{\underline{L}^2 \left( Q_r \right)} \leq C \left( \frac{r}{R} \right)^{1-\ep} \left\| u - \left( u \right)_{Q_R} \right\|_{\underline{L}^2 \left( Q_R \right)} + \int_{-R^2}^0 \sum_{x \in \Zd} e^{-c  \left(\ln \beta\right)  \left( R \vee |x| \right)} \left|  u(t , x) \right|^2 \, dt.
    \end{equation}
    \item The $C^{0,1-\ep}$-regularity estimate~\eqref{eq:C^1-epregcal} can be deduced from the estimate~\eqref{eq:TV11461} by the integral characterization of H\"older continuous functions due to Meyers~\cite{Mey}. The adaptation to the discrete setting being straightforward and we omit the details.
\end{itemize}

\medskip

\textit{Substep 1.} The proof can essentially be extracted from the first step of the proof of Proposition~\ref{prop:prop4.6}. We let $\bar u$ be the solution of the parabolic boundary value problem
\begin{equation*}
    \left\{ \begin{aligned}
    \partial_t \bar u - \frac{1}{2} \Delta \bar u = 0 ~\mbox{in}~ Q_r, \\
\bar u = u ~\mbox{on}~ \partial_{\sqcup} Q_r,
    \end{aligned} \right.
\end{equation*}
where the notation $\partial_{\sqcup} Q_r$ denotes the parabolic boundary of the cylinder $\partial_{\sqcup} Q_r$ (see Section~\ref{Secchap2parapb} of Chapter~\ref{Chap:chap2}). We then apply the proof of Step 1 of Proposition~\ref{prop:prop4.6} to obtain the result. There are two differences in the demonstration: we do not have a Laplacian in the $\phi$-variable as in~\eqref{def.wprop4.2} and the problem is not elliptic but parabolic, nevertheless the extension of the proof Proposition~\ref{prop:prop4.6} requires a mostly notational modification of the argument so we omit the details.

\medskip

\textit{Substep 2.} The strategy is similar to the one presented in Proposition~\ref{prop:prop4.6} and follows standard arguments; we only give a sketch of the proof. We choose the inverse temperature $\beta$ large enough so that the estimate~\eqref{eq:TV11101} holds with the parameters $\delta : = \delta_{\frac\ep2}$, where $\delta_{\frac\ep2}$ is the parameter which appears in the statement of Lemma~\ref{lem:lemma4.6} associated to the regularity exponent $\ep /2$. We apply Lemma~\ref{lem:lemma4.6} with the function $\phi(r) := \left\| \nabla u \right\|_{\underline{L}^2 \left( Q_r\right)}$ to obtain that, for each pair of radii $r , R \geq 1$ such that $R \geq r \geq \left( \ln R \right)^2$,
\begin{equation} \label{eq:TV10491}
    \left\| \nabla u  \right\|_{\underline{L}^2 \left( Q_r \right)} \leq C \left( \frac{R}{r} \right)^{\frac \ep2} \left\| \nabla u \right\|_{\underline{L}^2 \left( Q_R \right)} + \int_{-R^2}^0 \sum_{x \in \Zd \setminus B_R} e^{-c \left(\ln \beta\right) |x|} \left| \nabla u(t , x) \right|^2 \, dt.
\end{equation}
We estimate the terms on the left and right side of the inequality thanks to the Caccioppoli inequality for uniformly elliptic parabolic equations and the Poincar\'e inequality for solutions of parabolic equations (see for instance~\cite{Str, GiSt}). These estimates should be adapted to the specific setting of the infinite range operator considered here; this can be achieved by using the approximation arguments presented in Section~\ref{sec:section4.2}. We obtain the estimate
\begin{equation} \label{eq:TV11251}
    \left\|  u - \left( u \right)_{Q_r}  \right\|_{\underline{L}^2 \left( Q_r \right)} \leq C \left( \frac{r}{R} \right)^{ 1 - \frac \ep2} \left\|  u - \left( u \right)_{Q_R} \right\|_{\underline{L}^2 \left( Q_R \right)} + \int_{-R^2}^0 \sum_{x \in \Zd} e^{-c \left(\ln \beta\right) \left(R \vee |x|\right)} \left| u(t , x) \right|^2 \, dt.
\end{equation}
There remains to extend the inequality~\eqref{eq:TV11251} to the small radii $r \in [1 , \left( \ln R \right)^2]$. We use the inequality
\begin{equation*}
    \left\|  u - \left( u \right)_{Q_r} \right\|_{\underline{L}^2 \left( Q_r \right)} \leq \left( \frac{ \left| Q_{\left( \ln R \right)^2}\right| }{\left| Q_r \right|} \right)^{\frac d2} \left\|  u - \left( u \right)_{Q_{\left( \ln R \right)^2}}  \right\|_{\underline{L}^2 \left( Q_{\left( \ln R \right)^2} \right)} \leq C \left( \ln R \right)^{d(d+2)}  \left\| u - \left( u \right)_{Q_{\left( \ln R \right)^2}} \right\|_{\underline{L}^2 \left( Q_{\left( \ln R \right)^2} \right)}.
\end{equation*}
We complete the argument by applying the inequality~\eqref{eq:TV11251} to estimate the $\underline{L}^2 \left( Q_{\left( \ln R \right)^2}\right)$-norm of $\nabla u$ and apply the estimate, valid under the assumption $r \leq \left( \ln R\right)^2$,
\begin{equation*}
    \left( \ln R \right)^{d(d+2)} \leq C \left( \frac{R}{r} \right)^{\frac{\ep}{2}}.
\end{equation*}
This completes the proof of Substep 2 and of the point (1).

\medskip

\textbf{Step 2. }We now treat the point (2); the objective is to deduce from the inequality~\eqref{eq:C^1-epregcal} the pointwise estimate
\begin{equation} \label{eq:L2Linfty33}
     \left\|  u \right\|_{L^\infty \left( Q_r \right)} \leq C \left\| u \right\|_{\underline{L}^2 \left( Q_{2r}\right)} + \int_{-r^2}^0 \sum_{x \in \Zd} e^{-c \left(\ln \beta\right) \left( r \vee |x|\right)} \left| u(t , x) \right|^2 \, dt.
\end{equation}
To prove the estimate~\eqref{eq:L2Linfty33}, we interpolate the space $L^\infty$ between the spaces $L^2$ and $C^{0 , 1-\ep}$ according to the formula, for any function $u : Q_r \to \R$,
\begin{equation*}
    \left\| u \right\|_{L^\infty \left( Q_{\frac{r}{2}} \right)} \leq C \left\| u \right\|_{L^2 \left(  Q_{\frac{r}{2}} \right)}^{\alpha} \left[ u \right]_{C^{1 - \ep},  Q_{\frac{r}{2}}}^{1-\alpha},
\end{equation*}
with the explicit value $\alpha = \frac{2(1-\ep)}{d+2 + 2(1 -\ep)} $. Applying the estimate~\eqref{eq:C^1-epregcal}, we obtain
\begin{align*}
     \left\| u \right\|_{L^\infty \left( Q_r \right)} &\leq C \left\| u \right\|_{L^2 \left( Q_r \right)}^{1 - \alpha} \left(  \left\| u \right\|_{\underline{L}^2 \left( Q_r \right)} + \int_{-r^2}^0 \sum_{x \in \Zd \setminus B_r} e^{-c \left(\ln \beta\right) \left( r \vee |x|\right)} \left|  u(t , x) \right|^2 \, dt \right)^{\alpha} \\
     & \leq C  \left\| u \right\|_{\underline{L}^2 \left( Q_r \right)} + \int_{-r^2}^0 \sum_{x \in \Zd} e^{-c \left( \ln \beta \right) \left( r \vee |x|\right)} \left|  u(t , x) \right|^2 \, dt.
\end{align*}
The proof of~\eqref{eq:L2Linfty33} is complete.

\medskip

\textbf{Step 3. }We now treat the point (3). The adjoint system of~\eqref{eq:TV11512} is given by
\begin{equation*}
      \partial_t v - \beta \mathcal{L}^{\phi_t}_{\mathrm{spat}} v  = 0,
\end{equation*}
so that a formal integration by parts leads to the identity
\begin{equation*}
    \iint \left( \partial_t + \beta \mathcal{L}^{\phi_t}_{\mathrm{spat}} \right) u \cdot v + \left( \partial_t - \beta \mathcal{L}^{\phi_t}_{\mathrm{spat}} \right) v \cdot u = 0.
\end{equation*}
For each point $x \in \Zd$ and each radius $r \geq 1$, we denote by $Q_r^*(x)$ the parabolic cylinder associated to the dual system
\begin{equation*}
    Q_r^* := (0 , r^2) \times B_r.
\end{equation*}
This operator $\left( \partial_t - \beta \mathcal{L}^{\phi_t}_{\mathrm{spat}} \right)$ is a perturbation of $\left( \partial_t - \frac 1{2} \Delta \right)$. One can thus apply the same arguments as the ones developed for the operator $\left( \partial_t + \beta \mathcal{L}^{\phi_t}_{\mathrm{spat}} \right)$ to prove the $L^\infty-L^2$-regularity estimate, for each function $v : Q_r^* \to \R$, solution of the parabolic system 
\begin{equation*}
    \left\| v \right\|_{L^\infty \left( Q_r^* \right)} \leq C \left\| v \right\|_{L^2 \left( Q_{2r}^* \right)} + C \int_{0}^{r^2} \sum_{x \in \Zd} e^{-c \left( \ln \beta \right) \left( r \vee |x|\right)} \left| \nabla v(t , x) \right|^2 \, dt.
\end{equation*}
This completes the proof of the point (3).

\medskip

\textbf{Step 4. }We now treat the point (4). We fix a Lipschitz function $\psi$ from $\Zd$ to $\R$. We denote by $\gamma$ the Lipschitz constant of the function $\psi$ and we always assume through the argument that $\gamma \leq 1$. We first record four inequalities. The first three estimates involve the discrete gradient of the function $\psi$. They read as follows: there exists a constant $C := C(d) < \infty$ such that
\begin{equation} \label{eq:propPhiK}
\left\{ \begin{aligned}
\left| \nabla e^{\psi} \right| & \leq C \gamma e^{\psi}, \\
\left| \nabla e^{-2 \psi} \right| & \leq C\gamma  e^{-2 \psi}\\
\forall A \subseteq \Zd, \hspace{2mm} &\sup_A e^{\psi} \leq e^{\diam A} \inf_A e^{\psi},
\end{aligned}
\right.
\end{equation}
where we used $\gamma \leq 1$ in the third inequality. The fourth, fifth and sixth ones pertain to the iteration of the discrete gradient $\nabla$ and is stated below. Since the iterated gradient is an operator which has range $n$, one has the estimate, for each integer $n \in \N$, each function $v : \Zd \to \R^{\binom d2}$ and each point $x \in \Zd$,
\begin{equation} \label{eq:propnablannn}
\left\{ \begin{aligned}
    \left| \nabla^n v(t,x) \right| & \leq C^{n-1} \sum_{y \in B(x , n)} \left| \nabla v (t,x) \right|, \\
    \left| \nabla^n v(x) \right| & \leq C^n \sum_{y \in B(x , n)} \left| v(x) \right|, \\
    \left| \nabla^n \left( e^{2 \psi} v\right) (x) -  e^{2 \psi(x)}(x) \nabla^n v(x)  \right| & \leq C^n \gamma \sum_{y \in B(x,n)} e^{2 \psi(y)} |v(y)|.
\end{aligned} \right.
\end{equation}
We now let $K$ be a large constant whose value is decided at the end of the proof and should only depend on the dimension $d$. Given a time $s \in \R$ and a point $y \in \Zd$, we let $\Gamma(\cdot, \cdot ; y, s) : (0 , \infty) \times \Zd \to \R^{\binom d2 \times \binom d2}$ be the parabolic Green's matrix, i.e., the solution of the parabolic system
\begin{equation} \label{def:defGamma}
    \left\{ \begin{aligned}
    \partial_t \Gamma(\cdot, \cdot ; y, s) + \beta \mathcal{L}^{\phi_\cdot}_{\mathrm{spat}} \Gamma(\cdot, \cdot ; y, s) & = 0 ~\mbox{in}~ (s , \infty) \times \Zd, \\
    \Gamma(s, \cdot ; y, s) & = \delta_y ~\mbox{in}~ \Zd.
    \end{aligned} \right.
\end{equation}
In particular we have the identity $\tilde P^{\phi_\cdot} = \Gamma \left( \cdot , \cdot ; \cdot , 0 \right)$. We introduce the notation $\Gamma$ because it gives some additional degrees of freedom regarding the starting time and point. We then let $Q_{s \to t}$ be the operator acting on compactly supported functions $f : \Zd \to \R^{\binom d2}$ according to the formula, for each $x \in \Zd$,
\begin{equation*}
    Q_{s \to t}^\psi f(x) := e^{-\psi(x)} \sum_{y \in \Zd} e^{\psi(y)} \Gamma \left(  x , t ; y  ,s\right) f(y),
\end{equation*}
and we note that the function $v(t ,x) := e^{\psi(x)} Q_{s \to t}^\psi f $ is solution of the parabolic system $\partial_t v + \beta \mathcal{L}_{\mathrm{spat}}^{\phi_\cdot} v = 0$. We compute
\begin{align} \label{eq:HKL2}
    \partial_t \left\| Q_{s \to t}^\psi f \right\|_{L^2\left( \Zd \right)}^2 & = \partial_t \sum_{x \in \Zd} \left| e^{\psi(x)} v(t , x) \right|^2  \\ & = 2 \sum_{x \in \Zd} e^{2\psi(x)} v(t , x) \partial_t v(t,x) \notag \\
    & = 2 \sum_{x \in \Zd} e^{2\psi(x)} v(t , x) \mathcal{L}^{\phi_\cdot}_{\mathrm{spat}} v(t,x) \notag \\
    & = \underbrace{- \sum_{x \in \Zd} \nabla \left( e^{2\psi(x)} v(t , x) \right) \nabla v(t,x)}_{\eqref{eq:HKL2}-(i)} - \underbrace{\sum_{n \geq 1} \frac{1}{\beta^\frac{n}{2}} \sum_{x \in \Zd}\nabla^{n+1} \left( e^{2 \psi(x)} v(t , x) \right) \nabla^{n+1} v(t,x)}_{\eqref{eq:HKL2}-(ii)} \notag \\
    & \qquad + \underbrace{\beta \sum_{q \in \mathcal{Q}} \nabla_q \left( e^{2\psi(x)} v(t , x) \right) \a_q \nabla_q v(t , x)}_{\eqref{eq:HKL2}-(iii)}. \notag
\end{align}
We estimate the three terms~\eqref{eq:HKL2}-(i),~\eqref{eq:HKL2}-(ii),~\eqref{eq:HKL2}-(iii) separately. For the term~\eqref{eq:HKL2}-(i), we expand the gradient of the product, use the inequality~\eqref{eq:propPhiK} and Young's inequality. We obtain
\begin{align} \label{eq:estNAfinalLaplacian}
    - \sum_{x \in \Zd} \nabla \left( e^{2 \psi(x)} v(t , x) \right) \nabla v(t,x) & =  -  \sum_{x \in \Zd} e^{2 \psi(x)} \left| \nabla v(t,x) \right|^2 - \sum_{x \in \Zd} \left( \nabla e^{2 \psi} \right)(x) v(t , x) \nabla v(t,x) \\
    & \leq  -  \sum_{x \in \Zd} e^{2 \psi(x)} \left| \nabla v(t,x) \right|^2 + C\gamma \sum_{x \in \Zd} e^{2 \psi(x)}  \left|v(t , x)\right| \left|\nabla v(t,x)  \right|  \notag \\
    & \leq - \frac{1}{2} \sum_{x \in \Zd} e^{2 \psi(x)} \left| \nabla v(t,x) \right|^2 +C\gamma^2 \sum_{x \in \Zd} e^{2 \psi(x)} \left| v(t,x) \right|^2 \notag \\
    & \leq - \frac{1}{2} \sum_{x \in \Zd} e^{2 \psi(x)} \left| \nabla v(t,x) \right|^2 + C\gamma^2\left\| Q_{s \to t}^\psi f \right\|_{L^2\left( \Zd \right)}^2. \notag
\end{align}
For the term~\eqref{eq:HKL2}-(ii), we use the inequality~\eqref{eq:propnablannn} and obtain
\begin{equation} \label{eq:tv22320.5}
        \sum_{x \in \Zd}\nabla^n \left( e^{2 \psi} v(t , \cdot ) \right)(x) \nabla^n v(t,x) \leq \sum_{x \in \Zd} e^{2 \psi(x)} \left| \nabla^n v(t , x) \right|^2 + C^n \gamma \sum_{x \in \Zd} \sum_{y \in B(x , n)} e^{2 \psi(y)}  \left| v(t , y) \right| \left| \nabla^n v(t,x) \right|.
\end{equation}
We use the inequalities~\eqref{eq:propnablannn} a second time, the property~\eqref{eq:propPhiK} and Young's inequality. We obtain
\begin{equation} \label{eq:TV224355}
    \sum_{x \in \Zd}\nabla^n \left( e^{2 \psi} v(t , \cdot ) \right)(x) \nabla^n v(t,x) \leq C^n \sum_{x \in \Zd} e^{2 \psi(x)} \left|\nabla v(t , x) \right|^2 + C^{n}\gamma^2 \sum_{x \in \Zd} e^{2 \psi(x)} \left|v(t , x)\right|^2.
\end{equation}
We then multiply the inequality~\eqref{eq:TV224355} by $\beta^{-\frac n2}$ and sum over the integers $n \in \N$. We obtain
\begin{multline*}
  \sum_{n \geq 1} \frac{1}{\beta^{\frac{n}{2}}}\sum_{x \in \Zd}\nabla^{n+1} \left( e^{2 \psi} v(t , \cdot ) \right)(x) \nabla^{n+1} v(t,x)\\ \leq \left(\sum_{n \geq 1} \left(\frac{C}{\sqrt{\beta}} \right)^n \right) \sum_{x \in \Zd} e^{2 \psi(x)} \left|\nabla v(t , x) \right|^2 + \left(\sum_{n \geq 1} \left(\frac{C}{\sqrt{\beta}} \right)^n \right) \gamma^2 \sum_{x \in \Zd} e^{2 \psi(x)} \left|v(t , x) \right|^2.
\end{multline*}
By choosing $\beta$ larger than the square of the constant $C$, we deduce
\begin{equation} \label{eq:TV0733}
  \sum_{n \geq 1} \frac{1}{\beta^{\frac{n}{2}}}\sum_{x \in \Zd}\nabla^{n+1} \left( e^{2 \psi} v(t , \cdot ) \right)(x) \nabla^n v(t,x) \leq \frac{C}{\beta^\frac12} \sum_{x \in \Zd} e^{2 \psi(x)} \left|\nabla v(t , x) \right|^2 + \frac{C \gamma^2}{\beta^\frac12}  \sum_{x \in \Zd} e^{2 \psi(x)} \left|v(t , x) \right|^2.
\end{equation}
For the term~\eqref{eq:HKL2}-(iii), we fix a charge $q \in \mathcal{Q}$. We recall the bound $\left| \a_q \right| \leq e^{-c \sqrt{\beta} \left\| q \right\|_1}$, the conventional notation for the constant $C_q$ and the estimates~\eqref{ineqcharge} of Chapter~\ref{Chap:chap2}. We compute
\begin{align*}
    \nabla_q \left( e^{2 \psi} v(t , \cdot) \right) \a_q \nabla_q v(t , \cdot) & = \left( \di^* \left( e^{2 \psi} v(t , \cdot)\right) , n_q \right) \a_q \nabla_q v(t , \cdot) \\
    & \leq C_q e^{-c \sqrt{\beta} \left\| q \right\|_1} \left\| \nabla \left( e^{2\psi} v(t  ,\cdot) \right) \right\|_{L^2
     \left( \supp n_q \right)} \left\| \nabla v(t  ,\cdot) \right\|_{L^2
     \left( \supp n_q \right)}.
\end{align*}
We then expand the gradient and use the properties~\eqref{eq:propPhiK}
\begin{align*}
    \nabla_q \left( e^{2 \psi} v(t , \cdot) \right) \a_q \nabla_q v(t , \cdot) & \leq C_q e^{-c \sqrt{\beta} \left\| q \right\|_1} \left\| \nabla \left( e^{2\psi} \right) v(t  ,\cdot)  \right\|_{L^2
     \left( \supp n_q \right)} \left\| \nabla v(t  ,\cdot) \right\|_{L^2
     \left( \supp n_q \right)} \\ & \qquad + C_q e^{-c \sqrt{\beta} \left\| q \right\|_1} \left\| e^{2\psi} \nabla v(t  ,\cdot)  \right\|_{L^2
     \left( \supp n_q \right)} \left\| \nabla v(t  ,\cdot) \right\|_{L^2
     \left( \supp n_q \right)} \\
     & \leq C_q e^{-c \sqrt{\beta} \left\| q \right\|_1}  \gamma \left(\sup_{\supp n_q} e^{\psi}  \right)  \left\| e^{\psi} v(t  ,\cdot)  \right\|_{L^2
     \left( \supp n_q \right)} \left\| \nabla v(t  ,\cdot) \right\|_{L^2
     \left( \supp n_q \right)} \\
     & \qquad + C_q e^{-c \sqrt{\beta} \left\| q \right\|_1} \left(\sup_{\supp_{n_q}} e^\psi \right) \left\| e^{\psi} \nabla v(t  ,\cdot)  \right\|_{L^2
     \left( \supp n_q \right)} \left\|  \nabla v(t  ,\cdot) \right\|_{L^2
     \left( \supp n_q \right)}.
\end{align*}
We use the property~\eqref{eq:propPhiK} (with the set $A = \supp n_q$). We obtain
\begin{align*}
    \nabla_q \left( e^{2 \psi} v(t , \cdot) \right) \a_q \nabla_q v(t , \cdot) & \leq C_q C^{\diam n_q} e^{-c \sqrt{\beta} \left\| q \right\|_1} \gamma \left\| e^{\psi} v(t  ,\cdot)  \right\|_{L^2
     \left( \supp n_q \right)} \left\| e^{\psi} \nabla v(t  ,\cdot) \right\|_{L^2
     \left( \supp n_q \right)} \\
     & \qquad + C_q C^{\diam n_q} e^{-c \sqrt{\beta} \left\| q \right\|_1} \left\| e^{\psi} \nabla v(t  ,\cdot)  \right\|_{L^2
     \left( \supp n_q \right)} \left\|  e^{\psi} \nabla v(t  ,\cdot) \right\|_{L^2
     \left( \supp n_q \right)}.
\end{align*}
We choose the inverse temperature $\beta$ large enough (depending only on the dimension $d$) so that the constants $C_q$ and $C^{\diam n_q}$ can be absorbed by the exponential term $e^{-c \sqrt{\beta} \left\| q \right\|_1 }$ and apply the Young's inequality. We obtain
\begin{equation*}
    \nabla_q \left( e^{2 \psi} v(t , \cdot) \right) \a_q \nabla_q v(t , \cdot) \leq C e^{-c \sqrt{\beta} \left\| q \right\|_1}\gamma^2 \left\| e^{\psi} v(t, \cdot) \right\|_{L^2 \left( \Zd \right)}^2 + C e^{-c \sqrt{\beta}\left\| q \right\|_1} \left\| e^{\psi} \nabla v(t, \cdot) \right\|_{L^2 \left( \Zd \right)}^2.
\end{equation*}
Summing over all the charges $q \in \mathcal{Q}$ and using the inequality, for each point $x \in \Zd$,
\begin{equation*}
    \sum_{x \in \Zd} e^{-c \sqrt{\beta} \left\| q \right\|_1} \indc_{\{ x \in \supp n_q \}} \leq C e^{-c \sqrt{\beta}},
\end{equation*}
we obtain the estimate
\begin{align} \label{eq:estNAchargesfinal}
    \beta \sum_{q\in \mathcal{Q}} \nabla_q \left( e^{2 \psi} v(t , \cdot) \right) \a_q \nabla_q v(t , \cdot) & \leq C \beta e^{-c \sqrt{\beta} }\gamma^2 \left\| e^{\psi} v(t, \cdot) \right\|_{L^2 \left( \Zd \right)}^2 + C \beta e^{-c \sqrt{\beta}} \left\| e^{\psi} \nabla v(t, \cdot) \right\|_{L^2 \left( \Zd \right)}^2 \\
    & \leq  C e^{-c \sqrt{\beta} }\gamma^2 \left\| e^{\psi} v(t, \cdot) \right\|_{L^2 \left( \Zd \right)}^2 + C  e^{-c \sqrt{\beta}} \left\| e^{\psi} \nabla v(t, \cdot) \right\|_{L^2 \left( \Zd \right)}^2, \notag
\end{align}
where we have absorbed the coefficient $\beta$ into the exponential terms $e^{-c \sqrt{\beta}}$ in the second line. We combine the estimates~\eqref{eq:HKL2},~\eqref{eq:estNAfinalLaplacian},~\eqref{eq:TV0733},~\eqref{eq:estNAchargesfinal} and choose $\beta$ large enough so that the term $C\beta^{-\frac 12}$ in the right side of~\eqref{eq:TV0733} and the term $C e^{-c \sqrt{\beta}}$ in the right side of~\eqref{eq:estNAchargesfinal} are both smaller than $\frac{1}{8}$. We obtain the estimate
\begin{align} \label{eq:ODENA}
    \partial_t \left\| Q_{s \to t}^\psi f \right\|_{L^2\left( \Zd \right)}^2 & \leq - \frac{1}{4} \sum_{x \in \Zd} e^{2 \psi(x)} \left| \nabla v(t,x) \right|^2 + C\gamma^2\left\| Q_{s \to t}^\psi f \right\|_{L^2\left( \Zd \right)}^2 \\
    & \leq C\gamma^2\left\| Q_{s \to t}^\psi f \right\|_{L^2\left( \Zd \right)}^2. \notag
\end{align}
By integrating the equation~\eqref{eq:ODENA} between the times $s$ and $t$, we obtain the inequality
\begin{equation*}
    \left\| Q_{s \to t}^\psi f \right\|_{L^2 \left( \Zd \right)} \leq e^{C \gamma^2(t - s)} \left\| f \right\|_{L^2 \left( \Zd \right)}.
\end{equation*}
The adjoint of the operator $Q_{s \to t}$ is given by the formula, for each compactly supported function $g : \Zd \to \R^{\binom d2}$,
\begin{equation*}
    \left(Q_{s \to t}^\psi\right)^* g(y) = e^{-\psi(y)} \sum_{x \in \Zd}  e^{\psi(x)}\Gamma^* \left(  y , s ; x  ,t\right) g(x),
\end{equation*}
where $\Gamma^* \left(  x , t ; y  ,s\right)$ is the fundamental solution of the dual operator $\partial_t - \beta\mathcal{L}^{\phi_\cdot}_{\mathrm{spat}}$. By  similar computation, we obtain the estimate
\begin{equation*}
     \left\| \left(Q_{s \to t}^\psi \right)^* g \right\|_{L^2 \left( \Zd \right)} \leq e^{C\gamma^2(t - s)} \left\| g \right\|_{L^2 \left( \Zd \right)}.
\end{equation*}
Considering the specific function $\psi = 0$ (in that case $\gamma = 0$), we obtain the $L^2$-estimates
\begin{equation*}
    \left\| Q_{s \to t}^0 f \right\|_{L^2 \left( \Zd \right)} \leq \left\| f \right\|_{L^2 \left( \Zd \right)} ~\mbox{and}~\left\| \left(Q_{s \to t}^0\right)^* g \right\|_{L^2 \left( \Zd \right)} \leq \left\| g \right\|_{L^2 \left( \Zd \right)}.
\end{equation*}
We then set $u(t , x) = Q_{s \to t}^0 f$, use the $L^\infty- L^2$ regularity estimate~\eqref{eq:L2Linfty33} in the parabolic cylinder $Q_{\frac{\sqrt{t - s}}{2}}(t , x)$ and the boundedness of the discrete gradient in $L^2\left( \Zd \right)$. We obtain
\begin{align*}
    \left| u(t , x)\right| & \leq \frac{C}{\left( t - s \right)^{1 + \frac{d}2}} \int_s^t  \sum_{y \in B(x , \sqrt{t - s} )} \left| u(t' , y) \right|^2 + \sum_{y \in \Zd} e^{-c \left(\ln \beta\right) \left( \sqrt{t - s} \vee |y-x|\right)} \left| u(t' , y) \right|^2   \, dt' \\
    & \leq \frac{C}{\left( t - s \right)^{1 + \frac{d}2}} \int_{s}^t \left\| u(t' , \cdot) \right\|_{L^2 \left( \Zd \right)}^2 \, dt' \\
    & \leq \frac{C}{\left( t - s \right)^{ \frac{d}2}} \left\| f \right\|_{L^2 \left( \Zd \right)}^2.
\end{align*}
Since the previous inequality is valid for any point $x \in \Zd$, we have obtained the following $L^\infty-L^2$-inequality
\begin{equation*}
    \left\| Q_{s \to t}^0 f \right\|_{L^\infty \left( \Zd \right)} \leq \frac{C}{\left( t - s \right)^{ \frac{d}2}} \left\| f \right\|_{L^2 \left( \Zd \right)}^2.
\end{equation*}
With the same computation, we obtain the $L^\infty-L^2$-estimate for the dual operator $\left( Q_{s \to t}^0 \right)^*$
\begin{equation} \label{eq:estLinftyL2fordualQ}
    \left\| \left( Q_{s \to t}^0 \right)^* g \right\|_{L^\infty \left( \Zd \right)} \leq \frac{C}{\left( t - s \right)^{ \frac{d}2}} \left\| g \right\|_{L^2 \left( \Zd \right)}^2.
\end{equation}
In the general case, we apply the regularity estimate~\eqref{eq:L2Linfty33} with the function $ e^{2 \psi(x)} Q_{s \to t}^\psi f$ and the radius $r = \frac{\sqrt{t - s}}{2}$. We obtain
\begin{multline} \label{eq:TVsam1055}
    e^{2 \psi(x)} \left|Q_{s \to t}^\psi f(t , x)\right|^2 \\ \leq \frac{C}{\left( t - s \right)^{1 + \frac{d}2}} \int_s^t  \sum_{y \in B(x , \sqrt{t - s} )} e^{2 \psi(y)} \left|Q_{s \to t}^\psi f(y)\right|^2 + \sum_{y \in \Zd} e^{-c \left( \ln \beta \right) \left( \sqrt{t - s} \vee |y - x|\right)} e^{2 \psi(y)} \left|Q_{s \to t}^\psi f( y)\right|^2  \, dt'.
\end{multline}
We then multiply each side of the inequality~\eqref{eq:TVsam1055} by $e^{-2\psi(x)}$ and note that we have the estimate, for each $y \in \Zd$,
\begin{equation*}
    e^{-2 \psi(x)} e^{2 \psi(y)} \leq \exp \left( 2 \gamma |x - y| \right).
\end{equation*}
We obtain the estimate
\begin{multline} \label{eq:TV1116sam}
    \left|Q_{s \to t}^\psi f( x)\right|^2 \leq \frac{C}{\left( t - s \right)^{1 + \frac{d}2}} \int_s^t  \sum_{y \in B(x , \sqrt{t - s} )} e^{2 \gamma |x-y|} \left|Q_{s \to t}^\psi f( y)\right|^2 \\ + \sum_{y \in \Zd} e^{-c  \left( \ln \beta \right)  \left( \sqrt{t - s} \vee |y - x |\right) + 2\gamma |y - x|} \left|Q_{s \to t}^\psi f( y)\right|^2  \, dt'.
\end{multline}
We assume that the inverse temperature $\beta$ is chosen large enough so that $c \ln \beta \geq 2$, where $c$ is the constant which appears in the exponential term $e^{-c \left( \ln \beta \right)  \left( \sqrt{t - s} \vee |y - x |\right)}$ in the right side of the inequality~\eqref{eq:TV1116sam}. Using this assumption $\gamma \leq 1$, we obtain the estimate
\begin{align} \label{eq:samTV1141}
    \left|Q_{s \to t}^\psi f(x)\right|^2 & \leq \frac{C}{\left( t - s \right)^{1 + \frac{d}2}} \int_s^t  \sum_{y \in B(x , \sqrt{t - s} )} e^{2\gamma \sqrt{t - s}} \left|Q_{s \to t}^\psi f(y)\right|^2 + \sum_{y \in \Zd \setminus B \left( x , \sqrt{t - s}\right) } \left|Q_{s \to t}^\psi f(y)\right|^2  \, dt'  \\
    & \leq \frac{C e^{2 \gamma \sqrt{s - t}}}{\left( t - s \right)^{1 + \frac{d}2}} \int_s^t  \left\|Q_{s \to t'}^\psi f\right\|^2_{L^2 \left( \Zd \right)}   \, dt' \notag \\
    & \leq \frac{C e^{2 \gamma \sqrt{s - t}}}{\left( t - s \right)^{1 + \frac{d}2}} \int_s^t  e^{C\gamma^2 (t' - s)} \left\|f\right\|^2_{L^2 \left( \Zd \right)}   \, dt' \notag\\
    & \leq \frac{C e^{2 \gamma\sqrt{s - t}}}{\gamma^2\left( t - s \right)^{1 + \frac{d}2}} e^{C\gamma^2 (t - s)}\left\|f\right\|^2_{L^2 \left( \Zd \right)}. \notag
\end{align}
Since the previous estimate is valid for any point $x \in \Zd$, we have obtained the following $L^\infty-L^2$ estimate for the operator $Q_{s \to t}^\psi$,
\begin{equation} \label{eq:TV13141}
    \left\| Q_{s \to t}^\psi f \right\|_{L^\infty \left( \Zd \right)}^2 \leq \frac{C}{\gamma^2 \left( t - s \right)^{\frac 12 + \frac{d}4}} \exp \left(\gamma \sqrt{s - t} + C\gamma^2 (t - s)\right)\left\|f\right\|^2_{L^2 \left( \Zd \right)}.
\end{equation}
A similar argument applies for the dual operator $\left( Q_{s \to t}^\psi\right)^*$ and we obtain
\begin{equation}\label{eq:TV13142}
    \left\| \left( Q_{s \to t}^\psi \right)^* g \right\|_{L^\infty \left( \Zd \right)} \leq \frac{C\beta }{ \gamma^2 \left( t - s \right)^{\frac 12 + \frac{d}4}} \exp \left(\gamma \sqrt{s - t} + C\gamma^2 (t - s)\right) \left\|g\right\|^2_{L^2 \left( \Zd \right)}.
\end{equation}
By duality the estimates~\eqref{eq:estLinftyL2fordualQ} and~\eqref{eq:TV13141} implies the inequalities
\begin{equation*}
     \begin{aligned}
    \left\| \left( Q_{s \to t}^0 \right) f \right\|_{L^2 \left( \Zd \right)} &\leq \frac{C}{\left( t - s \right)^{ \frac{d}4}} \left\|f\right\|^2_{L^1 \left( \Zd \right)}, \\
     \left\| \left( Q_{s \to t}^\psi \right) f \right\|_{L^2 \left( \Zd \right)} &\leq \frac{C }{\gamma^2\left( t - s \right)^{\frac12 + \frac{d}4}} \exp \left( \gamma \sqrt{s - t} + C\gamma^2 (t - s)\right) \left\|f\right\|^2_{L^1 \left( \Zd \right)}.
     \end{aligned}
\end{equation*}
We then set $\tau = \frac{t + s}{2}$ and use the semigroup property $Q_{s \to t}^\psi = Q_{s \to \tau}^\psi \circ Q_{\tau \to t}^\psi$. We obtain the estimate
\begin{equation} \label{eq:TV14033}
     \begin{aligned}
    \left\| \left( Q_{s \to t}^0 \right) f \right\|_{L^\infty \left( \Zd \right)} & \leq \frac{C}{\left( t - s \right)^{ \frac{d}2}} \left\|f\right\|^2_{L^1 \left( \Zd \right)}, \\
     \left\| \left( Q_{s \to t}^\psi \right) f \right\|_{L^\infty \left( \Zd \right)} & \leq \frac{C }{\gamma^2\left( t - s \right)^{1 + \frac{d}2}} \exp \left(\gamma \sqrt{s - t} + C\gamma^2 (t - s)\right) \left\|f\right\|^2_{L^1 \left( \Zd \right)}.
     \end{aligned}
\end{equation}
The estimate~\eqref{eq:TV14033} implies the upper bounds on the Green's matrix $\Gamma \left( t , x ; s , y \right)$
\begin{equation} \label{eq:TV17.5211}
    \begin{aligned}
    \left| \Gamma \left( t , x ; s , y \right) \right| &\leq \frac{C}{(t-s)^{\frac d2}},\\
    e^{\psi(x) - \psi(y)} \left| \Gamma \left( t , x ; s , y \right) \right|& \leq \frac{C \exp \left( \gamma \sqrt{ (t-s)} + C\gamma^2 \left(t - s \right) \right)}{\gamma^2 \left( t-s \right)^{1 + \frac d2}}.
    \end{aligned}
\end{equation}
We choose the function $\psi$ according to the formula $\psi(z) = \gamma |z - y|$. The estimate~\eqref{eq:TV17.5211} becomes
\begin{equation*}
    \left| \Gamma \left( t , x ; s , y \right) \right| \leq \frac{C_0}{(t - s)^\frac d2} \min\left( 1 , \frac{\exp \left( \gamma \sqrt{(t-s)} +C_0 \gamma^2 \left(t - s \right) - \gamma \left| x - y \right|\right)}{\gamma^2 (t-s)} \right).
\end{equation*}
We select the value of the coefficient $\gamma$: we let $C_0 := C_0(d) < \infty$ be a large constant and set 
\begin{equation*}
    \gamma := \left\{ \begin{aligned}
    \frac{|x - y|}{C_0 (t - s)} \hspace{3mm} &\mbox{if} \hspace{3mm} |x - y| \leq (t - s), \\
    \gamma = \frac{1}{C_0}  \hspace{3mm} &\mbox{if} \hspace{3mm}  |x - y| \geq (t - s).
    \end{aligned} \right.
\end{equation*}
By choosing  the constant $C_0$ large enough depending on the dimension $d$, we obtain the following result. There exists a constant $C := C(d) < \infty$ such that
\begin{equation*}
    \left| \Gamma \left( t , x ; s , y \right) \right| \leq \left\{\begin{aligned}
    \frac{C}{(t - s)^\frac d2} \exp \left( - \frac{|x - y|^2}{C(t-s)}\right) ~\mbox{if}~ |x - y| \leq t - s, \\
    C \exp \left( - \frac{|x - y|}{C}\right) ~\mbox{if}~ |x - y| \geq t - s.
    \end{aligned} \right.
\end{equation*}
Choosing the specific values $y =0$ and $s = 0$ completes the proof of Proposition~\ref{prop:prop4.8}.
\end{proof}

\subsection{Regularity theory for the heat kernel in dynamic environment}

This section is devoted to the proof of Proposition~\ref{prop:prop4.9} which establishes pointwise bounds on the gradient and mixed derivative of the heat kernel in terms of the regularity exponent $\ep$ and the function $\Phi_C$.

\begin{proposition}[$C^{0,1-\ep}$-regularity for the heat kernel] \label{prop:prop4.9}
For each exponent $\ep > 0$, there exists an inverse temperature $\beta_0 := \beta_0 \left( d , \ep \right) < \infty$ such that the following result holds. For any $\beta \geq \beta_0$, there exists a constant $C := C(d , \ep) < \infty$ such that for each triplet $(t , x, y) \in (0, \infty) \times \Zd \times \Zd$ and any time-continuous coefficient field $\phi : \R \times \Zd \to \R$, one has the estimate
\begin{equation} \label{eq:gradparareg}
\left| \nabla_x P^{\phi_\cdot} (t , x ; y) \right| \leq C \left(\frac{ \beta}{t} \right)^{\frac 12 + \frac d2 - \ep} \Phi_C \left(\frac{t}{ \beta} ,x-y \right),
\end{equation}
and
\begin{equation} \label{eq:gradpararegreg}
    \left| \nabla_x \nabla_y P^{\phi_\cdot} (t , x ; y) \right| \leq  C \left(\frac{ \beta}{t} \right)^{1 + \frac d2 - \ep}\Phi_C \left(\frac{t}{ \beta} ,x-y \right).
\end{equation}
\end{proposition}

\begin{proof}
By performing the change of variable $t \to \frac{t}{\beta}$, it is sufficient to prove that the heat kernel $\tilde P^{\phi_\cdot}$ defined in~\eqref{eq:TV13592} satisfies the estimates
\begin{equation*}
        \left| \nabla_x \tilde P^{\phi_\cdot} (t , x ; y) \right| \leq  \frac{C}{t^{\frac 12 + \frac d2 - \ep}}\Phi_C (t,x-y) \hspace{3mm} \mbox{and} \hspace{3mm}
    \left| \nabla_x \nabla_y \tilde P^{\phi_\cdot} (t , x ; y) \right| \leq  \frac{C }{t^{  1 + \frac d2 - \ep}}\Phi_C (t,x-y).
\end{equation*}

We fix an exponent $\ep > 0$, a time $t \geq 0$ and a point $x \in \Zd$. We let $\beta_0 := \beta_0 \left( d , \ep\right) <\infty$ be the inverse temperature above which the the Gaussian bound~\eqref{eq:TV1482} and the regularity estimate~\eqref{eq:C^1-epregcal} are valid. We apply the inequality~\eqref{eq:C^1-epregcal} with the function $u = \tilde P^{\phi_\cdot}(\cdot , \cdot ; y)$ in the parabolic cylinder $Q_{\sqrt{\frac{t}{2}}}(t , x)$ and obtain the estimate
\begin{multline} \label{eq:TV185922}
     \left[ \tilde P^{\phi_\cdot} \right]_{C^{0, 1-\ep}\left(Q_{\sqrt{\frac{t}{8}}}(t , x )\right)} \leq \frac{C}{t^{\frac 12 - \ep}} \left\| \tilde P^{\phi_\cdot} - \left(\tilde P^{\phi_\cdot} \right)_{Q_{\sqrt{\frac{t}{2}}}(t , x)} \right\|_{\underline{L}^2 \left( Q_{\sqrt{\frac{t}{2}}}(t , x) \right)} \\ + \int_{\frac {t}{2}}^t \sum_{z \in \Zd} e^{-c \left( \ln \beta \right) \left(\sqrt{\frac{t}{2}} \vee   |z| \right)} \left| \tilde P^{\phi_\cdot}(t' , z ; y) \right|^2 \, dt'.
\end{multline}
We then use the Gaussian bound~\eqref{eq:TV1482} to estimate the right side of~\eqref{eq:TV185922}. We obtain
\begin{equation} \label{eq:TV185923}
     \left[ \tilde P^{\phi_\cdot} \right]_{C^{0, 1-\ep}\left(Q_{\sqrt{\frac{t}{8}}}(t , x )\right)} \leq \frac{C}{t^{\frac 12-\ep}}\Phi_C (t , x-y).
\end{equation}
In the discrete setting, assuming that $t \geq 8$, we can write
\begin{equation} 
\label{e.missing}
    \left| \nabla \tilde P^{\phi_\cdot}(t , x) \right| \leq \sum_{y \sim x } \left| \tilde P^{\phi_\cdot}(t , y) - \tilde P^{\phi_\cdot}(t , x) \right|  \leq C \osc_{Q_1(t,x)} \tilde P^{\phi_\cdot}  \leq C \left[ \tilde P^{\phi_\cdot} \right]_{C^{0, 1-\ep}\left(Q_{\sqrt{\frac{t}{8}}}(t,x)\right)}.
\end{equation}
A combination of the estimates~\eqref{eq:TV185923} and~\eqref{e.missing} completes the proof of the estimate~\eqref{eq:gradparareg}. We note that the same argument gives the more general bound involving the parabolic Green's matrix $\Gamma$ defined in~\eqref{def:defGamma}: for each pair of times $0 \leq s < t < \infty$ and each pair of points $x, y \in \Zd$,
\begin{equation} \label{eq:estgradGamma}
    \left| \nabla_x \Gamma \left(t , x ; s , y \right) \right| \leq \frac{C}{(t-s)^{\frac 12 - \ep}} \Phi_C \left( t-s , x-y \right).
\end{equation}

To prove the estimate~\eqref{eq:gradpararegreg} we use that the function $(s , y) \mapsto \nabla_x \Gamma \left( t , x ; s , y \right)$ is solution of the dual equation $\partial_s \nabla_x \Gamma \left( t , x ; \cdot , \cdot \right)- \beta \mathcal{L}_{\mathrm{spat},y}^{\phi_\cdot} \nabla_x \Gamma \left( t , x ; \cdot  , \cdot \right) = 0$. We can thus apply the estimate~\eqref{eq:C^1-epregcal} (since this $C^{0 , 1-\ep}$-regularity estimate also holds for the dual parabolic problem by the same perturbation argument) and the arguments used in the proof of the inequality~\eqref{eq:gradparareg}. We obtain
\begin{equation*}
    \left| \nabla_x \nabla_y \Gamma (t , x ; s , y) \right| \leq \frac{C}{t^{\frac 12 - \ep}} \left\| \nabla_x \Gamma \left( t , x ; \cdot , \cdot \right) \right\|_{\underline{L}^2 \left( Q^*_{\sqrt{\frac{t - s}2}  }(s , y)\right)} + \int_{s}^{\frac{s + t}{2}} \sum_{z \in \Zd} e^{-c \left( \ln \beta \right)\left( \sqrt{\frac{t - s}2} \vee |z| \right)} \left| \nabla_x \Gamma(t , x ; t' ,z) \right|^2 \, dt'.
\end{equation*}
We use the estimate~\eqref{eq:estgradGamma} on the gradient of the function $\Gamma$ to obtain
\begin{equation} \label{eq:TV19555}
    \left| \nabla_x \nabla_y \Gamma (t , x ; s , y) \right| \leq \frac{C}{(t-s)^{1-2\ep}} \Phi_C \left(t-s , x-y \right).
\end{equation}
There is an exponent $2\ep$ instead of $\ep$ in the right side of~\eqref{eq:TV19555}. Since this estimate is valid for any exponent $\ep > 0$ (by increasing the inverse temperature $\beta$ if necessary), one can fix this issue by rewriting the proof with the exponent $\frac{\ep}{2}$ instead of $\ep$ to obtain the desired upper bound. Finally the estimate~\eqref{eq:TV19555} implies the inequality~\eqref{eq:gradpararegreg} by setting $s = 0$.
\end{proof}

\subsection{Upper bounds and regularity theory for the elliptic Green's matrix \texorpdfstring{$\G_{\f}$}{4}} \label{sec:section4.4}.
In this section, we consider the elliptic Green's matrix associated to the operator $\mathcal{L}$ defined in \eqref{e.Green} in Chapter~\ref{section3.4}. We fix an exponent $p \in [1 , \infty ]$, and a function $\f \in L^p \left( \mu_\beta \right)$. We consider the map $\Pa_\f$ introduced in Section~\ref{sec:section4.3} and define the elliptic Green's matrix $\G_\f:  \Zd \times \Omega \times \Zd \mapsto \R$ by the formula, for each $(x, \phi , y) \in \Zd \times \Omega \times \Zd$,
\begin{equation} \label{eq:formulaGreens}
\G_\f (x , \phi ; y) := \int_0^\infty \Pa_\f \left( t , x , \phi ; y \right) \, dt.
\end{equation}
As a consequence of the Feynman-Kac formula, this function can be equivalently characterized as the unique solution of the equation (see \eqref{e.Green} in Chapter~\ref{section3.4}).
\begin{equation*}
\mathcal{L} \G_\f(x  , \phi ; y) = \f( \phi) \delta_y(x) ~\mbox{in}~ \Zd \times \Omega,
\end{equation*}
such that $\left\| \G_\f(x ,\cdot ; y) \right\|_{L^p \left( \mu_\beta \right)}$ tends to $0$ as $x$ tends to infinity.

Using the equivalent characterization by Feynman-Kac, and bounds on the mixed gradients of the heat kernel established in the previous section, we obtain asymptotic estimates on the $L^p\left( \mu_\beta \right)$-norm of the Green's matrix, its gradient and its mixed derivative. This proves Proposition \ref{prop.prop4.11chap4} of Chapter~\ref{section3.4}.

\begin{proof}[Proof of Proposition \ref{prop.prop4.11chap4} of Chapter~\ref{section3.4}]
The proof is obtained by using the formula~\eqref{eq:formulaGreens} and integrating the estimates~\eqref{eq:GaussianP},~\eqref{eq:gradGaussianP} and~\eqref{eq:gradgradGaussianP} on the heat kernel over time.
\end{proof}

\section{Regularity estimate for the differentiated Helffer-Sj{\"o}strand equation} \label{sec.section4.5} In this section, we study the differentiated Helffer-Sj{\"o}strand obtained by the following procedure. We consider a function $u \in C_c^\infty (\Zd \times \Omega)$ and we denote by $G := \mathcal{L} u$. We apply the operator $\partial_{x}$ to both the left and right hand sides of the identity $\mathcal{L} u = G$. We obtain the identity
\begin{equation} \label{eq:derivHS1}
 \partial_{x} \Delta_\phi u   - \frac{1}{2\beta} \Delta \partial_{x} u + \frac{1}{2\beta}\sum_{n \geq 1} \frac{1}{\beta^{ \frac n2}} (-\Delta)^{n+1}  \partial_{x } u + \partial_{x} \left( \sum_{q\in \mathcal{Q}} \nabla_q^* \cdot \a_q \nabla_q u \right) = \partial_x G. 
\end{equation}
To go further in the computation, we introduce the following notations:
\begin{itemize}
\item We define the function $v ,h : \Zd \times \Zd \times \Omega \mapsto \R^{\binom d2 \times \binom d2}$ by the formulas, for each for $(x , y , \phi) \in \Zd \times \Zd \times \Omega$, $v(x , y, \phi) = \partial_x u(y,\phi)$ and $h(x , y , \phi) = \partial_x G(y , \phi).$ \medskip
\item Given a map $h : \Zd \times \Zd \times \Omega \mapsto \R^{\binom d2 \times \binom d2}$, we denote by $\Delta_x$ the spatial Laplacian in the first variable and by $\di^*_y$ the Laplacian in the second variable. We also denote by $\sum_{q_x\in \mathcal{Q}}  \nabla_{q_x}^* \cdot \a_{q_x} \nabla_{q_x} h $ and by $\sum_{q_y \in \mathcal{Q}}  \nabla_{q_y}^* \cdot \a_{q_y} \nabla_{q_y} h $ the operators
\begin{equation*}
\sum_{q_x \in \mathcal{Q}}  \nabla_{q_x}^* \cdot \a_{q_x} \nabla_{q_x} h : (x , y , \phi) \mapsto \sum_{ q \in \mathcal{Q}}  \a_{q}(\phi) \left( h\left( \cdot , y , \phi \right) , q \right)  q(x)
\end{equation*}
and
\begin{equation*}
\sum_{q_y \in \mathcal{Q}}  \nabla_{q_y}^* \cdot \a_{q_y} \nabla_{q_y} h : (x , y , \phi) \mapsto \sum_{q \in \mathcal{Q}}  \a_{q}(\phi) \left( h\left( x , \cdot , \phi \right) , q \right)  q(y).
\end{equation*}
\item Finally, we denote by $\mathcal{L}_{\mathrm{spat} , x }$ and $\mathcal{L}_{\mathrm{spat} , y }$ the operators
\begin{equation*}
\mathcal{L}_{\mathrm{spat} , x } := - \frac{1}{2\beta} \Delta_x u + \frac{1}{2\beta}\sum_{n \geq 1} \frac{1}{\beta^{ \frac n2}} (-\Delta_x)^{n+1} u + \sum_{q_x \in \mathcal{Q}} \nabla_{q_x}^* \cdot \a_{q_x} \nabla_{q_x} u,
\end{equation*}
and
\begin{equation*}
\mathcal{L}_{\mathrm{spat} , y } :=  - \frac{1}{2\beta} \Delta_y u + \frac{1}{2\beta}\sum_{n \geq 1} \frac{1}{\beta^{ \frac n2}} (-\Delta_y)^{n+1} u + \sum_{q_y\in \mathcal{Q}} \nabla_{q_y}^* \cdot \a_{q_y} \nabla_{q_y} u,
\end{equation*}
\end{itemize}
The term $\partial_{\cdot} \Delta_\phi u$ can be computed by using the same strategy as the one used to derive the Helffer-Sj{\"o}strand equation in Section~\ref{secchap3HSPDE} of Chapter~\ref{chap:chap3} and we obtain, for each $(x , y , \phi) \in \Zd \times \Zd \times \Omega$,
\begin{align} \label{eq:derivHS2}
\partial_{x} \Delta_\phi u (y,\phi) & = \Delta_\phi  v(x , y , \phi) - \frac{1}{2\beta} \Delta_x v(x , y , \phi)  + \frac{1}{2\beta}\sum_{n \geq 1} \frac{1}{\beta^{ \frac n2}} (-\Delta_x)^{n+1}  v (x , y , \phi) +  \sum_{q_x\in \mathcal{Q}} \nabla_{q_x}^* \cdot \a_{q_x}  \nabla_{q_x} v(x , y , \phi) \\
						& =  \Delta_\phi  v(x , y , \phi)  +  \mathcal{L}_{\mathrm{spat} ,x } v(x , y , \phi)  \notag
\end{align}
The term $\partial_{x} \left( \sum_{q\in \mathcal{Q}} \nabla_q^* \cdot \a_q \nabla_q u \right)$ can be computed by the exact formula stated in~\eqref{eq:TV18290702}:
\begin{align} \label{eq:derivHS3}
\partial_{x} \left( \sum_{q\in \mathcal{Q}} \nabla_q^* \cdot \a_q \nabla_q u (y,\phi) \right) & = \partial_{x} \left( \sum_{q\in \mathcal{Q}}  \a_q \left( u, q \right) q (y) \right) \\
													& = \sum_{q\in \mathcal{Q}}  \partial_{x} \a_q  \left( u, q \right) q(y) + \sum_{q\in \mathcal{Q}}     \a_q \left( v (x , \cdot, \phi ),  q \right) q \notag \\
													& = \sum_{q\in \mathcal{Q}} 2\pi z \left( \beta , q\right) \cos 2\pi\left( \phi , q \right) \left( u, q \right) q(x) \otimes q(y) + \sum_{q_y \in \mathcal{Q}} \nabla^*_{q_y} \a_{q_y} \nabla_{q_y} v(x , y , \phi) . \notag
\end{align}
Combining the identities~\eqref{eq:derivHS1},~\eqref{eq:derivHS2} and~\eqref{eq:derivHS3}, we obtain that the map $v$ solves the equation
\begin{equation} \label{eq:TV19370302}
\Delta_\phi v (x , y , \phi) + \mathcal{L}_{\mathrm{spat}, x } v (x , y , \phi) + \mathcal{L}_{\mathrm{spat}, y} v (x , y , \phi)  = -  \sum_{q\in \mathcal{Q}} 2\pi z \left( \beta , q\right) \cos 2\pi\left( \phi , q \right) \left( u, q \right) q(x) q(y) + \partial_x G(y , \phi).
\end{equation}
This equality can be rigorously justified using the arguments of Section~\ref{secchap3HSPDE} of Chapter~\ref{chap:chap3} and~\cite{NS, GOS}, we omit the details. The identity~\eqref{eq:TV19370302} motivates the definition of the differentiated Helffer-Sj{\"o}strand operator
\begin{equation}
\label{e.Lder}
\mathcal{L}_{\mathrm{der}} := \Delta_\phi  + \mathcal{L}_{\mathrm{spat}, x } + \mathcal{L}_{\mathrm{spat}, y} .
\end{equation}
It is natural to consider the Green's function associated to the differentiated Helffer-Sj{\"o}strand equation
\begin{equation}
\label{e.Greender}
\mathcal{L}_{\mathrm{der}} \G_{\mathrm{der},\f} =\f \delta_{(x,y)} ~\mbox{in}~ \Zd \times \Zd \times \Omega.  
\end{equation}
 Notice that we can solve \eqref{e.Greender} variationally, by applying the Gagliardo-Nirenberg-Sobolev inequality as in Lemma \ref{l.Greenvariation} of Chapter~\ref{chap:chap3}.

\subsection{Gaussian bounds and regularity estimates for the heat kernel \texorpdfstring{$\Pa_{\mathrm{der}, \f}$}{5}}
As in Sections~\ref{sec:section4.3} and~\ref{sec:section4.4}, we wish to study the properties of the heat kernel and the Green's matrix associated to the differentiated Helffer-Sj\"{o}strand operator. Given a real number $p \in [1 , \infty)$, a function $\f \in L^p \left( \mu_\beta \right)$ and a point $(x , x_1) \in \Zd \times \Zd $, we denote by $\Pa_{\mathrm{der}, \f}(\cdot, \cdot , \cdot , \cdot ; y , y_1) : (0 , \infty) \times \Zd \times \Zd \times \Omega  \to \R^{\binom d2^4}$ the solution of the $2d$-dimensional parabolic system
\begin{equation*}
    \left\{\begin{aligned}
     \partial_t  \Pa_{\mathrm{der}, \f} \left(\cdot, \cdot , \cdot , \cdot ; y , y_1\right) + \L_{\mathrm{der}} \Pa_{\mathrm{der}, \f} \left(\cdot, \cdot , \cdot , \cdot ; y , y_1\right) & = 0 ~\mbox{in}~ (0 , \infty) \times \Zd \times \Zd \times \Omega, \\
     \Pa_{\mathrm{der}, \f} \left(0, \cdot, \cdot , \cdot ; y , y_1\right) & = \f \delta_{(y , y_1)} ~\mbox{in}~ \Zd \times \Zd \times \Omega. 
     \end{aligned} \right.
\end{equation*}
We then define the elliptic Green's matrix $\G_{\mathrm{der}, \f}$ according to the Duhamel principle by the formula, for each $(x, x_1, y , y_1) \in \left( \Zd\right)^4 $ and each field $\phi \in \Omega$,
\begin{equation*}
    \mathcal{G}_{\mathrm{der}, \f} \left(x , x_1 , \phi ; y , y_1\right) := \int_{0}^\infty \Pa_{\mathrm{der}, \f} \left(t , x , x_1  , \phi ; y , y_1\right) dt.
\end{equation*}
We denote by $\nabla_x $, $\nabla_y $, $\nabla_{x_1} $, $\nabla_{y_1} $ the gradient with respect to the first, second, third and fourth spatial variables of the maps  $ \Pa_{ \mathrm{der}, \f}$ and $\G_{\mathrm{der}, \f}$.

We will prove in Proposition~\ref{prop:prop4.13} below upper bounds on the heat kernel $\Pa_{\mathrm{der}, \f}$ and its derivatives. We then combine this with the Duhamel principle to deduce upper bounds on the elliptic Green's matrix $\G_{\mathrm{der}, \f}$ in Corollary \ref{cor:corollary4.14}.

Before stating the propositions, we make a few remarks about the results. If we let $\left(\phi_t\right)_{t \geq 0}$ be the diffusion process defined by the formula~\eqref{def.phi_t} and if we recall the notation $\E_\phi$ introduced in the paragraph following~\eqref{def.phi_t}, then one has the Feynman-Kac formula
    \begin{equation} \label{eq:FeynmanKacder}
        \mathcal{P}_{\mathrm{der}, \mathbf{f}} \left( t , x , x_1 , \phi ; y, y_1 \right) = \E_{\phi} \left[ \mathbf{f}(\phi_t) P^{\phi_\cdot}_{\mathrm{der}}(t , x , x_1 ; y , y_1) \right],
    \end{equation}
where $P^{\phi_\cdot}_{\mathrm{der}}(\cdot , \cdot \, ; y)$ is the solution of the system of equations,
    \begin{equation} \label{eq:defPhatphider}
        \left\{ \begin{aligned}
        \partial_t P^{\phi_\cdot}_{\mathrm{der}}\left(\cdot , \cdot , \cdot \, ; y , y_1\right) + \left(\mathcal{L}_{\mathrm{spat} , x}^{\phi_t} + \mathcal{L}_{\mathrm{spat} , y}^{\phi_t}\right)  P^{\phi_\cdot}_{\mathrm{der}}\left(\cdot , \cdot , \cdot \, ; y , y_1\right) & =0 ~\mbox{in}~ (0 , \infty) \times \Zd \times \Zd, \\
        P^{\phi_\cdot}_{\mathrm{der}} \left(0,\cdot , \cdot \, ;y , y_1 \right) & = \delta_{(y , y_1)} ~\mbox{in}~\Zd \times \Zd.
        \end{aligned} \right.
    \end{equation}

The operator $\mathcal{L}_{\mathrm{spat}, x }^{\phi_\cdot} + \mathcal{L}_{\mathrm{spat}, y}^{\phi_\cdot}$ is a uniformly elliptic operator on the $2d$-dimensional space $\Zd \times \Zd$. Additionally, if the inverse temperature $\beta$ is chosen large enough, then this operator is a perturbation of the $2d$-dimensional Laplacian $\Delta_x + \Delta_y$. Hence the same arguments as in Section~\ref{sec:section4.3} can be used to prove Gaussian bounds and $C^{0 , 1 - \ep}$-regularity estimates on the heat kernel $ \Pa_{\mathrm{der} , \f}$; the only difference is that the underlying space is $2d$-dimensional.

The result stated in Proposition~\ref{prop:prop4.13} is strictly stronger than the ones stated in Propositions~\ref{prop:prop4.8} and Proposition~\ref{prop:prop4.9} since we obtain estimates on the triple and quadruple gradients of the heat kernel $\mathcal{P}_{\mathrm{der}, \mathbf{f}}$. This results are obtained by making use of the specific structure of the problem as we now describe. The elliptic operators $ \mathcal{L}_{\mathrm{spat}, x }$ and $\mathcal{L}_{\mathrm{spat}, y}$ only acts on the $x$ and $y$ variables respectively; in particular \emph{they commute}. This remark implies that heat kernel $P^{\phi_\cdot}_{\mathrm{der}}$ defined in~\eqref{eq:defPhatphider} factorises, i.e., we have the identity
\begin{equation} \label{eq:identitysplitLxLy}
P^{\phi_\cdot}_{\mathrm{der}}\left(t , x , x_1 \, ; y , y_1\right) = P^{\phi_\cdot}\left(t , x \, ; y \right) \otimes P^{\phi_\cdot}\left(t ,  x_1 \, ; y_1\right),
\end{equation}
where $ P^{\phi_\cdot}$ denotes the $d$-dimensional heat kernel defined in~\eqref{eq:defPhatphi}.
Thanks to this property, one can obtain additional regularity; for instance applying the gradruple gradient  $ \nabla_{x} \nabla_{x_1}\nabla_{y}\nabla_{y_1} $ to the heat kernel gives
\begin{equation*}
\nabla_{x} \nabla_{y}\nabla_{x_1}\nabla_{y_1}  P^{\phi_\cdot}_{\mathrm{der}} (x , y ; x_1 , y_1) = \nabla_x \nabla_{y} P^{\phi_\cdot}\left(t , x \, ; y \right) \otimes \nabla_{x_1} \nabla_{y_1} P^{\phi_\cdot}\left(t , x_1 \, ; y_1 \right).
\end{equation*}
We can then apply the regularity estimate~\eqref{eq:gradpararegreg} proved in Proposition~\ref{prop:prop4.9}.

\begin{proposition} \label{prop:prop4.13}
For any regularity exponent $\ep > 0$, there exists an inverse temperature $\beta_0\left(d , \ep \right) < \infty$ such that the following statement holds. For any inverse temperature $\beta > \beta_0$, there exists a constant $C(d , \ep) <\infty$ such that for each $ (x , y , x_1 , y_1) \in \left(\Zd\right)^4$, one has the estimate
\begin{equation*}
\left\| \mathcal{P}_{\mathrm{der}, \mathbf{f}} \left( t , x , x_1 , \cdot ; y, y_1 \right) \right\|_{L^p \left( \mu_\beta \right)} \leq C \left\| \f \right\|_{L^p \left( \mu_\beta \right)} \Phi_C \left( \frac{t}{\beta} , x - x_1 \right) \Phi_C \left( \frac{t}{\beta} , y - y_1 \right) ,
\end{equation*}
and the $C^{0 , 1- \ep}$-regularity estimates: if we let $\nabla_1 , \nabla_2 , \nabla_3$ and $\nabla_4$ be any permutation of the set of gradients $ \nabla_{x},  \nabla_{x_1},  \nabla_{y} $ and $\nabla_{y_1}$, then one has the four inequalities
\begin{itemize}
\item[(i)] On the gradient of the heat kernel
\begin{equation*}
\left\| \nabla_{1}   \mathcal{P}_{\mathrm{der}, \mathbf{f}} \left( t , x , x_1 , \cdot ; y, y_1 \right) \right\|_{L^p \left( \mu_\beta \right)} \leq C \left\| \f \right\|_{L^p \left( \mu_\beta \right)} \left(\frac{\beta}{t}\right)^{\frac 12 - \ep} \Phi_C \left( \frac{t}{\beta} , x - x_1 \right) \Phi_C \left( \frac{t}{\beta} , y - y_1 \right);
\end{equation*} 
\item[(ii)] On the double gradient of the heat kernel
\begin{equation*}
\left\| \nabla_{1} \nabla_{2}    \mathcal{P}_{\mathrm{der}, \mathbf{f}} \left( t , x , x_1 , \cdot ; y, y_1 \right) \right\|_{L^p \left( \mu_\beta \right)} \leq C  \left\| \f \right\|_{L^p \left( \mu_\beta \right)} \left(\frac{\beta}{t}\right)^{1 - \ep} \Phi_C \left(  \frac{t}{\beta}, x - x_1 \right) \Phi_C \left( \frac{t}{\beta} , y - y_1 \right);
\end{equation*}
\item[(iii)] On the triple gradient of the heat kernel
\begin{equation*}
\left\| \nabla_{1} \nabla_{2}\nabla_{3}   \mathcal{P}_{\mathrm{der}, \mathbf{f}} \left( t , x , x_1 , \cdot ; y, y_1 \right) \right\|_{L^p \left( \mu_\beta \right)} \leq C \left\| \f \right\|_{L^p \left( \mu_\beta \right)} \left(\frac{\beta}{t}\right)^{\frac 32 - \ep} \Phi_C \left(  \frac{t}{\beta} , x - x_1 \right) \Phi_C \left(  \frac{t}{\beta} , y - y_1 \right);
\end{equation*} 
\item[(iv)] On the quadruple gradient of the heat kernel
\begin{equation*}
\left\| \nabla_{1} \nabla_{2}\nabla_{3}\nabla_{4}   \mathcal{P}_{\mathrm{der}, \mathbf{f}} \left( t , x , x_1 , \cdot ; y, y_1 \right) \right\|_{L^p \left( \mu_\beta \right)} \leq C  \left\| \f \right\|_{L^p \left( \mu_\beta \right)} \left(\frac{\beta}{t}\right)^{2 - \ep} \Phi_C \left(  \frac{t}{\beta} , x - x_1 \right) \Phi_C \left( \frac{t}{\beta} , y - y_1 \right).
\end{equation*} 
\end{itemize}
\end{proposition}

\begin{proof}[Proof of Proposition~\ref{prop:prop4.13}]
The proof is essentially given in the paragraph preceding Proposition~\ref{prop:prop4.13}. We use the Feynman-Kac formula~\eqref{eq:FeynmanKacder} together with the factorization formula~\eqref{eq:identitysplitLxLy} and the regularity estimates stated in Proposition~\ref{prop:prop4.9}.
\end{proof}

\subsection{Upper bounds and regularity estimates for the Green's function \texorpdfstring{$\G_{\mathrm{der}, \f}$}{6}} \label{sectionchap4.3}
From these estimates, we deduce the bounds on the elliptic Green's matrix and its gradient stated in the following proposition.

\begin{proposition} \label{cor:corollary4.14}
For any regularity exponent $\ep > 0$, there exists an inverse temperature $\beta_0\left(d , \ep \right) < \infty$ such that the following statement holds. For any inverse temperature $\beta > \beta_0$, there exists a constant $C(d  , \ep) <\infty$ such that for each $ (x , y , x_1 , y_1) \in \left(\Zd\right)^4$, one has the estimate
\begin{equation*}
\left\| \G_{\mathrm{der}, \f} (x , y, \cdot ; x_1 , y_1)\right\|_{L^p \left( \mu_\beta \right)} \leq \frac{C \beta \left\|\f  \right\|_{L^p \left( \mu_\beta \right)}}{|x - x_1|^{2d-2} + |y - y_1|^{2d-2}}.
\end{equation*}
Then, for any permutation $\nabla_1 , \nabla_2 , \nabla_3$ and $\nabla_4$ of the set of gradients $ \nabla_{x},  \nabla_{x_1},  \nabla_{y} $ and $\nabla_{y_1}$, one has the estimates:
\begin{itemize}
\item[(i)] On the gradient of the Green's matrix
\begin{equation*}
\left\| \nabla_{1}   \G_{\mathrm{der}, \f} (x , y, \cdot ; x_1 , y_1) \right\|_{L^p \left( \mu_\beta \right)} \leq \frac{C \beta\left\|\f  \right\|_{L^p \left( \mu_\beta \right)}}{|x - x_1|^{2d-1 - \ep} + |y - y_1|^{2d-1- \ep}};
\end{equation*} 
\item[(ii)] On the double gradient of the Green's matrix
\begin{equation*}
\left\| \nabla_{1} \nabla_{2}   \G_{\mathrm{der}, \f} (x , y, x_1 , y_1,  \cdot)  \right\|_{L^p \left( \mu_\beta \right)} \leq \frac{C \beta\left\|\f  \right\|_{L^p \left( \mu_\beta \right)}}{|x - x_1|^{2d- \ep} + |y - y_1|^{2d- \ep}};
\end{equation*}
\item[(iii)] On the triple gradient of the Green's matrix
\begin{equation*}
\left\| \nabla_{1} \nabla_{2}\nabla_{3} \G_{\mathrm{der}, \f} (x , y, \cdot ; x_1 , y_1)  \right\|_{L^p \left( \mu_\beta \right)} \leq  \frac{C\beta\left\|\f  \right\|_{L^p \left( \mu_\beta \right)}}{|x - x_1|^{2d + 1 - \ep} + |y - y_1|^{2d + 1 - \ep}};
\end{equation*} 
\item[(iv)] On the quadruple gradient of the Green's matrix
\begin{equation*}
\left\| \nabla_{1} \nabla_{2}\nabla_{3}\nabla_{4} \G_{\mathrm{der}, \f} (x , y, \cdot ; x_1 , y_1) \right\|_{L^p \left( \mu_\beta \right)} \leq \frac{C\beta\left\|\f  \right\|_{L^p \left( \mu_\beta \right)}}{|x - x_1|^{2d +2 - \ep} + |y - y_1|^{2d +2 - \ep}}.
\end{equation*} 
\end{itemize}
\end{proposition}

\begin{proof}[Proof of Proposition~\ref{cor:corollary4.14}]
The estimates on the elliptic Green's matrix are obtained by integrating the inequalities of Proposition~\ref{prop:prop4.13} over the times $t$ in $[0, \infty)$.
\end{proof}

\chapter{Quantitative convergence of the subadditive quantities} \label{section5}

In this chapter, we introduce two subadditive energy quantities related to the variational formulation associated to the Helffer-Sj{\"o}strand operator described in Chapter~\ref{chap:chap3}. The first one, denoted by $\nu(\cu , p)$, represents the energy of the minimizer associated to the Dirichlet problem in a cube $\cu$ with affine boundary condition $l_p(x) := p \cdot x$. The second one, denoted by $\nu^*(\cu , q)$, represents the energy of the minimizer associated to the Neumann problem with boundary flux $\nabla l_q$. These two quantities satisfy a subadditivity property with respect to the domain of integration and converges as the side length of the cube tends to infinity. Moreover, the quantities $\nu$ and $\nu^*$ are convex with respect to the slopes of the boundary condition $p$ and $q$ and are in some sense convex dual to each other. The main focus of this section is then to prove by a multiscale argument that as the size of the domains tends to infinity, these quantities converge to a pair of dual convex conjugate functions and to extract from the proof a quantification of the rate of convergence.

While the general strategy comes from the theory of quantitative stochastic homogenization presented in~\cite{AKM}, the adaptation of the techniques presented in this monograph requires to overcome three types of difficulties:
\begin{itemize}
    \item One needs to take into account the Laplacian with respect to the $\phi$-variable; 
    \item One needs to take into account the infinite range of the operator $\mathcal{L}$;
    \item We need to homogenize an elliptic system instead of an elliptic PDE.
\end{itemize}
While the first point has been successfully treated in \cite{AW} to study the $\nabla\phi$ model, the last two points are intrinsic to the Coulomb gas representation of the Villain model and will be treated in this chapter. 

In Section~\ref{sec:section5.4}, we introduce a finite-volume version of first-order corrector associated to the Hellfer-Sj\"{o}strand operator $\mathcal{L}$. We use the quantitative rate of convergence of the energy $\nu$ to establish quantitative sublinearity of the corrector and to prove a quantitative estimate on the weak norm of its flux. This function and its properties are crucial to prove the quantitative homogenization of the mixed derivative of the Green's matrix in Chapter~\ref{sec:section6}.

Throughout this entire chapter, we fix a regularity exponent $\ep$ which is small compared to $1$ and depends only on the dimension $d$. We assume that the inverse temperature $\beta$ is large enough so that all the results presented in Chapter~\ref{section:section4} hold with the regularity exponent $\ep$.

We complete this section by mentioning that in this chapter, the constants are only allowed to depend in the dimension $d$ as we need to be track their dependence in the inverse temperature $\beta$. The objective is to prove that the quantitative rate of convergence $\alpha$ obtained in Proposition~\ref{prop:mainpropsect5} and~\ref{prop:prop5.25} remains bounded away from $0$ as $\beta$ tends to infinity.

\section{Definition of the subadditive quantities and basic properties}

\subsection{Definition of the energy quantities}

Let $\cu \subseteq \Zd$ be a cube of $\Zd$, we define the energy functional $\mathbf{E}_\cu$ according to the formula, for each function $u \in H^1 \left(\Zd, \mu_\beta \right)$,
\begin{equation*}
\mathbf{E}_{\cu} \left[ u \right]  :=  \beta \sum_{y \in \Zd} \left\| \partial_y u \right\|_{L^2 \left( \cu , \mu_\beta \right)}^2 +  \frac1{2}  \left\| \nabla  u\right\|_{L^2 \left( \cu , \mu_\beta \right)}^2 + \frac{1}{2}\sum_{n \geq 1} \frac1{\beta^{\frac n2}}\left\| \nabla^{n+1}  u \right\|_{L^2 \left( \Zd, \mu_\beta \right)}^2   -  \beta \sum_{\supp q \cap \cu \neq \emptyset}  \left\langle \nabla_q u \cdot \a_q  \nabla_q u   \right\rangle_{\mu_{\beta}}.
\end{equation*}
We introduce the bilinear form associated to the energy $\mathbf{E}_\cu$ according to the formula, for each function $u \in H^1 \left(\Zd, \mu_\beta \right)$,
\begin{multline*}
\mathbf{B}_{\cu} \left[ u , v \right] := \beta \sum_{x \in \cu} \sum_{y \in \Zd} \left\langle \partial_y u(x , \cdot ), \partial_y v(x , \cdot ) \right\rangle_{\mu_\beta} +  \frac1{2}  \sum_{x \in \cu} \left\langle \nabla  u(x , \cdot ), \nabla  v(x , \cdot ) \right\rangle_{\mu_\beta} \\ + \frac1{2}  \sum_{n \geq 1} \sum_{x \in \Zd} \frac1{\beta^{\frac n2}}\left\langle \nabla^{n+1}  u(x , \cdot ), \nabla^{n+1}  v(x , \cdot ) \right\rangle_{\mu_\beta} -  \sum_{\supp q \cap \cu \neq \emptyset} \beta \left\langle \nabla_q u \cdot \a_q  \nabla_q v   \right\rangle_{\mu_{\beta}}.
\end{multline*}
One cannot consider the energy $\mathbf{E}_{\cu}$ of a function $v$ only defined in the cube $\cu$ since the infinite range of the operator $\mathcal{L}$ requires to know the value of the function on the entire space $\Zd$. To fix this issue, we need to remove a boundary layer from a given cube $\cu$. This is done in the definition below. 

\begin{definition}[Trimmed cube] \label{def:deftrimmed}
Given a cube $\cu := z + \left( - \frac{R}{2} , \frac{R}{2} \right)^d$, we defined the trimmed cube $\cu^-$ by the formula
\begin{equation*}
    \cu^{-} :=  z + \left( - \frac{R}{2} + \frac{\sqrt{R}}{10} , \frac{R}{2} - \frac{\sqrt{R}}{10} \right)^d.
\end{equation*}
\end{definition}

We define the energy $\mathbf{E}^*_\cu$ according to the formula
\begin{align*}
\mathbf{E}_{\cu}^* \left[ u \right]  = \beta \sum_{y \in \Zd} \left\| \partial_y u \right\|_{L^2 \left( \cu , \mu_\beta \right)}^2 +  \frac1{2}  \sum_{n \geq 0} \sum_{x\in\cu, \dist \left( x , \partial \cu \right) \geq n} \frac1{\beta^{\frac n2}}\left\| \nabla^{n+1}  u(x , \cdot ) \right\|_{L^2 \left( \mu_\beta \right)}^2  - \frac{1}{\beta^{\frac 14}} \left\| \nabla u \right\|_{L^2 \left( \cu \setminus \cu^-, \mu_\beta\right)}^2 \\ - \beta \sum_{\supp q \subseteq \cu}  \left\langle \nabla_q u \cdot \a_q  \nabla_q u \right\rangle_{\mu_{\beta}},
\end{align*}
as well as the corresponding bilinear form $\mathbf{B}_{\cu}^*$, for each $u , v \in H^1 \left(\cu, \mu_\beta \right)$,
\begin{multline*}
\mathbf{B}_{\cu}^* \left[ u , v \right]  :=  \beta \sum_{x \in \cu} \sum_{y \in \Zd} \left\langle \partial_y u(x , \cdot ), \partial_y v(x , \cdot ) \right\rangle_{\mu_\beta}  +  \frac1{2}  \sum_{n \geq 1} \sum_{x\in\cu, \dist \left( x , \partial \cu \right) \geq n} \frac1{\beta^{\frac n2}}\left\langle \nabla^{n+1}  u(x , \cdot ), \nabla^{n+1}  v(x , \cdot ) \right\rangle_{\mu_\beta}  \\ - \frac{1}{\beta^{\frac 14}} \sum_{x \in \cu \setminus \cu^-} \left\langle \nabla u(x , \cdot) , \nabla v(x  , \cdot)\right\rangle_{\mu_\beta} -  \beta \sum_{\supp q \subseteq \cu}  \left\langle \nabla_q u \cdot \a_q  \nabla_q v   \right\rangle_{\mu_{\beta}}.
\end{multline*}
Let us make a few remarks about the definition of the energy $\mathbf{E}_{\cu}^*$.

\begin{remark} 
The iterated Laplacian $\Delta^n$ has range $2n$; given a point $x \in \cu$, we only consider the iteration of the Laplacian until the integer $n :=  \dist(x , \partial \cu )$. This ensures that for any function $v \in H^1 \left(\cu , \mu_\beta \right)$, the quantity $\Delta^n v$ is well-defined.
\end{remark}
\begin{remark}
We only consider the charges $q$ whose support is included in the cube $\cu$, this ensures that for any function $v \in H^1 \left(\cu , \mu_\beta \right)$, the quantity $\nabla_q \cdot \a_q \nabla_q v$ is well-defined.
\end{remark}
\begin{remark}
We subtract an additional term in the boundary layer $\left\{ x \in \cu \, : \, \dist(x , \partial \cu ) \leq \frac{\sqrt{R}}{10} \right\}$. This term is a perturbative terms for two reasons: we are only summing on a small boundary layer of size $\frac{\sqrt{R}}{10}$ of the cube $\cu$ and the multiplicative factor $\beta^{-\frac 14}$ is much smaller than the leading order term of the energy $\mathbf{E}_\cu^*$, which is of order $1$. The reason justifying the presence of this term is that it is useful to deal with the infinite range of the operator $\mathcal{L}$; in particular, it is useful to prove the subadditivity of the energy functional $\nu^*$ (see Definition~\ref{def.defnunustar}) in Proposition~\ref{p.subaddnustar}. The specific choice for the exponent $\frac{1}4$ for the power of $\beta$ is arbitrary; we only need an exponent which is strictly between $0$ and $\frac 12$.
\end{remark}

By choosing the inverse temperature $\beta$ sufficiently large, one can prove that the energy $\mathbf{E}_\cu$ satisfies the following coercivity and boundedness properties: there exist constants $c(d) > 0$ and $C(d) < \infty$ such that, for each $u \in H^1_0 \left(\Zd , \mu_\beta \right)$,
\begin{equation} \label{eq:coerccontenergy}
 c \left\llbracket u \right\rrbracket_{H^1 \left( \cu , \mu_\beta \right)} \leq \mathbf{E}_\cu \left[ u \right] \leq C  \left\llbracket u \right\rrbracket_{H^1 \left( \cu , \mu_\beta \right)},
\end{equation} 
where we recall the notation $\left\llbracket u \right\rrbracket_{H^1 \left( \cu , \mu_\beta \right)}$ introduced in Section~\ref{sec:notGIbbsmeasure} of Chapter~\ref{Chap:chap2},
\begin{equation*}
\left\llbracket u \right\rrbracket_{H^1 \left( \cu , \mu_\beta \right)} := \left(\beta \sum_{y \in \Zd} \left\| \partial_y  u \right\|_{L^2 \left( \cu ,\mu_\beta \right)}^2\right)^\frac 12 +  \left\| \nabla u \right\|_{L^2 \left( \cu, \mu_\beta \right)}.
\end{equation*}
The same estimate holds for the energy functional $\mathbf{E}_\cu^*$: for each $u \in H^1 \left(\cu , \mu_\beta \right)$,
\begin{equation} \label{eq:coerccontenergystar}
 c\left\llbracket u \right\rrbracket_{H^1 \left( \cu , \mu_\beta \right)} \leq \mathbf{E}_\cu^* \left[ u \right] \leq C \left\llbracket u \right\rrbracket_{H^1 \left( \cu , \mu_\beta \right)}.
\end{equation}
We now proceed by giving the definitions of the subadditive quantities $\nu$ and $\nu^*$.

\begin{definition}[Linear functions and subadditive quantities] \label{def.defnunustar}
Given a vector $p \in  \R^{d \times\binom d2}$, we write $p = \left( p_1 , \ldots, p_{\binom d2} \right)$ where the components $ p_1 , \ldots, p_{\binom d2}$ belong to the space $\R^d$. We denote by $l_p$ the affine function defined by the formula
\begin{equation*}
    l_p := \left\{\begin{aligned}
    \Zd &\to \R^{\binom d2}, \\
    x &\to \left( p_1 \cdot x , \ldots,  p_{\binom d2} \cdot x \right).
    \end{aligned} \right.
\end{equation*}
For each cube $\cu \subseteq \Zd$ and each pair of vectors $p , p^* \in \R^{d \times \binom{d}{2}}$, we define the energies
\begin{equation} \label{def:defnu}
\nu \left( \cu , p \right) := \inf_{u \in l_p + H^1_0\left( \cu , \mu_\beta \right)} \frac{1}{2|\cu|}\mathbf{E}_{\cu}[u],
\end{equation}
and
\begin{equation} \label{def:defnustar}
\nu^* \left( \cu , p^* \right) := \sup_{v \in H^1 \left( \cu , \mu_\beta \right)} - \frac{1}{2|\cu|}\mathbf{E}^*_{\cu}[v] + \frac{1}{|\cu|}  \sum_{x \in \cu} p^* \cdot \left\langle \nabla v(x) \right\rangle_{\mu_\beta}.
\end{equation}
\end{definition}
It is clear from the estimate~\eqref{eq:coerccontenergy} that the energy quantities $\nu$ and $\nu$* are well-defined, quadratic in the variables $p$ and $p^*$ respectively and that they satisfy the upper and lower bounds, for each cube $\cu \subseteq \Zd$ and each pair of vectors $p , p^* \in \R^{d \times\binom d2}$,
\begin{equation} \label{eq:TV08234}
    c\left| p \right|^2 \leq \nu \left( \cu , p \right)  \leq C \left| p \right|^2 ~\mbox{and}~  c \left| p^* \right|^2 \leq \nu \left( \cu , p^* \right)  \leq C \left| p^* \right|^2.
\end{equation}
It follows from the standard argument of the calculus of variations that the minimizer in the variational definition~\eqref{def:defnu} exists and is unique. We denote it by $u\left( \cdot , \cu , p\right)$.

The maximizer of the variational formulation~\eqref{def:defnustar} exists and is unique up to additive constant. This property is not a direct consequence of the standard arguments. It requires to use the properties of the Helffer-Sj\"{o}strand equation and the regularity estimates established in Chapter~\ref{section:section4}. We postpone the proof of this result to Appendix~\ref{app.appB}. We denote by $v \left(\cdot , \cu , p^* \right)$ the unique maximizer which satisfies $\sum_{x \in \cu} \left\langle v \left( x , \cdot , \cu , p^* \right) \right\rangle_{\mu_\beta} = 0$. We record from Appendix~\ref{app.appB} that it satisfies the variance estimate, for each point $x \in \frac 13 \cu$ and each $p^* \in \R^{d \times \binom d2}$,
\begin{equation} \label{eq:TV17570501}
    \var \left[ v \left( x , \cdot , \cu_n , p^* \right) \right] \leq C \left|p^*\right|^2.
\end{equation}
The maps $p \mapsto u \left( \cdot , \cdot , \cu , p \right)$ and $p^* \mapsto v \left( \cdot , \cdot , \cu , p^* \right)$ are linear and that by the upper bound on the energies $\mathbf{E}$ and $\mathbf{E}^*$ stated in~\eqref{eq:coerccontenergy}, they satisfy the estimates
\begin{equation} \label{eq:uppboundoptimi}
    \left\| \nabla u \left( \cdot  , \cdot, \cu , p \right) \right\|_{\underline{L}^2 \left( \cu , \mu_\beta  \right)} \leq C \left| p \right| ~\mbox{and}~\left\| \nabla v \left( \cdot  , \cdot,  \cu , p^* \right) \right\|_{\underline{L}^2 \left( \cu , \mu_\beta  \right)} \leq C \left| p^* \right|.
\end{equation}

The goal of this section is to prove that, as the size of the cube $\cu$ tends to infinity, the two quantities $\nu$ and $\nu^*$ converge and to obtain an algebraic rate of convergence. We obtain a result along a specific sequence of cubes defined below.

\begin{figure}[h!]
\centering
\includegraphics[scale=0.1]{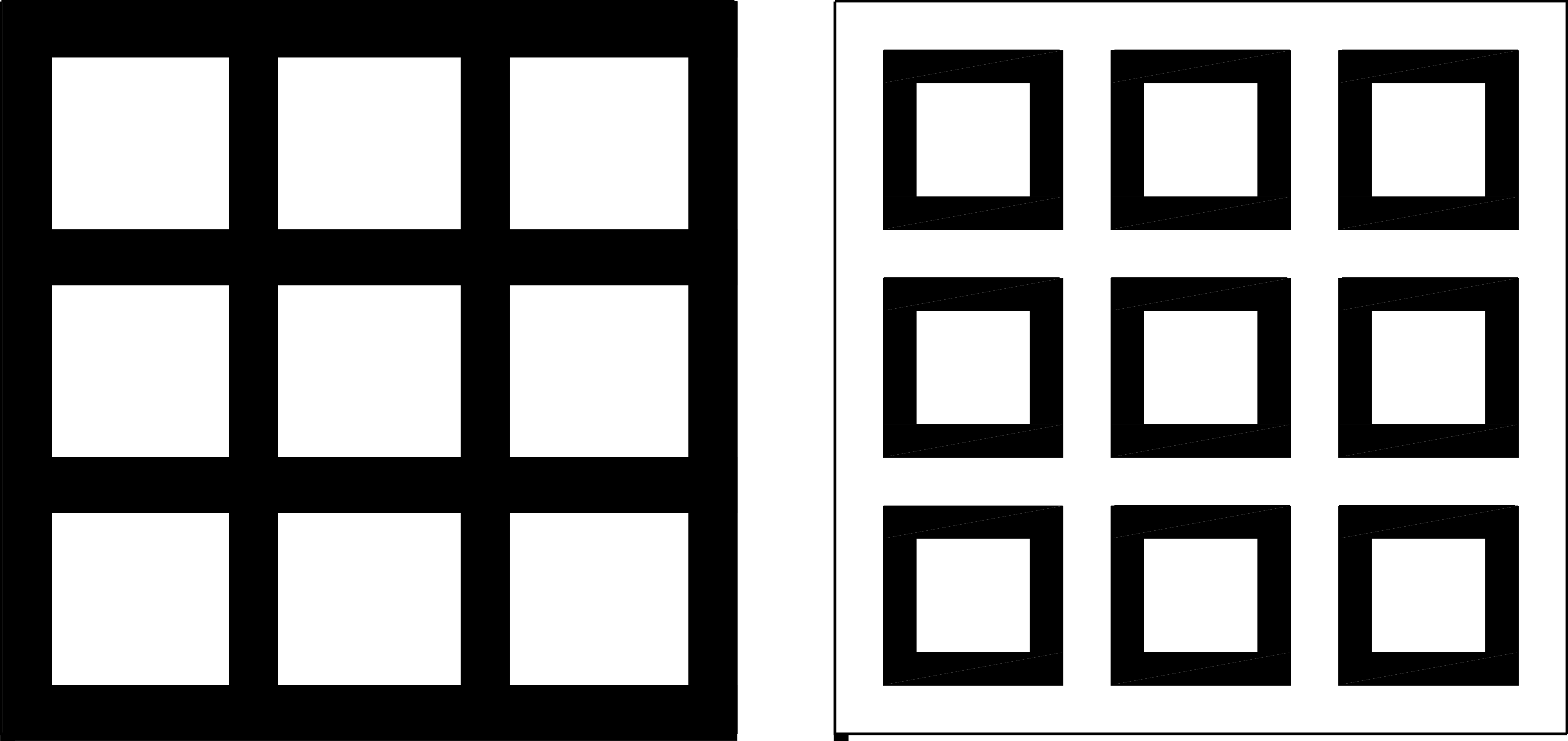}

\caption{The picture on the left represents the cube $\cu_{n+1}$, the white interior cubes are the cubes $(z + \cu_n)_{z \in \mathcal{Z}_n}$ and the set in black is the boundary layer $BL_n$. The picture on the right represents the same cube; the set $A_n$ is drawn in black.} \label{fig:boundarylayers}
\end{figure}

\begin{definition}[Triadic cube and $\mathcal{Z}_n$] \label{def:deftriadic}
We define the sequence $l_n$ of non-negative real numbers according to the induction formula
\begin{equation*}
l_0 = 1~ \mbox{and for each}~n \in \N, ~ l_{n+1} = 3 l_n + \sqrt{l_n}.
\end{equation*}
For each $n \in \N$, we define the cube $\cu_n := \left( - \frac{l_n}2 , \frac{l_n}2 \right)^d$. We denote by $\mathcal{Z}_{m,n} := l_{n}3^{m-n} \Zd \cap \cu_{n}$ and by $BL_{m,n}$ the mesoscopic boundary layer defined by the formula $BL_{m,n} := \cu_{n} \setminus \bigcup_{z \in \mathcal{Z}_{m,n}} \left( z + \cu_{m} \right)$. The cube $\cu_n$ can be partitioned according to the formula
\begin{equation*}
\cu_{n} := \bigcup_{z \in \mathcal{Z}_{m,n}} \left( z + \cu_m \right) \cup BL_{m,n}.
\end{equation*} 
We also introduce the notations $\mathcal{Z}_n := \mathcal{Z}_{n,m}$, $BL_n := BL_{n+1,n}$ and let  $A_n$ be the set
\begin{equation} \label{eq:TVdefAn}
   A_n := \bigcup_{z \in \mathcal{Z}_n}  \left( z + \cu_n \right) \setminus \left( z + \cu_n^- \right).
\end{equation}
We refer to Figure~\ref{fig:boundarylayers} for an illustration of these definitions. The reason we introduce the sets $BL_{m,n}$ and $A_n$ is to treat the infinite range of the operator $\mathcal{L}$.
\end{definition}

In the following remarks, we record without proof some properties pertaining the Definition~\ref{def:deftriadic}.

\begin{remark} \label{rmk:TV1414}
 There exists a universal constant $C$ such that, for each integer $n \in \N$, $3^n \leq l_n \leq C 3^n$.
 \end{remark}
 
 \begin{remark} \label{rmk:cardinality}
 The cardinality of $\mathcal{Z}_{m,n}$ is equal to $3^{d(n-m)}$.
 \end{remark}
 
 \begin{remark} \label{rem:volest}
 One has the volume estimate $\left| BL_{m,n} \right| \leq C 3^{-\frac m2} |\cu_n|$.
\end{remark}

\subsection{Statement of the main result} \label{sec:Chap6statmainres1014}

The main result obtained in this chapter is a quantitative rate of convergence for the two energy quantities $\nu$ and $\nu^*$; it is stated below.

\begin{proposition} \label{prop:mainpropsect5}
There exists an inverse temperature $\beta_0 := \beta_0 \left( d \right) < \infty$ such that the following statement holds. There exist constants $c := c(d)>0$, $C := C(d) < \infty$ and an exponent $\alpha := \alpha (d) > 0$ such that for each inverse temperature $\beta \geq \beta_0$ there exists a symmetric positive definite matrix $\ahom \in \R^{d \binom d2 \times d \binom d2}$
such that for each integer $n \in \N$, and each pair of vectors $p , p^* \in \R^{d \times\binom d2}$, one has the estimate
\begin{equation*}
    \left| \nu \left( \cu_n^- , p \right) - \frac 12 p \cdot \ahom p \right| \leq C  3^{- \alpha n} |p|^2 \hspace{5mm}\mbox{and} \hspace{5mm} \left| \nu^* \left( \cu_n , p^* \right) - \frac{1}{2}  p \cdot \ahom^{-1} p^* \right| \leq C 3^{- \alpha n} |p^*|^2.
\end{equation*}
\end{proposition}

\begin{remark} \label{remark1.1110320302}
Using the symmetries of the model, we can prove that the following properties. If we let $L^{2 , \di^*}$ be the linear map introduced in Section~\ref{sec:diffform=vectfct} of Chapter~\ref{Chap:chap2}, then there exists a coefficient $\bar \lambda_\beta:= \bar \lambda_\beta(d, \beta)$ which tends to $0$ as $\beta$ tends to infinity such that
\begin{equation} \label{eq:TV19240202}
   \left\{ 
   \begin{aligned}
     \ahom & = \frac{1}{2}I_d ~\mbox{in the space}~ \mathrm{Ker} \, L_{2 , \di^*}, \\ 
     \ahom & = \frac{(1 + \bar \lambda_\beta)}{2} I_d  ~\mbox{in the space}~ \left(\mathrm{Ker} \, L_{2 , \di^*}\right)^\perp.
    \end{aligned}
   \right.
\end{equation}
A direct consequence of~\eqref{eq:TV19240202} is the identity between the elliptic systems
\begin{equation*}
    - \nabla \cdot \ahom \nabla = \frac{1}{2} \left( \di^* \di + \left(1+\bar \lambda_\beta \right) \di \di^*\right).
\end{equation*}
These properties are a consequence of Property (3) of Proposition~\ref{prop:prop5.2}and Proposition~\ref{prop:mainpropsect5}.
\end{remark}

The proof of Proposition~\ref{prop:mainpropsect5} relies on ideas which were initially developed in~\cite{AS}, and follows the presentation given in~\cite{AKM}. The argument relies on the definition of the quantity
\begin{equation} \label{def:defJquan}
    J \left( \cu , p , p^* \right) := \nu \left( \cu^- , p \right) + \nu^* \left( \cu, p^* \right) - p \cdot p^*.
\end{equation}
By the estimate~\eqref{eq:onesidedconvexdual} below, we know that the quadratic form $J$ is is almost positive, in the sense that it satisfies the inequality, for each cube $\cu$ of size $R$ and each pair of slopes $p , p^* \in \R^{d \times\binom d2}$,
\begin{equation*}
    J \left( \cu , p , p^* \right) \geq - C R^{-\frac 12} \left( |p|^2 + |p^*|^2 \right).
\end{equation*}
To prove Proposition~\ref{prop:mainpropsect5}, we argue that the map $J \left( \cu , p , p^* \right)$ can be bounded from above in the following sense: for each vector $p \in \Rd$, there exists a vector $p^* \in \Rd$ such that
\begin{equation} \label{eq:Jsupbound}
    J \left( \cu , p , p^* \right) \leq C 3^{- \alpha n} |p|^2.
\end{equation}
Additionally, we prove that the vector $p^*$ is close to $\ahom p$. The quantitative rate of convergence stated in Proposition~\ref{prop:mainpropsect5} is then a relatively straightforward consequence of the estimate~\eqref{eq:Jsupbound}. The proof of~\eqref{eq:Jsupbound} relies on a hierarchical decomposition of space and requires to introduce the subadditivity defect at scale $l_n$,
\begin{align} \label{def.taun}
    \tau_n & := \sup_{p , p^* \in B_1} \left(\nu \left( \cu_{n}^-, p \right) - \nu \left( \cu_{n +1}^-, p \right) \right) + \left(\nu^* \left( \cu_{n}, p^* \right) - \nu^* \left( \cu_{n +1}, p^* \right) \right) \\
    & = \sup_{p , p^* \in B_1} J \left( \cu_{n}, p, p^* \right) - J \left( \cu_{n +1}, p , p^* \right). \notag
\end{align}
We then prove a series of propositions and lemmas (Propositions~\ref{p.subadd} and~\ref{p.subaddnustar}, Lemmas~\ref{lem:lemma5.6},~\ref{lemmvarest},~\ref{lem:lemma5.8} and~\ref{e.lemma4.3}), where various quantities are estimated in terms if the subadditivity defect $\tau_n$. From these results we deduce an inequality of the form: for each integer $n \in \N $ and each vector $p \in \R^{d \times\binom d2}$, there exists a vector $p^* \in \R^{d \times\binom d2}$ such that
\begin{equation*}
    J \left( \cu_{n+1} , p , p^* \right) \leq C \tau_n,
\end{equation*}
which can be rewritten
\begin{equation} \label{eq:TV11194}
    J \left( \cu_{n+1} , p , p^* \right) \leq \frac{C}{C+1}J \left( \cu_{n} , p , p^* \right).
\end{equation}
The estimate~\eqref{eq:TV11194} shows that, by passing from one scale to another, the energy quantity $J$ has to contract by a multiplicative factor strictly less than $1$. An iteration of the inequality~\eqref{eq:TV11194} yields the algebraic rate of convergence stated in the inequality~\eqref{eq:Jsupbound}.

\subsection{Basic properties}

We first record some basic properties of the energy quantities $\nu$ and $\nu^*$; they are analogous to \cite[Lemma 2.2]{AKM}.

\begin{proposition}[Basic properties of $\nu$ and $\nu^*$] \label{prop:prop5.2}
Fix a cube $\cu \subseteq \Zd$ and two parameters $p , p^* \in \R^{ d \times \binom{d}{2}}$. The energy quantity $\nu(\cu ,p)$ (resp. $\nu^* (\cu , p^*)$) and the minimizer $u(\cdot , \cu , p)$ (resp. maximizer $v(\cdot, \cu , p^*)$) satisfy the properties:
\begin{enumerate}
\item \textnormal{First variation.} The optimizing functions satisfy the following identities:
\begin{equation*}
\mathbf{B}_\cu [u (\cdot , \cu , p) , w ] = 0, ~\forall w \in H^1_0 \left( \cu , \mu_\beta \right),
\end{equation*}
and
\begin{equation*}
\mathbf{B}_\cu^* [v (\cdot , \cu , p^*) , w ] = \frac{1}{|\cu|} \sum_{x \in \cu} p^* \cdot \left\langle \nabla w(x , \cdot) \right\rangle_{\mu_\beta}  , ~\forall w  \in H^1 \left( \cu , \mu_\beta \right).
\end{equation*}
\item \textnormal{Second variation.} For each function $w \in l_p + H^1_0 (\cu , \mu_\beta),$
\begin{equation} \label{eq.secondvarnu}
\frac1{2|\cu|} \mathbf{E}_\cu \left[ w \right] - \nu(\cu , p) = \frac1{2|\cu|} \mathbf{E}_\cu \left[ u \left( \cdot , \cu , p \right) - w \right].
\end{equation}
For each $w \in H^1 (\cu , \mu_\beta)$,
\begin{equation}  \label{eq.secondvarnustar}
\nu^* \left( \cu, p^* \right)  + \frac 12 \mathbf{E}_\mathbf{\cu } \left[ w \right] - \frac{1}{|\cu|}  \sum_{x \in \cu} p^* \cdot \left\langle \nabla w(x) \right\rangle_{\mu_\beta}=   \frac1{2|\cu|} \mathbf{E}_\cu \left[ v \left( \cdot , \cu , p^* \right) - w \right].
\end{equation}
\item \textnormal{Quadratic representation.} We recall the definition linear map $L_{2, \di^*}:  \R^{d\times \binom d2} \to \R^{d }$ introduced in Section~\ref{sec:diffform=vectfct} of Chapter~\ref{Chap:chap2}. There exist two symmetric positive definite matrices $\a(\cu), \a_* (\cu) \in \R^{d \binom d2 \times d \binom d2}$ such that
\begin{equation} \label{eq:TV15214}
    \nu (\cu , p) = \frac 12 p \cdot \a(\cu) p \hspace{5mm} \mbox{and} \hspace{5mm}  \nu^* (\cu , p^*) = \frac 12 p^* \cdot \a^{-1}_* (\cu)  p^*.
\end{equation}
Additionally, there exist two coefficients $ \lambda_\cu$ and $ \lambda_\cu^*$ such that
\begin{equation} \label{eq:TV19250202}
   \left\{ 
   \begin{aligned}
     \ahom & = I_d ~\mbox{in the space}~ \mathrm{Ker} L_{2 , \di^*}, \\ 
     \ahom & = (1 + \lambda_\cu) I_d  ~\mbox{in the space}~ \left(\mathrm{Ker} L_{2 , \di^*}\right)^\perp.
    \end{aligned}
   \right.
\end{equation}
and
\begin{equation} \label{eq:TV19260202}
   \left\{ 
   \begin{aligned}
     \ahom & = I_d ~\mbox{in the space}~ \mathrm{Ker} L_{2 , \di^*}, \\ 
     \ahom & = (1 + \lambda_\cu^*) I_d  ~\mbox{in the space}~ \left(\mathrm{Ker} L_{2 , \di^*}\right)^\perp.
    \end{aligned}
   \right.
\end{equation}
 We denote by $L^{t}_{2, \di^*} : \R^{d} \to \R^{d \times \binom d2}$ its adjoint of the map $L_{2, \di^*}$. By differentiating the identities~\eqref{eq:TV15214} with respect to the parameters $p$ and $p^*$, we obtain the equalities
\begin{equation} \label{eq:derivpnu}
\frac{1}{|\cu|}\sum_{x \in \cu} \left( \frac1{2}  \left\langle \nabla  u(x , \cdot , \cu ,p ) \right\rangle_{\mu_\beta}  -  \beta  \sum_{\supp q \cap \cu \neq \emptyset}   \left\langle \a_q  \nabla_q u(\cdot , \cdot , \cu ,p )   \right\rangle_{\mu_{\beta}} L^{t}_{2, \di^*} \left( n_q(x) \right)  \right)= \a(\cu) p
\end{equation}
and
\begin{equation}  \label{eq:derivpnustar}
\frac{1}{|\cu|}\sum_{x \in \cu} \left\langle \nabla v(x, \cdot , \cu , p^*) \right\rangle_{\mu_\beta} = \a^{-1}_* (\cu) p^*.
\end{equation}
\item \textnormal{One-sided convex duality.} For each discrete cube $\cu \subseteq \Zd$ of sidelength $R$, we have the estimate
\begin{equation} \label{eq:onesidedconvexdual}
\nu \left( \cu^- , p \right) + \nu^* \left( \cu , p^* \right) - p \cdot p^* = \frac1{2|\cu|} \mathbf{E}_{\cu}^* \left[  v\left( \cdot , \cu , p^* \right) - u \left( \cdot, \cu^{-} , p \right)\right] + O \left( C |p|^2 R^{-\frac 12} \right),
\end{equation}
where we recall the notation $O$ introduced in Section~\ref{SecNotandprelim} of Chapter~\ref{Chap:chap2}: given two real numbers $X , Y$ and a non-negative real number $\kappa$, we write $X = Y + O\left( \kappa \right)$ if and only if $|X - Y| \leq \kappa$;
\end{enumerate}
\end{proposition}

\begin{proof}
The proof of the properties (1), (2) and (3) are straightforward and we refer to~\cite[Lemma 2.2]{AKM}. For the identity~\eqref{eq:TV15214}, the arguments of~\cite{AKM} give the following results: for each cube $\cu \subseteq \Zd$, there exist two positive definite matrices $\mathbf{a} (\cu), \mathbf{a}_* (\cu) \in \R^{d \binom d2 \times d \binom d2}$, such that, for each $p , p^* \in \R^{d \times \binom d2}$,
\begin{equation*}
     \nu (\cu , p) = \frac 12 p \cdot \mathbf{a} (\cu) p \hspace{5mm} \mbox{and} \hspace{5mm}  \nu^* (\cu , p^*) = \frac 12 p^* \cdot \mathbf{a}_* (\cu) p^*.
\end{equation*}
To prove the estimate~\eqref{eq:TV19250202}, we first use that any $p \in \mathrm{Ker} \, L_{2,\di^*}$, one has the identity $\di l_p = 0$. This implies that the minimizer in the energy $\nu \left( \cu , p \right)$ is attained by the map $l_p$, from which one obtains that the linear map $\mathbf{a}$ is equal to the identity on the space $\mathrm{Ker} \, L_{2,\di^*}$. The proof of the result on the orthogonal complement of the space $\mathrm{Ker} \, L_{2,\di^*}$ is a consequence of the rotation and symmetry invariance of the dual Villain model. The proof of~\eqref{eq:TV19260202} is identical.

For the identity~\eqref{eq:derivpnu}, by differentiating the equality~\eqref{eq.secondvarnu} with respect to the variable $p$, we obtain the identities, for each $p , p' \in \R^{d \times \binom d2}$,
\begin{align} \label{eq:TV15474}
    \a(\cu) p\cdot p' & = \frac{1}{|\cu|} \sum_{x \in \cu}  \frac1{2}  \left\langle \nabla  u(x , \cdot , \cu ,p ) \cdot p' \right\rangle_{\mu_\beta}  -  \beta  \sum_{\supp q \cap \cu \neq \emptyset}   \left\langle \a_q  \nabla_q u(\cdot , \cdot , \cu ,p )   \right\rangle_{\mu_{\beta}}  \left( n_q , \di^* l_{p'} \right)   \\
      & = \frac{1}{|\cu|}\sum_{x \in \cu} \frac1{2}  \left\langle \nabla  u(x , \cdot , \cu ,p ) \cdot p' \right\rangle_{\mu_\beta}  -  \beta  \sum_{\supp q \cap \cu \neq \emptyset}   \left\langle \a_q  \nabla_q u(\cdot , \cdot , \cu ,p )   \right\rangle_{\mu_{\beta}}  \left( n_q ,  L_{2, \di^*}  \left( \nabla l_{p'} \right) \right)   \notag \\
     & = \frac{1}{|\cu|} \sum_{x \in \cu} \frac1{2}  \left\langle \nabla  u(x , \cdot , \cu ,p ) \cdot p' \right\rangle_{\mu_\beta}  -  \beta  \sum_{\supp q \cap \cu \neq \emptyset}   \left\langle \a_q  \nabla_q u(\cdot , \cdot , \cu ,p )   \right\rangle_{\mu_{\beta}}  \left( n_q ,  L_{2, \di^*}  \left( p' \right) \right)   \notag \\
    & =  \frac{1}{|\cu|} \sum_{x \in \cu} \left( \frac1{2}  \left\langle \nabla  u(x , \cdot , \cu ,p ) \cdot p' \right\rangle_{\mu_\beta}  -  \beta  \sum_{\supp q \cap \cu \neq \emptyset}   \left\langle \a_q  \nabla_q u(\cdot , \cdot , \cu ,p )   \right\rangle_{\mu_{\beta}}  L^{t}_{2, \di^*} \left( n_q(x)\right) \cdot   p'  \right) . \notag
\end{align}
Using that the identity~\eqref{eq:TV15474} is valid for every vector $p' \in \R^{d \times \binom d2}$, we obtain the identity~\eqref{eq:derivpnu}. 

There only remains to prove the one-sided convex duality property stated in~\eqref{eq:onesidedconvexdual}. We apply the second variation formula~\eqref{eq.secondvarnustar}, with the function $u = u\left( \cdot, \cu^- , p\right)$ and use the identity $$ \frac{1}{|\cu|}  \sum_{x \in \cu} p^* \cdot \left\langle \nabla u(x , \cdot , \cu^- , p) \right\rangle_{\mu_\beta} = p \cdot p^*,$$ which is a consequence of the inclusion $\cu^- \subseteq \cu$ and the fact that the map $u$ belongs to the space $l_p + H^1_0 \left( \cu, \mu_\beta\right)$. We obtain
\begin{equation*} 
\nu^* \left( \cu, p^* \right)  + \frac 1{2\left| \cu \right|} \mathbf{E}_\mathbf{\cu}^* \left[ u \right] - p^* \cdot p=   \frac1{2|\cu|} \mathbf{E}_\cu^* \left[ v \left( \cdot , \cu , p^* \right) - u \left( \cdot , \cu^- , p^* \right) \right].
\end{equation*}
By definition of the function $u$, we have the equality $\nu \left( \cu^- , p \right) = \frac 1{2\left| \cu^- \right|} \mathbf{E}_\mathbf{\cu^- } \left[ u \right]$. To prove the inequality~\eqref{eq:onesidedconvexdual}, it is thus sufficient to prove
\begin{equation} \label{eq:approxucucumin}
     \left| \frac 1{\left| \cu^- \right|} \mathbf{E}_\mathbf{\cu^- } \left[ u \right] - \frac{1}{|\cu|} \mathbf{E}_\cu^* \left[ u \right] \right| \leq C R^{- \frac 12}.
\end{equation}
The rest of the argument is devoted to the proof of the inequality~\eqref{eq:approxucucumin}. We use the two estimates
\begin{equation*}
    \mathbf{E}_\mathbf{\cu^- } \left[ u \right] \leq C |p|^2  ~\mbox{and}~ \frac{ \left| \cu \setminus \cu^- \right| }{ \left| \cu \right|} \leq \frac{C}{\sqrt R}
\end{equation*}
to deduce
\begin{equation} \label{eq:approxucucumin22}
    \left| \frac 1{\left| \cu^- \right|} \mathbf{E}_\mathbf{\cu^- } \left[ u \right] - \frac{1}{|\cu|} \mathbf{E}_\mathbf{\cu^- } \left[ u \right] \right| \leq \frac{C |p|^2}{\sqrt R}.
\end{equation}
From the inequality~\eqref{eq:approxucucumin22}, we see that to prove~\eqref{eq:approxucucumin}, it is sufficient to prove the inequality
\begin{equation} \label{eq:approxucucumin222}
      \left| \mathbf{E}_\cu^* \left[ u \right] -  \mathbf{E}_\mathbf{\cu^- } \left[ u \right] \right| \leq C R^{d - \frac 12}.
\end{equation}
since the volume of the cube $\cu$ is equal to $R^d$. The rest of the argument is devoted to the proof of the estimate~\eqref{eq:approxucucumin222}. Using the definitions of the energies $\mathbf{E}_\mathbf{\cu^-}$ and $\mathbf{E}_\cu^*$ and noting that the term involving the sum over the boundary layer $\{ x \in \cu \, : \, \dist \left( x , \partial \cu \right) \leq R^{\frac 12} \}$ in the definition of $\mathbf{E}_\cu^*$ is negative, we have the inequality
\begin{multline*}
    \left| \mathbf{E}_\cu^* \left[ u \right] - \mathbf{E}_\mathbf{\cu^- } \left[ u \right] \right| \leq \beta  \sum_{y \in \Zd} \left\| \partial_y u \right\|_{L^2 \left( \cu \setminus \cu^-, \mu_\beta \right)}^2 + \frac1{2}  \left\| \nabla  u \right\|_{L^2 \left( \cu \setminus \cu^-, \mu_\beta \right)}^2 \\
     + \frac1{2}  \sum_{n \geq 1} \frac1{\beta^{\frac n2}}\left\| \nabla^{n+1}  u \right\|_{L^2 \left( \Zd\setminus \cu^n,  \mu_\beta \right)}^2   - \beta \sum_{q \in \mathcal{Q}^-_\cu}  \left\langle \nabla_q u \cdot \a_q  \nabla_q u \right\rangle_{\mu_{\beta}},
\end{multline*}
where we used the notation $\cu^n$ to denote the set $\cu^n := \left\{ x \in \cu \, : \, \dist \left( x , \partial \cu \right) \geq n  \right\}$ and the notation $\mathcal{Q}^-_\cu$ is used to denote the set of charges
\begin{equation*}
        \mathcal{Q}^-_\cu := \left\{ q \in \mathcal{Q} \, : \, \supp q \subseteq \cu \setminus \cu^- ~\mbox{or}~ \left( \supp q \cap \cu^- \neq \emptyset ~\mbox{and}~ \supp q \cap \left(\Zd \setminus \cu \right) \neq \emptyset \right) \right\}.
\end{equation*}
We then use the following ingredients:
\begin{itemize}
    \item The function $u$ is equal to the affine function $l_p$ outside the cube $\cu^-$. This implies the identities, for each point $x \in \cu \setminus \cu^-$, each point $y \in \Zd$ and each charge $q \in \mathcal{Q}$ such that $\supp q \subseteq \cu \setminus \cu^-$,
    \begin{equation*}
        \partial_y u(x , \cdot) = 0, \hspace{5mm} \nabla u(x , \cdot ) = p, \hspace{5mm} \nabla_q u = \left( \di^* l_p , n_q \right).
    \end{equation*}
    From these identities, we deduce that
    \begin{equation*}
        \sum_{y \in \Zd} \left\| \partial_y u \right\|_{L^2 \left( \cu \setminus \cu^- ,  \mu_\beta \right)}^2 = 0, \hspace{5mm} \left\| \nabla  u \right\|_{L^2 \left(\cu \setminus \cu^-, \mu_\beta \right)}^2 = \left| p \right|^2 |\cu \setminus \cu^-| \leq C |p|^2 R^{- \frac12} R^d.
    \end{equation*}
    Using the estimate $ \left| \a_q \right| \leq C e^{- c \sqrt{\beta} \left\| q \right\|_1}$, we obtain
    \begin{equation*}
        \left| \sum_{\supp q \subseteq \cu \setminus \cu^-} \left\langle \nabla_q u \cdot \a_q  \nabla_q u \right\rangle_{\mu_{\beta}} \right| \leq C \left| p\right|^2  \left| \cu \setminus \cu^- \right| \leq C \left| p\right|^2 R^{d - \frac 12}.
    \end{equation*}
    \item If a charge $q \in \mathcal{Q}$ is such that its support intersects both the cube $\cu^-$ and the set $\Zd \setminus \cu$, then its diameter must be larger than $R^{\frac12}$. This implies the inequality, for any such charge $q \in \mathcal{Q}$, $\left| \a_q \right| \leq e^{-c \sqrt{\beta} \left\| q \right\|_1} \leq e^{-c \sqrt{\beta} R^{\frac 12}}$. This implies
    \begin{align*}
        \left| \left\langle \nabla_q u \cdot \a_q  \nabla_q u \right\rangle_{\mu_{\beta}} \right| & \leq e^{- c \sqrt{\beta} \left\| q \right\|_1} \left\| n_q \right\|_2^2 \left\| \nabla u \right\|_{L^2 \left( \supp n_q , \mu_\beta \right)}^2 \\
        & \leq C_q e^{- c \sqrt{\beta} \left\| q \right\|_1} \left\| \nabla u \right\|_{L^2 \left( \cu , \mu_\beta \right)}^2.
    \end{align*}
    Summing over all the charges whose support intersects the cube $\cu^-$ and the set $\Zd \setminus \cu$, we obtain
    \begin{equation*}
        \sum_{q\in \mathcal{Q}} \left| \left\langle \nabla_q u \cdot \a_q  \nabla_q u \right\rangle_{\mu_{\beta}} \right| \leq \left(\sum_{q\in \mathcal{Q}} e^{- c \sqrt{\beta} \left\| q \right\|_1} \left\| n_q \right\|_2 \right) \left\| \nabla u \right\|_{L^2 \left( \cu \right)}^2 \leq e^{-c \sqrt{\beta R} } \left\| \nabla u \right\|_{L^2 \left( \cu , \mu_\beta \right)}^2.
    \end{equation*}
    We then use the upper bounds stated in~\eqref{eq:uppboundoptimi} to deduce
    \begin{equation*}
        \sum_{q\in \mathcal{Q}} \left| \left\langle \nabla_q u \cdot \a_q  \nabla_q u \right\rangle_{\mu_{\beta}} \right| \leq C e^{- c \sqrt{\beta R}} R^d \left\| \nabla u \right\|_{\underline{L}^2 \left( \cu , \mu_\beta \right)}^2 \leq e^{- c \sqrt{\beta R}} |p|^2,
    \end{equation*}
where we reduced the value of the constant $c$ in the second inequality to absorb the volume term $R^d$ in the exponential term $e^{- c \sqrt{\beta R}}$.
\item Using that the function $u$ is equal to the affine function outside the cube $\cu^-$, that for each integer $n$ larger than $2$, the iterated gradient of the affine function $\nabla^n l_p $ is equal to $0$ and the fact that the operator $\nabla^n$ has range $n$, we obtain the identity, for each point $x \in \Zd$ such that $\dist \left( x , \cu^- \right) \geq n$, $\nabla^n u(x) = 0$. From this identity, we deduce
\begin{equation} \label{eq:TV09291111}
     \sum_{n \geq 0} \frac1{\beta^{\frac n2}}\left\| \nabla^n u \right\|_{L^2 \left( \Zd \setminus \cu^{n}, \mu_\beta \right)}^2 \leq \sum_{n \geq \frac{\sqrt{R}}{2}}  \sum_{\dist \left( x , \cu^- \right) \leq n} \frac1{\beta^{\frac n2}}\left\| \nabla^n  u(x , \cdot ) \right\|_{L^2 \left( \mu_\beta \right)}^2.
\end{equation}
We then estimate the $L^2$-norm of the iterated gradient $\nabla^n$ according to the estimate, for each $x \in \Zd$,
\begin{align} \label{eq:TV092911}
    \left\| \nabla^n u (x , \cdot) \right\|_{L^2 \left( \mu_\beta \right)}^2 \leq C^n \left\| \nabla u  \right\|_{L^2 \left( B(x , n) \mu_\beta \right)}^2 & \leq C^n \left( \left\| \nabla u \right\|_{L^2 \left( B(x , n) \cap \cu , \mu_\beta \right)}^2 +  \left\| \nabla l_p \right\|_{L^2 \left( B(x , n) \cap \left( \Zd \setminus \cu \right) , \mu_\beta \right)}^2 \right) \\ & \leq  C^n \left( R^d \left\| \nabla u \right\|_{\underline{L}^2 \left( \cu , \mu_\beta \right)}^2 + n^d |p|^2 \right) \notag \\
    & \leq  C^n R^d  |p|^2, \notag
\end{align}
where we used the estimate~\eqref{eq:uppboundoptimi} in the last inequality. A combination of the inequalities~\eqref{eq:TV09291111} and~\eqref{eq:TV092911} with the upper bound~\eqref{eq:uppboundoptimi} shows
\begin{equation*}
    \sum_{n \geq 0} \frac1{\beta^{\frac n2}}\left\| \nabla^n  u \right\|_{L^2 \left( \Zd \setminus \cu^{n}, \mu_\beta \right)}^2  \leq \sum_{n \geq \frac{\sqrt{R}}{2}} \frac{\left| \{ x \in \Zd \, : \, \dist \left( x , \cu^- \right) \leq n \}  \right| C^n R^d}{\beta^{\frac n2}} |p|^2 .
\end{equation*}
The volume factor $\left| \{ x \in \Zd \, : \, \dist \left( x , \cu^- \right) \leq n \}  \right|$ can be estimated by the value $( R + n)^d$. Thus, by choosing the inverse temperature $\beta$ large enough, we obtain that there exists a constant $c := c(d) > 0$ such that
\begin{equation*}
    \sum_{n \geq 0} \frac1{\beta^{\frac n2}}\left\| \nabla^n  u \right\|_{L^2 \left( \Zd \setminus \cu^n \mu_\beta \right)}^2 \leq \left( \frac{C}{\sqrt{\beta}} \right)^{\frac{\sqrt{R}}{2}} R^d  \left| p \right|^2 \leq C e^{- c \left(\ln \beta\right) \sqrt{R}} \left| p \right|^2.
\end{equation*}
\end{itemize}
A combination of the three previous items completes the proof of completes the proof the inequality~\eqref{eq:onesidedconvexdual}.
\end{proof}

\section{Subadditivity for the energy quantities}

In this section, we prove a subadditivity property for the two energies $\nu$ and $\nu^*$. The result is quantified and we estimate the $H^1$-norm of the difference of the minimizer $u$ (resp. maximizer $v$) over two different scales in terms of the difference $\nu \left( \cu_m , p \right) - \nu \left( \cu_n , p \right)$ (resp. $\nu^* \left( \cu_m , p^* \right) - \nu^* \left( \cu_n , p^* \right)$).

\subsection{Subadditivity for the energy \texorpdfstring{$\nu$}{7}}
In this section, we prove that the energy quantity $\nu$ satisfies a subadditivity property with respect to the domain of integration and deduce from it the existence of the homogenized coefficient $\ahom$. The statement of Proposition~\ref{p.subadd} is quantified; we prove that the $H^1$-norm of the difference of the minimizer $u$ over two different scales in terms of the subadditivity defect for the energy $\nu$.

\begin{proposition}[Subadditivity for $\nu$]  \label{p.subadd}
There exists an inverse temperature $\beta_0 := \beta_0 \left( d \right) < \infty$ such that for each $\beta \geq \beta_0$ the following statement is valid. There exists a constant $C := C(d) < \infty$ such that for each pair of integers $(m,n) \in \N$ satisfying $n > m$ and each vector $p \in \R^{d \times \binom d2}$,
\begin{equation} \label{eq:subaddnu}
\frac{1}{\left| \mathcal{Z}_{m,n} \right|}\sum_{z \in \mathcal{Z}_{m,n}} \left\llbracket u(\cdot , \cu_{n} , p) - u(\cdot ,z + \cu_{m} , p)  \right\rrbracket_{\underline{H}^1(\cu_{n+1},\mu_\beta)}^2 \leq C \left( \nu \left( \cu_m , p \right) - \nu \left( \cu_{n} , p \right)  +  C 3^{-\frac m2} |p|^2 \right).
\end{equation}
\end{proposition}

\begin{remark}  \label{rem.subaddtrim}
Since it is useful in the rest of the proof, we note that the demonstration of Proposition~\ref{p.subadd} can be adapted to the case of trimmed cubes so as to obtain the estimate, for each pair of integers $m,n \in \N$ such that $m \leq n$,
\begin{equation*} \label{eq:subaddnutrim}
\frac{1}{\left| \mathcal{Z}_{m,n} \right|}\sum_{z \in \mathcal{Z}_{m,n}} \left\llbracket u(\cdot , \cu_{n}^- , p) - u(\cdot ,z + \cu_{m}^- , p)  \right\rrbracket_{\underline{H}^1(\cu_{n+1},\mu_L)}^2 \leq C \left( \nu \left( \cu_m^- , p \right) - \nu \left( \cu_{n}^- , p \right)  +  C 3^{-\frac m2} |p|^2 \right).
\end{equation*}
Since the proof is essentially the same as the proof of Proposition~\ref{p.subadd}; the details are left to the reader.
\end{remark}

Before proving Proposition~\ref{p.subadd}, we record an immediate corollary of the the subadditivity property for the energy $\nu$.

\begin{corollary} \label{cor:coro5.13}
There exists an inverse temperature $\beta_0 := \beta_0 \left( d \right) < \infty$ such that, for each $\beta \geq \beta_0$, there exists a non-negative real number $\ahom$ such that for each vector $p \in \R^{d \times \binom d2}$, one has
\begin{equation*}
    \nu \left( \cu_n , p \right) \underset{n \to \infty}{\longrightarrow} p \cdot \ahom p.
\end{equation*}
By Property (3) of Proposition~\ref{prop:prop5.2}, this statement can be rewritten equivalently as
\begin{equation*}
    \a \left( \cu_n \right)  \underset{n \to \infty}{\longrightarrow} \ahom.
\end{equation*}
Additionally, one deduces from~\eqref{eq:subaddnu} the lower bound estimate in the sense of symmetric positive definite matrices
\begin{equation} \label{eq:TV18244}
     \a \left( \cu_n \right) \geq \ahom - C 3^{-\frac n2}.
\end{equation}
\end{corollary}

\begin{remark}
By Remark~\ref{rem.subaddtrim}, the convergence also holds with the trimmed triadic cubes and we have, for each vector $p \in \R^{d \times \binom d2}$,
\begin{equation*}
        \nu \left( \cu_n^- , p \right) \underset{n \to \infty}{\longrightarrow} p \cdot \ahom p, \hspace{5mm} \a \left( \cu_n^- \right) \underset{n \to \infty}{\longrightarrow} \ahom \hspace{5mm} \mbox{and} \hspace{5mm} \forall n \in \N, ~\a \left( \cu_n^- \right) \geq \ahom - C 3^{-\frac n2}.
\end{equation*}
\end{remark}

\begin{proof} 
Since the left side of~\eqref{eq:subaddnu} is non-negative, we have the inequality, for each pair of integers $m, n \in \N$ such that  $n > m$
\begin{equation} \label{eq:TV18114}
     \nu \left( \cu_n , p \right) \leq  \nu \left( \cu_{m} , p \right)  +  C 3^{-\frac m2} |p|^2.
\end{equation}
Combining the inequality~\eqref{eq:TV18114} with the fact that the sequence $\left(\nu \left( \cu_n , p \right)\right)_{n \in \N}$ is non-negative implies that it converges with the estimate~\eqref{eq:TV18244}.
\end{proof}

We now focus on the proof of Proposition~\ref{p.subadd}.

\begin{proof}[Proof of Proposition~\ref{p.subadd}]
For the sake of simplicity, we only write the proof in the case when the difference between the integers $m$ and $n$ is equal to $1$: we consider the specific case of the pair $(n , n+1)$. The proof of the general case is similar. We assume without loss of generality that $|p|=1.$

We let $w$ be the function of $l_p + H^1_{0} \left(\cu_{n+1}, \mu_\beta \right)$ defined by the following construction:
\begin{itemize}
\item For each point $z \in \mathcal{Z}_{n+1}$, we set $w  := u(\cdot , z + \cu_{n} , p);$
\item On the mesoscopic boundary layer $BL_n$, we set $w := l_p$.
\end{itemize}
Applying the second variation formula~\eqref{eq.secondvarnu} and the coercivity of the energy functional $\mathbf{E}$ stated in~\eqref{eq:coerccontenergy} gives the inequality 
\begin{equation}  \label{eq:YK105326}
\left\llbracket u(\cdot , \cu_{n+1} , p) - w \right\rrbracket_{\underline{H}^1(\cu_{n+1},\mu_\beta)}^2 \leq C
\left( \frac1{2|\cu_{n+1}|} \mathbf{E}_{\cu_{n+1}} \left[ w \right] - \nu(\cu , p) \right).
\end{equation}
Using that, for each point $z \in \mathcal{Z}_n$, the function $w$ is equal to the minimizer $u(\cdot , z + \cu_{n} , p)$ in the cube $\left(z + \cu_n\right)$, we have the inequality
\begin{equation} \label{eq:TV105222}
    \sum_{z \in \mathcal{Z}_{n+1}} \left\llbracket u(\cdot , \cu_{n+1} , p) - u(\cdot, z + \cu_{n} , p) \right\rrbracket_{\underline{H}^1(\cu_{n+1},\mu_\beta)}^2 \leq \left\llbracket u(\cdot , \cu_{n+1} , p) - w \right\rrbracket_{\underline{H}^1(\cu_{n+1},\mu_\beta)}^2.
\end{equation}
By the estimates~\eqref{eq:YK105326} and~\eqref{eq:TV105222}, we see that to prove the inequality~\eqref{eq:subaddnu}, it is thus sufficient to prove
\begin{equation} \label{eq:estw2scale}
\frac1{2|\cu|} \mathbf{E}_{\cu_{n+1}} \left[ w \right]  \leq \nu \left( \cu_n , p \right) + C 3^{-\frac n2}.
\end{equation}
We now prove the inequality~\eqref{eq:estw2scale}. By definition of the energy $\mathbf{E}$, we have
\begin{multline} \label{eq:splitEn+1}
\mathbf{E}_{\cu_{n+1}} \left[ w \right] :=  \underbrace{\beta \sum_{y \in \Zd} \left\| \partial_y  w \right\|_{L^2 \left( \cu_{n+1}, \mu_\beta \right)}^2}_{\eqref{eq:splitEn+1}-(i)} + \underbrace{  \frac1{2}  \left\| \nabla   w \right\|_{L^2 \left( \cu_{n+1}, \mu_\beta \right)}^2}_{\eqref{eq:splitEn+1}-(ii)} \\
+ \underbrace{ \frac1{2}  \sum_{k \geq 1}  \frac1{\beta^{\frac k2}}\left\| \nabla^{k+1}   w \right\|_{L^2 \left( \Zd, \mu_\beta \right)}^2}_{\eqref{eq:splitEn+1}-(iii)}- \underbrace{ \beta \sum_{\supp q \cap \cu_{n+1} \neq \emptyset}  \left\langle \nabla_q  w\cdot \a_q  \nabla_q u   \right\rangle_{\mu_{\beta}}}_{\eqref{eq:splitEn+1}-(iv)}.
\end{multline}
We estimate the four terms on the right side separately. The term~\eqref{eq:splitEn+1}-(i) involving the derivative with respect to the field $\phi$ can be estimated by the following argument. Since the map $w$ is equal to the deterministic affine function $l_p$ in the boundary layer $BL_n$, we have the identity $\partial_y w(x , \cdot) = 0$ for any point $x \in BL_n$ and any point $y \in \Zd$. This implies the equality
\begin{equation} \label{eq:TVs21}
\sum_{y \in \Zd} \left\| \partial_y  w \right\|_{_{L^2 \left( \cu_{n+1} \mu_\beta \right)}}^2 = \sum_{z \in \mathcal{Z}_n} \sum_{y \in \Zd} \left\| \partial_y  u(\cdot , z + \cu_n ) \right\|_{L^2 \left( z + \cu_{n} \mu_\beta \right)}^2.
\end{equation} 
This completes the estimate of the term~\eqref{eq:splitEn+1}-(i). For the term~\eqref{eq:splitEn+1}-(ii), we use the same argument and note that $\nabla w(x , \cdot) = p$ for any point $x \in BL_n$. We obtain
\begin{align} \label{eq:TVS22}
    \frac{1}{\left|\cu_{n+1} \right|} \left\| \nabla   w \right\|_{L^2 \left( \cu_{n+1} \mu_\beta \right)}^2 & =  \frac{1}{|\cu_{n+1}|}\sum_{z \in \mathcal{Z}_n} \left\| \nabla   u(\cdot, z + \cu_n , p ) \right\|_{L^2 \left( z + \cu_{n}, \mu_\beta \right)}^2  + \frac{\left| BL_n \right|}{\left| \cu_{n+1} \right|} \\
    & \leq \frac{1}{|\cu_{n+1}|}\sum_{z \in \mathcal{Z}_n} \left\| \nabla   u( \cdot, z + \cu_n , p ) \right\|_{L^2 \left( z + \cu_n, \mu_\beta \right)}^2  + C 3^{-\frac n2}. \notag
\end{align}
To estimate the third term~\eqref{eq:splitEn+1}-(iii), we note that, by Remark~\ref{rmk:TV1414}, the boundary layer $BL_n$ has a width of order $c 3^{\frac n2}$, where $c$ is a universal constant. We split the sum over the integer $k$ at the value $c 3^{\frac n2}$ and write
\begin{equation} \label{eq:TV130209}
\sum_{ k   \geq 1} \frac1{\beta^{\frac k2}}\left\| \nabla^{k+1}   w \right\|_{L^2 \left( \Zd \mu_\beta \right)}^2 = 
\underbrace{\sum_{1 \leq k \leq  c3^{\frac n2}} \frac1{\beta^{\frac k2}}\left\| \nabla^{k+1}   w \right\|_{L^2 \left( \Zd, \mu_\beta \right)}^2}_{ \eqref{eq:TV130209}-(i)} + \underbrace{\sum_{k > c3^{\frac n2}} \frac1{\beta^{\frac k2}}\left\| \nabla^{k+1}   w \right\|_{L^2 \left( \Zd, \mu_\beta \right)}^2}_{\eqref{eq:TV130209}-(ii)},
\end{equation}
and estimate the two terms in the right side separately. For the term~\eqref{eq:TV130209}-(i), we use the following ingredients:
\begin{itemize}
    \item The boundary layer $BL_n$ has width $c 3^{\frac n2}$;
    \item For each integer $k \in \N$, the iterated gradient $\nabla^k$ has range $k$;
    \item For each integer $k \geq 2$, the $k$th-iterated gradient of the affine function $l_p$ is equal to $0$, i.e., $\nabla^k l_p =0$;
    \item The function $w$ is equal to the affine function $l_p$ in the boundary layer $BL_n$ and, for each point $z \in \mathcal{Z}_n$, is equal to the minimizer $u \left( \cdot , z + \cu_n , p \right)$ in the subcube $\left(z + \cu_n\right)$.
\end{itemize}
We obtain the identity
\begin{equation} \label{eq:TV16152}
    \sum_{1 \leq k \leq  c3^{\frac n2}} \frac1{\beta^{\frac k2}}\left\| \nabla^{k+1}   w \right\|_{L^2 \left( \Zd ,  \mu_\beta \right)}^2 = \sum_{z \in \mathcal{Z}_n} \sum_{1 \leq k \leq  c3^{\frac n2}} \frac1{\beta^{\frac k2}} \left\| \nabla^{k+1} u \left( \cdot , z + \cu_n , p \right) \right\|_{L^2 \left( \Zd ,  \mu_\beta \right)}^2.
\end{equation}
The term~\eqref{eq:TV130209}-(ii) is an error term which is small. Using that the the discrete gradient is a bounded operator which has range $k$ and that the function $w$ is equal to the affine function $l_p$ outside the cube $\cu_{n+1}$, we write
\begin{equation*}
    \sum_{k > c3^{\frac n2}} \frac1{\beta^{\frac k2}}\left\| \nabla^{k+1}   w \right\|_{L^2 \left( \Zd, \mu_\beta \right)}^2 \leq \sum_{k > c3^{-\frac n2}} \frac{C^k}{\beta^{\frac k2}} \left\| \nabla w \right\|_{L^2 \left( \cu_{n+1}, \mu_\beta \right)}^2 \leq \sum_{k > c3^{\frac n2}} \frac{C^k \left| \cu_{n+1} \right|}{\beta^{\frac k2}} \left\| \nabla w \right\|_{\underline{L}^2 \left( \cu_{n+1}, \mu_\beta \right)}^2.
\end{equation*}
We then use that the volume of the cube $\cu_{n+1}$ is smaller than the value $C3^{dn}$ and the upper bound $\left\| \nabla w \right\|_{\underline{L}^2 \left( \cu_{n+1}, \mu_\beta \right)}^2 \leq C$ which is a consequence of the upper bound~\eqref{eq:uppboundoptimi} and the the definition of the map $w$. This argument yields, if the inverse temperature $\beta$ is chosen large enough,
\begin{equation} \label{eq:TV16153}
    \sum_{k > c3^{\frac n2}} \frac1{\beta^{\frac k2}}\left\| \nabla^{k+1}   w \right\|_{L^2 \left( \Zd, \mu_\beta \right)}^2 \leq \left( \sum_{k > c3^{\frac n2}} \frac{C^k}{\beta^{\frac k2}} \right) 3^{dn} \leq  \left( \frac{C}{\sqrt{\beta}} \right)^{c3^{\frac n2}} 3^{dn} \leq C e^{-c \left( \ln \beta \right) 3^{- \frac n2}}.
\end{equation}
With the same argument, we obtain the estimate, for each point $z \in \mathcal{Z}_n$,
\begin{equation} \label{eq:TV16154}
    \sum_{k > c3^{\frac n2}} \frac{1}{\beta^{\frac k2}} \left\| \nabla^{k+1} u \left( \cdot , z + \cu_n , p \right) \right\|_{L^2 \left( \Zd , \mu_\beta \right)} \leq C e^{-c \left( \ln \beta \right) 3^{- \frac n2}}.
\end{equation}
By combining the estimates~\eqref{eq:TV16152},~\eqref{eq:TV16153} and~\eqref{eq:TV16154}, we have obtained the upper bound
\begin{equation} \label{eq:TVYK2037}
    \sum_{ k \geq 1} \frac1{\beta^{\frac k2}}\left\| \nabla^{k+1}   w \right\|_{L^2 \left( \Zd ,  \mu_\beta \right)}^2 \leq \sum_{z \in \mathcal{Z}_n} \sum_{ k \geq 1} \frac1{\beta^{\frac k2}}\left\| \nabla^{k+1} u \left( \cdot , z + \cu_n, p \right) \right\|_{L^2 \left( \Zd ,  \mu_\beta \right)}^2  + C e^{- c \left( \ln \beta \right) 3^{\frac n2}}.
\end{equation}
This completes the estimate of the term term~\eqref{eq:splitEn+1}-(iii).

The term~\eqref{eq:splitEn+1}-(iv) can be estimated similarly, we partition the set of charges $q$ whose support intersects the cube $\cu_{n+1}$ into three subsets:
\begin{itemize}
\item The set of charges whose support does not intersect any cube of the collection $(z + \cu_n)_{z \in \mathcal{Z}_n}$. We denote this set by $\mathcal{Q}_0$;
\item The set of charges $q$ whose support intersects exactly one cube of the collection $(z + \cu_n)_{z \in \mathcal{Z}_n}$. We denote this set by $\mathcal{Q}_1$ and for $z \in \mathcal{Z}_n$, we denote by $\mathcal{Q}_{1,z}$ the set of charges whose support only intersects the cube $\left( z + \cu_n \right)$;
\item The set of charges $q$ whose support intersects at least two cubes of the collection $(z + \cu_n)_{z \in \mathcal{Z}_n}$. We denote this set by $\mathcal{Q}_2$ and note that a charge belonging to this set must satisfy the property $\diam q \geq c 3^{- \frac n2}$.
\end{itemize} 
We then partition the sum
\begin{equation} \label{eq:TV091711}
\sum_{\supp q \cap \cu_{n+1} \neq \emptyset}  \left\langle \nabla_q  w\cdot \a_q  \nabla_q w   \right\rangle_{\mu_{\beta}} =  \underbrace{\sum_{ q \in \mathcal{Q}_0}  \left\langle \nabla_q  w\cdot \a_q  \nabla_q w   \right\rangle_{\mu_{\beta}}}_{\eqref{eq:TV091711}-(i)}  + \underbrace{\sum_{q \in \mathcal{Q}_1}  \left\langle \nabla_q  w\cdot \a_q  \nabla_q u \right\rangle_{\mu_{\beta}}}_{\eqref{eq:TV091711}-(ii)} + \underbrace{\sum_{q \in \mathcal{Q}_2}  \left\langle \nabla_q  w\cdot \a_q  \nabla_q w \right\rangle_{\mu_{\beta}}}_{\eqref{eq:TV091711}-(iii)}.
\end{equation}
We then estimate the three terms in the right side separately. For the term~\eqref{eq:TV091711}-(i), we have, by the definitions of the function $w$ and of the set $\mathcal{Q}_0$,
\begin{equation*}
    \sum_{ q \in \mathcal{Q}_0}  \left\langle \nabla_q  w\cdot \a_q  \nabla_q w   \right\rangle_{\mu_{\beta}} = \sum_{ q \in \mathcal{Q}_0}  \left\langle \nabla_q  l_p \cdot \a_q  \nabla_q l_p   \right\rangle_{\mu_{\beta}} \leq \sum_{q \in \mathcal{Q}_0} e^{- c \sqrt{\beta} \left\| q \right\|_1 } \left\| n_q \right\|_{L^1}^2.
\end{equation*}
Using that the volume of the boundary layer satisfies $\left| BL_n \right| \leq C 3^{-\frac n2} \left| \cu_{n+1} \right|$ and the estimate, for each $x \in \Zd$,
\begin{equation*}
    \sum_{q \in \mathcal{Q}_0} e^{- c \sqrt{\beta} \left\| q \right\|_1 } \left\| n_q \right\|_{L^1}^2 \indc_{\{ x \in \supp n_q \}}  \leq C, 
\end{equation*}
we obtain that
\begin{equation} \label{eq:TVYK13}
    \sum_{q \in \mathcal{Q}_0} e^{- c \sqrt{\beta} \left\| q \right\|_1 } \left\| n_q \right\|_{L^1}^2\leq C 3^{- \frac n2} \left| \cu_{n+1} \right|.
\end{equation}
We then estimate the term~\eqref{eq:TV091711}-(ii). By definition of the function $w$, we can write
\begin{equation} \label{eq:TVYK12}
    \sum_{ q \in \mathcal{Q}_1}  \left\langle \nabla_q  w\cdot \a_q  \nabla_q w   \right\rangle_{\mu_{\beta}} = \sum_{z \in \mathcal{Z}_n} \sum_{ q \in \mathcal{Q}_{1, z}}  \left\langle \nabla_q  u \left( \cdot, \cdot, z + \cu_n , p  \right)\cdot \a_q  \nabla_q u \left( \cdot, \cdot, z + \cu_n , p  \right)   \right\rangle_{\mu_{\beta}}.
\end{equation}
To estimate the term~\eqref{eq:TV091711}-(iii), we use the upper bound $ \left\| \nabla w \right\|_{
\underline{L}^2 \left( \cu_{n+1} , \mu_\beta \right)} \leq C |p|$ on the average $L^2$-norm of the gradient of $w$, the fact that the function $w$ is equal to the affine function $l_p$ outside the cube $\cu_{n+1}$, the estimate \eqref{e.aest} on the coefficients $\a_q$ and the property, for each charge $q \in \mathcal{Q}_3$, $\diam q \geq c 3^{\frac n2}$. We obtain
\begin{equation} \label{eq:TVYK11}
\left| \sum_{\supp q  \in \mathcal{Q}_3 }  \left\langle \nabla_q  w\cdot \a_q  \nabla_q w   \right\rangle_{\mu_{\beta}} \right| \leq e^{-c \sqrt{\beta} 3^{ \frac n2}}.
\end{equation}
With the same argument, we obtain the following estimate: for each $z \in \mathcal{Z}_n$,
\begin{equation} \label{eq:TVYK11111}
    \left| \sum_{\supp q \cap z + \cu_n \neq \emptyset, q \notin \mathcal{Q}_{1 , z}} \left\langle \nabla_q  u \left( \cdot , \cdot, z + \cu_n, p \right) \cdot \a_q  \nabla_q u \left( \cdot , \cdot, z + \cu_n, p \right)  \right\rangle_{\mu_{\beta}} \right| \leq e^{-c \sqrt{\beta} 3^{ \frac n2}}.
\end{equation}
Combining the identity~\eqref{eq:TV091711} with the inequalities~\eqref{eq:TVYK13},~\eqref{eq:TVYK12}~\eqref{eq:TVYK11} and~\eqref{eq:TVYK11111} shows the estimate
\begin{equation} \label{eq:TV203618}
\sum_{\supp q \cap \cu_{n+1} \neq \emptyset}  \left\langle \nabla_q  w\cdot \a_q  \nabla_q w \right\rangle_{\mu_{\beta}} \leq  \sum_{z \in \mathcal{Z}_n} \sum_{ \supp q \cap \left( z + \cu_n \right)}  \left\langle \nabla_q  u \left( \cdot, z + \cu_n , p  \right)\cdot \a_q  \nabla_q u \left( \cdot, z + \cu_n , p  \right)   \right\rangle_{\mu_{\beta}} + C 3^{-\frac n2} |\cu_{n+1}| .
\end{equation}
We finally combine the equality~\eqref{eq:splitEn+1}, the estimates~\eqref{eq:TVs21},~\eqref{eq:TVS22},~\eqref{eq:TVYK2037} ~\eqref{eq:TV203618} to obtain the inequality~\eqref{eq:estw2scale}. The proof of Proposition~\ref{p.subadd} is complete. 
\end{proof}

\subsection{Subadditivity for the energy \texorpdfstring{$\nu^*$}{9}}
In this section, we prove a similar statement for the energy~$\nu^*$.

\begin{proposition}[Subadditivity for $\nu^*$]  \label{p.subaddnustar}
There exists a constant $C := C(d) < \infty$ such that for each pair of integers $(n,m) \in \N$ such that $n> m$ and each vector $p^* \in \R^{d \times \binom d2}$,
\begin{equation} \label{eq:subaddnustar}
\frac{1}{\left| \mathcal{Z}_{m,n} \right|}\sum_{z \in \mathcal{Z}_{m,n}} \left\llbracket v(\cdot , \cdot, \cu_{n} , p^*) - v(\cdot , \cdot , z + \cu_{m} , p^*)  \right\rrbracket_{H^1(z + \cu_{m},\mu_\beta)}^2 \leq C \left( \nu^* \left( \cu_m , p \right) - \nu^* \left( \cu_{n} , p \right)  +  3^{-\frac n2}|p^*|^2 \right).
\end{equation}
\end{proposition}

As it was the case for the energy quantity $\nu$, we deduce from Proposition~\ref{p.subaddnustar} that the sequence $\left( \nu^* \left( \cu_n , p^* \right) \right)_{n \in \N}$ converges as $n$ tends to infinity.

\begin{corollary} \label{cor:coro5.15}
There exists an inverse temperature $\beta_0 := \beta_0 \left( d \right) < \infty$ such that for each $\beta \geq \beta_0$ the following statement is valid. There exists a non-negative real number $\ahom_*$ such that for each vector $p^* \in \R^{d \times \binom d2}$, one has
\begin{equation*}
    \nu^* \left( \cu_n , p^* \right) \underset{n \to \infty}{\longrightarrow} \ahom_*^{-1} \left| p^* \right|^2.
\end{equation*}
By the Property (3) of Proposition~\ref{prop:prop5.2}, this statement can be rewritten equivalently
\begin{equation*}
    \a_* \left( \cu_n \right)^{-1}  \underset{n \to \infty}{\longrightarrow} \ahom_*^{-1}.
\end{equation*}
We also have the lower bound, for each integer $n \in \N$,
\begin{equation*}
    \a_* \left( \cu_n \right)^{-1} \geq \ahom_*^{-1} - C 3^{-\frac n2}.
\end{equation*}
\end{corollary}

\begin{proof}[Proof of Proposition~\ref{p.subaddnustar}]
For the sake of simplicity, we only write the proof in the case when the difference between the integers $m$ and $n$ is equal to $1$. We consider the specific case of the pair $(n +1 , n)$. The proof of the general case is similar. We assume without loss of generality that $\left| p^* \right| = 1$.

We consider the function $v := v \left( \cdot , \cu_{n+1} , p^* \right)$ and, for $z \in \mathcal{Z}_n$, we restrict it to the cubes~$\left(z + \cu_n \right)$. We apply the second variation formula~\eqref{eq.secondvarnustar} and the coercivity of the energy functional $\mathbf{E}_{z + \cu_n}^*$. We obtain, for each point $z \in \mathcal{Z}_n$,
\begin{equation*}
\left\llbracket v(\cdot , \cu_{n+1} , p^*) - v(\cdot ,z + \cu_{n} , p^*)  \right\rrbracket_{\underline{H}^1( z + \cu_{n},\mu_\beta)}^2 \leq C \left( \nu^* \left( z+  \cu_n , p^* \right)  + \frac 1{2|\cu_n|} \mathbf{E}_{z + \cu_n}^* \left[ v \right] + \frac{1}{| \cu_n|}  \sum_{x \in z + \cu} p^* \cdot \left\langle \nabla v(x) \right\rangle_{\mu_\beta} \right).
\end{equation*}
Summing over the points $z \in \mathcal{Z}_n$ and dividing by the cardinality of $\mathcal{Z}_n$ shows
\begin{multline*}
\frac{1}{\left| \mathcal{Z}_n \right|}\sum_{z \in \mathcal{Z}_n} \left\llbracket v(\cdot , \cu_{n+1} , p^*) - v(\cdot ,z + \cu_{n} , p^*)  \right\rrbracket_{\underline{H}^1( z + \cu_{n},\mu_\beta)}^2  \\ \leq C \left(\nu^* \left( \cu_n , p^* \right)  + \sum_{z \in \mathcal{Z}_n} \frac 1{2\left|\mathcal{Z}_n \right| \cdot \left|\cu_n \right|} \mathbf{E}_{z + \cu_n}^* \left[ v \right]  +  \frac 1{\left|\mathcal{Z}_n \right| \cdot \left|\cu_n \right|}  \sum_{x \in \cu_n} p^* \cdot \left\langle \nabla v(x) \right\rangle_{\mu_\beta}\right).
\end{multline*}
The factor $\left| \mathcal{Z}_n \right| = 3^d$ on the left side depends only on the dimension $d$ and can thus be absorbed in the constant $C$ in the right side. We deduce that, to prove the inequality~\eqref{eq:subaddnustar}, it is sufficient to prove
\begin{multline} \label{eq:TV1654}
 \sum_{z \in \mathcal{Z}_n} \frac 1{\left|\mathcal{Z}_n \right| \cdot \left|\cu_n \right|} \left( \frac 12\mathbf{E}_{z + \cu_n}^* \left[ v \right]  +  \sum_{x \in z +\cu_n} p^* \cdot \left\langle \nabla v(x) \right\rangle_{\mu_\beta}\right) \\ \leq  \underbrace{\frac{1}{2|\cu_{n+1}|} \mathbf{E}_{\cu_{n+1}}^* \left[ v \right]}_{\eqref{eq:TV1654}-(i)} + \underbrace{\frac{1}{|\cu_{n+1}|}\sum_{x \in \cu_{n+1}} p^* \cdot \left\langle \nabla v(x) \right\rangle_{\mu_\beta}}_{\eqref{eq:TV1654}-(ii)} +  C 3^{-\frac n2}.
\end{multline}
We first estimate the term~\eqref{eq:TV1654}-(ii). We use the estimate~\eqref{eq:uppboundoptimi} on the $L^2$-norm of the gradient of the function~$v$, the Cauchy-Schwarz inequality and the volume estimate 
\begin{equation} \label{eq:TVvolest1625}
\left| \cu_{n+1} \right| - \left| \mathcal{Z}_n \right| \cdot \left| \cu_n \right| = \left| \cu_{n+1}  \setminus \bigcup_{z \in \mathcal{Z}_n } \left( z+ \cu_n \right) \right| = \left| BL_n \right| \leq C 3^{-\frac n2} \left| \cu_{n+1} \right|.
\end{equation}
We obtain
\begin{align} \label{eq:TV1653.2}
\sum_{z \in \mathcal{Z}_n} \frac 1{\left|\mathcal{Z}_n \right| \cdot \left|\cu_n \right|}  \sum_{x \in z +\cu_n} p^* \cdot \left\langle \nabla v(x, \cdot ) \right\rangle_{\mu_\beta} & \leq \frac 1{\left|\cu_{n+1} \right|} \sum_{z \in \mathcal{Z}_n} \sum_{x \in z +\cu_n} p^* \cdot \left\langle \nabla v(x) \right\rangle_{\mu_\beta} + \left(\frac{\left|BL_n \right| }{\left|\cu_{n+1} \right|} \right)^{\frac12} \left\| \nabla v\right\|_{\underline{L}^2 \left( \cu_{n+1}, \mu_\beta \right)} \\
				& \leq  \frac 1{\left|\cu_{n+1} \right|}  \sum_{x \in \cu_{n+1}} p^* \cdot \left\langle \nabla v(x, \cdot ) \right\rangle_{\mu_\beta} + \frac 1{\left|\cu_{n+1} \right|}  \sum_{x \in BL_n} p^* \cdot \left\langle \nabla v(x, \cdot ) \right\rangle_{\mu_\beta} + C3^{-\frac n4} \notag\\
				& \leq \frac 1{\left|\cu_{n+1} \right|}  \sum_{x \in \cu_{n+1}} p^* \cdot \left\langle \nabla v(x, \cdot ) \right\rangle_{\mu_\beta} + \left( \frac{|BL_n|}{|\cu_{n+1}|} \right)^{\frac 12} \left\| \nabla v\right\|_{\underline{L}^2 \left( \cu_{n+1}, \mu_\beta \right)}  + C 3^{- \frac n4} \notag \\
				& \leq  \frac 1{\left|\cu_{n+1} \right|}  \sum_{x \in \cu_{n+1}} p^* \cdot \left\langle \nabla v(x, \cdot) \right\rangle_{\mu_\beta}  +  C 3^{- \frac n4}. \notag
\end{align}
To estimate the term pertaining to the energy functional $\mathbf{E}^*$, we introduce two notations:
\begin{itemize}
    \item We let $\mathcal{Q}_{n+1}$ the set of charges whose support is included in the cube $\cu_{n+1}$ and intersects the boundary layer $BL_n$;
    \item For each integer $k \in \N$, we let $C_k$ be the set
    \begin{equation*}
        C_k := \left\{ x \in \cu_{n+1} \, : \, B(x , k) \subseteq \cu_{n+1}~\mbox{and}~
        B(x , k) \not\subseteq \bigcup_{z \in \mathcal{Z}_n} \left( z + \cu_n \right) \right\}.
    \end{equation*}
\end{itemize} 
We note that we have the inclusion of the boundary layers $\cu_{n+1} \setminus \cu_{n+1}^- \subseteq BL_n$. Thus, by the definition of the energy functional $\mathbf{E}_{\cu_{n+1}}^*$, we have the inequality
\begin{multline*}
    \mathbf{E}_{\cu_{n+1}}^* \left[ v \right]  \geq  \beta \sum_{y \in \Zd} \left\| \partial_y v \right\|_{L^2 \left( \cu , \mu_\beta \right)}^2 +  \left( \frac1{2} - \frac{1}{\beta^{\frac 14}} \right)  \left\| \nabla  v \right\|_{L^2 \left( BL_n, \mu_\beta \right)}^2 + \frac1{2} \sum_{z \in \mathcal{Z}_n} \left\| \nabla  v \right\|_{L^2 \left(z + \cu_n , \mu_\beta \right)}^2\\
    + \frac1{2}  \sum_{n \geq 1} \sum_{\dist \left( x , \partial \cu \right) \geq n} \frac1{\beta^{\frac n2}}\left\| \nabla^{n+1}  v(x , \cdot ) \right\|_{L^2 \left( \mu_\beta \right)}^2 - \beta \sum_{\supp q \subseteq \cu}  \left\langle \nabla_q v \cdot \a_q  \nabla_q v \right\rangle_{\mu_{\beta}}.
\end{multline*}
We choose the inverse temperature $\beta$ enough so that, we have the estimate $\frac1{2} - \beta^{-\frac 14} \geq \frac1{4}$. We recall the definition of the set $A_n$ stated in~\eqref{eq:TVdefAn} and use the definition of the energy $\mathbf{E}_{ z + \cu_{n}}^*$ to obtain the estimate
\begin{align} \label{eq:TV12481}
\mathbf{E}_{\cu_{n+1}}^* \left[ v \right] - \sum_{z \in \mathcal{Z}_n} \mathbf{E}_{z + \cu_n}^* \left[ v \right] & \geq \beta \sum_{y \in \Zd} \left\| \partial_y v  \right\|_{L^2 \left(BL_n, \mu_\beta \right)}^2 + \frac1{4}  \left\| \nabla   v \right\|_{L^2 \left(BL_n, \mu_\beta \right)}^2  +  \frac{1}{\beta^{\frac 14}}\left\| \nabla v \right\|_{L^2 \left( A_n \mu_\beta\right)}^2 \\
    & \quad  - \beta \sum_{\supp q \in \mathcal{Q}_{n+1}}  \left\langle \nabla_q v \cdot \a_q  \nabla_q v   \right\rangle_{\mu_{\beta}} + \frac{1}{2}  \sum_{k \geq 1}  \frac1{\beta^{\frac k2}}\left\| \nabla^{k+1}   v \right\|_{L^2 \left(C_k, \mu_\beta \right)}^2 . \notag \\
    & \geq \frac1{4}  \left\| \nabla   v \right\|_{L^2 \left(BL_n, \mu_\beta \right)}^2  +  \frac{1}{\beta^{\frac 14}}\left\| \nabla v \right\|_{L^2 \left( A_n \mu_\beta\right)}^2 \notag \\
    &\quad  - \underbrace{ \beta \sum_{\supp q \in \mathcal{Q}_{n+1}}  \left\langle \nabla_q v \cdot \a_q  \nabla_q v   \right\rangle_{\mu_{\beta}}}_{\eqref{eq:TV12481}-(i)} + \underbrace{\frac{1}{2}  \sum_{k \geq 1}  \frac1{\beta^{\frac k2}}\left\| \nabla^{k+1}   v \right\|_{L^2 \left(C_k, \mu_\beta \right)}^2}_{\eqref{eq:TV12481}-(ii)} . \notag
\end{align}
We first estimate the term~\eqref{eq:TV12481}-(i). To this end, we partition the set $\mathcal{Q}_{n+1}$ into two sets of charges $\mathcal{Q}_{n+1,1}$ and $\mathcal{Q}_{n+1,2}$ defined by the formulas
\begin{equation*}
\mathcal{Q}_{n+1,1} := \left\{ q \in \mathcal{Q}_{n+1} ~:~ \supp q \subseteq BL_n \cup A_n \right\} ~\mbox{and}~ \mathcal{Q}_{n+1,2} := \mathcal{Q}_{n+1} \setminus \mathcal{Q}_{n+1,1}.
\end{equation*}
We then use the two following arguments.
First, if a charge belongs to the set $\mathcal{Q}_{n+1,2}$, then its diameter has to be larger larger than $c 3^{-\frac n2}$. By the estimate $\left|\a_q \right| \leq C e^{-c \sqrt{\beta} \left\| q \right\|_1}$, we obtain
\begin{align*}
 \left| \sum_{ q \in \mathcal{Q}_{n+1,2}}  \left\langle \nabla_q  v\cdot \a_q  \nabla_q v   \right\rangle_{\mu_{\beta}} \right| & \leq  \sum_{ q \in \mathcal{Q}_{n+1,2}} e^{- c \sqrt{\beta} \left\| q\right\|_1} \left\| n_q\right\|_2^2 \left\| \nabla v \right\|_{L^2 \left( \supp n_q , \mu_\beta \right)}^2  \\ &
  \leq e^{- c \sqrt{\beta} 3^{\frac n2}} \left\| \nabla v \right\|_{L^2 \left( \cu_{n+1} , \mu_\beta \right)}^2 .
\end{align*}
We then use the estimate~\eqref{eq:uppboundoptimi} to bound the $L^2$-norm of the gradient of $v$. We obtain
\begin{equation} \label{eq:TV15311}
 \left| \sum_{\supp q \in \mathcal{Q}_{n+1,2}}  \left\langle \nabla_q  v \cdot \a_q  \nabla_q v   \right\rangle_{\mu_{\beta}} \right| \leq C \left| \cu_{n+1}\right| e^{- c 3^{\frac n2}} \leq C  e^{- c \sqrt{\beta} 3^{\frac n2}},
\end{equation}
by reducing the value of the constant $c$ in the second inequality.

Second, for the charges belonging to the set $\mathcal{Q}_{n+1,1}$, using the estimate $\left| \a_q \right| \leq C e^{-c \sqrt{\beta} \left\| q \right\|_1}$, we have
\begin{equation} \label{eq:TV153122}
\left| \sum_{q \in \mathcal{Q}_{n+1,1}}  \left\langle \nabla_q  v\cdot \a_q  \nabla_q v   \right\rangle_{\mu_{\beta}} \right| \leq C e^{- c \sqrt{\beta}} \left\| \nabla v \right\|_{L^2 \left(BL_n \cup A_n ,  \mu_\beta\right)}^2.
\end{equation}

By combining~\eqref{eq:TV15311} and~\eqref{eq:TV153122}, we obtain
\begin{equation} \label{eq:TV15313}
    \left| \sum_{q \in \mathcal{Q}}  \left\langle \nabla_q  v\cdot \a_q  \nabla_q v   \right\rangle_{\mu_{\beta}} \right| \leq  C e^{- c \sqrt{\beta}} \left\| \nabla v \right\|_{L^2 \left(BL_n \cup A_n ,  \mu_\beta\right)}^2 + C  e^{- c \sqrt{\beta} 3^{\frac n2}}.
\end{equation}

The term~\eqref{eq:TV12481}-(ii) can be estimated with similar arguments: we need to decompose over the integers $k$ such that $C_k \subseteq BL_n \cup A_n$ and the integers $k$ such that $C_k \not\subseteq BL_n \cup A_n$. Since the argument is almost the same, we omit it and only give the result. We obtain the inequality
\begin{equation}  \label{eq:TV15314}
    \frac{1}{2}  \sum_{k \geq 1}  \frac1{\beta^{\frac k2}}\left\| \nabla^k   v \right\|_{L^2 \left(C_k, \mu_\beta \right)}^2 \leq \frac{C}{\beta^{\frac 12}} \left\| \nabla v \right\|_{L^2 \left( BL_n \cup A_n, \mu_\beta \right)} + C e^{- c \left( \ln \beta \right) 3^{\frac{n}{2}}}.
\end{equation}
We now combine the estimates~\eqref{eq:TV15313} and~\eqref{eq:TV15314} and deduce that
\begin{equation*}
    \left| \mbox{\eqref{eq:TV12481}-(i)} \right| + \left|\mbox{\eqref{eq:TV12481}-(ii)}\right| \leq \left( \frac{C}{\beta^{\frac 12}} + C e^{- c \sqrt{\beta}} \right) \left\| \nabla v \right\|_{L^2 \left( BL_n \cup A_n, \mu_\beta \right)} + C \left(  e^{- c \sqrt{\beta} 3^{\frac n2}}+ e^{- c \left( \ln \beta \right) 3^{\frac{n}{2}}}\right).
\end{equation*}
As a consequence, if $\beta$ is chosen large enough depending only on the dimension $d$,  then we have
\begin{equation} \label{eq:TBB1547}
\left|\mbox{\eqref{eq:TV12481}-(i)}\right| + \left|\mbox{\eqref{eq:TV12481}-(ii)}\right| \leq \frac1{2}  \left\| \nabla   v \right\|_{L^2 \left(BL_n, \mu_\beta \right)}^2  +  \frac{1}{\beta^{\frac 12}}\left\| \nabla v \right\|_{L^2 \left( A_n \mu_\beta\right)}^2 +  Ce^{- c \left( \ln \beta \right) 3^{\frac{n}{2}}}.
\end{equation}
Combining the estimates~\eqref{eq:TV12481} and~\eqref{eq:TBB1547} shows
\begin{equation*}
\mathbf{E}_{\cu_{n+1}} \left[ v \right] - \sum_{z \in \mathcal{Z}_n} \mathbf{E}_{z + \cu_n} \left[ v \right]  \geq - C e^{- c \left( \ln \beta \right) 3^{\frac{n}{2}}}.
\end{equation*}
Dividing the previous display by $2\left|\mathcal{Z}_n \right| \cdot \left|\cu_n \right|$ shows the estimate
\begin{equation} \label{eq:TV1652}
 \sum_{z \in \mathcal{Z}_n} \frac 1{2\left|\mathcal{Z}_n \right| \cdot \left|\cu_n \right|} \mathbf{E}_{z + \cu_n}[v]  \leq \frac 1{2 \left|\mathcal{Z}_n \right| \cdot \left|\cu_n \right|} \mathbf{E}_{\cu_{n+1}}[v] + C e^{- c 3^{\frac n2}}.
\end{equation}
We use the volume estimate~\eqref{eq:TVvolest1625}, the bound on the average $L^2$-norm of the gradient of $v$ stated in~\eqref{eq:uppboundoptimi} and the coercivity of the energy functional $\mathbf{E}^*$ stated in~\eqref{eq:coerccontenergystar} to deduce
\begin{equation} \label{eq:TV1652.5}
 \frac 1{2\left|\mathcal{Z}_n \right| \cdot \left|\cu_n \right|} \mathbf{E}_{\cu_{n+1}}[v] \leq \frac 1{2\left| \cu_{n+1} \right|} \mathbf{E}_{\cu_{n+1}}[v]  +  C 3^{-\frac n2}.
\end{equation}
Combining the estimates~\eqref{eq:TV1652},~\eqref{eq:TV1652.5} and~\eqref{eq:TV1653.2} shows the inequality~\eqref{eq:TV1654} and completes the proof of Proposition~\ref{p.subaddnustar}.
\end{proof}

\section{Quantitative convergence of the subadditive quantities} \label{section5.3}

In this section, we prove an algebraic rate of convergence for the quantity $J$ defined in~\eqref{def:defJquan}. We recall the definition of the subadditivity defect $\tau_n$ given in~\eqref{def.taun} and we introduce the following notation: for each integer $n \in \N$, we denote by
\begin{equation} \label{defapproxhomogcoeff}
    \ahom_n := \a_*\left( \cu_n \right),
\end{equation}
and call the matrix $\ahom_n$ the \emph{approximate homogenized matrix}. We first prove a series of lemmas, estimating various quantities in terms of the subadditivity defect $\tau_n$ following the strategy described in Section~\ref{sec:Chap6statmainres1014}. 

Before starting the proofs, let us make the following remark: By Corollaries~\ref{cor:coro5.13} and~\ref{cor:coro5.15}, the subadditivity defect $\tau_n$ converges to $0$ as $n$ tends to infinity. In particular all the quantities which are bounded from above by the subadditivity defect $\tau_n$ tend to $0$ when $n$ tends to infinity.

\subsection{Control over the approximate homogenized coefficient}
The first lemma we prove establishes that the difference between the coefficients $\ahom_n$ over two different scales can be estimated in terms of the subadditivity defect $\tau_n$. 

\begin{lemma} \label{lem:lemma5.6}
There exists a constant $C := C(d) < \infty$ such that for any pair of integers $(m,n) \in \N^2$ with $m \leq n$, the following estimate holds
\begin{equation*}
\left| \ahom^{-1}_n  -\ahom^{-1}_m  \right|^2 \leq \sum_{k=m}^n \tau_k + C 3^{-\frac{m}2}.
\end{equation*}
\end{lemma}

\begin{proof}
Before starting the proof, we collect a few ingredients and notations used in the argument:
\begin{itemize}
    \item We recall the notation $O$ introduced in Section~\ref{SecNotandprelim} of Chapter~\ref{Chap:chap2}, given two real numbers $X , Y$ and a non-negative real number $\kappa$, we write $X = Y + O\left( \kappa \right)$ if and only if $|X - Y| \leq \kappa$; \smallskip
    \item By the formula~\eqref{eq:derivpnustar}, we have the identity $\sum_{x \in \cu_n} \left\langle \nabla v(x, \cdot , \cu_n , p^*) \right\rangle_{\mu_\beta} = \a_n^{-1} p^*$; \smallskip
    \item We recall the definition of the set $\mathcal{Z}_{m,n}  := l_{n} 3^{m-n} \Zd \cap \cu_{n}$, the definition of the boundary layer $BL_{m,n}$ given in Definition~\ref{def:deftriadic} and the volume estimate $\left| BL_{m,n} \right| \leq C 3^{- \frac m2} \left| \cu_n \right|$ stated in Remark~\ref{rem:volest}; \smallskip
    \item By definition of the subadditivity defect $\tau_k$, we have the identity, for each $p \in \R^{d \binom d2}$,
    \begin{equation*}
        \nu^* \left( \cu_{m} , p \right) - \nu^* \left( \cu_n , p \right) \leq |p|^2 \sum_{k = m}^n \tau_k.
    \end{equation*}
\end{itemize}
We fix a vector $p^* \in \R^{d \binom d2}$ such that $\left| p^*\right| = 1$ and use the formula~\eqref{eq:derivpnustar} to write
\begin{align} \label{eq:TV18195}
 \ahom_n^{-1} p^* & = \frac{1}{|\cu_n|}\sum_{x \in \cu_n} \left\langle \nabla v(x, \cdot , \cu_n , p^*) \right\rangle_{\mu_\beta} \\ & = \underbrace{ \frac{1}{|\cu_n|}\sum_{z \in \mathcal{Z}_{m,n}} \sum_{x \in z +\cu_m} \left\langle \nabla v(x, \cdot , \cu_n , p^*) \right\rangle_{\mu_\beta}}_{\eqref{eq:TV18195}-(i)} +   \underbrace{\frac{1}{|\cu_n|} \sum_{x \in BL_{m,n}} \left\langle \nabla v(x, \cdot , \cu_n , p^*) \right\rangle_{\mu_\beta}}_{\eqref{eq:TV18195}-(ii)}. \notag
\end{align}
The term~\eqref{eq:TV18195}-(ii) is the simplest one, we estimate it by the Cauchy-Schwarz inequality, the estimate on the $L^2$-norm of the gradient of $v$ stated in~\eqref{eq:uppboundoptimi} and the volume estimate $\left| BL_{m,n} \right| \leq C 3^{- \frac m2} \left| \cu_n \right|$. We obtain
\begin{equation} \label{eq:TV11447}
 \left| \frac{1}{\left| \cu_n \right|} \sum_{x \in BL_{m,n}} \left\langle \nabla v(x, \cdot , \cu_n , p^*) \right\rangle_{\mu_\beta} \right| \leq C3^{-\frac{m}2}.
\end{equation} 
To estimate the term~\eqref{eq:TV18195}-(i), we use the estimate~\eqref{eq:uppboundoptimi}, the identity $BL_{m,n} = \cu_n \setminus \bigcup_{z \in \mathcal{Z}_{m,n}}$ and the volume estimate~$\left| BL_{m,n} \right| \leq C 3^{- \frac m2} \left| \cu_n \right|$. We obtain
\begin{equation*}
 \frac{1}{|\cu_n|} \sum_{z \in \mathcal{Z}_n} \sum_{x \in z +\cu_m} \left\langle \nabla v(x, \cdot , \cu_n , p^*) \right\rangle_{\mu_\beta} = \frac{1}{\left| \mathcal{Z}_{m,n}\right|} \frac{1}{\left| z + \cu_m \right|} \sum_{z \in \mathcal{Z}_n}   \sum_{x \in z +\cu_m} \left\langle \nabla v(x, \cdot , \cu_n , p^*) \right\rangle_{\mu_\beta} + O \left( C 3^{-\frac m2} \right).
\end{equation*}
Applying the subadditivity estimate stated in Proposition~\ref{p.subadd}, we find that
\begin{align} \label{eq:TV111987}
     \lefteqn{ \frac{1}{\left| \mathcal{Z}_{m,n}\right|} \frac{1}{\left| z + \cu_m \right|} \sum_{z \in \mathcal{Z}_n}   \sum_{x \in z +\cu_m} \left| \left\langle \nabla v(x, \cdot , \cu_n , p^*) - \nabla v(x, \cdot , z + \cu_m , p^*) \right\rangle_{\mu_\beta} \right|} \qquad & \\ &
     \leq \frac{1}{\left| \mathcal{Z}_{m,n}\right|} \sum_{z \in \mathcal{Z}_{m,n}} \left\| \nabla v \left( \cdot , \cu_n, p^* \right) - \nabla v \left(  \cdot , z + \cu_m , p^* \right) \right\|_{\underline{L}^2 \left( z + \cu_m , \mu_\beta \right)} \notag \\
     & \leq \left( \frac{1}{\left| \mathcal{Z}_{m,n}\right|} \sum_{z \in \mathcal{Z}_{m,n}} \left\| \nabla v \left( \cdot , \cu_n, p^* \right) - \nabla v \left(  \cdot , z + \cu_m , p^* \right) \right\|_{\underline{L}^2 \left( z + \cu_m , \mu_\beta \right)}^2 \right)^{\frac12} \notag  \\
     & \leq  \left( \frac{1}{\left| \mathcal{Z}_{m,n}\right|} \sum_{z \in \mathcal{Z}_{m,n}} \left\llbracket \nabla v \left( \cdot , \cu_n, p^* \right) - \nabla v \left(  \cdot , z + \cu_m , p^* \right) \right\rrbracket_{\underline{H}^1 \left( z + \cu_m , \mu_\beta \right)}^2 \right)^{\frac12} \notag \\
     & \leq C \left( \sum_{k = m}^n \tau_k \right)^\frac 12 + C 3^{-\frac m2}. \notag
\end{align}
We use the inequality~\eqref{eq:TV111987}, the translation invariance of the measure $\mu_\beta$ and the identity $\sum_{x \in \cu_n} \left\langle \nabla v(x, \cdot , \cu_m , p^*) \right\rangle_{\mu_\beta} = \ahom_m^{-1} p^*$. We obtain
\begin{align} \label{eq:TV114271}
  \lefteqn{ \frac{1}{\left| \mathcal{Z}_{m,n}\right|} \frac{1}{\left| z + \cu_m \right|} \sum_{z \in \mathcal{Z}_n} \sum_{x \in z +\cu_m} \left\langle \nabla v(x, \cdot , \cu_n , p^*) \right\rangle_{\mu_\beta}} \qquad & \\ & =  \frac{1}{\left| \mathcal{Z}_{m,n}\right|}  \sum_{z \in \mathcal{Z}_{m,n}}  \frac{1}{\left| z + \cu_m \right|} \sum_{x \in z +\cu_m} \left\langle \nabla v(x, \cdot , z + \cu_m , p^*) \right\rangle_{\mu_\beta} +  O \left( C \left( \sum_{k=m}^n \tau_k \right)^\frac 12 + C 3^{-\frac m2}  \right) \notag \\
  &=  \frac{1}{\left|\cu_m \right|} \sum_{x \in \cu_m} \left\langle \nabla v(x, \cdot , \cu_m , p^*) \right\rangle_{\mu_\beta} + O \left( C \left( \sum_{k=m}^n \tau_k \right)^\frac 12 + C 3^{-\frac m2} \right) \notag \\
  &= \ahom_m^{-1} p^* +  O \left( C \left( \sum_{k=m}^n \tau_k \right)^\frac 12 + C 3^{-\frac m2}  \right). \notag
\end{align}
We then combine the identity~\eqref{eq:TV18195} with the estimates~\eqref{eq:TV11447} and~\eqref{eq:TV114271} to complete the proof of Lemma~\ref{lem:lemma5.6}.
\end{proof}

\subsection{Control over the variance of the spatial average of the maximizer \texorpdfstring{$v$}{10}}
The next step in the argument is to control the variance of the spatial average of the maximiser $v$. We prove that its variance contracts and obtain an algebraic rate of convergence. The proof relies on an explicit computation and makes use of the differentiated Helffer-Sj\"{o}strand equation introduced in Section~\ref{secchap3HSPDE} of Chapter~\ref{chap:chap3} to estimate the correlation between the random variables $ \phi \mapsto v(x , \phi , \cu_{n+1} , p)$ and $ \phi \mapsto v(x' , \phi , \cu_{n+1} , p)$ for a pair of points $x , x' \in \cu_{n+1}$ distant from one another.

\begin{lemma}[Variance estimate] \label{lemmvarest}
There exists a constant $C := C(d) < \infty$ such that for each $n \in \N$ and each $p^* \in \R^{d \times \binom d2}$,
\begin{equation} \label{eq:TV5.55}
\var_{\mu_\beta} \left[ \frac 1{|\cu_n|} \sum_{x \in \cu_n} \nabla v(x , \cdot , \cu_{n+1} , p^*) \right] \leq C 3^{-\left(d-\frac{5}{2}\right) n} |p^*|^2.
\end{equation}
For later purposes, we also record that the variance of the flux contracts
\begin{equation} \label{eq:TV09200}
    \var \left[ \frac 1{|\cu_n|} \sum_{x \in \cu_n}  \left( \frac 1{2} \nabla v \left( x , \cdot, \cu_{n+1} , p^* \right) + \beta \sum_{ q \in \mathcal{Q}} \a_q \nabla_q  v \left(\cdot, \cdot, \cu_{n+1} , p^* \right) n_q(x) \right) \right] \leq   C 3^{-\left(d-\frac{5}{2}\right) n} |p^*|^2.
\end{equation}
\end{lemma}

\begin{remark}
    The value of the coefficient $d - \frac{5}{2}$ is arbitrary; we can prove the result for any fixed number strictly smaller than $d - 2$ by choosing $\beta$ large enough accordingly.
\end{remark}

\begin{remark}
    The argument presented in the proof below can be adapted to prove the variance estimate, for each point $x \in \cu_n$,
    \begin{equation*}
        \var_{\mu_\beta} \left[  \nabla v(x , \cdot , \cu_{n+1} , p^*) \right] \leq C.
    \end{equation*}
    Since this estimate is an \emph{a priori} estimate, we use it in Appendix~\ref{app.appB} to prove the solvability of the Neumann problem.
\end{remark}

\begin{proof}
We fix an inverse temperature $\beta$ large enough so that all the regularity results of Chapter~\ref{section:section4} hold with the regularity exponent $\ep = \frac 12$. We decompose the argument into two steps. 

\smallskip

\textit{Step 1.} To ease the notation, we denote by $v:= v \left( \cdot ,\cdot, \cu_{n+1} , p^*\right)$. We assume without loss of generality that $|p^*|=1$. We first decompose the variance
\begin{equation} \label{eq:TV15547}
    \var_{\mu_\beta} \left[ \frac 1{|\cu_n|} \sum_{x \in \cu_n} \nabla v(x , \cdot) \right] = \frac{1}{\left| \cu_n \right|^2} \sum_{x , x' \in  \cu_n} \cov_{\mu_\beta} \left[ \nabla v \left( x, \cdot \right), \nabla v \left( x', \cdot \right) \right].
\end{equation}
We then prove the estimate, for each pair of points $x , x' \in \cu_n$,
\begin{equation} \label{eq:TV12517}
    \left|\cov_{\mu_\beta} \left[ \nabla v \left( x, \cdot \right), \nabla v \left( x', \cdot \right) \right] \right| \leq \frac{C3^{\frac{n}{2}}}{|x - x'|^{d-2}}.
\end{equation}
The estimate~\eqref{eq:TV5.55} can then be deduced from~\eqref{eq:TV12517} and~\eqref{eq:TV15547}; indeed we have
\begin{align*}
    \var_{\mu_\beta} \left[ \frac 1{|\cu_n|} \sum_{x \in \cu_n} \nabla v(x , \cdot) \right] &\leq \frac{1}{\left| \cu_n \right|^2} \sum_{x , x' \in  \cu_n} \cov_{\mu_\beta} \left[ \nabla v \left( x, \cdot \right), \nabla v \left( x', \cdot \right) \right] \\
    & \leq \frac{C3^{\frac{n}{2}}}{\left| \cu_n \right|^2} \sum_{x , x' \in  \cu_n} \frac{1}{|x - x'|^{d-2}}\\
    & \leq C 3^{-\left(d-\frac{5}{2}\right)n}.
\end{align*}
We now fix two points $x, x' \in \cu_n$ and focus on the proof of ~\eqref{eq:TV12517}. By applying the Helffer-Sj{\"o}strand formula, we write
\begin{equation} \label{eq:covXzXz'}
    \cov_{\mu_\beta} \left[ \nabla v \left( x, \cdot \right), \nabla v \left( x', \cdot\right) \right] = \sum_{y \in \Zd} \left\langle \partial_{y}  \nabla v \left( x, \cdot \right) \mathcal{H}_{x'}(y, \cdot) \right\rangle_{\mu_\beta},
\end{equation}
where $\mathcal{H}_{z'}$ is the solution of the Helffer-Sj{\"o}strand equation, for each $(y , \phi) \in \Zd \times \Omega$,
\begin{equation*} 
    \mathcal{L} \mathcal{H}_{x'}(y , \phi) = \partial_y \nabla v \left( x', \phi \right). 
\end{equation*}
We then decompose the function $\mathcal{H}_{x'}$ according to the collection of Green's matrices $\left( \mathcal{G}_{\partial_y \nabla v \left( x', \cdot \right)}\right)_{y \in \Zd}$, following the notation introduced in~\eqref{eq:formulaGreens} of Section~\ref{sec:section4.4} of Chapter~\ref{section:section4}. We obtain
\begin{equation*}
    \mathcal{H}_{x'}(y , \phi) = \sum_{y' \in \Zd} \mathcal{G}_{\partial_{y'}\nabla v \left( x', \cdot \right)} \left( y , \phi ;y' \right).
\end{equation*}
Using Proposition~\ref{prop.prop4.11chap4} of Chapter~\ref{chap:chap3}, we can estimate the $L^2(\mu_\beta)$-norm of the function $\mathcal{H}_{x'}$, for each point $y \in \Zd$,
\begin{equation} \label{eq:TV12607}
    \left\| \mathcal{H}_{x'}\left( y , \cdot \right) \right\|_{L^2 \left( \mu_\beta \right)} \leq C \sum_{y' \in \Zd} \frac{\left\| \partial_{y'} \nabla v \left( x', \cdot \right) \right\|_{L^2 \left( \mu_\beta \right)}}{|y - y'|^{d-2}}.
\end{equation}
We then claim that we have the estimates, for each pair of points $y , y' \in \Zd$,
\begin{equation} \label{eq:TV12597}
      \left\| \partial_y \nabla v \left(x , \cdot \right)\right\|_{L^2 \left( \mu_\beta \right)} \leq \frac{C3^{\frac{n}{4}}}{|y-x|^{d+\frac 34}}  ~ \mbox{and}~ \left\| \partial_{y'} \nabla v \left(x' , \cdot \right)\right\|_{L^2 \left( \mu_\beta \right)} \leq \frac{C3^{\frac{n}{4}}}{|y'-x'|^{d+\frac 34}}.
\end{equation}
The estimate~\eqref{eq:TV12597} is proved in Step 2 below. Combining the inequalities~\eqref{eq:TV12607},~\eqref{eq:TV12597} and the formula~\eqref{eq:covXzXz'}, we obtain
\begin{equation} \label{eq:TV08095}
    \cov_{\mu_\beta} \left[ \nabla v \left( x, \cdot \right), \nabla v \left( x', \cdot\right) \right]  \leq C 3 ^{\frac{n}{2}} \sum_{y , y' \in \Zd} \frac{1}{|y'-x'|^{d+\frac 34}} \times \frac{1}{|y-x|^{d+\frac 34}} \times \frac{1}{|y - y'|^{d-2}}.
\end{equation}
The sum in the right side of the inequality~\eqref{eq:TV08095} can be explicitly computed and we obtain the inequality~\eqref{eq:TV12517}.

\smallskip

\textit{Step 2. Proof of~\eqref{eq:TV12597}.} The argument relies on the differentiated Helffer-Sj{\"o}strand equation introduced in Section~\ref{sec.section4.5} of Chapter~\ref{section:section4} and on the reflection principle to solve the Neumann problem~\eqref{eq:TV054165} below. Given a cube $Q \subseteq \Zd$ of sidelength $R$, we recall the notation $\frac 12 Q$ to denote the cube which has the same center as $Q$ and sidelength $\frac{R}{2}$. We consider the specific cube $\cu := \left( 0 , l_{n+1} \right)^d$ and the function $v \left( \cdot , \cdot, \cu, p^* \right)$. Since the cube $\cu_{n+1}$ can be obtained from the cube $\cu$ by a translation and since the measure $\mu_\beta$ is translation invariant, we see that to prove the estimate~\eqref{eq:TV12597} it is sufficient to prove the inequality, for each point $y \in \frac 13 \cu$ and each point $z \in \Zd$,
\begin{equation} \label{eq:TV11357}
    \left\| \partial_z \nabla v \left(y , \cdot, \cu, p^*  \right)\right\|_{L^2 \left( \mu_\beta \right)} \leq \frac{C3^{\frac{n}{4}}}{|y-z|^{d+\frac 34}}.
\end{equation}
The reason justifying this specific choice of the cube $\cu$ will become clear later in the proof. Using the definition of the map $v := v \left( \cdot , \cdot , \cu , p^* \right)$  as a minimizer in the variational formulation of $\nu^*\left( \cu , p^* \right)$ stated in~\eqref{def:defnustar}, we see that it is a solution of the Neumann problem
\begin{equation} \label{eq:TV16343}
\left\{ \begin{aligned}
\Delta_\phi v + \mathcal{L}_{\cu} v & = 0 &~\mbox{in}~ \cu \times \Omega,\\
\mathbf{n} \cdot \nabla v & = \mathbf{n} \cdot p^* &~\mbox{on}~ \partial \cu \times \Omega,
\end{aligned} \right.
\end{equation}
where the operator $\mathcal{L}_{\cu}$ is the uniformly elliptic operator defined by the formula
\begin{equation*}
    \mathcal{L}_{\cu} := - \frac{1}{2 \beta} \Delta +  \frac 1{2\beta}\sum_{k \geq 1} \frac1{\beta^{\frac k2}} \nabla^{k+1} \cdot \left( \indc_{\cu^k} \nabla^{k+1} \right) + \frac{1}{\beta^{\frac 54}} \nabla \cdot \left(\indc_{\cu \setminus \cu^-} \nabla\right) + \sum_{\supp q \subseteq \cu} \nabla_q \cdot \a_q \nabla_q,
\end{equation*}
where we recall the notation $\cu^k := \left\{ x \in \cu \, : \, \dist (x , \partial \cu) \geq k \right\}$. The specific, technical formula of the operator $\mathcal{L}_{\cu}$ is not relevant in the proof; the important point of the argument is that the operator $\mathcal{L}_{\cu}$ is well-defined for functions which are only defined in the interior of the triadic cube $\cu$ and that, as it is the case for elliptic operator $\mathcal{L}_{\mathrm{spat}}$, it is uniformly elliptic and is a perturbation of the Laplacian $- \frac 1{2\beta} \Delta$. As a consequence, all the results stated in Chapter~\ref{section:section4} for the Helffer-Sj{\"o}strand operator $\mathcal{L}$ are also valid for the operator $\Delta_\phi +  \mathcal{L}_{\cu}$. In particular all the arguments stated in Section~\ref{sec.section4.5} about the differentiated Helffer-Sj{\"o}strand equation apply in this setting. By applying the partial derivative $\partial$ to the system~\eqref{eq:TV16343}, we obtain that, if we denote by $w(y , z , \phi) = \partial_z v \left( y , \phi \right)$, then the function $w$ is the solution of the system
\begin{equation} \label{eq:TV054165}
    \left\{ \begin{aligned}
    \Delta_\phi w +  \mathcal{L}_{\cu, y} w +  \mathcal{L}_{\mathrm{spat}, z} w & = \sum_{\supp q \subseteq \cu}  z \left( \beta , q\right) \sin 2\pi\left( \phi , q \right) \left( v, q \right) q_y \otimes q_z &~\mbox{in}~ \cu \times \Zd \times \Omega, \\
    \mathbf{n} \cdot \nabla_y w & = 0  &~\mbox{on}~ \partial \cu \times \Zd \times \Omega,
    \end{aligned} \right.
\end{equation}
where the subscripts $y$ (resp. $z$) in the notation $\mathcal{L}_{\cu, y}$ (resp. $\mathcal{L}_{\mathrm{spat}, z}$) means that the spatial operator $\mathcal{L}_{\cu_{n+1}}$ (resp. $\mathcal{L}_{\mathrm{spat}, z}$) only acts on the spatial variable $y$. We introduce the notation $\mathbf{f}$ to denote the function
\begin{equation*}
   \mathbf{f} := \left\{ \begin{aligned}
    \cu \times \cu \times \Omega & \to  \R^{d \times d}, \\
    (y , z , \phi) & \mapsto   \sum_{\supp q \subseteq \cu}  z \left( \beta , q\right) \cos 2\pi\left( \phi , q \right) \left( v, q \right) n_{q}(y) \otimes n_{q}(z).
    \end{aligned} \right.
\end{equation*}
Using this notation, the system~\eqref{eq:TV054165} becomes
\begin{equation} \label{eq:TV05416}
    \left\{ \begin{aligned}
    \Delta_\phi w +  \mathcal{L}_{\cu, y} w +  \mathcal{L}_{\mathrm{spat}, z} w & =\di_y \di_z \mathbf{f} &~\mbox{in}~ \cu \times \Zd \times \Omega, \\
    \mathbf{n} \cdot \nabla_y w & = 0  &~\mbox{on}~ \partial \cu \times \Zd \times \Omega.
    \end{aligned} \right.
\end{equation}
To solve the system~\eqref{eq:TV05416}, we use the reflection principle. To this end, we need to introduce a few definitions, notations and remarks. We fix a point $z \in \Zd$ and extend the elliptic operator $\mathcal{L}_{\cu}$, the functions $v$ and $\mathbf{f}\left( \cdot  , z \right)$, initially defined on the cube $\cu$, to the entire space according to a the following procedure. We let $\tilde \cu$ be the discrete cube $\left( - l_{n+1} , l_{n+1} \right)^d$. For each point $x = (x_1 , \ldots, x_d) \in \tilde \cu$, we let $N_x$ be the number of negative coordinates of the components of $x$ and define
\begin{equation*}
        \mathcal{L}_{\cu} \left( x \right) =  \mathcal{L}_{\cu} \left( \left| x_1 \right| , \ldots, \left|x_d \right| \right) \hspace{5mm} \mbox{and} \hspace{5mm} \mathbf{f} \left( x , z , \phi \right) = (-1)^{N_x} \mathbf{f} \left( \left| x_1 \right| , \ldots, \left|x_d \right| , z , \phi \right).
\end{equation*}
We extend the operator $\mathcal{L}_{\cu}$ and the function $\mathbf{f}$ periodically from the cube $\tilde \cu$ to $\Zd$ and let $\tilde w$ be the solution of the elliptic system
\begin{equation} \label{eq:systsoltildew}
    \Delta_\phi \tilde w +  \mathcal{L}_{\cu, y} \tilde w +  \mathcal{L}_{\mathrm{spat}, z} \tilde w =\di_y \di_z \tilde{\mathbf{f}} \hspace{5mm}\mbox{in}~ \Zd \times \Zd \times \Omega.
\end{equation}
It is straightforward to verify that with this construction, the restriction of the function $\tilde w$ to the subcube $\cu$ satisfies the elliptic system~\eqref{eq:TV05416}; it is thus equal to the function $w$. We now study the function $\tilde w$. We denote by $\tilde{\mathcal{G}}_{\mathrm{der}}$ the Green's matrix associated to the operator $\Delta_\phi +  \mathcal{L}_{\cu, y} + \mathcal{L}_{\mathrm{spat}, z}$. As was already mentioned, the operator $\mathcal{L}_{\cu}$ is a perturbation of the Laplacian $\frac 1{2\beta} \Delta$; as a consequence, one can apply the same proofs as the ones written in Chapter~\ref{section:section4} and obtain the same results. In particular the statement of Proposition~\ref{cor:corollary4.14} of Chapter~\ref{section:section4} holds for the differentiated Green's matrix $\tilde{\mathcal{G}}_{\mathrm{der}}$. Using that the function $\tilde{w}$ solves the system~\eqref{eq:systsoltildew}, we obtain the explicit formula
\begin{equation*}
    \nabla_y \tilde w(y , z , \phi) = \sum_{y_1 , z_1 \in \Zd} 
    \nabla_y \di^*_{y_1} \di^*_{z_1} \tilde{\mathcal{G}}_{\mathrm{der}, \mathbf{f}\left( y_1 , z_1 , \cdot \right) }\left( y , z , \phi ; y_1 , z_1 \right).
\end{equation*}
Using the statement of Proposition~\ref{cor:corollary4.14}, Chapter~\ref{section:section4}, we obtain the estimate on the $L^2\left( \mu_\beta \right)$-norm of the function $\tilde w$,
\begin{equation} \label{eq:TV16436}
    \left\| \nabla_y  \tilde w(y , z , \cdot ) \right\|_{L^2 \left( \mu_\beta \right)} \leq C \sum_{y_1 , z_1 \in \Zd} \frac{\left\| \mathbf{f}\left( y_1 , z_1 , \cdot \right) \right\|_{L^2 \left( \mu_{\beta} \right)}}{\left| y_1 - y \right|^{2d +\frac 34} + \left| z_1 - z \right|^{2d +\frac 34}}.
\end{equation}
To compute~\eqref{eq:TV16436}, we prove the estimate, for each pair of points $y_1 , z_1 \in \Zd$,
\begin{equation} \label{eq:TV15376}
    \left\| \mathbf{f} \left( y_1 , y_1 + z_1, \cdot \right) \right\|_{L^2 \left( \mu_\beta \right)} \leq C e^{-c \sqrt{\beta} |z_1|} \sum_{y_0 \in \Zd} e^{-c \sqrt{\beta}|y_0 - y_1| } \left\| \nabla v (y_0, \cdot) \right\|_{L^2 \left( \mu_\beta \right)}.
\end{equation}
Let us make a comment about the estimate~\eqref{eq:TV15376}. Due to the exponential decay $\left| z \left( \beta , q \right)\right| \leq Ce^{-c \sqrt{\beta} \left\| q \right\|_1}$, the function $\mathbf{f}$ decays exponentially fast outside the diagonal $y = z$ of $\Z^{2d}$. This phenomenon can be observed in the inequality~\eqref{eq:TV15376}: the exponential term $e^{-c \sqrt{\beta} |z|}$ is small when the norm of $z$ is large, i.e., when the point $(y_1 , y_1 + z_1)$ is far from the diagonal $\left\{ (y , y) \in \Zd \times \Zd  \right\}$. Furthermore, on the diagonal, the term $\left\| \mathbf{f} \left( y_1 , y_1, \cdot \right) \right\|_{L^2 \left( \mu_\beta \right)}$ is approximately equal to the value $\left\| \nabla v(y_1, \cdot) \right\|_{L^2 \left( \mu_\beta \right)}$; but again the sum over all the charges needs to be taken into consideration and explains the sum over all the radii in the right side of~\eqref{eq:TV15376} with the exponential decay $e^{-c \sqrt{\beta} r}$.

 We now prove the estimate~\eqref{eq:TV15376}. We start from the inequality, for each pair of points $y_1 , z_1 \in \Zd$,
\begin{equation}  \label{eq:TV16226}
        \left\| \mathbf{f} \left( y_1 , y_1 + z_1, \cdot \right) \right\|_{L^2 \left( \mu_\beta \right)} \leq \sum_{q \in \mathcal{Q}} \sum_{y \in \supp n_q} e^{-c \sqrt \beta \left\| q \right\|_1} \left\| \nabla v(y, \cdot) \right\|_{L^2 \left( \mu_\beta \right)} \left\| n_q \right\|_{L^\infty} \left| n_q(y_1) \right| \left| n_q(y_1 + z_1) \right|.
\end{equation}
We then note that if a charge $q$ is such that the two points $y_1 $ and $y_1 + z_1$ belong to the support of $n_q$, then the diameter of $n_q$ is larger than $|z_1|$ and thus the diameter of $q$ has to be larger than $c |z_1|$. From this remark, we deduce that
\begin{equation} \label{eq:TV16236}
    \sum_{q \in \mathcal{Q}} e^{-c \sqrt \beta \left\| q \right\|_1}  \left\| n_q \right\|_{L^\infty} \left| n_q(y_1) \right| \left| n_q(y_1 + z_1) \right| \leq C e^{-c \sqrt{\beta} |z_1|}.
\end{equation}
Similarly, if a charge $q$ is such that the three points $y_1 $ and $y_1 + z_1$ and $y$ belong to the support of $n_q$, then the diameter of $n_q$ is larger than $\max \left( |z_1| , |y - y_1| \right) \geq \frac{|z_1| + |y - y_1|}{2}$. This implies that the diameter of $q$ has to be larger than $c \left(|z_1| + |y - y_1|\right)$ and we deduce that
\begin{equation} \label{eq:TV16246}
    \sum_{q \in \mathcal{Q}} e^{-c \sqrt \beta \left\| q \right\|_1} \indc_{\{ y \in \supp n_q \}} \left\| n_q \right\|_{L^\infty} \left| n_q(y_1)\right| \left| n_q(y_1 + z_1)\right| \leq C e^{-c \sqrt{\beta} \left( |z_1| + |y - y_1| \right)}. 
\end{equation}
Combining the estimates~\eqref{eq:TV16226},~\eqref{eq:TV16236} and~\eqref{eq:TV16246}, we obtain
\begin{align*}
    \left\| \mathbf{f} \left( y_1 , y_1 + z_1, \cdot \right) \right\|_{L^2 \left( \mu_\beta \right)} & \leq \sum_{q \in \mathcal{Q}} \sum_{y_0 \in \supp n_q} e^{-c \sqrt \beta \left\| q \right\|_1} \left\| \nabla v(y_0, \cdot) \right\|_{L^2 \left( \mu_\beta \right)} \left\| n_q \right\|_{L^\infty} n_q(y_1) n_q(y_1 + z_1) \\
    & \leq   \sum_{y \in \Zd} \sum_{q \in \mathcal{Q}} e^{-c \sqrt \beta \left\| q \right\|_1} \left\| \nabla v(y_0, \cdot) \right\|_{L^2 \left( \mu_\beta \right)} \indc_{\{y_0 \in \supp n_q\}} \left\| n_q \right\|_{L^\infty} n_q(y_1) n_q(y_1 + z_1) \\
    & \leq C e^{-c \sqrt{\beta} |z_1|} \sum_{y_0 \in \Zd} e^{-c \sqrt{\beta}|y_0 - y_1| } \left\| \nabla v (y_0, \cdot) \right\|_{L^2 \left( \mu_\beta \right)}
\end{align*}
and we have proved the inequality~\eqref{eq:TV15376}.

\smallskip

We now come back to the estimate~\eqref{eq:TV16436}, fix a point $y \in \cu$ and use the estimate~\eqref{eq:TV15376}. We obtain
\begin{equation} \label{eq:TV19146}
    \left\| \nabla_y \tilde w(y , z , \phi) \right\|_{L^2 \left( \mu_\beta \right)} \leq  C \sum_{y_0, y_1 , z_1 \in \Zd} \frac{ e^{-c \sqrt{\beta} \left(|z_1 - y_1| + |y_0 - y_1|\right)} \left\| \nabla v\left( y_0 , \cdot \right) \right\|_{L^2 \left( \mu_\beta \right)}}{\left| y_1 - y \right|^{2d +\frac{3}{4}} + \left| z_1 - z \right|^{2d +\frac{3}{4}}}.
\end{equation}
We focus on the sum over the variable $y_1$ and $z_1$. The exponential decay of the terms $ e^{-c \sqrt{\beta} |z_1 - y_1|}$ and $ e^{-c \sqrt{\beta} |y_0 - y_1|}$ forces the sum to contract on the points $y_1 = y_0$ and $z_1 = y_0$. We have the inequality,
\begin{equation*}
    \sum_{ y_1, z_1 \in \Zd} \frac{  e^{-c \sqrt{\beta} \left(|z_1 - y_1| + |y_0 - y_1|\right)}}{\left| y_1 - y \right|^{2d +\frac{3}{4}} + \left| z_1 - z \right|^{2d +\frac{3}{4}}} \leq  \frac{C}{\left| y_0 - y \right|^{2d +\frac{3}{4}} + \left| y_0 - z \right|^{2d +\frac{3}{4}}}.
\end{equation*}
Using the previous estimate, we can simplify the inequality~\eqref{eq:TV19146} and we obtain
\begin{equation*}
    \left\| \nabla_y \tilde w(y , z , \phi) \right\|_{L^2 \left( \mu_\beta \right)} \leq C \sum_{y_0  \in \Zd} \frac{ \left\| \nabla v(y_0 , \cdot) \right\|_{L^2 \left(\mu_\beta \right)}}{\left| y_0 - y \right|^{2d +\frac{3}{4}} + \left| y_0 - z \right|^{2d +\frac{3}{4}}}.
\end{equation*}
We then truncate the sum, depending on whether the point $y_0$ belongs to the cube $\frac 12 \cu$. We write
\begin{equation} \label{eq:TV192767}
    \left\| \nabla_y \tilde w(y , z , \phi) \right\|_{L^2 \left( \mu_\beta \right)} \leq C  \underbrace{\sum_{y_0  \in \frac 12 \cu} \frac{  \left\| \nabla v \left( y_0 , \cdot \right) \right\|_{L^2 \left(  \mu_\beta \right)}}{\left| y_0 - y \right|^{2d +\frac{3}{4}} + \left| y_0 - z \right|^{2d +\frac{3}{4}}}}_{\eqref{eq:TV192767}-(i)} + C \underbrace{\sum_{y_0  \in \Zd \setminus  \frac 12 \cu} \frac{  \left\| \nabla v \left( y_0, \cdot \right) \right\|_{L^2 \left(  \mu_\beta \right)}}{\left| y_0 - y \right|^{2d +\frac{3}{4}} + \left| y_0 - z \right|^{2d +\frac{3}{4}}}}_{\eqref{eq:TV192767}-(ii)}.
\end{equation}
We treat the two terms in the right side of~\eqref{eq:TV192767} separately. For the term~\eqref{eq:TV192767}-(i), we use that the map $v$ is a solution of the Helffer-Sj\"{o}strand equation~\eqref{eq:TV16343} in the cube $\cu$ and apply Proposition~\ref{prop:prop4.6} of Chapter~\ref{section:section4} with the regularity exponent $\ep = \frac 14$. We obtain, for each point $y_0 \in \frac 12 \cu$,
\begin{equation} \label{eq:TV10426}
    \left\| \nabla v \left( y_0, \cdot \right) \right\|_{L^2 \left(  \mu_\beta \right)} \leq C \left(l_{n+1}\right)^{\frac 12} \left\| \nabla v \right\|_{\underline{L}^2 \left( \cu ,  \mu_\beta \right)} \leq C 3^{\frac n2},
\end{equation}
where we used Remark~\ref{rmk:TV1414} and the inequality~\eqref{eq:uppboundoptimi} in the second inequality. Using the estimate~\eqref{eq:TV10426}, we can compute the term~\eqref{eq:TV192767}-(i)
\begin{align} \label{eq:TV110477}
    \sum_{y_0  \in \frac 12 \cu} \frac{  \left\| \nabla v \left( y_0, \cdot \right) \right\|_{L^2 \left(  \mu_\beta \right)}}{\left| y_0 - y \right|^{2d +\frac{3}{4}} + \left| y_0 - z \right|^{2d +\frac{3}{4}}} & \leq C 3^{\ep n} \sum_{y_0  \in \frac 12 \cu} \frac{ 1}{\left| y_0 - y \right|^{2d +\frac{3}{4}} + \left| y_0 - z \right|^{2d +\frac{3}{4}}} \\
    & \leq C 3^{\ep n} \sum_{y_0  \in \Zd} \frac{ 1}{\left| y_0 - y \right|^{2d +\frac{3}{4}} + \left| y_0 - z \right|^{2d +\frac{3}{4}}} \notag \\
    & \leq   C 3^{\ep n} \sum_{y_0  \in \Zd} \frac{ 1}{\left| y_0 \right|^{2d +\frac{3}{4}} + \left| y_0 + y - z \right|^{2d +\frac{3}{4}}} \notag \\
    & \leq  C \frac{3^{\ep n}}{\left| y - z \right|^{d + \frac{3}{4}}}, \notag
\end{align}
where we used Proposition~\ref{propappCp97} of Appendix~\ref{app.appC} in the last inequality. We now treat the term~\eqref{eq:TV192767}-(ii). In that case, we use the estimate $|y - y_0| \geq c |y_0|$ valid for any point $y_0 \in \Zd \setminus \frac{1}{2} \cu$ and any point $y \in \frac13 \cu$. We obtain the inequality
\begin{equation*}
\sum_{y_0  \in \Zd \setminus  \frac12 \cu} \frac{  \left\| \nabla v \left( y_0,\cdot \right) \right\|_{L^2 \left(  \mu_\beta \right)}}{\left| y_0 - y \right|^{2d +\frac{3}{4}} + \left| y_0 - z \right|^{2d +\frac{3}{4}}} \leq \sum_{y_0  \in \Zd \setminus  \frac12 \cu} \frac{  \left\| \nabla v \left( y_0, \cdot \right) \right\|_{L^2 \left(  \mu_\beta \right)}}{\left| y_0 \right|^{2d +\frac{3}{4}} + \left| y_0 - z \right|^{2d +\frac{3}{4}}}.
 \end{equation*}
 To estimate the previous inequality, we partition the space into cubes. We consider the set $\mathcal{K} :=  \frac 12 l_{n+1} \Zd$ and note that the collection of cubes $\left( \kappa + \frac 12 \cu\right)_{\kappa \in \mathcal{K}}$ is a partition of $\Zd$. We note that, for each point $\kappa \in \mathcal{K}\setminus \{ 0 \}$, each point $y_0 \in \left( \kappa + \frac 12 \cu \right)$ and each point $z \in \Zd$, one has the inequalities
 \begin{equation} \label{eq:TV10467}
      c \left( \left| y_0 \right|^{2d +\frac{3}{4}} + \left| y_0 - z \right|^{2d +\frac{3}{4}}\right) \leq  \left| \kappa \right|^{2d +\frac{3}{4}} + \left| \kappa - z \right|^{2d +\frac{3}{4}} \leq  C\left( \left| y_0 \right|^{2d +\frac{3}{4}} + \left| y_0 - z \right|^{2d +\frac{3}{4}}\right).
 \end{equation}
Using the extension of the function $v$ from the cube $\cu$ to the entire space $\Zd$ stated in Step 2 of the proof of Lemma~\ref{lemmvarest}, the volume identity $\left| \frac{1}{2} \cu \right| = \left( \frac{1}{2} \right)^d \left| \cu \right| $ and the inequality~\eqref{eq:uppboundoptimi}, we obtain the estimate, for each $\kappa \in \mathcal{K}$,
\begin{equation} \label{eq:TV12336}
    \left\| v \right\|_{\underline{L}^2 \left( \kappa + \frac 12 \cu , \mu_\beta \right)} \leq C.
\end{equation}
A combination of the inequalities~\eqref{eq:TV10467} and~\eqref{eq:TV12336} yields
 \begin{align} \label{eq:TV10457}
     \sum_{y_0  \in \Zd \setminus  \frac 12 \cu} \frac{  \left\| \nabla v \left( y_0, \cdot \right) \right\|_{L^2 \left(  \mu_\beta \right)}}{\left| y_0 \right|^{2d +\frac{3}{4}} + \left| y_0 - z \right|^{2d +\frac{3}{4}}} & \leq  \sum_{\kappa \in \mathcal{K} \setminus \{ 0 \}} \sum_{y_0  \in \kappa + \frac 12 \cu} \frac{  \left\| \nabla v \left( y_0, \cdot \right) \right\|_{L^2 \left(  \mu_\beta \right)}}{\left| y_0 \right|^{2d +\frac{3}{4}} + \left| y_0 - z \right|^{2d +\frac{3}{4}}} \\
     & \leq \sum_{\kappa \in \mathcal{K} \setminus \{ 0 \}} \sum_{y_0  \in \kappa + \frac 12 \cu} \frac{  \left\| \nabla v \left( y_0, \cdot \right) \right\|_{L^2 \left(  \mu_\beta \right)}}{\left| \kappa \right|^{2d +\frac{3}{4}} + \left| \kappa - z \right|^{2d +\frac{3}{4}}} \notag \\
     & \leq C \sum_{\kappa \in \mathcal{K} \setminus \{ 0 \}}  \frac{ \left| \frac12 \cu \right|}{\left| \kappa \right|^{2d +\frac{3}{4}} + \left| \kappa - z \right|^{2d +\frac{3}{4}}}. \notag
\end{align}
To estimate the sum in the right side of~\eqref{eq:TV10457}, we use the estimate~\eqref{eq:TV10467} a second time and write
\begin{align} \label{eq:TV13546}
\sum_{y_0  \in \Zd \setminus  \frac 12 \cu} \frac{  \left\| \nabla v \left( y_0, \cdot \right) \right\|_{L^2 \left(  \mu_\beta \right)}}{\left| y_0 \right|^{2d +\frac{3}{4}} + \left| y_0 - z \right|^{2d +\frac{3}{4}}} 
     & \leq C \sum_{\kappa \in \mathcal{K} \setminus \{ 0 \}}  \frac{ \left| \frac 12 \cu \right|}{\left| \kappa \right|^{2d +\frac{3}{4}} + \left| \kappa - z \right|^{2d +\frac{3}{4}}} \\
     & \leq C \sum_{\kappa \in \mathcal{K} \setminus \{ 0 \}} \sum_{y_0  \in \kappa + \frac 12 \cu} \frac{1}{\left| \kappa \right|^{2d +\frac{3}{4}} + \left| \kappa - z \right|^{2d +\frac{3}{4}}} \notag \\
     & \leq  C \sum_{\kappa \in \mathcal{K} \setminus \{ 0 \}} \sum_{y_0  \in \kappa + \frac 12 \cu} \frac{1}{\left| y_0 \right|^{2d +\frac{3}{4}} + \left| y_0 - z \right|^{2d +\frac{3}{4}}} \notag \\
     & \leq C \sum_{y_0  \in \Zd \setminus  \frac 12 \cu} \frac{1}{\left| y_0 \right|^{2d +\frac{3}{4}} + \left| y_0 - z \right|^{2d +\frac{3}{4}}} \notag \\
     & \leq \frac{C}{\max \left(|z|, 3^n \right)^{d+\frac 34}}. \notag
\end{align}
where we used Remark~\ref{propappCp97rem} of Appendix~\ref{app.appC} in the last inequality. We finally slightly modify the result: using that the point $y$ belongs to the cube $\frac 13 \cu$, we have the inequality
\begin{equation} \label{eq:TV13556}
    \frac{1}{\left( |z| \vee 3^n \right)^{d+\frac 12}} \leq \frac{C}{|z - y|^{d+\frac 12}}.
\end{equation}
We use the computation~\eqref{eq:TV13546} and the inequality~\eqref{eq:TV13556} to obtain
\begin{equation}\label{eq:TV182616}
    \sum_{y_0  \in \Zd \setminus  \frac 12 \cu} \frac{  \left\| \nabla v \left( y_0 , \cdot \right) \right\|_{L^2 \left(  \mu_\beta \right)}}{\left| y_0 \right|^{2d +\frac{3}{4}} + \left| y_0 - z \right|^{2d +\frac{3}{4}}}  \leq \frac{C}{|z - y|^{d+\frac 12}}.
\end{equation}
By combining the estimates~\eqref{eq:TV192767},~\eqref{eq:TV110477} and~\eqref{eq:TV182616}, we deduce that
\begin{equation} \label{eq:TV11287}
    \left\| \nabla_y \tilde w(y , z , \cdot ) \right\|_{L^2 \left( \mu_\beta \right)} \leq \frac{C 3^{\frac{n}{4}}}{\left| z - y  \right|^{d+\frac{3}{4}}}.
\end{equation}
We complete the argument by recalling that for each $y \in \cu$ and each $z \in \Zd$, the function $\tilde w$ is defined so that we have $ \nabla_y \tilde w \left( y , z , \cdot  \right) = \partial_z \nabla v \left( y , \cdot , \cu \right)$. The inequality~\eqref{eq:TV11287} can thus be rewritten
\begin{equation*}
    \left\| \partial_z \nabla v \left( y , \cdot , \cu , p^* \right) \right\| \leq \frac{C 3^{\frac{n}{4}}}{|y - z|^{d+\frac 34}}.
\end{equation*}
The proof of the inequality~\eqref{eq:TV11357} and thus of Step 2 is complete.
\end{proof}

\subsection{Control over the \texorpdfstring{$L^2$}{11}-norms of the functions \texorpdfstring{$u - l_p$}{12} and \texorpdfstring{$v - \a_* \left( \cu_n \right)^{-1} l_{ p^*}$}{13}.} The objective of this section is to prove that the optimizers $u$ and $v$ are close in the $\underline{L}^2\left( \cu_n , \mu_\beta \right)$-norm to affine functions. The result relies on the multiscale Poincar\'e inequality stated in Appendix~\ref{section:multiscPoinc} and is quantified in terms of the subadditivity defect $\tau_n$.

\begin{lemma}[$L^2$ estimate for the optimizers $u$ and $v$] \label{lem:lemma5.8}
There exist an inverse temperature $\beta_0 := \beta_0(d)< \infty$ and a constant $C := C(d) < \infty$ such that for each $\beta > \beta_0$, each integer $n \in \N$, and each pair of vectors $p,p^* \in \R^{d \times \binom{d}{2}}$,
\begin{equation} \label{mainest.lamma4.2u}
\left\| u (\cdot, \cdot, \cu_{n+1}^- , p) -  l_p  \right\|_{\underline L^2(\cu_{n+1} , \mu_{\beta}) }^2 \leq  C |p|^2 3^{2n} \left( 3^{- \frac n2}+ \sum_{m = 0}^n 3^{- \frac{m -n}{2}} \tau_m \right),
\end{equation}
and
\begin{equation}  \label{mainest.lamma4.2v}
\left\| v (\cdot, \cdot, \cu_{n+2} , p^*) -  l_{\ahom_n^{-1} p^*}  - \left( v \right)_{\cu_{n+1}, \mu_\beta}\right\|_{\underline L^2(\cu_{n+1} , \mu_{\beta}) }^2 \leq  C |p^*|^2 3^{2n} \left( 3^{-\frac n2}+ \sum_{m = 0}^{n+1} 3^{-\frac{(m -n)}{2}} \tau_m \right).
\end{equation}
\end{lemma}

\begin{proof}
We assume without loss of generality that $|p|=1$ and $|p^*|=1$. The strategy of the proof relies on two ingredients:
\begin{itemize}
    \item First, we need to estimate the spatial averages of the gradients of the functions $u - l_p$ and $v -  l_{\a_* \left( \cu_n \right)^{-1} p^*}$ and prove that they are small. To be more precise, we estimate these spatial averages in terms of the subadditivity defects $\tau_n$. The proof relies on different arguments depending on which function we consider:
    \begin{itemize}
        \item For the function $u$ associated to the energy quantity $\nu$, we use the subadditivity property stated in Proposition~\ref{p.subadd} and the following fact: for any discrete cube $\cu \subseteq \Zd$ and any function $f : \cu \to \R$ which is equal to $0$ on the boundary of the cube $\cu$, one has the identity
        \begin{equation*}
            \sum_{x \in \cu} \nabla f(x) = 0;
        \end{equation*}
        \item For the function $v$ associated to the energy quantity $\nu^*$, we use the subadditivity property stated in Proposition~\ref{p.subadd} and Lemma~\ref{lemmvarest} to control the variance of the spatial average of its gradients.
    \end{itemize}
    \item The multiscale Poincar\'e inequality, which is stated in Proposition~\ref{prop:multiscPoin} in Appendix~\ref{section:multiscPoinc}. This inequality allows to estimate the $L^2$-norm of a function in terms of the spatial averages of its gradient.
\end{itemize}
We first focus on the function $u :=  u (\cdot, \cdot, \cu_{n+1}^- , p)$ and prove the inequality~\eqref{mainest.lamma4.2u}. We first recall that the function $u$ is extended by the affine function $l_p$ outside the cube $\cu_{n+1}^-$. We thus have
\begin{equation*}
    \left\| u (\cdot, \cdot,  \cu_{n+1}^- , p) -  l_p  \right\|_{L^2(\cu_{n+1} , \mu_{\beta}) }^2 = \left\| u (\cdot, \cdot,  \cu_{n+1}^- , p) -  l_p  \right\|_{L^2(\cu_{n+1}^- , \mu_{\beta}) }^2.
\end{equation*}
By the multiscale Poincar\'e inequality stated in Proposition~\ref{prop:multiscPoin} of Appendix~\ref{section:multiscPoinc}, we have
\begin{multline} \label{multcspoincare1}
\left\| u (\cdot, \cdot, \cu_{n+1}^- , p) -l_p \right\|_{\underline L^2(\cu_{n+1}^- , \mu_{\beta}) }^2  \\ \leq \underbrace{C \left\|\nabla  u (\cdot, \cdot, \cu_{n+1}^- , p)  - p\right\|_{\underline L^2(\cu_{n+1} , \mu_{\beta})}^2}_{\eqref{multcspoincare1}-(i)} + \underbrace{C 3^n \sum_{m = 0}^n \frac{3^m}{\left|\mathcal{Z}_{m,n} \right|} \left\langle \left( \frac{1}{\left| z + \cu_m \right|}\sum_{x \in z + \cu_m}  \nabla  u (\cdot, \cdot, \cu_{n+1}^- , p)  - p \right)^2 \right\rangle_{\mu_\beta}}_{\eqref{multcspoincare1}-(ii)}.
\end{multline}
We bound the first term~\eqref{multcspoincare1}-(i) using the estimate~\eqref{eq:uppboundoptimi}. We obtain the inequality
\begin{equation} \label{term1multsc}
  \left\|\nabla  u (\cdot ,\cdot, \cu_{n+1}^- , p)  - p\right\|_{\underline L^2(\cu_{n+1} , \mu_{\beta})}^2   \leq 2 \left\|\nabla  u (\cdot , \cdot, \cu_{n+1} , p) \right\|_{\underline L^2(\cu_{n+1}^- , \mu_{\beta})}^2 + 2 |p^2|   \leq C  |p^2|.
\end{equation}
To estimate the term~\eqref{multcspoincare1}-(ii), we use the two following ingredients:
\begin{itemize}
\item The subadditivity of the energy $\nu$ which is stated in Proposition~\ref{p.subadd} and Remark~\ref{rem.subaddtrim}. It reads, for each integer $m \in \left\{ 1 , \ldots , n \right\}$,
\begin{align*}
\left| \mathcal{Z}_{m,n}\right|^{-1} \sum_{z \in \mathcal{Z}_{m,n}} \left\llbracket u(\cdot , \cdot , \cu_{n+1}^- , p) - u(\cdot, \cdot ,z + \cu_{m}^- , p)  \right\rrbracket_{\underline{H}^1(z + \cu_{m}^-,\mu_\beta)}^2 & \leq C \left( \nu \left( \cu_m^- , p \right) - \nu \left( \cu_{n+1}^- , p \right)  +  3^{-\frac m2} |p|^2 \right) \\
& \leq C \left( \sum_{k = m}^{n} \tau_k +  3^{-\frac m2} |p|^2 \right).
\end{align*}
\item For each point $z \in \mathcal{Z}_{m,n}$, the function $ u(\cdot ,z + \cu_{m} , p)$ belongs to the space $l_p + H^1_0 \left( z + \cu_m , \mu_\beta \right)$. This implies that, for each realization of the field $\phi \in \Omega$,
\begin{equation} \label{eq:TV17394}
\frac{1}{\left| z + \cu_m^- \right|}\sum_{x \in z + \cu_m}  \nabla  u (x , \phi, z + \cu_{m}^- , p)  = p.
\end{equation}
\end{itemize}
We deduce the inequality, for each integer $m \in \left\{ 1 , \ldots , n \right\}$,
\begin{equation} \label{eq:estspatav2241}
 \sum_{z \in \mathcal{Z}_{m,n}}\frac{1}{\left|\mathcal{Z}_{m,n} \right|} \left\langle \left( \frac{1}{\left| z + \cu_m \right|}\sum_{x \in z + \cu_m}  \nabla  u (x , \cdot, \cu_{n+1}^- , p)  - p \right)^2 \right\rangle_{\mu_\beta} \leq C \left( \sum_{k = m}^{n} \tau_k +  3^{-\frac m2} |p|^2 \right).
\end{equation}
Combining the estimates~\eqref{multcspoincare1},~\eqref{term1multsc} and~\eqref{eq:estspatav2241} completes the proof of the estimate~\eqref{mainest.lamma4.2u}. 

\smallskip

We now prove the inequality~\eqref{mainest.lamma4.2v}. By the multiscale Poincar\'e inequality, we have
\begin{multline} \label{multcspoincare2}
\left\| v (\cdot, \cdot, \cu_{n+2} , p^*) -  l_{\ahom_n^{-1} p^*} - \left( v (\cdot, \cdot, \cu_{n+2} , p^*) -   l_{\ahom_n^{-1} p^*} \right)_{\cu_{n+1}}  \right\|_{\underline L^2(\cu_{n+1} , \mu_{\beta}) }^2 \\ \leq \underbrace{C \left\|\nabla  v(\cdot, \cu_{n+2} , p^*)  - \ahom_n^{-1} p^*\right\|_{\underline L^2(\cu_{n+1} , \mu_{\beta})}^2}_{\eqref{multcspoincare2}-(i)} + \underbrace{C 3^n \sum_{m = 0}^n \frac{3^m}{\left|\mathcal{Z}_{m,n} \right|} \left\langle \left( \frac{1}{\left| z + \cu_m \right|}\sum_{x \in z + \cu_m}  \nabla  v (\cdot, \cu_{n+2} , p^*)  - \ahom_n^{-1} p^* \right)^2 \right\rangle_{\mu_\beta}}_{\eqref{multcspoincare2}-(ii)}.
\end{multline}
We first treat the term on the left side. Since the average value of a linear map on a cube centered at $0$ is equal to $0$, we have that
\begin{equation*}
    \left( v (\cdot, \cdot, \cu_{n+2} , p^*) -   l_{\ahom_n^{-1} p^*} \right)_{\cu_{n+1}} = \frac{1}{\left| \cu_{n+1} \right|}\sum_{x \in \cu_{n+1}} v (x, \cdot, \cu_{n+2} , p^*) -  l_{\ahom_n^{-1}p^*}(x) = \frac{1}{\left| \cu_{n+1} \right|}\sum_{x \in \cu_{n+1}} v (x, \cdot, \cu_{n+2} , p^*).
\end{equation*}
We then use the estimate~\eqref{eq:TV17570501} and the inclusion $\cu_{n+1} \subseteq \frac 13 \cu_{n+2}$. We obtain
\begin{align} \label{onetermmesomultscter}
    \left\| \left( v (\cdot, \cdot, \cu_{n+2} , p^*) -   l_{\ahom_n^{-1} p^*} \right)_{\cu} - \left( v (\cdot, \cdot, \cu_{n+2} , p^*) \right)_{\cu_{n+1}, \mu_\beta} \right\|_{L^2 \left( \mu_\beta \right)}^2 & = \var_{\mu_\beta} \left[ \left( v (\cdot, \cdot, \cu_{n+2} , p^*)  \right)_{\cu_{n+1}} \right] \\ 
    &\leq \frac{C}{\left| \cu_n \right|}\sum_{x \in  \cu_{n+1}} \var_{\mu_\beta} \left[ v (x, \cdot, \cu_{n+2} , p^*)  \right] \notag
   \\  & \leq \frac{C}{\left| \cu_n \right|}\sum_{x \in \frac{1}{3} \cu_{n+2}} \var \left[ v (x, \cdot, \cu_{n+2} , p^*)  \right] \notag \\
   & \leq C |p^*|. \notag
\end{align}
The first term~\eqref{multcspoincare2}-(i) can be estimated with the same argument as in the inequality~\eqref{term1multsc}. We obtain
\begin{equation} \label{onetermmesomultscbis}
    \left\|\nabla  v(\cdot, \cdot, \cu_{n+2} , p^*)  - \ahom_n^{-1} p^*\right\|_{\underline L^2(\cu_{n+1} , \mu_{\beta})}^2 \leq C.
\end{equation}
To estimate the term~\eqref{multcspoincare2}-(ii), we prove that, for each integer $m \in \left\{ 1 , \ldots , n \right\}$,
\begin{equation} \label{onetermmesomultsc}
\frac{1}{\left|\mathcal{Z}_{m,n} \right|} \left\langle \left( \frac{1}{\left| z + \cu_m \right|}\sum_{x \in z + \cu_m}  \nabla  v (\cdot, \cdot, \cu_{n+2} , p^*)  - \ahom_n^{-1} p^* \right)^2 \right\rangle_{\mu_\beta} \leq C 3^{-  \frac m2} + C \sum_{k = m}^n \tau_k.
\end{equation}
To this end, we decompose the left side of~\eqref{onetermmesomultsc} and write
\begin{align} \label{eq:TV16479}
\lefteqn{\frac{1}{\left|\mathcal{Z}_{m,n} \right|} \sum_{z \in \mathcal{Z}_{m,n}}\left\langle \left( \frac{1}{\left| z + \cu_m \right|}\sum_{x \in z + \cu_m}  \nabla  v (\cdot, \cdot, \cu_{n+2} , p^*)  - \ahom_n^{-1} p^* \right)^2 \right\rangle_{\mu_\beta}  } \qquad & \\ &
							\leq 3 \left| \mathcal{Z}_{m,n}\right|^{-1} \sum_{z \in \mathcal{Z}_{m,n}} \left\llbracket v(\cdot, \cdot , \cu_{n+2} , p^*) - v(\cdot , \cdot, z + \cu_{m} , p^*)  \right\rrbracket_{\underline{H}^1(z + \cu_{m},\mu_\beta)}^2 \notag \\
							& \quad + 3\left| \ahom_n^{-1}p^*  -\ahom_m^{-1} p^*  \right|^2 \notag \\
							& \quad  +  3 \left| \mathcal{Z}_{m,n}\right|^{-1} \sum_{z \in \mathcal{Z}_{m,n}}\left\langle \left( \frac{1}{\left| z + \cu_m \right|}\sum_{x \in z + \cu_m}  \nabla  v (\cdot,\cdot, z + \cu_{m} , p^*)  - \ahom_m^{-1} p^* \right)^2 \right\rangle_{\mu_\beta}. \notag
\end{align}
We estimate the first term on the right side by Proposition~\ref{p.subaddnustar}, the second term by Lemma~\ref{lem:lemma5.6}. We obtain
\begin{equation} \label{eq:TV16552}
\left| \mathcal{Z}_{m,n}\right|^{-1} \sum_{z \in \mathcal{Z}_{m,n}} \left\llbracket v(\cdot ,\cdot, \cu_{n+2} , p^*) - v(\cdot , \cdot, z + \cu_{m} , p^*)  \right\rrbracket_{\underline{H}^1(z + \cu_{m},\mu_{\beta})}^2 + \left| \ahom_n^{-1}p^*  -\ahom_m^{-1} p^*  \right|^2 \leq C 3^{-  \frac m2} + C \sum_{k = m}^n \tau_k.
\end{equation}
There remains to estimate the third term in the right side of~\eqref{eq:TV16479}. We first recall the identity, for each integer $m \in \N$,
\begin{equation*}
\frac{1}{|\cu_m|}\sum_{x \in \cu_m} \left\langle  \nabla v(x, \cdot , \cu_m , p^*) \right\rangle_{\mu_\beta} = \ahom^{-1}_m p^*.
\end{equation*} 
We use the translation invariance of the measure $\mu_\beta$ and Lemma~\ref{lemmvarest}. To ease the notation, we note that in dimension larger than $3$, we have the estimate $d - \frac 52 \geq \frac 12$. We obtain
\begin{align} \label{eq:TV165522}
 \left| \mathcal{Z}_{m,n}\right|^{-1} \sum_{z \in \mathcal{Z}_{m,n}}\left\langle \left( \frac{1}{\left| z + \cu_m \right|}\sum_{x \in z + \cu_m}  \nabla  v (x , \cdot, z + \cu_{m} , p^*)  - \ahom_m^{-1} p^* \right)^2 \right\rangle_{\mu_\beta} & = \left\langle \left( \frac{1}{\left| \cu_m \right|}\sum_{x \in \cu_m}  \nabla  v (x , \cdot, \cu_{m} , p^*)  - \ahom_m^{-1} p^* \right)^2 \right\rangle_{\mu_\beta} \\
				& = \var_{\mu_\beta} \left[ \frac 1{|\cu_m|} \sum_{x \in \cu_n} \nabla v(x, \cdot , \cu_m, p^*) \right] \notag \\
				& \leq C \left( 3^{- \frac m2} + \tau_m \right). \notag
\end{align}
Combining the estimates~\eqref{onetermmesomultscter},~\eqref{onetermmesomultscbis},~\eqref{eq:TV16479},~\eqref{eq:TV165522} and~\eqref{eq:TV16552} completes the proof of~\eqref{mainest.lamma4.2v}.
\end{proof}

\subsection{Control over the energy \texorpdfstring{$J$}{15}}
In this section, we deduce from the previous results and the Caccioppoli inequality a control over the energy quantity $J \left( \cu_n, p , \ahom_n p \right)$. The argument needs to take into account the infinite range of the Helffer-Sj{\"o}strand operator and the specific forms of the energies $\mathbf{E}$ and $\mathbf{E}^*$ which causes some technicalities in the argument. The result is stated in the lemma below.

\begin{lemma} \label{e.lemma4.3}
There exist an inverse temperature $\beta_0 := \beta_0(d) < \infty$ and a constant $C := C(d) < \infty$ such that for each $\beta \geq \beta_0$, each integer $n \in \N$ and each $p \in \R^{d \times \binom d2}$,
\begin{equation} \label{eq:Lemma4.3}
\nu  \left( \cu_n^-, p\right) + \nu^* \left( \cu_n , \ahom_n p \right) - \ahom_n | p|^2 \leq C|p|^2 \left( 3^{-\frac{n}{2}} + \sum_{m = 0}^{n+1} 3^{- \frac{n-m}{2}} \tau_m \right).
\end{equation} 
\end{lemma}

\begin{proof}
The strategy of the proof relies on three ingredients: the Caccioppoli inequality stated in Proposition~\ref{Caccio.ineq} of Chapter~\ref{section:section4}, the one-sided convex duality formula~\eqref{eq:onesidedconvexdual} stated in Proposition~\ref{prop:prop5.2} and the $L^2$-norm estimate on the optimizers $u$ and $v$ stated in Lemma~\ref{lem:lemma5.8}.

We fix a slope $p \in \Rd$ and assume without loss of generality that $|p|=1$. By Proposition~\ref{prop:prop5.2}, we have the identity
\begin{equation*}
\nu  \left( \cu_n^-, p\right) + \nu^* \left( \cu_n , \ahom_n p \right) - \ahom_n |p|^2 = \mathbf{E}_{\cu_n}^* \left[ u\left(\cdot , \cdot , \cu_n^- , p \right) -  v\left(\cdot , \cdot , \cu_n , \ahom_n p \right) \right] + O \left( C 3^{-\frac m2}\right).
\end{equation*}
To prove the estimate~\eqref{eq:Lemma4.3}, it is thus sufficient to prove the estimate
\begin{equation} \label{eq:TV12183}
    \mathbf{E}_{\cu_n}^* \left[ u\left(\cdot , \cdot , \cu_n^- , p \right) -  v\left(\cdot , \cdot , \cu_n , \ahom_n p \right) \right]  \leq C \left( 3^{-\frac{n}{2}} + \sum_{m = 0}^{n+1} 3^{- \frac{n-m}{2}} \tau_m \right).
\end{equation}
Using the coercivity of the energy $\mathbf{E}_{\cu_n}^*$ stated in~\eqref{eq:coerccontenergystar}, we see that to prove the inequality~\eqref{eq:TV12183}, it is sufficient to prove the estimate
\begin{equation} \label{eq:TV14093}
    \left\llbracket  u\left(\cdot, \cdot , \cu_n^- , p \right) -  v\left(\cdot,\cdot , \cu_n , \ahom_n p \right) \right\rrbracket_{\underline{H}^1 \left( \cu_n , \mu_{\beta} \right)}^2 \leq C \left( 3^{-\frac{n}{2}} + \sum_{m = 0}^{n+1} 3^{- \frac{n-m}{2}} \tau_m \right),
\end{equation}
and by Propositions~\ref{p.subadd} and~\ref{p.subaddnustar}, we see that to prove~\eqref{eq:TV14093} it is sufficient to prove
\begin{equation} \label{eq:TV14103}
    \left\llbracket  u\left(\cdot,\cdot , \cu_{n+1}^- , p \right) -  v\left(\cdot,\cdot , \cu_{n+2} , \ahom_n p \right) \right\rrbracket_{\underline{H}^1 \left( \cu_n , \mu_{\beta} \right)}^2 \leq C \left( 3^{-\frac{n}{2}} + \sum_{m = 0}^{n+1} 3^{- \frac{n-m}{2}} \tau_m \right).
\end{equation}
We now focus on the proof of~\eqref{eq:TV14103}. In the rest of the proof, we make use of the notations $u := u\left(\cdot, \cdot , \cu_{n+1}^- , p \right)$ and $v := v\left(\cdot, \cdot , \cu_{n+2} , \ahom_n p \right) - \left( v\left(\cdot, \cdot , \cu_{n+2} , \ahom_n p \right) \right)_{\cu_{n+1}, \mu_\beta}$. By Lemma~\ref{lem:lemma5.8}, we have the $\underline L^2 \left( \cu_{n+1} , \mu_\beta \right)$-estimate
\begin{equation} \label{eq:TV18283}
    \left\| u- v \right\|_{\underline L^2 \left( \cu_{n+1} , \mu_\beta \right)}^2 \leq 2 \left\| u - l_p \right\|_{\underline L^2 \left( \cu_{n+1} , \mu_\beta \right)}^2 + 2 \left\| v - l_p \right\|_{ \underline L^2 \left( \cu_{n+1} , \mu_\beta \right)}^2 \leq  C 3^{2n} \left( 3^{-\frac{n}{2}}+ \sum_{m = 0}^{n+1} 3^{-\frac{m-n}{2}} \tau_m \right).
\end{equation}
We recall the following notation: for each integer $k \in \N$, we denote by $\cu_{n+2}^k$ the interior cube $\cu_{n+2}^k := \left\{ x \in \cu_{n+2} \, : \, \dist  \left(x , \partial \cu_{n+2} \right) \geq k \right\}$. By the first variation formula stated in Proposition~\ref{prop:prop5.2}, the maps $u$ and $v$ are solutions of the equations
\begin{equation*}
\mathcal{L} u = 0 ~\mbox{in}~ \cu_{n+2}^- \times \Omega \hspace{5mm} \mbox{and} \hspace{5mm} \mathcal{L}_{\cu_{n+1}, *} v = 0 ~\mbox{in}~ \cu_{n+2} \times \Omega,
\end{equation*}
where we recall the definition of the Helffer-Sj{\"o}strand operator $\mathcal{L}_{\cu_{n+2}, *}$
\begin{equation*}
    \mathcal{L}_{\cu_{n+2}, *} := \Delta_\phi - \frac{1}{2 \beta} \Delta +  \frac 1{2\beta}\sum_{k \geq 1} \frac{(-1)^{k+1}}{\beta^{\frac k2}} \nabla^{k+1} \cdot \left( \indc_{\cu_{n+2}^k} \nabla^{k+1} \right) - \frac{1}{\beta^{\frac 54}} \nabla \cdot \left(\indc_{\cu \setminus \cu^-} \nabla\right) + \sum_{\supp q \subseteq \cu_{n+1}} \nabla_q \cdot \a_q \nabla_q.
\end{equation*}

One can adapt the proof of the Caccioppoli inequality (Proposition~\ref{Caccio.ineq} of Chapter~\ref{section:section4}) to the operator $\mathcal{L}_{\cu_{n+2}, *}$ and obtain the following statement. There exists a constant $C := C(d) < \infty$ such that for any vector fields $F : \cu_{n+2} \times \Omega \to \R^{d \times \binom d2}$ and $G : \cu_{n+2} \times \Omega \to \R^{d}$, any ball $B(x , r)$ such that $B(x , 2r)$ is included in the cube $\cu_{n+2}$ and every solution $w : \cu_{n+2} \times \Omega \to \R^{\binom d2}$ of the equation
\begin{equation*}
    \mathcal{L}_{\cu_{n+2}, *} w = \nabla \cdot F + \di G ~\mbox{in}~ B(x , 2r) \times \Omega,
\end{equation*}
one has the estimate
\begin{multline} 
 \left\llbracket w \right\rrbracket_{\underline{H}^1 \left( B_r(x) , \mu_{\beta} \right)}  \leq \frac{C}{R} \left\| w \right\|_{\underline{L}^2 \left( B_{2r}(x) , \mu_\beta \right)} \\  + \left\| F \right\|_{\underline{L}^2\left( B_{2r}(x) , \mu_\beta \right) } + \left\| G \right\|_{\underline{L}^2\left( B_{2r}(x) , \mu_\beta \right) }  + \sum_{y \in \cu_{n+2} \setminus B_{2r}(x)} e^{-c \left( \ln \beta \right) |y -x|}   \left\| w (y , \cdot) \right\|_{L^2\left( \mu_\beta \right) }.
\end{multline}
We then note that, by the definition of the operator $\mathcal{L}_{\cu_{n+2}, *}$, the function $u$ satisfies the equation
\begin{equation*}
    \mathcal{L}_{\cu_{n+2}, *}  u = \nabla \cdot F + \di G ~\mbox{in}~ \cu_{n+2}^- \times \Omega,
\end{equation*}
where the vector fields $F$ and $G$ are defined by the formulas
\begin{equation*}
    F :=  - \frac 1{2\beta}\sum_{k \geq \dist \left( x , \partial \cu_{n+2} \right)} \frac1{\beta^{\frac k2}} (-\Delta)^k \nabla u \hspace{5mm} \mbox{and} \hspace{5mm} G := \sum_{\supp q \not\subseteq \cu_{n+2}} \a_q \left( \nabla_q u \right) n_q.
\end{equation*}
We estimate the $L^2\left( \cu_{n+2}^- , \mu_\beta \right)$-norm of the functions $F$ and $G$. We first note that every point $x$ in the cube $\cu_{n+2}^-$ satisfies the inequality $\dist (x , \partial \cu_{n+2}) \geq c 3^{\frac n2}$. Using the boundedness of the discrete Laplacian operator, the upper bound on the $L^2$-norm of the gradient of the function $u$ stated in~\eqref{eq:uppboundoptimi} and choosing the inverse temperature $\beta$ large enough, we have
\begin{align} \label{eq:TV18083}
    \left\| F  \right\|_{\underline{L}^2 \left( \cu_{n+2}^- ,  \mu_\beta \right)} & = \left\| \sum_{k \geq \dist \left( x , \partial \cu_{n+2} \right)} \frac1{\beta^{\frac k2}}(- \Delta)^k \nabla u  \right\|_{\underline{L}^2 \left( \cu_{n+2}^- ,  \mu_\beta \right)} \\
    & \leq   \sum_{k \geq \dist \left( x , \partial \cu_{n+2} \right)} \frac1{\beta^{\frac k2}} \left\|\Delta^k \nabla u  \right\|_{\underline{L}^2 \left( \cu_{n+2}^- ,  \mu_\beta \right)} \notag \\
    & \leq   \sum_{k \geq \dist \left( x , \partial \cu_{n+2} \right)} \frac{C^k}{\beta^{\frac k2}} \left\| \nabla u  \right\|_{\underline{L}^2 \left( \cu_{n+2}^- ,  \mu_\beta \right)} \notag \\
    & \leq \frac{1}{1 - \frac{C}{\beta^{\frac 12}}}\left( \frac{C}{\beta^{\frac12}} \right)^{c3^{\frac n2}} \left\| \nabla u \right\|_{\underline{L}^2 \left(\cu_{n+2}^- , \mu_\beta \right)} \notag \\ & \leq C e^{-c \left( \ln \beta \right) 3^{\frac n2}}. \notag
\end{align}
Using a similar argument, we note that for each point $x$ in the interior cube $\cu_{n+1}^-$, if a charge $q \in \mathcal{Q}$ is such that its support is not included in the cube $\cu_{n+1}$ and such that the point $x$ belongs to the support of $n_q$, then its diameter must be larger than $c 3^{\frac n2}$. We can then use the estimate on the coefficient $\a_q$ stated in \eqref{e.aest} and the estimate~\eqref{eq:uppboundoptimi} to obtain
\begin{equation} \label{eq:TV18084}
    \left\| G \right\|_{\underline L^2 \left( \cu_{n+2}^- , \mu_\beta \right)} = \left\| \sum_{\supp q \not\subseteq \cu_{n+2}} \a_q \left( \nabla_q u \right) n_q \right\|_{\underline L^2 \left( \cu_{n+2}^- , \mu_\beta \right)} \leq C e^{- c \sqrt{\beta} 3^{\frac n2}} \left\| \nabla u \right\|_{\underline L^2 \left( \cu_{n+2}^- , \mu_\beta \right)} \leq  C e^{- c \sqrt{\beta} 3^{\frac n2}}.
\end{equation}
We now apply the Caccioppoli inequality ~\eqref{est:Caccio.HS}, Chapter~\ref{section:section4}, to the function $w := u - v$ which is solution of the equation $\mathcal{L}_{\cu_{n+2}, *} \left( u - v \right) = \nabla \cdot F + \di G$ in the set $\cu_{n+1}^- \times \Omega$. We obtain
\begin{align} \label{eq:TV18343}
    \lefteqn{\beta \sum_{y \in \Zd} \left\| \partial_y \left( u - v  \right) \right\|_{\underline{L}^2 \left(\cu_n , \mu_\beta \right)} + \left\|\nabla \left( u - v  \right) \right\|_{\underline{L}^2\left( \cu_n , \mu_\beta \right) } } \hspace{15mm} & \\ & \leq \underbrace{ C3^{-2n} \left\| u - v \right\|_{\underline{L}^2\left( \cu_{n+1}^- , \mu_\beta \right) }^2}_{\eqref{eq:TV18343}-(i)}  + \underbrace{\left\| F \right\|_{\underline{L}^2\left( \cu_{n+1}^- , \mu_\beta \right) }^2 + \left\| G \right\|_{\underline{L}^2\left( \cu_{n+1}^- , \mu_\beta \right) }^2 }_{\eqref{eq:TV18343}-(ii)} \notag \\ & \quad  + \underbrace{ \left(\sum_{x \in \cu_{n+1}^- \setminus \cu_n } e^{-c \left( \ln \beta \right) |x|}   \left\| u(x , \cdot) - v (x , \cdot) \right\|_{L^2\left( \mu_\beta \right) }\right)^2}_{\eqref{eq:TV18343}-(iii)}. \notag
\end{align}
We estimate the term~\eqref{eq:TV18343}-(i) thanks to the inequality~\eqref{eq:TV18283}. We obtain
\begin{equation} \label{eq:TV18353}
    C3^{-2n} \left\| u - v \right\|_{\underline{L}^2\left( \cu_{n+1}^- , \mu_\beta \right) }^2  \leq  C  \left( 3^{-\frac{n}{2}}+ \sum_{m = 0}^{n+1} 3^{-\frac{m -n}{2}} \tau_m \right).
\end{equation}
We estimate the term~\eqref{eq:TV18343}-(ii) by the inequalities~\eqref{eq:TV18083} and~\eqref{eq:TV18084}. We obtain
\begin{equation} \label{eq:TV1835344}
    \left\| F \right\|_{\underline{L}^2\left( \cu_{n+1}^- , \mu_\beta \right) }^2 +  \left\| G \right\|_{\underline{L}^2\left( \cu_{n+1}^- , \mu_\beta \right)} \leq C e^{-c \left( \ln \beta \right) 3^{\frac n2}}.
\end{equation}
For the term~\eqref{eq:TV18343}-(iii), we use the estimate~\eqref{eq:TV18283}, the observation $\tau_n \leq C$ and note that if a point $x$ lies outside the cube $\cu_{n}$ then its norm must be larger than $c3^n$. We obtain
\begin{align}  \label{eq:TV18443}
    \sum_{x \in \cu_{n+1} \setminus \cu_n } e^{-c \sqrt{\beta} |x|}   \left\|(u - v) (x , \cdot) \right\|_{L^2\left( \mu_\beta \right) } & \leq C e^{-c \sqrt \beta 3^n}\sum_{x \in \cu_{n+1}  }   \left\| \nabla (u - v) (x , \cdot) \right\|_{L^2\left( \mu_\beta \right) } \\ 
    & \leq C e^{-c \sqrt \beta 3^n} 3^{\frac{dn}{2}} \left\| u - v \right\|_{\underline L^2\left( \cu_{n+1}, \mu_\beta \right) } \notag \\
    & \leq  C e^{-c \sqrt \beta 3^n} 3^{n \left(\frac{d}{2} + 1 \right)} \notag \\
    & \leq C e^{-c \sqrt \beta 3^n}. \notag
\end{align}
Combining the estimates~\eqref{eq:TV18343},~\eqref{eq:TV18353},~\eqref{eq:TV1835344} and~\eqref{eq:TV18443} completes the proof of Lemma~\ref{e.lemma4.3}.
\end{proof}

\subsection{Quantitative rate of convergence for the energy \texorpdfstring{$J$}{16}}
In this section, we use Lemma~\ref{e.lemma4.3} together with an iteration argument to obtain an algebraic rate of convergence for the quantity $J \left( \cu_n, p , \ahom_n p \right)$. The strategy implemented in the proof is essentially the one described in the paragraph following Proposition~\ref{prop:mainpropsect5} up to a technical difficulty: the term in the right side of the estimate~\eqref{eq:Lemma4.3} of Lemma~\ref{e.lemma4.3} is not the subadditivity defect $\tau_n$ but a weighted average the subadditivity defects. This additional technicality requires to make use of a weighted quantity denoted by $\tilde F_n$ in the proof below.

\begin{proposition} \label{prop:prop5.10}
There exist a constant $C := C(d) < \infty$ and an exponent $\alpha := \alpha (d) > 0$ such that for each integer $n \in \N$ and each $p \in \R^{d \times \binom d2}$,
\begin{equation*} 
 \nu  \left( \cu_n^-, p\right) + \nu^* \left( \cu_n , \ahom_n p \right) - \ahom_n |p|^2 \leq C |p|^2 3^{-\alpha n}.
\end{equation*}
\end{proposition}

We record, as a corollary, that the quantitative rate of convergence established in Proposition~\ref{prop:prop5.10} implies a quantitative estimate on the subadditivity defect $\tau_n$. The result is stated below.

\begin{corollary}  \label{cor:cor5.23}
There exist a constant $C := C(d) < \infty$ and an exponent $\alpha := \alpha (d) > 0$ such that for each integer $n \in \N$,
\begin{equation} \label{eq:quantesttaun}
 -C 3^{-\frac n2} \leq \tau_n \leq C 3^{-\alpha n}.
\end{equation}
\end{corollary}

\begin{proof}[Proof of Proposition~\ref{prop:prop5.10} and Corollary~\ref{cor:cor5.23}]
For $k\in\N$, we let $B_1(\R^k)$ be the unit ball in $\R^k$. We denote by $C_0$ the constant which appears in the right side of the identity~\eqref{eq:onesidedconvexdual} and define, for each integer $n \in \N$, 
\begin{equation*}
 F_n := \sup_{p \in B_1 \left(\R^{d \times \binom d2} \right)}  \nu  \left( \cu_n^-, p\right) + \nu^* \left( \cu_n , \ahom_n p \right) -  \ahom_n  |p|^2 + C_0 |p|^2 3^{-\frac n2}.
\end{equation*}
We note that by the inequality~\eqref{eq:TV08234}, we have the upper bound, for each integer $n \in \N$, $F_n \leq C$. By Proposition~\ref{prop:prop5.2} and Lemma~\ref{e.lemma4.3}, we have for each integer $n \in \N$,
\begin{equation} \label{eq:TV08244}
 0 \leq F_n \leq C \left(3^{-\frac{n}{2}} + \sum_{m = 0}^{n+1} 3^{- \frac{n- m}{2}} \tau_m \right).
\end{equation}
Additionally, we obtain from the subadditivity properties stated in Propositions~\ref{p.subadd} and~\ref{p.subaddnustar} the inequality 
\begin{equation} \label{eq:TV18300501}
F_{n+1} \leq F_n + C 3^{-\frac n2}.
\end{equation}
Combining the estimates~\eqref{eq:TV08244} and~\eqref{eq:TV18300501} implies that
\begin{equation*}
    0 \leq F_{n+1} \leq C \left(3^{-\frac{n+1}{2}} + \sum_{m = 0}^{n+1} 3^{- \frac{n- m}{2}} \tau_m \right).
\end{equation*}
By definition of the subadditivity defect $\tau_n$, and the fact that the maps $p \to  \nu  \left( \cu_n, p\right) - \nu  \left( \cu_{n+1},p\right) + C |p|^2 3^{-\frac n2}$ and $p^* \to  \nu  \left( \cu_n, p^*\right) - \nu  \left( \cu_{n+1},p^*\right) + C \left|p^*\right|^2 3^{-\frac n2}$ are quadratic and non-negative, we have
\begin{align} \label{eq:TV08254}
\tau_n & \leq C \sum_{k=1}^d \left( \nu  \left( \cu_n, e_k\right) - \nu  \left( \cu_{n+1},e_k\right) + \nu^*  \left( \cu_n, e_k\right) - \nu^* \left( \cu_{n+1},e_k\right) \right) + C 3^{- \frac n2}\\
		& \leq C \left( F _n -  F_{n+1}  + 3^{-\frac{n}{2}}\right). \notag
\end{align}
We then define
$
\tilde F_n := 3^{-\frac{n}{4}} \sum_{k=0}^n 3^{ \frac{k}4}  F_k.
$
From the estimates~\eqref{eq:TV08244} and~\eqref{eq:TV08254} and the inequality $F_0 \leq C$, we deduce
\begin{align} \label{eq:TV10054}
\tilde F_{n} - \tilde F_{n+1}  = 3^{-\frac{n}4 } \sum_{k = 0}^n 3^{\frac k4 } \left( F_{k} - F_{k+1} \right) - 3^{-\frac{(n + 1)}4 } F_0 
& \geq 3^{-\frac{n}4} \sum_{k = 0}^n 3^{\frac k4 } \left( \frac{1}{C} \tau_k - 3^{-\frac k2}\right) - C 3^{-\frac{n}4 }   \\
& \geq  \frac{1}{C} \sum_{k = 0}^n 3^{- \frac{(n -k)}4 } \tau_k - \sum_{k = 0}^n 3^{- \frac{(n -k)}4 } 3^{-\frac k2} - C 3^{-\frac{n}4 } \notag \\
    & \geq \frac{1}{C} \sum_{k = 0}^n 3^{- \frac{(n -k)}4 } \tau_k - C 3^{-\frac{n}4}. \notag
\end{align}
We then compute by using the inequalities~\eqref{eq:TV18300501} and~\eqref{eq:TV08244}
\begin{align*} 
    \tilde F_{n+1}  = 3^{-  \frac{n+1}4 } \sum_{k=0}^{k+1} 3^{ \frac k4}  F_k  = 3^{-  \frac n4} \sum_{k=0}^n 3^{ \frac k4}  F_{k+1} +  3^{-  \frac {n+1}4} F_0 
        & \leq 3^{-  \frac n4} \sum_{k=0}^n 3^{ \frac k4}  \left( F_{k} +  C 3^{-\frac n2}\right) +  C 3^{-  \frac n4} \\
        & \leq \tilde F_n + C 3^{-\frac n4}.
\end{align*}
We then use the estimate~\eqref{eq:TV08244} and write
\begin{align} \label{eq:TV10064}
    \tilde F_{n+1}  \leq  3^{-  \frac n4} \sum_{k=0}^n 3^{ \frac k4}  F_{k} + C 3^{-\frac n4} 
    & \leq  3^{-  \frac n4} \sum_{k=0}^n 3^{ \frac k4} \left(C 3^{-\frac k2} + \sum_{m = 0}^k 3^{- \frac{k- m}{2}} \tau_m \right) + C 3^{-\frac n4}  \\
    & \leq C 3^{-  \frac n4} \sum_{k=0}^n 3^{ - \frac k4} +  3^{-  \frac n4} \sum_{k=0}^n 3^{ - \frac k4} \sum_{m = 0}^k 3^{- \frac m2} \tau_m + C 3^{-\frac n4} \notag \\
    & \leq C  \sum_{k=0}^n 3^{ - \frac{n-k}4} \tau_k + C 3^{-\frac n4}. \notag
\end{align}
By combining the estimates~\eqref{eq:TV10054} and~\eqref{eq:TV10064}, we have obtained
\begin{equation*}
     \tilde F_{n+1} \leq C \left( \tilde F_{n} - \tilde F_{n+1} \right) + C 3^{-\frac n4}.
\end{equation*}
The previous inequality can be rewritten
\begin{equation} \label{eq:TV10254}
    \tilde F_{n+1} \leq \frac{C}{C+1} \tilde F_n + C 3^{-\frac n4}.
\end{equation}
We set $\alpha_0 := \frac{1}{\ln 3} \ln \frac{C}{C+1}$ so that we have $3^{\alpha_0} = \frac{C}{C+1}$ and define the exponent $\alpha := \min \left(\alpha_0, \frac{1}8 \right)$. We iterate the inequality~\eqref{eq:TV10254} and note that the inequality $F_0 \leq C$ implies the inequality $\tilde F_0 \leq C$. We obtain
\begin{equation*}
     \tilde F_{n} \leq 3^{- \alpha_0 n} \tilde{F}_0 + C \sum_{k = 0}^n 3^{- \alpha_0 k} 3^{-\frac{n-k}{4}} \leq C 3^{-\alpha n} .
\end{equation*}
Finally, by the definition of the weighted sum $\tilde F_n$, we have the inequality $F_n \leq \tilde F_n$. The proof of Proposition~\ref{prop:prop5.10} is complete.

There only remains to prove Corollary~\ref{cor:cor5.23}. The lower bound in~\eqref{eq:quantesttaun} is a direct consequence subadditivity properties stated in Propositions~\ref{p.subadd} and~\ref{p.subaddnustar}. For the upper bound, we use the inequality~\eqref{eq:TV08254} together with the estimates $F_n \leq C3^{-\alpha n}$ and $F_{n+1} \geq 0$.
\end{proof}

\subsection{Quantitative rate of convergence for the subadditive quantities \texorpdfstring{$\nu$}{17} and \texorpdfstring{$\nu^*$}{18}} In this section, we deduce Proposition~\ref{prop:mainpropsect5} from Proposition~\ref{prop:prop5.10}.

\begin{proof}[Proof of Proposition~\ref{prop:mainpropsect5}]
Before starting the proof, we collect some ingredients which were proved in this chapter:
\begin{itemize}
\item By Proposition~\ref{prop:prop5.2} and Definition~\ref{defapproxhomogcoeff}, we have the identities, for each integer $n \in \N$ and each $p, p^* \in \Rd$,
\begin{equation} \label{eq:TV11254}
    \nu \left( \cu_n^- , p \right) = \frac 12 p \cdot \a \left( \cu_n^- \right) p \hspace{5mm} \mbox{and} \hspace{5mm}  \nu^* \left( \cu_n , p^* \right) = \frac 12 p^* \cdot \ahom_n^{-1} p^*;
\end{equation}
\item By Property (4) of Proposition~\ref{prop:prop5.2}, there exist two strictly positive constants $c , C$ depending only on the dimension $d$ such that, for every cube $\cu \subseteq \Zd$,
\begin{equation} \label{eq:TV11304}
 c \leq \a(\cu), \a_* (\cu) \leq C;
 \end{equation}
\item By Corollaries~\ref{cor:coro5.13} and~\ref{cor:coro5.15}, we have the convergences
\begin{equation} \label{eq:TV11264}
         \a \left( \cu_n^- \right) \underset{n \to \infty}{\longrightarrow} \ahom \hspace{5mm} \mbox{and} \hspace{5mm} \ahom_n^{-1} \underset{n \to \infty}{\longrightarrow} \ahom_*^{-1};
\end{equation}
\item By the one sided convex duality estimate~\eqref{eq:onesidedconvexdual} and Proposition~\ref{prop:prop5.10}, we have the inequalities, for each $p \in \R^{d \times \binom d2}$,
\begin{equation*}
     -C |p|^2 3^{-\frac n2} \leq \nu  \left( \cu_n, p\right) + \nu^* \left( \cu_n , \ahom_n p \right) - \ahom_n |p|^2 \leq C |p|^2 3^{-\alpha n},
\end{equation*}
which can be rewritten, by using~\eqref{eq:TV11254},
\begin{equation}  \label{eq:TV11274}
    -C 3^{-\frac n2} \leq \a \left( \cu_n^- \right) - \ahom_n \leq C 3^{-\alpha n};
\end{equation}
    \item By Lemma~\ref{lem:lemma5.6} and Corollary~\ref{cor:cor5.23}, we have the inequality, for each pair of integers $(m,n) \in \N$ such that $m \leq n$,
    \begin{equation} \label{eq:TV11284}
        \left| \ahom^{-1}_n  -\ahom^{-1}_m  \right|^2  \leq \sum_{k=m}^n \tau_k + C 3^{-\frac{m}2}
    \leq \sum_{k=m}^n C 3^{-\alpha k}+ C 3^{-\frac{m}2} \leq C 3^{-\alpha m}.
    \end{equation}
\end{itemize}
We now combine the four previous results to complete the proof of Proposition~\ref{prop:mainpropsect5}. First by sending $n$ to infinity in the inequality~\eqref{eq:TV11274} and using the convergence~\eqref{eq:TV11264}, we obtain the identity $\ahom = \ahom_*^{-1}$. Then by sending $n$ to infinity in the inequality~\eqref{eq:TV11284}, we obtain the inequality, for each integer $m \in \N$,
\begin{equation} \label{eq:TV12114}
    \left| \ahom^{-1}_m  -\ahom^{-1}  \right| \leq C 3^{-\alpha m}.
\end{equation}
We then combine the inequality~\eqref{eq:TV11304} with the inequality~\eqref{eq:TV12114} to obtain
\begin{equation} \label{eq:TV12364}
    \left| \ahom_m  -\ahom  \right| \leq C 3^{-\alpha m}.
\end{equation}
Combining the estimate~\eqref{eq:TV12364} with the estimate~\eqref{eq:TV11274} and using that the exponent $\alpha$ is smaller than $\frac 12$, we deduce that, for each integer $n \in \N$,
\begin{equation} \label{eq:TV12354}
    \left| \ahom \left( \cu_n^- \right) - \ahom \right| \leq \left| \ahom \left( \cu_n^- \right) - \ahom_n \right| + \left| \ahom_n - \ahom \right| \leq C 3^{-\alpha n}.
\end{equation}
Proposition~\ref{prop:prop5.2} is then a consequence of the estimates~\eqref{eq:TV12364},~\eqref{eq:TV12354} and the representation formulas~\eqref{eq:TV11254}.
\end{proof}

\section{Definition of the first-order  corrector and quantitative sublinearity} \label{sec:section5.4}
An important ingredient to prove the quantitative homogenization of the mixed derivative of the Green's matrix associated to the Helffer-Sj{\"o}strand operator $\mathcal{L}$ (which is the subject of Section~\ref{sec:section6}) is the first-order corrector. The objective of this section is to introduce this function and to deduce from the algebraic rate of convergence on the energy $\nu$ established in Proposition~\ref{prop:mainpropsect5} two properties on this map:
\begin{itemize}
    \item The quantitative sublinearity of the corrector, this is stated in the equation~\eqref{eq:quantsublin};
    \item A quantitative estimate on the $H^{-1}$-norm of the flux of the corrector, this is stated in the estimate~\eqref{eq:fluxsublin}.
\end{itemize}
The corrector which is introduced in this section is a finite-volume version of the corrector (see Definition~\ref{def.finivolcorr}), the reason justifying this choice is that it is is simpler to construct from the subadditive energy $\nu$ than the infinite-volume corrector and allows the arguments developed in Chapter~\ref{sec:section6} to work. We do not try to construct the infinite-volume corrector as it would require to prove a quantitative homogenization theorem and establish a large-scale regularity theory (following the techniques of~\cite[Chapter 3]{AKM}) and the development of this technology is unnecessary to prove Theorem~\ref{t.main}. Nevertheless, the specific structure of the problem (and the strong regularity properties established in Chapter~\ref{section:section4}) allows to define the gradient of the infinite-volume corrector; the construction is carried out in Proposition~\ref{prop5.26}.

\subsection{Finite-volume corrector}

This section is devoted to the definition and the study of the finite-volume corrector.

\begin{definition}[Finite-volume corrector] \label{def.finivolcorr}
For each integer $m \in \N$, and each slope $p \in \R^{d \times \binom{d}{2}}$, we define the finite-volume corrector at scale $3^n$ to be the function $\chi_{n,p} : \Zd \times \Omega \to \R^{\binom d2}$ defined by the formula
\begin{equation*}
    \chi_{n,p} := u \left(\cdot, \cdot,  \cu_n^-, p \right) - l_p.
\end{equation*}
We recall that the corrector is equal to $0$ outside the trimmed cube $\cu_n^-$. Given two integers $i \in \{ 1, \ldots , d \}$ and $j \in \{ 1, \ldots , \binom d2 \}$, we denote by $e_{ij} \in \R^{d \times \binom d2}$ the vector
\begin{equation*}
    e_{ij} = \left( 0 , \ldots, e_i, \ldots, 0 \right),
\end{equation*}
where the vector $e_i \in \Rd$ appears at the $j$-th position. We note that the collection of vectors $(e_{ij})_{1 \leq i \leq d, \, 1 \leq j \leq \binom d2}$ is a basis of the vector space $\R^{d \times \binom d2}$. We frequently refer to the corrector $\chi_{n, e_{ij}}$ by the notation $\chi_{n, ij}$.
\end{definition}

\begin{remark} \label{remark4.20302}
The finite volume corrector $\chi_{n,p}$ is the solution of the equation
\begin{equation*}
    \left\{ \begin{aligned}
     \Delta_\phi \chi_{n , p} - \frac{1}{2\beta} \Delta \chi_{n , p} + \frac{1}{2\beta}\sum_{n \geq 1} \frac{1}{\beta^{ \frac n2}} \left(-\Delta\right)^{n+1} \chi_{n , p} + \sum_{q \in \mathcal{Q}} \nabla_q^* \cdot \a_q \nabla_q \left(l_p + \chi_{n , p} \right) &= 0 ~\mbox{in}~ \cu_n^{-}, \\
     \chi_{n,p} & = 0 ~\mbox{on}~ \partial \cu_n^{-}.
    \end{aligned} \right.
\end{equation*}
By the identity $\nabla_q \left(l_p + \chi_{n , p} \right)$, we see that the corrector depends only on the value of $\di^* l_p$. In particular, if $\di^* l_p = 0$ then $\chi_{n , p} = 0$.
\end{remark}

The following proposition establishes quantitative sublinearity of the corrector and provides a quantitative estimate for the $H^{-1}$-norm of its flux.

\begin{proposition}[Quantitative sublinearity] \label{prop:prop5.25}
There exist a constant $C := C(d)$, an exponent $\alpha(d) > 0$ and an inverse temperature $\beta_0(d) < \infty$ such that for every inverse temperature $\beta > \beta_0$ and every vector $p \in \R^{d \times \binom d2}$, the finite-volume corrector satisfies the following estimates
\begin{equation} \label{eq:quantsublin}
    \left\| \chi_{n,p} \right\|_{\underline{L}^2 \left( \cu_n^-, \mu_\beta \right)} \leq C |p| 3^{(1- \alpha) n}
\end{equation}
and 
\begin{equation} \label{eq:fluxsublin}
    \left\| \frac 1{2} \left( p +  \nabla \chi_{n,p} \right) + \beta \sum_{q \in \mathcal{Q}} \a_q \nabla_q \left( l_p + \chi_{n,p} \right) L^{t}_{2, \di^*} \left( n_q \right) - \ahom p \right\|_{\underline{H}^{-1}\left( \cu_n^- , \mu_\beta \right)  } \leq C|p| 3^{(1 - \alpha) n}.
\end{equation}
\end{proposition}

\begin{proof}
The estimate~\eqref{eq:quantsublin} is obtained by combining Lemma~\ref{lem:lemma5.8} and Corollary~\ref{cor:cor5.23}. The proof of the estimate~\eqref{eq:fluxsublin} regarding the flux is more involved and we split the argument into two steps. The argument requires to take into account the infinite range of the sum over the charges (by using the boundary layer $BL_n$ and the exponential decay of the coefficient $\a_q$), which makes the proof technical. Since similar technicalities have already been treated in the previous sections and the analysis does not contain any new arguments, we omit some of the details and only write a (detailed) sketch of the proof. 

\medskip

\textit{Step 1.} In this step, we prove that, to prove~\eqref{eq:fluxsublin} is is sufficient to prove the estimate, for each $p^* \in \R^{d\times \binom d2}$,
\begin{equation} \label{eq:TV13357}
    \left\| \frac 1{2} \nabla v \left( \cdot , \cdot, \cu_n , p^* \right) +  \beta \sum_{\supp q \subseteq \cu_n} \a_q \nabla_q  v \left( \cdot , \cdot, \cu_n , p^* \right) L^{t}_{2, \di^*} \left( n_q \right) - p^* \right\|_{\underline{H}^{-1}\left( \cu_n^- , \mu_\beta \right)  } \leq C|p^*| 3^{(1 - \alpha) n}.
\end{equation}
We fix a vector $p^* \in \R^{d \times \binom 2d}$ and recall that, by definition of the first order corrector, $l_p + \chi_{n,p} = u \left( \cdot , \cdot , \cu_n^- , p \right)$. To ease the notation, we denote by $u := u \left( \cdot , \cdot , \cu_n^- , p \right)$ and by $v := v\left( \cdot , \cdot , \cu_n, \ahom p \right)$. First, we note that Proposition~\ref{prop:prop5.2} implies the inequality $\left| \a\left( \cu_m^- \right) - \ahom \right| \leq C 3^{-\alpha m}$. Combining this result with the estimate~\eqref{eq:uppboundoptimi}, we obtain the inequality, for each vector $p \in \R^{d \times \binom d2}$,
\begin{equation} \label{eq:TV13429}
    \left\| \nabla v \left( \cdot , \cdot , \cu_n , \ahom p \right) - \nabla v \left( \cdot , \cdot , \cu_n , \ahom_n p \right) \right\|_{\underline{L}^2 \left( \cu_n , \mu_\beta \right)} = \left\| \nabla v \left( \cdot , \cdot , \cu_n , \ahom p - \ahom_n p \right) \right\|_{\underline{L}^2 \left( \cu_n , \mu_\beta \right)} \leq C 3^{-\alpha n} |p|.
\end{equation}
We use the inequality~\eqref{eq:TV13429} with the estimate~\eqref{eq:TV14093} stated in the proof of Proposition~\ref{e.lemma4.3} and Corollary~\ref{cor:cor5.23}. We deduce that
\begin{align} \label{eq:TV18157}
     \left\| \nabla u - \nabla v  \right\|_{\underline{L}^2 \left( \cu_n , \mu_\beta \right)} &  \leq \left\| \nabla u -  \nabla v \left( \cdot , \cdot , \cu_n , \ahom_n p \right) \right\|_{\underline{L}^2 \left( \cu_n , \mu_\beta \right)} + \left\| \nabla v \left( \cdot , \cdot , \cu_n , \ahom_n p \right) -  \nabla v \right\|_{\underline{L}^2 \left( \cu_n , \mu_\beta \right)} \\ & \leq C 3^{-\alpha n}|p|. \notag
\end{align}
Using the estimate~\eqref{eq:TV18157}, we can write
\begin{align*}
    \lefteqn{\left\| \frac 1{2} \nabla u +  \beta \sum_{q \in \mathcal{Q}} \a_q \left( \nabla_q  u \right)  L^{t}_{2, \di^*} \left( n_q \right) - \ahom p \right\|_{\underline{H}^{-1}\left( \cu_n^- , \mu_\beta \right)}} \qquad & \\ &  \leq  \left\| \frac 1{2} \nabla v  +  \beta \sum_{q \in \mathcal{Q}} \a_q \left( \nabla_q  v\right)   L^{t}_{2, \di^*} \left( n_q \right) - \ahom p \right\|_{\underline{H}^{-1}\left( \cu_n^- , \mu_\beta \right)} \\ & \quad + \left\| \frac 1{2 } \nabla \left( u - v \right)  + \beta \sum_{q \in \mathcal{Q}} \a_q \left( \nabla_q \left( u -  v \right)\right)   L^{t}_{2, \di^*} \left( n_q \right)  \right\|_{\underline{H}^{-1}\left( \cu_n , \mu_\beta \right)} \\ 
    & \leq  \left\| \frac 1{2} \nabla v  +  \beta \sum_{q \in \mathcal{Q}} \a_q \left( \nabla_q  v\right)   L^{t}_{2, \di^*} \left( n_q \right) - p \right\|_{\underline{H}^{-1}\left( \cu_n , \mu_\beta \right)} \\ & \quad + CR \left\| \frac 1{2} \nabla \left( u - v \right)  +  \beta \sum_{q \in \mathcal{Q}} \a_q \left( \nabla_q \left( u -  v \right)\right)   L^{t}_{2, \di^*} \left( n_q \right) - p \right\|_{\underline{L}^{2}\left( \cu_n , \mu_\beta \right)}.
\end{align*}
Using the estimate \eqref{e.aest} on the coefficient $\a_q$, we see that
\begin{equation*}
   \left\| \frac 1{2} \nabla \left( u - v \right)  +  \beta \sum_{q \in \mathcal{Q}} \a_q \left( \nabla_q \left( u -  v \right)\right)   L^{t}_{2, \di^*} \left( n_q \right) - p \right\|_{\underline{L}^{2}\left( \cu_n , \mu_\beta \right)} \leq C \left\| \nabla \left( u - v \right)  \right\|_{L^2 \left( \cu_n , \mu_\beta  \right)} \leq C 3^{-\alpha n} |p|.
\end{equation*}
A combination of the two previous displays shows
\begin{multline} \label{eq:TV14309}
    \left\| \frac 1{2} \nabla u +  \beta \sum_{q \in \mathcal{Q}} \a_q \left( \nabla_q  u \right) n_q - p \right\|_{\underline{H}^{-1}\left( \cu_n , \mu_\beta \right)} \\ \leq  \left\| \frac 1{2 } \nabla v  + \beta \sum_{q \in \mathcal{Q}} \a_q \left( \nabla_q  v\right)  L^{t}_{2, \di^*} \left( n_q \right) - \ahom p \right\|_{\underline{H}^{-1}\left( \cu_n , \mu_\beta \right)} + C3^{(1-\alpha) n} |p|.
\end{multline}
The estimate~\eqref{eq:TV14309} implies that to prove the inequality~\eqref{eq:fluxsublin}, it is sufficient to prove~\eqref{eq:TV13357}.

\medskip

\textit{Step 2. Proving the estimate~\eqref{eq:TV13357}.} The argument is similar to the proof presented in Lemma~\ref{lem:lemma5.8}. To ease the notation, we denote by $v :=  v \left( \cdot , \cdot, \cu_n , p^* \right)$ and by $v_{z , m} := v \left( \cdot , \cdot, z + \cu_m , p^* \right)$ and assume without loss of generality that $\left| p^* \right| =1$. We use the $H^{-1}$-version of the multiscale Poincar\'e inequality stated in Proposition~\ref{prop:multiscPoin} of Appendix~\ref{section:multiscPoinc}. We obtain
\begin{align} \label{eq:TV20439}
    \lefteqn{\left\| \frac 1{2} \nabla v  +  \beta \sum_{ q \in \mathcal{Q}} \a_q \nabla_q  v  L^{t}_{2, \di^*} \left( n_q \right) - p^* \right\|_{\underline{H}^{-1}\left( \cu_n^- , \mu_\beta \right)  }} \qquad & \\ & \leq C \left\|  \frac 1{2} \nabla v  +  \beta \sum_{q \in \mathcal{Q}} \a_q \nabla_q  v   L^{t}_{2, \di^*} \left( n_q \right) - p^* \right\|_{\underline{L}^2 \left( \cu_n^- , \mu_\beta \right)} \notag \\
    & \quad + C 3^n \sum_{m = 0}^n \sum_{z \in \mathcal{Z}_{m,n}}\frac{3^m}{\left|\mathcal{Z}_{m,n} \right|} \left\langle \left( \frac{1}{\left| z + \cu_m \right|}\sum_{x \in z + \cu_m}  \frac 1{2} \nabla v(x , \cdot) +  \beta \sum_{q \in \mathcal{Q}} \a_q \nabla_q  v  L^{t}_{2, \di^*} \left( n_q(x) \right) - p^* \right)^2 \right\rangle_{\mu_\beta}.  \notag
\end{align}
The first term in the right side of~\eqref{eq:TV20439} can be estimated by the estimate~\eqref{eq:uppboundoptimi}. We obtain
\begin{equation} \label{eq:TV21299}
     \left\|  \frac 1{2} \nabla v  +  \beta \sum_{q \in \mathcal{Q}} \a_q \nabla_q  v   L^{t}_{2, \di^*} \left( n_q \right) - p^* \right\|_{\underline{L}^2 \left( \cu_n^- , \mu_\beta \right)} \leq C.
\end{equation}
To estimate the second term in the right side of~\eqref{eq:TV20439}, we proceed as in Lemma~\ref{lem:lemma5.8} and use the subadditivity estimate stated in Proposition~\ref{p.subaddnustar} and Corollary~\ref{cor:coro5.15}. We obtain
\begin{align} \label{eq:TV21309}
  \lefteqn{\sum_{z \in \mathcal{Z}_{m,n}}\frac{1}{\left|\mathcal{Z}_{m,n} \right|} \left\langle \left( \frac{1}{\left| z + \cu_m \right|}\sum_{x \in z + \cu_m}  \frac 1{2} \nabla v(x , \cdot)  +  \beta \sum_{q \in \mathcal{Q}} \a_q \nabla_q  v   L^{t}_{2, \di^*} \left( n_q \right)- p^* \right)^2 \right\rangle_{\mu_\beta} } \qquad & \\ & \leq \sum_{z \in \mathcal{Z}_{m,n}}\frac{1}{\left|\mathcal{Z}_{m,n} \right|} \left\langle \left( \frac{1}{\left| z + \cu_m \right|}\sum_{x \in z + \cu_m}  \frac 1{2} \nabla v_{z , m}(x , \cdot)  +  \beta \sum_{q \in \mathcal{Q}} \a_q \nabla_q  v_{z , m}  L^{t}_{2, \di^*} \left( n_q(x) \right) - p^* \right)^2 \right\rangle_{\mu_\beta}  + C 3^{- \alpha m }  .\notag
\end{align}
We then use the two following results:
\begin{itemize}
    \item One has the identity, for each point $z \in \mathcal{Z}_{m,n}$,
    \begin{equation*}
        \left\langle \frac{1}{\left| z + \cu_m \right|}\sum_{x \in z + \cu_m} \left(  \frac 1{2 } \nabla v_{z , m}(x , \cdot) + \beta \sum_{q \in \mathcal{Q}} \a_q \nabla_q  v_{z  ,m}  L^{t}_{2, \di^*} \left( n_q(x)\right) \right)   \right\rangle_{\mu_\beta} = p^*;
    \end{equation*}
    \item By Lemma~\ref{lemmvarest}, the inequality $d - \frac 52 \geq \frac 12$ valid in dimension larger than $3$ and the translation invariance of the measure $\mu_\beta$, one has the variance estimate
    \begin{equation*}
        \var \left[ \frac{1}{\left| z + \cu_m \right|}\sum_{x \in z + \cu_m} \left( \frac 1{2} \nabla v_{z ,m}(x , \cdot) +  \beta \sum_{q \in \mathcal{Q}} \a_q \nabla_q  v_{z , m}   L^{t}_{2, \di^*} \left( n_q(x) \right) \right) \right] \leq C  3^{-\frac m2}.
    \end{equation*}
\end{itemize}
We obtain the estimate
\begin{equation} \label{eq:TV21269}
    \left\langle \left( \frac{1}{\left| z + \cu_m \right|}\sum_{x \in z + \cu_m}  \frac 1{2} \nabla v_{z , m}(x , \cdot)  +  \beta \sum_{q \in \mathcal{Q}} \a_q \nabla_q  v_{z , m}  L^{t}_{2, \di^*} \left( n_q(x) \right) - p^* \right)^2 \right\rangle_{\mu_\beta} \leq C 3^{-\frac m2}.
\end{equation}
Combining the estimates~\eqref{eq:TV20439},~\eqref{eq:TV21299},~\eqref{eq:TV21309} and~\eqref{eq:TV21269}, we have obtained
\begin{equation*}
    \left\| \frac 1{2} \nabla v +  \beta \sum_{ q \in \mathcal{Q}} \a_q \nabla_q  v L^{t}_{2, \di^*} \left( n_q \right) - p^* \right\|_{\underline{H}^{-1}\left( \cu_n^- , \mu_\beta \right)}  \leq C 3^{(1-\alpha)n}.
\end{equation*}
The proof of Proposition~\ref{prop:prop5.25} is complete. 
\end{proof}

\subsection{Gradient of the infinite-volume corrector}

The next proposition establishes the existence and stationarity of the spatial gradient of the infinite-volume corrector.

\begin{proposition}[Existence of the gradient of the infinite-volume corrector and stationarity] \label{prop5.26}
There exists a stationary random field $\nabla \chi : \Zd \times \Omega \to \R$ satisfying the following property, for each $p \in \Rd$ and each integer $n \in \N$,
\begin{equation*}
    \left\| \nabla \chi_{n,p} - \nabla \chi_p \right\|_{\underline{L}^2 \left( \cu_m , \mu_\beta \right)} \leq C 3^{-n \alpha}.
\end{equation*}
\end{proposition}

\begin{remark} \label{remark09110302}
The property stated in Remark~\ref{remark4.20302} about the finite volume corrector also applies to the infinite volume corrector: the function $\nabla \chi_p$ depends only on the value of $\di^* l_p$. AS the vectors $\di^* l_p$ belong to the space $\Rd$, the collection of correctors $\left( \chi_p \right)_{p \in \R^{d \times \binom d2}}$ forms a $d$-dimensional vector space from which we extract a basis: for each integer $i \in \{ 1 , \ldots, d \}$, we select a vector $p \in \R^{d \times \binom d2}$ such that $\di^* l_p = e_i$ and denote by $\nabla \chi_{i} = \nabla \chi_{p_i}$.
\end{remark}

Let us first present the main idea of the argument. By assuming that the inverse temperature is large enough, one has $C^{0 , 1- \ep}$-regularity estimates for the solutions of the Helffer-Sj{\"o}strand equation, following the arguments given in Section~\ref{sec:section4.2} of Chapter~\ref{section:section4}. By Proposition~\ref{prop:mainpropsect5}, one also has an algebraic rate of convergence for the subadditive energy $\nu$ with exponent $\alpha$. The exponent $\ep$ depends on the inverse temperature $\beta$ and tends to $0$ as $\beta$ tends to infinity while the exponent $\alpha$ depends only on the dimension and remains unchanged by sending the inverse temperature to infinity. In other words, as the inverse temperature $\beta$ tends to infinity, the regularity exponent $\ep$ tends to $0$ and the exponent $\alpha$ remains bounded away from $0$. It is thus possible to choose $\beta$ sufficiently large so that the exponent $\ep$ is smaller than the exponent $\frac\alpha2$ and to leverage on this property, the $C^{0,1-\ep}$-regularity estimate presented in Proposition~\ref{prop:prop4.5} of Chapter~\ref{section:section4} and the Caccioppoli inequality to prove the existence of the gradient of the infinite-volume corrector.

\begin{proof}
We fix a vector $p \in \R^{d \times \binom d2}$ and assume without loss of generality that $|p|=1$. We decompose the proof into two steps. In the first step, we prove that for each point $x \in \Zd$, the sequence $\left( \nabla \chi_{n , p}(x) \right)_{n \in \N}$ is Cauchy in the space $L^2 \left( \mu_\beta \right)$. This implies that it converges and we define the gradient of the infinite-volume corrector to be its limit. In the second step we prove that the function $\nabla \chi_p$ is stationary.

\medskip

\textit{Step 1.} We prove the inequality, for each point $x \in \Zd$ integer $n \in \N$ such that $x \in \cu_n^-$,
\begin{equation} \label{eq:TV22044}
    \left\| \nabla \chi_{n,p}(x, \cdot) - \nabla \chi_{n+1,p}(x,\cdot) \right\|_{\underline{L}^2 \left(  \mu_\beta \right)} \leq C 3^{- \frac\alpha2 n }.
\end{equation}
We now fix a point $x \in \Zd$ and prove the estimate~\eqref{eq:TV22044}.
By the definition of the correctors stated in Definition~\ref{def.finivolcorr}, and the definition of the function $u$ as the minimizer in the definition of the energy quantity $\nu$ given in~\eqref{def:defnu}, we see that for each integer $n \in \N$, the functions $\chi_n$ and $\chi_{n+1}$ are solutions of the Helffer-Sj{\"o}strand equations
    \begin{equation*}
        \mathcal{L} \left( l_p + \chi_{n,p} \right) = 0 ~\mbox{in}~ \cu_{n}^- \times \Omega ~\mbox{and}~\mathcal{L} \left( l_p + \chi_{n+1,p} \right) = 0 ~\mbox{in}~ \cu_{n+1}^- \times \Omega .
    \end{equation*}
    In particular, the difference $\chi_{n+1,p} - \chi_{n,p}$ is solution of the equation $\mathcal{L} \left( \chi_{n+1,p} - \chi_{n,p} \right) = 0$ in the set $ \cu_m^- \times \Omega $. We can thus apply Proposition~\ref{prop:prop4.6} of Chapter~\ref{section:section4} to obtain, for each integer $n \in \N$ such that $x \in \cu_{n}^-$,
    \begin{align} \label{eq:TV22364}
        \left\| \nabla \chi_{n,p}(x, \cdot) - \nabla \chi_{n+1 , p}(x , \cdot) \right\|_{L\left( \mu_\beta \right)} & \leq \sup_{y \in \cu_{n}^-} \left\| \nabla \chi_{n,p} \left( y , \cdot \right) - \nabla \chi_{n+1 , p}  \left( y , \cdot \right) \right\|_{L^2(\mu_\beta)}  \\
        & \leq C3^{(\ep-1)n} \left\|  \chi_{n,p} - \chi_{n+1 , p} - \left( \chi_{n,p} - \chi_{n+1 , p} \right)_{\cu_n^-} \right\|_{\underline{L}^2 \left( \cu_n^- , \mu_\beta \right)} \notag \\
        & \leq  C3^{(\ep-1)n} \left\|  \chi_{n,p} - \chi_{n+1 , p} \right\|_{\underline{L}^2 \left( \cu_n^- , \mu_\beta \right)}. \notag
    \end{align}
By combining the estimate~\eqref{eq:TV22364} and Proposition~\ref{prop:prop5.25}, we obtain the estimate, for each pair of integers $n \in \N$ such that $x \in \cu_n^-$,
\begin{equation*}
    \left\| \nabla \chi_{n,p} \left( x , \cdot \right) - \nabla \chi_{n+1 , p} \left( x , \cdot \right) \right\|_{L^2\left( \mu_\beta \right)} \leq C 3^{(\ep - \alpha) n} . 
\end{equation*}
Using the assumption $\ep \leq \frac\alpha2$, we obtain
\begin{equation} \label{eq:TV22414}
    \left\| \nabla \chi_{n,p} \left( x , \cdot \right) - \nabla \chi_{n+1 , p} \left( x , \cdot \right) \right\|_{L^2\left( \mu_\beta \right)} \leq C 3^{ - \frac\alpha2 n}.
\end{equation}
The inequality~\eqref{eq:TV22414} implies that, the sequence $\left( \nabla \chi_{n,p}  \right)_{n \in \N}$ is Cauchy in the space $L^2\left(\mu_\beta\right)$. This implies that it converges in the space $L^2\left( \mu_\beta\right)$. We define the gradient of the corrector $\nabla \chi_p(x)$ to be the limiting object. 

From the estimate~\eqref{eq:TV22414}, we also deduce that it satisfies the inequality, for each pair of integers $n \in \N$,
\begin{equation*}
    \left\| \nabla \chi_{n,p} \left( x , \cdot \right) - \nabla \chi_{p} \left( x , \cdot \right) \right\|_{L^2\left( \mu_\beta \right)} \leq C 3^{ - \frac\alpha2 n}.
\end{equation*}
The proof of Step 1 is complete. \medskip

\textit{Step 2.} In this step, we prove the stationarity of the infinite-volume gradient corrector. For $z \in \Zd$, we recall the notation $\tau_z$ for the translation of the field introduced in Chapter~\ref{Chap:chap2}. We prove the identity, for each $(x, \phi) \in \Zd \times \Omega$,
\begin{equation} \label{eq:sttphipro}
    \nabla \chi_p \left(  x , \phi \right) = \nabla \chi_p \left( z + x , \tau_{z} \phi \right).
\end{equation}
To prove the equality~\eqref{eq:sttphipro}, we first note that, by the definition of the function $u$, we have the equality, for each point $z \in \Zd$, each cube $\cu \subseteq \Zd$, and each pair $(x , \phi) \in \left( y + \cu \right) \times \Omega$,
\begin{equation} \label{eq:TV8115}
    u \left( x , \phi , y + \cu , p\right) = u \left( x - y , \tau_{-y} \phi , \cu , p  \right).
\end{equation}
Using the identity~\eqref{eq:TV8115}, the result established in Step 1 and the translation invariance of the measure $\mu_\beta$, we obtain that the sequence $\left( \nabla u \left( x , \cdot , y + \cu_n , p\right) - p \right)_{n \in \N}$ converges in the space $L^2 \left( \mu_\beta \right)$ to the random variable $\phi \to \nabla \chi_p \left( x-y , \tau_{-y} \phi  \right)$. Thus to prove the identity~\eqref{eq:sttphipro}, it is sufficient to prove that the sequence $\left( \nabla u \left( x , \cdot , y + \cu_n , p\right) - p \right)_{n \in \N}$ also converges in $L^2 \left(\mu_\beta \right)$ to the gradient of the corrector $\phi \to \nabla \chi_p(x, \phi)$. This is what we now prove.

We first note that the proof Proposition~\ref{prop:prop5.2} can be adapted so as to have the following result. For each $y \in \Zd$ and each integer $n$ such that $3^{\frac n2} \geq 2 |y|$, one has the estimate
\begin{equation} \label{eq:TV10145}
    \sum_{z \in \mathcal{Z}_n} \left\|  \nabla u \left( \cdot  , \cdot , y + z + \cu_n , p\right) - \nabla u \left( \cdot  , \cdot , \cu_{n+1} , p\right)   \right\|_{\underline{L}^2 \left( y + z + \cu_n , \mu_\beta \right)}^2 \leq C \left( \nu \left( \cu_n,p \right) - \nu \left( \cu_{n+1}, p \right) + 3^{-\frac n2} \right).
\end{equation}
The proof is identical; indeed under the assumption $3^{\frac n2} \geq 2|y|$, one can partition the triadic cube $\left( y + \cu_{n+1}\right)$ into the collection of triadic cubes $\left(y + z + \cu_n \right)_{z \in \mathcal{Z}_n}$ and a boundary layer of width of size $3^{\frac n2}$. One can then rewrite the proof of Proposition~\ref{prop:prop5.2} to obtain the estimate~\eqref{eq:TV10145}. We then use Proposition~\ref{prop:prop5.2} (or more precisely Corollary~\ref{cor:cor5.23}) and obtain the inequality
\begin{equation*}
    \left\|  \nabla u \left( \cdot  , \cdot , y + \cu_n , p\right) - \nabla  u \left( \cdot  , \cdot , \cu_{n+1} , p\right) \right\|_{\underline{L}^2 \left( y + \cu_n , \mu_\beta \right)}^2 \leq C 3^{- \alpha n}.
\end{equation*}
Using the $C^{1-\ep}$-regularity estimate stated in Proposition~\ref{prop:prop4.6}, the assumption $\ep \leq \frac \alpha2$ and an argument similar to the one presented in Step 1, we obtain, for each integer $n \in \N$ such that $3^{\frac n2} \geq 2|y|$ and each point $x \in \cu_{n}^-$,
\begin{equation*}
    \left\|  \nabla u \left( x  , \cdot , y + \cu_n , p\right) - \nabla u \left( x  , \cdot , \cu_{n+1} , p \right) \right\|_{L^2 \left( \mu_\beta \right)}^2 \leq C  3^{-\frac \alpha2 n}.
\end{equation*}
Using the definition of the finite-volume corrector given in Definition~\ref{def.finivolcorr} and the inequality~\eqref{eq:TV22044}, we deduce that
\begin{equation*}
     \left\|  \nabla u \left( x  , \cdot , y + \cu_n , p\right) - p - \nabla \chi_p \left( x  , \cdot \right) \right\|_{L^2 \left( \mu_\beta \right)}^2 \leq C  3^{-\frac \alpha2 n}.
\end{equation*}
The previous inequality implies that the sequence $\left( \nabla u \left( x , \cdot , y + \cu_n , p\right) - p \right)_{n \in \N}$ converges in the space $L^2 \left( \mu_\beta \right)$ to the random variable $\phi \to \nabla \chi_p \left( x ,\phi  \right)$. The proof of Proposition~\ref{prop5.26} is complete.
\end{proof}

\chapter{Quantitative homogenization of the Green's matrix} \label{sec:section6}

\section{Statement of the main result}
The objective of this chapter is to prove the homogenization of the mixed gradient of the Green's matrix. We first introduce the notation $\ahom_\beta := \frac{\ahom}{\beta}$ and the Green's matrix associated to the homogenized operator $\nabla \cdot \ahom_\beta \nabla$: we denote by $\bar G : \Zd \to \R^{\binom d2 \times \binom d2}$ the fundamental solution of the elliptic system
\begin{equation} \label{def.barGTV145288}
    -\nabla \cdot \ahom_\beta \nabla  \bar G = \delta_0 ~\mbox{in}~\Zd.
\end{equation}
The matrix $\ahom_\beta$ is a small perturbation of the matrix $\frac{1}{2\beta} I_d$ and the size of the perturbation is of order $\beta^{-\frac 32} \ll \beta^{-1}$. The solvability of the equation is thus ensured by the arguments of Chapter~\ref{section:section4}; more specifically, a Nash-Aronson estimate holds for the heat-kernel associated to the operator $-\nabla \cdot \ahom_\beta \nabla$ (which implies the solvability by an integration in time of the heat in dimension larger than $3$).

The following theorem is the main result of this chapter.

\begin{theorem}[Homogenization of the mixed derivative of the Green's matrix] \label{thm:homogmixedder}
Fix a charge $q_1 \in \mathcal{Q}$ such that $0$ belongs to the support of $n_{q_1}$ and let $\mathcal{U}_{q_1}$ be the solution of the Helffer-Sj{\"o}strand equation
\begin{equation} \label{eq:defmathcalU}
\L \mathcal{U}_{q_1} = \cos 2\pi\left( \phi , q_1\right) q_1 ~\mbox{in}~\Zd \times \Omega.
\end{equation}
for each integer $k \in \left\{ 1 , \ldots , \binom d2 \right\}$, we define the function $\bar G_{q_1, k} : \Zd \to \R$ by the formula
\begin{equation*}
    \bar G_{q_1, k} = \sum_{1 \leq i \leq d} \sum_{1 \leq j\leq \binom d2} \left\langle \cos 2\pi\left( \phi , q_1\right) \left( n_{q_1} , \di^* l_{e_{ij}} + \di^* \chi_{ij} \right)   \right\rangle_{\mu_\beta} \nabla_i \bar G_{jk}.
\end{equation*}
Then there exist an inverse temperature $\beta_0 := \beta_0(d) <\infty$, an exponent $\gamma := \gamma(d) >0$ and a constant $C_{q_1}$ which satisfies the estimate
$\left| C_{q_1} \right| \leq  C \left\| q_1 \right\|_1^{k}$ for some $C := C(d, \beta) <\infty$ and $k := k(d) < \infty$, such that for each $\beta \geq \beta_0$ and each radius $R \geq 1$, one has the inequality
\begin{equation} \label{eq:TV19253}
    \left\| \nabla \mathcal{U}_{q_1} - \sum_{1 \leq i \leq d} \sum_{1 \leq j\leq \binom d2} \left( e_{ij} + \nabla \chi_{ij}\right) \nabla_i \bar G_{q_1, j}\right\|_{\underline{L}^2 \left( A_R , \mu_\beta \right)} \leq \frac{C_{q_1}}{R^{d + \gamma}}.
\end{equation}
\end{theorem}

\begin{remark}
Since the codifferential $\di^*$ is a linear functional of the gradient, the map $\di^* \chi_{ij}$ is well-defined even if we have only defined the gradient of the infinite-volume corrector: we have the formula $\di^* \chi_{ij} :=  L_{2, \di^*} \left( \nabla \chi_{ij} \right)$.
\end{remark}

\begin{remark}
We recall that in this chapter, the constants are allowed to depend on the dimension $d$ and on the inverse temperature $\beta$.
\end{remark}

\begin{remark}
We recall the definition of the annulus $A_R := B_{2R} \setminus B_R$. The volume of the annulus $A_R$ is of order $R^d$.
\end{remark}

\begin{remark}
    The double sum $\sum_{1 \leq i\leq d} \sum_{1 \leq j\leq \binom d2}$ appears frequently in the proofs of this chapter; to ease the notation, we denote it by $\sum_{i,j}$.
\end{remark}

\begin{remark}
Since the the form $q_1$ can be written $\di n_{q_1}$, we expect the two gradients $\nabla \mathcal{U}_{q_1}$ and $\nabla \bar G_{q_1}$ to behave like mixed derivatives of the Green's matrix, i.e., they should be of order $R^{-d}$ in the annulus $A_R$. The proposition asserts that the difference between the two terms $\nabla \mathcal{U}_{q_1}$ and $\sum_{i,j} \left( e_{ij} + \nabla \chi_{ij} \right) \nabla_i \bar G_{q_1 , j}$ is quantitatively smaller than the typical size of the two terms considered separately.
\end{remark}

\section{Outline of the argument}

The strategy of the proof of Theorem~\ref{thm:homogmixedder} relies on a classical strategy in homogenization: the two-scale-expansion. The proofs presented in the chapter make essentially use of two ingredients established in Chapters~\ref{section:section4} and~\ref{section5}:
\begin{itemize}
    \item The quantitative sublinearity of the finite-volume corrector and the estimate on the $H^{-1}$-norm of the flux stated in Proposition~\ref{prop:prop5.25} of Chapter~\ref{section5};
    \item The $C^{0 , 1-\ep}$-regularity theory established in Chapter~\ref{section:section4}.
\end{itemize}
We now give an outline of the proof of Theorem~\ref{thm:homogmixedder}. The argument is split into two sections:
\begin{itemize}
    \item In Section~\ref{sec:section6.1}, we perform the two-scale expansion and obtain a result of homogenization for the gradient of the Green's matrix as stated in Proposition~\ref{prop:prop6.1};
    \item In Section~\ref{sec:section6.2}, we use the result of Proposition~\ref{prop:prop6.1} and perform the two-scale expansion a second time to obtain the quantitative homogenization of the mixed derivative of the Green's matrix stated in Theorem~\ref{thm:homogmixedder}.
\end{itemize}

\subsection{Homogenization of the gradient of the Green's matrix}
In Section~\ref{sec:section6.1}, we prove the homogenization of the gradient of the Green's matrix stated in Proposition~\ref{prop:prop6.1} below.

\begin{proposition}[Homogenization of the Green's matrix] \label{prop:prop6.1}
Let $\mathcal{G}: \Zd \times \Omega \to \R^{\binom d2 \times \binom d2}$ be the Green's matrix associated to the Helffer-Sj{\"o}strand equation
\begin{equation} \label{def:GreenGf19}
    \mathcal{L}  \mathcal{G} =  \delta_0 \hspace{3mm} \mbox{in} \hspace{3mm} \Zd \times \Omega.
\end{equation}
Then there exist an inverse temperature $\beta_0(d) < \infty$, an exponent $\gamma := \gamma(d) > 0$ and a constant $C:=C(d) < \infty$ such that for any $\beta > \beta_0$, any radius $R \geq 1$, for any of integer $k \in \left\{  1 , \ldots , \binom d2 \right\}$,
\begin{equation} \label{eq:TVmainprop6.2}
\left\| \nabla \mathcal{G}_{\cdot k} - \sum_{i,j} \left( e_{ij} + \nabla \chi_{ij} \right) \nabla_i \bar G_{jk} \right\|_{\underline{L}^2 \left(A_R,  \mu_\beta\right)} \leq \frac{C}{R^{ d-1 + \gamma}}.
\end{equation}
\end{proposition}

The argument relies on a two-scale expansion. To set up the argument, we first select an inverse temperature $\beta$ large enough, depending only on the dimension $d$, such that the quantitative sublinearity of the finite-volume corrector and of its flux stated in Proposition~\ref{prop:prop5.25} of Chapter~\ref{section5} hold with exponent $\alpha > 0$. Following the argument explained at the beginning of Section~\ref{sec:section5.4} there, we can choose the parameter $\beta$ large enough so that all the results presented in Chapter~\ref{section:section4} pertaining to the $C^{0,1-\ep}$-regularity theory for the Helffer-Sj{\"o}strand operator $\mathcal{L}$ are valid with a regularity exponent $\ep$ which is small compared to the exponent $\alpha$ (we assume for instance that the ratio between $\ep$ and $\alpha$ is smaller than $100d^2$). We also fix an exponent $\delta$ which is both larger than $\ep$ and smaller than $\alpha$ and corresponds to the size of a mollifier exponent which needs to be taken into account in the argument (we assume for instance that the ratios between the exponents $\alpha$ and $\delta$ and between the exponents $\ep$ and $\delta$ are both smaller than $10d$). To summarize, we have three exponents in the argument, which can be ordered by the following relations 
\begin{equation} \label{eq:TVexpord}
    0 < \underbrace{\ep}_{\mathrm{regularity}} \ll \underbrace{\delta}_{\mathrm{mollifier}\, \mathrm{exponent}}  \ll  \underbrace{\alpha}_{\mathrm{homogenization}} \ll 1.
\end{equation}
We additionally assume that the exponents $\ep$, $\delta$ and $\alpha$ are chosen in a way that they depend only on the dimension $d$. The exponent $\gamma$ in the statement of Proposition~\ref{prop:prop6.2} depends only $\ep$, $\delta$ and $\alpha$ (and thus only on the dimension $d$).

We now give an outline of the proof of the inequality~\eqref{eq:TVmainprop6.2}; the details can be found in Step 1 of the proof of Proposition~\ref{prop:prop6.2}. The first step of the argument is to approximate the Green's matrices $\mathcal{G}$ and $\bar G$; the main issue is that the spatial Dirac function $\delta_0$ in the definitions of the Green's matrices $\mathcal{G}$ in~\eqref{def:GreenGf19} and $\bar G$ in~\eqref{def.barGTV145288} is too singular and causes some problems in the analysis. To remedy this issue, we replace the Dirac function $\delta_0$ by a smoother function and make use of the boundary layer exponent $\delta$: we let $\rho_\delta$ be a discrete function from $\Zd$ to $\R^{\binom d2 \times \binom d2}$, we denote its components by $\left( \rho_{\delta, ij} \right)_{1 \leq i , j \leq \binom d2}$ and assume that they satisfy the four properties
\begin{equation} \label{prop.rhodelta}
\supp \rho_\delta \subseteq B_{R^{1 - \delta}}, ~ 0 \leq \rho_{\delta, ij}  \leq C R^{-(1-\delta) d } , ~\sum_{x \in \Zd} \rho_{\delta, ij}(x) = \indc_{\{i=j\}}, ~\mbox{and} ~\forall k \in \N, ~\left| \nabla^k \rho_{\delta, ij} \right| \leq \frac{C}{R^{(d + k)(1 - \delta)}},
\end{equation}
and define the functions $ \mathcal{G}_\delta : \Zd \times \Omega \to \R^{\binom d2 \times \binom d2}$ and $\bar G_\delta:  \Zd  \to \R^{\binom d2 \times \binom d2}$ to be the solution of the systems, for each integer $k \in \left\{1 , \ldots , \binom d2 \right\}$
\begin{equation} \label{def:GdeltaGhomdelta}
    \L \mathcal{G}_{\delta, \cdot k} = \rho_{\delta, \cdot k} ~\mbox{in} ~\Omega \times \Zd, \hspace{5mm} - \nabla \cdot \left( \ahom_\beta \nabla  \bar G_{\delta, \cdot k} \right) =  \rho_{\delta, \cdot k} ~\mbox{in} ~\Zd,
\end{equation}
We then prove, by using the $C^{0,1-\ep}$-regularity theory established in Section~\ref{section3.4} of Chapter~\ref{section:section4}, that the functions $\mathcal{G}_\delta$, $\bar G_\delta$ are good approximations of the functions
$\mathcal{G}$, $\bar G$. This is the subject of Lemma~\ref{prop:prop6.2} where we prove that for $\beta$ sufficiently large, there exists an exponent $\gamma := \gamma(d , \beta , \delta, \ep) > 0$ such that 
\begin{equation} \label{eq:TV0421}
    \left\| \nabla \mathcal{G}_\delta - \nabla \mathcal{G} \right\|_{L^\infty \left( A_R , \mu_\beta \right)} \leq CR^{1 - d - \gamma} \hspace{5mm} \mbox{and} \hspace{5mm} \left\| \nabla \bar G_\delta - \nabla \bar G \right\|_{L^\infty \left( A_R , \mu_\beta \right)} \leq CR^{1 - d - \gamma}.
\end{equation}
By the estimates~\eqref{eq:TV0421}, we see that to prove Proposition~\ref{prop:prop6.2} it is sufficient to prove the inequality, for each integer $k \in \{1 , \ldots , \binom d2 \}$,
\begin{equation} \label{eq:TV1013}
    \left\| \nabla \mathcal{G}_{\delta, \cdot k} - \sum_{i,j} \left( e_{ij} + \nabla \chi_{ij} \right) \nabla_i \bar G_{\delta, jk} \right\|_{\underline{L}^2 \left( A_R , \mu_\beta \right)} \leq C R^{1 - d - \gamma}.
\end{equation}
We now sketch the proof of the inequality~\eqref{eq:TV1013}. We let $m$ be the integer uniquely defined by the inequalities $3^m \leq R^{1 + \delta} < 3^{m+1}$, consider the collection finite-volume correctors $\left( \chi_{m , ij} \right)_{1 \leq i \leq d, 1 \leq j \leq \binom d2}$. We then define the two-scale expansion $\mathcal{H}_\delta : \Zd \times \Omega \to \R^{\binom d2 \times \binom d2}$ according to the formula, for each integer $k \in \left\{ 1 , \ldots , \binom d2 \right\}$,
\begin{equation}  \label{def.Ho2sc}
    \mathcal{H}_{\delta,\cdot k} := \bar G_{\delta,\cdot k} + \sum_{i,j} \left( \nabla_i \bar G_{\delta, j k} \right) \chi_{m , ij}.
\end{equation}
We fix an integer $k \in \left\{ 1 , \ldots , \binom d2 \right\}$. The strategy is to compute the value of $\mathcal{L} \mathcal{H}_{\delta, \cdot k}$ by using the explicit formula on $\mathcal{H}_{\delta, \cdot k}$ stated in~\eqref{def.Ho2sc} and to prove that it is quantitatively close to the map $\rho_{\delta, \cdot k}$ in the correct functional space; precisely, we prove the $H^{-1}$-estimate, for each integer $k \in \left\{1 , \ldots, \binom d2 \right\}$
\begin{equation} \label{eq:insertineq}
    \left\| \mathcal{L} \mathcal{H}_{\delta, \cdot k} - \rho_{\delta, \cdot k} \right\|_{\underline{H}^{-1} \left( B_{R^{1+\delta} , \mu_\beta} \right)} \leq C R^{1-d  - \gamma}.
\end{equation}
Obtaining this result relies on the quantitative behavior of the corrector and of the flux established in Proposition~\ref{prop:prop5.25} of Chapter \ref{section5}. Once one has a good control over the $H^{-1}$-norm of $\mathcal{L} \mathcal{H}_{\delta, \cdot k} - \rho_{\delta, \cdot k}$, the inequality~\eqref{eq:TV1013} can be deduced from the following two arguments:
\begin{itemize}
    \item We use that the function $\mathcal{G}_{\delta, \cdot k}$ satisfies the equation $\mathcal{L} \mathcal{G}_{\delta, \cdot k} = \rho_{\delta, \cdot k}$ to obtain that the $H^{-1}$-norm of the term $\mathcal{L} \left( \mathcal{H}_{\delta, \cdot k} - \mathcal{G}_{\delta, \cdot k} \right)$ is small. We then introduce a cutoff function $\eta : \Zd \to \R$ which satisfies:
    \begin{equation*}
    \supp \eta \subseteq A_R, ~ 0 \leq \eta \leq 1 , ~\eta = 1 ~\mbox{on}~ \left\{ x \in \Zd \, : \, 1.1 R \leq |x| \leq 1.9 R \right\}, ~\mbox{and} ~\forall k \in \N, ~\left| \nabla^k \eta \right| \leq \frac{C}{R^k},
\end{equation*}
    and use the function $\eta \left( \mathcal{H}_{\delta, \cdot k} - \mathcal{G}_{\delta, \cdot k} \right)$ as a test function in the definition of the $H^{-1}$-norm of the inequality~\eqref{eq:insertineq}. We obtain that the $L^2$-norm of the difference $\left( \nabla \mathcal{H}_{\delta, \cdot k} - \nabla \mathcal{G}_{\delta, \cdot k} \right)$ is small (the cutoff function is used to ensure that the function $\eta \left( \mathcal{H}_{\delta, \cdot k} - \mathcal{G}_{\delta, \cdot k} \right)$ is equal to $0$ on the boundary of the ball $B_{R^{1+\delta}}$ and can thus be used as a test function). The precise estimate we obtain is written below
    \begin{equation} \label{eq:TV15488de}
        \left\| \nabla \mathcal{H}_{\delta, \cdot k} - \nabla \G_{\delta, \cdot k} \right\|_{L^2 \left( \Zd , \mu_\beta \right)} \leq CR^{\frac d2 + 1 - d - \gamma};
    \end{equation}
    \item By using the formula~\eqref{def.Ho2sc}, we can compute an explicit formula for the gradient of the two-scale expansion $\mathcal{H}_{\delta, \cdot k}$. We then use the quantitative sublinearity of the corrector stated in Proposition~\ref{prop:prop5.25}, Chapter~\ref{section5} and the property of the gradient of the infinite volume corrector stated in Proposition~\ref{prop5.26} to deduce that, for each integer $k \in \{1, \ldots, \binom d2 \}$, the $L^2$-norm of the difference $\nabla \mathcal{H}_{\delta , \cdot k} - \sum_{i,j} ( e_{ij} + \nabla \chi_{ij}) \nabla_i \bar G_{jk}$ is small; the precise result we obtain is the following
    \begin{equation} \label{eq:TV09109}
        \left\| \nabla \mathcal{H}_{\delta , \cdot k} - \sum_{i,j} ( e_{ij} + \nabla \chi_{ij}) \nabla_i \bar G_{jk} \right\|_{L^2 \left( \Zd  \mu_\beta \right)} \leq C R^{\frac d2 + 1 - d - \gamma}.
    \end{equation}
\end{itemize}
The inequality~\eqref{eq:TV1013} is then a consequence of the inequalities~\eqref{eq:TV15488de} and~\eqref{eq:TV09109}.

\subsection{Homogenization of the mixed derivative of the Green's matrix} \label{sec3.2chap60851}
In Section~\ref{sec:section6.2}, we use Proposition~\ref{prop:prop6.1} to prove Theorem~\ref{thm:homogmixedder}. The argument is decomposed into three steps:
\begin{itemize}
    \item In Step 1, we use Proposition~\ref{prop:prop6.1} and the symmetry of the Helffer-Sj{\"o}strand operator $\mathcal{L}$ to prove the inequality in expectation
    \begin{equation} \label{eq:TV15057}
    \left( R^{-d} \sum_{z\in A_R} \left| \left\langle \mathcal{U}_{q_1}(z, \cdot ) \right\rangle_{\mu_\beta} - \bar G_{q_1}(z) \right|^2 \right)^\frac 12 \leq \frac{C}{R^{d - 1+\gamma}}.
    \end{equation}
    \item In Step 2, we prove the variance estimate, for each point $z \in \Zd$,
    \begin{equation} \label{eq:TV15087}
         \var \left[ \mathcal{U}_{q_1}(z, \cdot) \right] \leq \frac{C_{q_1}}{|z|^{2d-2\ep}}.
    \end{equation}
    Since we expect the function $z \mapsto \mathcal{U}_{q_1}(z)$ to decay like $|z|^{1-d}$; its variance should be of order $|z|^{2-2d}$. The estimate~\eqref{eq:TV15087} states that the variance of the random variable $\phi \to \mathcal{U}_{q_1}(z, \phi)$ is (quantitatively) smaller than its size; this means that the random variable $\mathcal{U}_{q_1}(z)$ concentrates on its expectation. We then use the result established in Step 1 to refine the result: since by~\eqref{eq:TV15057}, one knows that the expectation of $\mathcal{U}_{q_1}(z)$ is close to the function $\bar G_{q_1}$, one deduces that the function $\mathcal{U}_{q_1}$ is close to the function $\bar G_{q_1}$ in the $\underline{L}^2(A_R, \mu_\beta)$-norm. The precise estimate we obtain is the following
    \begin{equation} \label{eq:TV08289}
    \left\| \mathcal{U}_{q_1} - \bar G_{q_1} \right\|_{\underline{L}^2 \left( A_R , \mu_\beta \right)} \leq \frac{C_{q_1}}{R^{d-1-\gamma}}.
    \end{equation}
    The proof of the inequality~\eqref{eq:TV15087} does not rely on tools from stochastic homogenization; we appeal to the Brascamp-Lieb inequality and used the differentiated Helffer-Sj{\"o}strand introduced in Section~\ref{sec.section4.5} of Chapter \ref{section:section4}.
    \item In Step 3, we prove the estimate~\eqref{eq:TV19253}, the proof is similar to the argument presented in the proof of Proposition~\ref{prop:prop6.1} and relies on a two-scale expansion. The argument is split into three substeps. We first define the two-scale expansion $\mathcal{H}_{q_1}$ by the formula
    \begin{equation} \label{eq:defHq12sc5}
    \mathcal{H}_{q_1} := \bar G_{q_1} + \sum_{i,j} \nabla_i \bar G_{q_1,j} \chi_{m,ij}.
    \end{equation}
    We can then use that the function $\bar G_{q_1}$ is the solution to the homogenized equation $\ahom \Delta \bar G_{q_1} = 0$ in the annulus $A_R$ to prove that the $\underline{H}^{-1}\left(A_R , \mu_\beta \right)$-norm of the term $ \L \mathcal{H}_{q_1}$ over the annulus $A_R$ is small; this is the subject of Substep 3.1 where we prove
    \begin{equation} \label{eq:TV10438}
        \left\| \L \mathcal{H}_{q_1} \right\|_{\underline{H}^{-1}\left(A_R , \mu_\beta \right)} \leq \frac{C_{q_1}}{R^{d + \gamma}}.
    \end{equation}
    The proof is essentially a notational modification of the proof of the estimate~\eqref{eq:insertineq} (and is even simpler since we do not have to take into account the exponent $\delta$ and the function $\rho_\delta$). Once we have proved that the inequality~\eqref{eq:TV10438}, we use that the function $\mathcal{U}_{q_1}$ satisfies the identity $\L \mathcal{U}_{q_1} = 0$ in the set $A_R \times \Omega$ to deduce that the $\underline{H}^{-1}\left(A_R , \mu_\beta \right)$-norm of the term $\L \left( \mathcal{H}_{q_1} - \mathcal{U}_{q_1}\right) = \L  \mathcal{H}_{q_1}$ is small. We then introduce the annulus $A_R^1 := \left\{ x \in \Zd \, : \, 1.1 R \leq |x| \leq 1.9 R \right\}$ and a cutoff function $\eta$ satisfying the properties:
\begin{equation*}
    \supp \eta \subseteq A_R, ~ 0 \leq \rho \leq 1 , ~\eta = 1 ~\mbox{on}~ A_R^1, ~\mbox{and} ~\forall k \in \N, ~\left| \nabla^k \eta \right| \leq \frac{C}{R^k}.
\end{equation*}
We note that the choice of the values $1.1$ and $1.9$ in the definition of the annulus $A_R^1$ is arbitrary; any pair of real numbers belonging to the interval $(1,2)$ would be sufficient for our purposes. We then use the function $\eta \left( \mathcal{H}_{q_1} - \mathcal{U}_{q_1} \right)$ as a test function in the definition of the $\underline{H}^{-1}\left(A_R , \mu_\beta \right)$-norm of the term $\L \left( \mathcal{H}_{q_1} - \mathcal{U}_{q_1}\right)$ and use the $\underline{L}^2 \left( A_R , \mu_\beta \right)$-estimate~\eqref{eq:TV08289} to obtain that the $\underline{L}^2 \left( A_R^1 , \mu_\beta \right)$-norm of the difference $\nabla \mathcal{H}_{q_1} - \nabla \mathcal{U}_{q_1}$ is small. This is the subject of Substep 3.2 where we prove
\begin{equation} \label{eq:TV09249V}
    \left\| \nabla \mathcal{U}_{q_1} - \nabla \mathcal{H}_{q_1} \right\|_{\underline{L}^2 \left( A_R^1 , \mu_\beta \right)} \leq \frac{C_{q_1}}{R^{d+\gamma}}.
\end{equation}
\item Step 4 is the conclusion of the argument, we use the explicit formula for the two-scale expansion $\mathcal{H}_{q_1}$ given in~\eqref{eq:defHq12sc5}, the quantitative sublinearity of the corrector stated in Proposition~\ref{prop:prop5.25} of Chapter~\ref{section5} and the quantitative estimate for the difference of the finite and infinite-volume gradient of the corrector stated in Proposition~\ref{prop5.26} there to prove the estimate
\begin{equation} \label{eq:TV12347V}
    \left\| \nabla \mathcal{H}_{q_1} - \sum_{i,j} \left( e_{ij} + \nabla \chi_{ij} \right) \nabla_i \bar G_{q_1,j} \right\|_{\underline{L}^2 \left( A_R , \mu_\beta \right)} \leq \frac{C_{q_1}}{R^{d+ \gamma}}.
\end{equation}
The argument is a notational modification of the one used to prove~\eqref{eq:defHq12sc5}. We finally combine the estimates~\eqref{eq:TV09249V} and~\eqref{eq:TV12347V} to obtain the estimate~\eqref{eq:TV19253} and complete the proof of Theorem~\ref{thm:homogmixedder}.
\end{itemize}

\medskip

\section{Two-scale expansion and homogenization of the gradient of the Green's matrix} \label{sec:section6.1}

This section is devoted to the proof of Proposition~\ref{prop:prop6.1}. We collect some preliminary results in Section~\ref{premest3.1chap6} and prove Theorem~\ref{thm:homogmixedder} in Sections~\ref{sec3.2chap608511},~\ref{sec3.3chap60851} and~\ref{sec3.4chap60851} following the outline given in Section~\ref{sec3.2chap60851}.

\smallskip

\subsection{Preliminary estimates} \label{premest3.1chap6} In this section, we collect some preliminary properties which are used in the proof of Proposition~\ref{prop:prop6.1}.

\subsubsection{Notations for the exponent $\gamma$} \label{notexpgamma} We first introduce some notations for the exponent $\gamma$. As was already mentioned, this exponent depends on the parameters $\alpha, \delta$ and $\ep$; in the argument, we need to keep track of its typical size and proceed as follows:
\begin{itemize}
    \item We use the notation $\gamma_1$ when the exponent is of order $1$; a typical example is the exponent $\gamma_\alpha := 1 - c_0 \alpha - c_1 \delta - c_2 \ep$ for some constants $c_0 , c_1, c_2$ depending only on the dimension $d$;
    \item We use the notation $\gamma_{\alpha}$ when the exponent is of order $\alpha$; a typical example is the exponent $\gamma_\alpha := \alpha - c_0 \delta - c_1 \ep$ for some constants $c_0 , c_1$ depending only on the dimension $d$;
    \item We use the notation $\gamma_\delta$ when the exponent is of order $\delta$; a typical example is the exponent $\gamma_\delta := \delta -  c_0 \ep$ for some constant $c_0$ depending only on the dimension $d$.
\end{itemize}
We always have the ordering
\begin{equation*}
    0 < \gamma_\ep \ll \gamma_\delta \ll \gamma_\alpha \ll \gamma_1.
\end{equation*}
We also allow the value of the exponents $\gamma_\ep, \gamma_\delta, \gamma_\alpha, \gamma_1$ to vary from line to line in the argument as long as the order of magnitude is preserved. In particular, we may write
\begin{equation*}
    \gamma_1 = \gamma_1  - \alpha, ~\gamma_\alpha  = \gamma_\alpha - \delta ~\mbox{and}~ \gamma_\delta  = \gamma_\delta - \ep.
\end{equation*}

\subsubsection{Regularity estimates} In this section, we record some regularity estimates pertaining to the Green's matrices $\G$, $\G_\delta$, $\bar G$ and $\bar G_\delta$.

\begin{proposition} \label{prop:prop6.2}
The following properties hold:
\begin{itemize}
\item There exists an exponent $\gamma_\delta > 0$ such that one has the estimates
\begin{equation} \label{eq:TV20383}
\left\|\nabla \G (x, \cdot) - \nabla \G_\delta(x , \cdot ) \right\|_{L^\infty \left(A_R,  \mu_\beta \right)} \leq \frac{C}{R^{d-1 + \gamma_\delta}} \hspace{5mm} \mbox{and} \hspace{5mm}  \left\|\nabla \bar G - \nabla \bar G_\delta \right\|_{L^\infty (A_R)} \leq \frac{C}{R^{d-1 + \gamma_\delta}};
\end{equation}
\item The Green's matrix $\G_\delta$ satisfies the following $L^\infty$-estimates
\begin{equation} \label{eq:TV090191}
    \left\| \G_\delta  \right\|_{L^\infty \left( \Zd , \mu_\beta  \right)} \leq  \frac{C}{R^{(1-\delta) (d-2) }} ~\mbox{and}~ \left\| \nabla \G_\delta  \right\|_{L^\infty \left( \Zd , \mu_\beta  \right)} \leq \frac{C}{R^{(1-\delta) (d-1 - \ep) }},
\end{equation}
as well as the estimates
\begin{equation} \label{eq:TVdes13370}
    \left\| \G_\delta \right\|_{L^\infty \left( A_{R^{1+\delta}}, \mu_\beta  \right)} \leq \frac{C}{R^{(1+\delta)(d-2)}}  ~\mbox{and}~  \left\| \nabla \G_\delta \right\|_{L^\infty \left( A_{R^{1+\delta}}, \mu_\beta  \right)} \leq \frac{C}{R^{(1+\delta)(d-1- \ep)}};
\end{equation}
\item The homogenized Green's matrix $\bar G_\delta$ satisfies the regularity estimate, for each integer $k \in \N$,
    \begin{equation} \label{eq:spatregbarGgeq3}
         \left\| \nabla^k \bar G_\delta \right\|_{L^\infty \left( \Zd , \mu_\beta \right)} \leq \frac{C}{R^{(1-\delta)(d-2+k)}},
    \end{equation}
as well as the estimate
        \begin{equation} \label{eq:spatregbarGgeq34}
         \left\| \nabla^k \bar G_\delta \right\|_{L^\infty \left( A_{R^{1+\delta}} , \mu_\beta \right)} \leq \frac{C}{R^{(1+\delta)(d-2+k)}}.
    \end{equation}
\end{itemize}
\end{proposition}

\begin{proof}[Proof of Proposition~\ref{prop:prop6.2}]
The proof relies on the regularity estimates established in Chapter~\ref{section:section4}. We first note that, by definitions of the functions $\G $ and $\G_\delta$, we have the identities 
\begin{equation} \label{eq:TV13219}
    \G \left( x , \phi \right) = \G_1 \left( x , \phi ;0 \right) \hspace{5mm} \mbox{and} \hspace{5mm} \G_\delta \left( x , \phi \right) = \sum_{y \in B_{R^{1-\delta}}} \G_1\left( x , \phi ; y \right)  \rho_\delta(y),
\end{equation}
where the product in the right side of~\eqref{eq:TV13219} is the standard matrix product between $\G_1\left( x , \phi ; y \right)$ and $\rho_\delta(y)$.
Using that the map $\rho_\delta$ has total mass $1$ and the regularity estimate on the Green's matrix stated in Proposition~\ref{cor:corollary4.14} of Chapter \ref{section:section4}, we obtain, for each point $x$ in the annulus $A_R$,
\begin{align*}
    \left\| \nabla_x \G (x , \phi ; 0) - \nabla_x \G_\delta (x , \phi ; y) \right\|_{L^\infty \left( \mu_\beta \right)}  & \leq  \sum_{y \in B_{R^{1-\delta}}} \rho_\delta (y) \left\|  \nabla_x \G_1 (x ,  \phi ; 0 ) -  \nabla_x  \G_1 (x , \phi ; y)  \right\|_{L^\infty \left( \mu_\beta \right)} \\
    & \leq R^{1 - \delta} \sup_{y \in B_{R^{1- \delta}}} \left\| \nabla_x \nabla_y \G (x , \phi ; y ) \right\|_{L^\infty \left( \mu_\beta \right)} \\
    & \leq R^{1 - \delta} \sup_{y \in B_{R^{1-\delta}}}|x - y|^{- d - \ep}\\
    & \leq R^{1 - \delta} R^{- d - \ep}.
\end{align*}
 This computation implies the estimate~\eqref{eq:TV20383} with the exponent $\gamma_\delta = \delta - \ep$ which is strictly positive by the assumption~\eqref{eq:TVexpord}.
 
The estimate on the homogenized Green's matrix is similar and even simpler since we only have to work with the Green's matrix associated to the discrete Laplacian on $\Zd$ (and not the Helffer-Sj{\"o}strand operator $\L$); we omit the details.

The proof of the inequality~\eqref{eq:TV090191} relies on the estimates on the Green's matrix and its gradient established in Corollary~\ref{cor:cor5.23} of Chapter~\ref{section:section4}. We use the identity~\eqref{eq:TV13219} and write, for each point $x \in \Zd$,
\begin{align*}
    \left\| \G_\delta \left( x , \cdot \right) \right\|_{L^\infty \left( \mu_\beta \right)}  = \sum_{y \in B_{R^{1-\delta}}} \left| \rho_\delta(y)\right| \left\| \G_1 \left( x , \phi ; y \right)  \right\|_{L^\infty \left( \mu_\beta \right)} \leq \frac{1}{R^{(1-\delta) d }} \sum_{y \in B_{R^{1-\delta}}} \frac{C}{|x- y|^{d-2}} 
    & \leq  \frac{1}{R^{(1-\delta) d }} \sum_{y \in B_{R^{1-\delta}}} \frac{C}{| y|^{d-2}} \\
    & \leq \frac{1}{R^{(1-\delta) (d-2) }}.
\end{align*}
A similar computation shows the bound for the gradient of the Green's matrix and the bounds~\eqref{eq:TVdes13370} in the annulus $A_{R^{1+\delta}}$.

To prove the regularity estimate~\eqref{eq:spatregbarGgeq3}, we use the definition of the map $\bar G_\delta$ given in~\eqref{def:GdeltaGhomdelta} and note that
\begin{equation*}
    -\nabla \cdot  \ahom_\beta \nabla  \left(\nabla^k \bar G_\delta \right)  = \nabla^k \rho_\delta ~\mbox{in}~\Zd.
\end{equation*}
We then use the the properties of the function $\rho_\delta$ stated in~\eqref{prop.rhodelta} and the standard estimates on the homogenized Green's matrix $\bar G$. We obtain, for each point $x \in \Zd$,
\begin{equation*} 
    \left| \nabla^k \bar G_\delta(x ) \right| \leq \sum_{y \in B_{R^{1-\delta}}} \left|\nabla^k  \rho_\delta(y) \right| \left| \bar G \left( x - y \right) \right|
     \leq \frac{C}{R^{(d+k)(1-\delta)}} \sum_{y \in B_{R^{1-\delta}}} \frac{1}{|x - y|^{d-2}}
     \leq \frac{C}{R^{(1-\delta)(d-2 + k)}}. \notag
\end{equation*}
There only remains to prove the estimate~\eqref{eq:spatregbarGgeq34}. We select a point $x \in A_{R^{1+\delta}}$ and write
\begin{align*} 
    \left| \nabla^k \bar G_\delta(x ) \right| = \left| \sum_{y \in B_{R^{1-\delta}}}    \nabla^k \bar G \left( x - y \right) \rho_\delta(y) \right|
     \leq \sum_{y \in B_{R^{1-\delta}}} \frac{\left|\rho_\delta (y)\right|}{|x - y|^{d-2+k}}
     &\leq \frac{C}{R^{(d-2+k)(1+\delta)}} \sum_{y \in B_{R^{1-\delta}}} \left|\rho_\delta (y)\right| \notag \\
    & \leq \frac{C}{R^{(1+\delta)(d-2 + k)}}. \notag
\end{align*}
\end{proof}

We have now collected all the necessary preliminary ingredients for the proof of Proposition~\ref{prop:prop6.1} and devote the rest of Section~\ref{sec:section6.1} to its demonstration. 

\smallskip

\subsection{Estimating the weak norm of \texorpdfstring{$\L \mathcal{H}_{\delta} - \rho_\delta$}{20}} \label{sec3.2chap608511}
In this section, we fix an integer $k \in \left\{1 , \ldots, \binom d2 \right\}$, let $\mathcal{H}_{\delta, \cdot k}$ be the two-scale expansion introduced in~\eqref{def.Ho2sc} and prove that there exists an exponent $\gamma_{\alpha} > 0$ such that
\begin{equation} \label{eq:TV19578}
    \left\| \L \mathcal{H}_{\delta, \cdot k} - \rho_{\delta, \cdot k} \right\|_{\underline{H}^{-1}\left(B_{R^{1+\delta}} , \mu_\beta \right)} \leq \frac{C}{R^{d-1+\gamma_{\alpha}}}.
\end{equation}

The strategy is to use the explicit formula for the map $\mathcal{H}_{\delta, \cdot k}$ to compute the value of the term $\mathcal{L} \mathcal{H}_{\delta, \cdot k}$. We then prove that its $\underline{H}^{-1}\left(B_{R^{1+\delta}} , \mu_\beta \right)$-norm is small by using the quantitative properties of the corrector stated in Proposition~\ref{prop:prop5.25} of Chapter~\ref{section5}. 
We first write
\begin{equation} \label{eq:TV0957897}
    \L \mathcal{H}_{\delta, \cdot k} = \underbrace{\Delta_\phi \mathcal{H}_{\delta, \cdot k}}_{\textnormal{Substep 1.1}} + \underbrace{\frac1{2\beta} \sum_{n \geq 1} \frac{1}{\beta^{\frac n2}}(- \Delta)^{n+1} \mathcal{H}_{\delta, \cdot k}}_{\textnormal{Substep 1.2}} - \underbrace{\frac1{2\beta} \Delta \mathcal{H}_{\delta, \cdot k} + \sum_{q \in \mathcal{Q}} \nabla_q^* \cdot \a_q \nabla_q \mathcal{H}_{\delta, \cdot k}}_{\textnormal{Substep 1.3}}.
\end{equation}
We treat the three terms in the right side in three distinct substeps.

\medskip

\textit{Substep 1.1} 
In this substep, we treat the term $\Delta_\phi \mathcal{H}_{\delta, \cdot k}$. Since the homogenized Green's matrix $\bar G_{\delta, \cdot k}$ does not depend on the field $\phi$, we have the formula
\begin{equation} \label{eq:mainsub1.1hom}
    \Delta_\phi \mathcal{H}_{\delta, \cdot k} =  \sum_{i,j} \nabla_i \bar G_{\delta , jk} \left( \Delta_{\phi} \chi_{m,ij}\right).
\end{equation}

\medskip

\textit{Sustep 1.2.} In this substep, we study the iteration of the Laplacian of the two-scale expansion. We prove the identity
\begin{equation} \label{eq:TV14557}
 	 \sum_{n \geq 1} \frac{1}{\beta^{\frac n2}}(- \Delta)^{n+1} \mathcal{H}_{\delta, \cdot k} =  \sum_{i,j} \sum_{n \geq 1}  \frac{1}{\beta^{\frac n2}} \nabla_i \bar G_{\delta , jk} (-\Delta)^{n+1} \chi_{m,ij} + R_{\Delta^n},
\end{equation}
where $R_{\Delta^n}$ is an error term which satisfies the $\underline{H}^{-1}\left( B_{R^{1+\delta}} , \mu_\beta \right)$-estimate
\begin{equation}  \label{eq:TV145677}
\left\| R_{\Delta^n} \right\|_{\underline{H}^{-1}\left( B_{R^{1+\delta}}, \mu_\beta \right)} \leq \frac{C}{R^{d-1+\gamma_{\alpha}}}.
\end{equation}
We use the following identity for the iteration of the Laplacian on a product of functions: given two smooth functions $f , g \in \C^\infty \left( \Rd \right)$, we have the identity
\begin{equation} \label{eq:expandLap}
    \Delta^n \left( f g \right) = \sum_{r=0}^{n} \sum_{l = 0}^r \binom{n- r}{l} \left( \nabla^{r} \Delta^l  f  \right) \cdot \left( \nabla^{r} \Delta^{n- r - l} g \right).
\end{equation}
We note that this formula is valid for continuous functions (with the continuous Laplacian), it can be adapted to the discrete setting by taking into considerations translations of the functions $f$ and $g$. Since this adaptation does not affect the overall strategy of the proof, we ignore this technical difficulty in the rest of the argument and apply the formula~\eqref{eq:expandLap} to the two-scale expansion $\mathcal{H}_{\delta, \cdot k}$ as such. We obtain
\begin{equation} \label{eq:TV11087}
\Delta^n \mathcal{H}_{\delta, \cdot k} = \Delta^n \bar G_{\delta, \cdot k} + \sum_{i,j} \sum_{r=0}^{n} \sum_{l = 0}^r \binom{n- r}{l} \left( \nabla^{r} \Delta^l  \nabla_i  \bar G_{\delta , j k}  \right) \cdot \left( \nabla^{r} \Delta^{n- r - l} \chi_{m,ij} \right) .
\end{equation}
We first focus on the term $\Delta^n \bar G_{\delta, \cdot k}$ in the identity~\eqref{eq:TV11087} and prove that it is small in the $\underline{H}^{-1} \left( B_{R^{1+\delta}} , \mu_\beta \right)$-norm. Using the regularity estimate~\eqref{eq:spatregbarGgeq3}, we have, for each integer $n \geq 2$,
\begin{align} \label{eq:TV14407}
    \left\| \Delta^n \bar G_{\delta, \cdot k} \right\|_{\underline{H}^{-1} \left( B_{R^{1+\delta}} , \mu_\beta \right)}  \leq C R^{1+\delta}\left\| \Delta^n \bar G_{\delta, \cdot k}  \right\|_{\underline{L}^{2} \left( B_{R^{1+\delta}} , \mu_\beta \right)} 
    & \leq  C R^{1+\delta}\left\| \Delta^n \bar G_{\delta, \cdot k}  \right\|_{L^{\infty} \left( B_{R^{1+\delta}} \right)} \\
    & \leq  \frac{C^{2n} R^{1+\delta}}{R^{(1-\delta)(d - 2 + 2n)}} \notag \\
    & \leq \frac{C^{2n} R^{1+\delta}}{R^{(1-\delta)(d - 2 + 4)}} \notag \\
    & \leq \frac{C^{2n} }{R^{d-1 + \gamma_1}}, \notag
\end{align}
where we have set $\gamma_1 := 2 + \delta (d+1) > 0$.

Using the regularity estimate~\eqref{eq:spatregbarGgeq3} a second time, we can estimate the terms of the right side of the identity~\eqref{eq:TV11087} with more than 3 derivatives on the homogenized Green's matrix $\bar G_\delta$. We obtain the following inequality: for each pair of integers $(i,j) \in \{ 1 , \ldots, d\} \times \{ 1 , \ldots, \binom d2\}$ and each pair of integers $(r, l) \in \{ 1 , \ldots, d \}^2$ such that $l \leq n-k$ and $k + 2 l \geq 2$,
\begin{align} \label{eq:TV11047}
\lefteqn{ \left\| \left( \nabla^{r} \Delta^l  \nabla_i \bar G_{\delta , j k}  \right) \cdot \left( \nabla^{r} \Delta^{n- r - l} \chi_{m,ij} \right) \right\|_{\underline{H}^{-1}(B_{R^{1+\delta}}, \mu_\beta)} } \qquad & \\  & \leq C R^{1+\delta} \left\| \left( \nabla^{r} \Delta^l  \nabla_i \bar G_{\delta , j k} \right) \cdot \left( \nabla^{r} \Delta^{n-r - l} \chi_{m,ij} \right) \right\|_{\underline{L}^2(B_{R^{1+\delta}}, \mu_\beta)}\notag \\
 & \leq  C R^{1+\delta} \left\|  \nabla^{r} \Delta^l  \nabla_i \bar G_{\delta , j k} \right\|_{L^\infty \left( B_{R^{1+\delta}}  \right)} \times \left\|  \nabla^{r} \Delta^{n-r - l} \chi_{m,ij}  \right\|_{\underline{L}^2(B_{R^{1+\delta}}, \mu_\beta)} \notag \\
 & \leq \frac{C^{r + 2l} R^{1+\delta}}{R^{(1-\delta)(d-1+2l + r)}} \left\|  \nabla^{r} \Delta^{n-r - l} \chi_{m,ij}  \right\|_{\underline{L}^2(B_{R^{1+\delta}}, \mu_\beta)}. \notag
 \end{align}
 We use that the discrete operator $ \nabla^{r} \Delta^{n- r - l}$ is bounded in the space $L^2 \left( B_{R^{1+\delta}} \right)$ and Proposition~\ref{prop:prop5.25} of Chapter~\ref{section5} to estimate the $L^2$-norm of the corrector. We obtain
 \begin{equation} \label{eq:TV11037}
      \left\|  \nabla^{r} \Delta^{n- r - l} \chi_{m,ij}  \right\|_{\underline{L}^2(B_{R^{1+\delta}}, \mu_\beta)} \leq C^{2n - 2l} \left\|   \chi_{m,ij}  \right\|_{\underline{L}^2(B_{R^{1+\delta}}, \mu_\beta)} \leq  C^{2n - 2l} R^{(1+\delta)(1-\alpha)}.
 \end{equation}
 Putting the estimate~\eqref{eq:TV11047} and~\eqref{eq:TV11037} together and using the inequality $3 \leq 2l+r \leq 2n$, we deduce that
 \begin{equation} \label{eq:TV14387}
       \left\| \left( \nabla^{r} \Delta^l  \nabla_i \bar G_{\delta , j k}  \right) \cdot \left( \nabla^{r} \Delta^{n- r - l} \chi_{m,ij} \right) \right\|_{\underline{H}^{-1}(B_{R^{1+\delta}}, \mu_\beta)} \leq  \frac{C^{2n} R^{1+\delta}}{R^{(1-\delta)(d+2)}} R^{(1+\delta)(1-\alpha)} \leq \frac{C^{2n}}{R^{d-1 + \gamma_{1}}},
 \end{equation}
where we have set $\gamma_{1} := 1 + \alpha - \alpha \delta + \delta \left( d-1 \right) + \delta >0$.

We then estimate the $H^{-1}$-norm of the terms corresponding to the parameters $r=1$ and $l=0$ in the sum in the right side of the identity~\eqref{eq:TV11087}. To estimate it, we select a function $h \in H^1_0 \left(B_{R^{1+\delta}}, \mu_\beta \right)$ such that $\left\| h \right\|_{\underline{H}^1 \left( B_{R^{1+\delta}}, \mu_\beta \right)}\leq 1$. We use the function $h$ as a test function, perform an integration by parts in the first line, use the Cauchy-Schwarz inequality in the second line and the continuity of the discrete Laplacian in the third line. We obtain 
\begin{align} \label{eq:TV14097}
\lefteqn{\frac{1}{R^{(1+\delta)d}}\sum_{x \in B_{R^{1+\delta}}} \left\langle \left( \nabla \nabla_i \bar G_{\delta , jk} \left( x , \cdot \right)  \right) \cdot \left( \nabla \Delta^{n- 1} \chi_{m,ij} (x , \cdot ) \right)  h(x , \cdot)  \right\rangle_{\mu_\beta} } \qquad & \\ 
			& = \frac{1}{R^{(1+\delta)d}} \sum_{x \in B_{R^{1+\delta}}} \left\langle \chi_{m,ij} (x , \cdot )   \nabla \cdot \Delta^{n- 1} \left( \left( \nabla \nabla_i \bar G_{\delta , j k} \left( x , \cdot \right)  \right) h  (x , \cdot )\right) \right\rangle_{\mu_\beta} \notag \\
			& \leq  \left\| \chi_{m,ij} \right\|_{L^2 \left( B_{R^{1+\delta}}, \mu_\beta \right) }   \left\| \nabla \cdot \Delta^{n- r}\left( \nabla \nabla_i \bar G_{\delta , j k}  h \right) \right\|_{\underline{L}^2 \left( B_{R^{1+\delta}} , \mu_\beta \right)} \notag \\
			& \leq C^{n} \left\| \chi_{m,ij} \right\|_{L^2 \left( B_{R^{1+\delta}}, \mu_\beta \right) }   \left\| \nabla \cdot \left( \nabla \nabla_i \bar G_{\delta , j k}  h \right) \right\|_{\underline{L}^2 \left( B_{R^{1+\delta}} , \mu_\beta \right)} .\notag
\end{align}
Using the regularity estimate for the homogenized Green's matrix stated in~\eqref{eq:spatregbarGgeq3} and the inequality $\left\| h \right\|_{\underline{L}^2 \left(B_{R^{1+\delta}} , \mu_\beta \right)} \leq C R^{1+\delta}$ (which is a consequence of the assumption $\left\| h \right\|_{\underline{H}^1 \left(B_{R^{1+\delta}} , \mu_\beta \right)} \leq 1$ and the Poincar\'e inequality), we obtain
\begin{align} \label{eq:TV14047}
    \left\| \nabla \cdot \left(\left( \nabla \nabla_i \bar G_{\delta , j k} \right) h \right) \right\|_{\underline{L}^2 \left( B_{R^{1+\delta}} , \mu_\beta \right)} & \leq \left\| \nabla^3 \bar G_{\delta , jk} h \right\|_{\underline{L}^2 \left( B_{R^{1+\delta}} , \mu_\beta \right)} + \left\| \nabla^2 \bar G_{\delta , j k}  \nabla h \right\|_{\underline{L}^2 \left( B_{R^{1+\delta}} , \mu_\beta \right)} \\
    & \leq  \frac{CR^{1+\delta}}{R^{\left( 1-\delta \right)\left(d + 1 \right)}}  +  \frac{C}{R^{\left( 1-\delta \right)d}} \notag \\ &
    \leq \frac{C}{R^{d - \delta (d+2)}}. \notag
\end{align}
We then combine the estimate~\eqref{eq:TV14097} with the inequality~\eqref{eq:TV14047} and the quantitative sublinearity of the corrector stated in Proposition~\ref{prop:prop5.25} of Chapter~\ref{section5}. We obtain
\begin{equation} \label{eq:TV14397}
    \frac{1}{R^{(1+\delta)d}}\sum_{x \in B_{R^{1+\delta}}} \left\langle \left( \nabla \nabla_i \bar G_{\delta , j k} \left( x \right)  \right) \cdot \left( \nabla \Delta^{n- 1} \chi_{m,ij} (x , \cdot ) \right)  h(x , \cdot)  \right\rangle_{\mu_\beta} \leq \frac{C^n R^{\left(1+\delta\right)(1 - \alpha) }}{R^{d - \delta (d+2)}} \leq  \frac{C^n}{R^{d - 1 + \gamma_{\alpha} }},
\end{equation}
where we have set $\gamma_{\alpha} :=\alpha (1+\delta) - \delta (d+3) > 0$.

\smallskip

By combining the identity~\eqref{eq:TV11087} with the estimates~\eqref{eq:TV14407},~\eqref{eq:TV14387},~\eqref{eq:TV14397} and choosing the inverse temperature $\beta$ large enough so that the series $\left( \frac{C^n}{\beta^{\frac n2}}\right)_{n \in \N}$ is summable, we obtain the main result~\eqref{eq:TV14557} and~\eqref{eq:TV145677} of this substep.

\medskip

\textit{Substep 1.3.}  In this substep, we study the term pertaining to the charges in the identity~\eqref{eq:TV0957897}. We prove the expansion
\begin{equation} \label{eq:mainsub1.3hom}
 \frac{1}{2\beta} \Delta \mathcal{H}_{\delta, \cdot k} + \sum_{q\in \mathcal{Q}} \nabla_q^* \cdot \a_q \nabla_q \mathcal{H}_{\delta, \cdot k}  = \ahom_\beta  \Delta \bar G_{\delta, \cdot k} + \sum_{i,j} \frac{1}{2\beta} \nabla_i  \bar G_{\delta , j k} \Delta \chi_{m,ij} +   \sum_{i,j}  \sum_{q\in \mathcal{Q}} \nabla_i  \bar G_{\delta , j k} \nabla_q^* \cdot \a_q \nabla_q \chi_{m,ij}+ R_{Q}.
\end{equation}
where $R_{Q}$ is an error term which satisfies the $\underline{H}^{-1}\left( B_{R^{1+\delta}} , \mu_\beta \right)$-norm estimate
\begin{equation}  \label{eq:TV14567}
\left\| R_{Q} \right\|_{\underline{H}^{-1}\left( B_{R^{1+\delta}} \mu_\beta \right)} \leq \frac{C}{R^{d-1+\gamma_{\alpha}}}.
\end{equation}
We first compute the gradient and the Laplacian of the two-scale expansion $\mathcal{H}
_{\delta, \cdot k}$ using the notation of~\eqref{eq:TV16450801} of Chapter~\ref{Chap:chap2}. We obtain the formulas
\begin{equation} \label{eq:formnablaH2sc}
\nabla \mathcal{H}_{\delta, \cdot k} = \nabla  \bar G_{\delta, \cdot k}  + \sum_{i,j} \left[ \nabla \nabla_i  \bar G_{\delta , j k} \otimes \chi_{m,ij} + \nabla_i  \bar G_{\delta , jk}   \nabla \chi_{m,ij} \right],
\end{equation}
and
\begin{equation} \label{eq:formnablaH2scLap}
    \Delta \mathcal{H}_{\delta, \cdot k}  = \Delta \bar G_{\delta, \cdot k} + \sum_{i,j} \nabla \cdot \left( \nabla  \nabla_i  \bar G_{\delta , j k} \otimes \chi_{m,ij} \right) + \left( \nabla \nabla_i  \bar G_{\delta , j k}  \right) \cdot  \left( \nabla \chi_{m,ij} \right)  + \left(  \nabla_i  \bar G_{\delta , j k}  \right) \Delta \chi_{m,ij}.
\end{equation}
We first treat the term $\Delta \mathcal{H}_{\delta, \cdot k}$ and use the two following ingredients:
\begin{enumerate}
    \item[(i)] First we introduce the notation $R_{Q,1} := \sum_{i,j}\nabla \cdot \left( \nabla  \nabla_i  \bar G_{\delta , j k} \otimes \chi_{m,ij} \right) $. By using the regularity estimate~\eqref{eq:spatregbarGgeq3} on the homogenized Green's matrix and the quantitative sublinearity of the corrector stated in 
    Proposition~\ref{prop:prop5.25} of Chapter~\ref{section5}, we prove that this term is an error term and estimate its $\underline{H}^{-1} \left( B_{R^{1+\delta}} , \mu_\beta \right)$-norm according to the following computation
    \begin{align*}
        \left\| R_{Q,1} \right\|_{\underline{H}^{-1} \left( B_{R^{1+\delta}} , \mu_\beta \right)} & \leq C \left\|  \sum_{i,j}   \nabla  \nabla_i  \bar G_{\delta , j k} \otimes \chi_{m,ij} \right\|_{\underline{L}^{2} \left( B_{R^{1+\delta}} , \mu_\beta \right)} \\
        & \leq \sum_{i,j} C \left\|   \nabla \nabla_i  \bar G_{\delta , j k}  \right\|_{L^{\infty} \left( B_{R^{1+\delta}} , \mu_\beta \right)} \left\|  \chi_{m,ij} \right\|_{\underline{L}^{2} \left( B_{R^{1+\delta}} , \mu_\beta \right)}\\
        & \leq \frac{C R^{\left( 1 + \delta \right)\left(1-\alpha\right)}}{R^{(1-\delta) (d)}} \\
        & \leq \frac{C}{R^{d-1 + \gamma_\alpha}},
    \end{align*}
    where we have set $\gamma_\alpha :=\alpha (1+\delta) - \delta (d+1) > 0$.
    \smallskip
    \item[(ii)] Second, we use the identity $\Delta \bar G_{\delta, \cdot k} = \nabla \cdot \nabla \bar G_{\delta, \cdot k} =  \sum_{i,j} \nabla \cdot \left( \nabla_i \bar G_{\delta , j k}  e_{ij} \right)$.
\end{enumerate}
We obtain
\begin{equation} \label{eq:TV21228}
     \Delta \mathcal{H}_{\delta, \cdot k}  = \nabla \cdot \left( \sum_{i,j}  \nabla_i \bar G_{\delta , j k} \left( e_{ij} + \nabla \chi_{m,ij} \right) \right) + \sum_{i,j} \left(  \nabla_i  \bar G_{\delta , jk}  \right) \Delta \chi_{m,ij} + R_{Q,1}.
\end{equation}
We then treat the term pertaining to the charges; the objective is to prove the identity 
\begin{equation} \label{eq:TV19558}
\sum_{q\in \mathcal{Q}} \nabla_q^* \cdot \a_q \nabla_q \mathcal{H}_{\delta, \cdot k} = \sum_{i,j} \nabla \nabla_i \bar G_{\delta , j k}  \sum_{q \in \mathcal{Q}}   \a_q \nabla_q \left( l_{e_{ij}} + \chi_{m,ij} \right) L^{t}_{2, \di^*} \left( n_q \right) + \sum_{q \in \mathcal{Q}} \nabla_i  \bar G_{\delta , j k} \nabla_q^* \cdot \a_q \nabla_q \left( l_{e_{ij}} + \chi_{m,ij} \right) + R_{Q,2},
\end{equation}
where $R_{Q,2}$ is an error term which satisfies the estimate
\begin{equation} \label{eq:TV19568}
    \left\| R_{Q,2} \right\|_{\underline{H}^{-1}\left(B_{R^{1+\delta}} , \mu_\beta \right)} \leq \frac{C}{R^{d-1-\gamma_{\alpha}}}.
\end{equation}
To prove this result, we select a test function $h : \Zd \to \R^{\binom d2}$ which belongs to the space $H^1_0 \left(B_{R^{1+\delta}} , \mu_\beta \right)$ and satisfies the estimate $\left\| h \right\|_{\underline{H}^1 \left(B_{R^{1+\delta}} , \mu_\beta \right)} \leq 1$. For each charge $q \in \mathcal{Q}$, we select a point $x_q$ which belongs to the support of the charge $q$ arbitrarily. We then write
\begin{equation} \label{eq:TV21447}
    \sum_{q\in \mathcal{Q}} \a_q \nabla_q \mathcal{H}_{\delta, \cdot k} \nabla_q h =  \sum_{q\in \mathcal{Q}}  \a_q \left( n_q , \di^* \mathcal{H}_{\delta, \cdot k} \right) \left( n_q , \di^* h \right).
\end{equation}
We use the exact formula for $\mathcal{H}_{\delta}$ and apply the codifferential. We obtain
\begin{align} \label{eq:TV21357}
    \di^* \mathcal{H}_{\delta, \cdot k}  = L^{t}_{2, \di^*}\left( \nabla \mathcal{H}_{\delta, \cdot k} \right) & = L_{2, \di^*} \left( \nabla  \bar G_{\delta, \cdot k}  + \sum_{i,j} \left[ \nabla \nabla_i  \bar G_{\delta , j k} \otimes  \chi_{m,ij} + \nabla_i  \bar G_{\delta , j k}   \nabla \chi_{m,ij} \right] \right) \\
    & =  \di^* \bar G_{\delta, \cdot k} + \sum_{i,j}   \nabla_i \bar G_{\delta, jk} \di^* \chi_{m,ij} +  \sum_{i,j}  L_{2, \di^*} \left( \nabla \nabla_i \bar G_{\delta, jk} \otimes \chi_{m,ij}\right).\notag
\end{align}
We record the following formula
\begin{align} \label{eq:TV16162212}
        \di^* \bar G_{\delta, \cdot k} =  L_{2, \di^*}  \left( \nabla \bar G_{\delta, \cdot k} \right)  =  L_{2, \di^*}  \left( \sum_{i,j} \nabla_i \bar G_{\delta,jk} e_{ij} \right) & =  L_{2, \di^*}  \left( \sum_{i,j} \nabla_i \bar G_{\delta,jk} \nabla l_{e_{ij}} \right)  \\
        & =    \sum_{i,j} \nabla_i \bar G_{\delta,jk} L_{2, \di^*} \left(\nabla l_{e_{ij}} \right) \notag \\
        & =  \sum_{i,j} \nabla_i \bar G_{\delta,jk} \di^* l_{e_{ij}}. \notag
\end{align}
Putting the identities~\eqref{eq:TV21357} and~\eqref{eq:TV16162212} back into~\eqref{eq:TV21447}, we obtain
\begin{multline} \label{eq:TV22007}
     \sum_{q \in \Qa} \a_q \nabla_q \mathcal{H}_{\delta, \cdot k} \nabla_q h = \underbrace{\sum_{i,j} \sum_{q \in \Qa}  \a_q \left( n_q , \nabla_i \bar G_{\delta, jk} \left( \di^* l_{e_{ij}} + \di^* \chi_{m,ij} \right) \right) \left( n_q , \di^* h \right)}_{\eqref{eq:TV22007}-(i)} \\ +  \underbrace{\sum_{q \in \Qa}  \a_q \left( n_q , L_{2, \di^*} \left( \nabla \nabla_i \bar G_{\delta, jk} \otimes \chi_{m,ij}\right) \right) \left( n_q , \di^* h \right)}_{\eqref{eq:TV22007}-(ii)} .
\end{multline}
The second term~\eqref{eq:TV22007}-(ii) is an error term which is small can be estimated thanks to the regularity estimate~\eqref{eq:spatregbarGgeq3} and Young's inequality. We obtain
\begin{align*}
     \lefteqn{ \left| \left\langle \sum_{q \in \Qa}  \a_q \left( n_q , L_{2, \di^*} \left( \nabla \nabla_i \bar G_{\delta, jk} \otimes \chi_{m,ij}\right) \right) \left( n_q , \di^* h \right) \right\rangle_{\mu_\beta} \right|} \qquad & \\ &  \leq \sum_{q \in \Qa} e^{-c \sqrt{\beta} \left\| q \right\|_1} \left\| n_q \right\|_2^2 \left\| \nabla^2 \bar G_{\delta, jk} \right\|_{L^\infty \left( \Zd , \mu_\beta \right)} \left\| \chi_{m,ij} \right\|_{L^2 \left( \supp n_q , \mu_\beta \right)} \left\| \nabla h \right\|_{L^2 \left( \supp n_q , \mu_\beta \right)} \\
     & \leq \frac{C}{R^{(1-\delta)(d+1)}}  \sum_{q \in \Qa} e^{-c \sqrt{\beta} \left\| q \right\|_1} \left\| n_q \right\|_2^2 \left\| \chi_{m,ij} \right\|_{L^2 \left( \supp n_q , \mu_\beta \right)} \left\| \nabla h \right\|_{L^2 \left( \supp n_q , \mu_\beta \right)} \\
     & \leq \frac{C}{R^{(1-\delta)(d+1)}}  \sum_{q \in \Qa} e^{-c \sqrt{\beta} \left\| q \right\|_1} \left\| n_q \right\|_2^2 \left( R^{-1 + \alpha} \left\| \chi_{m,ij} \right\|_{L^2 \left( \supp n_q , \mu_\beta \right)}^2 + R^{1 - \alpha} \left\| \nabla h \right\|_{L^2 \left( \supp n_q , \mu_\beta \right)}^2 \right).
\end{align*}
We then use the inequality, for each point $x \in \Zd$,
\begin{equation} \label{eq:TV09127}
    \sum_{q \in \Qa} e^{-c \sqrt{\beta} \left\| q \right\|_1} \left\| n_q \right\|_2^2 \indc_{\{ x \in \supp n_q\}} \leq C.
\end{equation}
We deduce that
\begin{multline*}
    \left| \left\langle \sum_{q \in \Qa}  \a_q \left( n_q , L_{2, \di^*} \left( \nabla \nabla_i\bar G_{\delta, jk} \otimes \chi_{m,ij}\right) \right) \left( n_q , \di^* h \right) \right\rangle_{\mu_\beta} \right| \\ \leq \frac{CR^{-1 + \alpha}}{R^{(1-\delta)(d+1)}}  \left\| \chi_{m,ij} \right\|_{L^2 \left(B_{R^{1+\delta}, \mu_\beta } \right)}^2 + \frac{CR^{1-\alpha}}{R^{(1-\delta)(d+1)}} \left\| \nabla h \right\|_{L^2 \left(B_{R^{1+\delta}, \mu_\beta } \right)}^2.
\end{multline*}
We then use Proposition~\ref{prop:prop5.25} of Chapter \ref{section5} and the assumption $\left\| \nabla h \right\|_{\underline{L}^2 \left(B_{R^{1+\delta}, \mu_\beta } \right)} \leq 1$. We obtain
\begin{equation} \label{eq:TV1854812}
    \left| \frac{1}{R^{(1+\delta)d}} \left\langle \sum_{q \in \Qa}  \a_q \left( n_q , L_{2, \di^*} \left( \nabla \nabla_i \bar G_{\delta, jk} \otimes \chi_{m,ij}\right) \right) \left( n_q , \di^* h \right) \right\rangle_{\mu_\beta} \right| \leq \frac{C R^{1-\alpha}}{R^{(1-\delta)(d+1)}} \leq \frac{C}{R^{d-1 + \gamma_{\alpha}}},
\end{equation}
where we have set $\gamma_{\alpha} = \alpha - \delta \left( d+1 \right)$.

To treat the term~\eqref{eq:TV22007}-(i), we make use of the point $x_q$ and write
\begin{align*}
     \lefteqn{\sum_{q\in \mathcal{Q}}  \a_q \left( n_q , \nabla_i \bar G_{\delta, jk} \left( \di^* l_{e_{ij}} + \di^* \chi_{m,ij} \right) \right) \left( n_q , \di^* h \right) } \qquad & \\ & = \sum_{q\in \mathcal{Q}}   \a_q \left( n_q ,  \left( \di^* l_{e_{ij}} + \di^* \chi_{m,ij} \right) \right) \left( n_q , \nabla_i \bar G_{\delta, jk} \di^* h \right) \\
     & + \sum_{q\in \mathcal{Q}} \a_q \left( n_q ,\left( \nabla_i \bar G_{\delta, jk} - \nabla_i \bar G_{\delta, jk} (x_q) \right)  \left( \di^* l_{e_{ij}} + \di^* \chi_{m,ij} \right) \right) \left( n_q , \di^* h \right) \\
     & +  \sum_{q\in \mathcal{Q}} \a_q \left( n_q   \left( \di^* l_{e_{ij}} + \di^* \chi_{m,ij} \right) \right) \left( n_q , \left( \nabla_i \bar G_{\delta, jk} - \nabla_i \bar G_{\delta, jk} (x_q) \right) \di^* h \right).
\end{align*}
The terms on the second and third lines are error terms which are small, they can be estimated by the regularity estimate~\eqref{eq:spatregbarGgeq3} on the gradient of the homogenized Green's matrix and Young's inequality. We obtain
\begin{align*}
    \lefteqn{\left| \left\langle \sum_{q\in \mathcal{Q}} \a_q \left( n_q ,\left( \nabla_i \bar G_{\delta, jk}- \nabla_i \bar G_{\delta, jk} (x_q) \right)  \left( \di^* l_{e_{ij}} + \di^* \chi_{m,ij} \right) \right) \left( n_q , \di^* h \right) \right\rangle_{\mu_\beta} \right|  } \qquad & \\ & \leq \sum_{q\in \mathcal{Q}} e^{-c \sqrt{\beta} \left\| q \right\|_1} \left\| n_q \right\|_2 \left\| \nabla \bar G_{\delta, jk} - \nabla \bar G_{\delta, jk} (x_q)\right\|_{L^\infty \left( \supp n_q , \mu_\beta \right)} \left\|  \nabla \chi_{m,ij} \right\|_{L^2 \left( \supp n_q , \mu_\beta \right)} \left\| \nabla h \right\|_{L^2 \left( \supp n_q , \mu_\beta \right)} \\
    & \leq \frac{C}{R^{(1-\delta)d}} \sum_{q\in \mathcal{Q}} e^{-c \sqrt{\beta} \left\| q \right\|_1} \left\| n_q \right\|_2  \left( \left\|  \nabla \chi_{m,ij} \right\|_{L^2 \left( \supp n_q , \mu_\beta \right)}^2 +  \left\| \nabla h \right\|_{L^2 \left( \supp n_q , \mu_\beta \right)}^2 \right). 
\end{align*}
We then apply the estimate~\eqref{eq:TV09127}, the bound $\left\| \nabla \chi_{m,ij} \right\|_{\underline{L}^2 \left( B_{R^{1+\delta}} , \mu_\beta \right)} \leq C$ on the gradient of the corrector and the assumption $\left\| \nabla h \right\|_{\underline{L}^2 \left( B_{R^{1+\delta}} , \mu_\beta \right)} \leq 1$ to conclude that
\begin{equation} \label{eq:TV18538}
    \left| \frac{1}{R^{(1+\delta)d}} \left\langle \sum_{q\in \mathcal{Q}} \a_q \left( n_q ,\left( \nabla_i \bar G_{\delta, jk} - \nabla_i \bar G_{\delta, jk}(x_q) \right)  \left( \di^* l_{e_{ij}} + \di^* \chi_{m,ij} \right) \right) \left( n_q , \di^* h \right) \right\rangle_{\mu_\beta} \right|  \leq \frac{C}{R^{(1-\delta)d}} \leq \frac{C}{R^{(d-1) + \gamma_1}},
\end{equation}
where we have set $\gamma_1 = 1 - \delta >0$.
The same argument proves the inequality
\begin{equation} \label{eq:TV18528}
     \left|  \frac{1}{R^{(1+\delta)d}} \left\langle \sum_{q\in \mathcal{Q}} \a_q \left( n_q   \left( \di^* l_{e_{ij}} + \di^* \chi_{m,ij} \right) \right) \left( n_q , \left( \nabla_i \bar G_{\delta,jk} - \nabla_i \bar G_{\delta,jk}(x_q) \right) \di^* h \right) \right\rangle_{\mu_\beta} \right| \leq \frac{C}{R^{(d-1) + \gamma}},
\end{equation}
with the same exponent $\gamma_1 >0$. Combining the identity~\eqref{eq:TV22007} with the estimates~\eqref{eq:TV1854812},~\eqref{eq:TV18538},~\eqref{eq:TV18528}, we have obtained the following result: for each function $h \in H^1_0 \left( B_{R^{1+\delta}} , \mu_\beta \right)$ such that $\left\| h \right\|_{\underline{H}^1 \left(B_{R^{1+\delta}} , \mu_\beta \right)} \leq 1$, one has the estimate
\begin{equation*}
     \frac{1}{R^{(1+\delta)d}} \sum_{q \in \Qa} \a_q \nabla_q \mathcal{H}_{\delta, \cdot k} \nabla_q h =   \frac{1}{R^{(1+\delta)d}} \sum_{q\in \mathcal{Q}}   \a_q \left( n_q ,  \left( \di^* l_{e_{ij}} + \di^* \chi_{m,ij} \right) \right) \left( n_q , \nabla_i \bar G_{\delta, jk} \di^* h \right) + O \left( \frac{C}{R^{d - 1 + \gamma_\alpha}} \right).
\end{equation*}
We then use the identity $\nabla_i \bar G_{\delta,jk} \di^* h = \di^*  \left(\nabla_i \bar G_{\delta,jk} h \right) -  L_{2, \di^*} \left( \nabla \nabla_i \bar G_{\delta, jk} \otimes h \right)$ which is established in~\eqref{eq:TV16162212}. We deduce that
\begin{multline} \label{eq:TV19548}
    \sum_{q \in \Qa} \a_q \nabla_q \mathcal{H}_{\delta, \cdot k} \nabla_q h =  \sum_{q\in \mathcal{Q}}   \a_q \left( n_q ,  \left( \di^* l_{e_{ij}} + \di^* \chi_{m,ij} \right) \right) \left( n_q , \di^* \left( \nabla_i \bar G_{\delta, jk}  h\right) \right) \\ + \sum_{q\in \mathcal{Q}}   \a_q \left( n_q ,  \left( \di^* l_{e_{ij}} + \di^* \chi_{m,ij} \right) \right) \left( n_q ,  L_{2, \di^*} \left( \nabla \nabla_i \bar G_{\delta, jk} \otimes h \right) \right) + O \left( \frac{C}{R^{d - 1 + \gamma_\alpha}} \right).
\end{multline}
This implies the identity~\eqref{eq:TV19558} and the estimate~\eqref{eq:TV19568}.

We now complete the proof of~\eqref{eq:mainsub1.3hom}. To prove this identity, it is sufficient, in view of~\eqref{eq:TV21228} and~\eqref{eq:TV19558}, to prove the estimate
\begin{equation} \label{eq:07499}
    \frac1{2\beta}\sum_{i,j}  \left( \nabla \nabla_i  \bar G_{\delta, jk}   \right) \cdot \left( e_{ij} + \nabla \chi_{m,ij} \right) + \sum_{i,j} \left( \nabla \nabla_i \bar G_{\delta, jk}  \right) \sum_{q \in \Qa}   \a_q \nabla_q \left( l_{e_{ij}} + \chi_{m,ij} \right) L_{2, \di^*}^t \left( n_q\right) = \nabla \cdot \left( \ahom_\beta \nabla  \bar G_{\delta, \cdot k} \right) + R_{Q,3},
\end{equation}
where the term $R_{Q,3}$ satisfies the estimate
\begin{equation} \label{eq:07509}
    \left\| R_{Q,3} \right\|_{\underline{H}^{-1}\left( B_{R^{1+\delta}} , \mu_\beta\right)} \leq \frac{C}{R^{1-d+ \gamma_\alpha}}.
\end{equation}
The proof relies on the quantitative estimate for the $\underline{H}^{-1}\left( B_{R^{1+\delta}} , \mu_\beta\right)$-norm of the flux corrector stated in Proposition~\ref{prop:prop5.25} of Chapter~\ref{section5} and the regularity estimate~\eqref{eq:spatregbarGgeq3} on the homogenized matrix $\bar G_\delta$. We select a function $h: \Zd \to \R^{\binom d2}$ which belongs to the space $H^1_0 \left( B_{R^{1+\delta}} , \mu_\beta \right)$ and such that $\left\| h \right\|_{\underline{H}^1 \left( B_{R^{1+\delta}} , \mu_\beta \right)} \leq 1$. We use it as a test function and write
\begin{align} \label{eq:TV07399}
    \lefteqn{\frac{1}{R^{(1+\delta)d}} \left| \left\langle  \sum_{x \in B_{R^{1+\delta}}}\sum_{i,j} \frac 1{2\beta} \left( \nabla \nabla_i  \bar G_{\delta,jk}(x , \cdot )  \right) \cdot \left( e_{ij} + \nabla \chi_{m,ij}(x , \cdot) \right) h(x , \cdot)  \right. \right.} \qquad & \\ & \left. \left. + \sum_{i,j} \sum_{q \in \mathcal{Q}}   \a_q \nabla_q \left( l_{e_{ij}} + \chi_{m,ij} \right) \left( n_q, L_{2, \di^*}\left( \nabla \nabla_i \bar G_{\delta, j k} \otimes h\right) \right) - \sum_{x \in B_{R^{1+\delta}}} \sum_{i = 1}^d \nabla \cdot \left( \ahom_\beta \nabla   \bar G_{\delta, \cdot k}(x ) \right) h(x , \cdot) \right\rangle_{\mu_\beta}\right| \notag \\ & \leq C \sum_{i,j}  \left\| \frac 1{2 \beta} \left(e_{ij} +  \nabla \chi_{m,ij}\right) + \sum_{q \in \mathcal{Q}} \a_q \nabla_q \left( l_{e_{ij}} + \chi_{m,ij} \right) L_{2, \di^*}^t \left(n_q\right) - \ahom_\beta e_{ij} \right\|_{\underline{H}^{-1}\left( B_{R^{1+\delta}} , \mu_\beta \right)} \left\| \nabla \nabla_i \bar G_{\delta,j k} \otimes  h \right\|_{\underline{H}^1 \left(B_{R^{1+\delta}} , \mu_\beta \right)}. \notag
\end{align}
We then use Proposition~\ref{prop:prop5.25} of Chapter~\ref{section5} to write
\begin{equation} \label{eq:TV07379}
    \sum_{i,j}  \left\| \frac 1{2 \beta} \left( e_{ij} +  \nabla \chi_{m,ij}\right) + \sum_{q \in \mathcal{Q}} \a_q \nabla_q \left( l_{e_{ij}} + \chi_{n,i} \right) L_{2, \di^*}^t \left(n_q\right) - \ahom_\beta e_{ij} \right\|_{\underline{H}^{-1}\left( B_{R^{1+\delta}} , \mu_\beta \right)} \leq CR^{(1+\delta)(1-\alpha)},
\end{equation}
and the regularity estimate~\eqref{eq:spatregbarGgeq3} to write 
\begin{align} \label{eq:TV07299}
    \left\| \nabla \nabla_i \bar G_{\delta, j k}  h \right\|_{\underline{H}^1 \left(B_{R^{1+\delta}} , \mu_\beta \right)} & \leq  \frac{1}{R^{1+\delta}}\left\|  \nabla^2 \bar G_{\delta, \cdot k}  \right\|_{L^\infty \left( B_{R^{1+\delta}} , \mu_\beta \right)}\left\|   h \right\|_{\underline{L}^2 \left( B_{R^{1+\delta}} , \mu_\beta \right)} + \left\|  \nabla^3 \bar G_{\delta, \cdot k}  \right\|_{L^\infty \left( B_{R^{1+\delta}} , \mu_\beta \right)}\left\|  h \right\|_{\underline{L}^2 \left( B_{R^{1+\delta}} , \mu_\beta \right)}  \\ & \qquad + \left\|  \nabla^2 \bar G_{\delta, \cdot k}  \right\|_{L^\infty \left( B_{R^{1+\delta}} , \mu_\beta \right)}\left\| \nabla h \right\|_{\underline{L}^2 \left( B_{R^{1+\delta}} , \mu_\beta \right)} \notag \\
    & \leq C \left(\frac{1}{R^{d(1-\delta)}} + \frac{1}{R^{(d-1)(1-\delta)}} + \frac{R^{1+\delta}}{R^{(d+1)(1-\delta)}} \right) \notag\\
    & \leq \frac{1}{R^{d - \delta\left( d+2 \right)}}. \notag
\end{align}
Combining the estimates~\eqref{eq:TV07399},~\eqref{eq:TV07379} and~\eqref{eq:TV07299}, we have obtained that, for each function $h \in H^1_0 \left( B_{R^{1+\delta}} , \mu_\beta \right)$ such that $\left\| h \right\|_{\underline{H}^1 \left( B_{R^{1+\delta}} , \mu_\beta \right)} \leq 1$,
\begin{align} \label{eq:07489}
    \lefteqn{\frac{1}{R^{(1+\delta)d}} \left| \left\langle  \sum_{x \in B_{R^{1+\delta}}}\sum_{i,j}  \left( \nabla \nabla_i  \bar G_{\delta, jk}(x )  \right) \cdot \left( e_{ij} + \nabla \chi_{m,ij}(x , \cdot) \right)  \right. \right.} \qquad & \\ & \left. \left. \qquad + \sum_{i,j} \sum_{q \in \mathcal{Q}}   \a_q \nabla_q \left( l_{e_{ij}} + \chi_{m,ij} \right) \left( n_q, L_{2, \di^*} \left( \nabla \nabla_i \bar G_{\delta,jk} \otimes h\right) \right) - \sum_{x \in B_{R^{1+\delta}}} \ahom \Delta  \bar G_{\delta, \cdot k}(x ) \cdot h(x , \cdot) \right\rangle_{\mu_\beta}\right| \notag \\ & \leq \frac{CR^{(1+\delta)(1-\alpha)}}{R^{d - \delta\left( d+2 \right)}} \notag \\
    & \leq \frac{C}{R^{d-1 + \gamma_{\alpha}}}, \notag
\end{align}
where we have set $\gamma := \alpha (1 + \delta) - \delta (d+3)$. Since the inequality~\eqref{eq:07489} is valid for any function $h \in H^1_0 \left( B_{R^{1+\delta}} , \mu_\beta \right)$ satisfying $\left\| h \right\|_{\underline{H}^1 \left( B_{R^{1+\delta}} , \mu_\beta \right)} \leq 1$, the estimate~\eqref{eq:07489} is equivalent to the identity~\eqref{eq:07499} and the $H^{-1}\left( B_{R^{1+\delta}} , \mu_\beta \right)$-norm estimate~\eqref{eq:07509}. The proof of~\eqref{eq:07499}, and thus of~\eqref{eq:mainsub1.3hom}, is complete. 

\medskip

\textit{Substep 1.4} In this substep, we conclude Step 1 and prove the estimate~\eqref{eq:TV19578}. We use the identity~\eqref{eq:TV0957897} and the identities~\eqref{eq:mainsub1.1hom} proved in Substep 1,~\eqref{eq:TV14557} proved in Substep 2 and~\eqref{eq:mainsub1.3hom} proved in Substep 3. We obtain
\begin{align} \label{eq:TV08399}
    \L \mathcal{H}_{\delta, \cdot k} & = \nabla \cdot \ahom_\beta \nabla \bar G_{\delta, \cdot k} \\ & \quad + \sum_{i,j}  \nabla_i  \bar G_{\delta, j k} \left(  \Delta_{\phi} \chi_{m,ij}  + \frac{1}{2\beta} \Delta \chi_{m,ij} + \sum_{q\in \mathcal{Q}}  \nabla_q^* \cdot \a_q \nabla_q \left( l_{e_{ij}} + \chi_{m,ij} \right)  + \frac 1{2\beta} \sum_{n \geq 1}  \frac{1}{\beta^{\frac n2}} (-\Delta)^{n+1} \chi_{m,ij} \right) \notag \\ & \quad + R_{Q}  + R_{\Delta^n}. \notag
\end{align}
We then treat the three lines of the previous display separately. For the first line, we use the identity
\begin{equation} \label{eq:TV09239}
     - \nabla \cdot \ahom_\beta  \nabla \bar G_{\delta, \cdot k} =\rho_{\delta, \cdot k} ~\mbox{in}~\Zd.
\end{equation}
For the second line, we use that, by the definition of the finite-volume corrector given in Definition~\ref{def.finivolcorr} of Chapter \ref{section5}, this map is a solution of the Helffer-Sj{\"o}strand equation $\L \left( l_{e_{ij}} + \chi_{m,ij} \right) = 0$ in the set $B_{R^{1+\delta}} \times \Omega$. We obtain
\begin{align} \label{eq:TV09249}
    \sum_{i,j}  \nabla_i  \bar G_{\delta, jk} \left(  \Delta_{\phi} \chi_{m,ij}  + \frac{1}{2\beta} \Delta \chi_{m,ij} + \sum_{q\in \mathcal{Q}}  \nabla_q^* \cdot \a_q \nabla_q \left( l_{e_{ij}} + \chi_{m,ij} \right)+   \frac1{2\beta} \sum_{n \geq 1}  \frac{1}{\beta^{\frac n2}} (-\Delta)^{n+1} \chi_{m,ij} \right) & =  \sum_{i,j}  \nabla_i  \bar G_{\delta,jk} \L \left( l_{e_{ij}} + \chi_{m,ij} \right) \\& = 0. \notag
    \end{align}
For the third line, we use the estimates~\eqref{eq:TV145677} and~\eqref{eq:TV14567} on the error terms $R_{Q}$ and $R_{\Delta^n}$ respectively. We obtain
\begin{equation} \label{eq:TV09259}
    \left\| R_{Q}+R_{\Delta^n} \right\|_{\underline{H}^{-1}\left(B_{R^{1+\delta}} , \mu_\beta \right)} \leq \frac{C}{R^{d-1+\gamma_\alpha}}.
\end{equation}
A combination of the identities~\eqref{eq:TV08399},~\eqref{eq:TV09239},~\eqref{eq:TV09249} and the estimate~\eqref{eq:TV09259} proves the inequality
\begin{equation*}
    \left\| \mathcal{L} \mathcal{H}_{\delta, \cdot k} - \rho_{\delta, \cdot k} \right\|_{\underline{H}^{-1}\left(B_{R^{1+\delta}} , \mu_\beta \right)} \leq \frac{C}{R^{d-1+\gamma_\alpha}}.
\end{equation*}
The proof of the estimate~\eqref{eq:TV19578} is complete.

\medskip

\subsection{Estimating the \texorpdfstring{$L^2$}{21}-norm of the term \texorpdfstring{$\nabla \G_\delta - \nabla \mathcal{H}_{\delta}$}{22}} \label{sec3.3chap60851}
The objective of this section is to prove that the gradient of the Green's matrix $\nabla \G_\delta$ and the gradient of the two-scale expansion $\nabla \mathcal{H}_\delta$ are close in the $\underline{L}^2 \left( A_R , \mu_\beta \right)$-norm. More specifically, we prove that there exists an exponent $\gamma_\delta > 0$ such that one has the estimate 
\begin{equation} \label{eq:TV1004020}
    \left\| \nabla \G_\delta - \nabla \mathcal{H}_\delta  \right\|_{\underline{L}^2 \left( A_R , \mu_\beta \right)} \leq \frac{C}{R^{(d-1) + \gamma_\delta}}.
\end{equation}
To prove this inequality, we work on the larger set $B_{R^{1+\delta}/2}$ and prove the estimate
\begin{align} \label{eq:TV10039}
    \left\| \nabla \G_\delta - \nabla \mathcal{H}_\delta  \right\|_{\underline{L}^2 \left( B_{R^{1+\delta}/2} , \mu_\beta \right)} \leq  \frac{C}{R^{(1+\delta)(d-1 - \ep/2)}}.
\end{align}
The inequality~\eqref{eq:TV1004020} implies~\eqref{eq:TV10039}; indeed by using that the annulus $A_R$ is included in the ball $B_{R^{1+\delta}}$ we can compute
\begin{align*}
    \left\| \nabla \G_\delta - \nabla \mathcal{H}_{\delta}  \right\|_{\underline{L}^2 \left( A_R , \mu_\beta \right)} & \leq \left(\frac{\left| B_{R^{1+\delta}/2}\right|}{\left| A_R \right|} \right)^\frac 12 \left\| \nabla \G_\delta - \nabla \mathcal{H}_{\delta}  \right\|_{\underline{L}^2 \left( B_{R^{1+\delta}/2} , \mu_\beta \right)} \\ 
    & \leq C \left( \frac{R^{d(1+\delta)}}{R^d}\right)^\frac12 \frac{C}{R^{(1+\delta)(d-1 - \ep/2)}} \\
    & \leq \frac{C}{R^{d-1 + \gamma_\delta}},
\end{align*}
where we have set $\gamma_\delta := \delta (\frac d2-1 - \ep/2)$.
We now focus on the proof of the estimate~\eqref{eq:TV10039}. The strategy is to first fix an integer $k \in \left\{1 , \ldots, \binom d2 \right\}$ and use the identity $\mathcal{L} \G_{\delta, \cdot k} =  \rho_{\delta, \cdot k}$ to rewrite the estimate~\eqref{eq:TV19578} in the following form
\begin{equation} \label{eq:TV10569}
    \left\| \L \left( \mathcal{H}_{\delta, \cdot k} -  \G_{\delta, \cdot k} \right)  \right\|_{\underline{H}^{-1} \left( B_{R^{1+\delta}} , \mu_\beta \right)} \leq \frac{C}{R^{d-1+\gamma_\alpha}}.
\end{equation}
We then use the function $\G_{\delta, \cdot k} - \mathcal{H}_{\delta, \cdot k}$ as a test function in the definition of the $H^{-1}$-norm in the inequality~\eqref{eq:TV10569} to obtain the $H^1$-estimate stated in~\eqref{eq:TV10039}, as described in the outline of the proof at the beginning of this chapter. The overall strategy is relatively straightforward; however, one has to deal with the following technical difficulty. By definition of the $H^{-1}$-norm, one needs to use a function in $H^1_0 \left( B_{R^{1+\delta}} , \mu_\beta \right)$ as a test function; in particular the function must be equal to $0$ outside the ball $B_{R^{1+\delta}}$. This condition is not verified by the function $\G_{\delta, \cdot k} - \mathcal{H}_{\delta, \cdot k}$ which is thus not a suitable test function. To overcome this issue, we introduce a cutoff function $\eta: \Zd \to \R$ supported in the ball $B_{R^{1+ \delta}}$ which satisfies the properties
\begin{equation} \label{eq:propeta}
    0 \leq \eta \leq \indc_{B_{R^{ 1 + \delta }}}, ~ \eta = 1 ~\mbox{in}~ B_{\frac{R^{ 1 + \delta }}{2}} , ~\mbox{and}~ \forall k \in \N, \, \left| \nabla^k \eta \right|\leq \frac{C}{R^{ (1 + \delta ) k }},
\end{equation}
and use the function $\eta \left( \G_{\delta, \cdot k} - \mathcal{H}_{\delta, \cdot k} \right)$ as a test function. The main difficulty is thus to treat the cutoff function. Nevertheless, this difficulty is similar to the one treated in the proof of the Caccioppoli inequality stated in Proposition~\ref{Caccio.ineq} of Chapter~\ref{section:section4}. We will thus omit some of the details of the argument and refer the reader to the proof of the Caccioppoli inequality  for the missing elements of the proof.

We use the function $\eta \left( \G_{\delta, \cdot k} - \mathcal{H}_{\delta, \cdot k} \right)$ as a test function in the inequality~\eqref{eq:TV10569}. We obtain
\begin{align} \label{eq:TV11239}
    \frac{1}{R^{(1+\delta)d}}\sum_{x \in B_{R^{1+\delta}}} \left\langle \eta \left( \G_{\delta, \cdot k} - \mathcal{H}_{\delta, \cdot k}\right) \L \left( \G_{\delta, \cdot k} - \mathcal{H}_{\delta, \cdot k} \right)  \right\rangle_{\mu_\beta} & \leq \left\| \L \left( \G_{\delta, \cdot k} - \mathcal{H}_{\delta, \cdot k} \right) \right\|_{\underline{H}^{-1} \left( B_{R^{1+\delta}} , \mu_\beta \right)} \left\|  \eta \left( \G_{\delta, \cdot k} - \mathcal{H}_{\delta, \cdot k} \right)  \right\|_{\underline{H}^1 \left( B_{R^{1+\delta}} , \mu_\beta \right)} \\
    & \leq \frac{C}{R^{d-1 + \gamma_{\alpha}}} \left\| \eta \left( \G_{\delta, \cdot k} - \mathcal{H}_{\delta, \cdot k} \right) \right\|_{\underline{H}^1 \left( B_{R^{1+\delta}} , \mu_\beta \right)}. \notag
\end{align}
We then treat the terms in the left and right sides of the inequality~\eqref{eq:TV11239} separately. Regarding the left side, we prove the estimate
\begin{equation} \label{eq:TV15049}
    \left\| \eta \left( \G_{\delta, \cdot k} - \mathcal{H}_{\delta, \cdot k} \right) \right\|_{\underline{H}^1 \left( B_{R^{1+\delta}} , \mu_\beta \right)} \leq \frac{C}{R^{d-1 - \delta d}}.
\end{equation}
The proof relies on the properties of the cutoff function $\eta$ stated in~\eqref{eq:propeta}, the regularity estimate on the Green's matrix stated in Proposition~\ref{prop:prop6.2} of Chapter~\ref{section:section4}, the $L^\infty$-bound on the homogenized Green's matrix $\G_{\mathrm{hom}}^\delta$ stated in~\eqref{eq:spatregbarGgeq3} and the bounds on the corrector and its gradient recalled below
\begin{equation*}
    \left\| \chi_{m,ij} \right\|_{\underline{L}^2 \left( B_{R^{1+\delta}} , \mu_\beta  \right)} \leq C R^{(1+\delta)(1-\alpha)},~ \left\| \nabla \chi_{m,ij} \right\|_{\underline{L}^2 \left( B_{R^{1+\delta}} , \mu_\beta  \right)} \leq C  ~\mbox{and}~  \sum_{x \in \Zd} \left\| \partial_x \chi_{m,ij} \right\|_{\underline{L}^2 \left( B_{R^{1+\delta}} , \mu_\beta  \right)}^2 \leq C  .
\end{equation*}
We first write
\begin{multline} \label{eq:TV11579}
    \left\| \eta \left( \G_{\delta, \cdot k} - \mathcal{H}_{\delta, \cdot k} \right) \right\|_{\underline{H}^1 \left( B_{R^{1+\delta}} , \mu_\beta \right)} \leq \underbrace{\frac{1}{R^{1+\delta}} \left\| \eta \left( \G_{\delta, \cdot k} - \mathcal{H}_{\delta, \cdot k} \right) \right\|_{\underline{L}^2 \left( B_{R^{1+\delta}} , \mu_\beta \right)}}_{\eqref{eq:TV11579}-(i)} + \underbrace{ \left\| \nabla \eta \left( \G_{\delta, \cdot k} - \mathcal{H}_{\delta, \cdot k} \right) \right\|_{\underline{L}^2 \left( B_{R^{1+\delta}} , \mu_\beta \right)} }_{\eqref{eq:TV11579}-(ii)} \\ + \underbrace{ \left\| \eta \left( \nabla \G_{\delta, \cdot k} - \nabla \mathcal{H}_{\delta, \cdot k} \right) \right\|_{\underline{L}^2 \left( B_{R^{1+\delta}} , \mu_\beta \right)}}_{\eqref{eq:TV11579}-(iii)}+ \underbrace{\beta \sum_{x \in \Zd}  \left\| \eta \left( \partial_x \G_{\delta, \cdot k} - \partial_x \mathcal{H}_{\delta, \cdot k} \right) \right\|_{\underline{L}^2 \left( B_{R^{1+\delta}} , \mu_\beta \right)}}_{\eqref{eq:TV11579}-(iv)},
\end{multline}
and treats the four terms in the right side separately. For the term~\eqref{eq:TV11579}-(i), we use that the function $\eta$ is non-negative and smaller than $1$ to write
\begin{equation*}
    \frac{1}{R^{1+\delta}} \left\| \eta \left( \G_{\delta, \cdot k} - \mathcal{H}_{\delta, \cdot k} \right) \right\|_{\underline{L}^2 \left( B_{R^{1+\delta}} , \mu_\beta \right)} \leq \frac{1}{R^{1+\delta}} \left( \left\| \G_{\delta, \cdot k}  \right\|_{\underline{L}^2 \left( B_{R^{1+\delta}} , \mu_\beta \right)} + \left\| \mathcal{H}_{\delta, \cdot k} \right\|_{\underline{L}^2 \left( B_{R^{1+\delta}} , \mu_\beta \right)} \right).
\end{equation*}
We then estimate the $L^2$-norm of the Green's matrix $ \G_\delta$ thanks to the estimate
\begin{equation*}
    \left\| \G_{\delta, \cdot k}  \right\|_{\underline{L}^2 \left( B_{R^{1+\delta}} , \mu_\beta \right)} \leq \left\| \G_{\delta, \cdot k} \right\|_{L^\infty \left( \Zd , \mu_\beta \right)} \leq \frac{C}{R^{(1-\delta)(d-2)}}.
\end{equation*}
The $L^2$-norm of the two-scale expansion $\mathcal{H}$ can be estimated according to the following computation
\begin{align*}
     \left\| \mathcal{H}_{\delta, \cdot k}  \right\|_{\underline{L}^2 \left( B_{R^{1+\delta}} , \mu_\beta \right)} & \leq  \left\| \bar G_{\delta , \cdot k}  \right\|_{\underline{L}^2 \left( B_{R^{1+\delta}} , \mu_\beta \right)} + \sum_{i,j} \left\| \nabla_i \bar G_{\delta , jk}  \right\|_{L^\infty \left( B_{R^{1+\delta}} , \mu_\beta \right)} \left\| \chi_{m,ij} \right\|_{\underline{L}^2 \left( B_{R^{1+\delta}} , \mu_\beta \right)} \\
     & \leq \frac{C}{R^{(1-\delta)(d-2)}} + \frac{C R^{(1 + \delta)(1-\alpha)}}{R^{(1-\delta)(d-1)}} \\
     & \leq \frac{C}{R^{(1-\delta)(d-2)}},
\end{align*}
where we have used the inequality $\alpha \gg \delta$ in the third inequality.
A combination of the three previous displays shows the estimate
\begin{equation} \label{eq:TV15019}
    \frac{1}{R^{1+\delta}} \left\| \eta \left( \G_{\delta, \cdot k} - \mathcal{H}_{\delta, \cdot k}  \right) \right\|_{\underline{L}^2 \left( B_{R^{1+\delta}} , \mu_\beta \right)} \leq \frac{C}{R^{1+\delta} \times R^{(1-\delta)(d-2)}} \leq \frac{C}{ R^{d-1 - \delta (d-3)}}.
\end{equation}
The proof of the term~\eqref{eq:TV11579}-(ii) is identical, we use the estimate $\left| \nabla \eta \right| \leq \frac{C}{R^{1+\delta}}$ and apply the estimate obtained for the term~\eqref{eq:TV11579}-(ii). We obtain
\begin{equation} \label{eq:TV15029}
     \left\| \nabla \eta \left( \G_{\delta, \cdot k} - \mathcal{H}_{\delta, \cdot k} \right) \right\|_{\underline{L}^2 \left( B_{R^{1+\delta}} , \mu_\beta \right)} \leq \frac{C}{ R^{d-1 - \delta (d-3)}}.
\end{equation}
For the term~\eqref{eq:TV11579}-(iii), we first write
\begin{equation*}
     \left\| \eta \left( \nabla \G_{\delta, \cdot k} - \nabla \mathcal{H}_{\delta, \cdot k} \right) \right\|_{\underline{L}^2 \left( B_{R^{1+\delta}} , \mu_\beta \right)} \leq \left\| \nabla \G_{\delta, \cdot k}  \right\|_{L^\infty \left(\Zd , \mu_\beta \right)} + \left\| \nabla \mathcal{H}_{\delta , \cdot k}  \right\|_{\underline{L}^2 \left( B_{R^{1+\delta}} , \mu_\beta \right)}.
\end{equation*}
The $L^\infty$-norm of the Green's matrix $\nabla \G_{\delta, \cdot k} $ is estimated by Proposition~\ref{prop:prop6.2}. We have
\begin{equation*}
    \left\| \nabla \G_{\delta, \cdot k}  \right\|_{L^\infty \left(\Zd , \mu_\beta \right)} \leq \frac{C}{R^{(1-\delta)(d-1-\ep)}}.
\end{equation*}
For the $L^2$-norm of the two-scale expansion $\mathcal{H}$, we use the formula~\eqref{eq:formnablaH2sc} and write
\begin{align*}
    \left\| \nabla \mathcal{H}_{\delta , \cdot k}  \right\|_{\underline{L}^2 \left( B_{R^{1+\delta}} , \mu_\beta \right)} & \leq  \left\| \nabla  \bar G_{\delta , \cdot k} \right\|_{L^\infty \left( \Zd  \right)}  + \sum_{i,j} \left\| \nabla \nabla_i  \bar G_{\delta,jk} \right\|_{L^\infty \left( \Zd  \right)} \left\| \chi_{m,ij} \right\|_{\underline{L}^2 \left( B_{R^{1+\delta}} , \mu_\beta  \right)} \\ & \quad + \sum_{i,j} \left\| \nabla_i  \bar G_{\delta, jk} \right\|_{L^\infty \left( \Zd  \right)}  \left\| \nabla \chi_{m,ij} \right\|_{\underline{L}^2 \left( B_{R^{1+\delta}} , \mu_\beta  \right)} \\
    & \leq \frac{C}{R^{(1-\delta)(d-1-\ep)}} + \frac{CR^{(1+\delta)(1-\alpha)}}{R^{(1-\delta)(d-\ep)}} + \frac{C}{R^{(1-\delta)(d-1-\ep)}} \\
    & \leq \frac{C}{R^{d-1 - \ep - \delta (d-1-\ep)}}.
\end{align*}
A combination of the three previous displays together with the inequality $\delta \gg \ep$ yields the estimate
\begin{equation} \label{eq:TV15039}
    \left\| \eta \left( \nabla \G_{\delta, \cdot k} - \nabla \mathcal{H}_{\delta, \cdot k}  \right) \right\|_{\underline{L}^2 \left( B_{R^{1+\delta}} , \mu_\beta \right)}  \leq \frac{C}{R^{d-1- \delta d}}.
\end{equation}
There remains to estimate the term~\eqref{eq:TV11579}-(iv). We first write
\begin{equation} \label{eq:TVde10120}
    \beta \sum_{x \in \Zd}  \left\| \eta \left( \partial_x \G_{\delta, \cdot k} - \partial_x \mathcal{H}_{\delta, \cdot k}  \right) \right\|_{\underline{L}^2 \left( B_{R^{1+\delta}} , \mu_\beta \right)} \leq \underbrace{\beta \sum_{x \in \Zd}  \left\| \eta \partial_x \G_{\delta, \cdot k}  \right\|_{\underline{L}^2 \left( B_{R^{1+\delta}} , \mu_\beta \right)}}_{\eqref{eq:TVde10120}-(i)} +  \underbrace{\beta \sum_{x \in \Zd} \left\| \eta \partial_x \mathcal{H}_{\delta}  \right\|_{\underline{L}^2 \left( B_{R^{1+\delta}} , \mu_\beta \right)}}_{\eqref{eq:TVde10120}-(ii)}
\end{equation}
and estimate the two terms in the right side separately. For the term~\eqref{eq:TVde10120}-(i), we use that the map $\G_\delta$ solves the equation $\mathcal{L} \G_{\delta, \cdot k} = \rho_\delta$ and use the map $\eta^2 \G_\delta$ as a test function. We obtain
\begin{multline*}
\beta \sum_{x \in \Zd}  \left\| \eta \partial_x \G_{\delta, \cdot k}  \right\|_{L^2 \left( B_{R^{1+\delta}} , \mu_\beta \right)}^2 = - \frac 1{2}\sum_{x \in \Zd}  \left\langle \nabla \G_{\delta, \cdot k} (x, \cdot) \cdot \nabla \left( \eta^2 \G_{\delta, \cdot k} \right)(x, \cdot) \right\rangle_{\mu_\beta} -   \beta \sum_{q \in \mathcal{Q}}  \left\langle \nabla_q \G_{\delta, \cdot k} \cdot \a_q \nabla_q \left( \eta^2 \G_{\delta, \cdot k} \right) \right\rangle_{\mu_\beta} \\ -  \frac 1{2} \sum_{n \geq 1} \sum_{x \in \Zd}  \frac{1}{\beta^{ \frac n2}} \left\langle  \nabla^{n+1} \G_{\delta , \cdot k}(x, \cdot) \cdot \nabla^{n+1} \left( \eta^2 \G_{\delta, \cdot k} \right)(x,\cdot) \right\rangle_{\mu_\beta} + \beta \sum_{x \in \Zd} \rho_{\delta , \cdot k}(x) \eta^2(x) \cdot \left\langle \G_{\delta , \cdot k}(x , \cdot) \right\rangle_{\mu_\beta}.
\end{multline*}
We then estimate the four terms in the right sides using the pointwise estimates on the function $\G_\delta$ and its gradient stated in Proposition~\ref{prop:prop6.2}, the properties on the functions $\rho_\delta$ and $\eta$ stated in~\eqref{prop.rhodelta} and~\eqref{eq:propeta} respectively. We omit the technical details and obtain the estimate
\begin{equation} \label{eq:TVde10220}
    \sum_{x \in \Zd}  \left\| \eta \partial_x \G_{\delta, \cdot k}  \right\|_{\underline{L}^2 \left( B_{R^{1+\delta}} , \mu_\beta \right)} \leq \frac{C}{R^{2(1-\delta)(d-1-\ep)}}.
\end{equation}

The term~\eqref{eq:TVde10120}-(ii) involving the two-scale expansion is the easiest one to estimate; using the explicit formula for the map $\mathcal{H}_{\delta, \cdot k} $ and the fact that the function $\bar G_\delta$ does not depend on the field $\phi$, we have the identity
\begin{equation*}
    \partial_x \mathcal{H}_{\delta , \cdot k} := \sum_{i,j} \nabla_i \bar G_{\delta, jk} \partial_x \chi_{m,ij}.
\end{equation*}
We deduce that
\begin{align} \label{eq:TVdec10340}
    \sum_{x \in \Zd} \left\| \eta \partial_x \mathcal{H}_{\delta , \cdot k}  \right\|_{\underline{L}^2 \left( B_{R^{1+\delta}} , \mu_\beta \right)} & \leq \sum_{x \in \Zd} \left\|  \partial_x \mathcal{H}_{\delta , \cdot k}  \right\|_{\underline{L}^2 \left( B_{R^{1+\delta}} , \mu_\beta \right)} \\
    & \leq C  \sum_{i,j} \left\| \nabla \bar G_{\delta , \cdot k} \right\|_{L^{\infty \left( \Zd \right)}} \sum_{x \in \Zd} \left\| \partial_x \chi_{m,ij} \right\|_{\underline{L}^2 \left( B_{R^{1+\delta}} , \mu_\beta \right)} \notag \\
    & \leq \frac{C}{R^{(1-\delta)(d - 1 - \ep)}}. \notag
\end{align}
Combining the inequalities~\eqref{eq:TVde10120},~\eqref{eq:TVde10220} and~\eqref{eq:TVdec10340} yields
\begin{equation} \label{eq:TVdec1034045}
    \sum_{x \in \Zd}  \left\| \eta \left( \partial_x \G_{\delta, \cdot k} - \partial_x \mathcal{H}_{\delta , \cdot k} \right) \right\|_{\underline{L}^2 \left( B_{R^{1+\delta}} , \mu_\beta \right)} \leq \frac{C}{R^{d-1 - \ep - \delta \left( d - 1 - \ep \right)}} \leq \frac{C}{R^{d-1 -  \delta  d }}.
\end{equation}

The inequality~\eqref{eq:TV15049} is then obtained by combining the estimates~\eqref{eq:TV15019}, \eqref{eq:TV15029},~\eqref{eq:TV15039} and~\eqref{eq:TVdec1034045}. We then put the inequality back into the inequality~\eqref{eq:TV11239} and deduce that
\begin{equation} \label{eq:TV06520}
     \frac{1}{R^{(1+\delta)d}}\sum_{x \in B_{R^{1+\delta}}} \left\langle \eta \left( \G_{\delta, \cdot k} - \mathcal{H}_{\delta , \cdot k}\right) \L \left( \G_{\delta, \cdot k} - \mathcal{H}_{\delta , \cdot k} \right)  \right\rangle_{\mu_\beta} \leq \frac{C}{R^{d-1 + \gamma_\alpha} \times R^{d-1 - \delta d}} \leq  \frac{C}{R^{2d-2 + \gamma_\alpha}},
\end{equation}
where we have used in the second inequality that the exponent $\gamma_\alpha$ is of order $\alpha$; it is thus much larger than the value $\delta d$ and we may write $\gamma_\alpha - \delta d = \gamma_\alpha$ following the conventional notation described at the beginning of Section~\ref{sec:section6.1}.

In the rest of this step, we treat the left side of~\eqref{eq:TV06520} and prove the inequality
\begin{equation} \label{eq:TV11321}
     \left\| \nabla \G_{\delta, \cdot k} - \nabla \mathcal{H}_{\delta , \cdot k} \right\|_{\underline{L}^2 \left( B_{\frac{R^{1+\delta}}{2}} , \mu_\beta \right)} \leq \frac{1}{R^{(1+\delta)d}}\sum_{x \in B_{R^{1+\delta}}} \left\langle \eta \left( \G_{\delta, \cdot k} - \mathcal{H}_{\delta , \cdot k}\right) \L \left( \G_{\delta, \cdot k} - \mathcal{H}_{\delta , \cdot k} \right)  \right\rangle_{\mu_\beta} + \frac{C}{R^{(1+\delta)(2d-2 - \ep) }} .
\end{equation}
First, by definition of the Helffer-Sj{\"o}strand operator $\mathcal{L}$, we have the identity
\begin{align} \label{eq:TV08101}
\lefteqn{ \sum_{x \in \Zd} \left\langle \eta \left( \G_{\delta, \cdot k} - \mathcal{H}_{\delta , \cdot k}\right) \L \left( \G_{\delta, \cdot k} - \mathcal{H}_{\delta , \cdot k} \right)  \right\rangle_{\mu_\beta} } \qquad & \\ & = \sum_{x , y \in \Zd} \eta(x) \left\langle \left( \partial_y \G_{\delta, \cdot k} (x , \cdot)  - \partial_y \mathcal{H}_{\delta , \cdot k} (x , \cdot) \right)^2 \right\rangle_{\mu_\beta} \\ & + \frac 1{2\beta}\sum_{x \in \Zd}  \left\langle  \left( \nabla \G_{\delta, \cdot k}  - \nabla \mathcal{H}_{\delta , \cdot k}\right)(x , \cdot) \cdot \nabla \left( \eta \left( \G_{\delta , \cdot k} -  \mathcal{H}_{\delta , \cdot k} \right)\right)(x , \cdot)  \right\rangle_{\mu_\beta} \notag \\ & +  \sum_{q \in \mathcal{Q}}  \left\langle \nabla_q \left(  \G_{\delta, \cdot k} -  \mathcal{H}_{\delta , \cdot k}\right) \cdot \a_q \nabla_q \left( \eta \left(  \G_{\delta, \cdot k}  - \mathcal{H}_{\delta , \cdot k} \right) \right) \right\rangle_{\mu_\beta} \notag \\ & +  \frac 1{2\beta} \sum_{n \geq 1} \sum_{x \in \Zd}  \frac{1}{\beta^{ \frac n2}} \left\langle  \nabla^{n+1} \left(  \G_{\delta, \cdot k}  -  \mathcal{H}_{\delta , \cdot k} \right)(x , \cdot) \cdot \nabla^{n+1} \left( \eta \left(  \G_{\delta, \cdot k}  -  \mathcal{H}_{\delta , \cdot k} \right) \right)(x,\cdot) \right\rangle_{\mu_\beta}. \notag
\end{align}
We then estimate the four terms on the right side separately. For the first one, we use that it is non-negative
\begin{equation*}
    \sum_{x , y \in \Zd} \eta(x)^2 \left\langle \left( \partial_y \G_{\delta, \cdot k} (x , \cdot)  - \partial_y \mathcal{H}_{\delta, \cdot k} (x , \cdot) \right)^2 \right\rangle_{\mu_\beta}  \geq 0.
\end{equation*}
For the second one, we expand the gradient of the product $\eta \left( \G_{\delta, \cdot k}  - \mathcal{H} \right)$ and write
\begin{align} \label{eq:TV09490}
    \lefteqn{\sum_{x \in \Zd}  \left\langle  \left( \nabla \G_{\delta, \cdot k}  - \nabla \mathcal{H}_{\delta , \cdot k}\right)(x , \cdot) \cdot \nabla \left( \eta \left( \G_{\delta , \cdot k} -  \mathcal{H}_{\delta , \cdot k} \right)\right)(x , \cdot)  \right\rangle_{\mu_\beta}} \qquad & \\ &
    = \sum_{x \in \Zd}  \eta(x)  \left\langle  \left( \nabla \G_{\delta, \cdot k}  - \nabla \mathcal{H}_{\delta}\right)(x , \cdot) \cdot   \nabla \left( \G_{\delta , \cdot k} -  \mathcal{H}_{\delta , \cdot k} \right)(x , \cdot)  \right\rangle_{\mu_\beta} \notag \\
    & \quad + \sum_{x \in \Zd}  \left\langle  \left( \nabla \G_{\delta, \cdot k}  - \nabla \mathcal{H}_{\delta , \cdot k}\right)(x , \cdot) \cdot  \nabla \eta(x)\left( \G_{\delta , \cdot k}-  \mathcal{H}_{\delta , \cdot k} \right)(x , \cdot)  \right\rangle_{\mu_\beta}. \notag
\end{align}
we divide the identity~\eqref{eq:TV09490} by the volume factor $R^{(1+\delta)d}$ and use the properties of the function $\eta$ stated in~\eqref{eq:propeta}. In particular, we use that the gradient of $\eta$ is supported in the annulus $A_{R^{1+\delta}} := B_{R^{1+\delta}} \setminus B_{\frac{R^{1+\delta}}{2}}$ and obtain
\begin{multline} \label{eq:TV07591}
    \frac{1}{R^{(1+\delta)d}} \sum_{x \in \Zd}  \left\langle  \left( \nabla \G_{\delta, \cdot k}  - \nabla \mathcal{H}_{\delta, \cdot k} \right)(x , \cdot) \cdot \nabla \left( \eta \left( \G_{\delta, \cdot k}-  \mathcal{H}_{\delta, \cdot k}  \right)\right)(x , \cdot)  \right\rangle_{\mu_\beta} \geq c \left\| \eta \left( \nabla \G_{\delta, \cdot k}  - \nabla \mathcal{H}_{\delta, \cdot k}  \right) \right\|_{\underline{L}^2 \left( B_{R^{1+\delta}} , \mu_\beta \right)}^2 \\ -  \frac{C}{R^{1+\delta}} \left\| \nabla \G_{\delta, \cdot k}  \ - \nabla \mathcal{H}_{\delta, \cdot k}  \right\|_{\underline{L}^2 \left( A_{R^{1+\delta}} , \mu_\beta \right)} \left\|  \G_{\delta, \cdot k}  - \mathcal{H}_{\delta, \cdot k}  \right\|_{\underline{L}^2 \left( A_{R^{1+\delta}} , \mu_\beta \right)}.
\end{multline}
By a computation similar to the one performed for the term ~\eqref{eq:TV11579}-(iii), but using the estimates~\eqref{eq:TV20383} for the Green's matrices in the distant annulus $A_{R^{1+\delta}}$, instead of the $L^\infty$-estimates~\eqref{eq:TV090191} and~\eqref{eq:spatregbarGgeq3}. We obtain
\begin{equation} \label{eq:TV07601}
    \left\| \nabla \G_{\delta, \cdot k}  \ - \nabla \mathcal{H}_{\delta, \cdot k}  \right\|_{\underline{L}^2 \left( A_{R^{1+\delta}} , \mu_\beta \right)} \leq \frac{C}{R^{(1+\delta)(d -1 - \ep)}}~\mbox{and}~\left\|  \G_{\delta, \cdot k}  - \mathcal{H}_{\delta, \cdot k}  \right\|_{\underline{L}^2 \left( A_{R^{1+\delta}}, \mu_\beta \right)} \leq \frac{C}{R^{(1+\delta)(d-2)}}.
\end{equation}
A combination of the inequalities~\eqref{eq:TV07591} and~\eqref{eq:TV07601} proves the estimate
\begin{multline} \label{eq:TV11111}
    \frac{1}{R^{(1+\delta)d}} \sum_{x \in \Zd}  \left\langle  \left( \nabla \G_{\delta, \cdot k}  - \nabla \mathcal{H}_{\delta, \cdot k} \right)(x , \cdot) \cdot \nabla \left( \eta \left( \G_\delta-  \mathcal{H}_{\delta, \cdot k}  \right)\right)(x , \cdot)  \right\rangle_{\mu_\beta} + \frac{C}{R^{(1+\delta)(2d-2 - \ep)}} \\ \geq c  \left\| \eta \left( \nabla \G_{\delta, \cdot k}  - \nabla \mathcal{H}_{\delta, \cdot k} \right) \right\|_{\underline{L}^2 \left( B_{R^{1+\delta}} , \mu_\beta \right)}^2.
\end{multline}
The other terms in the right side of the identity~\eqref{eq:TV08101} involving the sum over the iteration of the Laplacian and over the charges $q \in \mathcal{Q}$ are treated similarly. Instead of expanding the gradient as it was done in~\eqref{eq:TV09490}, we use the commutation estimates (see~\eqref{eq:Caccio1925} and~\eqref{eq:TV1549C} in the proof of Proposition~\ref{Caccio.ineq} in Chapter~\ref{section:section4}), for each integer $n \in \N$, and each pair $(x , \phi) \in \Zd \times \Omega$,
\begin{equation}
\left| \nabla^n \left( \eta \left ( \G_{\delta, \cdot k}  - \mathcal{H}_{\delta, \cdot k}  \right) \right) (x , \phi)  -  \eta(x) \nabla^n\left ( \G_{\delta, \cdot k}  - \mathcal{H}_{\delta, \cdot k}  \right)(x , \phi)  \right| \leq \frac{C^n}{R^{1+\delta}} \sum_{z \in B(x,n)} | \G_{\delta, \cdot k} ( z, \phi)  - \mathcal{H}_{\delta, \cdot k} ( z, \phi)|.
\end{equation}
and for each charge $q \in \mathcal{Q}$, each point $x$ in the support of the charge $q$ and each field $\phi \in \Omega$,
\begin{equation*}
\left| \nabla_q \left( \eta \left ( \G_{\delta, \cdot k}  - \mathcal{H}_{\delta, \cdot k}  \right) \right) - \eta(x) \nabla_q  \left( \G_{\delta, \cdot k}  - \mathcal{H}_{\delta, \cdot k}  \right) \right| \leq \frac{C}{R^{1+\delta}} \left(\diam q \right) \left\|q \right\|_{1} \sum_{z \in \supp q} | \G_{\delta, \cdot k}  - \mathcal{H}|.
\end{equation*}
The details of the argument are similar to the one presented in the proof of the Caccioppoli inequality (Proposition~\ref{Caccio.ineq} of Chapter~\ref{section:section4}) and are omitted. The results obtained are stated below
\begin{equation} \label{eq:TV11121}
    \frac{1}{R^{(1+\delta)d}}  \sum_{q \in \mathcal{Q}}  \left\langle \nabla_q \left(  \G_{\delta, \cdot k} -  \mathcal{H}_{\delta, \cdot k} \right) \cdot \a_q \nabla_q \left( \eta \left(  \G_{\delta, \cdot k}  - \mathcal{H}_{\delta, \cdot k}  \right) \right) \right\rangle_{\mu_\beta}  + \frac{C}{R^{(1+\delta)(2d-2 - \ep) }} \geq -C e^{-c \sqrt{\beta}} \left\| \eta \left( \nabla \G_{\delta, \cdot k}  - \nabla \mathcal{H}_{\delta, \cdot k}  \right) \right\|_{L^2 \left( \Zd , \mu_\beta \right)}^2
\end{equation}
and
\begin{align} \label{eq:TV11131}
    \lefteqn{ \frac{1}{R^{(1+\delta)d}}  \sum_{n \geq 1} \sum_{x \in \Zd}  \frac{1}{\beta^{ \frac n2}} \left\langle  \nabla^{n+1} \left(  \G_{\delta, \cdot k}  -  \mathcal{H}_{\delta, \cdot k}  \right)(x , \cdot) \cdot \nabla^{n+1} \left( \eta \left(  \G_{\delta, \cdot k}  -  \mathcal{H}_{\delta, \cdot k}  \right) \right)(x,\cdot) \right\rangle_{\mu_\beta} + \frac{C}{R^{(1+\delta)(2d-2 - \ep) }} }  \qquad \qquad\qquad\qquad \qquad\qquad\qquad & \\ & \geq   \sum_{n \geq 1} \sum_{x \in \Zd}  \frac{1}{\beta^{ \frac n2}} \left\langle \eta(x) \left| \nabla^{n+1} \left(  \G_{\delta, \cdot k}  -  \mathcal{H}_{\delta, \cdot k}  \right)(x , \cdot) \right|^2 \right\rangle_{\mu_{\beta}} \notag \\
    & \geq 0. \notag
\end{align}
We then combine the identity~\eqref{eq:TV08101} with the estimates~\eqref{eq:TV11111},~\eqref{eq:TV11121} and~\eqref{eq:TV11131} and assume that the inverse temperature $\beta$ is large enough. We obtain
\begin{align*}
     \frac{1}{R^{(1+\delta)d}} \sum_{x \in \Zd} \left\langle \eta \left( \G_{\delta, \cdot k} - \mathcal{H}\right) \L \left( \G_{\delta, \cdot k} - \mathcal{H}_{\delta, \cdot k}  \right)  \right\rangle_{\mu_\beta} +  \frac{C}{R^{(1+\delta)(2d-2 - \ep) }} & \geq c \left\| \eta \left( \nabla \G_{\delta, \cdot k}  - \nabla \mathcal{H}_{\delta, \cdot k}  \right) \right\|_{\underline{L}^2 \left( \Zd , \mu_\beta \right)}^2 \\ & \geq c \left\|  \nabla \G_{\delta, \cdot k}  - \nabla \mathcal{H}_{\delta, \cdot k}   \right\|_{\underline{L}^2 \left( B_{R^{1+\delta}/2}, \mu_\beta \right)}^2. \notag
\end{align*}
The proof of the inequality~\eqref{eq:TV11321} is then complete. To complete the proof of Step 2, we combine the estimates~\eqref{eq:TV06520} and~\eqref{eq:TV11321}. We obtain
\begin{equation} \label{eq:TV17540801}
     \left\| \nabla \G_{\delta, \cdot k} - \nabla \mathcal{H}_{\delta, \cdot k}  \right\|_{L^2 \left( B_{\frac{R^{1+\delta}}{2}} , \mu_\beta \right)}^2 \leq \frac{C}{R^{2d-2 + \gamma_\alpha}} + \frac{C}{R^{(1+\delta)(2d-2 - \ep) }} \leq \frac{C}{R^{(1+\delta)(2d-2 - \ep)}},
\end{equation}
where the last inequality is a consequence of the fact that $\gamma_\alpha$ is of order $\alpha$ and of the ordering $\alpha \gg \delta \gg \ep$. Since the inequality~\eqref{eq:TV17540801} is valid for any integer $k \in \left\{1 , \ldots , \binom d2 \right\}$, the proof of the estimate~\eqref{eq:TV10039} is complete.

\medskip

\subsection{Homogenization of the gradient of the Green's matrix} \label{sec3.4chap60851}
In this section, we post-process the conclusion~\eqref{eq:TV10039} of Section~\ref{sec3.3chap60851} and prove that the gradient of the Green's matrix $\nabla \mathcal{G}_{\cdot k}$ is close to the map $\sum_{i,j} \left( e_{ij} + \nabla \chi_{ij} \right) \nabla_i \bar G_{jk}$. The objective is to prove that there exists an exponent $\gamma_\delta > 0$ such that
\begin{equation} \label{eq:TV14282}
    \left\| \nabla \G_{\cdot k} - \sum_{i,j} \left( e_{ij} + \nabla \chi_{ij} \right) \nabla_i \bar G_{jk} \right\|_{\underline{L}^2 \left( A_R , \mu_\beta \right)}  \leq \frac{C}{R^{d-1 + \gamma_{\delta}}}.
\end{equation}
We first use the regularity estimates stated in Proposition~\ref{prop:prop6.2} and the $L^2$-bound on the gradient of the infinite-volume corrector, for each $x \in \Zd$, each pair of integers $(i,j) \in \{ 1 , \ldots , d \} \times \{1 , \ldots, \binom d2 \}$, $\left\| \nabla \chi_{ij}(x , \cdot) \right\|_{L^2 \left( \mu_\beta \right)} \leq C$. We write
\begin{align} \label{eq:TV14272}
    \lefteqn{\left\| \nabla \left( \G_{\cdot k} - \G_{\delta,\cdot k} \right) - \sum_{i,j} \left( e_{ij} + \nabla \chi_{ij} \right) \nabla_i \left( \bar G_{\delta, jk} - \bar G_{jk} \right) \right\|_{\underline{L}^2 \left( A_R , \mu_\beta \right)}} \qquad & \\ & \leq \left\| \nabla \left( \G_{\cdot k} - \G_{\delta, \cdot k} \right) \right\|_{\underline{L}^2 \left( A_R , \mu_\beta \right)} + \sum_{i,j} \left\| \left( e_{ij} + \nabla \chi_{ij} \right) \right\|_{\underline{L}^2 \left( A_R , \mu_\beta \right)} \left\| \nabla_i \left( \bar G_{\delta,jk} - \bar G_{jk} \right) \right\|_{L^\infty \left( A_R , \mu_\beta \right)}  \notag \\
    & \leq \frac{C}{R^{d-1+\gamma_\delta}}. \notag
\end{align}
Using the inequality~\eqref{eq:TV14272}, we see that to prove~\eqref{eq:TV14282} it is sufficient to prove the estimate 
\begin{equation} \label{eq:TV15172}
 \left\| \nabla \G_{\delta, \cdot k} - \sum_{i,j} \left( e_{ij} + \nabla \chi_{ij} \right) \nabla_i \bar G_{\delta,jk} \right\|_{\underline{L}^2 \left( A_R , \mu_\beta \right)}  \leq \frac{C}{R^{d-1 + \gamma_\delta}}.
 \end{equation}
We then use the main estimate~\eqref{eq:TV10039} and deduce that, to prove the inequality~\eqref{eq:TV15172}, it is sufficient to prove
\begin{equation} \label{eq:TV15182}
    \left\| \nabla \mathcal{H}_{\delta, \cdot k} - \sum_{i,j} \left( e_{ij} + \nabla \chi_{ij} \right) \nabla_i \bar G_{\delta,jk} \right\|_{\underline{L}^2 \left( A_R , \mu_\beta \right)} \leq \frac{C}{R^{d-1 + \gamma_\delta}}.
\end{equation}
The rest of the argument of this step is devoted to the proof of~\eqref{eq:TV15182}. We first use the explicit formula for the gradient of the two-scale expansion $\nabla \mathcal{H}_{\delta, \cdot k}$ stated in~\eqref{eq:formnablaH2sc} and write 
\begin{multline*}
\left\| \nabla \mathcal{H}_{\delta, \cdot k} - \sum_{i,j} \left( e_{ij} + \nabla \chi_{m,ij} \right) \nabla_i G_{\delta,jk} \right\|_{\underline{L}^2 \left( A_R , \mu_\beta \right)} \\ \leq \sum_{i,j} \left\|  \nabla \nabla_i  \bar G_{\delta,jk}  \chi_{m,ij} \right\|_{\underline{L}^2 \left( A_R , \mu_\beta \right)}+ \left\| \left(\nabla_i  \bar G_{\delta,jk} \right) \left( \nabla \chi_{m,ij} - \nabla \chi_{ij} \right) \right\|_{\underline{L}^2 \left(A_R , \mu_\beta \right)}.
\end{multline*}
We then use the regularity estimate~\eqref{eq:spatregbarGgeq3}, the quantitative sublinearity of the corrector stated in Proposition~\ref{prop:prop5.25} and Proposition~\ref{prop5.26} of Chapter \ref{section5} to quantify the $L^2$-norm of the difference between the gradient of finite-volume corrector and the gradient of the infinite-volume corrector. We obtain
\begin{align} \label{eq:TV13422}
    \left\| \nabla \mathcal{H}_{\delta, \cdot k} - \sum_{i,j} \left( e_{ij} + \nabla \chi_{m,ij} \right) \nabla_i \bar G_{\delta,jk} \right\|_{\underline{L}^2 \left( A_R , \mu_\beta \right)} & \leq \left( \frac{\left| A_R \right|}{\left| B_{R^{1+\delta}} \right|} \right)^\frac 12 \frac{C R^{1-\alpha}}{R^{d - \ep}} + \left( \frac{\left| A_R \right|}{\left| B_{R^{1+\delta}} \right|} \right)^\frac 12 \frac{C R^{-\alpha}}{R^{d - 1 - \ep}} \\
    & \leq \frac{C}{R^{d-1 + \gamma_\alpha}}, \notag
\end{align}
where we have set $\gamma_\alpha := \alpha - \ep - \frac{d \delta}{2}$. Using that the exponent $\gamma_\alpha$ is larger than the exponent $\gamma_\delta$ completes the proof of the estimate~\eqref{eq:TV14282}.

\section{Homogenization of the mixed derivative of the Green's matrix} \label{sec:section6.2}

In this section, we use Proposition~\ref{prop:prop6.1} to prove Theorem~\ref{thm:homogmixedder}. We fix a charge $q_1 \in \mathcal{Q}$ and recall the definitions of the maps $\mathcal{U}_{q_1}$ and $\bar G_{q_1}$ given in the statement of Theorem~\ref{prop:prop6.1}. The proof is decomposed into four sections and follows the outline of the proof given in Section~\ref{sec3.2chap60851}.

\subsection{Preliminary estimates}

In this section, we record some properties pertaining to the functions $\mathcal{U}_{q_1}$ and $\bar G_{q_1}$ which are used in the argument.

\begin{proposition} \label{prop:prop6.5}
There exists an inverse temperature $\beta_0 := \beta_0(d) < 0$ and a constant $C_{q_1}$ which satisfies the estimate $C_{q_1} \leq C \left\| q_1 \right\|_1^k$, for some $C(d) < \infty$ and $k(d) < \infty$, such that the following statement holds. For each pair of points $y \in \Zd$ and each integer $k \in \N$, one has the estimates
\begin{equation*}
\left\| \nabla \mathcal{U}_{q_1} \left( y, \cdot \right)\right\|_{L^\infty \left(\mu_\beta \right)} \leq \frac{C_{q_1}}{|y|^{d-\ep}}, \hspace{3mm} \left\| \mathcal{U}_{q_1} \left( y , \cdot \right)\right\|_{L^\infty \left(\mu_\beta \right)} \leq \frac{C_{q_1}}{|y|^{d-1-\ep}} \hspace{3mm} \mbox{and} \hspace{3mm} \left| \nabla^k \bar G_{q_1} (y) \right| \leq \frac{C_{q_1}}{|y |^{d-1 + k}}.
\end{equation*}
\end{proposition}

\begin{proof}
The proof is a consequence of the regularity estimates given in Proposition~\ref{prop.prop4.11chap4} of Chapter~\ref{chap:chap3} and the identity $q = \di n_q$.
\end{proof}

\subsection{Exploiting the symmetry of the Helffer-Sj{\"o}strand operator} \label{sec4.2explsymmHS} The objective of this section is to use Proposition~\ref{prop:prop6.2} and the symmetry of the Helffer-Sj{\"o}strand operator $\mathcal{L}$ to prove the following estimate of the expectations
\begin{equation} \label{eq:TV07495}
    \left( R^{-d} \sum_{z\in A_R} \left| \left\langle \mathcal{U}_{q_1}(z, \cdot) \right\rangle_{\mu_\beta} - \bar G_{q_1}(z) \right|^2 \right)^\frac 12 \leq \frac{C}{R^{d+\gamma_\delta}}.
\end{equation}
We start from the formula, for each integer $k \in \left\{ 1, \ldots , \binom d2\right\}$
\begin{equation} \label{eq:TV22283}
\left\| \di^* \mathcal{G}_{\cdot k} - \sum_{i,j} \left( \di^* l_{e_{ij}} + \di^* \chi_{ij} \right) \nabla_i \bar G_{jk} \right\|_{\underline{L}^2 \left(A_R,  \mu_\beta\right)} \leq \frac{C}{R^{ d-1 + \gamma_\delta}},
\end{equation}
which is a direct consequence of Proposition~\ref{prop:prop6.1} since the codifferential is a linear functional of the gradient.
Using the estimate~\eqref{eq:TV22283}, we deduce that
\begin{multline*}
    R^{-d} \sum_{x \in A_R} \left| \left\langle \cos 2\pi\left( \phi , q_1(x + \cdot) \right)  \left( n_{q_1}\left( x + \cdot \right), \di^* \mathcal{G}_{\cdot k} \right) \right\rangle_{\mu_\beta}- \sum_{i,j} \left\langle \cos 2\pi\left( \phi , q_1(x + \cdot) \right) \left( n_{q_1}\left( x + \cdot \right) \left(\di^* l_{e_{ij}} + \di^* \chi_{ij}\right) \right)  \right\rangle_{\mu_\beta }\nabla_i \bar G_{jk}(x) \right| \\ \leq \frac{C_{q_1}}{R^{d - 1+\gamma_\delta}}.
\end{multline*}
By the translation invariance of the measure $\mu_\beta$ and the stationarity of the gradient of the infinite-volume corrector, we deduce that
\begin{align*}
    \lefteqn{\sum_{i,j} \left\langle \cos 2\pi\left( \phi , q_1(x + \cdot) \right) \left( n_{q_1}\left( x + \cdot \right) \left(\di^* l_{e_{ij}} + \di^* \chi_{ij}\right) \right)  \right\rangle_{\mu_\beta }\nabla_i \bar G_{jk}(x)} \qquad & \\ & = \sum_{i,j} \left\langle \cos 2\pi\left( \phi , q_1 \right) \left( n_{q_1} \left( \di^* l_{e_{ij}} + \di^* \chi_{ij} \right) \right)  \right\rangle_{\mu_\beta }\nabla_i \bar G_{jk}(x) \\
    & = \bar G_{q_1} \left( x \right).
\end{align*}
We now claim that we have the identity, for each point $x \in A_R$,
\begin{equation*}
    \left\langle \cos 2\pi\left( \phi , q_1(x + \cdot) \right)  \left( n_{q_1}\left( x + \cdot \right), \di^* \mathcal{G} \right) \right\rangle_{\mu_\beta} = \left\langle \mathcal{U}_{q_1(x + \cdot)} (0 , \cdot) \right\rangle_{\mu_\beta}.
\end{equation*}
The proof of this result is a consequence of the symmetry of the Helffer-Sj{\"o}strand operator $\L$. To argue this, we use the computation
\begin{align*}
     \left\langle \cos 2\pi\left( \phi , q_1(x + \cdot) \right)  \left( n_{q_1}\left( x + \cdot \right), \di^* \mathcal{G} \right) \right\rangle_{\mu_\beta} & =  \left\langle   \left( \cos 2\pi\left( \phi , q_1(x + \cdot) \right) q_1\left( x + \cdot \right),  \mathcal{G} \right) \right\rangle_{\mu_\beta} \\
     & = \left\langle   \left( \cos 2\pi\left( \phi , q_1(x + \cdot) \right) q_1\left( x + \cdot \right),  \L^{-1}\delta_0 \right) \right\rangle_{\mu_\beta} \\
     & = \left\langle   \left( \L^{-1} \cos 2\pi\left( \phi , q_1(x + \cdot) \right) q_1\left( x + \cdot \right), \delta_0 \right) \right\rangle_{\mu_\beta} \\
     & = \left\langle \mathcal{U}_{q_1(x + \cdot)} (0, \cdot) \right\rangle_{\mu_\beta}.
\end{align*}
A combination of the four previous displays implies
\begin{equation} \label{eq:TV8194}
    R^{-d}\sum_{x\in A_R} \left| \left\langle \mathcal{U}_{q_1 \left( x + \cdot \right)}(0, \cdot) \right\rangle_{\mu_\beta} - \bar G_{q_1}( x) \right| \leq \frac{C}{R^{d+\gamma_\delta}}.
\end{equation}
We then use the translation invariance of the measure $\mu_\beta$ and the definition of the map $\mathcal{U}_{q_1}$ as the solution of the Helffer-Sj{\"o}strand equation~\eqref{eq:defmathcalU} to write
\begin{equation} \label{eq:TV21154}
    \left\langle \mathcal{U}_{q_1 \left( x + \cdot \right)}(0, \cdot) \right\rangle_{\mu_\beta} = \left\langle \mathcal{U}_{q_1}(x, \cdot) \right\rangle_{\mu_\beta}.
\end{equation}
Combining the inequality~\eqref{eq:TV8194} with the identity~\eqref{eq:TV21154}, we obtain
\begin{equation*}
    R^{-d}\sum_{x\in A_R} \left| \left\langle \mathcal{U}_{q_1}(x, \cdot) \right\rangle_{\mu_\beta} - \bar G_{q_1}( x) \right| \leq \frac{C}{R^{d+\gamma_\delta}}.
\end{equation*}
We finally upgrade the $L^1$-inequality stated in~\eqref{eq:TV8194} into an $L^2$-inequality: by using Proposition~\ref{prop:prop6.5}, we write
\begin{align*}
    \lefteqn{R^{-d}\sum_{x\in A_R} \left| \left\langle \mathcal{U}_{q_1}(x, \cdot) \right\rangle_{\mu_\beta} - \bar G_{q_1}(x) \right|^2} \qquad & \\ & \leq \left(R^{-d}\sum_{x\in A_R} \left| \left\langle \mathcal{U}_{q_1}(x, \cdot) \right\rangle_{\mu_\beta} - \bar G_{q_1}(x) \right|\right) \left(  \left\| \mathcal{U}_{q_1}(x, \cdot) \right\|_{L^\infty \left(A_R, \mu_\beta\right)}  + \left\| \bar G_{q_1} \right\|_{L^\infty \left(A_R \right)}\right) \\
    & \leq \frac{C_{q_1} }{R^{d-1+\gamma_\delta} \times R^{d-1-\ep}} \\
    & \leq \frac{C_{q_1} }{R^{2d-2+\gamma_\delta}},
\end{align*}
where we have used the convention notation described at the beginning of Section~\ref{sec:section6.1} to absorb the exponent $\ep$ into the exponent $\gamma_\delta$ in the third inequality.

\medskip

\subsection{Contraction of the variance of \texorpdfstring{$\mathcal{U}_{q_1}$}{25}} \label{SEctiocontracvar4.3chap7} In this section, we prove that the random variable $\mathcal{U}_{q_1}$ contracts around its expectation. To this end, we prove the variance estimate, for each point $z \in \Zd$,
\begin{equation} \label{eq:TV08015}
    \var \left[ \mathcal{U}_{q_1}(z, \cdot) \right] \leq \frac{C_{q_1}}{|z|^{2d-2\ep}}.
\end{equation}
Let us make a comment about the result: since the size of the random variable $\mathcal{U}_{q_1}(z, \cdot)$ is of order $|z|^{1-d}$ (since it behaves like the gradient of a Green's function), we would expect its variance to be of order $|z|^{2-2d}$. The inequality~\eqref{eq:TV08015} asserts that it is in fact of order $|z|^{2d - 2\ep}$ which is smaller than the typical size of the random variable $\mathcal{U}_{q_1}(z, \cdot)$ by an algebraic factor.

Once this estimate is established, we can combine it with the main result~\eqref{eq:TV07495} of Section~\ref{sec4.2explsymmHS} to prove that the map $\mathcal{U}_{q_1}$ is close to the (deterministic) Green's function $\bar G_{q_1}$ in the $\underline{L}^2 \left( A_R , \mu_\beta \right)$-norm: we obtain the inequality
\begin{equation} \label{eq:TV16256}
    \left\| \mathcal{U}_{q_1} - \bar G_{q_1} \right\|_{\underline{L}^2 \left( A_R , \mu_\beta \right)} \leq \frac{C}{R^{d-1-\gamma_\delta}}.
\end{equation}
We now prove of the variance estimate~\eqref{eq:TV08015}. We first apply the Brascamp-Lieb inequality and write
\begin{equation} \label{eq:TV08435}
     \var \left[ \mathcal{U}_{q_1}(z, \cdot) \right] \leq C \sum_{y,y_1 \in \Zd} \left\| \partial_y \mathcal{U}_{q_1}(z, \cdot)  \right\|_{L^2 \left( \mu_\beta \right)} \frac{C}{|y - y_1|^{d-2}} \left\| \partial_{y_1} \mathcal{U}_{q_1}(z, \cdot)  \right\|_{L^2 \left( \mu_\beta \right)}.
\end{equation}
A consequence of the inequality~\eqref{eq:TV08435} is that to estimate the variance of the random variable $\mathcal{U}_{q_1}(z,\cdot)$, it is sufficient to understand the behavior of the mapping $y \mapsto \partial_y \mathcal{U}_{q_1}(z, \cdot)$. To this end, we appeal to the differentiated Helffer-Sj{\"o}strand equation: following the arguments developed in Section~\ref{sec.section4.5} of Chapter \ref{section:section4}, the map $u : (y , z , \phi) \mapsto \partial_y \mathcal{U}_{q_1}(z , \phi)$ is solution of the equation
\begin{equation*}
    \L_{\mathrm{der}} u (x , y , \phi) = -  \sum_{q \in \mathcal{Q}} 2\pi z \left( \beta , q\right) \cos 2\pi\left( \phi , q \right) \left( \mathcal{U}_{q_1}, q \right) q(x) \otimes q(y) + 2\pi\sin 2\pi\left( \phi , q_1 \right) q_1 (x) \otimes q_1(y) \hspace{3mm} \mbox{in} \hspace{3mm} \Zd \times \Zd \times \Omega.
\end{equation*}
The function $u$ can be expressed in terms of the Green's function $\mathcal{G}_{\mathrm{der}}$ and we write, for each triplet $(x , y , \phi) \in \Zd \ \times \Zd \times \Omega$,
\begin{align*}
     u (x , y , \phi) & = \sum_{q \in \mathcal{Q}}2\pi z \left( \beta , q \right) \sum_{x_1 , y_1 \in \Zd} \di^*_{x_1} \di^*_{y_1} \G_{\mathrm{der}, \cos 2\pi\left( \phi , q \right) \left( \mathcal{U}_{q_1}, q \right)} \left( x , y , \phi ; x_1 , y_1 \right) n_{q}(x_1) \otimes n_{q}(y_1)  \\
     & \quad + \sum_{x_1 , y_1 \in \Zd} 2\pi \di^*_{x_1} \di^*_{y_1} \G_{\mathrm{der}, \sin 2\pi\left( \phi , q_1 \right)} \left( x , y , \phi ; x_1 , y_1 \right)  n_{q_1} (x_1) \otimes n_{q_1}(y_1).
\end{align*}
We use the regularity estimates on the Green's function stated in Proposition~\ref{cor:corollary4.14} of Chapter~\ref{section:section4} to obtain, for each pair of points $(x , y) \in \Zd \times \Zd$,
\begin{align} \label{eq:TV10395}
    \left\| u (x , y , \cdot) \right\|_{L^\infty \left( \mu_\beta \right)} & \leq \underbrace{C \sum_{q \in \mathcal{Q}} e^{-c \sqrt{\beta} \left\| q \right\|_1} \sum_{x_1 , y_1 \in \Zd} \frac{\left| n_q(x_1) \right| \left| n_q(y_1) \right| \left\| \left( \di^* \mathcal{U}_{q_1} , n_q \right) \right\|_{L^\infty \left( \mu_\beta \right)} }{|x - x_1|^{2d - \ep} + |y - y_1|^{2d - \ep}}}_{\eqref{eq:TV10395}-(i)} \\
    & \quad + \underbrace{\sum_{x_1 , y_1 \in \Zd} \frac{\left| n_{q_1}(x_1) \right| \left| n_{q_1}(y_1) \right| }{|x - x_1|^{2d - \ep} + |y - y_1|^{2d - \ep}}}_{\eqref{eq:TV10395}-(ii)}. \notag
\end{align}
We then estimate the two terms~\eqref{eq:TV10395}-(i) and~\eqref{eq:TV10395}-(ii) separately. We first focus on the term~\eqref{eq:TV10395}-(i) and prove the inequality
\begin{equation} \label{eq:TV16095}
    \eqref{eq:TV10395}-(i) \leq  \frac{C_{q_1}}{|x-y|^{d - \ep} \max \left( \left| x \right|, \left| y \right|\right)^{d-1}}
\end{equation}
To prove the estimate~\eqref{eq:TV16095}, we first decompose the set of charges $\mathcal{Q}$ according to the following procedure. For each $z \in \Zd$, we denote by $\mathcal{Q}_z$ the set of charges $q \in \mathcal{Q}$ such that the point $z$ belongs to the support of $n_q$, i.e., $\mathcal{Q}_z := \left\{ q \in \mathcal{Q} \, : \, z \in \supp n_q  \right\}$. We note that we have the equality $\mathcal{Q} := \bigcup_{z \in \Zd} \mathcal{Q}_z$ but the collection $\left( \mathcal{Q}_z\right)_{z \in \Zd}$ is not a partition of $\mathcal{Q}$. We first prove that, for each point $z \in \Zd$,
\begin{equation} \label{eq:TV16585}
    \sum_{q \in \mathcal{Q}_z} e^{-c\sqrt{\beta} \left\| q \right\|_1}\sum_{x_1 , y_1 \in \Zd} \frac{\left| n_q(x_1) \right| \left| n_q(y_1) \right| \left\| \left( \di^* \mathcal{U}_{q_1} , n_q \right) \right\|_{L^\infty \left( \mu_\beta \right)}}{|x - x_1|^{2d - \ep} + |y - y_1|^{2d - \ep}} \leq \frac{C_{q_1}}{\left(|x - z|^{2d-\ep} + |y - z|^{2d-\ep}\right) \times |z|^{d-\ep} }
\end{equation}
To prove the estimate~\eqref{eq:TV16585}, we first use Proposition~\ref{prop:prop6.5} to estimate the term $\left\| \left( \di^* \mathcal{U}_{q_1} , n_q \right) \right\|_{L^\infty \left( \mu_\beta \right)}$. We write, for each charge $q \in \mathcal{Q}_z$, 
\begin{align} \label{eq:TV17545}
    \left\| \left( \di^* \mathcal{U}_{q_1} , n_q \right) \right\|_{L^\infty \left( \mu_\beta \right)} & \leq \left\| \nabla \mathcal{U}_{q_1} \right\|_{L^\infty \left( \supp n_q , \mu_\beta \right)} \left\| n_q \right\|_{L^1} \\
    & \leq C_{q_1,q} \sup_{z_1 \in \supp n_q} \frac{1}{|z_1|^{d-\ep}}. \notag
\end{align}
Since we have assumed that the point $z$ belongs to the support of the charge $n_q$, we have the inequality
\begin{equation} \label{eq:TV17555}
    \sup_{z_1 \in \supp n_q} \frac{1}{|z_1|^{d-\ep}} \leq \frac{\left| \diam n_q\right|^{d - \ep}}{|z|^{d-\ep}} \leq \frac{C_q}{|z|^{d-\ep}}.
\end{equation}
Combining the estimates~\eqref{eq:TV17545},~\eqref{eq:TV17555}, we deduce that
\begin{equation} \label{eq:TV18015}
    \left\| \left( \di^* \mathcal{U}_{q_1} , n_q \right) \right\|_{L^\infty \left( \mu_\beta \right)} \leq \frac{C_{q_1,q}}{|z|^{d - \ep}}.
\end{equation}
Putting the inequality~\eqref{eq:TV18015} back into the left side of the estimate~\eqref{eq:TV16585}, we obtain
\begin{align} \label{eq:TV19005}
    \lefteqn{\sum_{q \in \mathcal{Q}_z} e^{-c\sqrt{\beta} \left\| q \right\|_1}\sum_{x_1 , y_1 \in \Zd} \frac{\left| n_q(x_1) \right| \left| n_q(y_1) \right| \left\| \left( \di^* \mathcal{U}_{q_1} , n_q \right) \right\|_{L^\infty \left( \mu_\beta \right)}}{|x - x_1|^{2d - \ep} + |y - y_1|^{2d - \ep}}} \qquad & \\ &
    \leq \sum_{q \in \mathcal{Q}_z} \frac{Ce^{-c\sqrt{\beta}   \left\| q \right\|_1} C_{q,q_1}}{|z|^{d - \ep}} \sum_{x_1 , y_1 \in \supp n_q}  \frac{1}{|x - x_1|^{2d - \ep} + |y - y_1|^{2d - \ep}}. \notag
\end{align}
Using that, for a given charge $q \in \mathcal{Q}_z$, the point $z$ belongs to the support of $n_q$, we write
\begin{equation} \label{eq:TV19015}
     \frac{1}{|x - x_1|^{2d-\ep} + |y - y_1|^{2d - \ep}} \leq \frac{\left| \diam n_{q}\right|^{2d}}{|x - z|^{2d-\ep} + |y - z|^{2d - \ep}} \leq \frac{C_q}{|x - z|^{2d-\ep} + |y-z|^{2d - \ep}}.
\end{equation}
We then combine the estimates~\eqref{eq:TV19005} and~\eqref{eq:TV19015} and use the exponential decay of the coefficient $e^{-c \sqrt{\beta}\left\| q \right\|_1}$ to absorb the algebraic growth of the constant $C_{q_1 , q}$ in the parameter $\left\| q_1 \right\|_1$. We obtain
\begin{equation*}
    \sum_{q \in \mathcal{Q}_z} e^{-c\sqrt{\beta} \left\| q \right\|_1}\sum_{x_1 , y_1 \in \Zd} \frac{\left| n_q(x_1) \right| \left| n_q(y_1) \right| \left\| \left( \di^* \mathcal{U}_{q_1} , n_q \right) \right\|_{L^\infty \left( \mu_\beta \right)}}{|x - x_1|^{2d - \ep} + |y - y_1|^{2d - \ep}} \leq \frac{C_{q_1}}{|z|^{d-\ep}\times \left( |x - z|^{2d-\ep} + |y - z|^{2d-\ep}  \right)}.
\end{equation*}
The proof of the estimate~\eqref{eq:TV16585} is complete. We then use the identity $\mathcal{Q} := \bigcup_{z \in \Zd} \mathcal{Q}_z$ to write
\begin{align} \label{eq:TV08086}
    \eqref{eq:TV10395}-(i) & =  \sum_{q \in \mathcal{Q}} e^{-c\sqrt{\beta} \left\| q \right\|_1}\sum_{x_1 , y_1 \in \Zd} \frac{\left| n_q(x_1) \right| \left| n_q(y_1) \right| \left\| \left( \di^* \mathcal{U}_{q_1} , n_q \right) \right\|_{L^\infty \left( \mu_\beta \right)}}{|x - x_1|^{2d - \ep} + |y - y_1|^{2d - \ep}} \\
    & \leq \sum_{z \in \Zd}  \sum_{q \in \mathcal{Q}_z} e^{-c\sqrt{\beta} \left\| q \right\|_1}\sum_{x_1 , y_1 \in \Zd} \frac{\left| n_q(x_1) \right| \left| n_q(y_1) \right| \left\| \left( \di^* \mathcal{U}_{q_1} , n_q \right) \right\|_{L^\infty \left( \mu_\beta \right)}}{|x - x_1|^{2d - \ep} + |y - y_1|^{2d - \ep}} \notag \\
    & \leq  \sum_{z \in \Zd}  \frac{C_{q_1}}{|z|^{d-\ep} \times \left( |x - z|^{2d-\ep} + |y - z|^{2d-\ep}  \right)}. \notag
\end{align}
We estimate the sum in the right side of the estimate~\eqref{eq:TV08086}. To this end, we note that, for each triplet $(x,y,z) \in \Zd \times \Zd \times \Zd$,
\begin{align} \label{eq:TV08576}
    \left( |x - z|^{2d-\ep} + |y - z|^{2d-\ep}  \right) \geq c \dist_{\Z^{2d}} \left( (x,y) , (z,z) \right)^{^{2d-\ep}} & \geq c \left( \left| \frac{x-y}{2} \right|^2 + \left| \frac{x+y}{2} - z \right|^2 \right)^{\frac{2d-\ep}{2}} \\ & \geq c \left( \left| x-y \right|^{2d-\ep} + \left| \frac{x+y}{2} - z \right|^{2d-\ep} \right), \notag
\end{align}
where the notation $\dist_{\Z^{2d}}\left( (x,y) , (z,z) \right)$ is used to denote the euclidean distance in the lattice $\Z^{2d}$ between the points $(x,y)$ and $(z,z)$, the second inequality is obtained by computing the orthogonal projection of the point $(x,y) \in \Z^{2d}$ on the diagonal $\left\{ (z,z) \in \Z^{2d} \, : \, z \in \Zd \right\}$ and the third inequality is obtained by reducing the value of the constant $c$. Using the estimate~\eqref{eq:TV08576}, we deduce that
\begin{align} \label{eq:TV10006}
    \sum_{z \in \Zd}  \frac{1}{|z|^{d-\ep} \times \left( |x - z|^{2d-\ep} + |y - z|^{2d-\ep}  \right)} & \leq \sum_{z \in \Zd} \frac{1}{|z|^{d-\ep}} \times \frac{1}{ \left| x-y \right|^{2d-\ep} + \left| \frac{x+y}{2} - z \right|^{2d-\ep}} \\
    & \leq \sum_{z \in \Zd} \frac{1}{|z|^{d-\ep}} \times \frac{1}{ \left| x-y \right|^{2d-\ep} + \left| \frac{x+y}{2} - z \right|^{2d-\ep}} \notag \\
    & \leq \frac{C}{|x-y|^{d} \max \left( \left| x \right|, \left| y \right|\right)^{d-2\ep}}, \notag
\end{align}
where the computation in the third line is performed in Proposition~\ref{propappClign6572} of Appendix~\ref{app.appC}.
Combining the estimates~\eqref{eq:TV08086} and~\eqref{eq:TV10006} completes the proof of the estimate~\eqref{eq:TV16095}.

To estimate the term~\eqref{eq:TV10395}-(ii), we use that $0$ belongs to the support of the charge $n_{q_1}$ to write, for each pair of points $x_1,y_1 \in \supp n_{q_1}$,
\begin{equation} \label{eq:TV14175}
    \frac{1}{|x - x_1|^{2d-\ep} + |y - y_1|^{2d - \ep}} \leq \frac{\left| \diam n_{q_1}\right|^{2d}}{|x|^{2d-\ep} + |y|^{2d-\ep}} \leq \frac{C_{q_1}}{|x|^{2d-\ep} + |y|^{2d - \ep}}.
\end{equation}
From the inequality~\eqref{eq:TV14175}, we deduce
\begin{equation} \label{eq:TV19375}
    \eqref{eq:TV10395}-(ii) = \sum_{x_1 , y_1 \in \Zd} \frac{\left| n_{q_1}(x_1) \right| \left| n_{q_1}(y_1) \right| }{|x - x_1|^{2d - \ep} + |y - y_1|^{2d - \ep}} \leq \frac{C_{q_1}}{|x|^{2d - \ep} + |y|^{2d-\ep}}.
\end{equation}
We then combine the estimates~\eqref{eq:TV10395},~\eqref{eq:TV16095},~\eqref{eq:TV19375} to deduce the inequality, for each pair of points $x , y \in \Zd$,
\begin{align} \label{eq:TV10186}
    \left\| u (x , y , \cdot) \right\|_{L^\infty \left( \mu_\beta \right)} & \leq \frac{C_{q_1}}{|x-y|^{d - \ep} \max \left( \left| x \right|, \left| y \right|\right)^{d-\ep}} + \frac{C_{q_1}}{|x|^{2d - \ep} + |y|^{2d-\ep}} \\
    & \leq \frac{C_{q_1}}{|x-y|^{d - \ep} \max \left( \left| x \right|, \left| y \right|\right)^{d-\ep}}. \notag
\end{align}
We then use this inequality to estimate the variance of the random variable $\mathcal{U}_{q_1}(x , \cdot)$ by using the formula~\eqref{eq:TV08435}. We obtain
\begin{equation*}
    \var \left[ \mathcal{U}_{q_1}(z, \cdot) \right] \leq C \sum_{y,y_1 \in \Zd} \frac{C_{q_1}}{|z-y|^{d - \ep} \max \left( \left| z \right|, \left| y \right|\right)^{d-1}} \cdot \frac{C}{|y - y_1|^{d-2}} \cdot \frac{C_{q_1}}{|z-y_1|^{d - \ep}  \max \left( \left| z \right|, \left| y_1 \right|\right)^{d-\ep}}.
\end{equation*}
We then use that the terms $\max \left( \left| z \right|, \left| y_1 \right|\right)$ and $\max \left( \left| z \right|, \left| y \right|\right)$ are both larger than the value $|z|$ to deduce that
\begin{equation*}
    \var \left[ \mathcal{U}_{q_1}(z, \cdot) \right] \leq \frac{C_{q_1}}{|z|^{2 d - 2 \ep}} \sum_{y,y_1 \in \Zd} \frac{1}{|z-y|^{d - \ep}} \cdot \frac{1}{|y - y_1|^{d-2}} \cdot \frac{1}{|z-y_1|^{d - \ep}} \leq \frac{C_{q_1}}{|z|^{2 d - 2 \ep}}.
\end{equation*}
The proof of the estimate~\eqref{eq:TV08015} is complete.

\medskip

\subsection{Homogenization of the mixed derivative of the Green's matrix} We fix a radius $R > 1$ and let $m$ be the smallest integer such that the annulus $A_R$ is included in the cube $\cu_m$. The proof relies on a two-scale expansion following the outline described in Section~\ref{sec3.2chap60851}. We define the function $\mathcal{H}_{q_1}$ by the formula
\begin{equation} \label{eq:TVdefHq12scexp}
    \mathcal{H}_{q_1} := \bar G_{q_1} + \sum_{i,j} \nabla_i \bar G_{q_1,j} \chi_{m,ij}.
\end{equation}
We decompose the argument into three Steps.

\medskip

\textit{Step 1.} In this step, we prove that the $\underline{H}^{-1}\left( A_R , \mu_\beta \right)$-norm of the term $\mathcal{L} \mathcal{H}_{q_1}$ is small, more specifically, we prove that there exists an exponent $\gamma_\alpha > 0$ such that one has the estimate
\begin{equation} \label{eq:TV15356}
    \left\| \mathcal{L} \mathcal{H}_{q_1} \right\|_{\underline{H}^{-1}\left( A_R , \mu_\beta \right)} \leq \frac{C_{q_1}}{R^{d - \gamma_\alpha}}.
\end{equation}
The proof is essentially identical to the argument presented in Section~\ref{sec3.2chap608511}: we use the exact formula for the two-scale expansion $\mathcal{H}_{q_1}$ given in~\eqref{eq:TVdefHq12scexp} to compute the value of $\L \mathcal{H}_{q_1}$ and then use the quantitative properties of the corrector stated in Proposition~\ref{prop:prop5.25} of Chapter \ref{section5} to prove that the $\underline{H}^{-1}\left( A_R , \mu_\beta \right)$-norm of the term $ \mathcal{L} \mathcal{H}_{q_1}$ satisfies the estimate~\eqref{eq:TV15356}. Since the proof is rather long due to the technicalities caused by the specific structure of the operator $\mathcal{L}$ (iterations of the Laplacian, sum over all the charges $q \in \mathcal{Q}$), we do not rewrite it but only point out the main differences:
\begin{itemize}
    \item We work in the annulus $A_R$ and not in the ball $B_{R^{1+\delta}}$, this difference makes the proof simpler since we do not have to take the additional parameter $\delta$ into considerations;
    \item We can always assume that the diameter of the charge $q_1$ is smaller than $R/2$, otherwise the constant $C_{q_1}$ is larger than $R^k$ for some large number $k := k(d)$ (since it is allowed to have an algebraic growth in the parameter $\left\| q_1 \right\|_1$) and the estimate~\eqref{eq:TV15356} is trivial in this situation. Under the assumption $\diam q_1 \leq R/2$, we use the identity $- \ahom \Delta \bar G_{q_1} = 0$ in the annulus $A_R$ instead of the identity $- \ahom \Delta \bar G_\delta = \rho_\delta$ in the ball $B_{R^{1+\delta}}$;
    \item We use the regularity estimates on the function $\bar G_{q_1}$ stated in Proposition~\ref{prop:prop6.5} instead of the estimates on the Green's function $\bar G$ stated in Proposition~\ref{prop:prop6.2}. Since the map $\bar G_{q_1}$ scales like the gradient of the Green's function (in particular it decays like $|x|^{1-d}$), we obtain an additional factor $R$ in the right side of~\eqref{eq:TV15356} compared to~\eqref{eq:TV19578}, i.e., we obtain
    \begin{equation*}
        \left\| \mathcal{L} \mathcal{H}_{q_1} \right\|_{\underline{H}^{-1}\left( A_R , \mu_\beta \right)} \leq \frac{C_{q_1}}{R^{d - \gamma_\alpha}} \hspace{3mm} \mbox{instead of}  \hspace{3mm} \left\| \mathcal{L} \mathcal{H}_{\delta, \cdot k} - \rho_{\delta, \cdot k} \right\|_{\underline{H}^{-1}\left( A_R , \mu_\beta \right)} \leq \frac{C}{R^{d - 1 - \gamma_\alpha}}.
    \end{equation*}
\end{itemize}

\medskip

\textit{Step 2.} In this step, we use the main result~\eqref{eq:TV15356} of Substep 3.1 to prove that the gradient of the Green's function $\nabla \mathcal{U}_{q_1}$ is close to the gradient of the two-scale expansion $\nabla \mathcal{H}_{q_1}$ in the $\underline{L}^2\left(A_R , \mu_\beta \right)$-norm. We prove the estimate
\begin{equation} \label{eq:TV17086}
    \left\| \nabla \mathcal{U}_{q_1} - \nabla \mathcal{H}_{q_1} \right\|_{\underline{L}^2 \left( A_R , \mu_\beta \right)} \leq \frac{C_{q_1}}{R^{d+\gamma_\alpha}}.
\end{equation}

To simplify the rest of the argument, we do not prove the estimate~\eqref{eq:TV17086} directly. We slightly reduce the size of the annulus $A_R$ and define the set $A_R^1$ to be the annulus $A_R^1 := \left\{ x \in \Zd \, : \, 1.1 R \leq |x| \leq 1.9R \right\}$. We note that we have the inclusion, for each radius $R \geq 1$, $A_R^1 \subseteq A_R$. In this substep, we prove the inequality
\begin{equation} \label{eq:TV17096}
    \left\| \nabla \mathcal{U}_{q_1} - \nabla \mathcal{H}_{q_1} \right\|_{\underline{L}^2 \left( A_R^1 , \mu_\beta \right)} \leq \frac{C_{q_1}}{R^{d+\gamma_\delta}}.
\end{equation}
The inequality~\eqref{eq:TV17086} can then be deduced from~\eqref{eq:TV17096} by a covering argument.

The argument is similar to the one presented in Section~\ref{sec3.3chap60851} except that, instead of making use of the mollifier exponent $\delta$ to prove that the $H^1$-norm is of the difference $\left( \nabla \mathcal{H}_\delta - \G_\delta \right)$ is small, as it was done in the estimates~\eqref{eq:TV07591} and~\eqref{eq:TV07601}, we use the main result~\eqref{eq:TV16256} of Section~\ref{SEctiocontracvar4.3chap7}. We first let $\eta$ be a cutoff function which satisfies the properties:
\begin{equation} \label{eq:propetebis}
    0 \leq \eta \leq 1, ~\supp \eta \subseteq A_R,~ \eta = 1 ~\mbox{in}~A_R^1,~\forall k \in \N, \, \left| \nabla^k \eta \right| \leq \frac{C}{R^k}.
\end{equation}
We then use the function $\eta \left( \mathcal{U}_{q_1} - \mathcal{H}_{q_1}\right)$ as a test function in the definition of the $\underline{H}^{-1} \left(A_R, \mu_\beta \right)$ of the inequality~\eqref{eq:TV15356} and use the identity $\L \mathcal{U}_{q_1} = 0$ in $A_R \times \Omega$. We obtain
\begin{align} \label{eq:TV10172112}
    \frac{1}{R^{d}}\sum_{x \in A_R} \left\langle \eta \left( \mathcal{U}_{q_1} - \mathcal{H}_{q_1}\right) \L \left( \mathcal{U}_{q_1} -  \mathcal{H}_{q_1} \right)  \right\rangle_{\mu_\beta} & \leq \left\| \L \left( \mathcal{U}_{q_1} - \mathcal{H}_{q_1} \right) \right\|_{\underline{H}^{-1} \left( B_{R^{1+\delta}} , \mu_\beta \right)} \left\|  \eta \left( \mathcal{U}_{q_1} -  \mathcal{H}_{q_1} \right)  \right\|_{\underline{H}^1 \left( A_R , \mu_\beta \right)} \\
    & \leq \frac{C}{R^{d + \gamma_\alpha}} \left\| \eta \left( \mathcal{U}_{q_1} - \mathcal{H}_{q_1} \right) \right\|_{\underline{H}^1 \left( A_R , \mu_\beta \right)}. \notag
\end{align}
We then estimate the $\underline{H}^1 \left( A_R , \mu_\beta \right)$-norm of the function $\mathcal{U}_{q_1} - \mathcal{H}_{q_1}$ with similar arguments as the one presented in the proof of the inequality~\eqref{eq:TV15049}, the only difference is that we use the regularity estimates stated in Proposition~\ref{prop:prop6.5} instead of the regularity estimates for the functions $\G_\delta$ and $\mathcal{H}$. We obtain
\begin{equation} \label{eq:TV10182112}
    \left\| \eta \left( \mathcal{U}_{q_1} - \mathcal{H}_{q_1} \right) \right\|_{\underline{H}^1 \left( A_R , \mu_\beta \right)} \leq \left\| \eta \mathcal{U}_{q_1} \right\|_{\underline{H}^1 \left( A_R , \mu_\beta \right)} + \left\| \eta  \mathcal{H}_{q_1} \right\|_{\underline{H}^1 \left( A_R , \mu_\beta \right)} \leq  \frac{C_{q_1}}{R^{d -1- \ep}}.
\end{equation}
For later use, we also note that the same argument yields to the inequality
\begin{equation} \label{eq:TV09577}
    \left\|\nabla  \mathcal{U}_{q_1} - \nabla \mathcal{H}_{q_1}  \right\|_{\underline{L}^2 \left( A_R , \mu_\beta \right)} \leq \left\|\nabla  \mathcal{U}_{q_1}  \right\|_{\underline{L}^2 \left( A_R , \mu_\beta \right)}  + \left\|\nabla  \mathcal{H}_{q_1} \right\|_{\underline{L}^2 \left( A_R , \mu_\beta \right)}  \leq  \frac{C_{q_1}}{R^{d - \ep}}.
\end{equation}
We then combine the inequalities~\eqref{eq:TV17096} and~\eqref{eq:TV10172112} and use that $\ep \ll \gamma_\alpha$ to deduce that
\begin{equation} \label{eq:TV12287}
    \frac{1}{R^{d}}\sum_{x \in A_R} \left\langle \eta \left( \mathcal{U}_{q_1} - \mathcal{H}_{q_1}\right) \L \left( \mathcal{U}_{q_1} -  \mathcal{H}_{q_1} \right)  \right\rangle_{\mu_\beta} \leq \frac{C_{q_1}}{R^{2d + \gamma_\alpha}}.
\end{equation}
Thus to prove the inequality~\eqref{eq:TV17096}, it is sufficient to prove the estimate
\begin{equation*}
    \left\| \nabla \mathcal{U}_{q_1} - \nabla \mathcal{H}_{q_1} \right\|_{\underline{L}^2 \left( A_R^1 , \mu_\beta \right)}^2 \leq \frac{1}{R^{d}}\sum_{x \in A_R} \left\langle \eta \left( \mathcal{U}_{q_1} - \mathcal{H}_{q_1}\right) \L \left( \mathcal{U}_{q_1} -  \mathcal{H}_{q_1} \right)  \right\rangle_{\mu_\beta}  + \frac{C_{q_1}}{R^{2d+\gamma_\delta}}.
\end{equation*}
First, by definition of the Helffer-Sj{\"o}strand operator $\mathcal{L}$, we have the identity
\begin{align} \label{eq:TV081011}
\sum_{x \in \Zd} \left\langle \eta \left( \mathcal{U}_{q_1} -  \mathcal{H}_{q_1}\right) \L \left( \mathcal{U}_{q_1} -  \mathcal{H}_{q_1} \right)  \right\rangle_{\mu_\beta} & = \sum_{x , y \in \Zd} \eta(x) \left\langle \left( \partial_y \mathcal{U}_{q_1} (x , \cdot)  - \partial_y \mathcal{H}_{q_1} (x , \cdot) \right)^2 \right\rangle_{\mu_\beta} \\ & + \frac 1{2\beta}\sum_{x \in \Zd}  \left\langle  \left( \nabla \mathcal{U}_{q_1} - \nabla \mathcal{H}_{q_1}  \right)(x , \cdot) \cdot \nabla \left( \eta \left( \mathcal{U}_{q_1} -  \mathcal{H}_{q_1} \right)\right)(x , \cdot)  \right\rangle_{\mu_\beta} \notag \\ & +  \sum_{q \in \mathcal{Q}}  \left\langle \nabla_q \left(  \mathcal{U}_{q_1} -  \mathcal{H}_{q_1}\right) \cdot \a_q \nabla_q \left( \eta \left(  \mathcal{U}_{q_1} -  \mathcal{H}_{q_1} \right) \right) \right\rangle_{\mu_\beta} \notag \\ & +  \frac 1{2\beta} \sum_{n \geq 1} \sum_{x \in \Zd}  \frac{1}{\beta^{ \frac n2}} \left\langle  \nabla^{n+1} \left(  \mathcal{U}_{q_1} -  \mathcal{H}_{q_1} \right)(x , \cdot) \cdot \nabla^{n+1} \left( \eta \left(  \mathcal{U}_{q_1} -  \mathcal{H}_{q_1} \right) \right)(x,\cdot) \right\rangle_{\mu_\beta}. \notag
\end{align}
We then estimate the four terms on the right side separately. For the first one, we use that it is non-negative
\begin{equation} \label{eq:TV12137}
    \sum_{x , y \in \Zd} \eta(x)^2 \left\langle \left( \partial_y \mathcal{U}_{q_1} (x , \cdot)  - \partial_y \mathcal{H}_{q_1} (x , \cdot) \right)^2 \right\rangle_{\mu_\beta}  \geq 0.
\end{equation}
For the second one, we expand the gradient of the product $\eta^2 \left( \mathcal{U}_{q_1}  - \mathcal{H}_{q_1}  \right)$ and write
\begin{align} \label{eq:TV094901}
    \lefteqn{\sum_{x \in \Zd}  \left\langle  \left( \nabla \mathcal{U}_{q_1}  - \nabla\mathcal{H}_{q_1} \right)(x , \cdot) \cdot \nabla \left( \eta \left(\mathcal{U}_{q_1}  - \mathcal{H}_{q_1} \right)\right)(x , \cdot)  \right\rangle_{\mu_\beta}} \qquad & \\ &
    = \sum_{x \in \Zd}  \eta(x)  \left\langle  \left( \nabla \mathcal{U}_{q_1}  -\nabla \mathcal{H}_{q_1}\right)(x , \cdot) \cdot   \nabla \left(\mathcal{U}_{q_1}  - \mathcal{H}_{q_1} \right)(x , \cdot)  \right\rangle_{\mu_\beta} \notag \\
    & \quad + \sum_{x \in \Zd}  \left\langle  \left( \nabla \mathcal{U}_{q_1}  - \nabla\mathcal{H}_{q_1}\right)(x , \cdot) \cdot  \nabla \eta(x)\left( \mathcal{U}_{q_1}  - \mathcal{H}_{q_1} \right)(x , \cdot)  \right\rangle_{\mu_\beta}. \notag
\end{align}
Dividing the identity~\eqref{eq:TV094901} by the volume factor $R^{d}$ and using the properties of the function $\eta$ stated in~\eqref{eq:propetebis}, we obtain
\begin{multline} \label{eq:TV075917}
    R^{-d}\sum_{x \in \Zd}  \left\langle  \left( \nabla \mathcal{U}_{q_1} (x , \cdot)  - \nabla \mathcal{H}_{q_1} \right)(x , \cdot) \cdot \nabla \left( \eta \left( \mathcal{U}_{q_1}-  \mathcal{H} \right)\right)(x , \cdot)  \right\rangle_{\mu_\beta} \geq c \left\| \eta \left( \nabla \mathcal{U}_{q_1}  - \nabla \mathcal{H}_{q_1} \right) \right\|_{\underline{L}^2 \left(A_R , \mu_\beta \right)}^2 \\ -  \frac{C}{R} \left\| \nabla \mathcal{U}_{q_1}  \ - \nabla \mathcal{H}_{q_1} \right\|_{\underline{L}^2 \left( A_{R} , \mu_\beta \right)} \left\|  \mathcal{U}_{q_1}  - \mathcal{H}_{q_1} \right\|_{\underline{L}^2 \left( A_{R} , \mu_\beta \right)}.
\end{multline}
We then use the inequality~\eqref{eq:TV16256} and the estimate~\eqref{eq:TV09577} and the quantitative sublinearity of the corrector to deduce that
\begin{equation} \label{eq:TV10537}
    \frac{1}{R} \left\| \nabla \mathcal{U}_{q_1}  \ - \nabla \mathcal{H}_{q_1} \right\|_{\underline{L}^2 \left( A_{R} , \mu_\beta \right)} \left\|  \mathcal{U}_{q_1}  - \mathcal{H}_{q_1} \right\|_{\underline{L}^2 \left( A_{R} , \mu_\beta \right)} \leq \frac{1}{R} \cdot \frac{C}{R^{d - \ep}} \cdot \frac{C}{R^{d - 1 + \gamma_\delta}} \leq \frac{C}{R^{2d + \gamma_\delta}}.
\end{equation}
We then combine the inequalities~\eqref{eq:TV075917} and~\eqref{eq:TV10537} to deduce that
\begin{equation} \label{eq:TV11427}
     R^{-d}\sum_{x \in \Zd}  \left\langle  \left( \nabla \mathcal{U}_{q_1} (x , \cdot)  - \nabla \mathcal{H}_{q_1} \right)(x , \cdot) \cdot \nabla \left( \eta \left( \mathcal{U}_{q_1}  - \nabla \mathcal{H}_{q_1} \right)\right)(x , \cdot)  \right\rangle_{\mu_\beta} + \frac{C_{q_1}}{R^{2d + \gamma_\delta}} \geq c \left\| \eta \left( \nabla \mathcal{U}_{q_1}  - \nabla \mathcal{H}_{q_1} \right) \right\|_{\underline{L}^2 \left(A_R , \mu_\beta \right)}^2.
\end{equation}
The two remaining terms in the right side of the estimate~\eqref{eq:TV081011} (involving the iteration of the Laplacian and the sum over the charges) are estimated following the ideas developed in in Section~\ref{sec3.3chap60851} (see~\eqref{eq:TV11121} and~\eqref{eq:TV11131}) or in the proof of the Caccioppoli inequality (Proposition~\ref{Caccio.ineq} of Chapter~\ref{section:section4}). We skip the details and write the result: we obtain
\begin{equation} \label{eq:TV111211}
    R^{-d}\sum_{q \in \mathcal{Q}}  \left\langle \nabla_q \left(  \mathcal{U}_{q_1}  - \nabla \mathcal{H}_{q_1}\right) \cdot \a_q \nabla_q \left( \eta \left(  \mathcal{U}_{q_1}  - \nabla \mathcal{H}_{q_1} \right) \right) \right\rangle_{\mu_\beta}  + \frac{C_{q_1}}{R^{(2d-\gamma_\delta) }} \geq -C e^{-c \sqrt{\beta}} \left\| \eta \left( \nabla \mathcal{U}_{q_1}  - \nabla \mathcal{H}_{q_1} \right) \right\|_{\underline{L}^2 \left(A_R , \mu_\beta \right)}^2
\end{equation}
and
\begin{equation} \label{eq:TV111311}
    R^{-d} \sum_{n \geq 1} \sum_{x \in \Zd}  \frac{1}{\beta^{ \frac n2}} \left\langle  \nabla^{n+1} \left(  \mathcal{U}_{q_1}  - \mathcal{H}_{q_1} \right)(x , \cdot) \cdot \nabla^{n+1} \left( \eta \left(   \mathcal{U}_{q_1}  - \mathcal{H}_{q_1} \right) \right)(x,\cdot) \right\rangle_{\mu_\beta} + \frac{C_{q_1}}{R^{(2d+\gamma_\delta) }} \geq 0.
\end{equation}
We then combine the estimates~\eqref{eq:TV12137},~\eqref{eq:TV11427},~\eqref{eq:TV111211} and~\eqref{eq:TV111311} with the identity~\eqref{eq:TV081011}, choose the inverse temperature $\beta$ large enough so that the right side of~\eqref{eq:TV111211} can be absorbed by the right side of~\eqref{eq:TV075917} and use that the cutoff function $\eta$ is equal to $1$ in the annulus $A_R^1$. We obtain
\begin{equation} \label{eq:TV12277}
     \left\| \nabla \mathcal{U}_{q_1}  - \nabla \mathcal{H}_{q_1} \right\|_{\underline{L}^2 \left( A_{R}^1 , \mu_\beta \right)} \leq \frac{C}{R^d} \sum_{x \in \Zd} \left\langle \eta \left( \mathcal{U}_{q_1} -  \mathcal{H}_{q_1}\right) \L \left( \mathcal{U}_{q_1} -  \mathcal{H}_{q_1} \right)  \right\rangle_{\mu_\beta} + \frac{C_{q_1}}{R^{d + \gamma_\delta}}.
\end{equation}
We then combine the inequality~\eqref{eq:TV12277} with the estimate~\eqref{eq:TV12287} to complete the proof of~\eqref{eq:TV17096}. Step 2 is complete.

\medskip

\textit{Step 3. The conclusion.} In this step, we prove the $L^2$-estimate
\begin{equation} \label{eq:TV12337}
    \left\| \nabla \mathcal{U}_{q_1} - \sum_{i,j} \left( e_{ij} + \nabla \chi_{ij} \right) \nabla \bar G_{q_1,j} \right\|_{\underline{L}^2 \left( A_R , \mu_\beta \right)} \leq \frac{C_{q_1}}{R^{d+ \gamma_\delta}}.
\end{equation}
In view of the estimate~\eqref{eq:TV17096} proved in Step 2, it is sufficient to prove the inequality
\begin{equation} \label{eq:TV12347}
    \left\| \nabla \mathcal{H}_{q_1} - \sum_{i,j} \left( e_{ij} + \nabla \chi_{ij} \right) \nabla \bar G_{q_1,j} \right\|_{\underline{L}^2 \left( A_R , \mu_\beta \right)} \leq \frac{C_{q_1}}{R^{d+ \gamma_\delta}}.
\end{equation}
The proof of~\eqref{eq:TV12347} relies on the regularity estimate on the function $\bar G_{q_1}$ stated in Proposition~\ref{prop:prop6.5}, the quantitative sublinearity of the corrector stated in Proposition~\ref{prop:prop5.25} of Chapter~\ref{section5} and the quantitative estimate for the difference of the finite and infinite-volume gradient of the corrector stated in Proposition~\ref{prop5.26} of Chapter~\ref{section5}. The argument is identical (and even simpler since we do not have to take into account the parameter $\delta$) to the argument given in Section~\ref{sec3.4chap60851} so we skip the details. The proof of Step 3, and thus of Theorem~\ref{thm:homogmixedder}, is complete.

\chapter{Proof of the estimates in Chapter~4} \label{section7}

In this chapter, we present the proofs of the technical lemmas which are used in Chapter~\ref{section3.4} to prove Theorem~\ref{t.main}. Most of the heuristic of the arguments are presented in this chapter and we refer to it for an overview of the results. As it may be useful to the reader, we record below the tools established in this article which are used in this chapter:
\begin{itemize}
    \item In Sections~\ref{sect:chap8.1},~\ref{sec:chap8.2} and~\ref{sec:chap8.3}, we study the correlation of random variables; this is achieved by using the Helffer-Sj{\"o}strand representation formula. We need to use the properties of the Green's function associated to the Helffer-Sj{\"o}strand operator stated in Proposition~\ref{prop.prop4.11chap4} of Chapter~\ref{chap:chap3};
    \item In Section~\ref{sec:chap8.3}, we need to study the correlation between a solution of a Helffer-Sj{\"o}strand equation and the random variables $X_x$ and $Y_0$. To this end, we appeal Helffer-Sj{\"o}strand representation formula and the differentiated Helffer-Sj{\"o}strand equation as well as to the properties of the Green's function associated to this operator stated in Proposition~\ref{cor:corollary4.14} of Chapter~\ref{section:section4};
    \item Sections~\ref{sec:chap8.4} and~\ref{sec:chap8.5} are devoted to the proofs of some properties of the discrete Green's function on the lattice $\Zd$; they can be read independently of the rest of the article.
\end{itemize}

\section{Removing the terms \texorpdfstring{$X_{\sin \cos},$}{30} \texorpdfstring{$X_{\cos \cos}$}{31} and \texorpdfstring{$X_{\sin \sin}$}{32}} \label{sect:chap8.1}

We recall the definitions of the values $Z_\beta (\sigma)$ and $Z_\beta(0)$ introduced in \eqref{e.Zsigma} of Chapter~\ref{chap:chap3}, the definitions of the random variables $Y_0$, $X_x$, $X_{\sin \cos}$, $X_{\cos \cos}$, $X_{\sin \sin}$ introduced in~\eqref{eq:moneq} of Chapter~\ref{section3.4} and the identity
\begin{equation} \label{eq:TV09121}
    \frac{Z_\beta (\sigma)}{Z_\beta(0)} = \left\langle Y_0 X_x X_{\sin \cos} X_{\cos \cos} X_{\sin \sin}  \right\rangle_{\mu_\beta}.
\end{equation}
Given a charge $q \in \mathcal{Q}$, we recall the conventional notation $C_q$ to mean the the constant $C$ depends on the variables $d , \beta$ and $q$ and that the dependence in the $q$ variable is at most algebraic, i.e., there exists an exponent $k := k \left( d \right)$ and a constant $C := C \left( d , \beta \right)$ such that $C_q \leq C \left\| q \right\|_1^k$. We also recall the notation, for each point $y \in \Zd$, $\mathcal{Q}_y := \left\{ q \in \mathcal{Q} \, : \, y \in \supp n_q \right\}$.

\begin{lemma} \label{lem.lemma7.1}
There exist constants $c := c(d, \beta)$ and $C := C(d , \beta) < \infty$ such that
\begin{equation} \label{eq:TV08511}
     \frac{Z_\beta (\sigma)}{Z_\beta(0)} = \left\langle Y_0 X_x \right\rangle_{\mu_\beta} + \frac{c \left\langle Y_0 X_x \right\rangle_{\mu_\beta}}{|x|^{d-2}} + O \left( \frac{C}{|x|^{d-1}} \right).
\end{equation}
As a consequence, the following statements are equivalent
\begin{multline} \label{eq:TV08491}
    \exists \gamma \in (0 , \infty), \, \exists c_1 , c_2 \in \R, \frac{Z_\beta (\sigma)}{Z_\beta(0)} = c_1 + \frac{c_2}{|x|^{d-2}} + O \left( \frac{C}{|x|^{d-2 + \gamma}} \right) \\ \iff  \exists \gamma \in (0 , \infty), \, \exists c_1, c_2 \in \R, \left\langle Y_0 X_x \right\rangle_{\mu_\beta} = c_1+ \frac{c_2}{|x|^{d-2}} + O \left( \frac{C}{|x|^{d-2 + \gamma}} \right).
\end{multline}
\end{lemma}

\begin{remark}
The values of the constants $c_1, c_2$ in the two sides of the equivalence~\eqref{eq:TV08491} are not necessarily equal; they are related through the constant $c$ which appears in~\eqref{eq:TV08511}. We use the same notation because we are not interested in their specific values but only in their existence.
\end{remark}

\begin{proof}
As is explained in Section~\ref{sec:chap4.1} of Chapter~\ref{section3.4}, the proof of estimate~\eqref{eq:TV08511} is based on the proof of the following estimates 
\begin{equation} \label{eq:estsmallprop3.1}
    \left\{ \begin{aligned}
    \left\| X_{\sin \cos} - 1 \right\|_{L^\infty} &\leq \frac{C}{|x|^{d-1}}, \\
    \left\| X_{\cos \cos} - 1 \right\|_{L^\infty} &\leq \frac{C}{|x|^{d-1}}, \\
    \var_{\mu_\beta} X_{\sin \sin} &\leq \frac{C}{|x|^{2d-2}}, \\
    \E \left[ X_{\sin \sin} \right] &= 1 + \frac{c}{|x|^{d-2}} + O \left( \frac{C}{|x|^{d-1}} \right).
    \end{aligned} \right.
\end{equation}

We first prove that~\eqref{eq:estsmallprop3.1} implies~\eqref{eq:TV08511}. From the identity~\eqref{eq:TV09121} and a direct computation involving the Cauchy-Schwarz inequality, we obtain the estimate
\begin{multline*}
    \left| \left\langle Y_0 X_x X_{\sin \cos} X_{\cos \cos} X_{\sin \sin}  \right\rangle_{\mu_\beta} - \left\langle Y_0 X_x \right\rangle_{\mu_\beta} \left\langle X_{\sin \sin}\right\rangle_{\mu_\beta} \right| \leq \left\| Y_0 X_x \right\|_{L^2 \left( \mu_\beta \right)} \left(\var_{\mu_\beta} X_{\sin \sin}\right)^\frac 12 \\ + \left\langle Y_0 X_x X_{\sin \sin} \right\rangle_{\mu_\beta} \left( \left\| X_{\sin \cos} - 1 \right\|_{L^\infty} +
    \left\| X_{\cos \cos} - 1 \right\|_{L^\infty} \right).
\end{multline*}
Using the Brascamp-Lieb inequality to estimate the $L^4 \left( \mu_\beta \right)$-norm of the random variables $Y_0 X_x$ and the estimates on the random variable $X_{\sin \sin}$ stated in~\eqref{eq:estsmallprop3.1}, we write
\begin{equation} \label{eq:TV08122}
    \left\| Y_0 X_x \right\|_{L^2 \left( \mu_\beta \right)} \leq \left\| Y_0 \right\|_{L^4 \left( \mu_\beta \right)} \left\| X_x \right\|_{L^4 \left( \mu_\beta \right)} \leq C
\end{equation}
and
\begin{equation*}
    \left\langle Y_0 X_x X_{\sin \sin} \right\rangle_{\mu_\beta} \leq  \left\| Y_0 X_x \right\|_{L^2 \left( \mu_\beta \right)}  \left\| X_{\sin \sin} \right\|_{L^2 \left( \mu_\beta\right)} \leq C \left\| Y_0 X_x \right\|_{L^2 \left( \mu_\beta \right)}  \left( \E \left[ X_{\sin \sin} \right] + \var^\frac 12 X_{\sin \sin} \right) \leq C.
\end{equation*}
We combine these inequalities with the estimates~\eqref{eq:estsmallprop3.1} to deduce the the estimate
\begin{equation*}
    \left| \left\langle Y_0 X_x X_{\sin \cos} X_{\cos \cos} X_{\sin \sin}  \right\rangle_{\mu_\beta} - \left\langle Y_0 X_x \right\rangle_{\mu_\beta} 
    \left(1 + \frac{c}{|x|^{d-2}} + O \left( \frac{C}{|x|^{d-1}} \right)\right) \right| \leq \frac{C}{|x|^{d-1}}.
\end{equation*}
The expansion~\eqref{eq:TV08511} is then a direct consequence of the identity~\eqref{eq:TV09121}, the estimate \eqref{eq:estsmallprop3.1} and the upper bound~\eqref{eq:TV08122}. 

It remains to prove the estimates stated in~\eqref{eq:estsmallprop3.1}; we first focus on the first two inequalities involving the random variables $X_{\sin \cos}$ and $X_{\cos \cos}$. The proof relies on the following ingredients:
\begin{itemize}
    \item For each point $y \in \Zd$ and each charge $q \in \mathcal{Q}_y$, we have the estimate
    \begin{align*}
        \left( \nabla G , n_q \right) \leq \left\| \nabla G \right\|_{L^\infty \left( \supp n_q \right)} \left\| n_q \right\|_{L^1} \leq \sup_{z \in \supp n_q} \frac{C}{|z|^{d-1}} \left\| n_q \right\|_{L^1} & \leq \sup_{|z| \leq \diam n_q} \frac{C}{|y + z|^{d-1}} \left\| n_q \right\|_{L^1} \\
        & \leq \frac{C \left( \diam n_q\right)^{d-1}}{|y|^{d-1}} \left\| n_q \right\|_{L^1} \\
        & \leq \frac{C_q}{|y|^{d-1}},
    \end{align*}
    where we used in the second inequality that, for each charge $q$ in the set $\mathcal{Q}_y$, the support of $n_q$ is included in the ball $B(y , \diam n_q)$. A similar computation shows the estimate
    \begin{equation*}
         \left( \nabla G_x , n_q \right) \leq \frac{C_q}{|y - x|^{d-1}};
    \end{equation*}
    \item The standard estimates, for each real number $a \in \R$, $|\sin a| \leq |a|$, $| \cos a - 1 | \leq \frac 12 |a|^2$ and the estimate, for each charge $q \in \mathcal{Q}$, $\left| z \left(\beta, q \right)\right| \leq e^{- c \sqrt{\beta} \left\| q \right\|_1}$.
\end{itemize}
We obtain the inequality
\begin{align} \label{eq:TV09082}
    \left| \sum_{q \in \mathcal{Q}} z(\beta , q) \sin2\pi(\phi , q)  \sin2\pi(\nabla G_x  , n_q) \left( \cos2\pi(\nabla G  , n_q) - 1 \right) \right| & \leq C \sum_{y \in \Zd} \sum_{q \in \mathcal{Q}_y} \frac{ e^{-c \sqrt{\beta} \left\| q \right\|_1}  C_q }{|y - x|^{d-1}} \frac{1}{|y|^{2d-2}} \\
    & \leq C \sum_{y \in \Zd} \frac{1}{|y - x|^{d-1}} \frac{1}{|y|^{2d-2}} \notag \\
    & \leq \frac{C}{|x|^{d-1}}, \notag
\end{align}
where we used the exponential decay of the term $e^{-c \sqrt{\beta} \left\| q \right\|_1}$ to absorb the algebraic growth of the constant $C_q$.
With a similar strategy, we obtain the two inequalities
\begin{equation} \label{eq:TV09092}
    \begin{aligned}
    \left| \sum_{q \in \mathcal{Q}} z(\beta , q) \sin2\pi(\phi , q)  \sin2\pi(\nabla G , n_q) \left( \cos2\pi(\nabla G_x  , q) - 1 \right) \right| \leq \frac{C}{|x|^{d-1}}, \\
    \left| \sum_{q \in \mathcal{Q}} z(\beta , q) \sin2\pi(\phi , q)  \frac12  \left( \cos (\nabla G_x , q) - 1 \right) \left(\cos (\nabla G , q) -1 \right) \right| \leq 
    \frac{C}{|x|^{d-1}}.
    \end{aligned} 
\end{equation}
We then combine the estimates~\eqref{eq:TV09082} and~\eqref{eq:TV09092} and use that the exponential function is Lipschitz on any compact subset of $\R$ to obtain, for each realization of the field $\phi \in \Omega$,
\begin{equation*}
    \left| X_{\sin \cos}(\phi) - 1 \right| \leq \frac{C}{|x|^{d-1}} \hspace{5mm} \mbox{and} \hspace{5mm}
    \left| X_{\cos \cos}(\phi) - 1 \right| \leq \frac{C}{|x|^{d-1}}.
\end{equation*}
This result implies the $L^\infty \left( \mu_\beta\right)$-estimates stated in \eqref{eq:estsmallprop3.1}. 

There remains to prove the estimates corresponding to the variance and the expectation of the random variable $X_{\sin \sin}$ in~\eqref{eq:estsmallprop3.1}. We first note that a computation similar to the one performed in~\eqref{eq:TV09082} gives the following $L^\infty(\mu_\beta)$-estimate: for each realization of the field $\phi \in \Omega$,
\begin{equation}
\label{e.inftysinsin}
    \left| \sum_{q \in \mathcal{Q}} z(\beta , q) \cos2\pi(\phi , q) \sin( \nabla G , n_q) \sin2\pi(\nabla G_x , n_q) \right| \leq C \sum_{y \in \Zd} \frac{1}{|y - x|^{d-1}} \frac{1}{|y|^{d-1}}  \leq \frac{C}{|x|^{d-2}}.
\end{equation}
By the estimate~\eqref{e.inftysinsin} and the Taylor expansion of the exponential, we obtain the bound
\begin{align*}
    \lefteqn{\left| X_{\sin \sin} - 1 - \sum_{q \in \mathcal{Q}} z(\beta , q) \cos2\pi(\phi , q) \sin2\pi(\nabla G , n_q) \sin2\pi(\nabla G_x , n_q) \right| } \qquad & \\ & \leq C \left( \sum_{q\in \mathcal{Q}} z(\beta , q) \cos2\pi(\phi , q) \sin2\pi(\nabla G , n_q) \sin2\pi(\nabla G_x
    , n_q) \right)^2 \\
    & \leq \frac{C}{|x|^{2d-4}}.
\end{align*}
Since the dimension $d$ is assumed to be larger than $3$, we have the inequality $2d-4 \geq d-1$. We deduce that to prove the estimates pertaining to the random variable $X_{\sin \sin}$ in~\eqref{eq:estsmallprop3.1}, it is sufficient to prove the inequalities
\begin{equation} \label{eq:TV09242}
    \var \left[ \sum_{q\in \mathcal{Q}} z(\beta , q) \cos2\pi(\phi , q) \sin2\pi(\nabla G , n_q) \sin2\pi(\nabla G_x , n_q) \right] \leq \frac{C}{|x|^{d-1}}
\end{equation}
    and the expansion
\begin{equation} \label{eq:TV09252}
    \E \left[ \sum_{q\in \mathcal{Q}} z(\beta , q) \cos2\pi(\phi , q) \sin2\pi(\nabla G , n_q) \sin2\pi(\nabla G_x , n_q) \right] = \frac{c}{|x|^{d-2}} + O \left( \frac{C}{|x|^{d-1}}
    \right) .
\end{equation}
The estimate~\eqref{eq:TV09242} involving the variance can be estimated by the Helffer-Sj{\"o}strand representation formula and the bounds on the Green's matrix $\mathcal{G}$ stated in Proposition~\ref{prop.prop4.11chap4} of Chapter~\ref{chap:chap3}. We first note that, for each point $y \in \Zd$,
\begin{multline} \label{eq:TV09462}
    \partial_y \left(  \sum_{q \in \mathcal{Q}} z(\beta , q) \cos2\pi(\phi , q) \sin2\pi(\nabla G , n_q) \sin2\pi(\nabla G_x , n_q) \right) \\ = - \sum_{q \in \mathcal{Q}}2\pi z(\beta , q) \sin2\pi(\phi , q) \sin2\pi(\nabla G , q) \sin2\pi(\nabla G_x , q) q(y).
\end{multline}
From the identity~\eqref{eq:TV09462}, we deduce that to compute the variance~\eqref{eq:TV09242}, one needs to solve the Helffer-Sj{\"o}strand equation
\begin{equation*}
    \mathcal{L} \mathcal{W}(y , \phi)= - \sum_{q \in \mathcal{Q}} z(\beta , q) \sin2\pi(\phi , q) \sin2\pi(\nabla G , q) \sin2\pi(\nabla G_x , q) q(y).
\end{equation*}
Using the notation $\mathcal{G}$ for the Green's function associated to the Helffer-Sj{\"o}strand operator $\mathcal{L}$ introduced in Section~\ref{sec:section4.4} of Chapter \ref{chap:chap3}, we have the identity
\begin{equation} \label{eq:TV11462}
    \mathcal{W} (y , \phi) = - \sum_{q \in \mathcal{Q}} z(\beta , q) \sin2\pi(\nabla G , n_q) \sin2\pi(\nabla G_x , n_q) \sum_{z \in \supp n_q} \di^*_z \mathcal{G}_{\sin2\pi(\phi , q)} \left( y , \phi ; z \right)  n_q(z).
\end{equation}
Taking the exterior derivative of the identity~\eqref{eq:TV11462} shows the equality
\begin{equation*}
    \di^* \mathcal{W} (y , \phi) = - \sum_{q \in \mathcal{Q}} z(\beta , q) \sin2\pi(\nabla G , n_q) \sin2\pi(\nabla G_x , n_q) \sum_{z \in \supp n_q} \di^*_y \di^*_z \mathcal{G}_{\sin2\pi(\phi , q)} \left( y , \phi ; z \right)  n_q(z).
\end{equation*}
Using the estimate on the Green's function proved in Proposition~\ref{prop.prop4.11chap4} of Chapter~\ref{chap:chap3}, and the fact that the codifferential $\di^*$ is a linear functional of the gradient, we deduce the estimate, for each pair of points $y , z \in \Zd$,
\begin{equation} \label{eq:TV11502}
    \left\| \di^*_y \di^*_z \mathcal{G}_{\sin2\pi(\phi , q)} \left( y , \phi ; z \right) \right\|_{L^\infty \left( \mu_\beta \right)} \leq \frac{C}{|y - z|^{d-\ep}}.
\end{equation}
Using the estimate~\eqref{eq:TV11502} and a computation similar to the one performed in~\eqref{eq:TV09082}, we obtain the inequality, for each point $y \in \Zd$,
\begin{align} \label{eq:TV12112}
    \left\| \di^* \mathcal{W} (y , \cdot) \right\|_{L^\infty \left( \mu_\beta \right)} & \leq \sum_{z \in \Zd} \sum_{q \in \mathcal{Q}_z} \frac{e^{-c \sqrt{\beta} \left\| q\right\|_1} C_q}{|z|^{d-1} |z - x|^{d-1}} \frac{1}{|y - z|^{d-\ep}} \\
    & \leq \sum_{z \in \Zd} \frac{C}{|z|^{d-1} |z - x|^{d-1}} \frac{1}{|y - z|^{d-\ep}}. \notag
\end{align}
Using the definition of the map $\mathcal{W}$, we apply the Helffer-Sj{\"o}strand representation formula and deduce that
\begin{multline*}
    \var \left[ \sum_{q \in \mathcal{Q}} z(\beta , q) \cos2\pi(\phi , q) \sin2\pi(\nabla G , n_q) \sin2\pi(\nabla G_x , n_q) \right] \\ \leq 4\pi^2 \sum_{y \in \Zd} \left\langle \left( \sum_{q \in \mathcal{Q}} z(\beta , q) \sin2\pi(\phi , q) \sin2\pi(\nabla G , n_q) \sin2\pi(\nabla G_x , q) n_q(y) \right) \di^* \mathcal{W}(y , \phi) \right\rangle_{\mu_\beta}.
\end{multline*}
Using the estimates~\eqref{eq:TV12112} and a computation similar to the one performed in~\eqref{eq:TV09082}, we deduce that 
\begin{align} \label{eq:TV16560901}
    \lefteqn{ \var \left[ \sum_{q \in \mathcal{Q}} z(\beta , q) \cos2\pi(\phi , q) \sin2\pi(\nabla G , n_q) \sin2\pi(\nabla G_x , n_q) \right]} \qquad & \\ & \leq \sum_{y \in \Zd} \sum_{q \in \mathcal{Q}_y} \frac{e^{-c \sqrt{\beta} \left\| q \right\|_1} C_q}{|y|^{d-1} |x-y|^{d-1}} \left\| \di^* \mathcal{W} (y , \cdot) \right\|_{L^\infty \left( \mu_\beta \right)} \notag \\  
    & \leq C \sum_{y, z \in \Zd} \frac{1}{|y|^{d-1} |x-y|^{d-1}} \times \frac{1}{|z|^{d-1} |z - x|^{d-1}} \times \frac{1}{|y - z|^{d-\ep}} \notag \\
    & \leq \frac{C}{|x|^{2d-2}}, \notag
\end{align}
where we used Proposition~\ref{propappClign6572} of Appendix~\ref{app.appC} in the last line. There only remains to prove the identity~\eqref{eq:TV09252}. To this end, we use the ideas presented in Section~\ref{sectionsection4.2789} of Chapter~\ref{section3.4}.
We first define an equivalence relation on the set $\mathcal{Q}$: one says that two charges $q$ and $q'$ are equivalent, and denote it by $q \sim q'$, if they are equal up to a translation, i.e.,
\begin{equation*}
    q \sim q' \iff \exists y \in \Zd ~\mbox{such that}~ q \left( \cdot + y \right) = q'.
\end{equation*}
We denote this quotient space by $\mathcal{Q}/\Zd$ and for each charge $q \in \mathcal{Q}$, we denote by $[q]$ its equivalence class. For each equivalence class $[q] \in \mathcal{Q}/\Zd$, we select a charge $q \in \mathcal{Q}$ such that $0$ belongs to the support of $n_q$ (if there is more than one candidate, we break ties by using an arbitrary criterion). We note that, for each charge $q \in \mathcal{Q}$, by the definition of the charge $n_q$ and the coefficient $z \left( \beta , q \right)$, we have the identities, for each point $z \in \Zd$,
\begin{equation} \label{eq:TV13485}
    z \left( \beta , q \right) = z \left( \beta , q(\cdot - z ) \right) \hspace{5mm} \mbox{and} \hspace{5mm} n_{q(\cdot - z)} = n_q (\cdot - z).
\end{equation}
Additionally, we can decompose the sum over the charges $q \in \mathcal{Q}$ along the equivalence classes, i.e., we can write, for any non-negative or summable (with respect to the counting measure on the set $\mathcal{Q}$) function $F : \mathcal{Q} \to \R$
\begin{equation} \label{eq:TV13495}
    \sum_{q \in \mathcal{Q}} F(q) = \sum_{[q] \in \mathcal{Q}/\Zd} \sum_{z \in \Zd} F(q \left( \cdot - z\right)).
\end{equation}
We can thus decompose the sum
\begin{multline*}
    \sum_{q \in \mathcal{Q}} z(\beta , q) \cos2\pi(\phi , q) \sin2\pi(\nabla G , n_q) \sin2\pi(\nabla G_x , n_q) \\ = \sum_{[q] \in \mathcal{Q}/\Zd} z(\beta , q) \sum_{y \in \Zd} \cos2\pi(\phi , q(y + \cdot)) \sin2\pi(\nabla G , n_q(y + \cdot)) \sin2\pi(\nabla G_x , n_q(y + \cdot)).
\end{multline*}
Taking the expectation, using the translation invariance of the measure $\mu_\beta$ and using the identities $(\nabla G , n_q(y + \cdot)) = (\nabla G(\cdot -y) , n_q) $ and $(\nabla G_x , n_q(y + \cdot)) = (\nabla G(\cdot -y) , n_q) $, we deduce that
\begin{multline} \label{eq:TV16582}
    \E \left[ \sum_{q \in \mathcal{Q}} z(\beta , q) \cos2\pi(\phi , q) \sin2\pi(\nabla G , n_q) \sin2\pi(\nabla G_x , n_q) \right] \\ = \sum_{[q] \in \mathcal{Q}/\Zd} z(\beta , q) \E \left[\cos2\pi(\phi , q)\right] \sum_{y \in \Zd} \sin2\pi(\nabla G(\cdot - y) , n_q) \sin2\pi(\nabla G_x(\cdot - y) , n_q).
\end{multline}
Fix an equivalence class $[q] \in \mathcal{Q}/\Zd$ and define the value $\left( n_q \right) := \sum_{z \in \Zd} n_q(z) \in \Rd $ (which only depends on the equivalence class of the charge $q$). We prove the expansion
\begin{equation} \label{eq:TV15312}
    \sum_{y \in \Zd} \sin2\pi(\nabla G (\cdot - y) , n_q) \sin2\pi(\nabla G_x(\cdot - y) , n_q) = 4\pi^2 \sum_{y \in \Zd} \nabla G(y) \cdot \left( n_q \right) \times \nabla G_x(y) \cdot \left( n_q \right) + O \left( \frac{C_q}{|x|^{d-1}} \right).
\end{equation}
With the same arguments as the ones presented in \eqref{eq:TV09466} of Chapter~\ref{section3.4}, we write the inequalities 
\begin{equation*}
    \left| \left( \nabla G(\cdot - y) , n_q  \right) - \nabla G (y) \cdot  \left( n_q \right)  \right| \leq \frac{C_q}{|y|^d} ~\mbox{and}~\left| \left( \nabla G_x(\cdot - y) , n_q  \right) -  \nabla G_x (y) \cdot  \left( n_q \right)  \right| \leq \frac{C_q}{|y+x|^d} .
\end{equation*}
Combining this result with the estimate, for each real number $a \in \R$, $\left| \sin a - a \right| \leq \frac{|a|^3}{6}$, we obtain the estimate
\begin{multline*}
    \left| \sin2\pi(\nabla G(\cdot - y) , n_q) \sin2\pi(\nabla G_x(\cdot - y) , n_q) - 4\pi^2 \nabla G(y) \cdot \left( n_q \right) \times  \nabla G(y + x) \cdot (n_q) \right| \\ \leq \frac{C_q }{|y|^d |x - y|^{d-1}} + \frac{C_q}{|y|^{d-1} |x - y|^{d}}.
\end{multline*}
Summing over the points $y \in \Zd$ completes the proof of the identity~\eqref{eq:TV15312}. It remains to prove that the expansion~\eqref{eq:TV15312} implies the estimate~\eqref{eq:TV09252}. To this end, we first note that, if we denote by $\left(n_q\right)_1, \ldots, \left(n_q\right)_d $ the $d$-coordinates of the vector $\left( n_q \right)$, then we have the equality
\begin{equation*}
    \sum_{y \in \Zd} \nabla G(y) \cdot \left( n_q \right) \times \nabla G_x(y) \cdot \left( n_q \right) = \sum_{i,j = 1}^d \left(n_q\right)_i \left(n_q\right)_j \sum_{y \in \Zd} \nabla_i G(y) \nabla_j G_x(y).
\end{equation*}
We sum over all the equivalence class $[q] \in \mathcal{Q}/\Zd$ and use the exponential decay of the term $z \left( \beta ,q \right)$ to absorb the (at most) algebraic growth of the various terms involving the charge $q$. We obtain
\begin{multline} \label{eq:TV16562}
    \sum_{[q] \in \mathcal{Q}/\Zd} z(\beta , q) \E \left[\cos2\pi(\phi , q)\right] \sum_{y \in \Zd} \sin 2\pi(\nabla G(\cdot - y) , n_q) \sin 2\pi(\nabla G_x(\cdot - y) , n_q) \\ = \sum_{i,j = 1}^d c_{ij} \sum_{y \in \Zd} \nabla_i G(y) \nabla_j G_x(y) + O \left( \frac{C}{|x|^{d-1}} \right),
\end{multline}
where the constants are defined by the formulas
\begin{equation*}
    c_{ij} = 4\pi^2 \sum_{[q] \in \mathcal{Q}/\Zd} z(\beta , q) \E \left[\cos2\pi(\phi , q)\right] \left(n_q\right)_i \left(n_q\right)_j.
\end{equation*}
By combining the estimates~\eqref{eq:TV16582} and~\eqref{eq:TV16562}, we have obtained the expansion
\begin{equation} \label{eq:TV170722}
    \E \left[ \sum_{q\in \mathcal{Q}} z(\beta , q) \cos 2\pi(\phi , q) \sin2\pi(\nabla G , n_q) \sin2\pi(\nabla G_x , n_q) \right] = \sum_{i,j = 1}^d c_{ij} \sum_{y \in \Zd} \nabla_i G(y) \nabla_j G_x(y) + O \left( \frac{C}{|x|^{d-1}} \right).
\end{equation}
The expansion~\eqref{eq:TV170722} is not exactly~\eqref{eq:TV09252}. To complete the argument, we appeal to the symmetry invariance of the dual Villain model following the argument presented in Section~\ref{chap4:sect4.1heuristic} of Chapter~\ref{section3.4} and the results proved in Section~\ref{sec:chap8.4} of this chapter. We let $H$ be the group of lattice-preserving maps introduced in Chapter~\ref{Chap:chap2}. Since the measure $\mu_\beta$ is invariant under the elements of the group $H$, the map $x \mapsto  \E \left[ \sum_{q\in \mathcal{Q}} z(\beta , q) \cos2\pi(\phi , q) \sin2\pi(\nabla G , n_q) \sin2\pi(\nabla G_x , n_q) \right]$ satisfies the same invariance property. We can thus apply Proposition~\ref{propsinginvaraincefss} of Section~\ref{sec:chap8.4} (whose proof is independent of the rest of the arguments developed in the article) to complete the proof of~\eqref{eq:TV09252}.

\end{proof}

\section{Removing the contributions of the cosines} \label{sec:chap8.2}

The goal of this section is to prove Lemma~\ref{p.removecos} of Chapter~\ref{section3.4}, which is restated below. We recall the definitions of the charges $Q_x$ and $n_{Q_x}$ stated in~\eqref{eq:TVdefcapitQ} and~\eqref{eq:TV14223} of Chapter~\ref{section3.4}: for each pair $(y , \phi) \in \Zd \times \Omega$,
\begin{equation} \label{eq:TV14050}
     Q_x(y, \phi) :=  \sum_{q \in \mathcal{Q}} 2\pi z(\beta , q) \cos 2\pi(\phi , q) \sin 2\pi(\nabla G_x  , n_q)  q(y)
\end{equation}
and
\begin{equation} \label{eq:TV14060}
n_{Q_x}(y, \phi) :=  \sum_{q \in \mathcal{Q}} 2\pi z(\beta , q) \cos 2\pi(\phi , q) \sin 2\pi(\nabla G_x  , n_q)  n_q(y).
\end{equation}
The statement of Lemma~\ref{p.removecos} is recalled below.

\begin{lemma}[Removing the contributions of the cosines]\label{p.removecos}
One has the identity
\begin{equation} \label{eq:TV14563V}
    \cov \left[ X_x , Y_0 \right] = \sum_{y \in \Zd} \left\langle X_x Q_x(y, \cdot) \mathcal{V}(y, \cdot ) \right\rangle_{\mu_\beta} + O \left( \frac{C}{|x|^{d - 1 - \ep}} \right),
\end{equation}
where $\mathcal{V}: \Zd \times \Omega \to \R^{\binom d2}$ is the solution of the Helffer-Sj{\"o}strand equation, for each $(y, \phi ) \in \Zd \times \Omega$,
\begin{equation} \label{eq:TV10003}
    \mathcal{L} \mathcal{V}(y , \phi) = Q_0(y) Y_0(\phi).
\end{equation}
\end{lemma}

\begin{proof}
We start from the Helffer-Sj{\"o}strand representation formula stated in~\eqref{eq:V07043} of Chapter~\ref{section3.4} and recalled below
\begin{equation} \label{eq:TV09013}
     \cov \left[ X_x , Y_0 \right] = \sum_{y \in \Zd} \left\langle \left(\partial_y  X_x\right) \mathcal{Y}(y, \cdot ) \right\rangle_{\mu_\beta},
\end{equation}
where $\mathcal{Y} : \Zd \times \Omega \to \R^{\binom d2}$ is the solution of the Helffer-Sj{\"o}strand equation, for each $(y , \phi) \in \Zd \times \Omega$,
\begin{equation} \label{def.wVill2}
    \mathcal{L}\mathcal{Y}(y , \phi) = \partial_y Y_0(\phi).
\end{equation}
Using the definition of the random variables $Y_0$ and $X_x$ stated in~\eqref{eq:moneq} of Chapter~\ref{section3.4}, we have the identities, for each $y \in \Zd$,
\begin{equation} \label{eq:TV07353}
    \partial_y Y_0(\phi) = - \left(Q_0(y, \phi)  + \frac12 2\pi \sum_{q\in \mathcal{Q}} z \left( \beta , q \right) \sin 2\pi(\phi, q) \left( \cos 2\pi(\nabla G , n_q) - 1 \right) q(y) \right) Y_0(\phi)
\end{equation}
and
\begin{equation} \label{eq:TV07363}
    \partial_y X_x(\phi) = - \left(Q_x(y, \phi)  +  \sum_{q\in \mathcal{Q}} \frac12 2\pi z \left( \beta , q \right) \sin 2\pi(\phi, q) \left( \cos 2\pi(\nabla G_x , n_q) - 1 \right) q(y) \right) X_x(\phi).
\end{equation}
The objective of the proof is to remove the terms involving the cosine in the right side of the identities~\eqref{eq:TV07353} and~\eqref{eq:TV07363}. The proof requires to use the following estimates established in~\eqref{eq:TV15583} and~\eqref{eq:TV14193} of Chapter~\ref{section3.4}: for each point $y \in \Zd$,
\begin{equation} \label{eq:TV09133}
        \left\| n_{Q_x}(y, \cdot) \right\|_{L^\infty \left( \mu_\beta \right)} \leq \frac{C}{|y- x|^{d-1}}
\end{equation}
and
\begin{equation} \label{eq:TV09143}
    \left| \sum_{q\in \mathcal{Q}} \frac12 z \left( \beta , q \right) \sin2\pi(\phi, q) \left( \cos 2\pi(\nabla G_x , n_q) - 1 \right) n_q(y) \right| \leq \frac{C}{|y - x|^{2d-2}}.
\end{equation}
The same arguments give the estimates
\begin{equation} \label{eq:TV08443}
        \left\| n_{Q_0}(y,\cdot) \right\|_{L^\infty \left( \mu_\beta \right)} \leq \frac{C}{|y|^{d-1}},
\end{equation}
and
\begin{equation} \label{eq:TV08453}
    \left| \sum_{q\in \mathcal{Q}} \frac12 z \left( \beta , q \right) \sin2\pi(\phi, q) \left( \cos 2\pi(\nabla G_0 , n_q) - 1 \right) n_q(y) \right| \leq \frac{C}{|y|^{2d-2}}.
\end{equation}
We split the argument into three steps:
\begin{itemize}
    \item In Step 1, we prove that the solution of the Helffer-Sj{\"o}strand equation $\mathcal{Y}$ satisfies the upper bound, for each $y \in \Zd$,
    \begin{equation} \label{eq:TV09163}
         \left\| \di^* \mathcal{Y}(y , \cdot ) \right\|_{L^2 \left( \mu_\beta \right)} \leq \frac{C}{|y|^{d- 1 - \ep}};
    \end{equation}
    \item In Step 2, we prove that the covariance between the random variables $X_x$ and $Y_0$ satisfies the expansion
    \begin{equation} \label{eq:TV09313}
        \cov \left[ X_x , Y_0 \right] = \sum_{y \in \Zd} \left\langle X_x Q_x(y,\cdot) \mathcal{Y}(y, \cdot ) \right\rangle_{\mu_\beta} + O \left( \frac{C}{|x|^{d - 1 - \ep}} \right);
    \end{equation}
    \item In Step 3, we use the symmetry of the Helffer-Sj{\"o}strand operator $\L$ to complete the proof of Lemma~\ref{p.removecos}.
\end{itemize}

\medskip

\textit{Step 1.} We first express the function $\mathcal{Y}$ in terms of the Green function associated to the Helffer-Sj{\"o}strand operator $\mathcal{L}$. From the equation~\eqref{def.wVill} of Chapter \ref{section3.4}, we deduce the formula, for each $(y, \phi) \in \Zd \times \Omega$,
\begin{multline} \label{eq:TV08333}
    \mathcal{Y} (y, \phi ) = \sum_{y_1 \in \Zd} \sum_{q \in \mathcal{Q}} 2\pi z(\beta , q) \bigg[ \sin2\pi(\nabla G  , n_q) \di^*_{y_1} \mathcal{G}_{\cos 2\pi \left( \cdot , q \right) Y_0}\left( y , \phi ; y_1\right) n_q(y_1) \\ + \frac12  \left( \cos 2\pi(\nabla G , n_q) - 1 \right) n_q(y_1) \di^*_{y_1} \mathcal{G}_{\sin2\pi(\cdot, q) Y_0}(y , \phi ; y_1) n_q(y_1) \bigg].
\end{multline}
From the identity~\eqref{eq:TV08333}, we deduce the following formula for the function $\di^* \mathcal{Y}$
\begin{multline*}
     \di^* \mathcal{Y} (y, \phi ) = 2\pi \sum_{y_1 \in \Zd} \sum_{q \in \mathcal{Q}} z(\beta , q) \sin 2\pi(\nabla G  , n_q) \di^*_y \di^*_{y_1} \mathcal{G}_{\cos 2\pi\left( \cdot , q \right) Y_0}\left( y , \phi ; y_1\right) n_q(y_1) \\ +  2\pi \sum_{y_1 \in \Zd} \sum_{q \in \mathcal{Q}} \frac12  \left( \cos 2\pi(\nabla G , n_q) - 1 \right) \di^*_y \di^*_{y_1} \mathcal{G}_{\sin2\pi(\cdot, q) Y_0}(y , \phi ; y_1) n_q(y_1).
\end{multline*}
Using the estimate on the Green's function proved in Proposition~\ref{prop.prop4.11chap4} of Chapter \ref{chap:chap3}, that the random variable $Y_0$ belongs to the space $L^2 \left( \mu_\beta \right)$ and that the codifferential $\di^*$ is a linear functional of the gradient, we obtain the estimate, for each pair of points $y,y_1 \in \Zd$,
\begin{equation*}
    \left\| \di^*_y \di^*_{y_1}\mathcal{G}_{\cos 2\pi\left( \cdot , q \right) Y_0}\left( y , \phi ; y_1\right)  \right\|_{L^2 \left( \mu_\beta \right)} \leq \frac{C \left\| \cos 2\pi\left( \phi , q \right) Y_0 \right\|_{L^2 \left( \mu_\beta \right)}}{|y - y_1|^{d-\ep}} \leq \frac{C \left\| Y_0 \right\|_{L^2 \left( \mu_\beta \right)}}{|y - y_1|^{d-\ep}} \leq  \frac{C}{|y - y_1|^{d-\ep}}
\end{equation*}
and, with the same argument,
\begin{equation*}
    \left\| \di^*_y \di^*_{y_1} \mathcal{G}_{\sin 2\pi\left( \cdot , q \right) Y_0}\left( y , \phi ; y_1\right)  \right\|_{L^2 \left( \mu_\beta \right)} \leq \frac{C}{|y - y_1|^{d-\ep}}.
\end{equation*}
We then combine the two previous inequalities with the estimates~\eqref{eq:TV08443} and~\eqref{eq:TV08453}. We obtain the inequality
\begin{align*}
    \left\| \di^* \mathcal{Y} (y, \phi ) \right\|_{L^2 \left( \mu_\beta \right)} & \leq \sum_{y_1 \in \Zd} \frac{C}{|y_1|^{d-1}} \left\| \di^*_y \di^*_{y_1}\mathcal{G}_{\cos 2\pi\left( \cdot , q \right) Y_0}\left( y , \phi ; y_1\right)  \right\|_{L^2 \left( \mu_\beta \right)} \\ & \quad + \sum_{y_1 \in \Zd} \frac{C}{|y_1|^{2d-2}} \left\| \di^*_y \di^*_{y_1}\mathcal{G}_{\sin 2\pi\left( \cdot , q \right) Y_0}\left( y , \phi ; y_1\right)  \right\|_{L^2 \left( \mu_\beta \right)} \\
    & \leq \sum_{y_1 \in \Zd} \frac{C}{|y_1|^{d-1} |y - y_1|^{d-\ep}} +  \frac{C}{|y - y_1|^{2d-2}|y - y_1|^{d-\ep}} \\
    & \leq \sum_{y_1 \in \Zd} \frac{C}{|y |^{d- 1 - \ep}}.
\end{align*}
The proof of Step 1 is complete.

\medskip

\textit{Step 2.} By the Helffer-Sj{\"o}strand formula~\eqref{eq:TV09013}, we have the identity
\begin{align} \label{eq:TV09073}
    \cov \left[ X_x , Y_0 \right]&  = \sum_{y \in \Zd} \left\langle \left(\partial_y  X_x\right) \mathcal{Y}(y, \cdot ) \right\rangle_{\mu_\beta} \notag \\
    & = \sum_{y \in \Zd}  \left\langle Q_x (y) X_x \mathcal{Y}(y , \cdot) \right\rangle_{\mu_\beta}  - \frac12 2\pi \left\langle \sum_{q \in \mathcal{Q}} z\left( \beta , q \right) \sin2\pi(\phi , q) \left( \cos 2\pi(\nabla G_x , n_q) - 1 \right)   n_q(y)  X_x \di^* \mathcal{Y}(y , \cdot) \right\rangle_{\mu_\beta}.
\end{align}
The objective of this step is to prove that the term involving the cosine in the right side of~\eqref{eq:TV09073} is of lower order; specifically we prove the inequality, for each $y \in \Zd$,
\begin{equation*}
    \left| \frac12  \sum_{q \in \mathcal{Q}} z\left( \beta , q \right) \left( \cos 2\pi (\nabla G_x , n_q) - 1 \right) n_q(y) \left\langle\sin 2\pi(\phi , q)  X_x \di^* \mathcal{Y}(y, \cdot ) \right\rangle_{\mu_\beta} \right| \leq \frac{C}{|y|^{d-1 - \ep}}.
\end{equation*}
The proof of the previous estimate relies on the three ingredients: the estimate~\eqref{eq:TV09143}, the $L^2\left( \mu_\beta \right)$-estimate $\left\| X_x \right\|_{L^2 \left( \mu_\beta \right)} \leq C$ and the estimate~\eqref{eq:TV09163} proved in Step 1. We obtain
\begin{align*}
    \lefteqn{\left| \frac12  \sum_{q \in \mathcal{Q}} z\left( \beta , q \right) \left( \cos 2\pi (\nabla G_x , n_q) - 1 \right) n_q(y) \left\langle\sin 2\pi(\phi , q)  X_x \di^* \mathcal{Y}(y, \cdot ) \right\rangle_{\mu_\beta} \right|} \qquad & \\ &
    \leq  \frac12  \sum_{q \in \mathcal{Q}} \left| z\left( \beta , q \right) \left( \cos 2\pi(\nabla G_x , n_q) - 1 \right) n_q(y) \right| \left\| \sin2\pi(\phi , q)  X_x \right\|_{L^2 \left( \mu_\beta \right)} \left\|\di^* \mathcal{Y}(y, \cdot ) \right\|_{L^2 \left( \mu_\beta\right)} \\
     & \leq \frac{C}{|y-x|^{2d-2}} \cdot \frac{1}{|y|^{d-1- \ep}}.
\end{align*}
Summing the inequality over all the points $y \in \Zd$ then shows
\begin{align*}
    \left| \sum_{y \in \Zd} \frac12 \left\langle \sum_{q \in \mathcal{Q}} z\left( \beta , q \right) \sin2\pi(\phi , q) \left( \cos 2\pi(\nabla G_x , n_q) - 1 \right)   n_q(y)  X_x \di^* \mathcal{Y}(y , \cdot) \right\rangle_{\mu_\beta} \right| & \leq C \sum_{y \in \Zd} \frac{1}{|y - x|^{2d -2}} \cdot \frac{1}{|y|^{d-1-\ep}} \\
    & \leq \frac{C}{|x|^{d-1-\ep}}.
\end{align*}
The proof of Step 2 is complete.

\medskip

\textit{Step 3. The conclusion.} We use the main result~\eqref{eq:TV09313} of Step 2 and the symmetry of the Helffer-Sj{\"o}strand operator to complete the proof of Lemma~\ref{p.removecos}. By the expansion~\eqref{eq:TV09313}, we see that to prove~\eqref{eq:TV14563V}, it is sufficient to prove the estimate
\begin{equation} \label{eq:TV10053}
     \sum_{y \in \Zd} \left\langle Q_x(y,\cdot) X_x \mathcal{V}(y, \cdot ) \right\rangle_{\mu_\beta} = \sum_{y \in \Zd} \left\langle Q_x(y,\cdot) X_x \mathcal{Y}(y, \cdot ) \right\rangle_{\mu_\beta} + O \left( \frac{C}{|x|^{d-1-\ep}} \right).
\end{equation}
Since the Helffer-Sj{\"o}strand operator $\mathcal{L}$ is symmetric, we can write
\begin{equation} \label{eq:TV10043}
    \sum_{y \in \Zd} \left\langle Q_x(y,\cdot) X_x \mathcal{Y}(y, \cdot ) 
    \right\rangle_{\mu_\beta} = \sum_{y \in \Zd} \left\langle \mathcal{X}_x \left( y , \cdot \right) \partial_y Y_0 \right\rangle_{\mu_\beta},
\end{equation}
where the mapping $\mathcal{X}_x : \Zd \times \Omega \to \R^{\binom d2}$ is the solution of the Helffer-Sj{\"o}strand equation,
\begin{align} \label{eq:TV18412601}
    \mathcal{L} \mathcal{X}_x & = Q_x X_x ~\mbox{in}~ \Zd \times \Omega.
\end{align}
The objective of this step is to prove the following expansion
\begin{equation} \label{eq:TV09543}
    \sum_{y \in \Zd} \left\langle \mathcal{X}_x \left( y , \cdot \right) \partial_y Y_0 \right\rangle_{\mu_\beta} = \sum_{y \in \Zd} \left\langle \mathcal{X}_x \left( y , \cdot \right) Q_0(y,\cdot) Y_0 \right\rangle_{\mu_\beta} + O \left( \frac{C}{|x|^{d-1-\ep}}\right).
\end{equation}
The proof is similar to the one written is Steps 1 and 2. With the same arguments as the ones developed in Step 1, one obtains the following upper bound for the function $\di^* \mathcal{X}$: for each $y \in \Zd$,
\begin{equation} \label{eq:TV09523}
    \left\| \di^* \mathcal{X}_x (y , \cdot) \right\|_{L^2 \left( \mu_\beta \right)} \leq \frac{C}{|y - x|^{d - 1 -\ep}}.
\end{equation}
Using the same arguments as the ones developed in Step 2, we obtain the inequality
\begin{align} \label{eq:TV09533}
     \left| \sum_{y \in \Zd} \sum_{q\in \mathcal{Q}} z \left( \beta , q \right) \left( \cos (\nabla G , n_q) - 1 \right) n_q(y,\phi) \left\langle \di^* \mathcal{X}_x(y, \phi ) \sin2\pi(\phi, q)  Y_0(\phi) \right\rangle_{\mu_\beta} \right| & \leq C \sum_{y \in \Zd} \frac{1}{|y - x|^{d-1 - \ep}} \cdot \frac{1}{|y|^{2d -2}}  \\
     & \leq  \frac{C}{|x|^{d-1 - \ep}}. \notag
\end{align}
Combining the inequalities~\eqref{eq:TV09523} and~\eqref{eq:TV09533} with the formula~\eqref{eq:TV07353} implies the expansion~\eqref{eq:TV09543}. We then use the symmetry of the Helffer-Sj{\"o}strand operator a second time to obtain the identity
\begin{equation} \label{eq:TV10013}
    \sum_{y \in \Zd} \left\langle \mathcal{X}_x \left( y , \cdot \right) Q_0(y,\cdot) Y_0 \right\rangle_{\mu_\beta} = \sum_{y \in \Zd} \left\langle Q_x(y,\cdot) X_x \mathcal{V} \left( y , \cdot \right) \right\rangle_{\mu_\beta},
\end{equation}
where the function $\mathcal{V}$ is defined as the solution of the Helffer-Sj{\"o}strand equation~\eqref{eq:TV10003}. Combining the identities~\eqref{eq:TV10013},~\eqref{eq:TV09543} and~\eqref{eq:TV10043}, we obtain the expansion~\eqref{eq:TV10053}. This completes the proof of Step 3 and of Lemma~\ref{p.removecos}.
\end{proof}

\section{Decoupling the exponentials} \label{sec:chap8.3}
The objective of this section is to remove the exponential terms $X_x$ and $Y_0$ from the computation. We prove the decorrelation estimate stated in Lemma~\ref{p.decoupleexp} below. The argument makes use of the bounds on the Green's function $\mathcal{G}$ obtained in Proposition~\ref{prop.prop4.11chap4} of Chapter~\ref{chap:chap3} and on the Green's function $\mathcal{G}_{\mathrm{der}, \f }$ associated to the differentiated Helffer-Sj{\"o}strand operator proved in Proposition~\ref{cor:corollary4.14} of Chapter~\ref{section:section4}. Before stating the lemma, we record two estimates which are used in its proof:
\begin{itemize}
    \item We recall the definition of the random variable $\mathcal{X}_x : \Zd \times \Omega \to \R^{\binom d2}$ defined as the solution of the Helffer-Sj{\"o}strand equation, for each $(z , \phi ) \in \Zd \times \Omega$, $\mathcal{L} \mathcal{X}_x(z , \phi) = \partial_z X_x$ in~\eqref{eq:TV18412601} of Section~\ref{sec:chap8.2}; by the inequality~\eqref{eq:TV09523}, it satisfies the $L^2 \left(\mu_\beta \right)$-estimate
    \begin{equation} \label{eq:TV08187}
         \left\| \di^* \mathcal{X}_x(z , \cdot) \right\|_{L^2 \left( \mu_\beta \right)} \leq \frac{C}{|z-x|^{d-1-\ep}}.
    \end{equation}
    For later purposes, we note that the same arguments lead to the estimate
    \begin{equation} \label{eq:TV08188456}
         \left\| \mathcal{X}_x(z , \cdot) \right\|_{L^2 \left( \mu_\beta \right)} \leq \frac{C}{|z-x|^{d-2-\ep}};
    \end{equation}
    \item The function $\mathcal{V}$ defined in the statement of Lemma~\ref{p.removecos}; by the estimate~\eqref{eq:TV16284} of Chapter~\ref{section3.4}, it satisfies the estimate
    \begin{equation} \label{eq:TV08197}
        \left\| \di^* \mathcal{V}(z , \cdot) \right\|_{L^2 \left( \mu_\beta \right)} \leq \frac{C}{|x|^{d-1-\ep}}.
    \end{equation}
\end{itemize}

\begin{lemma}[Decoupling the exponential terms] \label{p.decoupleexp}
One has the following estimate
\begin{equation} \label{eq:TV174233}
    \cov \left[ X_x , Y_0 \right]  = \left\langle Y_0 \right\rangle_{\mu_\beta} \left\langle X_0 \right\rangle_{\mu_\beta}  \sum_{y \in \Zd} \left\langle Q_x(y) \mathcal{U}(y, \cdot ) \right\rangle_{\mu_\beta} + O \left( \frac{C}{|x|^{d-1+\ep}} \right),
\end{equation}
where the function $\mathcal{U}: \Zd \times \Omega \to \R^{\binom d2}$ is the solution of the Helffer-Sj{\"o}strand equation 
\begin{equation} \label{eq:VER12128}
    \mathcal{L} \mathcal{U} = Q_0~ \mbox{in} ~\Zd \times \Omega.
\end{equation}
\end{lemma}

\begin{proof}
We recall the notations and results introduced in Remarks~\ref{rem:rem3.13},~\ref{rem:rem3.3} and~\ref{rem:rem3.4} of Chapter~\ref{section3.4} which will be used in the proof. We start from the result of Lemma~\ref{p.removecos} which reads
\begin{equation*}
    \cov \left[ X_x , Y_0 \right] = \sum_{y \in \Zd} \left\langle X_x Q_x(y,\cdot) \mathcal{V}(y, \cdot ) \right\rangle_{\mu_\beta} + O \left( \frac{C}{|x|^{d - 1 - \ep}} \right),
\end{equation*}
where $\mathcal{V}$ is the solution of the Helffer-Sj{\"o}strand equation, for each $(y, \phi ) \in \Zd \times \Omega$,
\begin{equation} \label{eq:TV10003V}
    \mathcal{L} \mathcal{V}(y , \phi) = Q_0(y,\phi) Y_0(\phi).
\end{equation}
We split the argument into two steps:
\begin{itemize}
    \item In Step 1, we prove the decorrelation estimate
\begin{equation} \label{eq:TV20114}
    \sum_{y \in \Zd} \left\langle X_x Q_x(y,\cdot) \mathcal{V}(y , \cdot) \right\rangle_{\mu_\beta}  = \left\langle X_x \right\rangle_{\mu_\beta}\sum_{y \in \Zd} \left\langle Q_x(y,\cdot)  \mathcal{V}(y , \cdot) \right\rangle_{\mu_\beta} + O \left( \frac{C}{|x|^{d-1-\ep}} \right).
\end{equation}
Note that since the measure $\mu_\beta$ is invariant under translations, the value $ \left\langle X_x \right\rangle_{\mu_\beta}$ does not depend on the point $x$.
    \item In Step 2, we prove the expansion
    \begin{equation} \label{eq:TV11153}
    \sum_{y \in \Zd} \left\langle Q_x(y,\cdot)  \mathcal{V}(y , \cdot) \right\rangle_{\mu_\beta} = \left\langle Y_0 \right\rangle_{\mu_\beta}\sum_{y \in \Zd} \left\langle Q_x(y,\cdot)  \mathcal{U}(y , \cdot) \right\rangle_{\mu_\beta} + O \left( \frac{C}{|x|^{d-1-\ep}} \right).
\end{equation}
\end{itemize}
The expansion~\eqref{eq:TV174233} is a consequence of~\eqref{eq:TV20114} and~\eqref{eq:TV11153}

\medskip

\textit{Step 1.} The expansion~\eqref{eq:TV20114} can be rewritten in terms of the covariance between the random variables $X_x$ and $Q_x(y)  \mathcal{V}(y , \cdot)$; it is equivalent to the expansion
\begin{equation} \label{eq:TV20174}
    \sum_{y \in \Zd} \cov \left[ X_x ,Q_x(y,\cdot)  \mathcal{V}(y , \cdot) \right] = O \left( \frac{C}{|x|^{d-1-\ep}} \right).
\end{equation}
To prove the expansion~\eqref{eq:TV20174}, we apply the Helffer-Sj{\"o}strand representation formula which reads, for each point $y \in \Zd$,
\begin{equation} \label{eq:TV19031}
    \cov \left[ X_x , Q_x(y,\cdot)  \mathcal{V}(y , \cdot) \right] = \sum_{z \in \Zd} \left\langle \mathcal{X}_x(z , \cdot) \partial_z \left( Q_x(y,\cdot)  \mathcal{V}(y , \cdot) \right) \right\rangle_{\mu_\beta}.
\end{equation}
Summing over the points $y \in \Zd$ and performing an integration by parts in the variable $y$, we deduce that
\begin{align*}
    \sum_{y \in \Zd} \cov \left[ X_x ,Q_x(y,\cdot)  \mathcal{V}(y , \cdot) \right] & = \sum_{y,z \in \Zd} \left\langle \mathcal{X}_x(z , \cdot) \partial_z \left(Q_x(y,\cdot)  \mathcal{V}(y , \cdot) \right) \right\rangle_{\mu_\beta} \\
    &   = \sum_{y,z \in \Zd} \left\langle \mathcal{X}_x(z , \cdot) \partial_z \left( n_{Q_x}(y,\cdot)  \di^* \mathcal{V}(y , \cdot) \right) \right\rangle_{\mu_\beta}.
\end{align*}
We split the proof into two substeps:
\begin{itemize}
    \item In Substep 1.1, we compute the value of $\partial_z \left( n_{Q_x}(y,\cdot)  \di^* \mathcal{V}(y , \cdot) \right)$. We prove the identity~\eqref{eq:TV20011} and the inequalities~\eqref{eq:TV20371};
    \item In Substep 1.2, we deduce the estimate~\eqref{eq:TV20114} from Substep 1.1.
\end{itemize}

\medskip

\textit{Substep 1.1.}
We first expand the derivative
\begin{equation} \label{eq:TV08195}
    \partial_z \left( n_{Q_x}(y,\cdot) \di^* \mathcal{V}(y , \cdot) \right) = \underbrace{\left( \partial_z  n_{Q_x}(y,\cdot) \right) \di^* \mathcal{V}(y , \cdot)}_{\eqref{eq:TV08195}-(i)} +  \underbrace{n_{Q_x}(y,\cdot) \partial_z \di^* \mathcal{V}(y , \cdot)  }_{\eqref{eq:TV08195}-(ii)}.
\end{equation}
The term~\eqref{eq:TV08195}-(i) can be computed explicitly from the definition of the charge $n_{Q_x}$ and the identity $q = \di n_q$. We obtain
\begin{align} \label{eq:HA08316}
    \left( \partial_z  n_{Q_x}(y,\phi) \right) \di^* \mathcal{V}(y , \phi) &=  \left( \sum_{q \in \mathcal{Q}}4\pi^2 z(\beta , q) \left( \sin2\pi(\phi , q) \sin2\pi(\nabla G_x  , n_q)  \right) n_q(y) \otimes q(z) \right) \di^* \mathcal{V}(y , \phi) \\
    & = \di_z \left( \left( \sum_{q \in \mathcal{Q}} 4\pi^2 z(\beta , q) \left( \sin2\pi(\phi , q) \sin2\pi(\nabla G_x  , n_q)  \right) n_q(y) \otimes n_q(z) \right) \di^* \mathcal{V}(y , \phi) \right). \notag
\end{align}
We then estimate the term in the right side of~\eqref{eq:HA08316}. To this end, we note that the sum over the charges $q \in \mathcal{Q}$ can be restricted to the set of charges $\mathcal{Q}_{y,z}$. Using the inequality $ \sum_{q \in \mathcal{Q}_{y , z}} e^{-c \sqrt{\beta} \left\| q \right\|_1} \leq e^{- c \sqrt{\beta} |y-z|}$ established in~\eqref{eq:sumcharges} of Chapter~\ref{Chap:chap2}.

\smallskip

We use the inequality on the sine of the gradient of the Green's function: for each charge $q \in \mathcal{Q}$ such that the point $y$ belongs to the support of $n_q$,
\begin{equation*}
    \left| \sin2\pi(\nabla G_x  , n_q) \right| \leq \left| 2\pi(\nabla G_x  , n_q) \right| \leq \frac{C_q}{|x-y|^{d-1}}.
\end{equation*}
We deduce that
\begin{align} \label{eq:TV08117}
   \left|  \sum_{q \in \mathcal{Q}} z(\beta , q) \sin2\pi(\phi , q) \sin2\pi(\nabla G_x  , n_q)  n_q(y) \otimes n_q(z) \right| & \leq \sum_{q \in \mathcal{Q}_{x,y}} e^{-c \sqrt{\beta} \left\| q \right\|_1} \frac{C_q}{|y - x|^{d-1}} \\ & \leq \sum_{q \in \mathcal{Q}_{x,y}} \frac{C \left\| q \right\|_1^k e^{-c \sqrt{\beta} \left\| q \right\|_1}}{|y - x|^{d-1}} \notag \\
   & \leq \sum_{q \in \mathcal{Q}_{x,y}} \frac{C  e^{-c \sqrt{\beta} \left\| q \right\|_1}}{|y - x|^{d-1}} \notag \\
   & \leq \frac{C  e^{-c \sqrt{\beta} |y - z|}}{|y - x|^{d-1}}, \notag
\end{align}
where we have reduced the value of the constant $c$ in the third inequality to absorb the algebraic growth of the term $\left\| q \right\|_1^k$ into the exponential term $e^{-c \sqrt{\beta} \left\| q \right\|_1}$. Combining the estimate~\eqref{eq:TV08117} with the inequality~\eqref{eq:TV08197} on the codifferential of the function $\mathcal{V}$, we obtain, for each pair of points $z , y \in \Zd$,
\begin{equation*}
     \left\| \left( \sum_{q \in \mathcal{Q}} z(\beta , q) \left( \sin 2\pi(\cdot , q) \sin2\pi(\nabla G_x  , n_q)  \right) n_q(y) \otimes n_q(z) \right) \di^* \mathcal{V}(y , \cdot) \right\|_{L^2 \left( \mu_\beta \right)} \leq \frac{C  e^{-c \sqrt{\beta} |y - z|}}{|y - x|^{d-1} \times |y|^{d-1- \ep}}.
\end{equation*}
We now treat the term~\eqref{eq:TV08195}-(ii). To estimate the $L^2 \left( \mu_\beta \right)$-norm of the map $\partial_z \di^* \mathcal{V}(y , \phi)$, we start from the definition of the map $\mathcal{V}$ as the solution of the Helffer-Sj{\"o}strand equation~\eqref{eq:TV10003V} and apply the derivative $\partial_z$ to both sides of the identity~\eqref{eq:TV10003V}. Following the arguments developed at the beginning of Section~\ref{sec.section4.5} of Chapter~\ref{section:section4}, we obtain that the map $\mathcal{V}_{\mathrm{der}} : (y , z , \phi) \to \partial_z \mathcal{V}(y , \phi)$ is the solution of the differentiated Helffer-Sj{\"o}strand equation
\begin{align} \label{eq:TV08555}
    \mathcal{L}_{\mathrm{der}} \mathcal{V}_{\mathrm{der}}(y , z , \phi) &  =   \left( \sum_{q \in \mathcal{Q}} 4\pi^2 z(\beta , q) \left( \sin2\pi(\phi , q) \sin2\pi(\nabla G  , n_q)  \right) q(y) \otimes q(z) \right) Y_0 \\ & \quad + \sum_{q \in \mathcal{Q}} 2\pi z\left( \beta , q \right) \sin 2\pi\left( \phi , q \right) \left( \di^* \mathcal{V} , n_q \right) q(y) \otimes q(z) \notag \\ & \quad - Q_0(y, \phi) \otimes \left(Q_0(z,\phi)  + \frac12 2\pi \sum_{q\in \mathcal{Q}} z \left( \beta , q \right) \sin2\pi(\phi, q) \left( \cos (\nabla G , n_q) - 1 \right) q(z) \right) Y_0. \notag
\end{align}
We decompose the function $\mathcal{V}_{\mathrm{der}}$ into three functions, $\mathcal{V}_{\mathrm{der},1}$, $\mathcal{V}_{\mathrm{der},2}$ and $\mathcal{V}_{\mathrm{der},3}$ according to the three terms in the right side of~\eqref{eq:TV08555}, i.e.,
\begin{equation} \label{eq:TV11470}
    \left\{ \begin{aligned}
    \mathcal{L}_{\mathrm{der}} \mathcal{V}_{\mathrm{der},1}(y,z,\phi) &=  \left( \sum_{q \in \mathcal{Q}} z(\beta , q)  \sin2\pi(\phi , q) \sin2\pi(\nabla G  , n_q)   q(y) \otimes q(z) \right) Y_0, \\ 
    \mathcal{L}_{\mathrm{der}} \mathcal{V}_{\mathrm{der},2}(y,z,\phi) &=  \sum_{q \in \mathcal{Q}} z\left( \beta , q \right) \sin 2\pi\left( \phi , q \right) \left( \di^* \mathcal{V} , n_q \right) q(y) \otimes q(z), \\
    \mathcal{L}_{\mathrm{der}} \mathcal{V}_{\mathrm{der},3}(y,z,\phi) &= - Q_0(y,\phi) \otimes \left(Q_0(z,\phi)  + \frac12 \sum_{q\in \mathcal{Q}} z \left( \beta , q \right) \sin2\pi(\phi, q) \left( \cos (\nabla G , n_q) - 1 \right) q(z) \right) Y_0.  \\
    \end{aligned} \right.
\end{equation}
We then estimate the three terms $\mathcal{V}_{\mathrm{der},1}$, $\mathcal{V}_{\mathrm{der},2}$ and $\mathcal{V}_{\mathrm{der},3}$ separately.

\smallskip

\textit{Estimate for the term $\mathcal{V}_{\mathrm{der},1}$.} We first express the function $\mathcal{V}_{\mathrm{der},1}$ in terms of the Green's matrix $\mathcal{G}_{\mathrm{der}}$ associated to the differentiated Helffer-Sj{\"o}strand operator $\mathcal{L}_{\mathrm{der}}$. We obtain, for each pair of points $y,z \in \Zd$ and each field $\phi \in \Omega$,
\begin{align} \label{eq:TV08447}
   \mathcal{V}_{\mathrm{der},1} (y , z , \phi) & = \sum_{q \in \mathcal{Q}} z(\beta , q) \sum_{y_1 , z_1 \in \Zd}  \mathcal{G}_{\mathrm{der}, \sin2\pi(\cdot , q) Y_0} \left( y , z , \phi ; y_1 , z_1 \right)     \sin2\pi(\nabla G  , n_q)   q(y_1) \otimes q(z_1)   \\
   & =\sum_{q \in \mathcal{Q}} z(\beta , q) \sum_{y_1 , z_1 \in \Zd}  \di^*_{y_1} \di^*_{z_1} \mathcal{G}_{\mathrm{der}, \sin2\pi(\cdot , q) Y_0} \left( y , z , \phi ; y_1 , z_1 \right)     \sin2\pi(\nabla G  , n_q)   n_q(y_1) \otimes n_q(z_1) . \notag
\end{align}
Taking the codifferential $\di^*$ with respect to the variable $y$ on both sides of the identity~\eqref{eq:TV08447}, we obtain the formula
\begin{equation*}
    \di^*_y \mathcal{V}_{\mathrm{der},1} (y , z,\phi)
    =\sum_{q \in \mathcal{Q}} z(\beta , q) \sum_{y_1 , z_1 \in \Zd}  \di^*_y \di^*_{y_1} \di^*_{y_2} \mathcal{G}_{\mathrm{der}, \sin2\pi(\cdot , q) Y_0} \left( y , z , \phi ; y_1 , z_1 \right) \sin2\pi(\nabla G  , n_q)   n_q(y_1) \otimes n_q(z_1) .
\end{equation*}
We then use the bound on the triple derivative of the Green's matrix $\mathcal{G}_{\mathrm{der}}$ stated in Proposition~\ref{cor:corollary4.14} of Chapter~\ref{section:section4}, the fact that the random variable $Y_0$ belongs to the space $L^2\left( \mu_\beta \right)$ and the pointwise estimate~\eqref{eq:TV08117}. We obtain
\begin{align} \label{eq:TV08487}
    \left\| \di^*_y \mathcal{V}_{\mathrm{der},1} (y , z , \cdot) \right\|_{L^2 \left( \mu_\beta \right)} & \leq C \sum_{y_1, z_1 \in \Zd} \frac{1}{|y - y_1|^{2d +1 - \ep} + |z - z_1|^{2d+ 1 - \ep}}  \frac{e^{-c \sqrt{\beta} |y_1 - z_1|}}{|y_1|^{d-1}} \\
    & \leq C \sum_{y_1 \in \Zd} \frac{1}{|y - y_1|^{2d +1- \ep} + |z - y_1|^{2d +1 - \ep}} \frac{1}{|y_1|^{d-1}}. \notag
\end{align}
The term in the right side of the inequality~\eqref{eq:TV08487} can be estimated as it was done in~\eqref{eq:TV08576} of Chapter \ref{sec:section6}; we note that we have the inequality, for each triplet $(x,y,z) \in \Zd \times \Zd \times \Zd$,
\begin{align*}
    \left( |y - y_1|^{2d +1-\ep} + |z - y_1|^{2d+1-\ep}  \right) \geq c \dist_{\Z^{2d}} \left( (y,z) , (y_1,y_1) \right)^{^{2d +1-\ep}} & \geq c \left( \left| \frac{y-z}{2} \right|^2 + \left| \frac{y+z}{2} - y_1 \right|^2 \right)^{\frac{2d+1-\ep}{2}} \\ & \geq c \left( \left| y-z \right|^{2d+1-\ep} + \left| \frac{y+z}{2} - z \right|^{2d+1-\ep} \right), \notag
\end{align*}
where the notation $\dist_{\Z^{2d}}\left( (y,z) , (y_1,y_1) \right)$ is used to denote the euclidean distance in the lattice $\Z^{2d}$ between the points $(y,z)$ and $(y_1,y_1)$, the second inequality is then obtained by computing the orthogonal projection of the point $(y,z) \in \Z^{2d}$ on the diagonal $\left\{ (y_1,y_1) \in \Z^{2d} \, : \, y_1 \in \Zd \right\}$ and the third inequality is obtained by reducing the value of the constant $c$. The right side of the inequality~\eqref{eq:TV08487} can be estimated by Proposition~\ref{propappClign6572} of Appendix~\ref{app.appC} and we obtain
\begin{align} \label{eq:TV09247}
   \sum_{y_1 \in \Zd} \frac{1}{|y - y_1|^{2d +1- \ep} + |z - y_1|^{2d +1 - \ep}} \frac{1}{|y_1|^{d-1}} & \leq \sum_{y_1 \in \Zd} \frac{1}{|y_1|^{d-1}} \times \frac{1}{ \left| y-z \right|^{2d+1-\ep} + \left| \frac{y+z}{2} - y_1 \right|^{2d+1-\ep}} \\
    & \leq \frac{C}{|y-z|^{d+1} \max \left( \left| y \right|, \left| z \right|\right)^{d-1-\ep}}. \notag
\end{align}
Plugging the estimate~\eqref{eq:TV09247} into the inequality~\eqref{eq:TV08487}, we deduce that
\begin{equation*}
    \left\| \di^*_y \mathcal{V}_{\mathrm{der},1} (y , z , \phi) \right\|_{L^2 \left( \mu_\beta \right)} \leq
     \frac{C}{|y-z|^{d+1} \max \left( \left| y \right|, \left| z \right|\right)^{d-1-\ep}}.
\end{equation*}
Multiplying the term by the value $n_{Q_x}(y)$ and applying the pointwise bound~\eqref{eq:TV09133} shows
\begin{equation*}
    \left\| n_{Q_x}(y,\cdot)  \di^*_y \mathcal{V}_{\mathrm{der},1} \left( y,z,\cdot\right) \right\|_{L^2 \left( \mu_\beta \right)} \leq \frac{C}{|x-y|^{d-1}\times |z-y|^{d +1} \times \max \left( \left| y \right|, \left| z \right|\right)^{d-1-\ep}}.
\end{equation*}
This inequality completes the estimate of the term $\mathcal{V}_{\mathrm{der},1}$.

\medskip

\textit{Estimate for the term $\mathcal{V}_{\mathrm{der},2}$.}  We first express the function $\mathcal{V}_{\mathrm{der},2}$ in terms of the Green function $\mathcal{G}_{\mathrm{der}}$. We obtain, for each $(x , y , \phi) \in \Zd \times \Zd \times \Omega$,
\begin{equation*}
\mathcal{V}_{\mathrm{der},2} \left(y , z, \phi \right) = - \sum_{y_1 , z_1 \in \Zd} \sum_{q \in \mathcal{Q}} 2\pi z\left( \beta , q \right) \di^*_{y_1} \di^*_{z_1} \mathcal{G}_{\mathrm{der},  \sin2\pi \left( \cdot , q \right) \left( \di^* \mathcal{V} , n_q \right)} \left( y , z , \phi ; y_1 , z_1 \right)  n_q(y_1) \otimes n_q(z_1) .
\end{equation*}
Taking the codifferential in the variable $y$ on both sides of the identity~\eqref{eq:TV08447}, we obtain the formula
\begin{equation*}
    \di^*_y \mathcal{V}_{\mathrm{der},2} (y , z , \phi)
    = -  \sum_{y_1 , z_1 \in \Zd} \sum_{q \in \mathcal{Q}} 2\pi z\left( \beta , q \right) \di^*_y \di^*_{y_1} \di^*_{z_1} \mathcal{G}_{\mathrm{der},  \sin 2\pi\left( \cdot , q \right) \left( \di^* \mathcal{V} , n_q \right)} \left( y , z , \phi ; y_1 , z_1 \right)  n_q(y_1) \otimes n_q(z_1).
\end{equation*}
Using Proposition~\ref{cor:corollary4.14} of Chapter~\ref{section:section4} (for the triple derivative of the Green's function $\mathcal{G}_{\mathrm{der}}$), we obtain the inequality
\begin{equation} \label{eq:TV09028}
    \left\| \di^*_y \di^*_{y_1} \di^*_{z_1} \mathcal{G}_{\mathrm{der},  \sin 2\pi\left( \cdot , q \right) \left( \di^* \mathcal{V} , n_q \right)} \left( y , z , \cdot ; y_1 , z_1 \right) \right\|_{L^2 \left( \mu_\beta \right)} \leq \frac{C \left\| \left( \di^* \mathcal{V} , n_q \right) \right\|_{L^2 \left( \mu_\beta \right)}}{|y - y_1|^{2d+1-\ep} + |z - z_1|^{2d+1-\ep}}.
\end{equation}
Using the estimate~\eqref{eq:TV08197} on the $L^2 \left( \mu_\beta \right)$-norm on the map $\di^* \mathcal{V}$, we obtain that, for each charge $q \in \mathcal{Q}_{y_1}$,
\begin{equation*}
    \left\| \left( \di^* \mathcal{V} , n_q \right) \right\|_{L^2 \left( \mu_\beta \right)} \leq  \frac{C_q}{|y_1|^{d-1-\ep}}.
\end{equation*}
Summing the estimate~\eqref{eq:TV09028} over all the charges $q \in \mathcal{Q}$ and using a computation similar to the one performed in~\eqref{eq:TV08117}, one obtains
\begin{align} \label{eq:TV09308}
    \left\| \di^* \mathcal{V}_{\mathrm{der},2} (y , z , \cdot) \right\|_{L^2 \left( \mu_\beta \right)} & \leq \sum_{y_1 , z_1 \in \Zd} \sum_{q \in \mathcal{Q}_{y_1 , z_1}}  \frac{C e^{-c \sqrt{\beta} \left\| q \right\|_1} \left\| \left( \di^* \mathcal{V} , n_q \right) \right\|_{L^2 \left( \mu_\beta \right)}  \left\| n_q\right\|_{L^\infty}^2 }{|y - y_1|^{2d+1-\ep} + |z - z_1|^{2d+1-\ep}}\\
    & \leq  \sum_{y_1 , z_1 \in \Zd} \sum_{q \in \mathcal{Q}_{y_1 , z_1}}  \frac{C_q e^{-c \sqrt{\beta} \left\| q \right\|_1} }{\left( |y - y_1|^{2d+1-\ep} + |z - z_1|^{2d+1-\ep} \right) \times |y_1|^{d-1-\ep}} \notag \\
    & \leq \sum_{y_1 , z_1 \in \Zd}  \frac{C e^{-c \sqrt{\beta} |y_1 - z_1|} }{\left( |y - y_1|^{2d+1-\ep} + |z - z_1|^{2d+1-\ep} \right) \times |y_1|^{d-1-\ep}}. \notag
\end{align}
Using the exponential decay of the term $e^{-c \sqrt{\beta} |y_1 - z_1|}$, we can estimate the sum over the variable $z_1$ of the right side of the inequality~\eqref{eq:TV09308}. We deduce that
\begin{equation} \label{eq:TV09348}
    \left\| \di^* \mathcal{V}_{\mathrm{der},2} (y , z , \cdot) \right\|_{L^2 \left( \mu_\beta \right)} \leq \sum_{y_1 \in \Zd}  \frac{C}{\left( |y - y_1|^{2d+1-\ep} + |z - y_1|^{2d+1-\ep} \right) \times |y_1|^{d-1-\ep}}.
\end{equation}
The right side of~\eqref{eq:TV09348} is almost identical to the right side of~\eqref{eq:TV08487} (the only difference is that there is an additional factor $\ep$ in the term $|y_1|^{d-1-\ep}$) and can be estimated with the same argument. We obtain
\begin{equation*}
     \left\| n_{Q_x}(y,\cdot)  \di^*_y \mathcal{V}_{\mathrm{der},2} \left( y,z,\cdot\right) \right\|_{L^2 \left( \mu_\beta \right)} \leq \frac{C}{|x-y|^{d-1-\ep}\times |z-y|^{d +1 -\ep} \times \max \left( \left| y \right|, \left| z \right|\right)^{d-1-\ep}}.
\end{equation*}

\textit{Estimate for the term $\mathcal{V}_{\mathrm{der},3}$.} This estimate is the most involved of the three terms. We prove that there exists a map $\mathcal{W}_{\mathrm{der} , 3} : \Zd \times \Zd \times \Omega \to \R^{\binom d2 \times d}$ which satisfies the identity, for each $(y , z , \phi) \in \Zd \times \Zd \times \Omega$,
\begin{equation} \label{eq:TV11180}
    \mathcal{V}_{\mathrm{der},3}(y , z , \phi) = \di_z  \mathcal{W}_{\mathrm{der},3}(y , z , \phi),
\end{equation}
as well as the upper bounds
\begin{equation} \label{eq:TV11190}
    \left\| \mathcal{W}_{\mathrm{der},3}(y , z , \cdot) \right\|_{L^2 \left( \mu_\beta \right)} \leq \frac{C}{|y|^{d-\frac32-\ep} \times |z|^{d-\frac 32-\ep}} \hspace{3mm} \mbox{and} \hspace{3mm} \left\| \di^*_y \mathcal{W}_{\mathrm{der},3}(y , z , \cdot) \right\|_{L^2 \left( \mu_\beta \right)} \leq \frac{C}{|y|^{d-1-\ep} \times |z|^{d-1-\ep}}.
\end{equation}
To prove the identity~\eqref{eq:TV11180} and the estimates~\eqref{eq:TV11190}, we appeal to the parabolic equation, following the strategy presented in Section~\ref{sec.section4.5} of Chapter~\ref{section:section4}. We recall the notations introduced in this section, and in particular the Feynman-Kac formula stated in~\eqref{eq:FeynmanKacder} and~\eqref{eq:defPhatphider} of Chapter~\ref{section:section4}. Applying this formula to the equation defining the map $\mathcal{V}_{\mathrm{der} , 3}$ stated in~\eqref{eq:TV11470}. We obtain the identity
\begin{multline} \label{eq:TV14080}
   \mathcal{V}_{\mathrm{der},3}(y , z , \phi)  =\sum_{y_1 , z_1 \in \Zd}\int_{0}^\infty  \E_{\phi} \left[  - Y_0(\phi_t) P^{\phi_\cdot}_{\mathrm{der}}(t , y , z ; y_1 , z_1) Q_0(y_1, \phi_t ) \otimes Q_0( z_1, \phi_t )    \right] \\
    - \frac 12 2\pi \sum_{y_1 , z_1 \in \Zd} \sum_{q\in \mathcal{Q}} z \left( \beta , q \right) \left( \cos (\nabla G , n_q) - 1 \right)\int_{0}^\infty  \E_{\phi} \left[  \sin2\pi(\phi_t, q) Y_0(\phi_t)  P^{\phi_\cdot}_{\mathrm{der}}(t , y , z ; y_1 , z_1) Q_0( y_1,\phi_t ) \otimes  q(z_1) \right],
\end{multline}
where, given a trajectory $(\phi_t)_{t \geq 0}$ of the Langevin dynamics, the map $P^{\phi_\cdot}_{\mathrm{der}}(\cdot , \cdot , \cdot \, ; y_1 , z_1): (0 , \infty) \times \Zd \times \Zd \to \R^{\binom d2^4}$ denotes the solution of the parabolic system of equations,
    \begin{equation} \label{eq:defPhatphiderbis}
        \left\{ \begin{aligned}
        \partial_t P^{\phi_\cdot}_{\mathrm{der}}\left(\cdot , \cdot , \cdot \, ; y_1 , z_1\right) + \left(\mathcal{L}_{\mathrm{spat} , x}^{\phi_t} + \mathcal{L}_{\mathrm{spat} , y}^{\phi_t}\right)  P^{\phi_\cdot}_{\mathrm{der}}\left(\cdot , \cdot , \cdot \, ; y_1 , z_1\right) & =0 ~\mbox{in}~ (0 , \infty) \times \Zd \times \Zd, \\
        P^{\phi_\cdot}_{\mathrm{der}} \left(0,\cdot , \cdot \, ;y_1 , z_1 \right) & = \delta_{(y_1 , z_1)} ~\mbox{in}~\Zd \times \Zd.
        \end{aligned} \right.
    \end{equation}
To ease the notation in the rest of the argument, we introduce the following definition. Given a pair of charges $q_1 , q_2 \in \mathcal{Q}$ and a trajectory of the Langevin dynamics $\left( \phi_t \right)_{t \geq 0}$, we let $P^{\phi_\cdot}_{q_1, q_2}(\cdot , \cdot , \cdot) : (0 , \infty) \times \Zd \times \Zd \to \R^{\binom d2^2}$ be the solution of the parabolic system
\begin{equation} \label{eq:TV20240}
        \left\{ \begin{aligned}
        \partial_t P^{\phi_\cdot}_{q_1 , q_2}+ \left(\mathcal{L}_{\mathrm{spat} , x}^{\phi_t} + \mathcal{L}_{\mathrm{spat} , y}^{\phi_t}\right)  P^{\phi_\cdot}_{ q_1 , q_2} & =0 ~\mbox{in}~ (0 , \infty) \times \Zd \times \Zd, \\
        P^{\phi_\cdot}_{ q_1 , q_2} \left(0,y , z\right) & = q_1(y) \otimes q_2(z).
        \end{aligned} \right.
\end{equation}
We note that since the operator $\mathcal{L}_{\mathrm{spat} , x}^{\phi_t}$  and $\mathcal{L}_{\mathrm{spat} , y}^{\phi_t}$ commutes, the solution of the equation~\eqref{eq:TV20240} factorizes; one can write $ P^{\phi_\cdot}_{ q_1 , q_2}(t , y , z) = P^{\phi_\cdot}_{ q_1} (t , y ) \otimes  P^{\phi_\cdot}_{ q_2} (t , z) $, where the maps $P^{\phi_\cdot}_{ q_1} $ and $ P^{\phi_\cdot}_{ q_2}$ are the solutions of the parabolic systems
\begin{equation} \label{eq:TV14531}
     \left\{ \begin{aligned}
        \partial_t P^{\phi_\cdot}_{q_1}+ \mathcal{L}_{\mathrm{spat}}^{\phi_t}  P^{\phi_\cdot}_{ q_1} & =0 ~\mbox{in}~ (0 , \infty) \times \Zd , \\
        P^{\phi_\cdot}_{ q_1 } \left(0,\cdot\right) & = q_1 ~\mbox{in}~ \Zd,
        \end{aligned} \right.
        \hspace{5mm}\mbox{and}  \hspace{5mm}
         \left\{ \begin{aligned}
        \partial_t P^{\phi_\cdot}_{q_2}+ \mathcal{L}_{\mathrm{spat}}^{\phi_t} P^{\phi_\cdot}_{ q_2} & =0 ~\mbox{in}~ (0 , \infty) \times \Zd, \\
        P^{\phi_\cdot}_{ q_2 } \left(0,y , z\right) & = q_2 ~\mbox{in}~ \Zd.
        \end{aligned} \right.
\end{equation}
We then use this notation and the definition of the random charge $Q_0$ stated in~\eqref{eq:TV14050} to rewrite the identity~\eqref{eq:TV14080}. We obtain
\begin{align} \label{eq:TV15501}
   \mathcal{V}_{\mathrm{der},3}(y , z , \phi) = & - \sum_{q_1 , q_2 \in \mathcal{Q}} z \left( \beta , q_1 \right) z \left( \beta , q_2 \right) \sin 2\pi(\nabla G , n_{q_1}) \sin 2\pi(\nabla G , n_{q_2}) \\ & \qquad \times \int_{0}^\infty   \E_{\phi} \left[  \cos2\pi(\phi_t, q_1) \cos2\pi(\phi_t, q_2)    Y_0(\phi_t)  P^{\phi_\cdot}_{q_1, q_2}(t , y , z) \right] dt \notag \\
   & + \frac 12 2\pi\sum_{q_1 , q_2 \in \mathcal{Q}} z \left( \beta , q_1 \right) z \left( \beta , q_2 \right) \sin 2\pi(\nabla G , n_{q_1}) \left( \cos 2\pi(\nabla G , n_{q_2}) - 1 \right) \notag \\ & \qquad \times \int_{0}^\infty   \E_{\phi} \left[  \cos2\pi(\phi_t, q_1) \cos2\pi(\phi_t, q_2)    Y_0(\phi_t)  P^{\phi_\cdot}_{q_1, q_2}(t , y , z) \right] dt. \notag
   \notag
\end{align}
We fix a trajectory $(\phi_t)_{t \geq 0}$ of the Langevin dynamics, two points $y_1, z_1 \in \Zd$, a pair of charges $(q_1, q_2) \in \mathcal{Q}_{y_1} \times \mathcal{Q}_{z_1}$ and study the map $P^{\phi_\cdot}_{q_1, q_2}$. More precisely, in view of the decomposition $P^{\phi_\cdot}_{q_1, q_2} = P^{\phi_\cdot}_{q_1} \otimes P^{\phi_\cdot}_{ q_2}$, we study the map $P^{\phi_\cdot}_{q}$ for a general charge $q \in \mathcal{Q}$ and prove the following results:
\begin{itemize}
    \item[(i)] There exist constants $C := C(d) < \infty$ and $C_{q} \leq C_0 e^{C_0 \left\| q \right\|_1 }$, for some $C_0 := C_0(d) < \infty$, such that each point $y \in \Zd$ and for each time $t \in (0 , \infty)$,
    \begin{equation} \label{eq:TV16480}
        \left| P^{\phi_\cdot}_{q}(t , y ) \right| \leq 
        C_{q} \left(\frac{ \beta}{t} \right)^{\frac 12 - \ep} \Phi_C \left( \frac{t}{\beta} , y - y_1\right) \hspace{4mm} \mbox{and} \hspace{4mm}  \left| \nabla P^{\phi_\cdot}_{q}(t , y) \right| \leq C_{q}
        \left( \frac{\beta}{t} \right)^{1 - \ep} \Phi_C \left( \frac{t}{\beta} , y - y_1\right).
    \end{equation}
    We note that here, contrary to the other results presented in the article, the constant $C_{q_1 , q_2}$ is allowed to grow exponentially fast in the parameter $ \left\| q \right\|_1$; this is caused by the exponential decay of the heat kernel. This growth does not cause problems in the analysis: since the constant $C_0$ depends only on the dimension~$d$, we can use the estimate $z \left( \beta , q \right) \leq e^{-c \sqrt{\beta} \left\| q\right\|_1}$ and set the inverse temperature $\beta$ large enough so that the exponent $c \sqrt{\beta}$ is strictly larger than the constant $C_0$ in order to absorb this term. In particular, from now on and until the estimate~\eqref{eq:TV1709} below, we assume that the constants are only allowed to depend on the dimension $d$ and keep track of their dependence in the inverse temperature $\beta$; \smallskip
    \item[(ii)] We prove that there exists a function $Q^{\phi_\cdot}_{q} : (0 , \infty) \times \Zd \to \R^{d}$ such that for each time $t \geq 0$ and each point $y \in \Zd$ one has the identity $P^{\phi_\cdot}_{q}(t,y) = \di Q^{\phi_\cdot}_{q}(t,y)$. Additionally, we prove that the function $Q^{\phi_\cdot}_{q}$ satisfies the estimates, for each point $y \in \Zd$ and for each time $t \geq \beta$,
    \begin{equation} \label{eq:TV16500}
        \left| Q^{\phi_\cdot}_{q}(t , y) \right| \leq 
        C_{q}\left( \frac{t}{\beta} \right)^{\ep} \Phi_C \left( \frac{t}{\beta} , y - y_1 \right).
    \end{equation}
\end{itemize}
To prove the estimate~\eqref{eq:TV16480}, we use the results established in Chapter~\ref{section:section4}. We first express the function $P^{\phi_\cdot}_{q}$ in terms of the heat kernel $P^{\phi_\cdot}$ (defined in~\eqref{eq:defPhatphiderbis}). We obtain
\begin{equation*}
    P^{\phi_\cdot}_{q}(t , y ) = \sum_{y' \in \Zd}  \di^*_{y'} P^{\phi_\cdot}_{\mathrm{der}}\left(t , y\, ; y'\right) n_{q}(y'),
\end{equation*}
and consequently
\begin{equation*}
\nabla P^{\phi_\cdot}_{q}(t , y ) = \sum_{y' \in \Zd}   \nabla_y \di^*_{y'} P^{\phi_\cdot}_{\mathrm{der}}\left(t , y  \, ; y'\right)n_{q}(y').
\end{equation*}
We use the Nash-Aronson estimate and the regularity theory for the heat kernel $P^{\phi_\cdot}_{\mathrm{der}}$ stated in Propositions~\ref{prop:prop4.8} and~\ref{prop:prop4.9} of Chapter~\ref{section:section4}. We obtain the upper bound
\begin{equation} \label{eq:TV17580}
    \left|P^{\phi_\cdot}_{q}(t , y )  \right| \leq  \left\| n_{q} \right\|_\infty \sum_{y'\in \supp n_{q}}  \left(\frac{ \beta}{t} \right)^{\frac 12 - \ep} \Phi_C \left( \frac{t}{\beta} , y - y'\right) .
\end{equation}
We use that, by assumption, the point $y_1$ belongs to the support of the charge $n_{q}$ and the inequality, for each point $y \in \Zd$, each point $y' \in \supp n_{q}$ and each time $t \in (0 , \infty)$,
\begin{align} \label{eq:TV17590}
    \Phi_C \left(t , y - y'\right)  \leq \Phi_C \left(t , y - y_1 \right) \exp \left( \frac{|y-y_1| }{C} \right)   \leq \Phi_C \left(t , y - y_1\right) \exp \left( \frac{\diam n_{q} }{C} \right)  & \leq \Phi_C \left(t , y - y_1 \right) \exp \left( \frac{\diam q}{C} \right) \\
    & \leq \Phi_C \left(t , y - y_1 \right) \exp \left( \frac{ \left\| q \right\|_1}{C} \right), \notag
\end{align}
where the first inequality is obtained by an explicit computation using the definition of the function $\Phi_C$. 
Combining the identity~\eqref{eq:TV17580} and the estimate~\eqref{eq:TV17590}, we obtain the inequality
\begin{align} \label{eq:TV09141}
     \left|P^{\phi_\cdot}_{q} \left( \frac{t}{\beta} , y , z \right)  \right| & \leq \left\| n_{q} \right\|_\infty  \sum_{y'\in \supp n_{q}} \left(\frac{ \beta}{t} \right)^{\frac 12 - \ep} \Phi_C \left(\frac{t}{\beta} , y - y_1 \right) \exp \left( \frac{ \left\| q\right\|_1 }{C} \right)
    \\  & \leq \left\| n_{q} \right\|_\infty  \left| \supp n_{q} \right| \left(\frac{ \beta}{t} \right)^{\frac 12 - \ep} \Phi_C \left( \frac{t}{\beta} , y - y_1 \right) \exp \left( \frac{ \left\| q_1 \right\|_1 }{C} \right) \notag \\
     & \leq C_{q} \left(\frac{ \beta}{t} \right)^{\frac 12 - \ep} \Phi_C \left( \frac{t}{\beta} , y - y_1\right). \notag
\end{align}
This completes the proof of the first estimate of~\eqref{eq:TV16480}. The second estimate follows similar lines, the only difference is that we use the regularity estimate stated in Proposition~\ref{prop:prop4.9} of Chapter~\ref{section:section4} instead of the Nash-Aronson type estimate stated in Proposition~\ref{prop:prop4.8} of Chapter~\ref{section:section4}.

We now focus on the existence of the function $Q^{\phi_\cdot}_{q}$ and the estimate~\eqref{eq:TV16500}. We first recall the explicit definition of the elliptic operator $\mathcal{L}_{\mathrm{spat}}$
\begin{equation*}
    \mathcal{L}_{\mathrm{spat}} = -\frac{1}{2\beta} \Delta + \frac{1}{2\beta}\sum_{n \geq 1} \frac{1}{\beta^{ \frac n2}} (-\Delta)^{n+1} + \sum_{q \in \mathcal{Q}} \nabla_{q}^* \cdot \a_{q} \nabla_{q}.
\end{equation*}
We define the function $Q^{\phi_\cdot}_{q}$ to be the solution of the parabolic system
\begin{equation} \label{eq:TV20150}
    \left\{ \begin{aligned}
        \partial_t Q^{\phi_\cdot}_{q} - \left( \frac{1}{2\beta} \Delta - \frac{1}{2\beta}\sum_{n \geq 1} \frac{1}{\beta^{ \frac n2}} (-\Delta)^{n+1} \right) Q^{\phi_\cdot}_{ q} & = - \sum_{q \in \mathcal{Q}}z \left( \beta , q \right) \cos 2\pi\left( \phi_t, q \right) \nabla_{q} P^{\phi_\cdot}_{q} n_{q}  &\mbox{in}~ (0 , \infty) \times \Zd,\\
        Q^{\phi_\cdot}_{ q} \left(0,\cdot\right) & =  n_{q} &\mbox{in}~\Zd.
        \end{aligned} \right.
\end{equation}
Applying the exterior derivative $\di$ to both sides of the equation~\eqref{eq:TV20150} and using that this operator commutes with the Laplacian $\Delta$, we obtain that
\begin{equation*}
    \left\{ \begin{aligned}
        \partial_t \di Q^{\phi_\cdot}_{q}-\left( \frac{1}{2\beta} \Delta - \frac{1}{2\beta}\sum_{n \geq 1} \frac{1}{\beta^{ \frac n2}} (-\Delta)^{n+1} \right)  \di Q^{\phi_\cdot}_{ q} & = - \sum_{q \in \mathcal{Q}}\nabla_{q}^* \a_{q} \nabla_{q} P^{\phi_\cdot}_{q} &\mbox{in}~ (0 , \infty) \times \Zd \times \Zd,\\
        \di Q^{\phi_\cdot}_{ q} \left(0,\cdot\right) & = q &~\mbox{in}~ \Zd.
        \end{aligned} \right.
\end{equation*}
Using the definition of the map $P^{\phi_\cdot}_{ q}$ given in~\eqref{eq:TV20240}, we see that the two maps $P^{\phi_\cdot}_{ q}$ and $\di Q^{\phi_\cdot}_{ q}$ solve the same parabolic equation with the same initial condition; this implies that they are equal.

It remains to prove the estimate~\eqref{eq:TV16500}. We denote by $K : (0, \infty) \times  \Zd \to \R^{\binom d2 \times \binom d2}$ the fundamental solution of the parabolic system
\begin{equation*}
    \left\{ \begin{aligned}
        \partial_t  K - \left( \frac{1}{2\beta} \Delta - \frac{1}{2\beta}\sum_{n \geq 1} \frac{1}{\beta^{ \frac n2}} (-\Delta)^{n+1} \right)  K  & = 0 &~\mbox{in}~ (0 , \infty) \times \Zd, \\
       K \left(0, \cdot \right) & = \delta_{0}  &~\mbox{in}~\Zd.
        \end{aligned} \right.
\end{equation*}
For the parabolic operator $\partial_t - \frac{1}{2\beta} \Delta+ \frac{1}{2\beta}\sum_{n \geq 1} \frac{1}{\beta^{ \frac n2}} (-\Delta)^{n+1}$, a Nash-Aronson estimate and a complete regularity theory is available: for each integer $k \in \N$, there exists a constant $C := C(k,d) < \infty$ such that
\begin{equation*}
    \left| \nabla^k K (t , y) \right| \leq C \left(\frac{\beta}{t} \right)^{\frac k2} \Phi_C \left( \frac{t}{\beta} , y\right).
\end{equation*}
We use the Duhamel principle to express the function $Q^{\phi_\cdot}_{ q}$ in terms of the kernel $K$. We obtain the formula, for each point $y \in \Zd$ and each time $t > 0$,
\begin{multline} \label{eq:TV08491852}
    Q^{\phi_\cdot}_{q} (t , y ) = \sum_{y'\in \Zd}  \di^* K \left(t , y - y' \right) n_{q}(y') \\ - \sum_{y' \in \Zd} \sum_{q \in \mathcal{Q}} \int_0^t  z \left( \beta , q \right) \cos 2\pi\left( \phi_s, q \right) \left( n_{q}, \di^* P^{\phi_\cdot}_{q}(s , \cdot) \right) K(t , y - y' ) n_q(y')\di s.
\end{multline}
To estimate the right side of~\eqref{eq:TV08491852}, we record two inequalities. The first one is obtained by the same computation as~\eqref{eq:TV09141} and reads
\begin{align} \label{eq:TV13541}
    \sum_{y' \in \Zd}  \left|   \di^*  K \left(t , y-y' \right) n_{q}(y') \right| & \leq \sum_{y'\in \Zd} C \left\| n_{q}\right\|_\infty \left(\frac{\beta}{t} \right)^{\frac 12 - \ep} \Phi_C \left( \frac{t}{\beta} ,  y - y' \right) \\
    & \leq C_{q} \left(\frac{\beta}{t} \right)^{\frac 12 - \ep} \Phi_C \left( \frac{t}{\beta} ,  y - y_1 \right), \notag
\end{align}
where the constant $C_{q}$ satisfies the same exponential growth in the parameter $\left\| q\right\|_1$ as the one in the right side of~\eqref{eq:TV09141}. For the second inequality, we fix a point $y'\in \Zd$, a time $s > 0$, use the Nash-Aronson estimate on the kernel $K$ and the estimate~\eqref{eq:TV16480} on the function $P^{\phi_\cdot}_{q}$. We obtain
\begin{align} \label{eq:TV11491}
    \lefteqn{ \sum_{q \in \mathcal{Q}} \left| z \left( \beta , q \right) \cos 2\pi\left( \phi_t, q \right) \left( n_{q}, \di^* P^{\phi_\cdot}_{q}(s , \cdot) \right) K(t-s , y- y' ) n_{q}(y') \right| } \qquad & \\ & \leq \sum_{q \in \mathcal{Q}_{y'}} C_q e^{-c \sqrt{\beta} \left\| q \right\|_1} \Phi_C \left( \frac{t-s}{\beta}, y-y'  \right)  \left\| \nabla P^{\phi_\cdot}_{q}(s ,  \cdot) \right\|_{L^2 \left( \supp n_q \right)} \notag \\ 
    & \leq \sum_{q \in \mathcal{Q}_{y'}} C_q e^{-c \sqrt{\beta} \left\| q \right\|_1} \Phi_C \left( \frac{t-s}{\beta}, y-y' \right)  \frac{\left\| \Phi_C \left( \frac{s}{\beta},\cdot \right) \right\|_{L^2 \left( \supp n_q \right)}}{\left( \frac{s}{\beta} \right)^{1 - \ep} \wedge 1}. \notag
\end{align}
With a computation similar to the one performed in~\eqref{eq:TV17590}, we obtain the inequality, for each charge $q \in \mathcal{Q}_{y'}$,
\begin{equation} \label{eq:TV11501}
    \left\| \Phi_C \left( \frac{s}{\beta}, \cdot \right) \right\|_{L^2 \left( \supp n_q \right)} \leq \left\| \Phi_C \left(  \frac{s}{\beta}, y' \right) \exp \left( \frac{\left\| q \right\|_1}{C} \right) \right\|_{L^2 \left( \supp n_q \right)} \leq C_q\exp \left( \frac{\left\| q \right\|_1}{C} \right)  \Phi_C \left( \frac{s}{\beta}, y' \right) .
\end{equation}
Combining the inequalities~\eqref{eq:TV11491} and~\eqref{eq:TV11501}, we have obtained
\begin{multline} \label{eq:TV11591}
     \sum_{q \in \mathcal{Q}} \left| z \left( \beta , q \right) \cos 2\pi\left( \phi_t, q \right) \left( n_{q}, \di^* P^{\phi_\cdot}_{q}( s , \cdot) \right) K(t-s, y-y' ) n_q(y')\right| \\ \leq   \frac{\Phi_C \left( \frac{t-s}{\beta}, y-y' \right) \Phi_C \left( \frac{s}{\beta}, y'\right)}{\left( \frac{s}{\beta} \right)^{1 - \ep} \wedge 1} \underbrace{\sum_{q \in \mathcal{Q}_{y'}} C_q e^{-c \sqrt{\beta} \left\| q \right\|_1}.}_{\eqref{eq:TV11591}-(i)}
\end{multline}
By choosing the inverse temperature $\beta$ large enough, we obtain that the sum~\eqref{eq:TV11591}-(i) is bounded from above by a constant $C$ depending only on $d$; we have proved
\begin{equation} \label{eq:TV12031}
     \left| \sum_{q \in \mathcal{Q}}  z \left( \beta , q \right) \cos 2\pi\left( \phi_t, q \right)\left( n_{q}, \di^* P^{\phi_\cdot}_{q}(s , \cdot) \right) K(t-s, y-y' ) n_q(y') \right| \\ \leq  C \frac{\Phi_C \left( \frac{t-s}{\beta}, y-y' \right) \Phi_C \left(  \frac{s}{\beta}, y'\right)}{s^{1 - \ep} \wedge 1}.
\end{equation}
We then use the inequality, for some constant $\tilde C > C$, 
\begin{equation} \label{eq:TV13481}
    \sum_{y' \in \Zd} \Phi_C \left(  \frac{t-s}{\beta}, y-y' \right) \Phi_C \left(  \frac{s}{\beta}, y'  \right) \leq \tilde C \Phi_{ \tilde C} \left( \frac{t}{\beta}, y  \right).
\end{equation}
The inequality~\eqref{eq:TV13481} is a convolution property for the discrete heat kernel and can be verified by an explicit computation using the formula for the map $\Phi_C$.
A combination of the inequalities~\eqref{eq:TV12031} and~\eqref{eq:TV13481} yields
\begin{equation} \label{eq:TV13561}
      \sum_{y'\in \Zd} \sum_{q \in \mathcal{Q}} \left| z \left( \beta , q \right) \cos 2\pi\left( \phi_t, q \right)  \left( n_{q}, \di^* P^{\phi_\cdot}_{q}(s , \cdot) \right) K(t-s, y-y' ) n_q(y') \right| \\ \leq  C \frac{\Phi_C \left( \frac{t}{\beta}, y \right) }{\left( \frac{s}{\beta} \right)^{1 - \ep} \wedge 1} .
\end{equation}
Integrating the inequality~\eqref{eq:TV13561} in the time interval $[0 , t]$ and using the assumption $t \geq \beta$, we obtain the estimate
\begin{align*}
    \int_0^t \sum_{y' \in \Zd}  \sum_{q \in \mathcal{Q}} \left| z \left( \beta , q \right) \cos 2\pi\left( \phi_t, q \right)  \left( n_{q}, \di^* P^{\phi_\cdot}_{q}(s , \cdot) \right) K(t-s , y-y') n_q(y')\right| & \leq \int_0^t C \frac{\Phi_C \left( \frac{t}{\beta}, y  \right) }{ \left( \frac{s}{\beta} \right)^{1 - \ep} \wedge 1} \, \di s \leq C \left( \frac{t}{\beta} \right)^\ep  \Phi_C \left( \frac{t}{\beta}, y \right). 
\end{align*}
This completes the proof of the estimate of~\eqref{eq:TV16500}.

\medskip

We are now in position to define the function $\mathcal{W}_{\mathrm{der}, 3}$ so that it satisfies the identity~\eqref{eq:TV11180} and the inequalities~\eqref{eq:TV11190}. We use the identity~\eqref{eq:TV15501} and define the map $\mathcal{W}_{\mathrm{der}, 3}$ by the formula
\begin{align} \label{eq:TV15521}
   \mathcal{W}_{\mathrm{der},3}(y , z , \phi) = & - \sum_{q_1 , q_2 \in \mathcal{Q}} z \left( \beta , q_1 \right) z \left( \beta , q_2 \right) \sin 2\pi(\nabla G , n_{q_1}) \sin 2\pi(\nabla G , n_{q_2}) \\ & \qquad \times \int_{0}^\infty   \E_{\phi} \left[  \cos2\pi(\phi_t, q_1) \cos2\pi(\phi_t, q_2)    Y_0(\phi_t)  P^{\phi_\cdot}_{q_1}(t , y) \otimes Q^{\phi_\cdot}_{q_2}(t , z)  \right] dt \notag \\
   & + \frac 12 2\pi \sum_{q_1 , q_2 \in \mathcal{Q}} z \left( \beta , q_1 \right) z \left( \beta , q_2 \right) \sin 2\pi(\nabla G , n_{q_1}) \left( \cos 2\pi(\nabla G , n_{q_2}) - 1 \right) \notag \\ & \qquad \times \int_{0}^\infty   \E_{\phi} \left[  \cos2\pi(\phi_t, q_1) \cos2\pi(\phi_t, q_2)    Y_0(\phi_t)  P^{\phi_\cdot}_{q_1}(t , y) \otimes Q^{\phi_\cdot}_{q_2}(t , z)  \right] dt. \notag
\end{align}
Applying the exterior derivative $\di$ in the $z$-variable and using the identity $ P^{\phi_\cdot}_{q_1} = \di Q^{\phi_\cdot}_{q_2}$, we obtain the identity~\eqref{eq:TV15501}. There only remains to prove the inequalities~\eqref{eq:TV11190}. To this end, we first take the $L^2(\mu_\beta)$-norm on both sides of the identity~\eqref{eq:TV15521} and use the following facts: the absolute value of the cosine is bounded by~$1$, the random variable $Y_0$ belongs to the space $L^2(\mu_\beta)$, the Langevin dynamics is invariant under the Gibbs measure $\mu_\beta$ and the functions $P^{\phi_\cdot}_{q_1}$ and $Q^{\phi_\cdot}_{q_2}$ are bounded from above by the right sides of~\eqref{eq:TV16480} and~\eqref{eq:TV16500}. We obtain
\begin{align} \label{eq:TV16001}
   \left\| \mathcal{W}_{\mathrm{der},3}(y , z , \cdot) \right\|_{L^2 \left( \mu_\beta \right)} \leq &  \sum_{y' , z' \in \Zd} \sum_{q_1 \in \mathcal{Q}_{y'} , q_2 \in \mathcal{Q}_{z'}} \left|z \left( \beta , q_1 \right) z \left( \beta , q_2 \right) \sin 2\pi(\nabla G , n_{q_1}) \sin 2\pi(\nabla G , n_{q_2}) \right| \\ & \qquad \times \int_{0}^\infty \left( \frac{t}{\beta} \right)^{-\frac 12 + 2 \ep} \Phi_C \left(\frac{t}{\beta} , y - y'\right) \Phi_C \left( \frac{t}{\beta} , z - z' \right) dt \notag \\
   & + \frac 12 2\pi \sum_{y' , z' \in \Zd}  \sum_{q_1 \in \mathcal{Q}_{y'} , q_2 \in \mathcal{Q}_{z'}} \left| z \left( \beta , q_1 \right) z \left( \beta , q_2 \right) \sin 2\pi(\nabla G , n_{q_1}) \left( \cos 2\pi(\nabla G , n_{q_2}) - 1 \right) \right| \notag \\ & \qquad \times \int_{0}^\infty \left( \frac{t}{\beta} \right)^{-\frac12 + 2 \ep} \Phi_C \left( \frac{t}{\beta} , y - y' \right) \Phi_C \left( \frac{t}{\beta} , z - z' \right) dt. \notag
\end{align}
We use the estimates, for each point $y' \in \Zd$ and each charge $q \in \mathcal{Q}_{y'}$
\begin{equation*}
        \left|  \sin 2\pi(\nabla G , n_{q}) \right| \leq \frac{C_q}{|y'|^{d-1}} \hspace{5mm} \mbox{and}  \hspace{5mm} \left|  \cos 2\pi(\nabla G , n_{q}) - 1 \right| \leq \frac{C_q}{|y'|^{2d-2}}.
\end{equation*}
Allowing the constants to depend on the inverse temperature $\beta$, we obtain that
\begin{align} \label{eq:TV1709}
     \left\| \mathcal{W}_{\mathrm{der},3}(y , z , \cdot) \right\|_{L^2 \left( \mu_\beta \right)} \leq &  \sum_{y' , z' \in \Zd} \sum_{q_1 \in \mathcal{Q}_{y'} , q_2 \in \mathcal{Q}_{z'}} \frac{e^{-c \sqrt{\beta} \left(\left\| q_1 \right\|_1 + \left\| q_2 \right\|_1 \right) } C_{q_1} C_{q_2} }{|y'|^{d-1} |z'|^{d-1} \left( |y - y'|^{2d-1 - 2\ep} + |z - z'|^{2d-1 - 2\ep} \right)}  \\ & \qquad + \sum_{y' , z' \in \Zd} \sum_{q_1 \in \mathcal{Q}_{y'} , q_2 \in \mathcal{Q}_{z'}} \frac{e^{-c \sqrt{\beta} \left(\left\| q_1 \right\|_1 + \left\| q_2 \right\|_1 \right) } C_{q_1} C_{q_2} }{|y'|^{d-1} |z'|^{2d-2} \left( |y - y'|^{2d - 1-2\ep} + |z - z'|^{2d -1- 2\ep} \right)} \notag \\
      & \leq \sum_{y' , z' \in \Zd} \frac{C}{|y'|^{d-1} |z'|^{d-1} \left( |y - y'|^{2d -1-2 \ep} + |z - z'|^{2d -1-2 \ep} \right)} \notag \\ & \qquad + \sum_{y' , z' \in \Zd} \frac{C}{|y'|^{d-1} |z'|^{2d-2} \left( |y - y'|^{2d -1-2 \ep} + |z - z'|^{2d -1-2 \ep} \right)}, \notag
\end{align}
where in the second inequality we used the exponential decays of the term $e^{-c \sqrt{\beta} \left(\left\| q_1 \right\|_1 + \left\| q_2 \right\|_1 \right) }$ to absorb the algebraic growth of the constants $C_{q_1}$ and $C_{q_2}$. The right hand side of the estimate~\eqref{eq:TV1709} is estimated by noting that $|z'|^{2d-2} \geq |z|^{d-1}$ and by using the inequality $a + b \geq 2 \sqrt{ab}$. We obtain
\begin{align*}
    \left\| \mathcal{W}_{\mathrm{der},3}(y , z , \cdot) \right\|_{L^2 \left( \mu_\beta \right)} & \leq \sum_{y' , z' \in \Zd} \frac{C}{|y'|^{d-1} |z'|^{d-1} \left( |y - y'|^{2d - 1 -2  \ep} + |z - z'|^{2d - 1 -2 \ep} \right)} \\
    & \leq \sum_{y' , z' \in \Zd} \frac{C}{|y'|^{d-1} |z'|^{d-1}  |y - y'|^{d - \frac 12 - \ep} |z - z'|^{d - \frac 12 - \ep}} \\
    & \leq \left( \sum_{y' \in \Zd} \frac{C}{|y'|^{d-1}  |y - y'|^{d -\frac 12 - \ep}} \right) \left( \sum_{ z' \in \Zd} \frac{C}{|z'|^{d-1} |z - z'|^{d - \frac 12 - \ep}} \right) \\ 
    & \leq \frac{C}{|y|^{d-\frac 32-\ep} |z|^{d- \frac 32 - \ep}}.
\end{align*}
This completes the estimate for the map $\mathcal{W}_{\mathrm{der},3}$ stated in~\eqref{eq:TV11190}. To estimate the second estimate~\eqref{eq:TV11190}, involving the codifferential $\di^*_y$ is similar and we only give an outline of the argument. First, by taking the codifferential on both sides of the identity~\eqref{eq:TV15521}, we obtain the formula
\begin{align*}
   \di^*_y \mathcal{W}_{\mathrm{der},3}(y , z , \phi) = & - \sum_{q_1 , q_2 \in \mathcal{Q}} z \left( \beta , q_1 \right) z \left( \beta , q_2 \right) \sin 2\pi(\nabla G , n_{q_1}) \sin 2\pi(\nabla G , n_{q_2}) \\ & \qquad \times \int_{0}^\infty   \E_{\phi} \left[  \cos2\pi(\phi_t, q_1) \cos2\pi(\phi_t, q_2)    Y_0(\phi_t)  \di^*_y P^{\phi_\cdot}_{q_1}(t , y) \otimes Q^{\phi_\cdot}_{q_2}(t , z)  \right] dt \notag \\
   & + \frac 12 2\pi \sum_{q_1 , q_2 \in \mathcal{Q}} z \left( \beta , q_1 \right) z \left( \beta , q_2 \right) \sin 2\pi(\nabla G , n_{q_1}) \left( \cos 2\pi(\nabla G , n_{q_2}) - 1 \right) \notag \\ & \qquad \times \int_{0}^\infty   \E_{\phi} \left[  \cos2\pi(\phi_t, q_1) \cos2\pi(\phi_t, q_2)    Y_0(\phi_t)  \di^*_y P^{\phi_\cdot}_{q_1}(t , y) \otimes Q^{\phi_\cdot}_{q_2}(t , z)  \right] dt. \notag
\end{align*}
One can then rewrite the same argument, and use the second estimate of~\eqref{eq:TV16480} on the map $\nabla P^{\phi_\cdot}_{q_1}$ (which provides an upper bound for the map $\di^* P^{\phi_\cdot}_{q_1}$ since the codifferential is a linear functional of the gradient) instead on the first estimate of~\eqref{eq:TV16480} on the map $P^{\phi_\cdot}_{q_1}$ to obtain the result. The estimate of the term $\mathcal{W}_{\mathrm{der},3}$ stated in~\eqref{eq:TV11190} is complete. We conclude the estimate of the term $\mathcal{V}_{\mathrm{der},3}$ by multiplying the term by the value $n_{Q_x}(y,\phi)$. We obtain
\begin{equation*}
    n_{Q_x}(y,\phi) \di^*_y \mathcal{V}_{\mathrm{der},3} (y , z , \phi)  =  n_{Q_x}(y,\phi) \di_z \di^*_y \mathcal{W}_{\mathrm{der},3} (y , z , \phi)  = \di_z \left( n_{Q_x}(y,\phi) \di^*_y \mathcal{W}_{\mathrm{der},3} (y , z , \phi)\right),
\end{equation*}
where in the second equality, we used that since the exterior derivative $\di_z$ only acts on the $z$-variable, it commutes with the codifferential $\di^*_y$ (which only acts on the $y$-variable). Applying the pointwise bound~\eqref{eq:TV09133} shows the inequality
\begin{equation*}
     \left\| n_{Q_x}(y,\cdot) \di^*_y \mathcal{W}_{\mathrm{der},3} (y , z , \cdot) \right\|_{L^2 \left( \mu_\beta \right)} \leq \left\| n_{Q_x}(y,\cdot) \right\|_{L^\infty \left( \mu_\beta \right)} \left\|  \di^*_y \mathcal{W}_{\mathrm{der},3} (y , z , \cdot) \right\|_{L^2 \left( \mu_\beta \right)}  \leq \frac{C}{|y - x|^{d-1} |y|^{d-1-\ep}|z|^{d-1-\ep}}.
\end{equation*}

\medskip

\textit{Conclusion of Substep 1.1.} Collecting the results proved in this step, we have obtained the identity, for each pair of points $(y , z) \in \Zd$,
\begin{align} \label{eq:TV20011}
    \partial_z \left( n_{Q_x}(y,\cdot) \di^* \mathcal{V}(y , \cdot) \right) & = \left( \partial_z  n_{Q_x}(y,\cdot) \right) \di^* \mathcal{V}(y,\cdot) +  n_{Q_x}(y,\cdot) \partial_z \di^* \mathcal{V}(y,\cdot) \\
    & = \left( \partial_z  n_{Q_x}(y,\cdot) \right) \di^* \mathcal{V}(y,\cdot)+  n_{Q_x}(y,\cdot) \di^*_y \mathcal{V}_{\mathrm{der},1}(y , z, \cdot) \notag \\ & \qquad  + n_{Q_x}(y,\cdot)  \di^*_y \mathcal{V}_{\mathrm{der},2}(y , z, \cdot) + \di_z \left(n_{Q_x}(y,\cdot) \di^*_y  \mathcal{W}_{\mathrm{der},3}(y , z, \cdot) \right), \notag
\end{align}
with the estimates
\begin{equation} \label{eq:TV20371}
    \left\{ \begin{aligned}
    \left\| \left( \partial_z  n_{Q_x}(y,\cdot) \right) \di^* \mathcal{V}(y , \cdot) \right\|_{L^2 \left( \mu_\beta \right)} &\leq \frac{C  e^{-c \sqrt{\beta} |y - z|}}{|y - x|^{d-1} \times |y|^{d-1- \ep}}, \\
    \left\| n_{Q_x}(y,\cdot) \di^*_y \mathcal{V}_{\mathrm{der},1}(y , z, \cdot) \right\|_{L^2 \left( \mu_\beta \right)} &\leq \frac{C}{|x-y|^{d-1-\ep}\times |z-y|^{d +1 -\ep} \times \max \left( \left| y \right|, \left| z \right|\right)^{d-1-\ep}}, \\
     \left\| n_{Q_x}(y,\cdot)  \di^*_y \mathcal{V}_{\mathrm{der},2}(y , z, \cdot) \right\|_{L^2 \left( \mu_\beta \right)} &\leq \frac{C}{|x-y|^{d-1-\ep}\times |z-y|^{d +1 -\ep} \times \max \left( \left| y \right|, \left| z \right|\right)^{d-1-\ep}},\\
    \left\| n_{Q_x}(y,\cdot) \di^*_y \mathcal{W}_{\mathrm{der},3} (y , z , \cdot) \right\|_{L^2 \left( \mu_\beta \right)} & \leq \frac{C}{|y - x|^{d-1} |y|^{d-1-\ep}|z|^{d-1-\ep}}.
    \end{aligned} 
    \right.
\end{equation}

\medskip

\textit{Substep 1.2.} We prove the covariance estimate~\eqref{eq:TV20174} rewritten below:
\begin{equation} \label{eq:TV20174bis}
    \sum_{y \in \Zd} \cov \left[ X_x ,Q_x(y,\cdot)  \mathcal{V}(y , \cdot) \right] = O \left( \frac{C}{|x|^{d-1-\ep}} \right).
\end{equation}
By the Helffer-Sj{\"o}strand representation formula we have, for each point $y \in \Zd$,
\begin{equation} \label{eq:TV19031bis}
    \cov \left[ X_x ,Q_x(y,\cdot)  \mathcal{V}(y , \cdot) \right] = \sum_{z \in \Zd} \left\langle \mathcal{X}_x(z , \cdot) \partial_z \left(Q_x(y,\cdot)  \mathcal{V}(y , \cdot) \right) \right\rangle_{\mu_\beta},
\end{equation}
where $\mathcal{X}_x$ is the solution of the Helffer-Sj{\"o}strand equation, for each pair $(z , \phi) \in \Zd \times \Omega$,
\begin{equation*}
    \mathcal{L} \mathcal{X}_x(z , \phi) = \partial_z X_x \left( \phi \right).
\end{equation*}
We recall the upper bounds~\eqref{eq:TV08187} and~\eqref{eq:TV08188456} on the $L^2(\mu_\beta)$-norm of the map $\X_x$ and the exterior derivative $\di^* \X_x$: for each point $z \in \Zd$,
\begin{equation} \label{eq:TV19231}
    \left\| \mathcal{X}_x(z , \cdot)  \right\|_{L^2 \left( \mu_\beta \right)} \leq \frac{C}{|z - x|^{d-2-\ep}}\hspace{5mm} \mbox{and} \hspace{5mm} \left\| \di^*\mathcal{X}_x(z , \cdot)  \right\|_{L^2 \left( \mu_\beta \right)} \leq \frac{C}{|z - x|^{d-1-\ep}}.
\end{equation}
Using the formula~\eqref{eq:TV19031bis}, we can rewrite the expansion~\eqref{eq:TV20174bis}
\begin{align} \label{eq:TV19581}
    \sum_{y \in \Zd} \cov \left[ X_x ,Q_x(y,\cdot)  \mathcal{V}(y , \cdot) \right] & = \sum_{y , z \in \Zd}  \left\langle \mathcal{X}_x(z , \cdot) \partial_z \left(Q_x(y,\cdot)  \mathcal{V}(y , \cdot) \right) \right\rangle_{\mu_\beta} \\ &= \sum_{y , z  \in \Zd}  \left\langle \mathcal{X}_x(z , \cdot) \partial_z \left( n_{Q_x}(y,\cdot)  \di^* \mathcal{V}(y , \cdot) \right) \right\rangle_{\mu_\beta}. \notag
\end{align}
We combine the identities ~\eqref{eq:TV20011} and~\eqref{eq:TV19581} and write
\begin{align*}
    \lefteqn{\sum_{y \in \Zd} \cov \left[ X_x ,Q_x(y,\cdot)  \mathcal{V}(y , \cdot) \right] } \qquad & \\ & = \sum_{y , z  \in \Zd}  \left\langle \mathcal{X}_x(z , \cdot) \left( \partial_z  n_{Q_x}(y) \right) \di^* \mathcal{V}(y , \cdot) \right\rangle_{\mu_\beta} + \sum_{y , z  \in \Zd}  \left\langle \mathcal{X}_x(z , \cdot)n_{Q_x}(y) \di^*_y \mathcal{V}_{\mathrm{der},1}(y , z, \cdot) \right\rangle_{\mu_\beta} \\ & \quad + \sum_{y , z  \in \Zd}  \left\langle \mathcal{X}_x(z , \cdot)  n_{Q_x}(y)  \di^*_y \mathcal{V}_{\mathrm{der},2}(y , z, \cdot) \right\rangle_{\mu_\beta} + \sum_{y , z  \in \Zd}  \left\langle \di^* \mathcal{X}_x(z , \cdot)  \left(n_{Q_x}(y) \di^*_y  \mathcal{W}_{\mathrm{der},3}(y , z ,\cdot) \right) \right\rangle_{\mu_\beta}.
\end{align*}
We use the estimates~\eqref{eq:TV19231} on the function $\mathcal{X}_x$ and the estimates~\eqref{eq:TV20371}. We obtain
\begin{align} \label{eq:TV20471}
     \sum_{y \in \Zd} \cov \left[ X_x ,Q_x(y,\cdot)  \mathcal{V}(y , \cdot) \right] & \leq \sum_{y , z  \in \Zd} \frac{C}{|z - x|^{d-2-\ep}} \frac{  e^{-c \sqrt{\beta} |y - z|}}{|y - x|^{d-1} \times |y|^{d-1- \ep}} \\
     & \quad + \sum_{y , z  \in \Zd} \frac{C}{|z - x|^{d-2-\ep}} \frac{1}{|x-y|^{d-1-\ep}\times |z-y|^{d +1 -\ep} \times \max \left( \left| y \right|, \left| z \right|\right)^{d-1-\ep}} \notag \\
     & \quad + \sum_{y , z  \in \Zd} \frac{C}{|z - x|^{d-1-\ep}} \frac{1}{|y - x|^{d-1} |y|^{d-1-\ep}|z|^{d-1-\ep}}. \notag
\end{align}
We estimate the three terms in the right side of~\eqref{eq:TV20471} separately. For the first term, we have
\begin{equation} \label{eq:TV11233}
      \sum_{y , z  \in \Zd} \frac{C}{|z - x|^{d-2-\ep}} \frac{C  e^{-c \sqrt{\beta} |y - z|}}{|y - x|^{d-1} \times |y|^{d-1- \ep}} \leq \sum_{y \in \Zd} \frac{C}{|y - x|^{2d-3- \ep} \times |y|^{d-1- \ep}} \leq \frac{C}{|x|^{d-1-2\ep}}.
\end{equation}
For the second term, we use the inequality $\max \left( \left| y \right|, \left| z \right|\right) \geq |y|$ and write
\begin{align} \label{eq:TV11223}
   \lefteqn{\sum_{y , z  \in \Zd} \frac{C}{|z - x|^{d-2-\ep}|x-y|^{d-1-\ep}\times |z-y|^{d +1 -\ep} \times \max \left( \left| y \right|, \left| z \right|\right)^{d-1-\ep}}} \qquad & \\ & \leq \sum_{y \in \Zd} \left[\frac{C}{|x-y|^{d-1-\ep}\left| y \right|^{d-1-\ep}} \sum_{ z  \in \Zd} \frac{1}{|z - x|^{d-2-\ep}|z-y|^{d +1 -\ep}} \right]\notag \\
   & \leq \sum_{y \in \Zd} \frac{C}{|x-y|^{d-1-\ep}\left| y \right|^{d-1-\ep}|x-y|^{d-2-\ep}} \notag \\
   & \leq \frac{C}{|x|^{d-1-3\ep}}. \notag
\end{align}
For the third term, we have
\begin{align} \label{eq:TV11213}
    \sum_{y , z  \in \Zd} \frac{C}{|z - x|^{d-1-\ep}} \frac{C}{|y - x|^{d-1} |y|^{d-1-\ep}|z|^{d-1-\ep}} & = \left( \sum_{y  \in \Zd} \frac{C}{|y - x|^{d-1} |y|^{d-1-\ep}} \right) \left( \sum_{ z  \in \Zd} \frac{C}{|z - x|^{d-1-\ep}|z|^{d-1-\ep}} \right) \\
    & \leq \frac{C}{|x|^{d-2-\ep}} \frac{C}{|x|^{d-2-\ep}} \notag \\
    & \leq \frac{C}{|x|^{d-1-2\ep}}, \notag
\end{align}
where we used the inequality $2d-4 \geq d - 1$, valid in dimension larger than $3$. Combining the inequalities~\eqref{eq:TV20471},~\eqref{eq:TV11233},~\eqref{eq:TV11223} and~\eqref{eq:TV11213} completes the proof of Step 1.

\medskip

\textit{Step 2.} To complete the proof of Lemma~\ref{p.decoupleexp}, there remains to prove the expansion~\eqref{eq:TV11153} which is restated below
\begin{equation} \label{eq:TV11383}
    \sum_{y \in \Zd} \left\langle Q_x(y,\cdot) \mathcal{V}(y , \cdot ) \right\rangle_{\mu_\beta} = \left\langle Y_0 \right\rangle_{\mu_\beta} \sum_{y \in \Zd} \left\langle Q_x(y,\cdot) \mathcal{U}(y , \cdot ) \right\rangle_{\mu_\beta} + O \left( \frac{C}{|x|^{d-1-\ep}} \right),
\end{equation}
where the functions $\mathcal{V}$ and $\mathcal{U}$ are respectively defined in~\eqref{eq:VER12128} and~\eqref{eq:TV10003V}. The strategy of the proof relies on the symmetry of the Helffer-Sj{\"o}strand operator $\mathcal{L}$; if we let $\mathcal{U}_x$ the solution of the equation $\mathcal{L} \mathcal{U}_x = Q_x$ in $\Zd \times \Omega$, then we have the identities
\begin{equation*}
    \sum_{y \in \Zd} \left\langle Q_x(y, \cdot) \mathcal{V}(y , \cdot ) \right\rangle_{\mu_\beta} = \sum_{y \in \Zd} \left\langle \mathcal{U}_x \left( y , \cdot \right) Q_0(y,\cdot) Y_0 \right\rangle_{\mu_\beta}
\end{equation*}
and
\begin{equation*}
    \sum_{y \in \Zd} \left\langle Q_x(y,\cdot) \mathcal{U}(y , \cdot ) \right\rangle_{\mu_\beta} = \sum_{y \in \Zd} \left\langle \mathcal{U}_x(y ,\cdot) Q_0(y,\cdot) \right\rangle_{\mu_\beta}.
\end{equation*}
Using these identities, we see that the expansion~\eqref{eq:TV11383} is equivalent to
\begin{equation*}
    \sum_{y \in \Zd} \left\langle \mathcal{U}_x \left( y , \cdot \right) Q_0(y,\cdot) Y_0 \right\rangle_{\mu_\beta}= \left\langle Y_0 \right\rangle_{\mu_\beta} \sum_{y \in \Zd} \left\langle \mathcal{U}_x \left( y , \cdot \right) Q_0(y,\cdot) \right\rangle_{\mu_\beta} + O \left( \frac{C}{|x|^{d-1-\ep}} \right).
\end{equation*}
The proof of this result is similar to the proof written in Step 1, and in fact simpler since we do not have to treat the term $\mathcal{V}_{\mathrm{der},3}$ in~\eqref{eq:TV11470}; we omit the details.
\end{proof}

\section{Using the symmetry and rotation invariance of the dual Villain model} \label{sec:chap8.4}

This section is devoted of some properties of the discrete convolution of the discrete Green's function on the lattice $\Zd$. We recall the definition of the group $H$ of the lattice-preserving maps introduced in Section~\ref{SecNotandprelim} of Chapter~\ref{Chap:chap2}.

\begin{lemma} \label{lem:lemma4.1chap8}
Fix four integers $j , j_1 , k , k_1 \in \left\{ 1 , \ldots, d \right\}$ and let $F: \Zd \to \R$ be the function
\begin{equation*}
    F_{j,k,j_1,k_1}(x) := \sum_{y , \kappa \in \Zd} \nabla_j G(y) \nabla_{k} G(x - y - \kappa) \nabla_{j_1} \nabla_{k_1} G(\kappa).
\end{equation*}
Then there exists a $(2-d)$-homogeneous map $J_{j , k , j_1 , k_1} : \Rd \setminus \{ 0 \} \to \R$ such that for any exponent $\ep > 0$, there exists a constant $C := C(d, \ep) <\infty$ satisfying
\begin{equation*}
     F_{j,k,j_1,k_1}(x) = J_{j , k , j_1 , k_1}(x) + O \left( \frac{C}{|x|^{d-1 - \ep}} \right).
\end{equation*}
The function $J_{i , j , i_1 , j_1}$ is given by the formula
\begin{equation*}
    J_{j , k , j_1 , k_1} (x) =  \int_{0}^\infty t^{-\frac{d+4}6} K_{j,k,j_1,k_1}\left(\frac{x }{t^{\frac 16}}\right) \, dt,
\end{equation*}
where $K : \Rd \to \R$ is defined as the inverse Fourier transform
\begin{equation*}
    K_{j,k,j_1,k_1}(x) = \frac{1}{\left( 2\pi\right)^d}\int_{\Rd} \xi_j \xi_k \xi_{j_1} \xi_{k_1} e^{- \left| \xi \right|^6 } e^{i x \cdot \xi} \, \di \xi.
\end{equation*}
\end{lemma}

\begin{remark} \label{rem:rem4.2chap8}
Fix two integers $j , j_1 \in \left\{1 , \ldots, d \right\}$ and let $F_{j,j_1} : \Zd \to \R$ be the map
\begin{equation*}
    F_{j,j_1}(x) =  \sum_{y \in \Zd} \nabla_j G(y) \nabla_{j_1} G(x - y).
\end{equation*}
An adaptation of the proof of Lemma~\ref{lem:lemma4.1chap8} shows the identity
\begin{equation*}
    F_{j,j_1}(x) = \int_{0}^\infty t^{-\frac{d+2}4} K_{j,j_1}\left(\frac{x }{t^{\frac 14}}\right) \, dt + O \left( \frac{C}{|x|^{d-1 - \ep}} \right),
\end{equation*}
where the map $K_{j , j_1}$ is defined as the inverse Fourier transform
\begin{equation*}
    K_{j,j_1}(x) =  \frac{1}{\left( 2 \pi\right)^d}\int_{\Rd} \xi_j  \xi_{j_1} e^{- \left| \xi \right|^4 } e^{i x \cdot \xi} \, \di \xi.
\end{equation*}
\end{remark}

\begin{proof}
The proof uses an explicit formula for the Green's function and relies on the discrete Fourier transform. Let us try to construct the Green's function $G$ by Fourier transform. Given a function $F: \Zd \to \R$ which decays sufficiently fast at infinity, we define its discrete Fourier transform according to the formula, for each $\xi = \left( \xi_1 , \ldots, \xi_d \right) \in \Rd$,
\begin{equation*}
    \hat F (\xi) = \sum_{z \in \Zd} F(z) e^{-i\xi \cdot z}.
\end{equation*}
Then we have $\widehat{\Delta F} (\xi) = 2 \sum_{l=1}^d \left( \cos \xi_l - 1 \right) \hat F(\xi)$. Applying this formula to the Green's function $G$, we formally obtain that $\hat G = \left(2 \sum_{l=1}^d \left( \cos \xi_l - 1 \right)\right)^{-1}$. Applying the inverse Fourier transform, we formally obtain the formula, for each $z \in \Zd$,
\begin{equation} \label{eq:TV13013}
    G (z) = \frac{1}{\left( 2\pi\right)^d}\int_{[-\pi,\pi]^d} \frac{e^{i\xi \cdot z}}{2 \sum_{l=1}^d \left( 1-\cos \xi_l \right)} \di \xi.
\end{equation}
One can then prove the formula~\eqref{eq:TV13013} rigorously. First, it is clear that the integral~\eqref{eq:TV13013} converges absolutely in dimension larger than $3$ and that it is bounded as a function of the variable $z$; one can then verify by an explicit computation that the discrete Laplacian in the variable $z$ of the integral~\eqref{eq:TV13013} is equal to the Dirac $\delta_0$. This argument provides a rigorous justification of the identity~\eqref{eq:TV13013}.

Once the formula~\eqref{eq:TV13013} is established, we obtain the identity for the gradient of the function $G$: for each integer $j \in \left\{ 1 , \ldots,d \right\}$,
\begin{equation*}
    \nabla_{j} G (z) = \frac{1}{\left( 2\pi\right)^d}\int_{[-\pi,\pi]^d} \frac{\left( e^{i \xi_j} - 1\right)e^{i\xi \cdot z}}{2 \sum_{l=1}^d \left( 1-\cos \xi_l  \right)} \di \xi.
\end{equation*}
Using that the discrete Fourier transform turns discrete convolution into products, we obtain the formula
\begin{equation} \label{eq:TV14531201}
    F_{j,k,j_1,k_1}(x) = \frac{1}{\left( 2\pi\right)^d}\int_{[-\pi, \pi]^d} \frac{\left( e^{i \xi_j} - 1\right)\left( e^{i \xi_k} - 1\right) \left( e^{i \xi_{j_1}} - 1\right) \left( e^{i \xi_{k_1}} - 1\right) e^{i\xi \cdot x}}{\left(2 \sum_{l=1}^d \left( 1-\cos \xi_l  \right)\right)^3} \di \xi.
\end{equation}
We use the identity, for any $\xi \in [-\pi , \pi]^d \setminus \{ 0 \}$,
\begin{equation*}
    \frac{1}{\left(2 \sum_{l=1}^d \left( \cos \xi_l - 1 \right)\right)^3} = \int_0^\infty e^{- t \left(2 \sum_{l=1}^d \left( 1-\cos \xi_l \right)\right)^3 } \di t
\end{equation*}
and apply Fubini's theorem to rewrite the identity~\eqref{eq:TV14531201},
\begin{equation*}
    F_{j,k,j_1,k_1}(x) = \frac{1}{\left( 2\pi\right)^d} \int_0^\infty \int_{[-\pi, \pi]^d} \left( e^{i \xi_j} - 1\right)\left( e^{i \xi_k} - 1\right) \left( e^{i \xi_{j_1}} - 1\right) \left( e^{i \xi_{k_1}} - 1\right) e^{i\xi \cdot x} e^{- t \left(2 \sum_{l=1}^d \left(1- \cos \xi_l \right)\right)^3 } \di \xi \, \di t.
\end{equation*}
To ease the notation, we denote by
\begin{equation} \label{eq:TV09291601}
    u(t , x) := \int_{[-\pi, \pi]^d} \left( e^{i \xi_j} - 1\right)\left( e^{i \xi_k} - 1\right) \left( e^{i \xi_{j_1}} - 1\right) \left( e^{i \xi_{k_1}} - 1\right) e^{i\xi \cdot x} e^{- t \left(2 \sum_{l=1}^d \left( 1-\cos \xi_l \right)\right)^3 } \di \xi.
\end{equation}
We prove the two following properties on the map $u$:
\begin{enumerate}
    \item For any exponent $\alpha > 0$, each integer $m \in \N$, there exists a constant $C_{m, \alpha} := C_{m , \alpha}(m , \alpha , d) < \infty$ such that for any $\lambda  \geq 1$, any time $t \leq 1$ and any point $x \in \Rd$ such that $|x| = 1$ and $\lambda x \in \Zd$,
    \begin{equation} \label{eq:TV09351901}
        \left| u(\lambda^{6 - \alpha} t , \lambda x) \right| \leq C_{m , \alpha} \lambda^{-m};
    \end{equation}
    \item For any time $t \in (0 , \infty)$, any point $x \in \Rd$ and any $\lambda \in (1 , \infty)$ such that $\sqrt[6]{t} \lambda \geq 1$,
\begin{equation} \label{eq:TV15411201}
    \lambda^{d+4} u( \lambda^{6} t , \lambda x) =  \int_{\Rd} \xi_j\xi_k\xi_{j_1}\xi_{k_1} e^{i\xi \cdot x} e^{- t \left| \xi \right|^6 } \di \xi + O \left( \frac{C}{\lambda t^{\frac{d+5}{6}}} \right).
\end{equation}
\end{enumerate} 
We first prove~\eqref{eq:TV09351901}. Fix a point $x \in \Rd$ such that $|x| = 1$ and denote by
\begin{equation} \label{eq:TV09571901}
    P(t , x) := \int_{[-\pi, \pi]^d} e^{i\xi \cdot x} e^{- t \left(2 \sum_{l=1}^d \left( 1-\cos \xi_l \right)\right)^3 } \di \xi.
\end{equation}
Note that 
\begin{equation*}
    u(t , x) = \nabla_{j} \nabla_{j_1} \nabla_{k} \nabla_{k_1} P(t , x).
\end{equation*}
Using that the discrete gradients are bounded operators, one sees that to prove~\eqref{eq:TV09351901} it is sufficient to prove the inequality
\begin{equation} \label{eq:TV09561901}
    \left| P(\lambda^{6 - \alpha} t , \lambda x) \right| \leq C_{k , \alpha} \lambda^{-k}.
\end{equation}
Consider the identity~\eqref{eq:TV09571901} and perform the change of variables $\xi := \lambda \xi$
\begin{equation} \label{eq:TV14501901}
    \lambda^{d} P( \lambda^{6-\alpha} t , \lambda x) = \int_{[-\lambda  \pi,  \lambda  \pi]^d}  e^{i\xi \cdot x} e^{- t \lambda^{6-\alpha} \left(2 \sum_{l=1}^d \left( 1-\cos \frac{\xi_l}{\lambda}  \right)\right)^3 } \di \xi.
\end{equation}
We write $x = \left( x_1 , \ldots, x_d \right)$ to denote the components of the vector $x$ and assume without loss of generality that $\left| x_1 \right| \geq \frac{1}{d}$ (since we have assumed $\left| x \right| = 1$). We let $g : \Rd \to \R$ and $g_{\lambda, t} : \Rd \to \R$ be the mappings defined by the formulas
\begin{equation*}
    g(\xi) := \left(2 \sum_{l=1}^d \left( 1-\cos \xi_l \right)\right)^3 \hspace{5mm} \mbox{and} \hspace{5mm}  g_{\lambda, t}(\xi) := t \lambda^{6-\alpha} \left(2 \sum_{l=1}^d \left( 1-\cos \frac{\xi_l}{\lambda} \right)\right)^3.
\end{equation*}
Note that the function $g$ is analytic and that the first order term of its Taylor expansion around the point $0$ is given by $g(\xi) = \left| \xi \right|^6 + O \left( C  \left| \xi \right|^7 \right)$. From these observations, one deduces the upper bound: for any integer $n \in \N$ there exists a constant $C_n := C_n (d , n) < \infty$ such that for any point $\xi \in [- \pi , \pi]^d$,
\begin{equation} \label{eq:TV14381901}
     \left| \partial_{1}^n g(\xi) \right| \leq C_n \left| \xi \right|^{\max \left( 0 , 6 - n \right) },
\end{equation}
where the symbol $\partial_{1}^n$ denotes the partial derivative with respect to the variable $\xi_1$ iterated $n$-times.
A consequence of the estimate~\eqref{eq:TV14381901} is the following inequality: for any integer $n \in \N_*$ and any $\xi \in [-\lambda \pi, \lambda \pi]^d$,
\begin{equation} \label{eq:TV14531901}
    \left| \partial_{1}^n g_{\lambda, t}(\xi) \right| \leq C_n \frac{t \lambda^{6-\alpha}}{\lambda^n} \left( \frac{\left| \xi \right|}{\lambda} \right)^{\max \left( 0 , 6 - n \right) }.
\end{equation}
For later use, we rewrite the inequality~\eqref{eq:TV14531901} in the following form
\begin{align} \label{eq:TV15181901}
    \left| \partial_{1}^n g_{\lambda, t}(\xi) \right| & \leq C_n \frac{\left( t \lambda^{-\alpha} \right)^{\min \left( 1 , \frac n6 \right) }}{\lambda^{\max \left( 0 , n - 6 \right)}} \left(  t^{\frac{1}{6}} \lambda^{-\frac{\alpha}{6}} \left| \xi \right| \right)^{\max \left( 0 , 6-n \right)} \\
    & \leq C_n \frac{1}{\lambda^{ \alpha \min\left( 1 , \frac{n}{6} \right) + \max \left( 0 , n - 6 \right)}} \left(  t^{\frac{1}{6}} \lambda^{-\frac{\alpha}{6}} \left| \xi \right| \right)^{\max \left( 0 , 6-n \right)} \notag \\
    & \leq C_n\lambda^{-\frac{\alpha n }{6}} \left(  t^{\frac{1}{6}} \lambda^{-\frac{\alpha}{6}} \left| \xi \right| \right)^{\max \left( 0 , 6-n \right)}. \notag
\end{align}
where in the second line we have used the inequality $t \leq 1$ and $\alpha \min\left( 1 , \frac{n}{6} \right) + \max \left( 0 , n - 6 \right) \geq \frac{\alpha n}{6}$ (since $\alpha \leq 1$) in the third line.
Using that the two maps $\xi \to e^{i x \cdot \xi}$ and $e^{- g_{\lambda, t} }$ are $2 \lambda \pi$-periodic (by the assumption $\lambda x \in \Zd$), we perform $k$ consecutive integration by parts with respect to the variable $\xi_1$ in the right side of~\eqref{eq:TV14501901}. We obtain
\begin{equation} \label{eq:TV17351901}
     \lambda^{d} P( \lambda^{6-\alpha} t , \lambda x) = \int_{[-\lambda  \pi,  \lambda  \pi]^d} \left( \frac{-1}{i x_1}\right)^k e^{i\xi \cdot x}  \partial_{1}^k e^{- g_{\lambda, t} } \left(\xi \right) \di \xi.
\end{equation}
Applying Fa\`a di Bruno's formula, we write
\begin{equation} \label{eq:TV11131901}
    \partial_{1}^k e^{- g_{\lambda , t} } \left(\xi \right) = \sum \frac{k!}{m_1!\,1!^{m_1}\,m_2!\,2!^{m_2}\,\cdots\,m_k!\,k!^{m_k}}\cdot e^{-g_{\lambda, t}(\xi)}\cdot \prod_{j=1}^k\left( - g^{(j)}_{\lambda,t}(\xi)\right)^{m_j},
\end{equation}
where the sum runs over over all $n$-tuples of nonnegative integers $(m_1, ..., m_k)$ satisfying the constraint
\begin{equation}\label{eq:TV16411901}
 m_1 + 2 m_2 + \ldots + k m_k = k.
\end{equation}
We combine the identity~\eqref{eq:TV11131901} with the estimate~\eqref{eq:TV15181901} and obtain
\begin{align} \label{eq:TV16401901}
    \left| \partial_{1}^k e^{- g } \left(\xi \right) \right| & \leq \sum \frac{k!}{m_1!\,1!^{m_1}\,m_2!\,2!^{m_2}\,\cdots\,m_k!\,k!^{m_k}}\cdot e^{-g_{\lambda, t}(\xi)} \cdot \prod_{j=1}^k \left| g^{(j)}(\xi)\right|^{m_j} \\
    & \leq \sum \frac{k!}{m_1!\,1!^{m_1}\,m_2!\,2!^{m_2}\,\cdots\,m_k!\,k!^{m_k}}\cdot e^{-g_{\lambda, t}(\xi)} \cdot \prod_{j=1}^k \left( C_{j} \lambda^{-\frac{\alpha j }{6}} \left(  t^{\frac{1}{6}} \lambda^{-\frac{\alpha}{6}} \left| \xi \right| \right)^{\max \left( 0 , 6- j \right)} \right)^{m_j} \notag \\
    & \leq \sum \frac{k! \prod_{j = 1}^k C_j^{m_j}} {m_1!\,1!^{m_1}\,m_2!\,2!^{m_2}\,\cdots\,m_k!\,k!^{m_k}}\cdot e^{-g_{\lambda, t}(\xi)} \lambda^{- \frac{\alpha}{6} \sum_{j = 1}^k j m_j}  \left(  t^{\frac{1}{6}} \lambda^{-\frac{\alpha}{6}} \left| \xi \right| \right)^{\sum_{j= 1}^k m_j \max \left( 0 , 6- j \right)} \notag 
\end{align}
To treat the right side of the inequality~\eqref{eq:TV16401901}, we use the identity~\eqref{eq:TV16411901} and note that since all the constants which appear in this term depend only on the integer $k$ and the dimension $d$, we may replace them by the notation $C_k$. We obtain
\begin{equation} \label{eq:TV16491901}
    \left| \partial_{1}^k e^{- g_{\lambda, t} } \left(\xi \right) \right| \leq C_k \lambda^{- \frac{\alpha k }{6} } \cdot e^{-g_{\lambda, t}(\xi)} \sum  \left(  t^{\frac{1}{6}} \lambda^{-\frac{\alpha}{6}} \left| \xi \right| \right)^{\sum_{j= 1}^k m_j \max \left( 0 , 6- j \right)}.
\end{equation}
By the identity~\eqref{eq:TV16411901}, we have
\begin{equation} \label{eq:TV16481901}
    \sum_{j= 1}^k m_j \max \left( 0 , 6- j \right) \leq 6 \sum_{j= 1}^k m_j \leq 6 \sum_{j= 1}^k j m_j \leq 6 k.
\end{equation}
Combining the inequality~\eqref{eq:TV16481901} with the identity $r^\kappa \leq 1 + r^{6k}$, valid for any $r \in [0 , \infty)$ and any $\kappa \in \left\{ 0 , \ldots , 6 k \right\}$, we can write
\begin{equation} \label{eq:TV16541901}
     \left(  t^{\frac{1}{6}} \lambda^{-\frac{\alpha}{6}} \left| \xi \right| \right)^{\sum_{j= 1}^k m_j \max \left( 0 , 6- j \right)} \leq 1 + \left(  t^{\frac{1}{6}} \lambda^{-\frac{\alpha}{6}} \left| \xi \right| \right)^{6k}.
\end{equation}
Using the estimate~\eqref{eq:TV16541901} and the fact that the sum in the right side of~\eqref{eq:TV16491901} contains at most $C_k$ terms, we deduce that
\begin{equation} \label{eq:TV17181901}
    \left| \partial_{1}^k e^{- g_{\lambda, t} } \left(\xi \right) \right| \leq  C_k \lambda^{- \frac{\alpha k }{6} } \cdot e^{-g_{\lambda, t}(\xi)} \left( 1 +  \left( t^{\frac{1}{6}} \lambda^{-\frac{\alpha}{6}} \left| \xi \right|  \right)^{6k} \right).
\end{equation}
We then use that there exists a universal constant $C$ such that, for any $r \in [-\pi , \pi]^d$, $1 - \cos r \leq \frac{r^2}{C}$ to obtain the following estimate on the map $e^{-g_{\lambda, t}}$: for any $\xi \in [- \lambda \pi , \lambda \pi]^d$,
\begin{equation*}
    e^{-g_{\lambda, t}(\xi)} \leq e^{- \frac{t \lambda^{-\alpha} \left| \xi \right|^6}{C}}.
\end{equation*}
Consequently, there exists a constant $C_k$ which depends only on the integer $k$ such that, for any $\xi \in [- \lambda \pi , \lambda \pi]^d$,
\begin{equation} \label{eq:TV17171901}
     \left( 1 +  \left( t^{\frac{1}{6}} \lambda^{-\frac{\alpha}{6}} \left| \xi \right|  \right)^{6k} \right) e^{-g_{\lambda, t}(\xi)} \leq \left( 1 +  \left( t^{\frac{1}{6}} \lambda^{-\frac{\alpha}{6}} \left| \xi \right|  \right)^{6k} \right) e^{-\frac{t \lambda^{-\alpha} \left| \xi \right|^6}{C}} \leq C_k e^{-\frac{t \lambda^{-\alpha} \left| \xi \right|^6}{2C}}.
\end{equation}
Combining~\eqref{eq:TV17181901} with~\eqref{eq:TV17171901}, we have obtained
\begin{equation} \label{eq:TV17231901}
    \left| \partial_{1}^k e^{- g_{\lambda, t} } \left(\xi \right) \right| \leq C_k \lambda^{- \frac{\alpha k }{6} }  e^{-\frac{t \lambda^{-\alpha} \left| \xi \right|^6}{2C}} \leq C_k \lambda^{- \frac{\alpha k }{6}}.
\end{equation}
We now complete the proof of~\eqref{eq:TV09561901} using~\eqref{eq:TV17231901}. We let $m$ be the integer which appears in~\eqref{eq:TV09561901} and $k$ be the smallest integer such that $\frac{\alpha k}{6} \geq m$. We consider the identity~\eqref{eq:TV17351901}, use the estimate~\eqref{eq:TV17231901} and the assumption $\left| x_1 \right| \geq \frac 1d$. We obtain
\begin{equation*}
    \left| P( \lambda^{6-\alpha} t , \lambda x) \right| \leq \lambda^{-d} \int_{[-\lambda  \pi,  \lambda  \pi]^d} d^k C_k \lambda^{- \frac{\alpha k }{6}} \leq C_k \lambda^{- \frac{\alpha k }{6}} \leq C_k \lambda^{- m},
\end{equation*}
since the integer $k$ is chosen such that it depends only on $m$ and $\alpha$, we have obtained~\eqref{eq:TV09561901}. The proof of~\eqref{eq:TV09351901} is complete.

It remains to prove~\eqref{eq:TV15411201}. Consider the identity~\eqref{eq:TV09291601} and perform the change of variable $\xi := \lambda t^{\frac 16} \xi$,
\begin{multline} \label{eq:TV10161601}
    \left( \sqrt[6]{t} \lambda\right)^{d+4} u( \lambda^6 t , \lambda x) \\ = \int_{[- \sqrt[6]{t} \lambda  \pi,  \sqrt[6]{t} \lambda  \pi]^d}  \left(\sqrt[6]{t} \lambda \right)^4 \left( e^{i \frac{\xi_j}{\sqrt[6]{t}\lambda}} - 1\right)\left( e^{i \frac{\xi_k}{ \sqrt[6]{t} \lambda}} - 1\right) \left( e^{i \frac{\xi_{j_1}}{ \sqrt[6]{t} \lambda}} - 1\right) \left( e^{i \frac{\xi_{k_1}}{ \sqrt[6]{t} \lambda}} - 1\right) e^{i\xi \cdot \frac{x}{\sqrt[6]{t}}} e^{- t \lambda^6 \left(2 \sum_{l=1}^d \left( 1-\cos \frac{\xi_l}{ \sqrt[6]{t} \lambda}  \right)\right)^3 } \di \xi.
\end{multline}
Note that, by using the assumption $t \geq 1$ and by substituting $\lambda$ by $\sqrt[6]{t} \lambda$ in~\eqref{eq:TV10161601}, it is sufficient, in order to prove~\eqref{eq:TV15411201}, to prove the following expansion: for any $\lambda \in (1 , \infty)$ and any point $ x \in \Rd$,
\begin{multline} \label{eq:TV11111601}
    \int_{[-\lambda  \pi,  \lambda  \pi]^d}  \lambda^4 \left( e^{i \frac{\xi_j}{\lambda}} - 1\right)\left( e^{i \frac{\xi_k}{ \lambda}} - 1\right) \left( e^{i \frac{\xi_{j_1}}{  \lambda}} - 1\right) \left( e^{i \frac{\xi_{k_1}}{  \lambda}} - 1\right) e^{i\xi \cdot x} e^{- \lambda^6 \left(2 \sum_{i=1}^d \left( 1-\cos \frac{\xi_i}{ \lambda}  \right)\right)^3 } \di \xi \\ =  \int_{\Rd} \xi_j\xi_k\xi_{j_1}\xi_{k_1} e^{i\xi \cdot x} e^{- \left| \xi \right|^6 } \di \xi + O \left( \frac{C}{\lambda} \right).
\end{multline}
We use a Taylor expansion of the exponential: for any $r \in \R$,
\begin{equation*}
    e^{i \frac{r}{\lambda}} = 1 + i \frac{r}{\lambda} + O \left( \frac{C r^2}{\lambda^2}\right).
\end{equation*}
Putting this estimate in the left side of~\eqref{eq:TV11111601} and using the identity $\left| e^{ir}\right| = 1$ valid for any $r \in \R$, we obtain
\begin{multline*}
    \left| \int_{[-\lambda  \pi,  \lambda  \pi]^d}  \left[ \lambda^4 \left( e^{i \frac{\xi_j}{\lambda}} - 1\right)\left( e^{i \frac{\xi_k}{ \lambda}} - 1\right) \left( e^{i \frac{\xi_{j_1}}{  \lambda}} - 1\right) \left( e^{i \frac{\xi_{k_1}}{  \lambda}} - 1\right) - \xi_{j} \xi_{k} \xi_{j_1} \xi_{k_1} \right] e^{i\xi \cdot x} e^{- \lambda^6 \left(2 \sum_{l=1}^d \left( 1-\cos \frac{\xi_l}{ \lambda}  \right)\right)^3 } \di \xi \right| \\ \leq \frac{C}{\lambda}  \int_{[-\lambda  \pi,  \lambda  \pi]^d} \left( \xi_{j}^2 +  \xi_{k}^2 + \xi_{j_1}^2 + \xi_{k_1}^2 \right) e^{- \lambda^6 \left(2 \sum_{l=1}^d \left( 1-\cos \frac{\xi_l}{ \lambda}  \right)\right)^3 } \di \xi.
\end{multline*}
Using that there exists a universal constant $C$ such that, for any $r \in [- \pi , \pi]$, $1 - \cos \left( r \right) \geq \frac{r^2}{C}$, we obtain the estimate
\begin{align} \label{eq:TV11521601}
\lefteqn{ \left| \int_{[-\lambda  \pi,  \lambda  \pi]^d}  \left[ \lambda^4 \left( e^{i \frac{\xi_j}{\lambda}} - 1\right)\left( e^{i \frac{\xi_k}{ \lambda}} - 1\right) \left( e^{i \frac{\xi_{j_1}}{  \lambda}} - 1\right) \left( e^{i \frac{\xi_{k_1}}{  \lambda}} - 1\right) - \xi_{j} \xi_{k} \xi_{j_1} \xi_{k_1} \right] e^{i\xi \cdot x} e^{- \lambda^6 \left(2 \sum_{i=1}^d \left( 1-\cos \frac{\xi_i}{ \lambda}  \right)\right)^3 } \di \xi \right|} \qquad \hspace{80mm} & \\ & \leq \frac{C}{\lambda}  \int_{[-\lambda  \pi,  \lambda  \pi]^d} \left( \xi_{j}^2 +  \xi_{k}^2 + \xi_{j_1}^2 + \xi_{k_1}^2 \right) e^{ -\frac{\left| \xi \right|^6}{C} } \di \xi  \notag \\ &
\leq \frac{C}{\lambda}. \notag
\end{align}
A consequence of~\eqref{eq:TV11521601} is the identity
\begin{multline*}
    \int_{[-\lambda  \pi,  \lambda  \pi]^d}  \lambda^4 \left( e^{i \frac{\xi_j}{\lambda}} - 1\right)\left( e^{i \frac{\xi_k}{ \lambda}} - 1\right) \left( e^{i \frac{\xi_{j_1}}{  \lambda}} - 1\right) \left( e^{i \frac{\xi_{k_1}}{  \lambda}} - 1\right) e^{i\xi \cdot x} e^{- \lambda^6 \left(2 \sum_{i=1}^d \left( 1-\cos \frac{\xi_i}{ \lambda}  \right)\right)^3 } \di \xi \\ = \int_{[-\lambda  \pi,  \lambda  \pi]^d}  \xi_{j} \xi_{k} \xi_{j_1} \xi_{k_1} e^{i\xi \cdot x} e^{- \lambda^6 \left(2 \sum_{i=1}^d \left( 1-\cos \frac{\xi_i}{ \lambda}  \right)\right)^3 } \di \xi + O \left( \frac{C}{\lambda} \right),
\end{multline*}
so that, in order to prove~\eqref{eq:TV11111601}, it is sufficient to prove
\begin{equation} \label{eq:TV13541601}
    \int_{[-\lambda  \pi,  \lambda  \pi]^d}  \xi_{j} \xi_{k} \xi_{j_1} \xi_{k_1} e^{i\xi \cdot x} e^{- \lambda^6 \left(2 \sum_{i=1}^d \left( 1-\cos \frac{\xi_i}{ \lambda}  \right)\right)^3 } \di \xi = \int_{\Rd}  \xi_{j} \xi_{k} \xi_{j_1} \xi_{k_1} e^{i\xi \cdot x} e^{- \left| \xi \right|^6 } \di \xi + O \left( \frac{C}{\lambda} \right).
\end{equation}
We now prove~\eqref{eq:TV13541601}. Using the inequality
\begin{equation*}
    \left| \int_{\Rd \setminus [-\lambda  \pi,  \lambda  \pi]^d}  \xi_{j} \xi_{k} \xi_{j_1} \xi_{k_1} e^{i\xi \cdot x} e^{- \left| \xi \right|^6 } \di \xi \right| \leq e^{- c \left| \xi \right|^6},
\end{equation*}
we see that to prove~\eqref{eq:TV13541601}, one needs to show
\begin{equation} \label{eq:TV13581601}
    \lambda \left| \int_{[-\lambda  \pi,  \lambda  \pi]^d}  \xi_{j} \xi_{k} \xi_{j_1} \xi_{k_1} e^{i\xi \cdot x} \left( e^{- \lambda^6 \left(2 \sum_{l=1}^d \left( 1-\cos \frac{\xi_l}{ \lambda}  \right)\right)^3 } - e^{- \left| \xi \right|^6} \right) \di \xi \right| \leq C.
\end{equation}
The strategy to prove~\eqref{eq:TV13581601} is to apply the dominated convergence theorem to prove the (stronger) statement
\begin{equation} \label{eq:TV14291601}
    \lambda \int_{[-\lambda  \pi,  \lambda  \pi]^d}  \xi_{j} \xi_{k} \xi_{j_1} \xi_{k_1} e^{i\xi \cdot x} \left[ e^{- \lambda^6 \left(2 \sum_{i=1}^d \left( 1-\cos \frac{\xi_i}{ \lambda}  \right)\right)^3 } - e^{- \left| \xi \right|^6} \right] \di \xi \underset{\lambda \to \infty}{\longrightarrow} 0.
\end{equation}
To apply the dominated convergence theorem to the left side of~\eqref{eq:TV14291601}, we verify its assumptions.

\smallskip

\textbf{Pointwise convergence.} By using the Taylor expansion of the cosine, we have, for any $\xi \in \Rd$,
    \begin{equation*}
        2 \sum_{i=1}^d \left( 1-\cos \frac{\xi_i}{ \lambda}  \right) = \frac{\left| \xi \right|^2}{\lambda^2} + O \left( \frac{C \left| \xi \right|^4}{\lambda^4} \right),
    \end{equation*}
    which yields
    \begin{equation} \label{eq:TV15521601}
        \lambda^6 \left( 2 \sum_{l=1}^d \left( 1-\cos \frac{\xi_l}{ \lambda}  \right) \right)^3 = \left| \xi \right|^6 + O \left( \frac{C \left| \xi \right|^8}{\lambda^2} \right).
    \end{equation}
    The expansion~\eqref{eq:TV15521601} implies the pointwise convergence, for any $\xi \in \Rd$,
    \begin{equation*}
        \lambda \left( e^{- \lambda^6 \left(2 \sum_{i=1}^d \left( 1-\cos \frac{\xi_i}{ \lambda}  \right)\right)^3 } - e^{- \left| \xi \right|^6} \right)  =  e^{- \left| \xi \right|^6} \lambda  \left( e^{- \lambda^6 \left(2 \sum_{l=1}^d \left( 1-\cos \frac{\xi_l}{ \lambda}  \right)\right)^3 + \left| \xi \right|^6 } - 1 \right) \underset{\lambda \to \infty}{\longrightarrow} 0.
    \end{equation*}

\textbf{Uniform upper bound.} We split this argument into two cases.

\smallskip

\textit{Case 1.} If $\left| \xi \right|^8 \leq \lambda^2$, then the expansion~\eqref{eq:TV15521601} implies the bound 
    \begin{align} \label{eq:TV16401601}
        e^{- \left| \xi \right|^6} \lambda  \left( e^{- \lambda^6 \left(2 \sum_{l=1}^d \left( 1-\cos \frac{\xi_l}{ \lambda}  \right)\right)^3 + \left| \xi \right|^6 } - 1 \right) & \leq e^{- \left| \xi \right|^6} \lambda  \left( e^{C \frac{\left| \xi \right|^8}{\lambda^2} } - 1 \right) \\
        & \leq  e^{- \left| \xi \right|^6} \lambda e^C  C \frac{\left| \xi \right|^8}{\lambda^2} \notag \\
        & \leq C \left| \xi \right|^8 e^{- \left| \xi \right|^6}, \notag
    \end{align}
    where we have used in the second line that for any constant $C>0$, the exponential is $e^C$-Lipschitz on the interval $[0 , C]$ and in the third line the inequality $\lambda \geq 1$.
    
\textit{Case 2.} If $\left| \xi \right|^8 \geq \lambda^2$, then we use the inequality $1 - \cos \left( r \right) \geq \frac{r^2}{C}$, valid for some universal constant $C$ and for any real number $r \in [- \pi , \pi]$, to obtain
\begin{align} \label{eq:TV16391601}
     \left| e^{- \left| \xi \right|^6} \lambda  \left( e^{- \lambda^6 \left(2 \sum_{l=1}^d \left( 1-\cos \frac{\xi_l}{ \lambda}  \right)\right)^3 + \left| \xi \right|^6 } - 1 \right) \right| & = \lambda  \left| \left( e^{- \lambda^6 \left(2 \sum_{l=1}^d \left( 1-\cos \frac{\xi_l}{ \lambda}  \right)\right)^3} - e^{- \left| \xi \right|^6}  \right) \right| \\ & \leq \lambda e^{- \frac{\left| \xi \right|^6}{C}}  \notag\\ 
     & \leq  C \left| \xi \right|^4 e^{- \frac{\left| \xi \right|^6}{C}}. \notag
\end{align}
A combination of~\eqref{eq:TV16401601} and~\eqref{eq:TV16391601} implies the upper bound, for any $\lambda \geq 1$ and any $\xi \in \left[ - \lambda \pi , \lambda \pi\right]^d$,
\begin{equation*}
     \left| \xi_{j} \xi_{k} \xi_{j_1} \xi_{k_1} e^{i\xi \cdot x} \lambda \left( e^{- \lambda^6 \left(2 \sum_{l=1}^d \left( 1-\cos \frac{\xi_l}{ \lambda}  \right)\right)^3 } - e^{- \left| \xi \right|^6} \right) \right| \leq  C \left| \xi \right|^{12} e^{- \frac{\left| \xi \right|^6}{C}}.
\end{equation*}
Since the map $\xi \mapsto \left| \xi \right|^{12} e^{- \frac{\left| \xi \right|^6}{C}}$ is integrable over $\Rd$, we can apply the dominated convergence theorem and conclude the proof of~\eqref{eq:TV14291601}. The proof of~\eqref{eq:TV15411201} is complete.

We use the properties~\eqref{eq:TV09351901} and~\eqref{eq:TV15411201} to complete the proof of Lemma~\ref{lem:lemma4.1chap8}. We fix two parameters $\lambda \in [1 , \infty]$ and $x \in \Rd$ such that $|x| =1$ and $\lambda x \in \Zd$. We let $\alpha \in (0 , 1)$ be an exponent whose value is decided later in the argument and shall depend only on the parameters $d$ and $\ep$. We write
\begin{align} \label{eq:TV20071201}
     \lambda^{d-2} F( \lambda x) & = \frac{1}{\left( 2\pi\right)^d}   \int_0^\infty \lambda^{d-2} u(t , \lambda x ) \, \di t \\
     & = \frac{1}{\left( 2\pi\right)^d}   \int_0^\infty \lambda^{d+4} u( \lambda^6 t , \lambda x ) \, \di t \notag \\
     & = \frac{1}{\left( 2\pi\right)^d}   \int_0^{\lambda^{-\alpha}} \lambda^{d+4} u( \lambda^6 t , \lambda x ) \, \di t + \frac{1}{\left( 2\pi\right)^d}   \int_{\lambda^{-\alpha}}^\infty \lambda^{d+4} u( \lambda^6 t , \lambda x ) \, \di t. \notag
\end{align}
We estimate the first term in the right side. We use the estimate~\eqref{eq:TV09351901} with the exponent $k = d+6$ and obtain
\begin{equation*}
    \frac{1}{\left( 2\pi\right)^d}   \int_0^{\lambda^{-\alpha}} \lambda^{d+4} \left| u( \lambda^6 t , \lambda x ) \right| \, \di t \leq \frac{C_{d+6 , \alpha} }{\left( 2\pi\right)^d} \lambda^{-\alpha} \lambda^{d+4} \lambda^{-d-6} \leq C \lambda^{-2}
\end{equation*}
We estimate the second term in the right side of~\eqref{eq:TV20071201}, we note that since $\lambda$ is chosen larger than $1$, we have $\lambda^{6 - \alpha} \geq 1$ and we can apply the inequality~\eqref{eq:TV15411201}. We obtain
\begin{equation*}
     \frac{1}{\left( 2\pi\right)^d}   \int_{\lambda^{-\alpha}}^\infty \lambda^{d+4} u( \lambda^6 t , \lambda x ) \, \di t = \frac{1}{\left( 2\pi\right)^d}   \int_{\lambda^{-\alpha}}^\infty \int_{\Rd} \xi_j\xi_k\xi_{j_1}\xi_{k_1} e^{i\xi \cdot x} e^{- t \left| \xi \right|^6 } \di \xi \di t + \int_{\lambda^{-\alpha}}^\infty O \left( \frac{C}{\lambda  t^{\frac{d+5}{6}} } \right) \, \di t. 
\end{equation*}
Using the estimate $\int_{\lambda^{-\alpha}}^\infty t^{-\frac{d+5}{6}} \leq C\lambda^{\frac{\alpha (d+1)}{6}}$ and setting $\alpha = \frac{6 \ep}{d+1}$, we deduce
\begin{equation*}
     \frac{1}{\left( 2\pi\right)^d}   \int_{\lambda^{-\alpha}}^\infty \lambda^{d+4} u( \lambda^6 t , \lambda x ) \, \di t = \frac{1}{\left( 2\pi\right)^d}   \int_{\lambda^{-\alpha}}^\infty \int_{\Rd} \xi_j\xi_k\xi_{j_1}\xi_{k_1} e^{i\xi \cdot x} e^{- t \left| \xi \right|^6 } \di \xi \di t + O \left( C\lambda^{\ep - 1} \right). 
\end{equation*}
By performing the change of variable $\xi := t^{\frac 16} \xi$, we see that
\begin{equation*}
    \int_{\Rd} \xi_j\xi_k\xi_{j_1}\xi_{k_1} e^{i\xi \cdot x} e^{- t \left| \xi \right|^6 } \di \xi = t^{-\frac{d + 4}{6}} \int_{\Rd} \xi_j\xi_k\xi_{j_1}\xi_{k_1} e^{i\xi \cdot x t^{-\frac{1}{6}}} e^{- t \left| \xi \right|^6 } \di \xi = t^{-\frac{d + 4}{6}} K \left( \frac{x}{t^{\frac 16}}\right)
\end{equation*}
A combination of the five previous displays shows
\begin{equation} \label{eq:TV09462001}
    \lambda^{d-2} F( \lambda x)  = \int_{\lambda^{-\alpha}}^\infty t^{-\frac{d + 4}{6}} K \left( \frac{x}{t^{\frac 16}}\right) + O \left( C\lambda^{\ep - 1} \right).
\end{equation}
To complete the argument, note that since the map $K$ belongs to the Schwartz space, it decays faster than any polynomials at infinity. We thus have the estimate $K\left( \frac{x}{t^{\frac 16}}\right) \leq C t^{\frac{d+4}{6} + \frac{1}{\alpha}}$, which implies
\begin{equation}  \label{eq:TV09472001}
    \int_{0}^{\lambda^{-\alpha}} t^{-\frac{d + 4}{6}} K \left( \frac{x}{t^{\frac 16}}\right) \leq C \int_{0}^{\lambda^{-\alpha}} t^{\frac{1}{\alpha} } \leq C \int_{0}^{\lambda^{-\alpha}} \lambda^{- 1} \leq  C\lambda^{-1}.
\end{equation}
Combining~\eqref{eq:TV09462001} and~\eqref{eq:TV09472001}, we have obtained
\begin{equation*}
    \lambda^{d-2} F( \lambda x)  = \int_{0}^\infty t^{-\frac{d + 4}{6}} K \left( \frac{x}{t^{\frac 16}}\right) + O \left( C\lambda^{\ep - 1} \right).
\end{equation*}
The proof of Lemma~\ref{lem:lemma4.1chap8} is complete.
\end{proof}

Lemma~\ref{lem:lemma4.1chap8} states that the large-scale behavior of the convolutions of gradients of the discrete Green's function $F_{j , k , j_1 , k_1}$ and $F_{j , j_1}$ is determined by the $(2-d)$-homogeneous maps $J_{j, k, j_1, k_1}$ and $J_{j , j_1}$. The rest of this section is devoted to the study of these maps. One of their properties is that, as $(2-d)$-homogeneous functions, they belong to the class of tempered distribution. Their Fourier transform turns out to be explicitly calculable; this is the purpose of the following lemma.

\begin{lemma} \label{eq:TV13071301}
    Fix four integers $ i , j , k ,l \in \{ 1 , \ldots, d \}$. One has the identities
    \begin{equation*}
        \widehat{J_{ i , j , k ,l}}(\xi) = \frac{\xi_i \xi_j \xi_{k} \xi_{l}}{\left| \xi \right|^6} \hspace{5mm} \mbox{and} \hspace{5mm} \widehat{J_{i,j}}(\xi) = \frac{\xi_i \xi_{j} }{\left| \xi \right|^4}.
    \end{equation*}
\end{lemma}

\begin{proof}
We only prove the result for the maps $J_{ i , j , k ,l}$. Fix a function $g: \Rd \to \R$ in the Schwartz space and prove the identity
\begin{equation*}
    \int_{\Rd} J_{i , j , k ,l} \left( x \right) g(x) \, \di x = \int_{\Rd} \frac{\xi_i \xi_j \xi_{k} \xi_{l}}{\left| \xi \right|^6} \hat{g}(\xi) \, \di \xi.
\end{equation*}
We first use the definition of the map $J_{ i , j , k ,l}$ and apply Fubini's theorem
\begin{align*}
     \int_{\Rd} J_{i , j , k ,l} \left( x \right) g(x) \, \di x & = \int_{\Rd} \int_0^\infty t^{-\frac{d+4}{6}} K_{i , j , k ,l} \left( \frac{x}{t^{\frac 16}}\right) g(x) \, \di x \\
     & = \int_0^\infty t^{-\frac{d+4}{6}} \int_{\Rd}  K_{i , j , k ,l} \left( \frac{x}{t^{\frac 16}}\right) g(x) \, \di x \\
     & = \int_0^\infty t^{-\frac{d+4}{6}} \int_{\Rd}  t^d \hat{K}_{i , j , k ,l} \left(  t^{\frac 16} x \right) \hat{g}(\xi) \, \di \xi.
\end{align*}
By definition of the map $K_{i , j , k ,l}$, we have the identity
\begin{equation*}
    \hat{K}(\xi) =  \xi_i\xi_j \xi_{k} \xi_{l} e^{- |\xi|^6}.
\end{equation*}
A combination of the two previous displays shows
\begin{equation*}
    \int_{\Rd} J_{i,j,k,l} \left( x \right) g(x) \, \di x = \int_0^\infty \int_{\Rd} \xi_i \xi_j \xi_{k} \xi_{l} e^{- t |\xi|^6} \hat{g}(\xi) \, \di \xi =  \int_{\Rd} \frac{\xi_i \xi_j \xi_{k} \xi_{l}}{|\xi|^6} \hat{g}(\xi) \, \di \xi ,
\end{equation*}
as claimed.
\end{proof}

The next proposition is used in the proof of Theorem~\ref{t.main}. It asserts that if a linear combination of the maps $F_{i,j, k,l}$ and $F_{i,j}$, with a specific structure given by the problem considered in this article, is invariant under the group $H$ lattice preserving maps, then it must satisfy the expansion given by~\eqref{eq:TV18052601}.

\begin{proposition} \label{propsinginvaraincefss}
Assume that there exist coefficients $\left(c_{ij}\right)_{1 \leq i ,  j \leq d}$ and $\left( K_{ij} \right)_{ 1 \leq i,j \leq d}$, an exponent $\alpha > 0$ and a map $U$ which is invariant under the group $H$ of the lattice-preserving maps such that
\begin{equation} \label{eq:TV11001301}
    U(x) = \sum_{i, j , k , l =1}^d c_{ij} c_{kl} F_{i,j, k,l} (x) + \sum_{i,  j = 1}^d K_{ij} F_{i,j}(x) + O \left( \frac{C}{|x|^{d- 2 + \alpha}} \right),
\end{equation}
then there exists a constant $c \in \R$ such that
\begin{equation} \label{eq:TV18052601}
     U(x) = \frac{c}{|x|^{d-2}} + O \left( \frac{C}{|x|^{d- 2 + \alpha}} \right).
\end{equation}
\end{proposition}

\begin{proof}
Applying Lemma~\ref{lem:lemma4.1chap8} and Remark~\ref{rem:rem4.2chap8}, the expansion~\eqref{eq:TV11001301} can be rewritten
\begin{equation*}
    U(x) = \sum_{i, j , k ,l =1}^d c_{ij} c_{kl} J_{i,j,k,l}(x) + \sum_{i , j = 1}^d K_{ij}  J_{i,j}(x) + O \left( \frac{C}{|x|^{d- 2 + \alpha}} \right).
\end{equation*}
Using that the maps $J_{i,j,k,l}$ and $J_{i,j}$ are $(2-d)$-homogeneous, we see that the assumption that $U$ is invariant under the lattice-preserving maps implies that the same property holds for the function $\sum_{i, j , k ,l =1}^d c_{ij} c_{kl} J_{i,j,k,l} + \sum_{i, j =1}^d K_{ij}  J_{i,j}$: for each $h \in H$ and each $x \in \Zd \setminus \{0 \}$, one has
\begin{equation*}
    \sum_{i, j , k ,l =1}^d c_{ij} c_{kl} J_{i,j,k,l}(h(x)) + \sum_{i, j =1}^d K_{ij}  J_{i,j}(h(x)) = \sum_{i, j , k ,l =1}^d c_{ij} c_{kl} J_{i,j,k,l}(x) + \sum_{i, j =1}^d K_{ij}  J_{i,j}(x).
\end{equation*}
Using the homogeneity of the maps $J_{i,j,k,l}$ and $J_{i,j}$, the result can be extended to each point of $\Rd \setminus \{ 0\}$: for each $h \in H$ and each $x \in \Rd \setminus \{0 \}$, one has
\begin{equation} \label{eq:TV13041301}
    \sum_{i, j , k ,l =1}^d c_{ij} c_{kl} J_{i,j,k,l}(h(x)) + \sum_{i, j =1}^d K_{ij}  J_{i,j}(h(x)) = \sum_{i, j , k ,l =1}^d c_{ij} c_{kl} J_{i,j,k,l}(x) + \sum_{i, j =1}^d K_{ij}  J_{i,j}(x).
\end{equation}
To ease the notation, we denote by $P$ the homogeneous polynomial
\begin{equation} \label{eq:TV10041501}
    P \left( \xi \right) = \left( \sum_{i,j = 1}^d c_{ij} \xi_i \xi_j  \right)^2 + \left| \xi \right|^2 \sum_{i,j = 1}^d K_{ij} \xi_i\xi_j,
\end{equation}
so that the Fourier transform of the map $\sum_{i, j , k ,l =1}^d c_{ij} c_{kl} J_{i,j,k,l} + \sum_{i, j =1}^d K_{ij}  J_{i,j}$ is equal to the function $\xi \mapsto \frac{P \left( \xi \right)}{\left| \xi \right|^6}$ by Lemma~\ref{eq:TV13071301}.

Taking the Fourier transform on both sides of the identity~\eqref{eq:TV13041301} and applying Lemma~\ref{eq:TV13071301}, we obtain the identity, for any $\xi \in \Rd $ and any $h \in H$,
\begin{equation} \label{eq:TV14301301}
    P \left( h\left( \xi \right) \right) =  P \left( \xi  \right).
\end{equation}
We now prove that the equality~\eqref{eq:TV14301301} implies that $P = a \left| \xi \right|^4$. We first use the following fact whose proof is omitted: if a polynomial $S \in \R \left[ X_1 , \ldots , X_d \right]$ is homogeneous of degree $4$ and is invariant under the lattice-preserving maps, then there exist $a , b \in \R$ such that
\begin{equation*}
    S = a \sum_{i = 1}^d X_i^4 + b \sum_{1 \leq i < j \leq d} X_i^2 X_j^2.
\end{equation*}
Applying this result to the polynomial $P$, we obtain that there exist $a , b \in \R$ such that, for any $\xi \in \Rd$,
\begin{equation} \label{eq:TV09591501}
    P \left( \xi \right) = a  \sum_{i = 1}^d \xi_i^4 +  b \sum_{1 \leq i < j \leq d} \xi_i^2 \xi_j^2.
\end{equation}
Using the equality
\begin{equation*}
    \left( \sum_{i=1}^d \xi_i^2 \right)^2 = \sum_{i = 1}^d \xi_i^4 +  2 \sum_{1 \leq i < j \leq d} \xi_i^2 \xi_j^2,
\end{equation*}
we can rewrite the identity~\eqref{eq:TV09591501}
\begin{equation*}
    P \left( \xi \right) = \left( a - \frac b2 \right) \sum_{i = 1}^d \xi_i^4 + \frac{b}{2} \left( \sum_{i=1}^d \xi_i^2 \right)^2.
\end{equation*}
The objective is thus to prove that $a - \frac{b}{2} = 0$ by using the specific structure of the polynomial $P$ stated in~\eqref{eq:TV10041501}.
We first write
\begin{equation} \label{eq:TV10211501}
     \left( \sum_{i,j = 1}^d c_{ij} \xi_i \xi_j  \right)^2 + \left| \xi \right|^2 \sum_{i,j = 1}^d K_{ij} \xi_i\xi_j =  \left( a - \frac b2 \right) \sum_{i = 1}^d \xi_i^4 + \frac{b}{2} \left| \xi \right|^4,
\end{equation}
which implies
\begin{equation} \label{eq:TV10271501}
    \left( \sum_{i,j = 1}^d c_{ij} \xi_i \xi_j  \right)^2 + \left| \xi \right|^2 \left( \sum_{i,j = 1}^d K_{ij} \xi_i\xi_j - \frac{b}{2} \left| \xi \right|^2 \right) = \left( a - \frac b2 \right) \sum_{i = 1}^d \xi_i^4.
\end{equation}
To prove that $a - \frac b2 = 0$, we argue by contradiction: if $a - \frac b2 \neq 0$, then, by~\eqref{eq:TV10211501}, there exist two real-valued polynomials $Q , R$ homogeneous of degree $2$ such that, for any $\xi \in \Rd$,
\begin{equation*}
    Q^2(\xi) + \left( \sum_{i = 1}^d \xi_i^2 \right) R(\xi) = \sum_{i = 1}^d \xi_i^4.
\end{equation*}
We now prove that these two polynomials do not exist. We first reduce the problem to the three dimensional case: if we denote by $Q_0 , R_0 : \R^3 \to \R$ the homogeneous polynomials of degree $2$ defined by the formulas, for each $\xi = \left( \xi_1 , \xi_2 , \xi_3 \right) \in \R^3$,
\begin{equation*}
    Q_0 (\xi) = Q \left( \left( \xi_1 , \xi_2 , \xi_3 , 0 , \ldots, 0 \right) \right) \hspace{5mm} \mbox{and}  \hspace{5mm} R_0 (\xi) = R \left( \left( \xi_1 , \xi_2 , \xi_3 , 0 , \ldots, 0 \right) \right),
\end{equation*}
then we have the identity, for each $\xi \in \R^3$,
\begin{equation} \label{eq:TV11081501}
    Q_0^2(\xi) + \left( \sum_{i = 1}^3 \xi_i^2 \right) R_0(\xi) = \sum_{i = 1}^3 \xi_i^4.
\end{equation}
It is thus sufficient to prove that the polynomials $Q_0$ and $R_0$ do not exist. The proof can be done by an explicit computation: if we denote by
\begin{equation*}
    Q_0 \left( \xi \right) = A \xi_1^2 + B \xi_2^2 + C \xi_3^2 + D \xi_1 \xi_2 + E \xi_1 \xi_3 + F \xi_2 \xi_3 ~\mbox{and}~ R_0 \left( \xi \right) = G \xi_1^2 + H \xi_2^2 + I \xi_3^2 + J \xi_1 \xi_2 + K \xi_1 \xi_3 + L \xi_2 \xi_3,
\end{equation*}
expand the left side of~\eqref{eq:TV11081501} and identify all the coefficients. We obtain a quadratic system of 15 equations and 12 variables (the parameters $A, B , C , D , E , F , G , H , I , J , K ,L$). This system is over-determined; it can be solved explicitly using a software of formal computation and it does not have any solution, ruling out the existence of the polynomials $Q_0$ and $R_0$. We have thus reached a contradiction and deduce that $a - \frac{b}{2} = 0$. The identity~\eqref{eq:TV09591501} can then be rewritten
\begin{equation*}
   P \left( \xi \right)  = a \left( \sum_{i = 1}^d \xi_i^2 \right)^2 = a \left| \xi \right|^4.
\end{equation*}
The equality~\eqref{eq:TV09591501} implies that the Fourier transform of the map $\sum_{i, j , k ,l =1}^d c_{ij} c_{kl} J_{i,j,k,l} + \sum_{i , j = 1}^d K_{ij}  J_{i,j}$ is equal to $\frac{a}{|\xi|^2}$, which implies, by taking the inverse Fourier transform, that there exists a constant $c$ such that, for any $x \in \Rd \setminus \{ 0 \}$,
\begin{equation} \label{eq:TV11311501}
    \sum_{i, j , k ,l =1}^d c_{ij} c_{kl} J_{i,j,k,l}(x) + \sum_{i , j = 1}^d K_{ij}  J_{i,j}(x) = \frac{c}{|x|^{d-2}}.
\end{equation}
Combining the identity~\eqref{eq:TV11311501} with the expansion~\eqref{eq:TV11001301}, we have obtained
\begin{equation*}
    U(x) = \frac{c}{|x|^{d-2}} + O \left(  \frac{C}{|x|^{d - 2 + \alpha}}\right).
\end{equation*}
The proof of Proposition~\ref{propsinginvaraincefss} is complete.
\end{proof}

\section{Treating the error term \texorpdfstring{$\mathcal{E}_{q_1,q_2}$}{40}} \label{sec:chap8.5}

This section is devoted to the treatment the error term $\mathcal{E}_{q_1,q_2}$. It is used in the proof of Theorem~\ref{t.main} in Section~\ref{sectionsection4.2789} of Chapter~\ref{section3.4}.

\begin{proposition} \label{prop:propB1}
Fix two exponents $\gamma, \ep \in ( 0, 1]$ such that $\ep \leq \frac{\gamma}{4(d-2)}$ and two charges $q_1,q_2 \in \mathcal{Q}$. Let $\mathcal{E}_{q_1, q_2} : \Zd \to \R$ be a function which satisfies the pointwise and $L^1$-estimates, for each point $\kappa \in \Zd$ and each radius $R \geq 1$,
\begin{equation} \label{app.assumptE}
    \left| \mathcal{E}_{q_1,q_2}(\kappa) \right| \leq \frac{C}{|\kappa|^{d- \ep}} ~\mbox{and}~ \sum_{\kappa \in B_{2R} \setminus B_R} \left| \mathcal{E}_{q_1 , q_2} \left( \kappa \right) \right| \leq C R^{-\gamma}.
\end{equation}
Then the constant $K_{q_1,q_2} := 4\pi^2\sum_{\kappa \in \Zd} \mathcal{E}_{q_1 , q_2} \left( \kappa \right) $ is well-defined in the sense that the sum converges absolutely and one has the expansion
\begin{multline} \label{eq:TV02504}
    4\pi^2\sum_{z_2 , \kappa  \in \Zd}  \nabla G(z_2) \cdot (n_{q_2}) \nabla G_x(z_2 + \kappa) \cdot (n_{q_1})  \mathcal{E}_{q_1,q_2}(\kappa) \\ = K_{q_1 , q_2} \sum_{z_2  \in \Zd}  \nabla G(z_2) \cdot (n_{q_2}) \nabla G(z_2-x) \cdot (n_{q_1})  + O \left(\frac{C_{q_1 , q_2}}{|x|^{d -2 + \frac{\gamma}{4(d-2)}}} \right).
\end{multline}
\end{proposition}

\begin{proof}
We first write
\begin{equation*}
    \nabla G(z_2) \cdot (n_{q_2}) = \sum_{i=1}^d \nabla_i G(z_2) (n_{q_2})_i \hspace{3mm}\mbox{and}\hspace{3mm} \nabla G(z_2 + \kappa - x) \cdot (n_{q_1}) = \sum_{j=1}^d \nabla_j G(z_2 + \kappa - x) (n_{q_1})_j
\end{equation*}
and note that to prove the expansion~\eqref{eq:TV02504} it is sufficient to prove that, for each pair of integers $i , j \in \{ 1 , \ldots , d \}$,
\begin{equation} \label{eq:TV02554}
    4\pi^2\sum_{z_2 , \kappa  \in \Zd}  \nabla_i G(z_2) \nabla_j G_x(z_2 + \kappa) \mathcal{E}_{q_1,q_2}(\kappa) = K_{q_1 , q_2} \sum_{z_2  \in \Zd}  \nabla_i G(z_2)  \nabla_j G(z_2-x)  + O \left(\frac{C_{q_1 , q_2}}{|x|^{d-2 + \frac{\gamma}{4(d-2)}}} \right).
\end{equation}
We fix two integers $i , j \in \{ 1 , \ldots , d \}$ and focus on the proof of~\eqref{eq:TV02554}.
Using the assumption on the function $\mathcal{E}_{q_1,q_2}$ stated in~\eqref{app.assumptE}, one has the estimate
\begin{equation*}
    \sum_{\kappa \in \Zd} \left| \mathcal{E}_{q_1 , q_2} \left( \kappa \right) \right| = \sum_{n = 0}^\infty  \sum_{\kappa \in B_{2^{n+1}} \setminus B_{2^n}}\left| \mathcal{E}_{q_1 , q_2} \left( \kappa \right) \right| \leq C \sum_{n = 0}^\infty 2^{-\gamma n} < \infty.
\end{equation*}
This proves that the sum $\sum_{\kappa \in \Zd}  \mathcal{E}_{q_1 , q_2} \left( \kappa \right)$ is absolutely convergent and that the constant $K_{q_1, q_2}$ is well-defined.
To ease the notation, we let $g$ be the function defined by the formula, for each $\kappa \in \Zd$,
\begin{equation*}
    g(\kappa) := \sum_{z_2  \in \Zd}  \nabla_i G(z_2) \nabla_j G(z_2 + \kappa),
\end{equation*}
so that the expansion~\eqref{eq:TV02554} can be rewritten in this notation
\begin{equation} \label{eq:TV03114}
4\pi^2\sum_{\kappa  \in \Zd}  g(\kappa + x) \mathcal{E}_{q_1,q_2}(\kappa) = K_{q_1 , q_2} g(x)  + O \left(\frac{C_{q_1 , q_2}}{|x|^{d -2+ \frac{\gamma}{4(d-2)}}} \right).
\end{equation}
Before proving the expansion~\eqref{eq:TV03114}, we record a property of the function $g$: by using standard estimates on the discrete Green's function $G$, it can be estimated by the formula
\begin{equation} \label{eq:TV04094}
    \left| g(\kappa) \right| \leq \sum_{z_2  \in \Zd} \left| \nabla_i G(z_2)  \nabla_j G(z_2 + \kappa) \right|  \leq C \sum_{z_2  \in \Zd} \frac{1}{|z_2|^{d-1}} \frac{1}{|z_2 + \kappa|^{d-1}}  \leq  \frac{C}{|\kappa|^{d-2}}.
\end{equation}
The gradient of $g$ can be bounded from above according to the following estimate
\begin{equation} \label{eq:TV03324}
    \left| \nabla g(\kappa) \right| \leq \sum_{z_2  \in \Zd} \left| \nabla_i G(z_2) \nabla \nabla_j G(z_2 + \kappa) \right|  \leq C \sum_{z_2  \in \Zd} \frac{1}{|z_2|^{d-1}} \frac{1}{|z_2 + \kappa|^{d}}  \leq  \frac{C \ln \left| x \right|}{|\kappa|^{d-1}}.
\end{equation}
To prove the expansion~\eqref{eq:TV03114}, it is sufficient to prove the estimate
\begin{equation} \label{eq:TV03284}
    \left| \sum_{\kappa \in \Zd} \left( g \left( x + \kappa \right) - g(x) \right)\mathcal{E}_{q_1 , q_2}(\kappa) \right| \leq \frac{C_{q_1,q_2}}{|x|^{d-2+\frac{\gamma}{4(d-2)}}}.
\end{equation}
To this end, We split the space into three regions.

\textit{Region 1. The ball of center $0$ and radius $|x|/2$.} We use that for each point $\kappa \in B(0 , |x|/2)$, $|\kappa + x| \geq |x|/2$. This implies, by~\eqref{eq:TV03324}, the inequality $\left| \nabla g(\kappa) \right| \leq C |x|^{d-1}$. From this estimate, we deduce that, for each $\kappa \in B(0 , |x|/2)$,
    \begin{equation*}
        \left|g \left( x+ \kappa \right) - g(x) \right| \leq |\kappa| \sup_{z \in B(0 , |x|/2)} \left| \nabla g(x + z) \right| \leq  \frac{C \left| \kappa \right| \ln \left| x \right|}{|x|^{d-1}} .
    \end{equation*}
    We then use this estimate to compute the sum
    \begin{align*}
        \left| \sum_{\kappa \in B \left( 0, \frac{|x|}{2}\right)} \left( g \left( x + \kappa \right) - g(x) \right)\mathcal{E}_{q_1 , q_2}(\kappa) \right| & \leq C \sum_{\kappa \in B \left( 0, \frac{|x|}{2}\right)} \frac{ |\kappa| \ln \left| x \right|}{|x|^{d-1}} \left| \mathcal{E}_{q_1 , q_2}(\kappa) \right|.
    \end{align*}
We partition the ball $B \left( 0, \frac{|x|}{2}\right)$ into dyadic annuli according to the inclusion $B \left( 0, \frac{|x|}{2}\right) \subseteq \cup_{n=0}^{\lfloor \ln_2\left( |x|/2\right) \rfloor} B_{2^{n+1}} \setminus B_{2^{n}}$, where we use the notation $\lfloor \ln_2\left( |x|/2\right) \rfloor$ to denote the floor of the real number $\ln_2\left( |x|/2\right)$. We additionally note that for each integer $n \in \N$ and each point $\kappa \in B_{2^{n+1}} \setminus B_{2^{n}}$, one has the estimate $|\kappa | \leq C 2^{n}$. Together with the estimate~\eqref{app.assumptE} on the error term $\mathcal{E}_{q_1 , q_2}$, we obtain
    \begin{align} \label{eq:TV04454}
        \left| \sum_{\kappa \in B \left( 0, \frac{|x|}{2}\right)} \left( G \left( x - \kappa \right) - G(x) \right)\mathcal{E}_{q_1 , q_2}(\kappa) \right| & \leq C \sum_{\kappa \in B \left( 0, \frac{|x|}{2}\right)} \frac{ \ln \left| x \right| |\kappa|}{|x|^{d-1}} \left| \mathcal{E}_{q_1 , q_2}(\kappa) \right| \\ 
        & \leq \frac{C \ln \left| x \right|}{  |x|^{d-1}} \sum_{n = 0}^{\lfloor \ln_2\left( |x|/2\right) \rfloor} \left| \kappa \right| \sum_{\kappa \in B_{2^{n+1}} \setminus B_{2^n} }  \left| \mathcal{E}_{q_1 , q_2}(\kappa) \right| \notag\\
        & \leq \frac{C \ln \left| x \right| }{ |x|^{d-1}} \sum_{n = 0}^{\lfloor \ln_2\left( |x|/2\right) \rfloor} 2^n 2^{-\gamma n}\notag \\
        & \leq \frac{C \ln \left| x \right|}{  |x|^{d-1}} 2^{(1-\gamma) \ln_2\left( |x|/2\right) } \notag\\
        & \leq \frac{C  }{|x|^{d-2+\frac{\gamma}{2}}}, \notag
    \end{align}
    where we have replaced the exponent $\gamma$ in the last line by $\frac{\gamma}{2}$ to absorb the logarithm.
    
\textit{Region 2. The ball of center $-x$ and of radius $|x|^{1-\frac{\gamma}{4(d-2)}}$.} In this region, we use the estimate~\eqref{eq:TV04094} on the function $g$ and the pointwise bound on the error term $\mathcal{E}_{q_1,q_2}$ stated in~\eqref{app.assumptE} to obtain
    \begin{align} \label{eq:TV04444}
         \left| \sum_{\kappa \in B \left(-x , |x|^{1-\frac{\gamma}{2(d-2)}} \right)} g \left( x + \kappa \right) \mathcal{E}_{q_1,q_2} \left( \kappa \right) \right| & \leq C \sum_{\kappa \in B \left(x , |x|^{1-\frac{\gamma}{2(d-2)}} \right)} \frac{1}{| \kappa - x|^{d-2}} \frac{1}{|\kappa|^{d- \ep}} \\ & \leq \frac{1}{|x|^{d- \ep}} \sum_{\kappa \in B \left(x , |x|^{1-\frac{\gamma}{2(d-2)}} \right)} \frac{1}{| \kappa - x|^{d-2}} \notag \\ & \leq \frac{C}{|x|^{d - 2 +\frac{\gamma}{2(d-2)} - \ep}} \notag\\
         & \leq \frac{C}{|x|^{d - 2 +\frac{\gamma}{4(d-2)}}}, \notag
    \end{align}
    where we used in the second inequality the lower bound $|\kappa| \geq c |x|$, valid for any point $\kappa \in B\left( x , |x|^{1-\frac{\gamma}{2}} \right)$, and the assumption $\ep \leq \frac{\gamma}{4(d-2)}$ in the fourth inequality.
    
\textit{Region 3. The set $C := \Zd \setminus \left( B\left(0, \frac{|x|}{2}\right)  \cup B \left(-x , |x|^{1-\frac{\gamma}{2(d-2)}} \right)\right)$.}
    By the inequality~\eqref{eq:TV04094}, we have the estimate, for each $\kappa \in C$,
    \begin{equation*}
        \left| g \left( x - \kappa \right) \right| \leq \frac{C}{|x|^{(1-\frac{\gamma}{2(d-2)})(d-2)}}.
    \end{equation*}
    Using the previous estimate and the inclusion $C \subseteq \cup_{n = \lceil \ln_2 |x|/2 \rceil}^\infty \left( B_{2^{n+1}} \setminus B_{2^n} \right)$, one obtains the inequality
        \begin{align} \label{eq:TV04324}
        \left| \sum_{\kappa \in C} \left( g \left( x + \kappa \right) - g(x) \right)\mathcal{E}_{q_1 , q_2}(\kappa) \right| & \leq C \sum_{\kappa \in C } \frac{1}{|x|^{(1-\frac{\gamma}{2(d-2)} )(d-2)}} \left| \mathcal{E}_{q_1 , q_2}(\kappa) \right| \\
        & \leq \frac{C}{|x|^{(1-\frac{\gamma}{2(d-2)} )(d-2)}} \sum_{n = \lceil \ln_2 |x|/2 \rceil}^\infty \sum_{\kappa \in  B_{2^{n+1}} \setminus B_{2^n}} \left| \mathcal{E}_{q_1 , q_2}(\kappa) \right| \notag \\
        & \leq \frac{C}{|x|^{(1-\frac{\gamma}{2(d-2)} )(d-1)}} \sum_{n = \lceil \ln_2 |x|/2 \rceil}^\infty 2^{-n\gamma} \notag \\
        & \leq \frac{C}{|x|^{(1-\frac{\gamma}{2(d-2)} )(d-2)+ \gamma}}. \notag
    \end{align}
    Computing the exponent in the last line of the inequality~\eqref{eq:TV04324} proves the estimate
    \begin{equation} \label{eq:TV04434}
        \left| \sum_{\kappa \in C} \left( G \left( x - \kappa \right) - G(x) \right)\mathcal{E}_{q_1 , q_2}(\kappa) \right| \leq \frac{C}{|x|^{(d-2)+ \frac{\gamma}{2}}}.
    \end{equation}
Combining the estimates~\eqref{eq:TV04454},~\eqref{eq:TV04444} and~\eqref{eq:TV04434} completes the proof of Lemma~\ref{prop:propB1}.

\end{proof}

\appendix

\chapter{Multiscale Poincar\'e inequality} \label{section:multiscPoinc}

\begin{proposition}[Multiscale Poincar\'e inequality] \label{prop:multiscPoin}
There exists a constant $C := C(d)$ such that for each cube integer $n \in \N$, the following statements hold:
\begin{enumerate}
\item For each function $f \in L^2 \left( \cu_n , \mu_\beta \right)$,
\begin{equation*}
\left\| f - \left( f \right)_{\cu_n} \right\|_{\underline{H}^{-1}(\cu_n , \mu_\beta)}^2 \leq C \left\| f \right\|_{\underline{L}^2 \left( \cu_n , \mu_\beta \right)}^2 + C 3^n \sum_{m = 0}^n \frac{1}{\left|\mathcal{Z}_{m,n} \right|} \left\langle \left( \frac{1}{\left| z + \cu_m \right|}\sum_{x \in z + \cu_m}   f (x, \cdot)  \right)^2 \right\rangle_{\mu_\beta};
\end{equation*}
\item For any function $f \in L^2 \left( \cu_n , \mu_\beta \right)$, one has the estimate
\begin{equation*}
\left\| f - \left( f \right)_\cu \right\|_{\underline{L}^{2}(\cu_n , \mu_\beta)}^2 \leq C \left\| \nabla f \right\|_{\underline{L}^2 \left( \cu_n , \mu_\beta \right)}^2 + C 3^n \sum_{m = 0}^n \frac{1}{\left|\mathcal{Z}_{m,n} \right|} \left\langle \left( \frac{1}{\left| z + \cu_m \right|}\sum_{x \in z + \cu_m}  \nabla f (x, \cdot)  \right)^2 \right\rangle_{\mu_\beta};
\end{equation*}
\item for each function $f \in L^2 \left( \cu_n , \mu_\beta \right)$ such that $f = 0$ on the boundary of the cube $\cu_n$
\begin{equation*}
\left\| f \right\|_{\underline{L}^{2}(\cu_n , \mu_\beta)}^2 \leq C \left\| \nabla f \right\|_{\underline{L}^2 \left( \cu_n , \mu_\beta \right)}^2 + C 3^n \sum_{m = 0}^n \frac{1}{\left|\mathcal{Z}_{m,n} \right|} \left\langle \left( \frac{1}{\left| z + \cu_m \right|}\sum_{x \in z + \cu_m}  \nabla f (x , \cdot)  \right)^2 \right\rangle_{\mu_\beta}.
\end{equation*}
\end{enumerate}
\end{proposition}

\begin{proof}
The proof is an almost immediate application of the multiscale Poincar\'e inequality proved in \cite[Proposition 1.7 and Lemma 1.8]{AKM}. We only treat the inequality (1); the other two estimates are similar. We consider a field $\phi \in \Omega$ and apply \cite[Proposition 1.7 and Lemma 1.8]{AKM} and a Cauchy-Schwarz inequality to the map $x \to f(x , \phi)$ (with a fixed field~$\phi$). We obtain
\begin{equation*}
    \left\| f(\cdot , \phi) - \left( f(\cdot, \phi) \right)_\cu \right\|_{\underline{H}^{-1}(\cu_n)}^2 \leq C \left\| f(\cdot, \phi) \right\|_{\underline{L}^2 \left( \cu_n  \right)}^2 + C 3^n \sum_{m = 0}^n \frac{1}{\left|\mathcal{Z}_{m,n} \right|} \left( \frac{1}{\left| z + \cu_m \right|}\sum_{x \in z + \cu_m}   f (x, \phi)  \right)^2.
\end{equation*}
Taking the expectation with respect to the field $\phi$ gives
\begin{equation*}
     \left\langle \left\| f - \left( f \right)_\cu \right\|_{\underline{H}^{-1}(\cu_n)}^2 \right\rangle_{\mu_\beta} \leq  C \left\| f \right\|_{\underline{L}^2 \left( \cu_n  \right)}^2 + C 3^n \sum_{m = 0}^n \frac{1}{\left|\mathcal{Z}_{m,n} \right|} \left\langle \left( \frac{1}{\left| z + \cu_m \right|}\sum_{x \in z + \cu_m}   f (x, \cdot)  \right)^2 \right\rangle_{\mu_\beta}.
\end{equation*}
We complete the proof by using the estimate
\begin{equation*}
\left\|  f - \left( f \right)_{\cu_n} \right\|_{\underline{H}^{-1}(\cu_n , \mu_\beta)}^2 \leq \left\langle \left\|   f - \left( f \right)_{\cu_n} \right\|_{ \underline{H}^{-1} \left( \cu_n \right)}^2 \right\rangle_{\mu_\beta},
\end{equation*}
which is a direct consequence of the definitions of the $\underline{H}^{-1}(\cu)$ and $\underline{H}^{-1}(\cu, \mu_\beta)$-norms stated in Chapter~\ref{Chap:chap2}.
\end{proof}

\chapter{Solvability of the Neumann problem} \label{app.appB}

In this appendix, we prove the existence and uniqueness of the maximizer in the variational formulation of the dual energy $\nu^*$ used in Chapter~\ref{section5}. We first recall a few definitions.

Given a discrete cube $\cu := \left( - \frac{R}{2} , \frac{R}{2} \right)^d \cap \Zd$, we recall the definition of the trimmed cube $\cu^-$
\begin{equation*}
    \cu^{-} := \left[ - \frac{R}{2} + \frac{\sqrt{R}}{10} , \frac{R}{2} - \frac{\sqrt{R}}{10} \right]^d.
\end{equation*}
We recall the definition of the dual energy $\mathbf{E}_\cu^*$, for each vector $p^* \in \R^{d \times \binom d2}$ and each map $u \in H^1\left( \cu , \mu_\beta \right)$,
\begin{multline*}
    \mathbf{E}_{\cu}^* \left[ v \right]  = \beta \sum_{y \in \Zd} \left\| \partial_y v \right\|_{L^2 \left( \cu , \mu_\beta \right)}^2 +  \sum_{n \geq 0} \sum_{\dist \left( x , \partial \cu \right) \geq n} \frac1{2\beta^{\frac n2}}\left\| \nabla^{n+1}  v(x , \cdot ) \right\|_{L^2 \left( \mu_\beta \right)}^2  \\ - \frac{1}{\beta^{\frac 14}} \left\| \nabla v \right\|_{L^2 \left( \cu \setminus \cu^-, \mu_\beta\right)}^2  - \beta \sum_{\supp q \subseteq \cu}  \left\langle \nabla_q v \cdot \a_q  \nabla_q v \right\rangle_{\mu_{\beta}}.
\end{multline*}
The energy $\mathbf{E}_{\cu}^*$ satisfies the coercivity and continuity estimates
\begin{equation*}
    c \left\llbracket v \right\rrbracket_{H^1(\cu ,\mu_\beta)} \leq  \mathbf{E}_{\cu}^* \left[ v \right] \leq C \left\llbracket v \right\rrbracket_{H^1(\cu ,\mu_\beta)}.
\end{equation*}
The dual subadditive quantity $\nu^*$ is defined by the formula
\begin{equation} \label{def:defnustarapp}
\nu^* \left( \cu , p^* \right) := \sup_{v \in H^1 \left( \cu , \mu_\beta \right)} - \frac{1}{2|\cu|}\mathbf{E}^*_{\cu}[v] + \frac{1}{|\cu|}  \sum_{x \in \cu} p^* \cdot \left\langle \nabla v(x) \right\rangle_{\mu_\beta}.
\end{equation}
We finally recall that the inverse temperature $\beta$ is chosen large enough so that all the results of Chapter~\ref{section:section4} hold with a regularity exponent $\ep \ll 1$ and that the constants are only allowed to depend on the dimension.

The following proposition states existence and uniqueness of the maximizer of the variational problem~\eqref{def:defnustarapp}.

\begin{proposition}[Solvability of the Neumann problem] \label{prop10282601}
For each $p^* \in \R^{d \times \binom d2}$, there exists a unique maximizer of the variational problem~\eqref{def:defnustarapp} up to a constant. We denote by $v (\cdot , \cdot , \cu , p^*)$ the unique maximizer which satisfies $\left( v (\cdot , \cdot , \cu , p^*) \right)_{\cu ,\mu_\beta} = 0$. Additionally, there exists a constant $C := C(d) > 0$ such that it satisfies the variance estimate
\begin{equation*}
    \sup_{x \in \frac 13 \cu} \var_{\mu_\beta} \left[ v (x , \cdot, \cu , p^*) \right] \leq C.
\end{equation*}
\end{proposition}

\begin{proof}
    The main difficulty in this proof is the absence of the Poincar\'e inequality in both the spatial and field variables. Indeed if one considers a maximizing sequence $(v_n)_{n \in \N}$ in the variational formulation~\eqref{def:defnustarapp}, then one can prove the upper bounds
    \begin{equation*}
        \left\llbracket v_n \right\rrbracket_{\underline{H}^1(\cu , \mu_\beta) } \leq C.
    \end{equation*}
Unfortunately, we do not have a Poincar\'e inequality of the form $\left\| v_n - \left( v_n \right)_{ \cu, \mu_\beta} \right\|_{L^2 \left( \cu, \mu_\beta \right)} \leq C \left\llbracket v_n \right\rrbracket_{\underline{H}^1(\cu , \mu_\beta) } $ as we need to integrate over both the field variable $\phi$ and the spatial variable $x$. This implies that we cannot prove boundedness of the $L^2 \left( \cu , \mu_\beta \right)$-norm of the map $v_n$ and eventually cannot prove the existence of the maximizer using this technique.

To overcome this issue, we introduce a massive term in the variational problem~\eqref{def:defnustarapp}: for each $\lambda > 0$, we define
\begin{equation} \label{def:defnustarapplambda}
    \nu^*_\lambda \left( \cu , p^* \right) := \sup_{v \in H^1 \left( \cu , \mu_\beta \right)} - \frac{1}{2|\cu|}\mathbf{E}^*_{\cu}[v] - \lambda \left\| v \right\|_{\underline{L}^2 \left( \cu , \mu_\beta \right)}^2 + \frac{1}{|\cu|}  \sum_{x \in \cu} p^* \cdot \left\langle \nabla v(x , \cdot) \right\rangle_{\mu_\beta}.
\end{equation}
In this case, it is clear that the maximizer of the variational problem~\eqref{def:defnustarapplambda} exists and is unique up to a constant. We denote this maximizer by $v_\lambda$. Additionally, it is clear that one has the estimate, for any $\lambda >0$,
\begin{equation} \label{eq:TV11220501}
     \sum_{x \in \cu} \left\langle v_\lambda \left( x , \cdot \right) \right\rangle_{\mu_\beta} = 0 \hspace{5mm} \mbox{and} \hspace{5mm} \left\llbracket v_\lambda \right\rrbracket_{\underline{H}^1(\cu ,\mu)} + \lambda \left\| v_\lambda \right\|_{\underline{L}^2 \left( \cu, \mu_\beta \right)} \leq C.
\end{equation}
The objective of the argument is to prove an upper bound on the variance of $v_\lambda (x , \cdot )$ uniform in $x \in \frac 13 \cu$ and $\lambda > 0$: we prove that there exists a constant $C := C(d)$ such that
\begin{equation} \label{eq:TV11230501bis}
    \sup_{\lambda > 0, x \in \frac 13 \cu} \var_{\mu_\beta} \left[ v_\lambda (x , \cdot ) \right] \leq C.
\end{equation}
The proof can thus be decomposed into two steps: proving the estimate~\eqref{eq:TV11230501bis} and proving that the estimate~\eqref{eq:TV11230501bis} implies Proposition~\ref{prop10282601}. We first focus on the second item of the list and show how~\eqref{eq:TV11230501bis} implies the existence and uniqueness of the maximizer $v$. We first use the estimates~\eqref{eq:TV11220501} and ~\eqref{eq:TV11230501bis} to verify that the collection of functions $\left( v_\lambda \right)_{\lambda >0}$ is uniformly bounded in the space $H^1( \cu , \mu_\beta)$. From the estimate~\eqref{eq:TV11220501}, we see that we only need to prove the $L^2\left( \cu , \mu_\beta \right)$-norm estimate
\begin{equation} \label{eq:TV16460501}
    \sup_{\lambda > 0 } \left\| v_\lambda\right\|_{L^2 \left( \cu , \mu_\beta \right)} < \infty.
\end{equation}
To prove the estimate~\eqref{eq:TV16460501}, we first decompose the $L^2 \left( \cu , \mu_\beta \right)$-norm
\begin{align} \label{eq:TV16580501}
    \left\| v_\lambda\right\|_{L^2 \left( \cu , \mu_\beta \right)} & \leq \underbrace{\left\| v_\lambda - \frac1{\left| \cu \right|} \sum_{x \in \cu}  v_\lambda \left( x , \cdot \right) \right\|_{L^2 \left( \cu , \mu_\beta \right)}}_{\eqref{eq:TV16580501}-(i)} + \underbrace{\left\| \frac1{\left| \cu \right|} \sum_{x \in \cu}  v_\lambda \left( x , \cdot \right) - v_\lambda (0 , \cdot)\right\|_{L^2 \left( \cu , \mu_\beta \right)}}_{\eqref{eq:TV16580501}-(ii)} \\
    & \quad + \underbrace{\left\| v_\lambda (0 , \cdot) - \left\langle v_\lambda (0 , \cdot) \right\rangle_{\mu_\beta} \right\|_{L^2 \left( \cu , \mu_\beta \right)}}_{\eqref{eq:TV16580501}-(iii)} + \underbrace{ \left| \left\langle v_\lambda (0 , \cdot) \right\rangle_{\mu_\beta} \right|}_{\eqref{eq:TV16580501}-(iv)}, \notag
\end{align}
and estimate the four terms in the right side separately. For the term~\eqref{eq:TV16580501}-(i), we apply the Poincar\'e inequality for each realization of the field $\phi$. We obtain
\begin{equation*}
    \left\| v_\lambda - \frac1{\left| \cu \right|} \sum_{x \in \cu}  v_\lambda \left( x , \cdot \right) \right\|_{L^2 \left( \cu , \mu_\beta \right)} \leq C R \left\| \nabla v_\lambda  \right\|_{L^2 \left( \cu , \mu_\beta \right)}  \leq CR.
\end{equation*}
For the term~\eqref{eq:TV16580501}-(ii), we let $g$ be the solution of the discrete Neumann problem
\begin{equation*}
     \left\{ \begin{aligned}
    -\Delta g = \delta_0 - \frac{1}{\left| \cu \right|} &~\mbox{in}~ \cu, \\
    \mathbf{n} \cdot \nabla g = 0 &~\mbox{on}~ \partial \cu.
    \end{aligned} \right.
\end{equation*}
The solvability is ensured by the identity $\sum_{x \in \cu} \left( \delta_0(x) - \frac{1}{\left|\cu \right|} \right) = 0$. Using the function $g$, we write
\begin{align*}
    \left\| \frac1{\left| \cu \right|} \sum_{x \in \cu}  v_\lambda \left( x , \cdot \right) - v_\lambda (0 , \cdot)\right\|_{L^2 \left( \cu , \mu_\beta \right)} & = \left\| \sum_{x \in \cu} \nabla v_\lambda \left( x , \cdot \right) \cdot \nabla g(x) \right\|_{L^2 \left( \cu , \mu_\beta \right)} \\
    & \leq \left\| \nabla v_\lambda  \right\|_{L^2 \left( \cu , \mu_\beta \right)} \left\| \nabla g \right\|_{L^2 \left( \cu , \mu_\beta \right)} \\
    & \leq C \left| \cu \right|^\frac 12 \left\| \nabla g \right\|_{L^2 \left( \cu , \mu_\beta \right)}.
\end{align*}
For the term~\eqref{eq:TV16580501}-(iii), we observe that
\begin{equation*}
    \left\| v_\lambda (0 , \cdot) - \left\langle v_\lambda (0 , \cdot) \right\rangle_{\mu_\beta} \right\|_{L^2 \left( \cu , \mu_\beta \right)} = \var_{\mu_\beta} \left[ v_\lambda (0 , \cdot) \right] \leq C.
\end{equation*}
For the term~\eqref{eq:TV16580501}-(iv), we use identity of~\eqref{eq:TV11220501} and the estimate for the term~\eqref{eq:TV16580501}-(ii) to write
\begin{align*}
    \left| \left\langle v_\lambda (0 , \cdot) \right\rangle_{\mu_\beta} \right| = \left| \left\langle v_\lambda (0 , \cdot) \right\rangle_{\mu_\beta} - \left\langle \frac{1}{\left| \cu \right|} \sum_{x \in \cu} v_\lambda (x , \cdot) \right\rangle_{\mu_\beta} \right| & \leq \left\| \frac1{\left| \cu \right|} \sum_{x \in \cu}  v_\lambda \left( x , \cdot \right) - v_\lambda (0 , \cdot)\right\|_{L^2 \left( \cu , \mu_\beta \right)} \\ 
    & \leq C \left| \cu \right|^\frac 12 \left\| \nabla g \right\|_{L^2 \left( \cu , \mu_\beta \right)}.
\end{align*}
A combination of the previous displays implies the estimate~\eqref{eq:TV16460501}.

Since the $H^1( \cu , \mu_\beta)$-norm of the family $\left( v_\lambda \right)_{\lambda > 0}$ is bounded uniformly in $\lambda$, we can extract a subsequence which converges weakly to a map $v \in H^1(\mu_\beta)$ as $\lambda$ goes to $0$. One can then verify that the map $v \in H^1(\mu, \beta)$ is a maximizer in the variational formulation~\eqref{def:defnustarapp} and satisfies the estimates
\begin{equation} \label{eq:TV19020501}
     \left( v \right)_{\cu, \mu_\beta} = 0, \hspace{5mm} \left\llbracket v \right\rrbracket_{\underline{H}^1(U ,\mu)} \leq C \hspace{5mm} \mbox{and} \hspace{5mm} \sup_{x \in \frac 13 \cu} \var_{\mu_\beta} \left[ v (x , \cdot) \right] \leq C.
\end{equation}
It only remains to prove~\eqref{eq:TV11230501bis}. The argument is similar to the proof of Lemma~\ref{lemmvarest} in Chapter~\ref{section5}. We fix a point $x \in \frac 13 \cu$ and apply the Brascamp-Lieb inequality (properly rescaled with respect to the inverse temperature $\beta$) to obtain
\begin{equation*}
    \var_{\mu_\beta} \left[ v_\lambda (x ,\cdot ) \right] \leq C \beta \sum_{y , z \in \Zd} \frac{\left\| \partial_y v_\lambda (x ,\cdot ) \right\|_{L^2 \left( \mu_\beta \right)} \left\| \partial_z v_\lambda (x ,\cdot )  \right\|_{L^2 \left( \mu_\beta \right)}}{|y - z|^{d-2}}.
\end{equation*}
Using the definition of $v_\lambda$ as the maximizer of the variational problem~\eqref{def:defnustarapplambda}, one sees that it is a solution to the Hellfer-Sj\"{o}strand equation
\begin{equation}
\label{e.HS.eqn.neubis}
\left\{ \begin{aligned}
\beta \Delta_\phi v_\lambda + \beta \mathcal{L}_{\cu} v_\lambda + \lambda v_\lambda & = 0 &~\mbox{in}~ \cu \times \Omega,\\
\mathbf{n} \cdot \nabla v_\lambda & = \mathbf{n} \cdot p^* &~\mbox{on}~ \partial \cu \times \Omega,
\end{aligned} \right.
\end{equation}
where the operator $\mathcal{L}_{\cu}$ is the uniformly elliptic operator defined by the formula 
\begin{equation*}
    \mathcal{L}_{\cu} := - \frac{1}{2 \beta} \Delta +  \frac 1{2\beta}\sum_{n \geq 1} \frac1{\beta^{\frac n2}} \nabla^{n+1} \cdot \left( \indc_{\cu} \nabla^{n+1} \right) + \frac{1}{\beta^{\frac 54}} \nabla \cdot \left(\indc_{\cu \setminus \cu^-} \nabla\right) + \sum_{\supp q \subseteq \cu} \nabla_q \cdot \a_q \nabla_q.
\end{equation*}
As is mentioned in the proof of Lemma~\ref{lemmvarest} of Chapter~\ref{section5}, the operator $\mathcal{L}_{\cu}$ is a perturbation of the Laplacian $\frac1{2\beta} \Delta$. Consequently, the same arguments as the ones developed in Chapter~\ref{section:section4} apply and the same regularity results hold.

Applying the operator $\partial$ to the equation~\eqref{e.HS.eqn.neubis} as it is done in the proof of Lemma~\ref{lemmvarest}, we obtain that the map $w_\lambda : (y , z , \phi) \to \partial_z v_\lambda(y , \phi) $ is a solution of the differentiated Helffer-Sj\"{o}strand equation
\begin{equation*}
    \left\{ \begin{aligned}
    \beta \Delta_\phi w_\lambda + \beta \mathcal{L}_{\cu, y} w_\lambda +  \beta \mathcal{L}_{\mathrm{spat}, z} w_\lambda + \lambda w_\lambda & = \beta \sum_{\supp q \subseteq \cu}  z \left( \beta , q\right) \sin 2\pi\left( \phi , q \right) \left( v_\lambda, q \right) q_y \otimes q_z &~\mbox{in}~ \cu \times \Zd \times \Omega, \\
    \mathbf{n} \cdot \nabla_y w_\lambda & = 0  &~\mbox{on}~ \partial \cu \times \Zd \times \Omega.
    \end{aligned} \right.
\end{equation*}
The proof is then almost identical to the proof of Lemma~\ref{lemmvarest} of Chapter~\ref{section5}; we use the reflection principle to express the function $w_\lambda$ in terms of the Green's function associated to the differentiated Hellfer-Sj\"{o}strand operator $\Delta_\phi +  \mathcal{L}_{\cu, y} +  \mathcal{L}_{\mathrm{spat}, z} + \lambda I_d$. There are two main differences in the argument which are listed below:
\begin{itemize}
    \item One needs to study the map $w_\lambda$ and not its gradient; this simplifies the computations.
    \item One needs to study the Green's function associated to the differentiated operator $\Delta_\phi +  \mathcal{L}_{\cu, y} +  \mathcal{L}_{\mathrm{spat}, z} + \lambda I_d$ and the weight $\lambda I_d$ has to be taken into account. This can be achieved with the same strategy as the one presented in Chapter~\ref{section:section4}. If we consider a function $\mathbf{f} \in L^2 \left( \mu_\beta \right)$, a pair of points $(x_1 , y_1) \in \Zd \times \Zd$ and let $\mathcal{G}_{\mathrm{der} , \f}^\lambda(\cdot , \cdot , \cdot ; x_1 , y_1)$ be the solution of the weighted equation
    \begin{equation*}
        \Delta_\phi \mathcal{G}_{\mathrm{der} , \f}^\lambda +  \mathcal{L}_{\cu, y} \mathcal{G}_{\mathrm{der} , \f}^\lambda +  \mathcal{L}_{\mathrm{spat}, z} \mathcal{G}_{\mathrm{der} , \f}^\lambda + \frac{\lambda}{\beta} \mathcal{G}_{\mathrm{der} , \f}^\lambda = \mathbf{f} \delta_0,
    \end{equation*}
    then an application of the Feynman-Kac formula shows the identity
    \begin{equation*}
        \mathcal{G}_{\mathrm{der} , \f}^\lambda(x , y , \phi; x_1 , y_1) = \beta^{-1} \int_0^\infty e^{-\frac{\lambda}{\beta} t} \E_\phi \left[ \mathbf{f}\left( \phi_t \right) P^{\phi_\cdot}_\cu\left(t , x \, ; x_1 \right) \otimes P^{\phi_\cdot}\left(t ,  y \, ; y_1\right)  \right] \, dt,
    \end{equation*}
    where we used the notations introduced at the begining of Chapter~\ref{section:section4} and where, given a realization of the dynamics $\left( \phi_t \right)_{t \geq 0}$, the map $P^{\phi_\cdot}_\cu(\cdot , \cdot ; y)$ is the solution of the parabolic equation
    \begin{equation*}
        \left\{\begin{aligned}
        \partial_t P^{\phi_\cdot}_\cu(\cdot , \cdot ; y) - \mathcal{L}_{\cu, y} P^{\phi_\cdot}_\cu(\cdot , \cdot ; y) = 0 &~ \mbox{in}~ (0 , \infty) \times \Zd, \\
        P^{\phi_\cdot}_\cu(0 , \cdot ; y) = \delta_{0} &~\mbox{on}~ \Zd. 
        \end{aligned} \right.
    \end{equation*}
    We then apply the estimate of Proposition~\ref{prop:prop4.7} of Chapter~\ref{section:section4}, which applies to both functions $P^{\phi_\cdot}_\cu(\cdot , \cdot ; y)$ and $P^{\phi_\cdot}(\cdot , \cdot ; y)$, to obtain the bound
    \begin{align} \label{eq:TV14510501}
        \left\| \mathcal{G}_{\mathrm{der} , \f}^\lambda(x , y , \cdot ; x_1 , y_1)  \right\|_{L^p \left( \mu_\beta \right)} & \leq C \beta^{-1} \left\|\mathbf{f} \right\|_{L^p \left( \mu_\beta \right)} \int_0^\infty e^{-\frac{\lambda}{\beta} t}  \Phi_C \left( \frac{t}{\beta},x-y \right) \Phi_C \left( \frac{t}{\beta},x_1-y_1 \right) \, dt \\
        & \leq C \beta^{-1} \left\|\mathbf{f} \right\|_{L^p \left( \mu_\beta \right)} \int_0^\infty \Phi_C \left( \frac{t}{\beta},x-y \right) \Phi_C \left( \frac{t}{\beta},x_1-y_1 \right) \, dt \notag \\
        & \leq \frac{C}{\left| x - y \right|^{2d-2} + \left| x_1 - y_1 \right|^{2d-2}}. \notag
    \end{align}
    In particular, the estimate~\eqref{eq:TV14510501} is uniform in the weight $\lambda$.
\end{itemize}
\end{proof}

\chapter{Basic estimates on discrete convolutions} \label{app.appC}

The objective of this appendix is to prove estimates on some discrete convolutions of functions decaying algebraically fast at infinity.

In the first proposition of this appendix, we consider two exponents $\alpha, \beta > 0$ such that $\alpha + \beta > d$ and prove estimates on the function 
\begin{equation*}
    y \to \sum_{x \in \Zd} \frac{1}{|x|^\alpha} \frac{1}{|x - y|^\beta}.
\end{equation*}
We distinguish different cases depending on the values of the exponents $\alpha$ and $\beta$; the results are collected in the following proposition.

\begin{proposition} \label{propappC}
    Given a pair of exponents $\alpha, \beta > 0$ such that $\alpha + \beta > d$ and a point $y \in \Zd$, one has the estimates
    \begin{itemize}
        \item[(i)] If $\alpha \in (0,d)$ and $\beta \in (0 , d)$,
        \begin{equation*}
            \sum_{x \in \Zd} \frac{1}{|x|^\alpha} \frac{1}{|x - y|^\beta} \leq \frac{C}{|y|^{\alpha + \beta - d}};
        \end{equation*}
        \item[(ii)] If $\alpha = d$ and $\beta \in (0 , d]$,
        \begin{equation*}
            \sum_{x \in \Zd} \frac{1}{|x|^\alpha} \frac{1}{|x - y|^\beta} \leq \frac{C\ln \left| y \right| }{|y|^{\beta}};
        \end{equation*}
        \item[(iii)] If $\alpha > d$ and $\beta \in (0 , \infty )$,
        \begin{equation*}
            \sum_{x \in \Zd} \frac{1}{|x|^\alpha} \frac{1}{|x - y|^\beta} \leq \frac{C}{|y|^{\min \left( \alpha, \beta \right)}}.
        \end{equation*}
    \end{itemize}
\end{proposition}

\begin{proof}
The proof of the points (ii) and (iii) can be found in ~\cite[Appendix]{MO16}. We only prove (i) and decompose the space into three regions according to the formula
\begin{equation} \label{eq:TV09011400}
    \sum_{x \in \Zd} \frac{1}{|x|^\alpha} \frac{1}{|x - y|^\beta} = \underbrace{\sum_{x \in B\left(\left|y \right|/2\right)} \frac{1}{|x|^\alpha} \frac{1}{|x - y|^\beta} }_{\eqref{eq:TV09011400}-(i)} +  \underbrace{\sum_{x \in B\left(y , \left| y \right|/2\right)}  \frac{1}{|x|^\alpha} \frac{1}{|x - y|^\beta} }_{\eqref{eq:TV09011400}-(ii)} +  \underbrace{\sum_{x \in \Zd \setminus \left(  B\left(\left| y \right| / 2\right) \cup B\left(y, \left| y \right| / 2\right) \right)}  \frac{1}{|x|^\alpha} \frac{1}{|x - y|^\beta}. }_{\eqref{eq:TV09011400}-(iii)}
\end{equation}
For the term~\eqref{eq:TV09011400}-(i), we use the inequality $|y - x| \geq \frac 12 |y|$, valid for any point $x \in B\left(\left|y \right|/2\right)$. We obtain
\begin{equation*}
    \sum_{x \in B\left(\left|y \right|/2\right)} \frac{1}{|x|^\alpha} \frac{1}{|x - y|^\beta}  \leq \frac{C}{|y|^\beta} \sum_{x \in B\left(\left|y \right|/2\right)} \frac{1}{|x|^\alpha} \leq \frac{C}{|y|^{\alpha + \beta -d}}
\end{equation*}
The term~\eqref{eq:TV09011400}-(ii) is estimated similarly, we use this time the inequality $|x| \geq \frac 12 |y|$, valid for any point $x \in B\left(y , \left|y \right|/2\right)$, and obtain
\begin{equation*}
    \sum_{x \in B\left(y , \left|y \right|/2\right)} \frac{1}{|x|^\alpha} \frac{1}{|x - y|^\beta}  \leq \frac{C}{|y|^\beta} \sum_{x \in B\left(y , \left|y \right|/2\right)} \frac{1}{|y -x|^\alpha} \leq \frac{C}{|y|^{\alpha + \beta -d}}.
\end{equation*}
For the term~\eqref{eq:TV09011400}-(iii), we use the inequality $\left| y - x \right| \geq c |x|$, valid for any point $x \notin B\left(\left| y \right| / 2\right) \cup B\left(y, \left| y \right| / 2\right)$, and obtain
\begin{align*}
    \sum_{x \in \Zd \setminus \left(  B\left(\left| y \right| / 2\right) \cup B\left(y, \left| y \right| / 2\right) \right)}  \frac{1}{|x|^\alpha} \frac{1}{|x - y|^\beta} & \leq C \sum_{x \in \Zd \setminus \left(  B\left(\left| y \right| / 2\right) \cup B\left(y, \left| y \right| / 2\right) \right)}  \frac{1}{|x|^\alpha} \frac{1}{|x|^\beta} \\ & \leq  C \sum_{x \in \Zd \setminus   B\left(\left| y \right| / 2\right)}  \frac{1}{|x|^{\alpha + \beta}} \\ & \leq \frac{C}{|y|^{\alpha+ \beta - d}}.
\end{align*}
\end{proof}

We record as a corollary three estimates which are used in~\eqref{eq:TV14460901} of Chapter~\ref{section3.4} and~\eqref{eq:TV08095} of Chapter~\ref{section5}.

\begin{corollary} \label{coro.appCcoro}
One has the estimates, for any point $x \in \Zd$,
\begin{equation} \label{eq:TV14480901}
         \sum_{z_1 , z_2 \in \Zd} \frac{1}{|x - z_1|^d}  \frac{1}{|z_1 - z_2|^{d - \ep}}   \frac{1}{|z_2|^{d-1}}  \leq \frac{C \ln \left| x \right|}{|x|^{d - 1 - \ep}}
\end{equation}
and
\begin{equation}  \label{eq:TV14470901}
     \sum_{z_1 , z_2 \in \Zd} \frac{1}{|x - z_1|^{d-1}} \times \frac{1}{|z_1 - z_2|^{d - \ep}} \frac{1}{|z_2|^{d}}  \leq \frac{C}{|x|^{d - 1 - \ep}}.
\end{equation}
For any pair of points $x , x' \in \Zd$, one has the estimate
\begin{equation} \label{eq:TV14530901}
     \sum_{y , y' \in \Zd} \frac{1}{|y'-x'|^{d+\frac 34}}  \frac{1}{|y-x|^{d+\frac 34}} \frac{1}{|y - y'|^{d-2}} \leq \frac{C}{|x - x'|^{d - 2 - \ep}}.
\end{equation}
\end{corollary}

\begin{proof}
To prove~\eqref{eq:TV14480901}, we apply Proposition~\ref{propappC} twice and obtain
\begin{equation*}
    \sum_{z_1 , z_2 \in \Zd} \frac{1}{|x - z_1|^d}  \frac{1}{|z_1 - z_2|^{d - \ep}}   \frac{1}{|z_2|^{d-1}}  \leq \sum_{z_1  \in \Zd} \frac{1}{|x - z_1|^d}  \frac{1}{|z_1 |^{d - 1 - \ep}} \leq \frac{C \ln \left| x\right|}{|x|^{d - 1 - \ep}}.
\end{equation*}
The proof of the estimate~\eqref{eq:TV14470901} is similar.
We now prove~\eqref{eq:TV14530901}. By the change of variables $y' := y' - x'$ and $y := y - x'$, one has the identity
\begin{equation*}
    \sum_{y , y' \in \Zd} \frac{1}{|y'-x'|^{d+\frac 34}}  \frac{1}{|y-x|^{d+\frac 34}} \frac{1}{|y - y'|^{d-2}} =  \sum_{y , y' \in \Zd} \frac{1}{|y'|^{d+\frac 34}}  \frac{1}{|y-x + x'|^{d+\frac 34}}  \frac{1}{|y - y'|^{d-2}}.
\end{equation*}
We apply Proposition~\ref{propappC} twice to obtain
\begin{equation*}
    \sum_{y , y' \in \Zd} \frac{1}{|y'|^{d+\frac 34}}  \frac{1}{|y-x + x'|^{d+\frac 34}}  \frac{1}{|y - y'|^{d-2}} \leq C \sum_{y' \in \Zd} \frac{1}{|y-x + x'|^{d+\frac 34}}  \frac{1}{|y - y'|^{d-2}} \leq \frac{C}{|x - x'|^{d - 2 - \ep}}.
\end{equation*}
\end{proof}

The following proposition is used in~\eqref{eq:TV10006} of Chapter~\ref{sec:section6} and~\eqref{eq:TV09247} of Chapter~\ref{section7}.

\begin{proposition} \label{propappClign6572}
One has the estimates, for each pair of points $y , z \in \Zd$,
\begin{equation} \label{eq:TV16041001}
    \sum_{y_1 \in \Zd} \frac{1}{\left| y_1 \right|^{d-1}} \frac{1}{|y - z|^{2d+1 - \ep} + \left| \frac{y + z}{2} - y_1 \right|^{2d + 1 - \ep}}  \leq \frac{C}{\left| y - z\right|^{d+1}\max \left( \left|y\right| , \left|z\right| \right)^{d -1 -\ep}}.
\end{equation}
and
\begin{equation*}
     \sum_{y_1 \in \Zd} \frac{1}{\left| y_1 \right|^{d- \ep}} \frac{1}{|y - z|^{2d - \ep} + \left| \frac{y + z}{2} - y_1 \right|^{2d - \ep}} \leq \frac{C}{\left| y - z\right|^{d}\max \left( \left|y\right| , \left|z\right| \right)^{d - 2 \ep}}
\end{equation*}
\end{proposition}

\begin{proof}
We only prove~\eqref{eq:TV16041001} and first show the estimate: for each real number $a \geq 1$ and each point $x \in \Zd$,
\begin{equation} \label{eq:TV14521001}
    \sum_{y_1 \in \Zd} \frac{1}{\left| y_1 \right|^{d-1}} \frac{1}{a^{2d + 1 - \ep} + \left| x - y_1 \right|^{2d + 1 - \ep}} \leq \frac{C}{a^{d+1}\max \left( |x| , a \right)^{d -1 -\ep}}.
\end{equation}
To prove the inequality~\eqref{eq:TV14521001}, we distinguish two cases.

\smallskip

\textbf{Case 1. $a \geq |x|$.} In that case, there exists a constant $c := c(d) > 0$ such that $a^{2d+1-\ep} + |y_1 - x|^{2d+1-\ep} \geq c \left( a^{2d+1-\ep} + |y_1|^{2d+1-\ep} \right)$ for any point $y_1 \in \Zd$. Using this inequality, we compute
\begin{align*}
    \sum_{y_1 \in \Zd} \frac{1}{\left| y_1 \right|^{d-1}} \frac{1}{a^{2d+1-\ep} + \left| x - y_1 \right|^{2d + 1 - \ep}} & \leq \sum_{y_1 \in \Zd} \frac{1}{\left| y_1 \right|^{d-1}} \frac{1}{a^{2d+1-\ep} + \left| x - y_1 \right|^{2d + 1 - \ep}} \\ & \leq \sum_{y_1 \in B(0 , a)}  \frac{1}{\left| y_1 \right|^{d-1}} \frac{1}{a^{2d+1-\ep} + \left| y_1 \right|^{2d + 1 - \ep}}  + \sum_{y_1 \in \Zd \setminus B(0, a)}  \frac{1}{\left| y_1 \right|^{d-1}} \frac{1}{a^{2d + 1 - \ep} + \left| y_1 \right|^{2d + 1 - \ep}}.
\end{align*}
We estimate the two terms in the right side separately. For the first term, we write
\begin{equation*}
    \sum_{y_1 \in B(0 , a)}  \frac{1}{\left| y_1 \right|^{d-1}} \frac{1}{a^{2d - 1 - \ep} + \left| y_1 \right|^{2d + 1 - \ep}}  \leq  \sum_{y_1 \in B(0 , a)}  \frac{1}{\left| y_1 \right|^{d-1}} \frac{2}{a^{2d + 1 - \ep}} \leq \frac{C}{a^{2d - \ep}}.
\end{equation*}
For the second term, we write
\begin{equation*}
    \sum_{y_1 \in \Zd \setminus B(0, a)}  \frac{1}{\left| y_1 \right|^{d-1}} \frac{1}{a^{2d + 1 - \ep} + \left| y_1 \right|^{2d + 1 - \ep}} \leq \sum_{y_1 \in \Zd \setminus B(0, a)}  \frac{1}{\left| y_1 \right|^{d-1}} \frac{1}{\left| y_1 \right|^{2d + 1 - \ep}} \leq  \frac{C}{a^{2d-\ep}}.
\end{equation*}
A combination of the three previous displays completes the proof of~\eqref{eq:TV14521001} in the case $a \geq |x|$.

\smallskip

\textbf{Case 2. $a \leq |x|$.} In that case, an application of Young's inequality yields the estimate $a^{2d + 1 - \ep} + \left| y_1 - x \right|^{2d + 1 - \ep} \geq c a^{d+1} \left| y_1 - x \right|^{d - \ep}$. We deduce that
\begin{equation*}
    \sum_{y_1 \in \Zd} \frac{1}{\left| y_1 \right|^{d-1}} \frac{1}{a^{2d - 1 - \ep} + \left| x - y_1 \right|^{2d - 1 - \ep}} \leq C \sum_{y_1 \in \Zd} \frac{1}{\left| y_1 \right|^{d-1}} \frac{1}{a^{d + 1 } \left| x - y_1 \right|^{d - \ep}}.
\end{equation*}
We apply Proposition~\ref{propappC} to obtain
\begin{equation*}
    \sum_{y_1 \in \Zd} \frac{1}{\left| y_1 \right|^{d-1}} \frac{1}{\left| x - y_1 \right|^{d - \ep}} \leq \frac{C}{\left| x \right|^{d-1 - \ep}}.
\end{equation*}
A combination of the two previous displays completes the proof of~\eqref{eq:TV14521001} in the case $a \leq |x|$.

\smallskip

We now prove the inequality~\eqref{eq:TV16041001}. Applying the inequality~\eqref{eq:TV14521001} with the choices $a = |y - z|$ and $x = \frac{y + z}{2}$, we obtain
\begin{equation*}
     \sum_{y_1 \in \Zd} \frac{1}{\left| y_1 \right|^{d-1}} \frac{1}{|y - z|^{2d+1 - \ep} + \left| \frac{y + z}{2} - y_1 \right|^{2d + 1 - \ep}}  \leq \frac{C}{\left| y - z\right|^{d+1}\max \left( \left| \frac{y + z}{2} \right| , \left| y - z \right| \right)^{d -1 -\ep}}.
\end{equation*}
We complete the proof of~\eqref{eq:TV16041001} by using the estimate $\max \left( \left| \frac{y + z}{2} \right| , \left| y - z \right| \right) \geq \frac 14 \max \left( |y| , |z|\right)$.

\end{proof}

The next proposition of this appendix is used in~\eqref{eq:TV16560901} of Chapter~\ref{section7}.
\begin{proposition}
One has the estimate, for any point $x \in \Zd$,
\begin{equation} \label{eq:TV14021001}
    \sum_{y,z \in \Zd} \frac{1}{|y|^{d-1} |x-y|^{d-1}} \frac{1}{|z|^{d-1}|z - x|^{d-1}} \frac{1}{|y - z|^{d- \ep}} \leq \frac{C}{|x|^{2d-2}}.
\end{equation}
\end{proposition}

\begin{proof}
We first fix a point $y \in \Zd$ and prove the estimate
\begin{equation} \label{eq:TV12291001}
    \sum_{z \in \Zd } \frac{1}{|z|^{d-1}|z - x|^{d-1}} \frac{1}{|y - z|^{d- \ep}} \leq 
    \left\{ \begin{aligned}
         \frac{C}{|x|^{d-1} \left| y \right|^{d - 1 -\ep}} &~\mbox{if}~ \left| y\right| \leq \frac{\left| x \right|}{4},\\
          \frac{C}{|x|^{d-1} \left| y - x \right|^{d - 1 -\ep}} &~\mbox{if}~ \left| y -x \right| \leq \frac{\left| x \right|}{4}\\
          \frac{C}{|x|^{2d-2-\ep}}, &~\mbox{otherwise}.
    \end{aligned} \right.
\end{equation}
In the case $\left| y \right| \leq \frac{\left| x \right|}{4}$, we split the sum according to the formula
\begin{align} \label{eq:TV090114001}
    \sum_{z \in \Zd } \frac{1}{|z|^{d-1}|z - x|^{d-1}} \frac{1}{|y - z|^{d- \ep}} & =  \underbrace{\sum_{z \in B\left(0,\left|x \right|/2\right)} \frac{1}{|z|^{d-1}|z - x|^{d-1}} \frac{1}{|y - z|^{d- \ep}} }_{\eqref{eq:TV090114001}-(i)} +  \underbrace{\sum_{z \in B\left(x , \left| x \right|/2\right)}  \frac{1}{|z|^{d-1}|z - x|^{d-1}} \frac{1}{|y - z|^{d- \ep}} }_{\eqref{eq:TV090114001}-(ii)} \\ & \quad +  \underbrace{\sum_{z \in \Zd \setminus \left(  B\left(0,\left| x \right| / 2\right) \cup B\left(x, \left| x \right| / 2\right) \right)} \frac{1}{|z|^{d-1}|z - x|^{d-1}} \frac{1}{|y - z|^{d- \ep}}. }_{\eqref{eq:TV090114001}-(iii)} \notag
\end{align}
For the term~\eqref{eq:TV090114001}-(i), we use that $\left| z - x \right| \geq \left| x \right| / 2$ if $z \in B\left(0,\left|x \right|/2\right)$. We obtain
\begin{equation*}
    \sum_{z \in B\left(0,\left|x \right|/2\right)} \frac{1}{|z|^{d-1}|z - x|^{d-1}} \frac{1}{|y - z|^{d- \ep}} \leq \frac{1}{|x|^{d-1}} \sum_{z \in B\left(0,\left|x \right|/2\right)} \frac{1}{|z|^{d-1}} \frac{1}{|y - z|^{d- \ep}} \leq \frac{1}{|x|^{d-1}} \sum_{z \in \Zd} \frac{1}{|z|^{d-1}} \frac{1}{|y - z|^{d- \ep}} \leq \frac{1}{|x|^{d-1} \left| y \right|^{d - 1 -\ep}},
\end{equation*}
where we used Proposition~\ref{propappC} in the third inequality.

For the term~\eqref{eq:TV090114001}-(ii), we use the inequalities $\left| z \right| \geq \left| x \right| / 2$ and $\left| z - y\right| \geq |x| / 4$ valid for any point $ z \in B(x , |x|/2)$ under the assumption $|y| \leq \frac{|x|}{4}$. We obtain
\begin{equation*}
    \sum_{z \in B\left(x , \left| x \right|/2\right)}  \frac{1}{|z|^{d-1}|z - x|^{d-1}} \frac{1}{|y - z|^{d- \ep}}  \leq \frac{C}{|x|^{2d - \ep}}\sum_{z \in B\left(x , \left| x \right|/2\right)}  \frac{1}{|z - x|^{d-1}} \leq \frac{C}{|x|^{2d -1 - \ep}}.
\end{equation*}
For the term~\eqref{eq:TV090114001}-(iii), we use the estimates $\left|z - x \right| \geq c \left| z\right|$ and $\left|y - x \right| \geq c \left| z\right|$, valid for any point $ z \in \Zd \setminus \left( B (|x|/2) \cup B(x , |x|/2) \right)$ under the assumption $|y| \leq \frac{|x|}{4}$. We obtain
\begin{align*}
    \sum_{z \in \Zd \setminus \left(  B\left(\left| x \right| / 2\right) \cup B\left(x, \left| x \right| / 2\right) \right)} \frac{1}{|z|^{d-1}|z - x|^{d-1}} \frac{1}{|y - z|^{d- \ep}} & \leq \sum_{z \in \Zd \setminus  B\left(\left| x \right| / 2\right) \cup B\left(x, \left| x \right| / 2\right) } \frac{1}{|z|^{3d-2-\ep}} \\ & \leq \sum_{z \in \Zd \setminus B\left(x, \left| x \right| / 2\right) } \frac{1}{|z|^{3d-2-\ep}} \\ & \leq \frac{C}{|x|^{2d-2 - \ep}}.
\end{align*}
A combination of the four previous displays and the assumption $|y| \leq \frac{\left| x\right|}{4}$ yields
\begin{equation*}
    \sum_{z \in \Zd} \frac{1}{|z|^{d-1}|z - x|^{d-1}} \frac{1}{|y - z|^{d- \ep}} \leq \frac{C}{|x|^{d-1} \left| y \right|^{d - 1 -\ep}},
\end{equation*}
which completes the proof of~\eqref{eq:TV12291001} in the case $|y| \leq |x| /4$. The proof of~\eqref{eq:TV12291001} in the case $|y - x| \leq |x| /4$ can be reduced to the case $|x| \leq |x| /4$ by performing the change of variable $z := z - y$.

There only remains to prove the estimate in the case~\eqref{eq:TV12291001} in the third case. We again split the sum into four terms 
\begin{align*}
    \sum_{z \in \Zd } \frac{1}{|z|^{d-1}|z - x|^{d-1}} \frac{1}{|y - z|^{d-\ep}}  & = \sum_{z \in B\left(0,\left|x \right|/8\right)} \frac{1}{|z|^{d-1}|z - x|^{d-1}} \frac{1}{|y - z|^{d- \ep}}  +  \sum_{z \in B\left(x , \left| x \right|/8\right)}  \frac{1}{|z|^{d-1}|z - x|^{d-1}} \frac{1}{|y - z|^{d- \ep}} \\ & \quad  +  \sum_{z \in  B\left(y , \left| x \right| /8\right) } \frac{1}{|z|^{d-1}|z - x|^{d-1}} \frac{1}{|y - z|^{d- \ep}}
    \\ & \quad +  \sum_{z \in \Zd \setminus \left(  B\left(0,\left| x \right| /8\right) \cup B\left(x, \left| x \right| /8\right) \cup B\left(y , \left| x \right| /8\right) \right)} \frac{1}{|z|^{d-1}|z - x|^{d-1}} \frac{1}{|y - z|^{d- \ep}}.
\end{align*}
We then estimate the four terms in the right side separately:
\begin{itemize}
    \item[(i)] For the first term, we use the inequalities $|y - z| \geq c |x| $ and $|z - x| \geq c |x|$;
    \item[(ii)] For the second term, we use the inequalities $| z| \geq c |x| $ and $|y - z| \geq c |x|$;
    \item[(iii)] For the third term, we use the inequalities $| z| \geq c |x| $ and $|z - x| \geq c |x|$;
    \item[(iv)] For the fourth term, we use the inequalities $|y - z| \geq c |x| $ and $|z - x| \geq c |z|$.
\end{itemize}
We obtain
\begin{align*}
    \sum_{z \in \Zd } \frac{1}{|z|^{d-1}|z - x|^{d-1}} \frac{1}{|y - z|^{d-\ep}}  & \leq  \frac{C}{| x|^{2d-1-\ep}}\sum_{z \in B\left(\left|x \right|/8\right)} \frac{1}{|z|^{d-1}} + \frac{1}{|x|^{2d-1-\ep}} \sum_{z \in B\left(x , \left| x \right|/8\right)}  \frac{1}{|z - x|^{d-1}}  \\ & +  \frac{1}{|x|^{2d-2}}\sum_{z \in  B\left(y , \left| x \right| /8\right) } \frac{1}{|y - z|^{d- \ep}}
    \\ & \quad + \frac{1}{|x|^{d- \ep} } \sum_{z \in \Zd \setminus \left(  B\left(\left| x \right| /8\right) \cup B\left(x, \left| x \right| /8\right) \cup B\left(y , \left| x \right| /8\right) \right)} \frac{1}{|z|^{2d-2}}  \\
    & \leq  \frac{C}{|x|^{2d-2-\ep}} + \frac{1}{|x|^{d- \ep}  } \sum_{z \in \Zd \setminus B\left(\left| x \right| /8\right)} \frac{1}{|z|^{2d-2}} \\
    & \leq \frac{C}{|x|^{2d-2-\ep}}.
\end{align*}
The proof of the estimate~\eqref{eq:TV12291001} is complete. We now complete the proof of~\eqref{eq:TV14021001}. By applying the estimate~\eqref{eq:TV12291001}, we obtain the inequality
\begin{align*}
     \sum_{y,z \in \Zd} \frac{1}{|y|^{d-1} |x-y|^{d-1}} \frac{1}{|z|^{d-1}|z - x|^{d-1}} \frac{1}{|y - z|^{d- \ep}} & \leq \frac{C}{|x|^{d-1}} \sum_{y \in B(0,|x|/4)}\frac{1}{|y|^{2d-2 -\ep} |x-y|^{d-1}} \\ & \quad +  \frac{C}{|x|^{d-1}} \sum_{y \in B(x , |x|/4)} \frac{1}{|y|^{d-1} |x-y|^{2d-2-\ep}} \\& \quad +  \frac{C}{|x|^{2d-2-\ep}} \sum_{y \in \Zd \setminus \left( B(0,|x|/4) \cup B(x , |x|/4) \right)} \frac{1}{|y|^{d-1} |x-y|^{d-1}}.
\end{align*}
We estimate each term on the right side by applying Proposition~\ref{propappC}. We obtain
\begin{equation*}
    \sum_{y,z \in \Zd} \frac{1}{|y|^{d-1} |x-y|^{d-1}} \frac{1}{|z|^{d-1}|z - x|^{d-1}} \frac{1}{|y - z|^{d- \ep}} \leq \frac{C}{|x|^{2d-2}}.
\end{equation*}
The proof of~\eqref{eq:TV14021001} is complete.

\end{proof}

The following estimate is used in~\eqref{eq:TV110477} and~\eqref{eq:TV13546} of Chapter~\ref{section5} with the exponent $\alpha = 2d + \frac 34$.

\begin{proposition} \label{propappCp97}
One has the estimate, for each point $y \in \Zd$ and each exponent $\alpha > d$,
\begin{equation} \label{eq:TV16440901}
    \sum_{y_0 \in \Zd} \frac{1}{\left| y_0 \right|^{\alpha} + \left| y_0 - y \right|^{\alpha}} \leq \frac{C}{\left| y\right|^{\alpha - d}},
\end{equation}
where the constant $C$ depends on the parameters $\alpha$ and $d$.
\end{proposition}

\begin{remark} \label{propappCp97rem}
A variation of the proof gives the following generalization of~\eqref{eq:TV16440901}: for every cube $\cu \subseteq \Zd$ of center $0$ and sidelength $R \geq 1$ and every point $y \in \Zd$,
\begin{equation*}
    \sum_{y_0 \in \Zd} \frac{1}{\left| y_0 \right|^{\alpha} + \left| y_0 - y \right|^{\alpha}} \leq \frac{C}{\max \left( R , \left| y\right| \right)^{\alpha - d}}.
\end{equation*}
\end{remark}

\begin{proof}
We split the space into two regions according to the formula
\begin{equation*}
    \sum_{y_0 \in \Zd} \frac{1}{\left| y_0 \right|^{\alpha} + \left| y_0 - y \right|^{\alpha}} = \sum_{y_0 \in  B\left(0,\left| y \right| / 2\right)} \frac{1}{\left| y_0 \right|^{\alpha} + \left| y_0 - y \right|^{\alpha}} + \sum_{y_0 \in \Zd \setminus B\left(0,\left| y \right| / 2\right)} \frac{1}{\left| y_0 \right|^{\alpha} + \left| y_0 - y \right|^{\alpha}}.
\end{equation*}
We estimate the two terms in the right side separately.

For the first term, we use the inequality $\left| y_0 \right|^{\alpha} + \left| y_0 - y \right|^{\alpha} \geq \frac{\left| y \right|^{\alpha}}{2^\alpha}$, valid for any point $y_0 \in B(\left| y \right|/2)$. We obtain
\begin{equation*}
    \sum_{y_0 \in  B\left(\left| y \right| / 2\right)} \frac{1}{\left| y_0 \right|^{\alpha} + \left| y_0 - y \right|^{\alpha}} \leq C \sum_{y_0 \in  B\left(\left| y \right| / 2\right)} \frac{1}{\left| y \right|^{\alpha}} \leq \frac{C}{\left| y \right|^{\alpha - d}}.
\end{equation*}
For the second term, we use the inequality $\left| y_0 \right|^{\alpha} + \left| y_0 - y \right|^{\alpha} \geq \left| y_0 \right|^{\alpha}$ and obtain
\begin{equation*}
    \sum_{y_0 \in \Zd \setminus B\left(\left| y \right| / 2\right)} \frac{1}{\left| y_0 \right|^{\alpha} + \left| y_0 - y \right|^{\alpha}}  \leq \sum_{y_0 \in \Zd \setminus B\left(\left| y \right| / 2\right)} \frac{1}{\left| y_0 \right|^{\alpha}} \leq \frac{C}{\left| y \right|^{\alpha - d}}.
\end{equation*}
\end{proof}

\small
\bibliographystyle{abbrv}
\bibliography{Villain}

\end{document}